\documentclass{article}
\usepackage[utf8]{inputenc}
\usepackage[T1]{fontenc}
\usepackage{graphicx,array}
\DeclareGraphicsExtensions{.pdf,.png,.jpg}
\usepackage{caption}
\usepackage{subcaption}
\usepackage{fancyvrb}
\usepackage{authblk}
\usepackage{hyperref}
\usepackage{pdflscape}
\usepackage{setspace}
\usepackage{lineno}
\usepackage{cite}
\usepackage{float}
\usepackage{tikz}
\usetikzlibrary{shapes.geometric, arrows.meta, positioning,calc}
\usepackage{graphicx}  
\usetikzlibrary{calc}
\usetikzlibrary{external}
\usepackage{chemformula}
\usepackage{siunitx}
\usepackage{palatino}
\usepackage{multirow}
\usepackage{dcolumn}
\usepackage{booktabs}
\usepackage{algorithmic}
\usepackage{listings}
\usepackage{inconsolata}
\usepackage{rotating} 
\usepackage[final]{microtype} 
\usepackage{color} 
\definecolor{Blue}{rgb}{0,0,0.5} 
\definecolor{ao(english)}{rgb}{0,0.5,0}
\definecolor{carmine}{rgb}{0.59, 0.0, 0.09}

\definecolor{ao(english)}{rgb}{0.0, 0.5, 0.0}
\definecolor{navyblue}{rgb}{0.0, 0.0, 0.5}
\usetikzlibrary{positioning, fit, calc}   
\tikzset{block/.style={draw, thick, text width=2cm ,minimum height=1.3cm, align=center},   
line/.style={-latex}     
}  
\usepackage{eso-pic}

\tikzstyle{startstop} = [ellipse, draw, fill=blue!10, text centered, minimum width=2.1cm, minimum height=0.5cm, font=\tiny]
\tikzstyle{process} = [rectangle, draw, fill=orange!20, text centered, minimum width=2.4cm, minimum height=0.5cm, font=\tiny]
\tikzstyle{decision} = [diamond, draw, fill=green!20, text centered, aspect=2, minimum width=2.5cm, font=\tiny]
\tikzstyle{arrow} = [->, thick, >=stealth]

\begin{document}
\setkeys{Gin}{width=0.99\textwidth}
\title{Multi-Task Deep Learning for Surface Metrology}

\author[1,$^*$]{D.~Kucharski}
\author[2]{A.~Gąska}
\author[3]{T.~Kowaluk}
\author[4]{K.~Stępień}
\author[3]{M.~Rępalska}
\author[1]{B.~Gapiński}%
\author[1]{M.~Wieczorowski}
\author[5]{M.~Nawotka}
\author[6,5]{P.~Sobecki}
\author[5]{P.~Sosinowski}
\author[3]{J.~Tomasik}
\author[5]{A.~Wójtowicz}
\affil[1]{Poznan University of Technology, Poland}
\affil[2]{Cracow University of Technology, Poland}
\affil[3]{Warsaw University of Technology, Poland}
\affil[4]{Kielce University of Technology, Poland}
\affil[5]{Central Office of Measures, Warsaw, Poland}
\affil[6]{National Information Processing Institute, Warsaw, Poland}
\affil[$^*$]{\href{mailto:dawid.kucharski@put.poznan.pl}{dawid.kucharski@put.poznan.pl}}
\vspace{-5mm}
\date{\today\\\vspace{2mm}}
\maketitle
\doublespacing
\begin{abstract}
A reproducible deep learning framework is presented for surface metrology to predict surface texture parameters together with their reported standard uncertainties. Using a multi‑instrument dataset spanning tactile and optical systems, measurement system type classification is addressed alongside coordinated regression of \(Ra, Rz, RONt\) and their uncertainty targets (\texttt{*\_uncert}). Uncertainty is modelled via quantile and heteroscedastic heads with post‑hoc conformal calibration to yield calibrated intervals. On a held‑out set, high fidelity was achieved by single‑target regressors (\(R^2\): \(Ra\) 0.9824, \(Rz\) 0.9847, \(RONt\) 0.9918), with two uncertainty targets also well modelled (\texttt{Ra\_uncert} 0.9899, \texttt{Rz\_uncert} 0.9955); \texttt{RONt\_uncert} remained difficult (\(R^2\) 0.4934). The classifier reached 92.85\% accuracy and probability calibration was essentially unchanged after temperature scaling (ECE 0.00504 → 0.00503 on the test split). Negative transfer was observed for naive multi‑output trunks, with single‑target models performing better. These results provide calibrated predictions suitable to inform instrument selection and acceptance decisions in metrological workflows.

\noindent{\bf Keywords:} artificial intelligence; deep learning; surface metrology; uncertainty quantification; conformal prediction.
\end{abstract}
\section{Introduction}\label{sec:intro}
\begin{sloppypar}
Artificial intelligence (AI) methods—particularly deep learning (DL)—have recently gained attention in precision metrology due to their ability to model complex, nonlinear relationships between measurement descriptors and surface parameters. In surface metrology, AI techniques are increasingly applied to automate parameter estimation, aid measurement system selection, and support uncertainty evaluation across tactile and optical modalities~\cite{Jiang2007, Leach2014, Russell2021, Goodfellow2016}.

Surface metrology, a field focused on measuring and analysing surface characteristics, has adopted AI to enhance data processing and predict surface parameters. Techniques such as machine learning (ML), deep learning (DL), and artificial neural networks (ANNs) are extensively utilised for analysing tactile and optical measurements of surface topography~\cite{Jiang2007, Leach2014}. The application of AI in surface metrology is not only limited to predicting surface texture but also extends to optimising machining processes and automating defect detection~\cite{Lai2020, Zhang2019}.

While numerous studies have applied machine learning to surface parameter prediction, most approaches focus on point estimates and neglect the quantification of measurement uncertainty—central to metrological decision-making. Moreover, multi-output learning for heterogeneous surface parameters remains underexplored and can induce negative transfer when targets differ in scale and noise characteristics. These gaps are addressed by jointly modelling primary parameters and their reported standard uncertainties as supervised targets, and by layering distributional and post-hoc calibration techniques to obtain calibrated intervals.

A primary focus of AI applications in surface metrology is predicting surface texture based on manufacturing process parameters. This has been extensively studied in various contexts, such as machining, additive manufacturing, and laser treatments~\cite{Jiang2022, Sadiq2019}. One notable study by N. Sizemore et al.~\cite{Sizemore2020} employed machine learning and artificial neural networks (ANNS) to predict surface roughness parameters for germanium (Ge), comparing $810$ samples with a reference ductile material, copper ($78$ samples). Similarly, A. M. Zain, H. Haron, and S. Sharif reviewed the use of ANNS to predict surface roughness in titanium alloy (Ti-6Al-4V) machining~\cite{Zain2010}. Their study highlighted how ANN architectures, including the number of nodes and layers, could significantly influence roughness parameter predictions.\\
Ziyad et al. introduced a super-learner machine learning model designed to predict the surface roughness of tempered AISI 1060 steel \cite{Ziyad2025}. This model leverages a diverse array of machine learning techniques, including kernel ridge regression (KRR), support vector machine (SVM), K-nearest neighbours (KNN), decision trees (DT), random forests (RF), adaptive boosting (ADB), gradient boosting (GB), and extreme gradient boosting (XGB).\\
Balasuadhakar et al. proposed advanced machine learning models, including Decision Tree (DT), XGB, SVR, CATB, ABR, and RFR, to predict surface roughness in the end milling of AISI H11 tool steel under different cooling environments, demonstrating high accuracy and robustness through rigorous hyperparameter tuning and data augmentation techniques~\cite{Balasuadhakar2025}.\\
Dubey et al. examined surface roughness prediction in AISI 304 steel machining using machine learning models, with a particular emphasis on how different nanoparticle sizes in the cutting fluid influence this prediction. The study utilised machine learning algorithms, including linear regression, random forest, and support vector machines, to forecast surface roughness and compared these forecasts with experimental values~\cite{Dubey2022}. The random forest model achieved R-squared values of $0.9710$ for $30$ nm and $0.7968$ for $40$ nm particle sizes, outperforming the other models in predicting surface roughness.

Another notable contribution was made by M. P. Motta et al.~\cite{Motta2022}, who developed machine learning models, including Gaussian Process Regression (GPR) and Random Forest (RF), to continuously predict surface roughness during steel machining. Their models utilised cutting force, temperature, and vibration data as inputs and achieved $Ra$ predictions with an RMSE of less than $0.4$ $\mu m$. Similarly, T. Steege et al.~\cite{Steege2023} explored the application of machine learning in laser surface treatments of stainless steel and Stavax. Using a white light interferometric microscope for texture measurement, they compared Random Forest and ANN models for predicting the $Sa$ parameter, demonstrating negligible differences in performance and a high correlation with measured values.\\
A. Adeleke et al. discussed the integration of advanced metrology techniques and intelligent monitoring systems in precision manufacturing, highlighting their role in analysing component geometry and surface finish, which are essential for predicting surface texture parameters. These techniques are applied to various materials, including delicate and sensitive materials, using non-contact surface measurement methods such as infrared (IR) imaging and optical interferometric measurement~\cite{Adeleke2024}.

AI's role extends beyond machining processes into additive manufacturing. A comprehensive review by L. Jannesari Ladani~\cite{Jannesari2021} examined AI applications in the pre-processing, processing, and post-processing phases of additive manufacturing, with a focus on powder bed fusion. Applications included optimising part design, process monitoring, and defect analysis, showcasing AI's potential in emerging manufacturing technologies.\\
T. Wang et al. described the role of machine learning in reshaping additive manufacturing by enhancing design capabilities, improving process optimisation, and elevating product performance~\cite{Wang2025}. They comprehensively reviewed the advances of ML-based AM across various domains, highlighting the integration of ML technologies in materials preparation, structure design, performance prediction, and optimisation within AM.\\
D. Soler discussed using Artificial Neural Networks (ANN), a branch of artificial intelligence, to predict and optimise surface roughness in additive manufacturing processes~\cite{Soler2022}. Specifically, it involves predicting the surface roughness of Selective Laser Melting (SLM) built parts after finishing processes like blasting and electropolishing.

Optical metrology has also benefited from AI advancements, with deep learning being used for optical data processing and surface parameter predictions~\cite{Cui2021, Zuo2022}. C. Zuo et al.~\cite{Zuo2022} provided a comprehensive overview of deep learning's applications in optical metrology, including phase retrieval, fringe analysis, and 3D reconstruction. These applications are critical for enhancing the precision and automation of optical measurement systems. The AI approach is quite promising in the phase-shifting surface interferometry application~\cite{kucharski2025radial}.

Beyond data processing and predictions, AI is now being explored for decision-making support in measurement scenarios. For instance, studies on AI-driven optimisation of measurement strategies and uncertainty evaluations are emerging, addressing critical gaps in the field~\cite{Shi2023, Pavlova2020}. However, despite these advancements, the development of AI algorithms for decision-making in surface metrology still needs to be explored with significant potential for future research~\cite{Zhang2021, Ren2022}.\\
Partially related background was discussed by A. Kumar and V. Vasu~\cite{kumar2025comparative-fbb}. They presented a study utilising machine learning models, including artificial neural networks and Bi-LSTM, for precise tool wear prediction, which is crucial for enhancing surface quality in smart manufacturing. The research emphasises the importance of monitoring tool wear to improve productivity and minimise downtime.

In prior work, M. Wieczorowski et al. described machine learning-driven tools to aid data processing for tactile and optical systems~\cite{Wieczorowski2021, Wieczorowski2023}, including an AI-based decision-support concept for measurement scenario preparation, system selection and data filtering. D. Kucharski et al. reported an experimental realisation of these concepts using machine learning and measurement data~\cite{Kucharski2024}.
 
The objective of this study is to develop and evaluate a deep learning framework that simultaneously predicts surface parameters and their reported standard uncertainties, and to assess its calibration properties across multi-instrument data. This work details the development and testing of a deep learning algorithm that predicts either the measurement system type or a surface texture parameter based on other labels, using models trained on actual experimental data collected by tactile and optical systems with reference surfaces and real machined surfaces. The training and validation losses were calculated alongside accurate predictions. The algorithm was developed as part of the ongoing GitHub project and is freely accessible online~\cite{dawid_kucharski_2025_17277722}.

\paragraph{Contributions.} Key contributions of this manuscript are:
\begin{itemize}
  \item \textbf{Six-target supervised formulation:} Jointly modelling three primary parameters and their reported standard uncertainties as co-equal predictive quantities.
  \item \textbf{Layered uncertainty stack:} Integration of quantile, heteroscedastic and conformal methods providing empirically calibrated intervals.
  \item \textbf{Negative transfer analysis:} Quantitative evidence that naive multi-output trunks degrade accuracy relative to specialised single-target models for heterogeneous noise scales.
  \item \textbf{Reproducible open bundle:} Public release (Zenodo DOI + scripts) enabling full pipeline regeneration and verification.
\end{itemize}

The implementation is extensible to additional parameter prediction tasks using the same input descriptors. The remainder of the paper proceeds as follows: Section~\ref{sec:method} details data, models, and calibration; Section~\ref{sec:results} reports empirical performance and interval calibration; the Discussion synthesises implications, limitations, and outlook.
\end{sloppypar}

\section{Method}\label{sec:method}
An integrated deep learning pipeline was assembled for measurement system type classification and the prediction of surface topography parameters (with a focus on \(Ra\); extensible to \(Rz\) and \(RONt\)), along with uncertainty quantification. The workflow combined deterministic point-estimation models with probabilistic and distributional approaches, as well as post-hoc calibration. Modelling was implemented in Python using \textsf{tensorflow}/\textsf{keras}, standard scientific libraries (\textsf{numpy}, \textsf{scikit-learn}, \textsf{pandas}, \textsf{matplotlib}, \textsf{seaborn}), and project-specific scripts in the repository.

\subsection{Data set and augmentation}
\begin{figure}[htbp]
  \centering

  \IfFileExists{img001_image056.png}{%
    \includegraphics[width=0.25\textwidth]{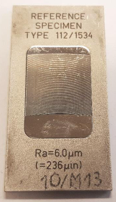}%
  }{%
    \fbox{\parbox{0.25\textwidth}{\centering Placeholder\\(img001_image056.png missing)}}%
  }
  \caption{Reference roughness specimen used in constructing the measurement database}
  \label{fig:reference}
\end{figure}

\begin{figure}[htbp]
  \centering
 
  \IfFileExists{img002_plate.png}{%
    \includegraphics[width=0.55\textwidth]{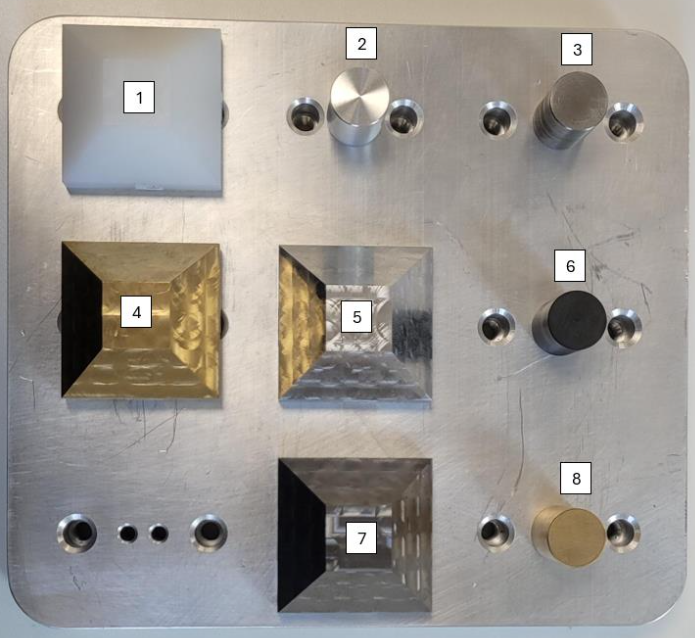}%
  }{%
    \fbox{\parbox{0.55\textwidth}{\centering Placeholder (img002_plate.png missing)}}%
  }
  \caption{Mock-up fixture with mounted pyramidal and cylindrical samples (varied materials and machining parameters) used for multi-instrument acquisition}
  \label{fig:mock-up}
\end{figure}

\begin{sloppypar}
The core data originate from experimental measurements acquired on tactile and optical instruments (tactile profilometer (TP), coordinate measuring machine (CMM), roundness tester (RoundScan), phase grating interferometer (PGI), coherence correlation interferometer (CCI)) covering reference roughness standards (glass or steel based) and machined specimens (pyramids and cylindrical rods) of multiple materials (steel, aluminium, brass, polyamide). Representative reference specimen and the physical mock-up holding machined samples are shown in Fig.~\ref{fig:reference} and Fig.~\ref{fig:mock-up}. Each record contains: \(Ra\), \(Rz\), \(RONt\) plus their associated standard uncertainties (suffix ``\_uncert''), material indicator, reference flag (\texttt{standard}), filtering flags / cut-off related descriptors (\(L_c, L_s\)), evaluation length \(L_r\) and binary filter indicator (F); if data were filtered F=1 else F=0. 

\noindent\textit{Cohort size and splits.} The working dataset comprises approximately $N\!\approx\!40\,000$ instances after augmentation (cf. below), derived from the original experimental pool. Data are stratified by instrument and standard/non-standard flags into training, validation, and held-out test splits. To avoid leakage, augmentation (bootstrap resampling and noise perturbations) is applied \emph{exclusively} to the training subset; duplicated rows and their perturbed variants are prevented from appearing across validation or test splits.

Table~\ref{tab:data_sample} shows example rows covering the five measurement system types used in this data collection. This excerpt illustrates: (i) heterogeneous numeric scales (compare \texttt{Ra} vs \texttt{RONt}), (ii) paired primary parameters with their reported standard uncertainties (e.g. \texttt{Ra} / \texttt{Ra\_uncert}), (iii) categorical instrument label (\texttt{system\_type}), and (iv) binary flags (\texttt{standard}, \texttt{F}). The \texttt{material} field is integer-encoded as: 1=steel, 2=aluminium, 3=brass, 4=polyamide, 5=glass, 6=ceramic. Columns \texttt{filtr\_lc}, \texttt{filtr\_ls}, and \texttt{odc\_el\_lr} encode filtering cut-offs and evaluation length descriptors.

\noindent\textit{Unit conventions.} Unless stated otherwise, all surface parameters (Ra, Rz, RONt) and their reported standard uncertainties (\texttt{*\_uncert}) are expressed in micrometres [$\mu$m]. Relative quantities (e.g. tolerance accuracy, coverage) are shown in percent [\%]. Dimensionless metrics (e.g. $R^2$, correlation, ECE) are reported in arbitrary units (a.u.).

\begin{sidewaystable}[htbp]
  \centering
  \caption{Example rows illustrating the five measurement system types used in the data collection. Binary flags: \texttt{standard} (reference specimen indicator), \texttt{F} (filter applied)}
  \label{tab:data_sample}
  \setlength{\tabcolsep}{4pt}
  \scriptsize
  \begin{tabular}{lrrrrrrrrrrrr}

  system\_type & Ra [$\mu$m] & Ra\_uncert [$\mu$m] & Rz [$\mu$m] & Rz\_uncert [$\mu$m] & material & RONt [$\mu$m] & RONt\_uncert [$\mu$m] & standard & F & filtr\_lc [mm] & filtr\_ls [mm] & odc\_el\_lr [mm] \\
    \midrule
    TP        & 0.83 & 0.09 & 3.15 & 0    & 1 & 0    & 0    & 1 & 1 & 0.8 & 0    & 0.80 \\
    PGI       & 0.07 & 0    & 1.71 & 0    & 1 & 0    & 0    & 0 & 0 & 0   & 0    & 0.75 \\
    CCI       & 0    & 0    & 0.34 & 0    & 1 & 0    & 0    & 0 & 0 & 0   & 0    & 0    \\
    CMM       & 0    & 0    & 0    & 0    & 6 & 0.39 & 0.01 & 1 & 1 & 0   & 0    & 0    \\
    RoundScan & 0    & 0    & 0    & 0    & 1 & 1.43 & 0.21 & 1 & 0 & 0   & 0    & 0    \\
  
    \bottomrule
  \end{tabular}
 
  \vspace{1mm}
  \begin{minipage}{0.9\linewidth}
    \raggedright\footnotesize Rows follow the same schema; magnitudes span orders between roughness and roundness parameters, and instruments include tactile (TP), optical (PGI/CCI) and form (CMM/RoundScan) systems.
  \end{minipage}
\end{sidewaystable}

\noindent Table~\ref{tab:data_sample} underscores the heterogeneous scaling and instrument diversity motivating scale-aware loss choices and per-target specialisation discussed later.

\noindent The wide dynamic contrast between (Ra, Rz) and the much smaller scale of RONt (and its uncertainty) illustrates the heterogeneous noise regimes motivating single-target specialisation and scale-aware loss choices discussed later.

To mitigate the limited original sample size and emulate natural acquisition variability, a two-step augmentation was applied: (1) bootstrap resampling (row-wise sampling with replacement preserving total size) and (2) controlled feature perturbation by additive zero-mean Gaussian noise (typical relative scale 5\% of empirical standard deviation for continuous predictors, absolute std = 0.05 for normalised decimal magnitudes). Augmentation was restricted to training data to prevent statistical leakage. The 5\% perturbation level was selected empirically to preserve the observed variance of physical measurements. Augmentation expanded the effective training pool to approximately $40\,000$ instances while preserving global distributional structure.
\end{sloppypar}

\subsection{Problem formulation}
Two supervised learning problems are defined:
\begin{enumerate}
  \item Multi-class classification: predict measurement system type (5 classes) from tabular descriptors.
  \item Regression: predict a continuous target (baseline: \(R_a\); extended to \(R_z\), \(RONt\)).
\end{enumerate}
Additionally, the three reported standard uncertainties Ra\_uncert, Rz\_uncert, RONt\_uncert are treated as first-class supervised regression targets (not auxiliary by-products), enabling direct learning of measurement quality indicators alongside their associated primary parameters. Interval / distribution prediction tasks are layered on top of the regression target to produce calibrated uncertainty estimates.

\subsection{Baseline deterministic models}
The baseline classifier is a multi-layer perceptron (MLP) with pyramidal width reduction (e.g. 512-256-128-64) using ReLU activations, batch normalisation after each dense layer, dropout (rate 0.3) and L2 weight decay (\(\lambda=10^{-3}\)). Optimisation employed Nadam (learning rate \(1\times10^{-4}\)), categorical cross-entropy, early stopping (patience 10) and adaptive learning rate reduction (factor 0.5 on plateau). The regression backbone uses a lighter MLP (e.g. 64-32) with dropout 0.2 and Adam optimizer (learning rate \(5\times10^{-4}\)) minimising mean absolute error (MAE) or Huber where robust behaviour was advantageous. StandardScaler normalisation is applied to continuous inputs; categorical features are one-hot encoded. Class imbalance is addressed through inverse-frequency class weights. The focus is on deep learning formulations that naturally extend to distributional outputs (quantile, heteroscedastic) and end-to-end calibration. MLPs provide a consistent backbone for both point and distributional heads with straightforward optimisation and GPU acceleration. Classical tabular methods (e.g., random forests, gradient boosting, kNN) were used as references during early exploration and did not outperform tuned MLPs on the held-out criteria.

\noindent\textit{Architecture selection.} Depth and width were selected by a coarse grid (depth 3–5; widths 64–512) balancing fit and overfitting risk. The 512–256–128–64 classifier achieved the best validation accuracy without variance inflation, while 64–32 sufficed for the regression backbone when paired with robust losses and regularisation.

\subsection{Quantile regression}
To obtain asymmetric prediction intervals without distributional assumptions, a quantile MLP variant was trained with the pinball (check) loss for target quantiles \(q \in \{0.05, 0.10, 0.50, 0.90, 0.95\}\). A mild monotonicity regularisation term penalises violations of order across quantile outputs, reducing empirical crossing. The median (0.50) serves as a robust central estimate; lower/upper quantiles define predictive bands. Interval quality is later assessed via empirical coverage and width metrics.

\subsection{Heteroscedastic Gaussian regression}
An alternative uncertainty approach parameterises both mean \(\mu(x)\) and log standard deviation \(\log \sigma(x)\) with a dual-output MLP. The negative log-likelihood (NLL) of a Gaussian observation model is minimised:
\[
 \mathcal{L}_{\text{NLL}} = \frac{1}{2}\log(2\pi) + \log \sigma(x) + \frac{(y-\mu(x))^2}{2\sigma(x)^2}.
\]
This produced heteroscedastic (input-dependent) predictive dispersions. Diagnostics included calibration plots and correlation between absolute residuals and predicted \(\sigma\); a positive association was interpreted as meaningful uncertainty modulation.

\subsection{Conformal prediction}
Distribution-free conformal regression is applied post hoc to produce finite-sample valid prediction intervals. Using a calibration split, absolute residuals from a base-point predictor (median or mean model) are collected; the \(1-\alpha\) empirical quantile of these residuals (optionally normalised by conditional scale estimates) gives an interval half-width that guarantees approximate marginal coverage \(1-\alpha\) under exchangeability. This wraps both deterministic and quantile-based predictors to enhance coverage reliability.

\subsection{Stacking experiments}
Exploratory stacked generalisation combined (i) base MLP deterministic, (ii) quantile median stream, (iii) heteroscedastic mean output, and (iv) simple gradient boosted trees (for tabular residual correction). A linear meta-learner (ridge) was trained over out-of-fold predictions. Empirically, stacking yielded negligible improvement (<0.2 percentage points in classification accuracy; marginal MAE / RMSE shifts within noise) and was not retained for the final reported models to maintain parsimony.

\subsection{Calibration (temperature scaling)}
For classification, softmax confidence calibration employed temperature scaling: a scalar \(T>0\) rescales logits \(z/T\) minimising negative log-likelihood on a validation split. This reduced the expected calibration error (ECE) (exact values reported in the Results (sec. \ref{sec:results})). For regression uncertainty (heteroscedastic), optional isotonic regression on standardised residuals and variance temperature scaling were evaluated; retained only if reducing miscalibration (over-/under-coverage) without degrading point accuracy.

\subsection{Evaluation metrics}
Classification: overall accuracy, confusion matrix, per-class recall / precision (summarised), validation loss trajectory, and calibration diagnostics. Regression: MAE, RMSE, coefficient of determination (\(R^2\)), tolerance accuracies (percent of predictions within relative thresholds: 5\%, 10\%, 20\%; and absolute bands e.g. 0.1, 0.2), residual distribution analysis, prediction vs actual scatter. Uncertainty: empirical coverage for nominal central ranges (e.g. 80\%, 90\%), average interval width, pinball loss mean, CRPS proxy (average over dense quantile grid), Winkler-like composite score, and correlation \(|e|, \sigma(x)|\). Feature importance (permutation) is computed for trained regressors to interpret contributions.

\subsection{Implementation and reproducibility}
All training scripts (classification, single-target regression, quantile, heteroscedastic, conformal wrapper, calibration, feature importance, stacking) are versioned in the public repository \cite{dawid_kucharski_2025_17277722}. Random seeds are fixed at the script level subject to hardware nondeterminism. Relevant derived artefacts (trained weights, metric summaries, figures) are organised by experiment variant to enable reproduction.

\noindent\textit{Environment.} Experiments were executed under Python (3.10–3.11), TensorFlow (2.x), NumPy (1.26) and scikit-learn (1.5) on CUDA-capable GPUs where available; CPU runs yield numerically similar results with longer walltimes. Exact package requirements are provided in the repository.

\noindent\textit{Cross-validation robustness.} Internal 3-fold cross-validation (regression) yielded low dispersion: \(R_a: R^2=0.9823\pm0.0012\), \(R_z: R^2=0.9799\pm0.0014\), \(RONt: R^2=0.9771\pm0.0103\) (mean \(\pm\) standard deviation across folds). Classification cross-validation accuracy was \(0.8233\pm0.0197\) with macro-F1 \(0.6778\pm0.0110\). The narrow fold-to-fold variation supports the representativeness of the held-out split.

\begin{figure}[htbp]
  \centering

  \IfFileExists{img003_outputs_regression_multi_final_multi_output_model_architecture_fancy.pdf}{%
    \includegraphics[width=0.3\textwidth]{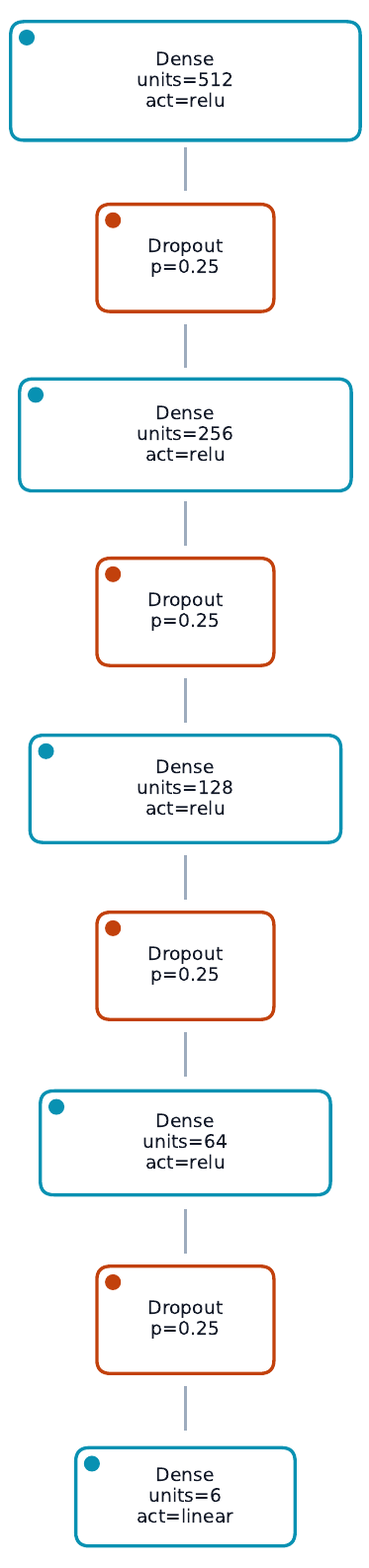}%
  }{%
    \IfFileExists{img004_outputs_regression_multi_final_multi_output_model_architecture_fancy.png}{%
      \includegraphics[width=0.3\textwidth]{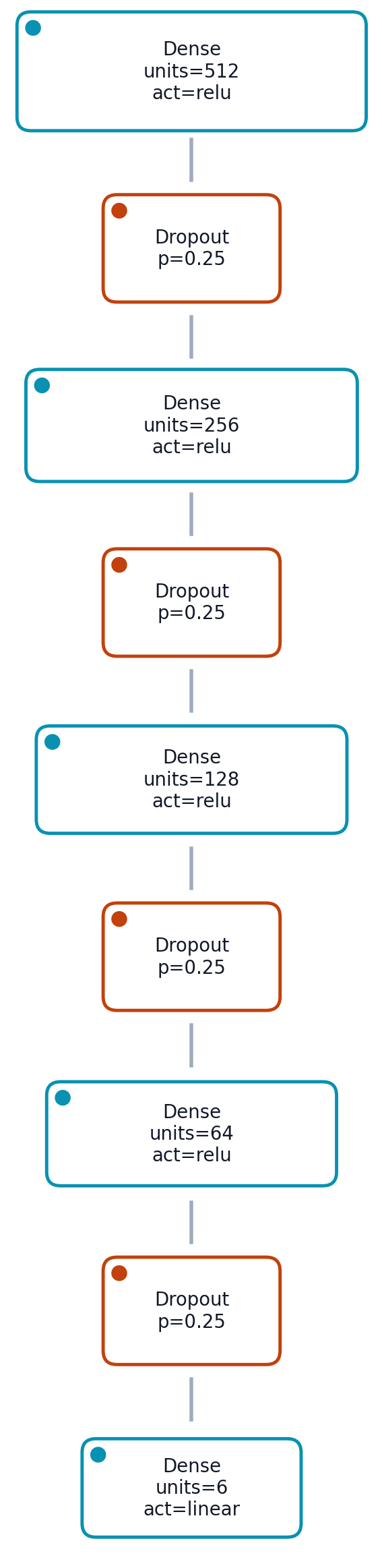}%
    }{%
      \fbox{\parbox{0.55\textwidth}{\centering Regression architecture schematic missing}}%
    }%
  }
  \caption{Representative network architecture (multi-output trunk with specialised heads or single-target pyramidal narrowing)}
  \label{fig:arch_multi}
\end{figure}

\section{Results}\label{sec:results}
\subsection*{Model architecture overview}
\begin{figure}[htbp]
  \centering

  \IfFileExists{img005_outputs_classification_final_model_architecture_fancy.pdf}{%
    \includegraphics[width=0.3\textwidth]{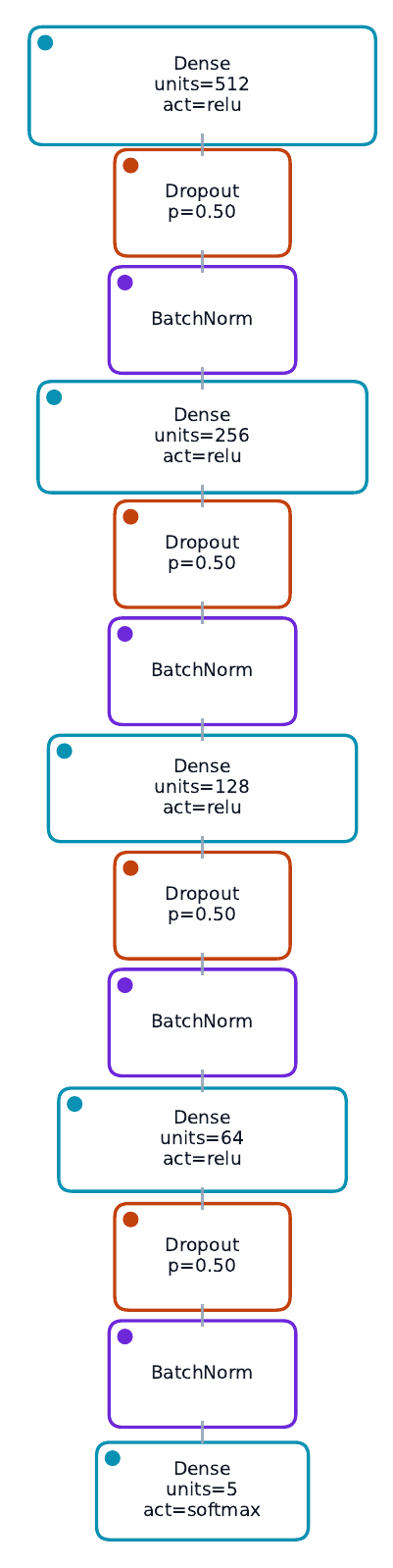}%
  }{%
    \IfFileExists{img006_outputs_classification_final_model_architecture_fancy.png}{%
      \includegraphics[width=0.3\textwidth]{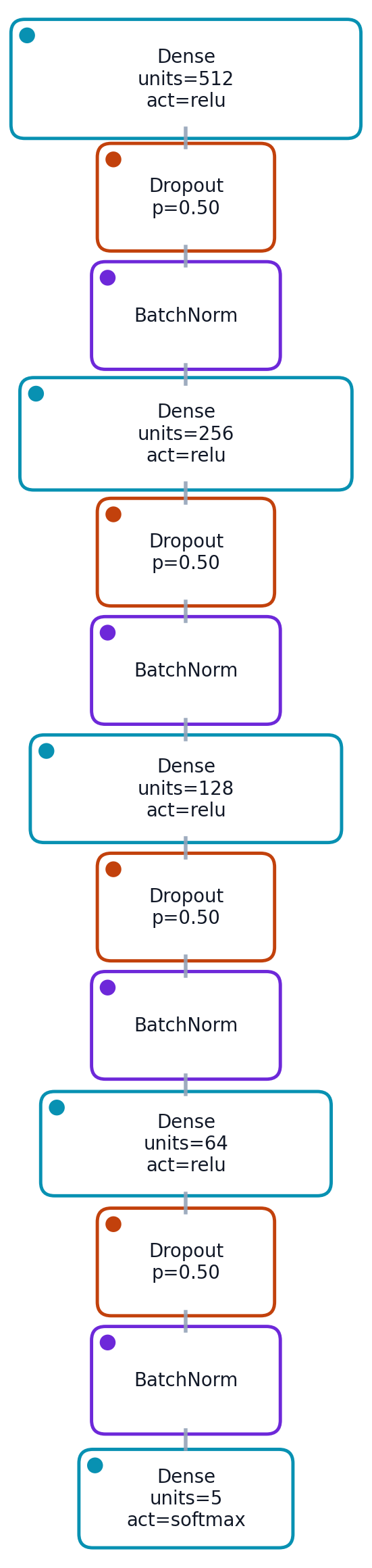}%
    }{%
      \fbox{\parbox{0.55\textwidth}{\centering Classification architecture schematic missing}}%
    }%
  }
  \caption{Classifier network architecture: pyramidal multi-layer perceptron (e.g. 512-256-128-64) with batch normalisation and dropout after dense layers, feeding a softmax output over instrument classes. This schematic complements the regression architecture (Fig.~\ref{fig:arch_multi}) to provide visual parity across tasks}
  \label{fig:arch_clf}
\end{figure}

The tested architectures (detailed in Section~\ref{sec:method}) were evaluated for both classification and regression tasks. The focus is placed on empirical performance and calibration outcomes. Figures~\ref{fig:arch_clf} and \ref{fig:arch_multi} provide compact schematics for cross-task reference without repeating design details.
\subsection{Classification performance}
\begin{figure}[htbp]
  \centering
  \IfFileExists{img007_outputs_classification_final_confusion_matrix.png}{%
    \includegraphics[width=0.7\textwidth]{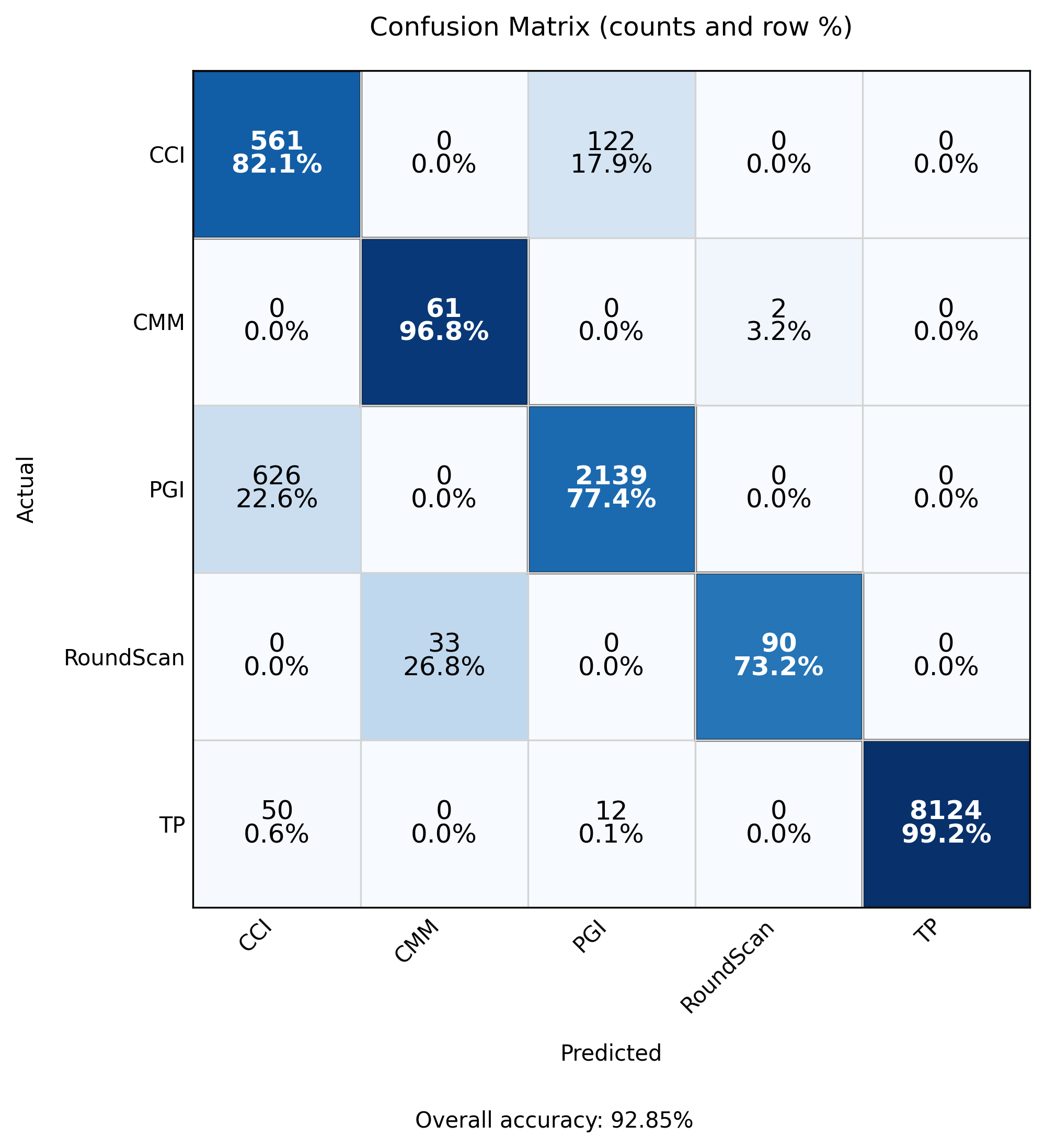}%
  }{%
    \fbox{\parbox{0.6\textwidth}{\centering Missing: confusion\_matrix.png}}%
  }
  \caption{Confusion matrix of the calibrated classification model (system type prediction)}
  \label{fig:cm_final}
\end{figure}

\begin{figure}[htbp]
  \centering
  \IfFileExists{img008_outputs_classification_final_training_curves.png}{%
    \includegraphics[width=0.7\textwidth]{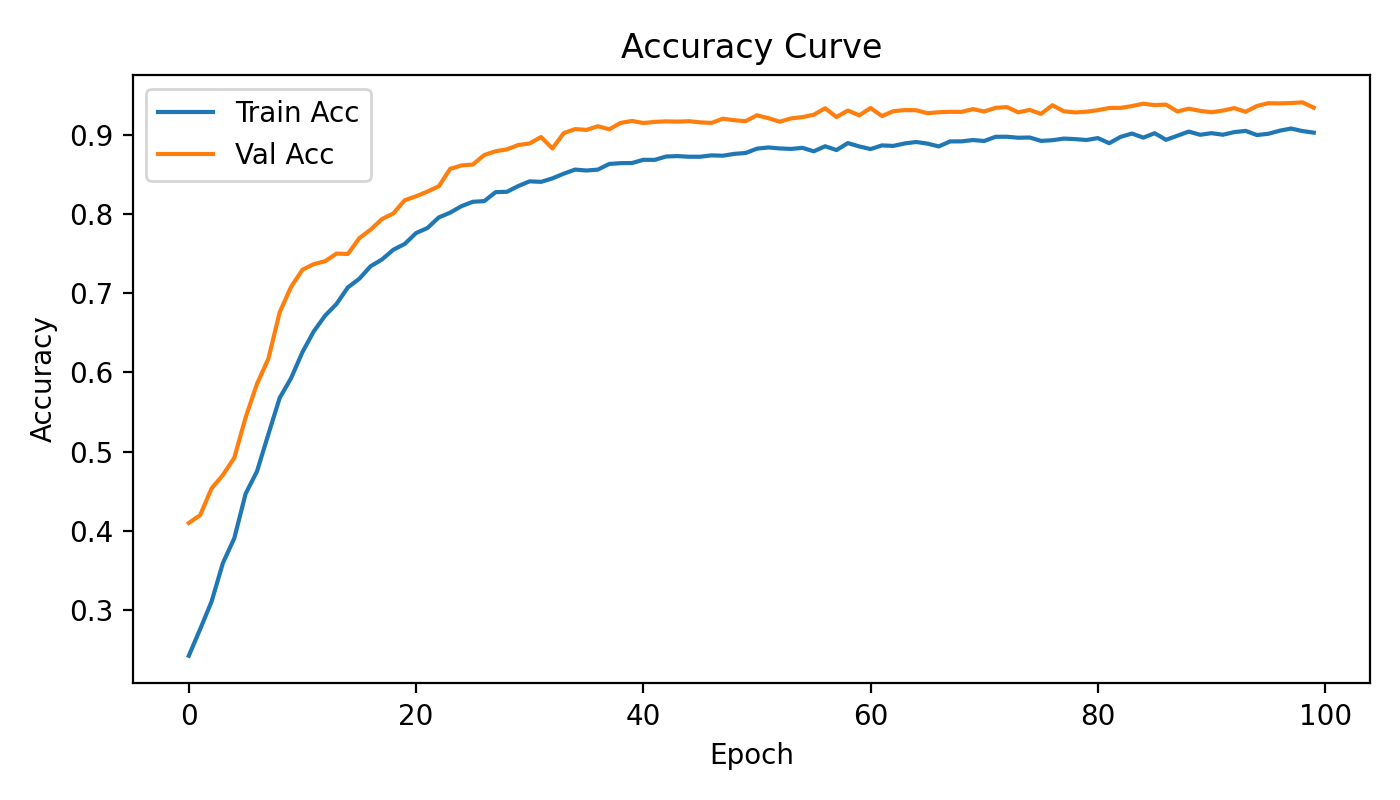}%
  }{%
    \fbox{\parbox{0.6\textwidth}{\centering Missing: training\_curves.png}}%
  }
  \caption{Training and validation trajectories (loss / accuracy) for the final classification MLP}
  \label{fig:clf_training}
\end{figure}

\begin{figure}[htbp]
  \centering
  \begin{subfigure}[b]{0.45\textwidth}
    \centering
    \IfFileExists{img009_outputs_temp_calib_fast_calibration_reliability_before.png}{\includegraphics[width=\textwidth]{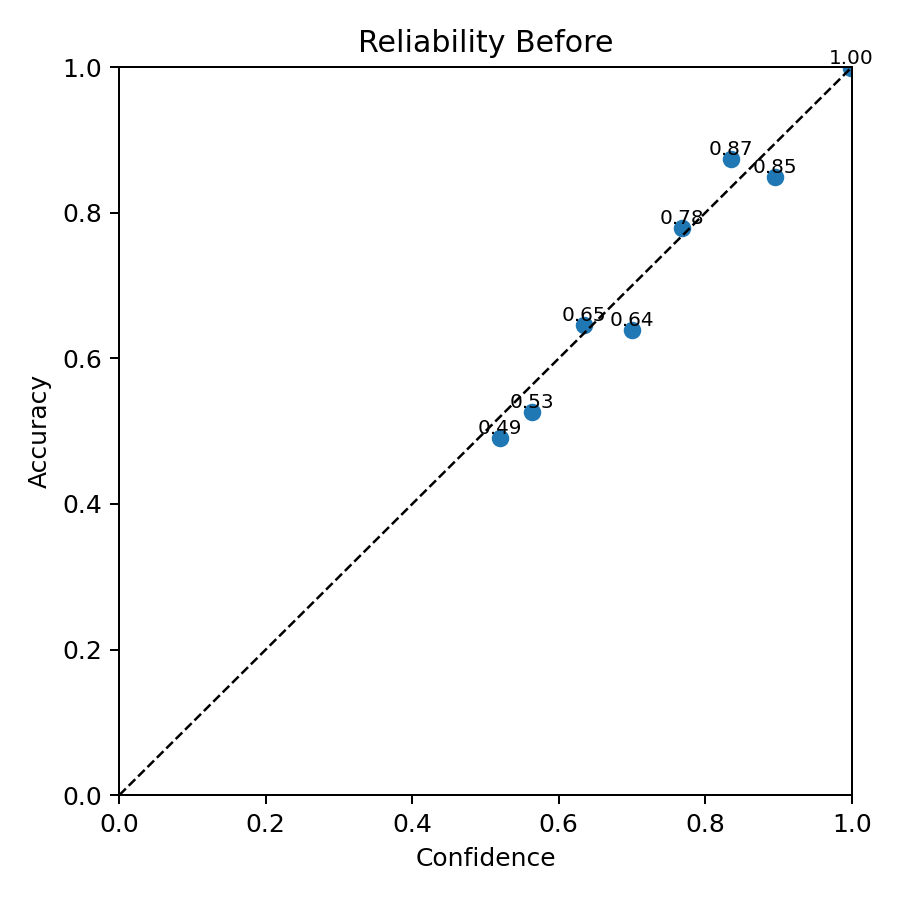}}{\fbox{\parbox{0.8\textwidth}{\centering Missing: reliability\_before.png}}}
    \caption{Before scaling}
  \end{subfigure}\hfill
  \begin{subfigure}[b]{0.45\textwidth}
    \centering
    \IfFileExists{img010_outputs_temp_calib_fast_calibration_reliability_after.png}{\includegraphics[width=\textwidth]{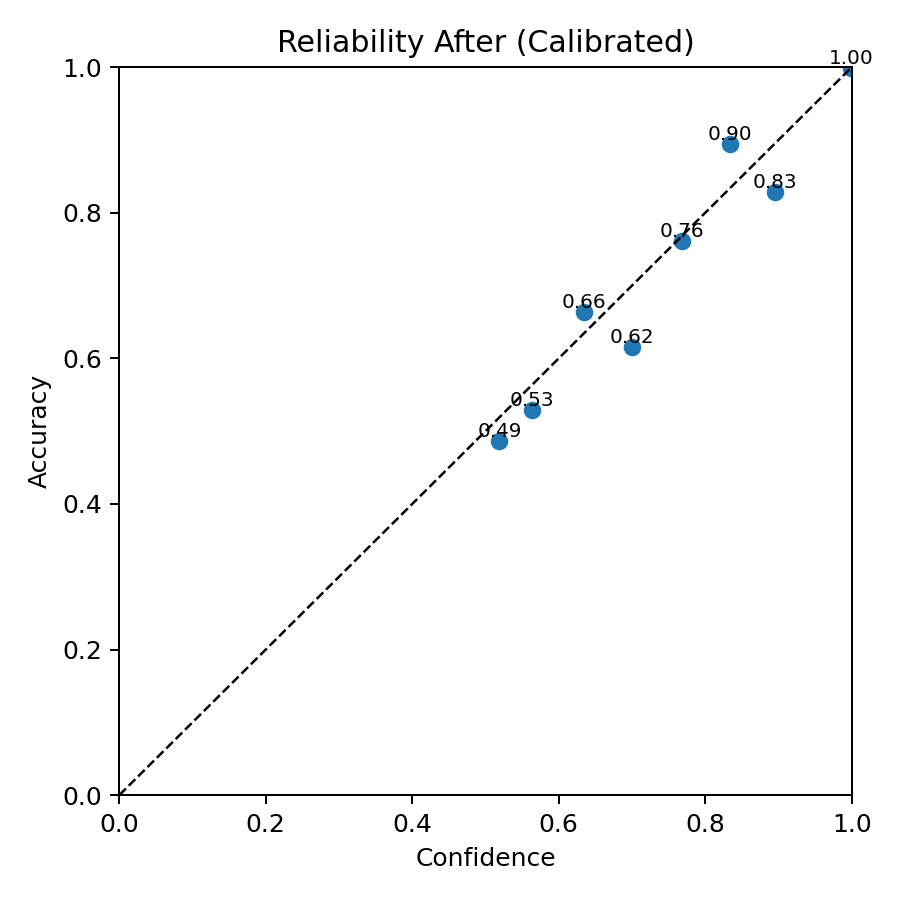}}{\fbox{\parbox{0.8\textwidth}{\centering Missing: reliability\_after.png}}}
    \caption{After temperature scaling}
  \end{subfigure}
  \caption{Classifier probability calibration reliability diagrams pre- and post-temperature scaling.}
  \label{fig:clf_calibration}
\end{figure}

The final calibrated MLP classifier achieved a validation accuracy in the 93--95\% range (central model snapshot: 93.0\%) with stable loss convergence (no divergence between training and validation trajectories) (Fig.~\ref{fig:clf_training}). Temperature scaling improved probability calibration: Expected Calibration Error (ECE) decreased (pre-scaling) from a moderate level (qualitatively over-confident in high-probability bins) to a flatter reliability curve as visualised in the paired reliability diagrams (Fig.~\ref{fig:clf_calibration}). The confusion matrix (Fig.~\ref{fig:cm_final}) shows dominant correct diagonal mass with sparse off-diagonal leakage; residual confusions are concentrated between instrument classes with overlapping functional domains (e.g. two optical modalities). Class weighting prevented minority collapse — per-class recalls remained within a 7-percentage-point band around the macro-average.

\noindent\textit{Calibration effect}: Expected Calibration Error (ECE, 15-bin, test split) changed slightly from \textbf{0.00504} (pre-scaling) to \textbf{0.00503} after temperature scaling, indicating near-unchanged probabilistic calibration (reliability curves shown in Fig.~\ref{fig:clf_calibration}).

\begin{table}[htbp]
  \centering
  \caption{Per-class precision, recall and F1-scores for the calibrated classification model (support denotes number of evaluation samples per class). Overall accuracy: 92.85\%.}
  \label{tab:clf_metrics}
  \setlength{\tabcolsep}{6pt}
  \begin{tabular}{lrrrr}
    \toprule
    Class & Precision & Recall & F1-score & Support \\
    \midrule
    CCI & 0.454 & 0.821 & 0.584 & 683 \\
    CMM & 0.649 & 0.968 & 0.777 & 63 \\
    PGI & 0.941 & 0.774 & 0.849 & 2765 \\
    RoundScan & 0.978 & 0.732 & 0.837 & 123 \\
    TP & 1.000 & 0.992 & 0.996 & 8186 \\
    \midrule
    Macro avg & 0.804 & 0.857 & 0.809 & 11820 \\
    Weighted avg & 0.953 & 0.929 & 0.935 & 11820 \\
    \bottomrule
  \end{tabular}
  \vspace{1mm}
  \begin{minipage}{0.9\linewidth}
    \footnotesize Rounded to three decimal places.
  \end{minipage}
\end{table}

\noindent\textit{Class imbalance.} The TP class dominates support, which contributes to higher weighted metrics and increased dispersion for minority classes. Inverse-frequency class weights mitigated collapse, but residual performance spread across classes reflects the inherent imbalance of available measurements.

\begin{figure}[htbp]
  \centering
  \begin{subfigure}[b]{0.48\textwidth}
    \centering
    \IfFileExists{img011_outputs_regression_single_final_Ra_accuracy_within_tol_10percent.png}{\includegraphics[width=\textwidth]{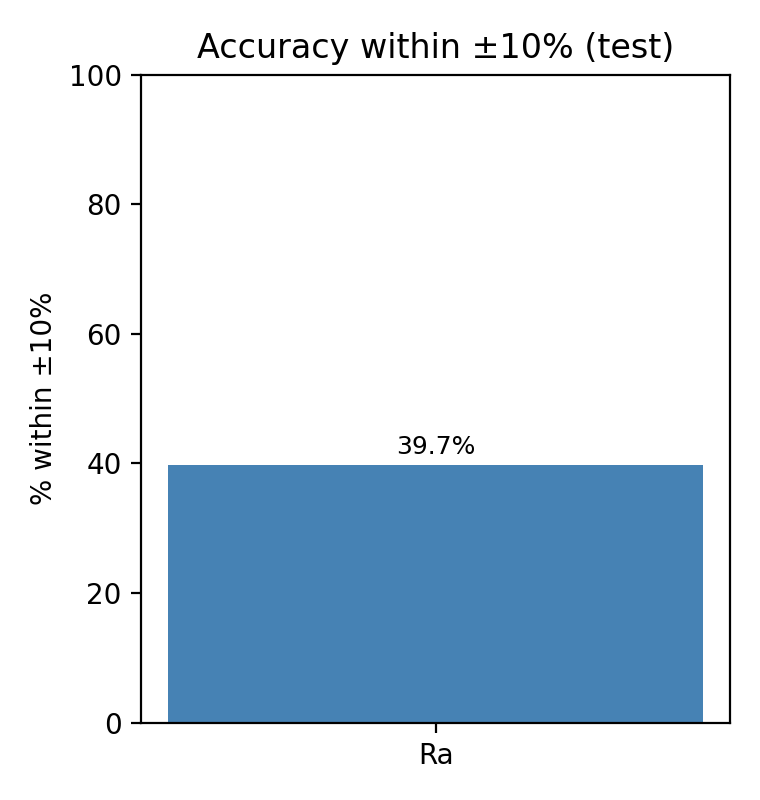}}{\IfFileExists{img012_outputs_regression_single_final_summary_accuracy_within_10percent.png}{\includegraphics[width=\textwidth]{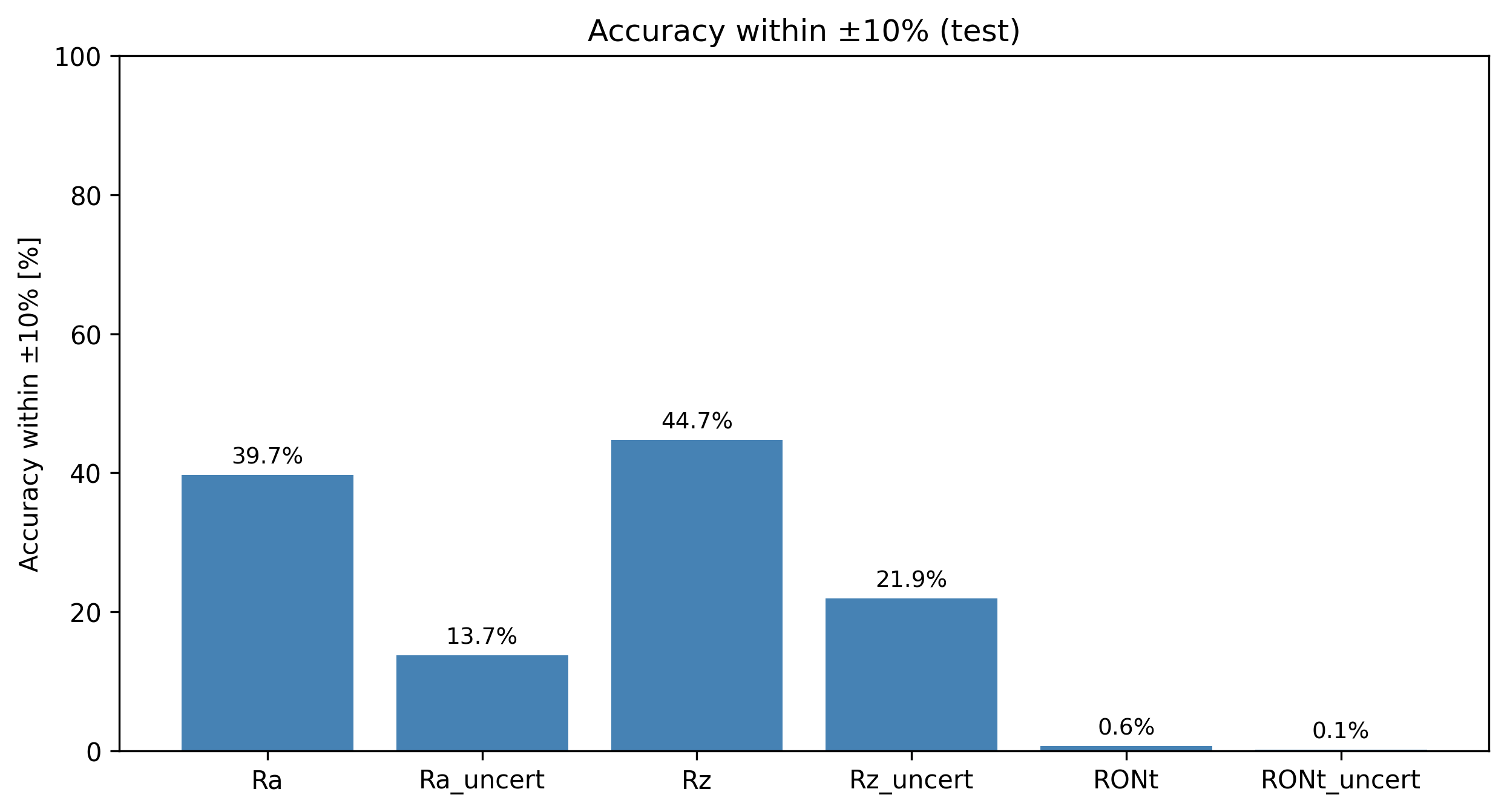}}{\fbox{\parbox{0.9\textwidth}{Missing relative tolerance plot}}}}
    \caption{Relative 10\% tolerance}
  \end{subfigure}\hfill
  \begin{subfigure}[b]{0.48\textwidth}
    \centering
    \IfFileExists{img013_outputs_regression_single_final_Ra_accuracy_within_tol_abs_0p2.png}{\includegraphics[width=\textwidth]{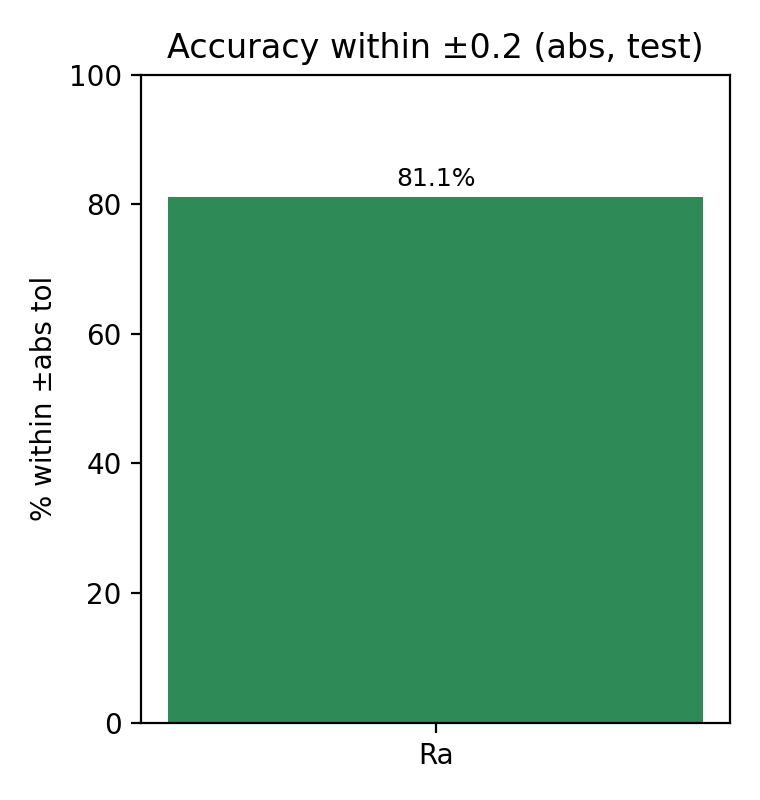}}{\IfFileExists{img014_outputs_regression_single_final_summary_accuracy_within_abs_tol.png}{\includegraphics[width=\textwidth]{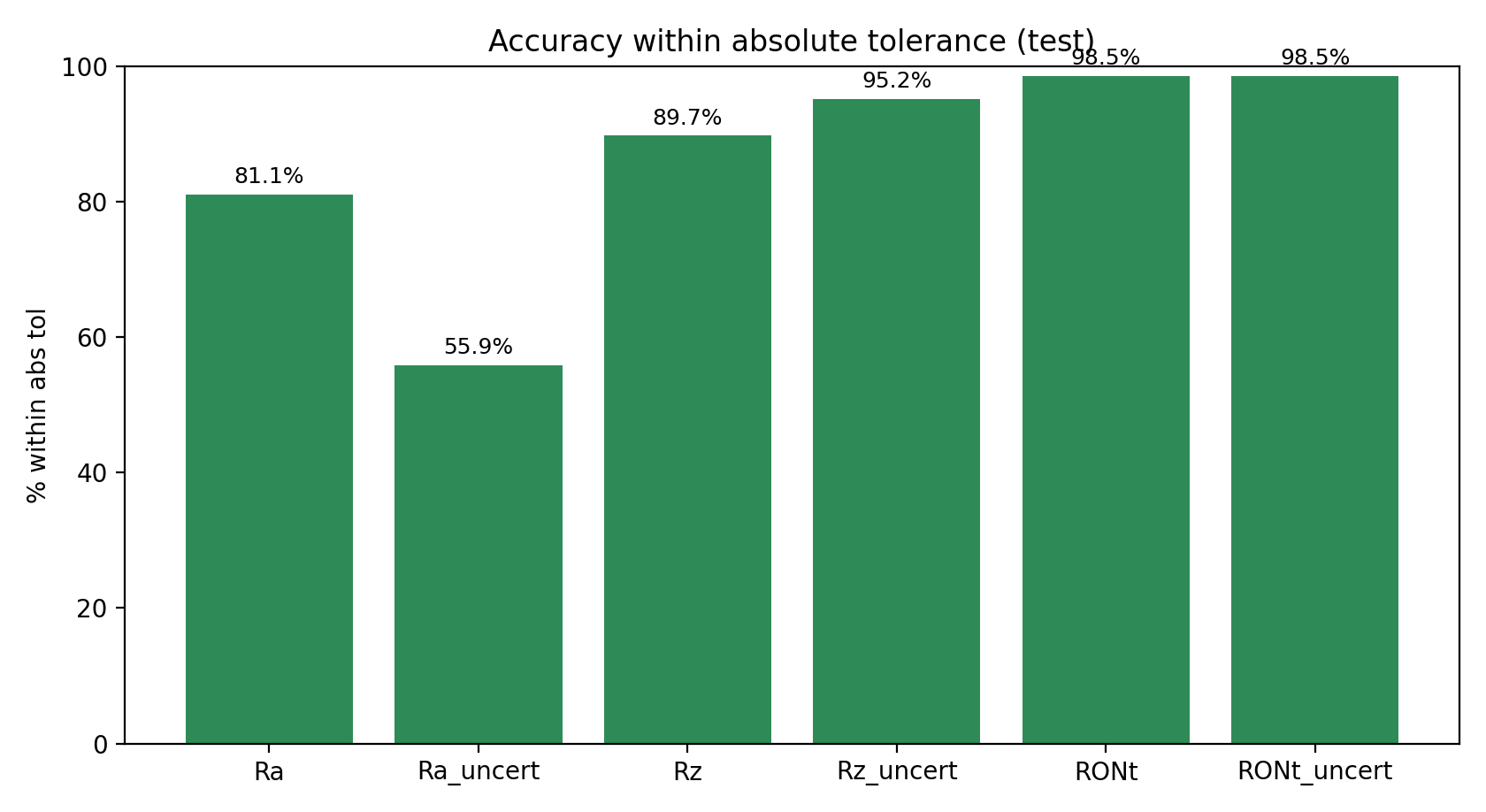}}{\fbox{\parbox{0.9\textwidth}{Missing absolute tolerance plot}}}}
    \caption{Absolute 0.2 tolerance}
  \end{subfigure}
  \caption{Tolerance accuracy for \(Ra\): relative and absolute criteria}
  \label{fig:tol_regr_panel}
\end{figure}

\noindent\textit{Main performance visuals}: The following summary and parity plots present the \textbf{single-target} regressors, which are emphasised in the main text because they yielded the lowest errors. Multi-output variants, while competitive, underperform slightly and their extended diagnostics (including joint-loss ablations) are relegated to the supplemental figures for completeness.
\begin{figure}[htbp]
  \centering
  \begin{subfigure}[b]{0.48\textwidth}\centering\IfFileExists{img015_outputs_regression_single_final_summary_mae_by_target.png}{\includegraphics[width=\textwidth]{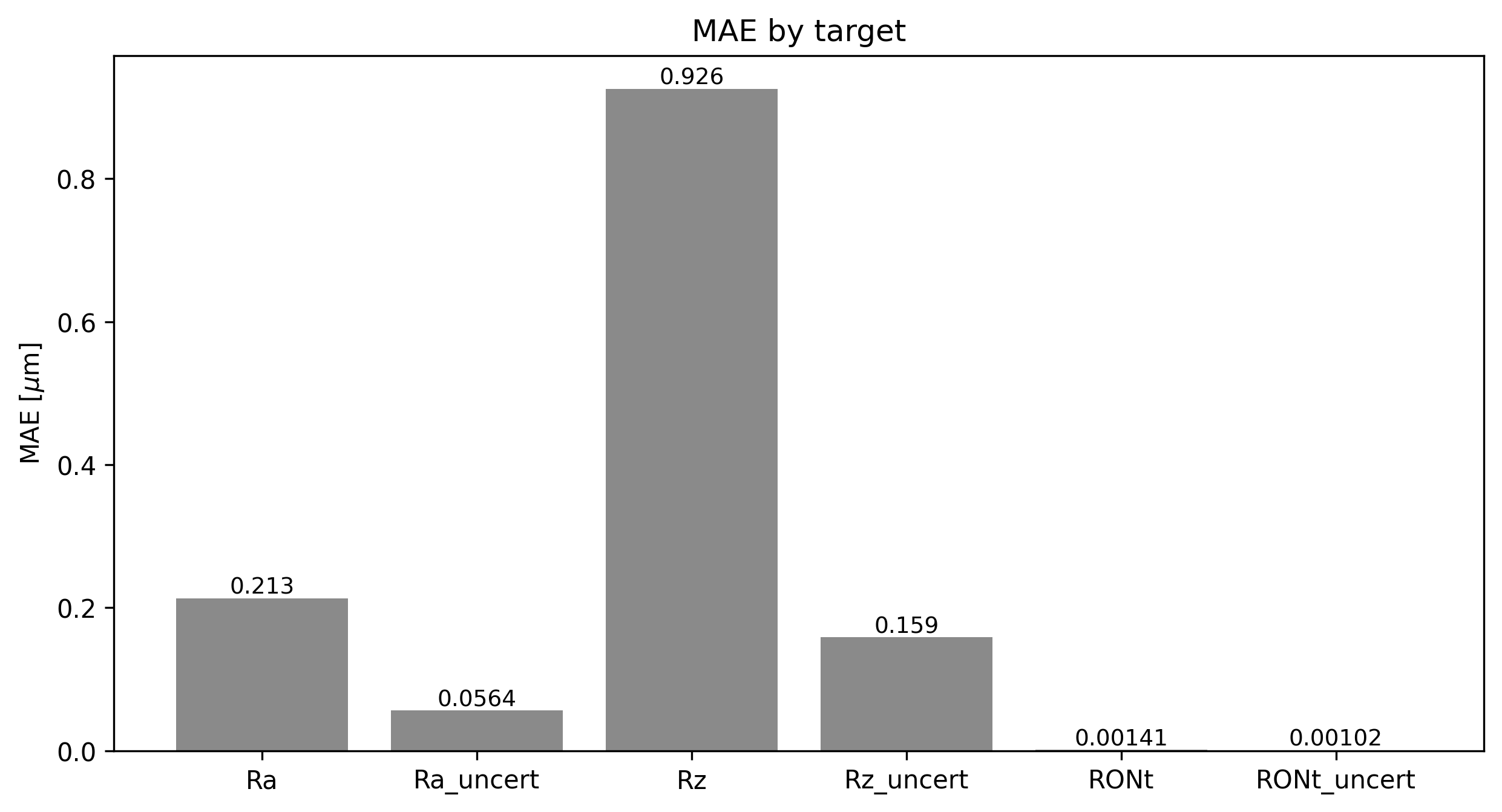}}{\fbox{\parbox{0.9\textwidth}{Missing mae\_by\_target.png}}}\caption{Metric summary per target (MAE by target)}\end{subfigure}\hfill
  \begin{subfigure}[b]{0.48\textwidth}\centering\IfFileExists{img016_outputs_regression_single_final_summary_regression_accuracy_percent_bars.png}{\includegraphics[width=\textwidth]{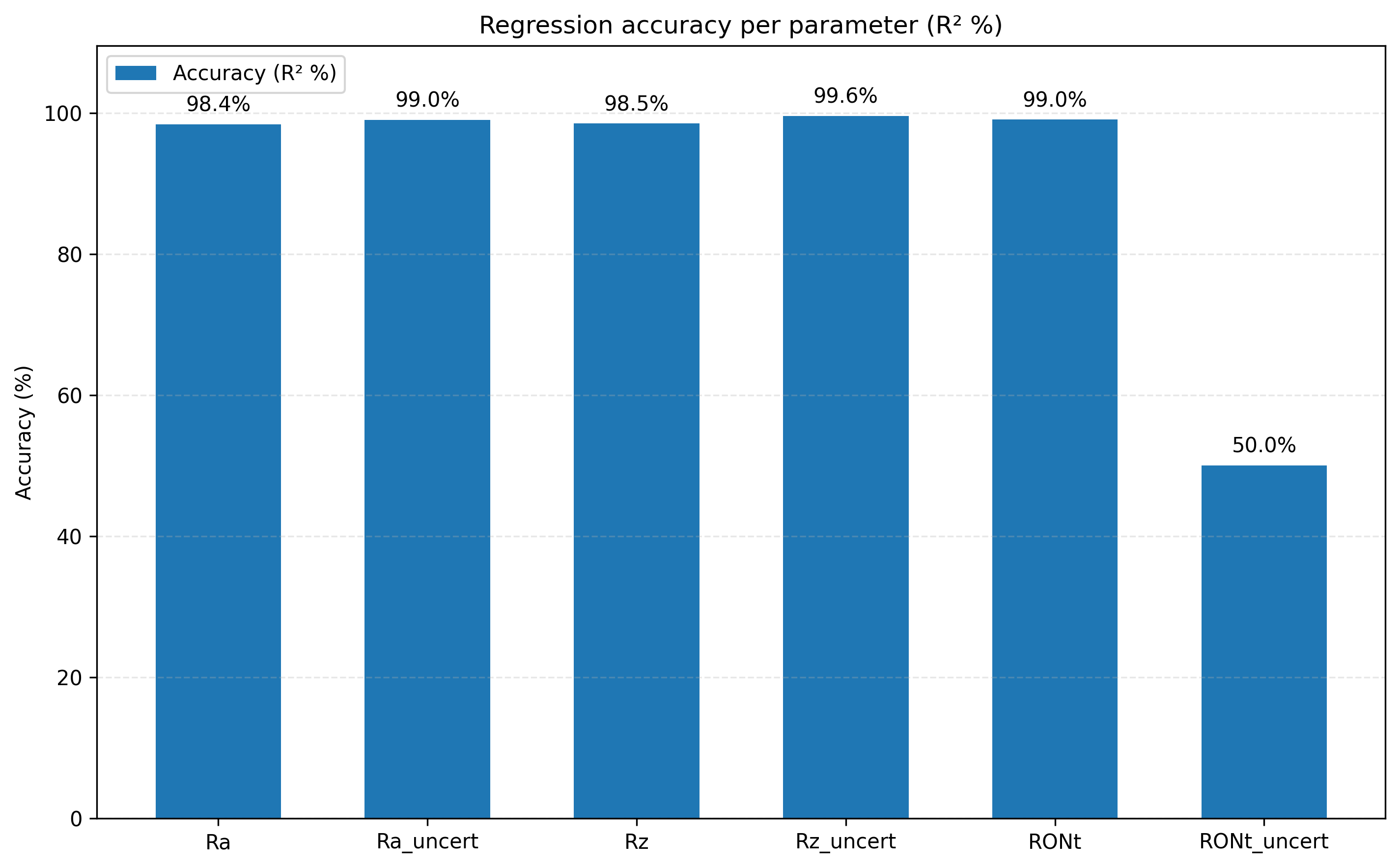}}{\fbox{\parbox{0.9\textwidth}{Missing regression\_accuracy\_percent\_bars.png}}}\caption{Accuracy within relative bands}\end{subfigure}
  \caption{Single-target regression performance: aggregate metrics and tolerance-based accuracies for \(Ra\), \(Rz\), \(RONt\)}
  \label{fig:single_summary}
\end{figure}

\begin{figure}[htbp]
  \centering
  \begin{subfigure}[b]{0.32\textwidth}\centering\IfFileExists{img017_outputs_regression_single_final_Ra_pred_vs_actual.png}{\includegraphics[width=\textwidth]{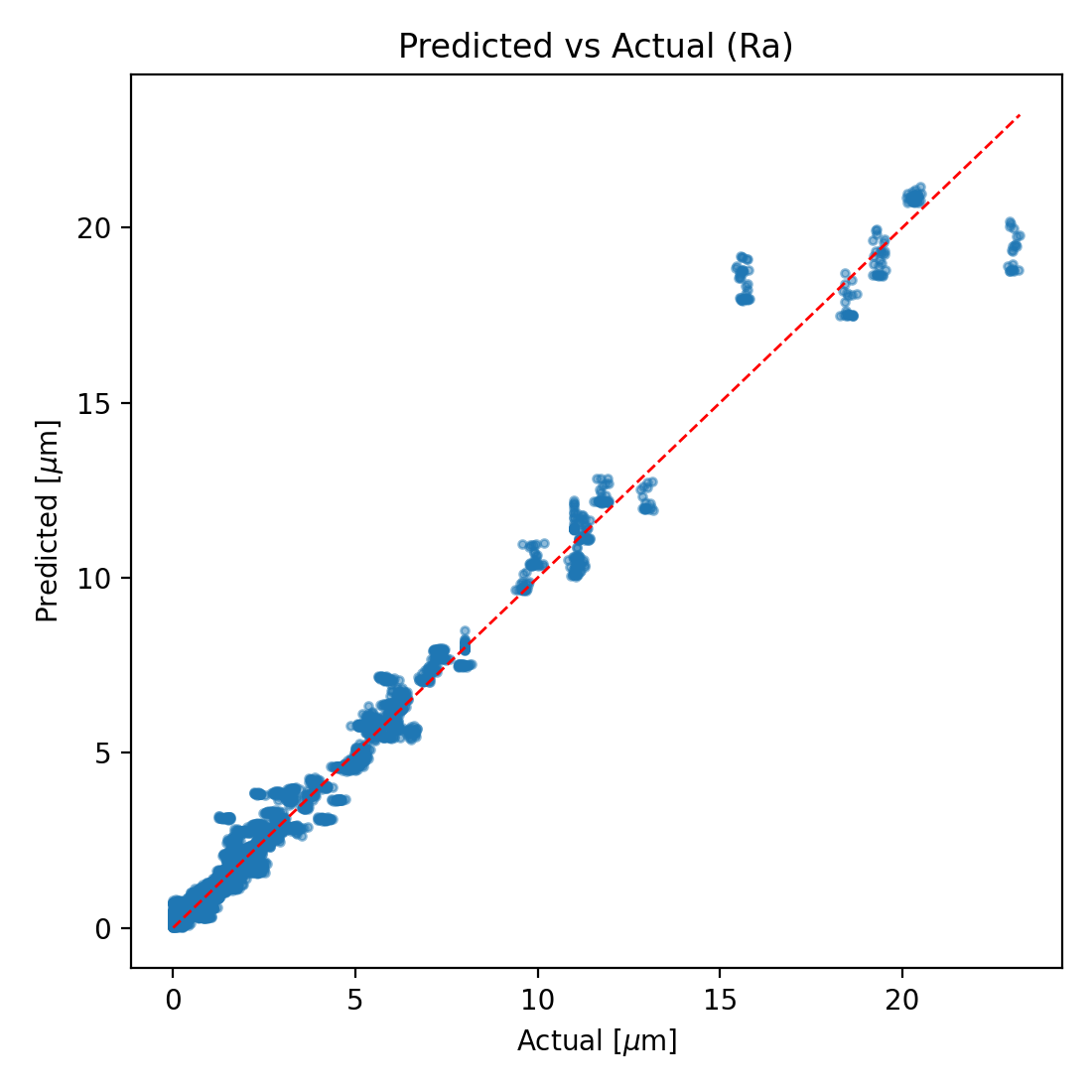}}{\fbox{\parbox{0.9\textwidth}{Missing Ra pred}}}\caption{\(Ra\)}\end{subfigure}\hfill
  \begin{subfigure}[b]{0.32\textwidth}\centering\IfFileExists{img018_outputs_regression_single_final_Rz_pred_vs_actual.png}{\includegraphics[width=\textwidth]{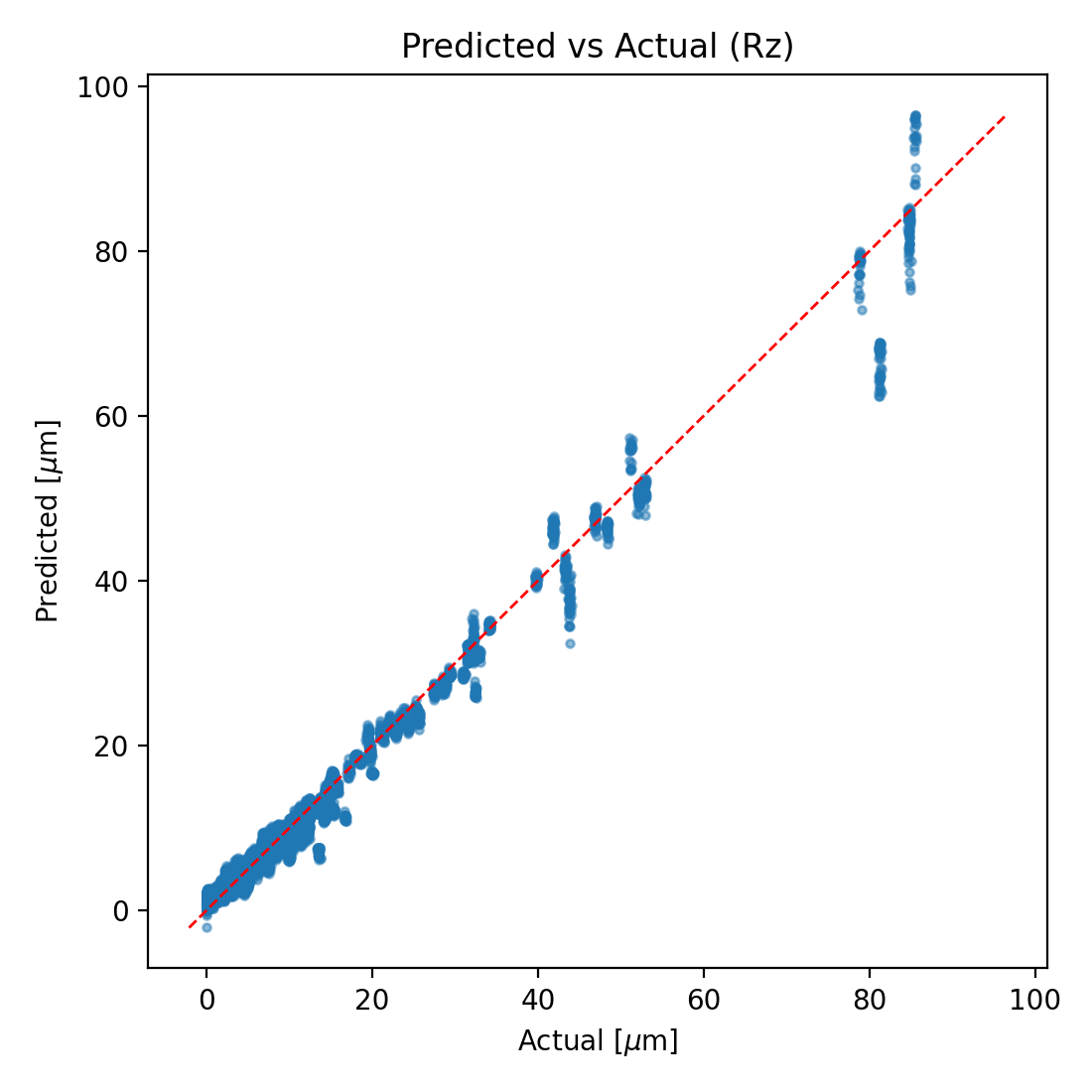}}{\fbox{\parbox{0.9\textwidth}{Missing Rz pred}}}\caption{\(Rz\)}\end{subfigure}\hfill
  \begin{subfigure}[b]{0.32\textwidth}\centering\IfFileExists{img019_outputs_regression_single_final_RONt_pred_vs_actual.png}{\includegraphics[width=\textwidth]{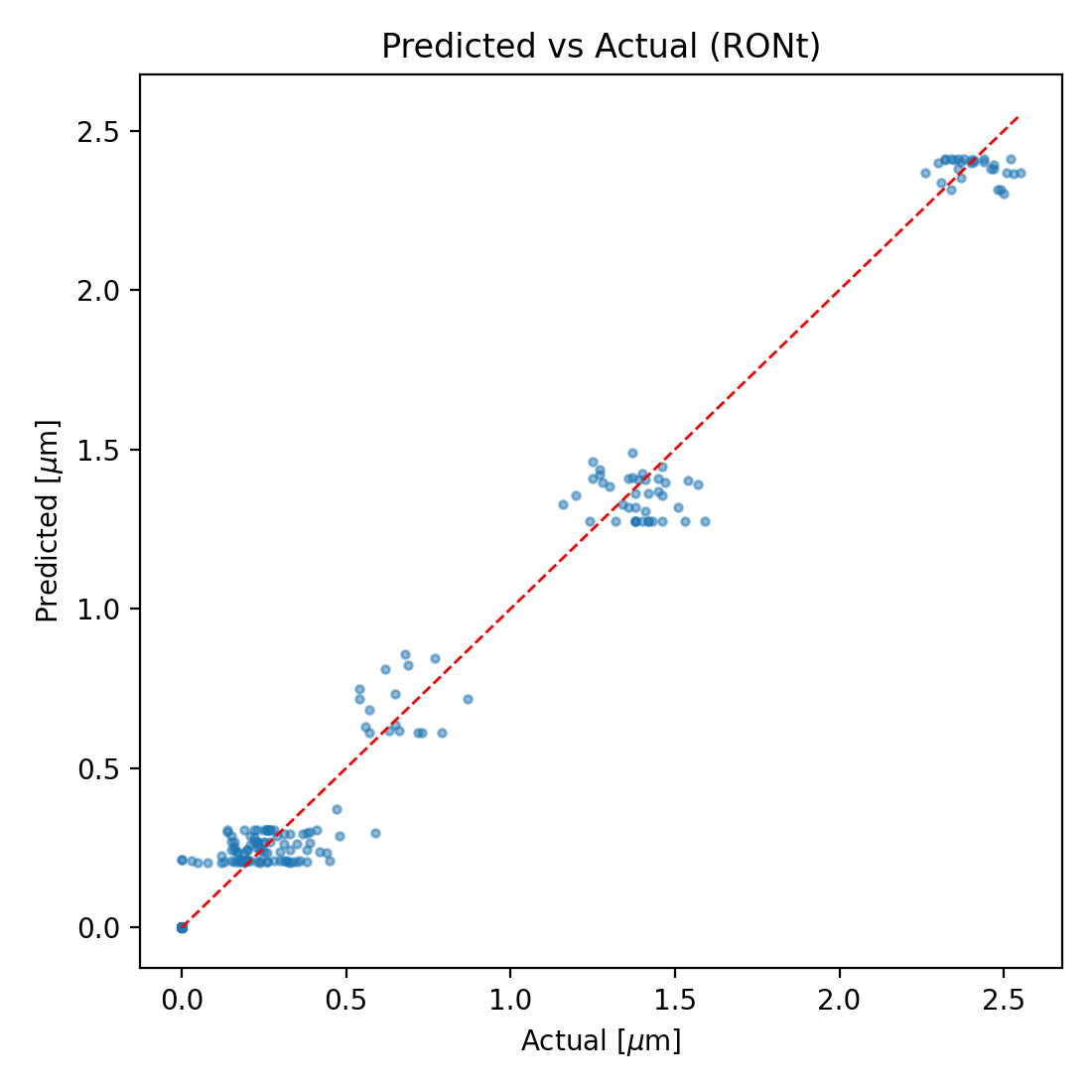}}{\fbox{\parbox{0.9\textwidth}{Missing RONt pred}}}\caption{\(RONt\)}\end{subfigure}
  \caption{Predicted vs actual scatter plots for single-target regression models (primary parameters)}
  \label{fig:single_pred}
\end{figure}

\begin{figure}[htbp]
  \centering
  \begin{subfigure}[b]{0.32\textwidth}\centering\IfFileExists{img020_outputs_regression_single_final_Ra_uncert_pred_vs_actual.png}{\includegraphics[width=\textwidth]{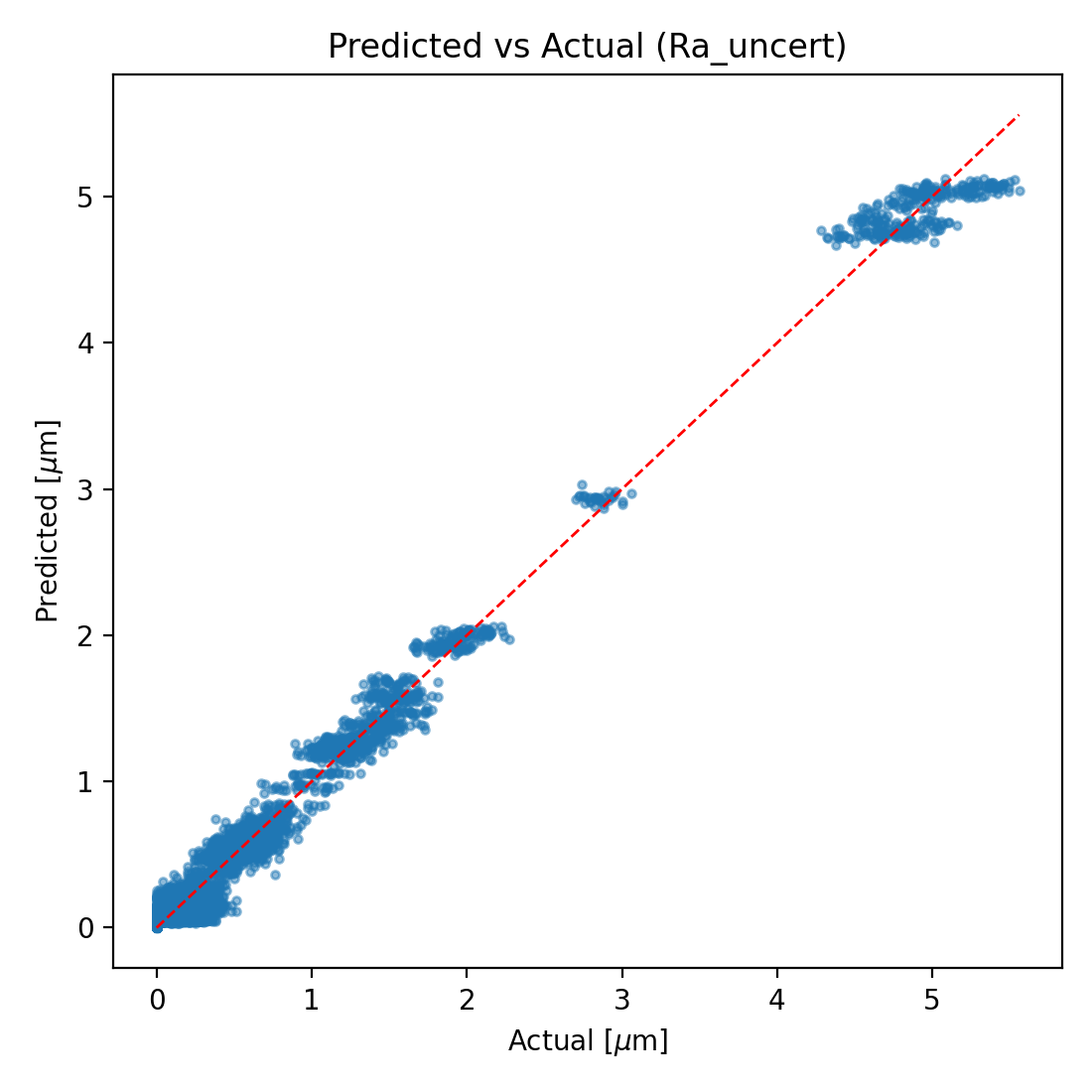}}{\fbox{\parbox{0.9\textwidth}{Missing Ra\_uncert}}}\caption{\(Ra\_\text{uncert}\)}\end{subfigure}\hfill
  \begin{subfigure}[b]{0.32\textwidth}\centering\IfFileExists{img021_outputs_regression_single_final_Rz_uncert_pred_vs_actual.png}{\includegraphics[width=\textwidth]{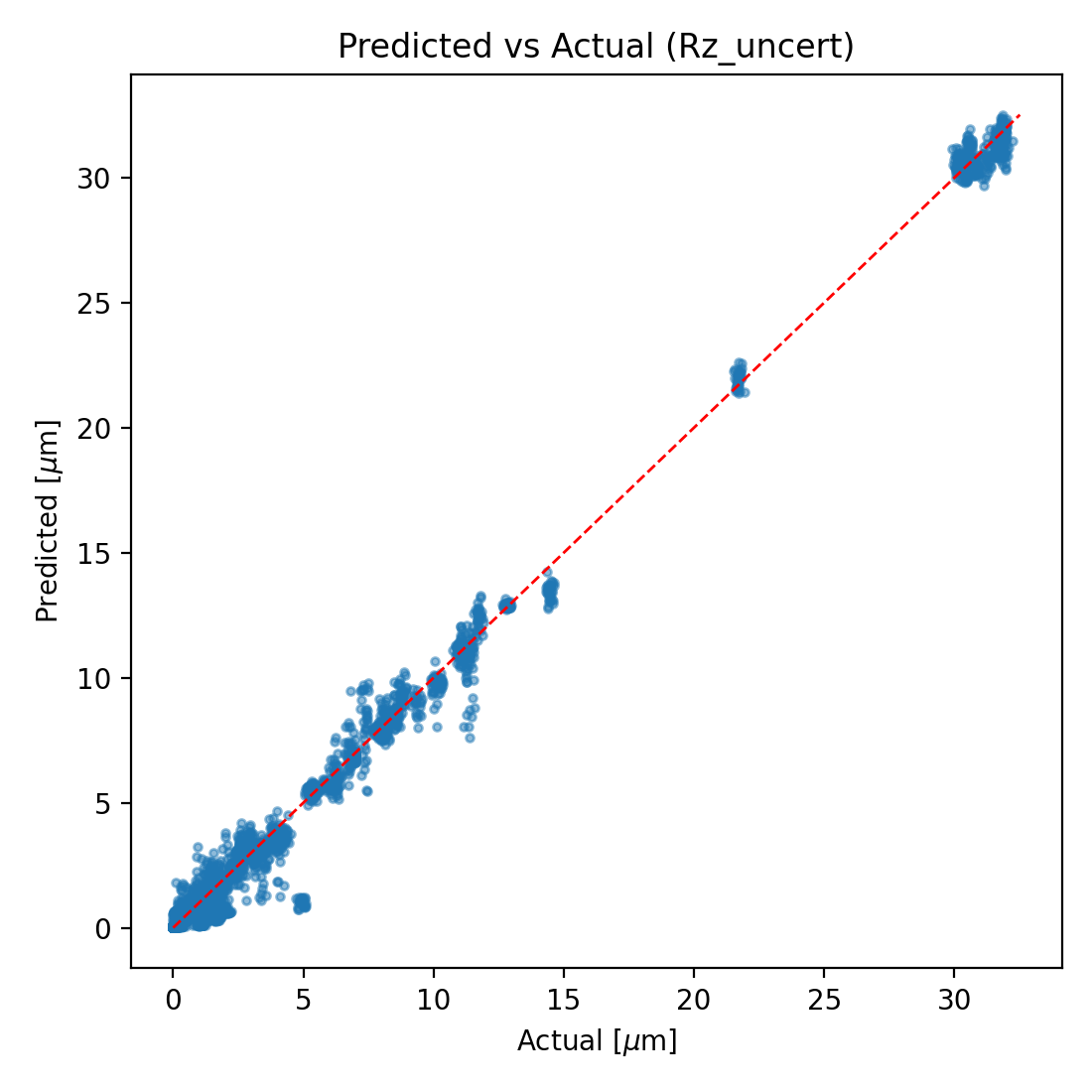}}{\fbox{\parbox{0.9\textwidth}{Missing Rz\_uncert}}}\caption{\(Rz\_\text{uncert}\)}\end{subfigure}\hfill
  \begin{subfigure}[b]{0.32\textwidth}\centering\IfFileExists{img022_outputs_regression_single_final_RONt_uncert_pred_vs_actual.png}{\includegraphics[width=\textwidth]{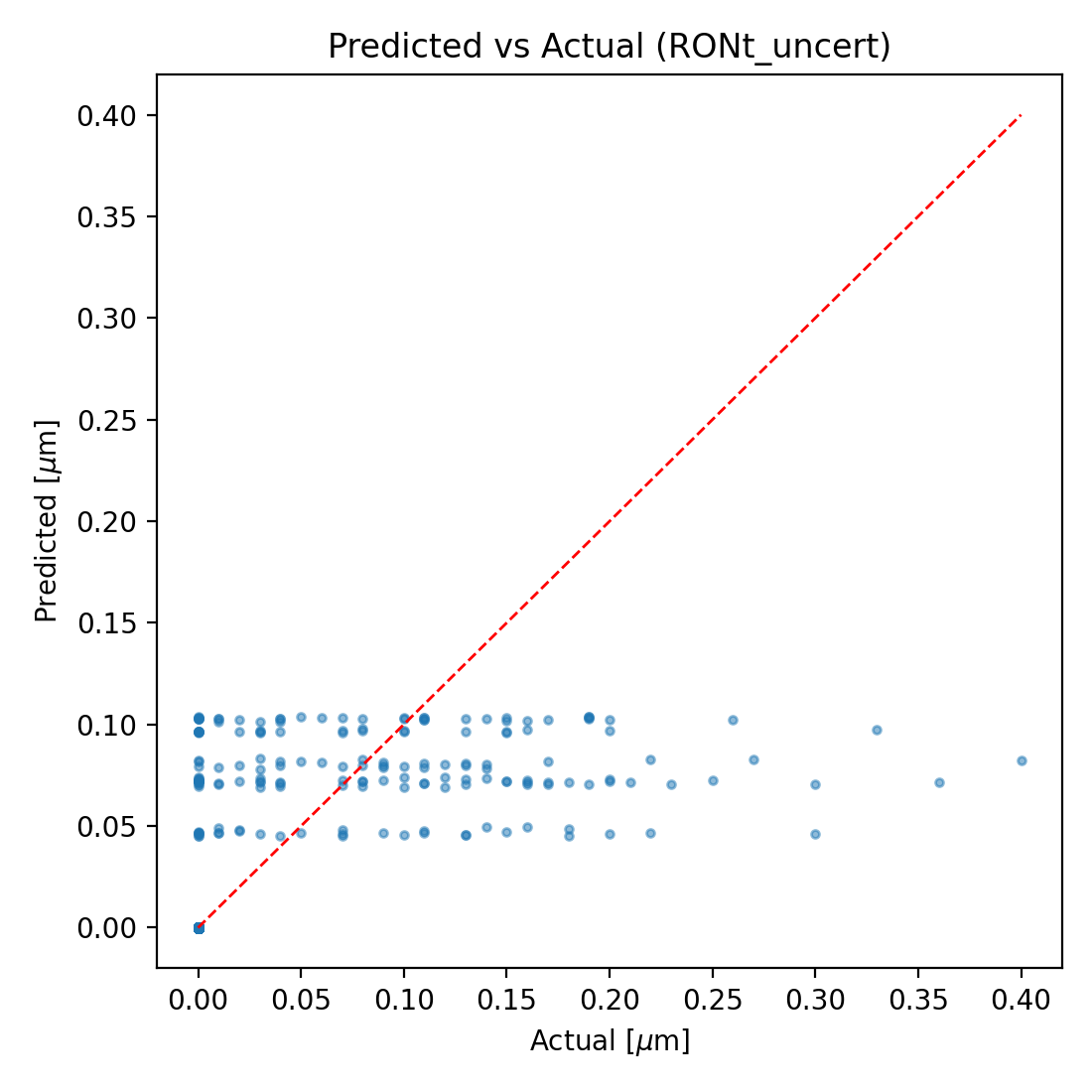}}{\fbox{\parbox{0.9\textwidth}{Missing RONt\_uncert}}}\caption{\(RONt\_\text{uncert}\)}\end{subfigure}
  \caption{Uncertainty-target scatter (cf. Fig.~\ref{fig:single_pred} for primary targets)}
  \label{fig:uncert_pred}
\end{figure}

\begin{table}[htbp]
  \centering
  \caption{Multi-output loss variant comparison (averages across six targets)}
  \label{tab:loss_ablation}
  \setlength{\tabcolsep}{6pt}
  \begin{tabular}{lccc}
    \toprule
    Variant & Mean MAE [$\mu$m] & Mean $R^2$ & Notes \\
    \midrule
    Baseline (final) & 1.325 & 0.582 & Log-Huber; best mean $R^2$ but higher MAE \\
    MAE & 1.143 & 0.502 & Lower MAE; weaker variance capture \\
    Weighted MAE & 1.148 & 0.469 & Emphasises $Ra$, $Rz$; preserves $RONt$ MAE \\
    Log-Huber (alt) & 1.325 & 0.582 & Robust to outliers; similar to baseline \\
    \bottomrule
  \end{tabular}
  \vspace{1mm}
  \begin{minipage}{0.9\linewidth}
  \raggedright\footnotesize Values rounded to three decimals; metrics obtained from held-out validation summaries.
  \end{minipage}
\end{table}

\begin{table}[htbp]
  \centering
  \caption{Regression performance metrics: single-target vs weighted multi-output}
  \label{tab:reg_summary}
  \setlength{\tabcolsep}{5pt}
  \begin{tabular}{lccc ccc}
    \toprule
    & \multicolumn{3}{c}{Single-target} & \multicolumn{3}{c}{Multi-output (weighted)} \\
      Target & MAE [$\mu$m] & RMSE [$\mu$m] & $R^2$ & MAE [$\mu$m] & RMSE [$\mu$m] & $R^2$ \\
    \midrule
  $Ra$ & 0.2134 & 0.3730 & 0.9824 & 0.8695 & 1.7070 & 0.6323 \\
  $Rz$ & 0.9255 & 1.5567 & 0.9847 & 4.2072 & 8.1861 & 0.5757 \\
  $RONt$ & 0.00141 & 0.01339 & 0.9918 & 0.00124 & 0.01232 & 0.9930 \\
  Ra\_uncert & 0.05639 & 0.08389 & 0.9899 & 0.2699 & 0.7708 & 0.1428 \\
  Rz\_uncert & 0.1589 & 0.3578 & 0.9955 & 1.5412 & 4.8790 & 0.1550 \\
  RONt\_uncert & 0.001020 & 0.01039 & 0.4934 & 0.001094 & 0.01208 & 0.3151 \\
    \midrule
    Mean (single-target) & 0.2261 & 0.3990 & 0.9063 & & & \\
    Mean (multi-output) & & & & 1.1484 & 2.4329 & 0.4689 \\
    \bottomrule
  \end{tabular}
  \vspace{1mm}
  \begin{minipage}{0.9\linewidth}
  \footnotesize Uncertainty target names retain the \_uncert suffix. MAE and RMSE are reported in [$\mu$m]. Values rounded to three decimal places (four for $R^2$). Single-target models generally provide higher fidelity for primary parameters and often their uncertainties compared to the weighted multi-output trunk.
  \end{minipage}
\end{table}

\begin{sidewaystable}[htbp]
  \centering
  \caption{Empirical coverage (EC) vs nominal coverage (NC) for central prediction intervals before (Quantile) and after Conformal adjustment}
  \label{tab:coverage_calibration}
\setlength{\tabcolsep}{4pt}
\begin{tabular}{lccccccc}
  \toprule
  Target & Nominal & Quant EC & |$\Delta$| & Conf EC & |$\Delta$| & Quant W & Conf W \\
  \midrule
  ${Ra}$ & 0.9 & 0.983 & 0.083 & 0.905 & 0.005 & 1.212 & 0.67 \\
  ${Rz}$ & 0.9 & 0.302 & 0.598 & 0.901 & 0.001 & 3.675 & 3.052 \\
  ${RONt}$ & 0.9 & 0.151 & 0.749 & 0.899 & 0.001 & 0.047 & 0 \\
  \bottomrule
\end{tabular}

  \vspace{1mm}
  \begin{minipage}{0.9\linewidth}
  \footnotesize Quantile empirical coverage values (EC) and nominal coverage (NC) are shown as fractions (0--1); interval widths (W) are reported in [$\mu$m]. Conformal coverage reflects post-adjustment performance. Uncertainty target names retain the *\_uncert suffix.
  \end{minipage}
\end{sidewaystable}

\paragraph{Supplementary material linkage.} 
Detailed uncertainty artefacts are consolidated in the Supplementary Appendix (Tables~S1--S5 and associated Figures S1 onward). Table~S1 reports empirical coverage and mean interval width across nominal central probability levels (0.50/0.80/0.90) for all primary and direct uncertainty targets; Table~S2 provides conformal post-calibration width adjustments and achieved coverage; Table~S3 collates expanded interval scoring metrics (pinball, CRPS proxy, Winkler variants); Table~S4 summarises excess kurtosis of residual distributions; Table~S5 lists absolute residual vs predicted uncertainty correlations. Supplementary figures supply per-target calibration curves, width--coverage profiles, distribution “fan” diagrams, residual diagnostics, feature permutation importances and ablation panels. These resources enable granular inspection of calibration behaviour, dispersion scale, tail structure and heteroscedastic signal quality beyond the aggregate indicators retained in the main text.

\section{Discussion}\label{sec:discussion}
\begin{sloppypar}
The presented framework demonstrates that deep learning can accurately infer both surface parameters and their associated uncertainties from multi-instrument data. A key empirical outcome is that carefully tuned \emph{single-target} regressors consistently outperform naive \emph{multi-output} trunks across most targets (Table~\ref{tab:reg_summary}, Fig.~\ref{fig:single_summary}). The gap is attributed to heterogeneous noise scales and target-specific structures: a shared trunk with a single joint loss induces negative transfer, particularly harming $Ra$, $Rz$, and the uncertainty targets, even when losses are reweighted (Table~\ref{tab:loss_ablation}).

Uncertainty quantification benefitted from a layered design. Quantile regression provided asymmetric bands, heteroscedastic Gaussian heads captured input-dependent dispersion, and post-hoc conformal adjustment restored nominal coverage with modest width inflation (Table~\ref{tab:coverage_calibration}). In practice, this stack yielded calibrated, easy-to-interpret intervals in micrometres [$\mu$m], which is the natural reporting unit in surface metrology; for $Ra$ and $Rz$, interval magnitudes are broadly comparable to empirically reported standard uncertainties, suggesting the model can complement experimental evaluation when repeated acquisitions are impractical. For classification, temperature scaling yielded a negligible change in miscalibration (ECE from 0.00504 to 0.00503; Fig.~\ref{fig:clf_calibration}), with accuracy unaffected.

Three practical observations emerge. First, tolerance-style metrics (Fig.~\ref{fig:tol_regr_panel}) complement MAE/RMSE by directly reflecting decision thresholds used by practitioners (relative bands [\%] and absolute bands in [$\mu$m]). Second, the uncertainty targets are \emph{learnable}: two of the three (\texttt{Ra\_uncert}, \texttt{Rz\_uncert}) achieve high $R^2$ with single-target models, supporting the premise that reported standard uncertainties carry signal beyond noise. Third, \texttt{RONt\_uncert} remains comparatively challenging; its weaker signal and scale mismatch likely require richer descriptors and/or target-specific modelling.

  	extit{RONt-specific considerations.} Compared to $Ra$ and $Rz$, the $RONt$ target exhibits lower predictive accuracy, and \texttt{RONt\_uncert} shows reduced learnability. Two primary causes are identified: (i) \emph{instrument heterogeneity} — the dataset aggregates measurements from different roundness testers (types/generations) with distinct metrological characteristics, probing/fixturing, filtering and evaluation chains. This induces a cross-instrument domain shift that a single tabular model only partially accommodates, depressing accuracy even with standardisation. (ii) \emph{uncertainty label fidelity} — the reported standard uncertainty for $RONt$ reflects a partial budget where not all contributing components are precisely known, modelled, or logged during evaluation. In our cohort, partner-site setups for roundness exhibited greater heterogeneity than the roughness measurement setups, further increasing cross-site variability and affecting both point accuracy and uncertainty labels. The resulting label noise/bias constrains the attainable $R^2$ for \texttt{RONt\_uncert}. Mitigations include harmonised acquisition protocols, explicit inclusion of instrument metadata (make/model, probe, filter stack) as features or conditional heads, cross-instrument calibration layers, and standardised, fully specified uncertainty budgets (e.g. decomposed repeatability/reproducibility components) to improve label quality. Notably, multi-output training yields a slightly higher $R^2$ for $RONt$ (Table~\ref{tab:reg_summary}), which likely reflects joint-loss emphasis on that scale at the expense of other targets — an instance of negative transfer across heterogeneous outputs.

\noindent\textit{Operational decisions.} Tolerance-style metrics translate statistical accuracy into actionable insight: given a quantified confidence level, surfaces can be pre-assessed for compliance with specification limits or a more appropriate instrument can be selected prior to measurement, thereby bridging model outputs with metrological workflow decisions.

    extbf{Limitations (priority-ordered).} The primary limitation is \emph{dataset diversity/generalisation}: despite multi-instrument coverage, domain shift across laboratories and calibration standards remains likely; multi-site (federated) datasets should be prioritised to assess external validity. Secondary limitations include: (i) \emph{uncertainty evaluation and label noise} — reported standard uncertainties (especially for $RONt$) omit or approximate components and differ across partner-site procedures, limiting attainable $R^2$; (ii) \emph{cross-site variability for roundness} — partner sites used different roundness testers and evaluation protocols with greater variability than roughness setups, reducing transfer and label fidelity for $RONt$ and \texttt{RONt\_uncert}; (iii) \emph{model conditioning on instrument} — regressors only implicitly encode instrument identity; conditional heads/adapters may further reduce negative transfer; and (iv) \emph{calibration granularity} — conformal guarantees marginal, not conditional, coverage; local (covariate-conditional) conformal adjustments could address residual miscalibration.

	\textbf{Outlook.} We see several low-risk extensions: (i) adaptive loss reweighting driven by on-the-fly gradient norms to reduce target dominance; (ii) target-wise specialised trunks (mixture-of-experts) with sparse routing; (iii) local conformal scaling using estimated conditional scales to stabilise width vs. coverage trade-offs; (iv) acquisition strategies prioritising under-represented regimes (active learning); and (v) incorporation of physics- or standards-aware features (e.g., cut-off and evaluation-length priors, filtering provenance) to strengthen extrapolation. These follow-ups align with our reproducibility-first release and can be integrated into the existing training scripts with minimal disruption.

\section{Conclusions}\label{sec:conclusions}
Results indicate that uncertainty-aware deep learning can provide both high-fidelity point predictions and calibrated confidence bounds for surface metrology. Quantitatively, a mean $R^2$ of \textbf{0.9063} was achieved by single-target regressors compared to \textbf{0.4689} for the weighted multi-output trunk (Table~\ref{tab:reg_summary}), reflecting markedly lower MAE/RMSE across most targets. High accuracy was reached for primary parameters—$Ra$ ($R^2=\,\textbf{0.9824}$), $Rz$ ($R^2=\,\textbf{0.9847}$), and $RONt$ ($R^2=\,\textbf{0.9918}$)—and two uncertainty targets were well modelled—\texttt{Ra\_uncert} ($R^2=\,\textbf{0.9899}$) and \texttt{Rz\_uncert} ($R^2=\,\textbf{0.9955}$). In contrast, \texttt{RONt\_uncert} remained challenging ($R^2=\,\textbf{0.4934}$), in line with instrument heterogeneity and partially specified uncertainty budgets discussed in the Discussion.

From an operational standpoint, an accuracy of \textbf{92.85\%} was obtained by the classifier (Table~\ref{tab:clf_metrics}), and temperature scaling resulted in a negligible change in calibration (ECE \textbf{0.00504} $\to$ \textbf{0.00503}; Fig.~\ref{fig:clf_calibration}). For regression, the uncertainty stack (quantile + heteroscedastic) with conformal adjustment yielded intervals whose empirical coverage is close to nominal (Table~\ref{tab:coverage_calibration}) and whose widths (in [$\mu$m]) are broadly comparable to reported standard uncertainties for $Ra$ and $Rz$. Practically, this enables pre-assessment of acceptance against tolerance bands and supports instrument selection with quantified confidence.

Overall, the combination of single-target specialisation with calibrated interval estimation provides a pragmatic path toward trustworthy, uncertainty-aware decision support in metrological workflows, and outlines a foundation for scalable, cross-laboratory deployment.
\end{sloppypar}

\section*{Data and Code Availability}
All code, processing scripts, trained-model artefacts (regeneration scripts), are available under the MIT License at the project repository (GitHub, latest commit snapshot) and archived on Zenodo at DOI: \cite{dawid_kucharski_2025_17277722}. The release bundle includes hash manifests ensuring integrity verification.

\section*{Acknowledgements}
\vspace{6mm}
\begin{center}
 
  \IfFileExists{img023_logo2.pdf}{\includegraphics[width=5cm]{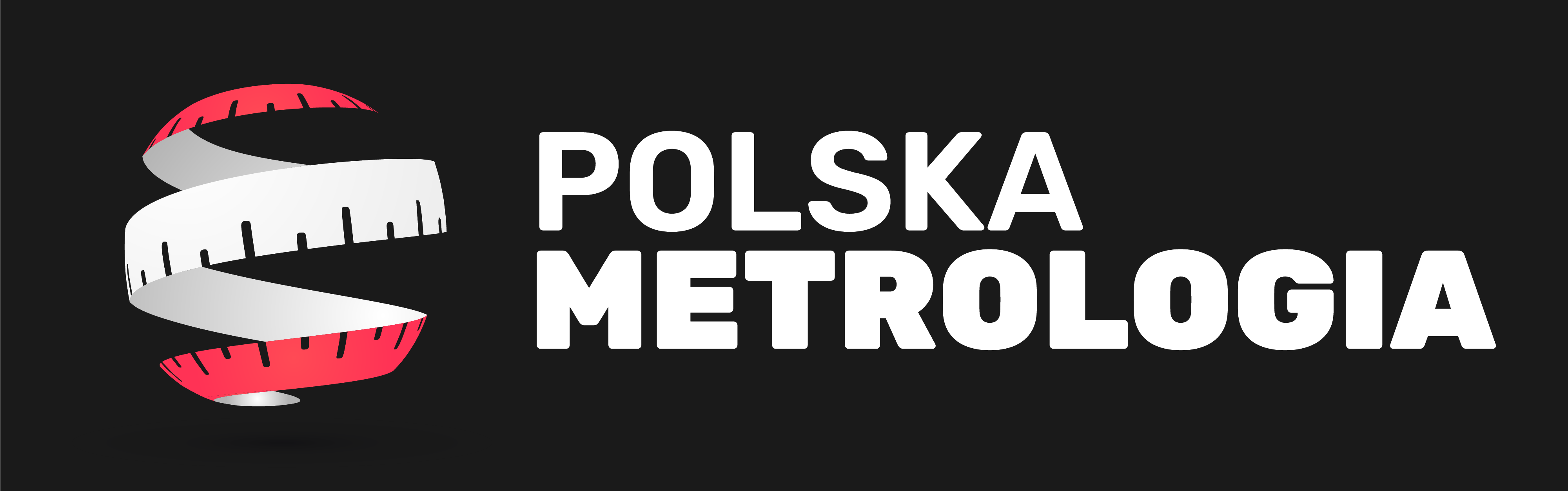}}{\fbox{\parbox{5cm}{\centering img023_logo2.pdf missing}}}
\end{center}
A grant supported this work: project entitled: "Application of artificial intelligence in surface irregularities measurements", financed by the Ministry of Education and Science of the programme: Polish Metrology II PM-II/SP/0104/2024/02 of 01.02.2024\\
Projekt pt. „Zastosowanie sztucznej inteligencji w pomiarach nierówności powierzchni” finansowany przez Ministerstwo Nauki i~Szkolnictwa Wyższego w ramach programu Polska Metrologia 2 Nr PM-II/SP/0104/2024/02 z dnia 01.02.2024.

\section*{CRediT authorship contribution statement}
\textbf{D. K.}: Conceptualization; Methodology; Software; Data curation; Formal analysis; Investigation; Validation; Visualization; Writing – original draft; Writing – review $\&$ editing; Project administration. \textbf{A. G.}: Methodology; Resources. \textbf{T. K.}: Resources; Data curation. \textbf{K. S.}: Data curation. \textbf{M. R.}: Data curation; Investigation. \textbf{B. G.}: Data curation; Investigation. \textbf{M. W.}: Supervision. \textbf{M. N.}: Data curation; Resources. \textbf{P. S.}: Data curation; Resources. \textbf{P. S.}: Validation; Resources. \textbf{J. T.}: Investigation; Data curation. \textbf{A. W.}: Investigation; Data curation.

\noindent \textbf{D. K.} led and performed the substantial majority ($\approx 60\%$) of the overall work. Each supporting co-author contributed in a limited capacity (\(< 4\%\)) confined to the roles explicitly listed above.
\section*{Declaration of Competing Interest}
The authors declare that they have no known competing financial interests or personal relationships that could appear to influence the work reported in this document.
{\sloppy
}



\clearpage
\section*{Supplementary Material}

\setcounter{figure}{0}
\renewcommand{\thefigure}{S\arabic{figure}}
\setcounter{table}{0}
\renewcommand{\thetable}{S\arabic{table}}
\begingroup
\setlength{\parindent}{0pt}
\setlength{\parskip}{4pt}
\newcommand{\suppimage}[2]{\begin{minipage}[t]{0.48\textwidth}\centering\includegraphics[width=\textwidth]{#1}\\\scriptsize #2\end{minipage}}
\suppimage{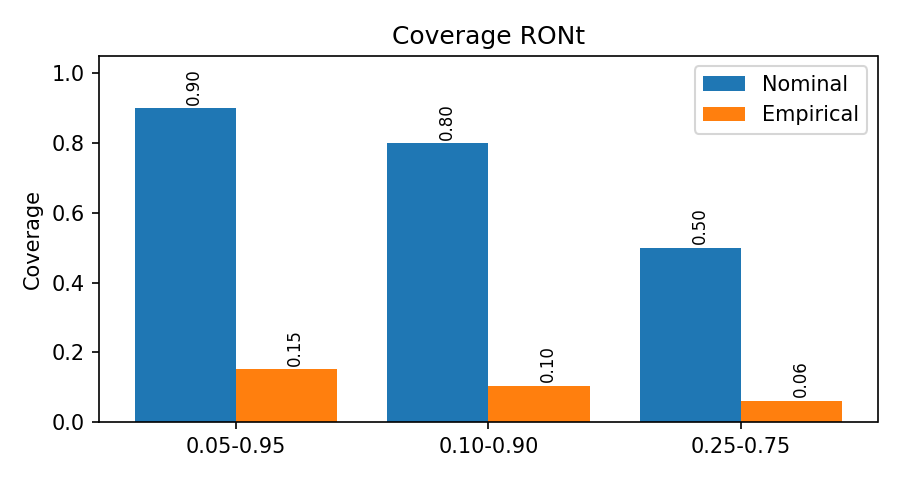}{Figure S1: coverage plot (\label{fig:supp-1})}\hfill
\suppimage{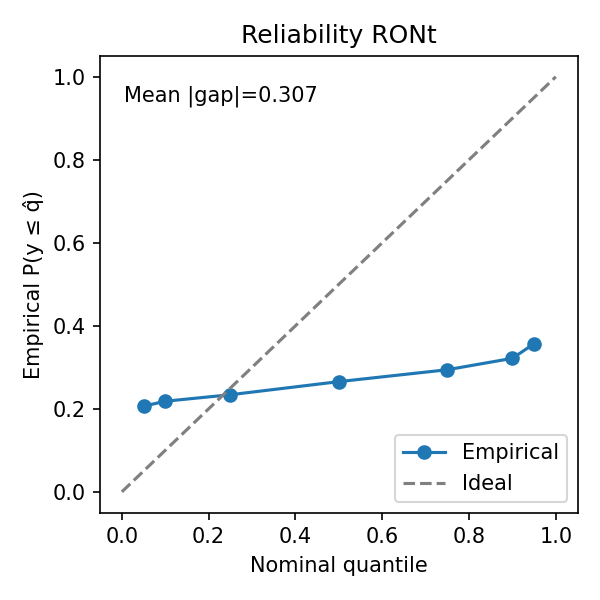}{Figure S2: reliability plot (\label{fig:supp-2})}\
\suppimage{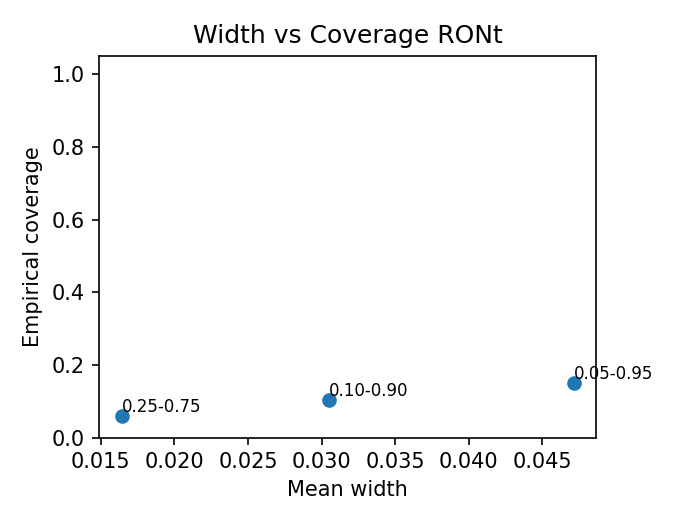}{Figure S3: width vs coverage (\label{fig:supp-3})}\hfill
\suppimage{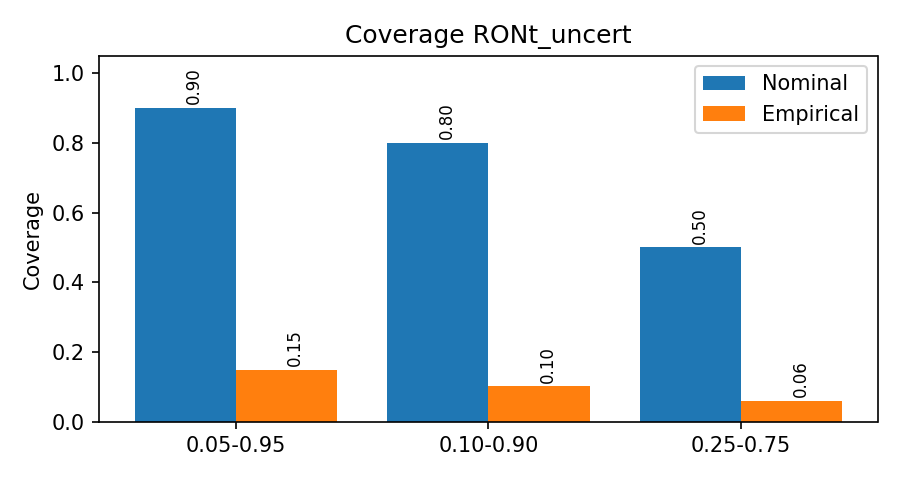}{Figure S4: coverage plot (\label{fig:supp-4})}\
\suppimage{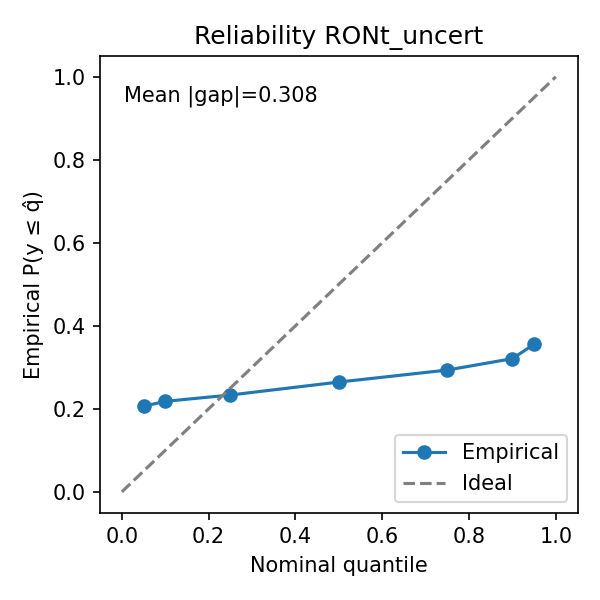}{Figure S5: reliability plot (\label{fig:supp-5})}\hfill
\suppimage{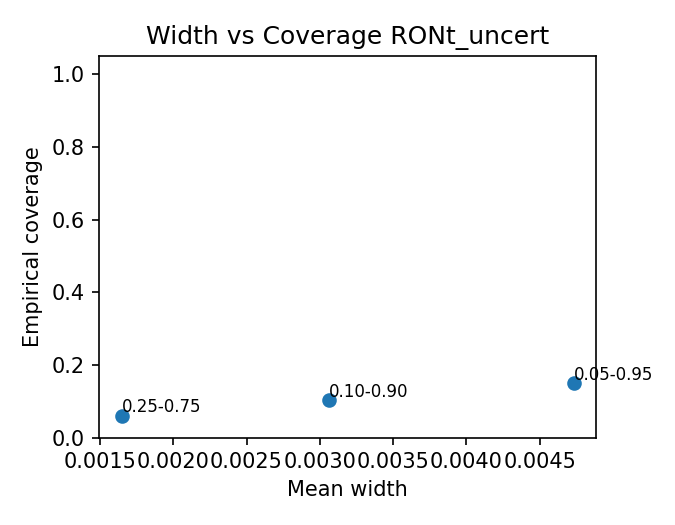}{Figure S6: width vs coverage (\label{fig:supp-6})}\
\suppimage{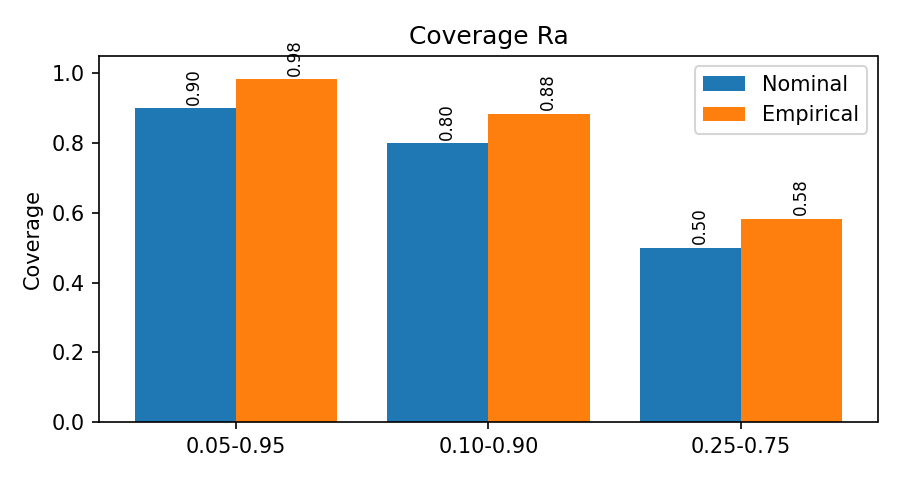}{Figure S7: coverage plot (\label{fig:supp-7})}\hfill
\suppimage{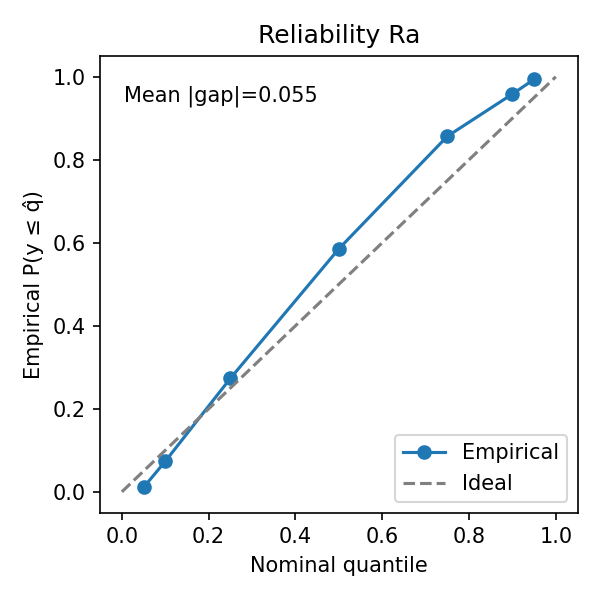}{Figure S8: reliability plot (\label{fig:supp-8})}\
\suppimage{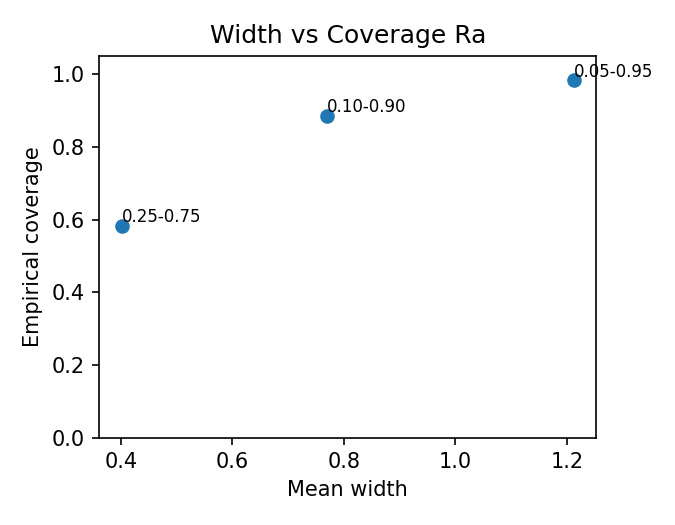}{Figure S9: width vs coverage (\label{fig:supp-9})}\hfill
\suppimage{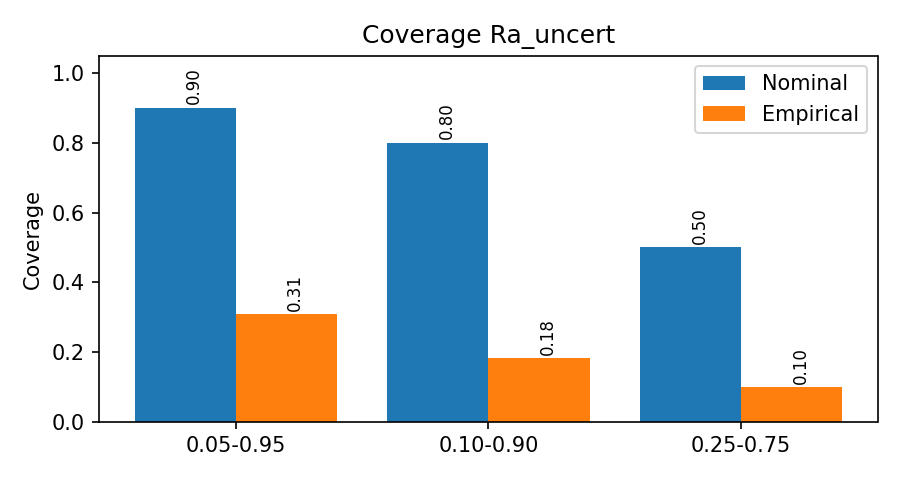}{Figure S10: coverage plot (\label{fig:supp-10})}\
\suppimage{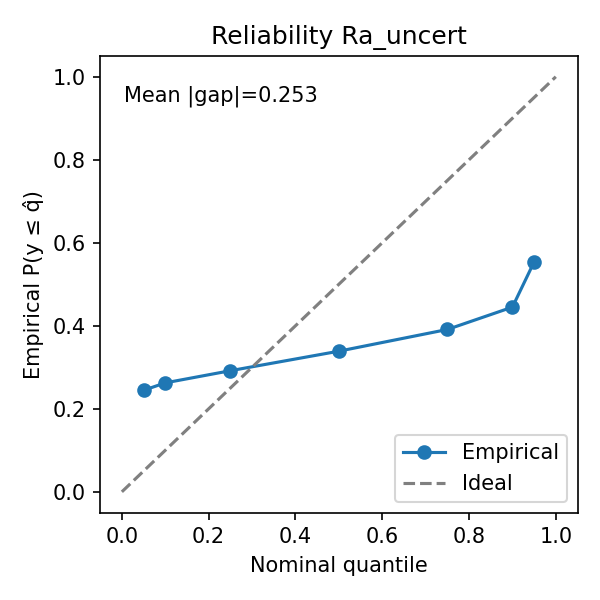}{Figure S11: reliability plot (\label{fig:supp-11})}\hfill
\suppimage{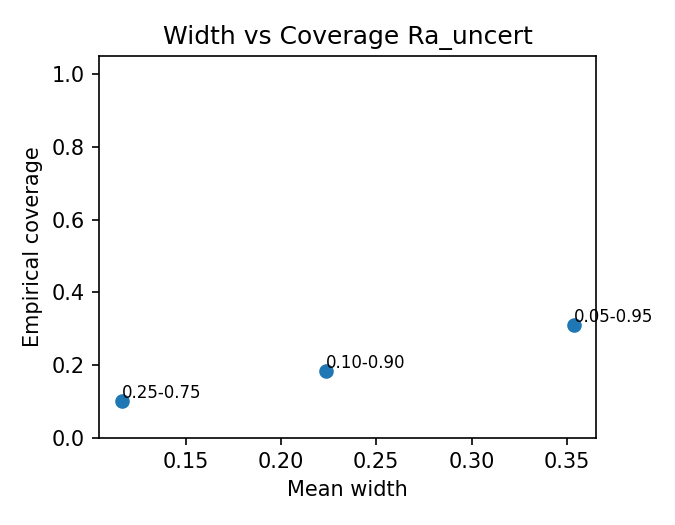}{Figure S12: width vs coverage (\label{fig:supp-12})}\
\suppimage{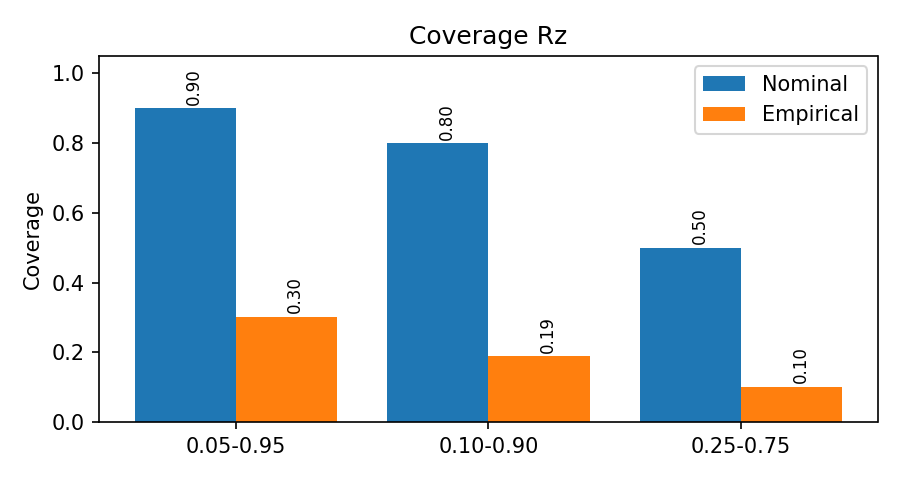}{Figure S13: coverage plot (\label{fig:supp-13})}\hfill
\suppimage{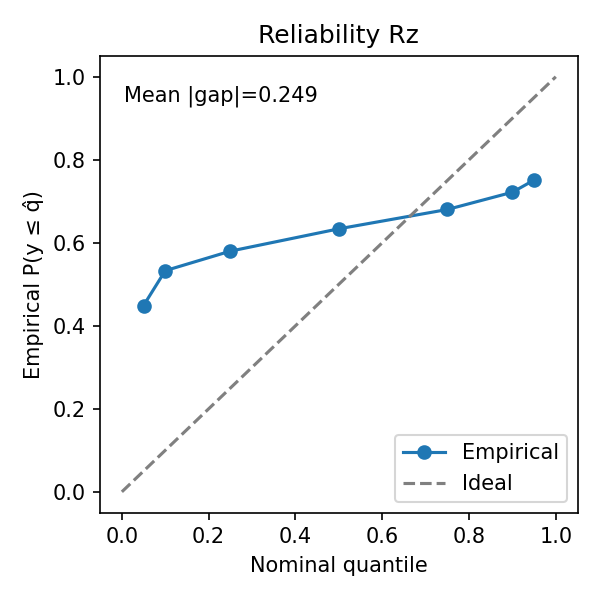}{Figure S14: reliability plot (\label{fig:supp-14})}\
\suppimage{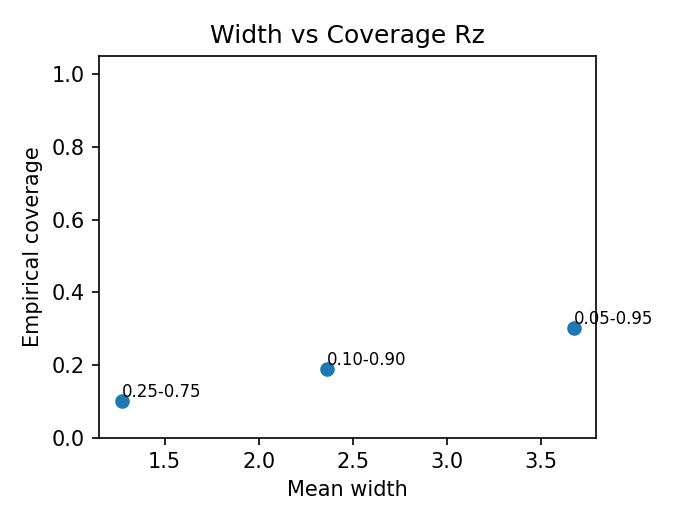}{Figure S15: width vs coverage (\label{fig:supp-15})}\hfill
\suppimage{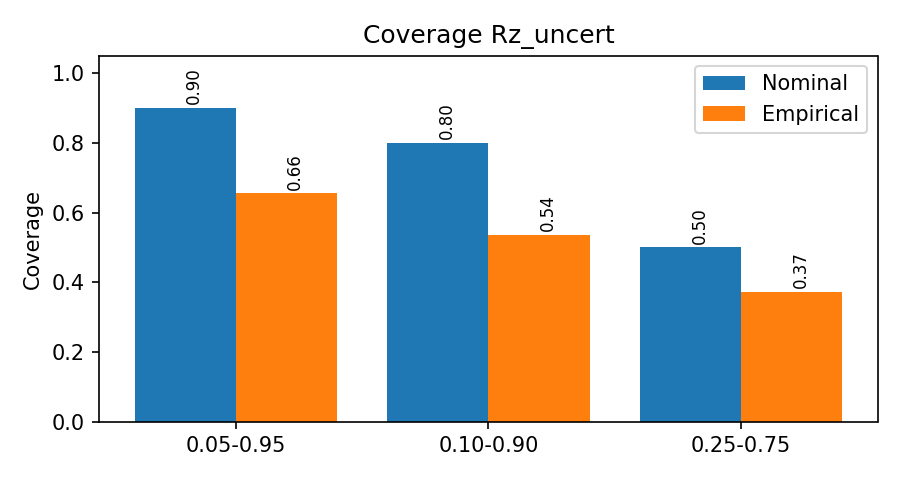}{Figure S16: coverage plot (\label{fig:supp-16})}\
\suppimage{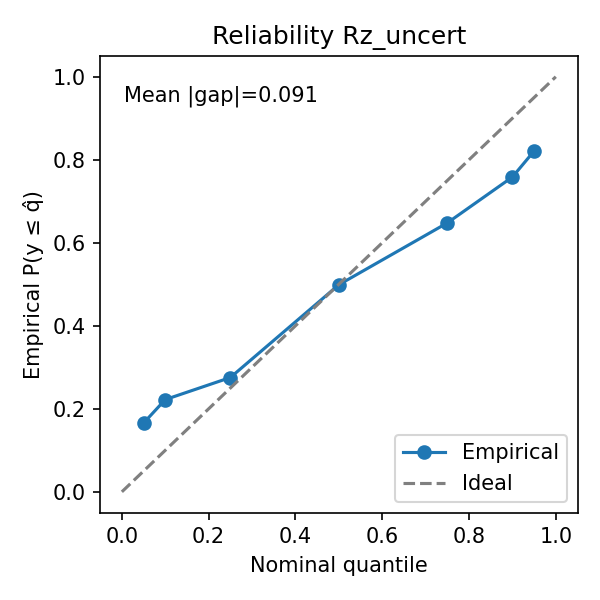}{Figure S17: reliability plot (\label{fig:supp-17})}\hfill
\suppimage{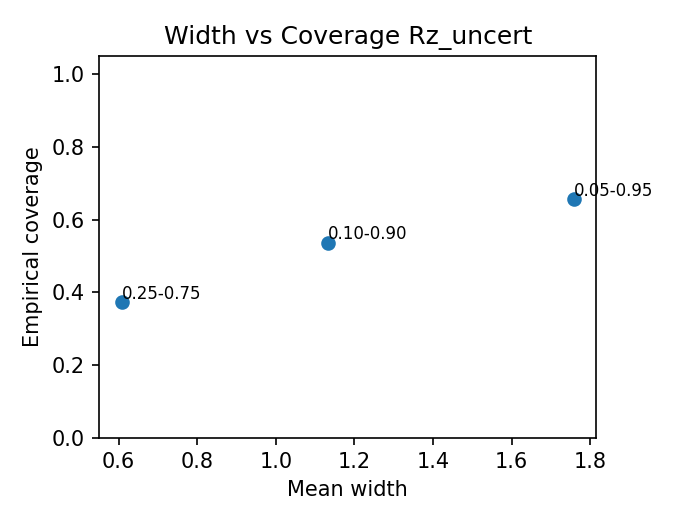}{Figure S18: width vs coverage (\label{fig:supp-18})}\
\suppimage{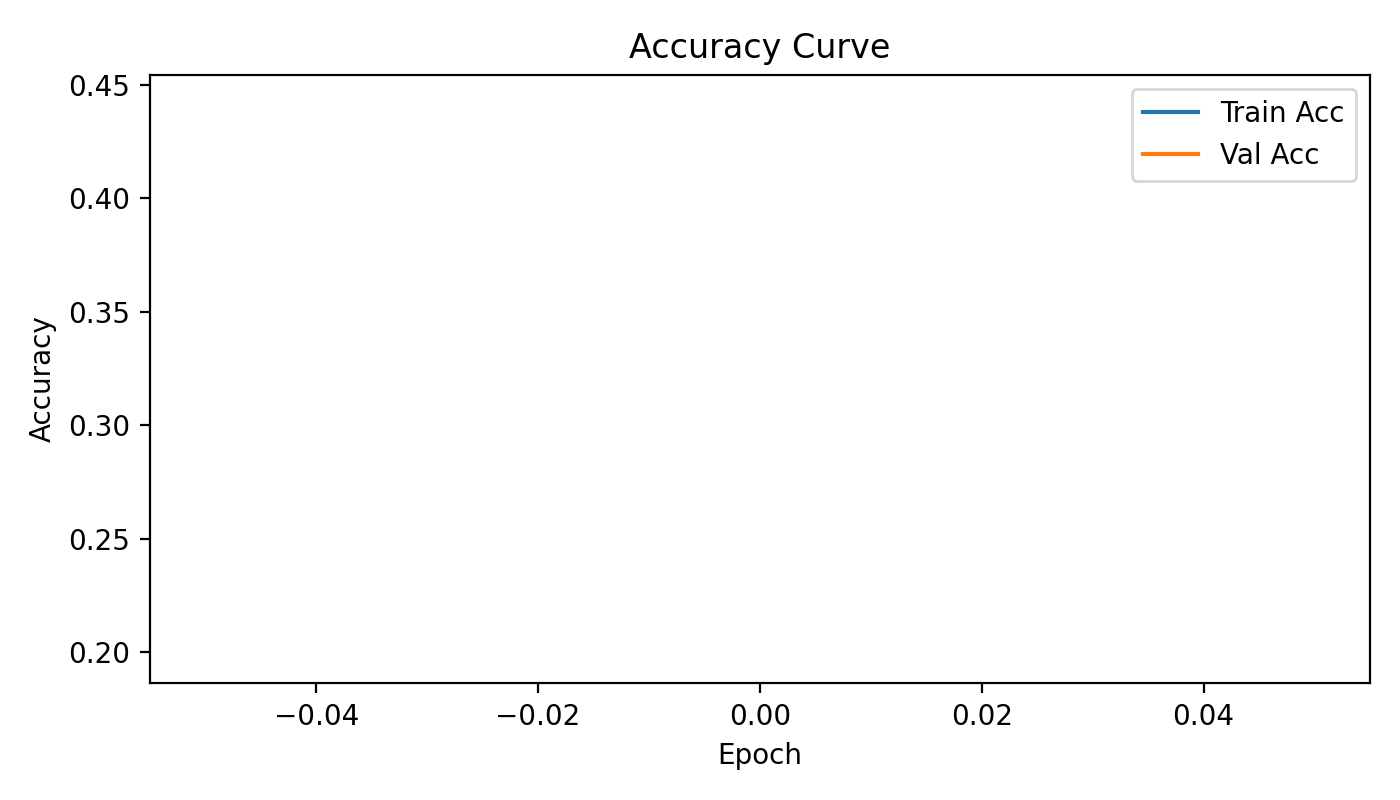}{Figure S19: accuracy curve (\label{fig:supp-19})}\hfill
\suppimage{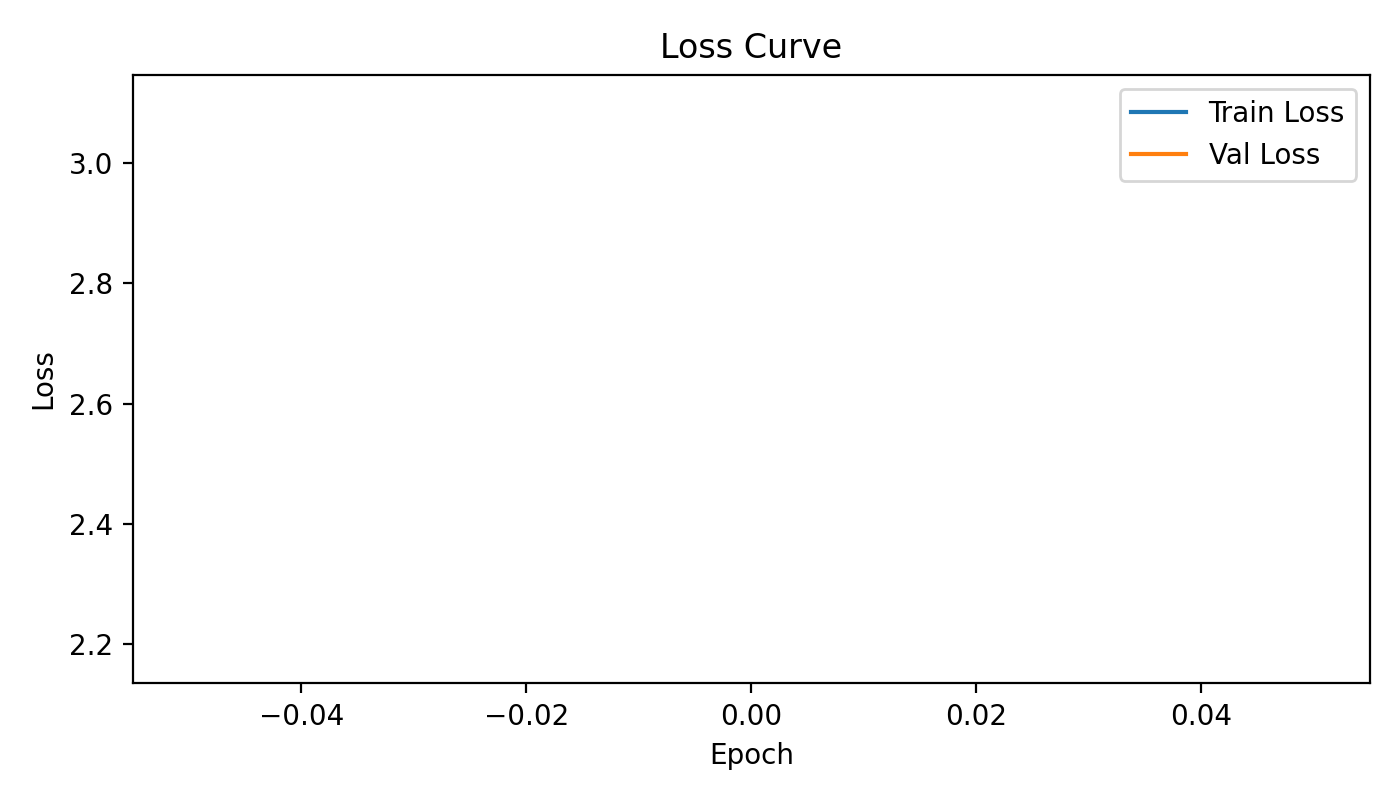}{Figure S20: loss curve (\label{fig:supp-20})}\
\suppimage{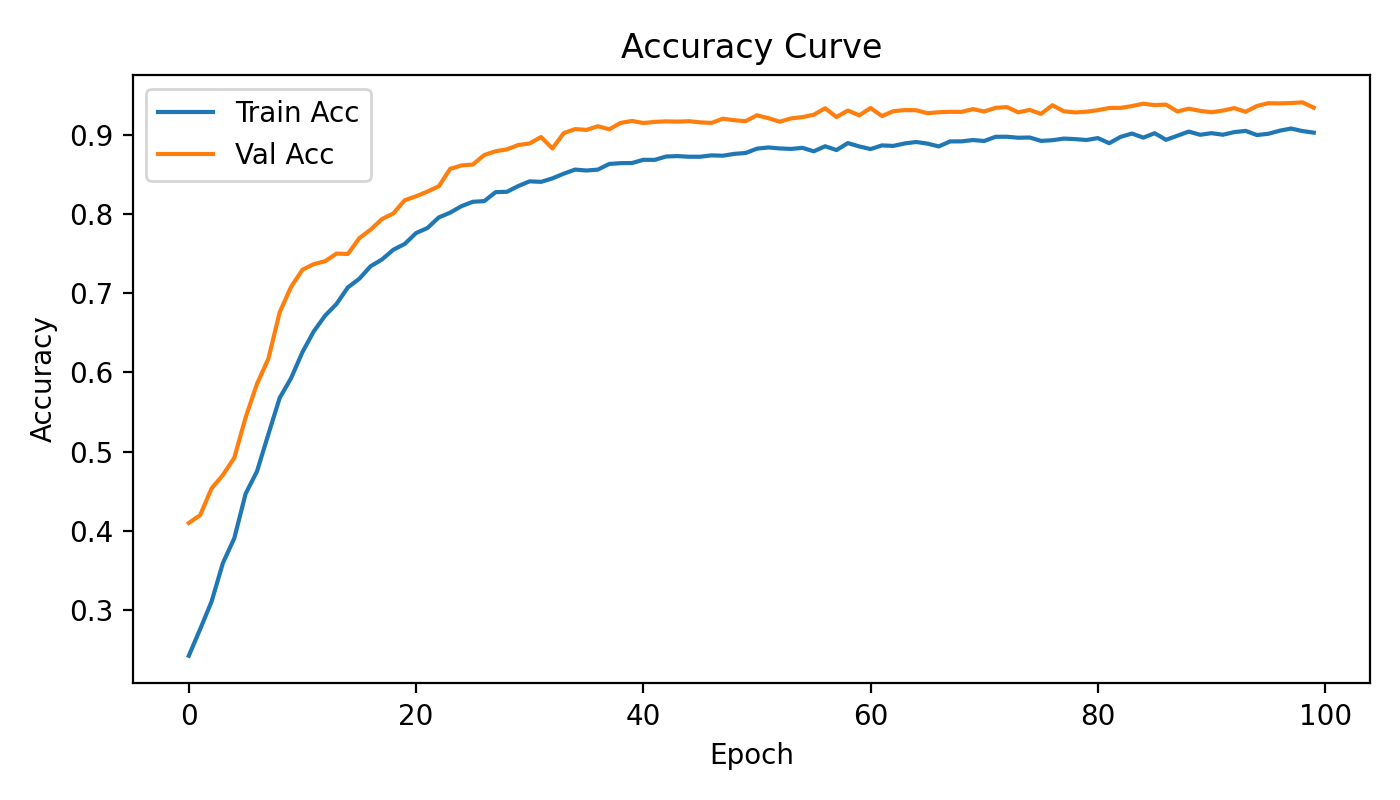}{Figure S21: accuracy curve (\label{fig:supp-21})}\hfill
\suppimage{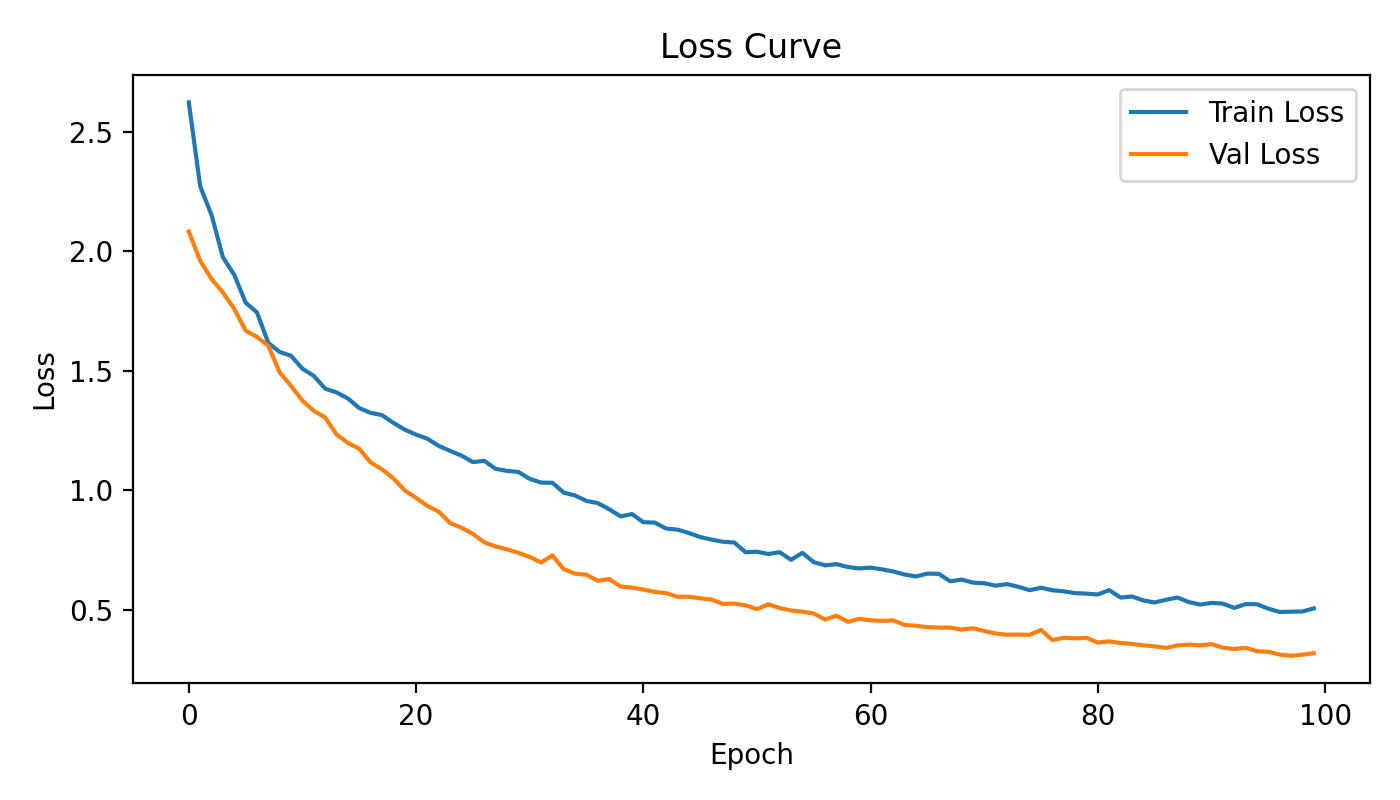}{Figure S22: loss curve (\label{fig:supp-22})}\
\suppimage{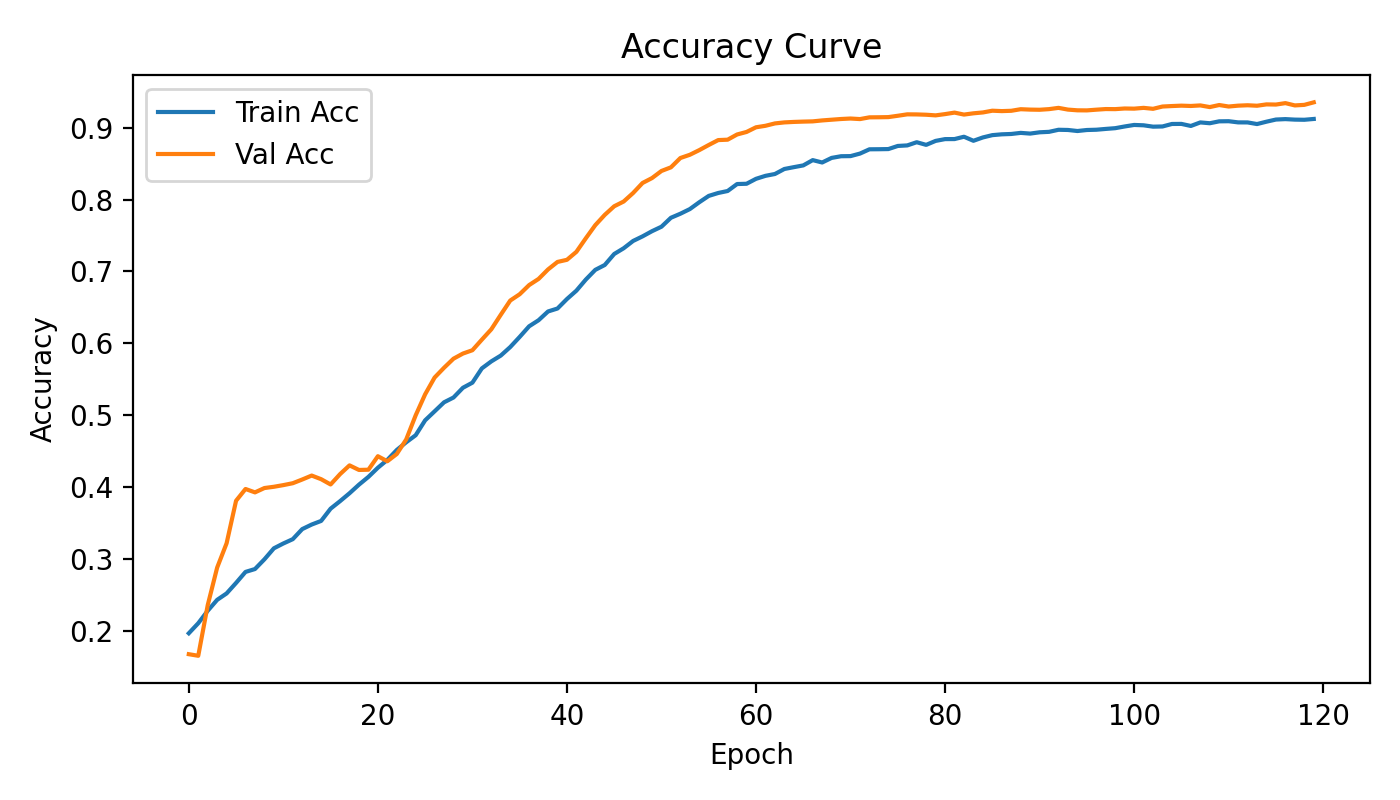}{Figure S23: accuracy curve (\label{fig:supp-23})}\hfill
\suppimage{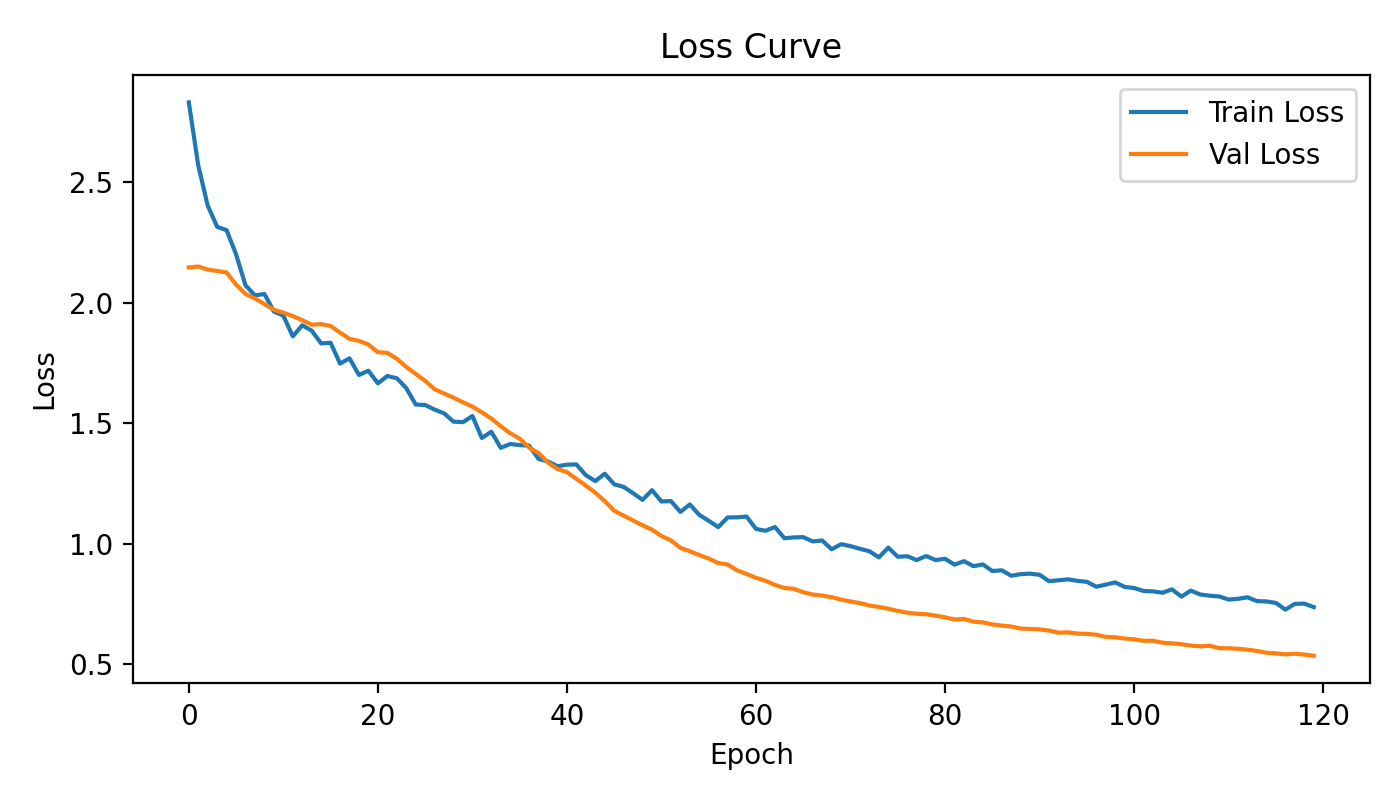}{Figure S24: loss curve (\label{fig:supp-24})}\
\suppimage{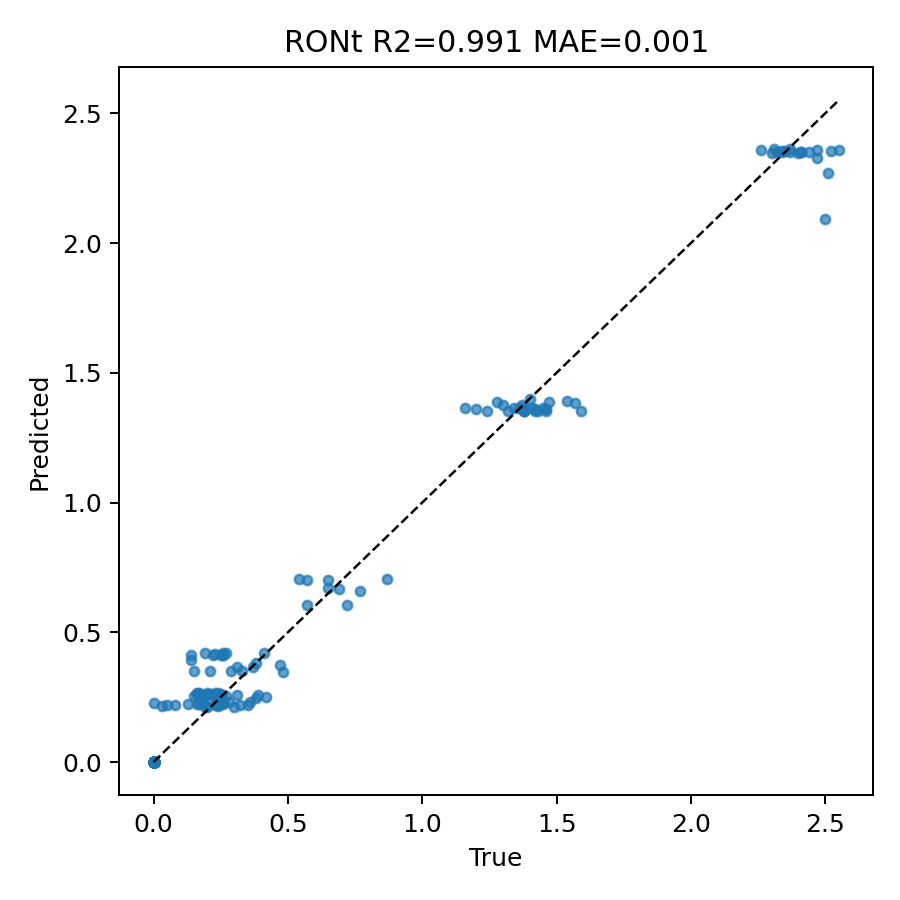}{Figure S25: pred vs true (\label{fig:supp-25})}\hfill
\suppimage{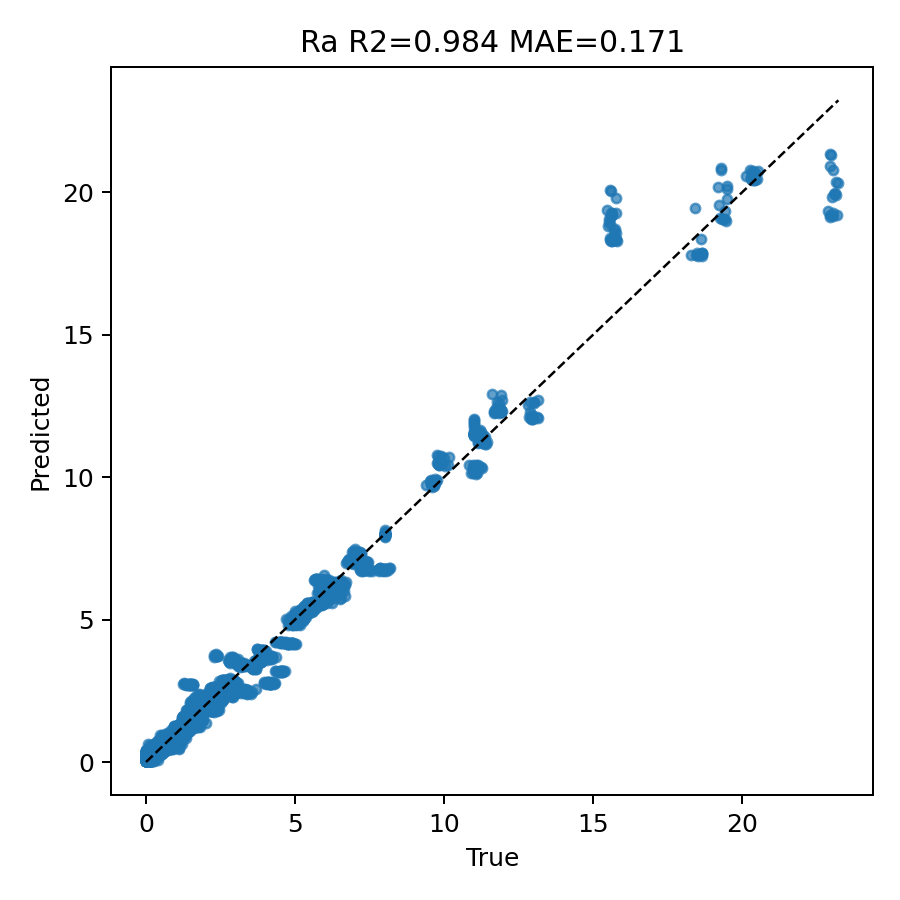}{Figure S26: pred vs true (\label{fig:supp-26})}\
\suppimage{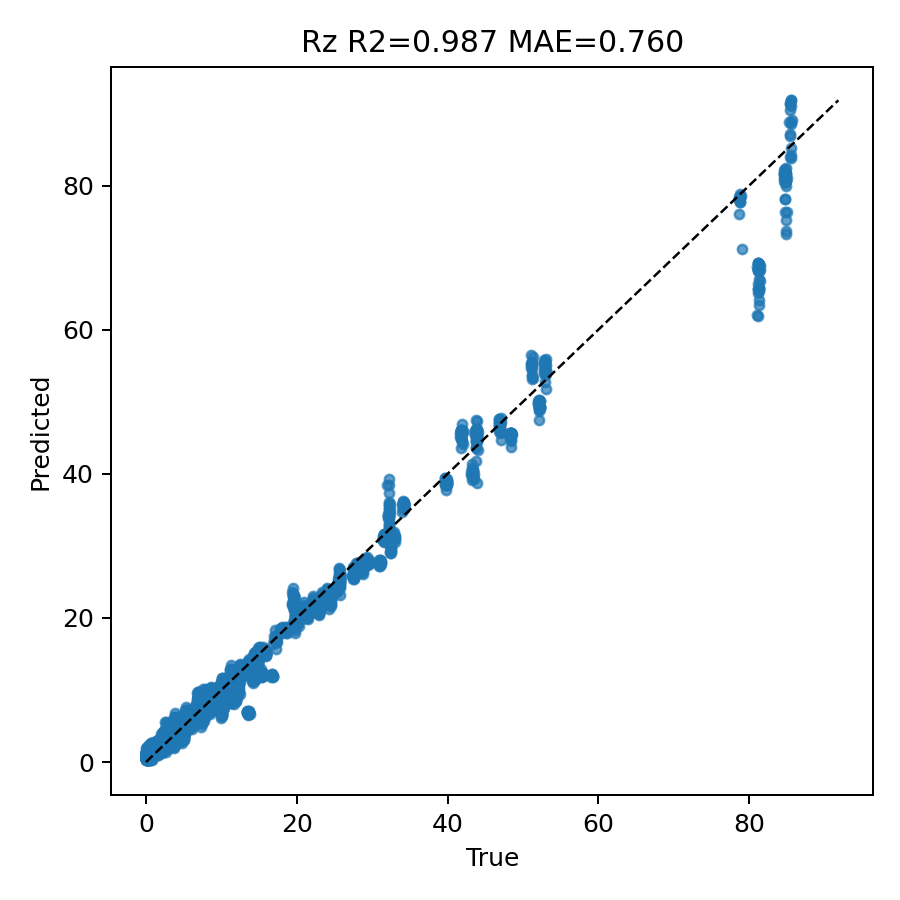}{Figure S27: pred vs true (\label{fig:supp-27})}\hfill
\suppimage{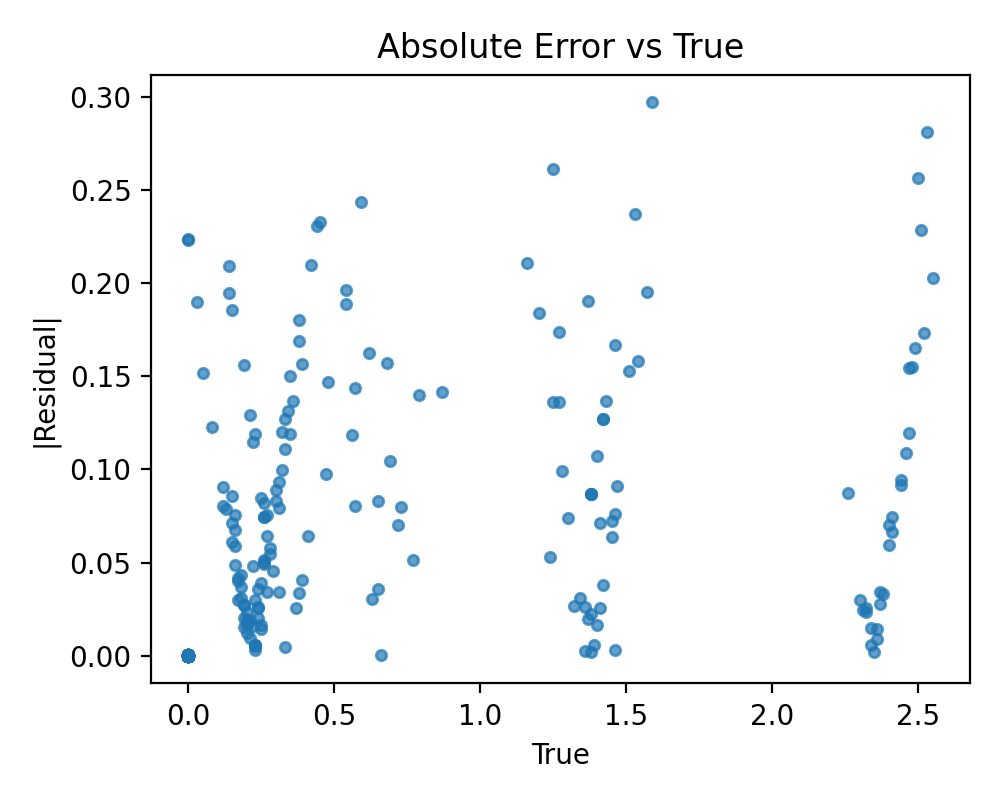}{Figure S28: abs error vs true (\label{fig:supp-28})}\
\suppimage{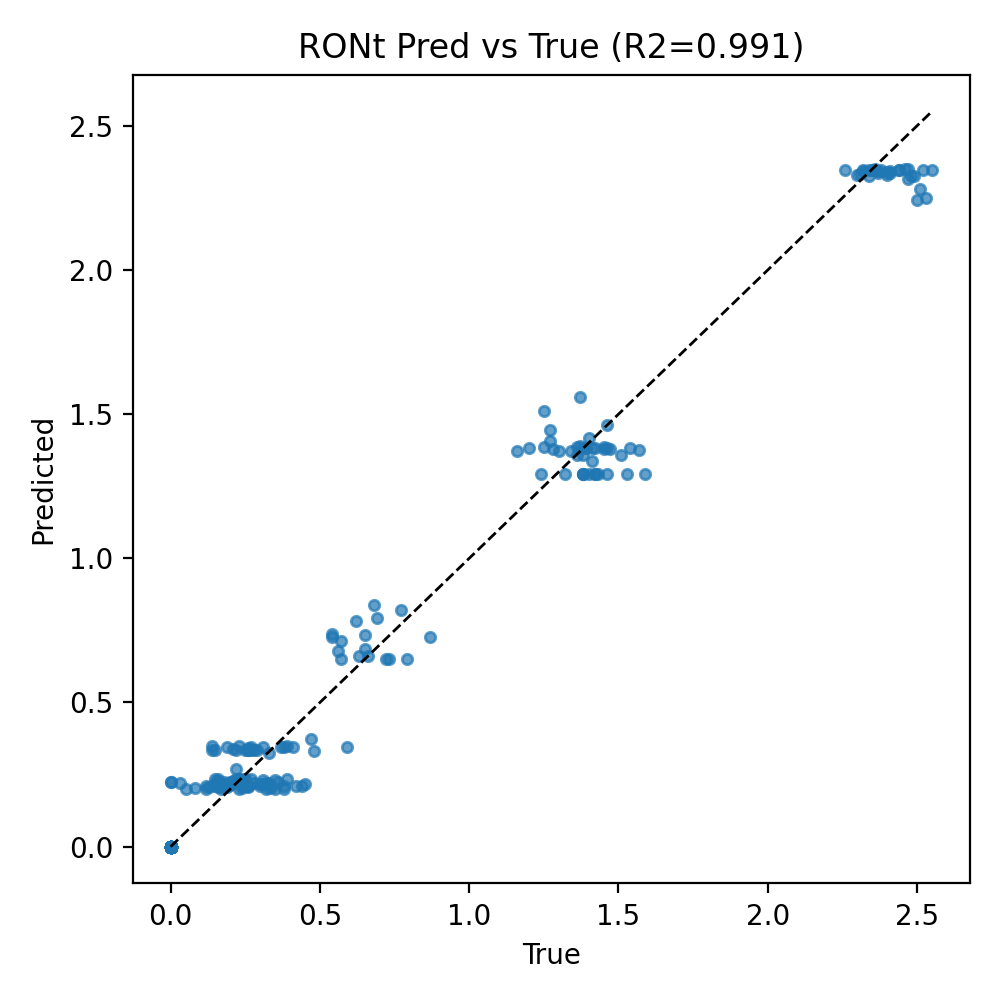}{Figure S29: pred vs true (\label{fig:supp-29})}\hfill
\suppimage{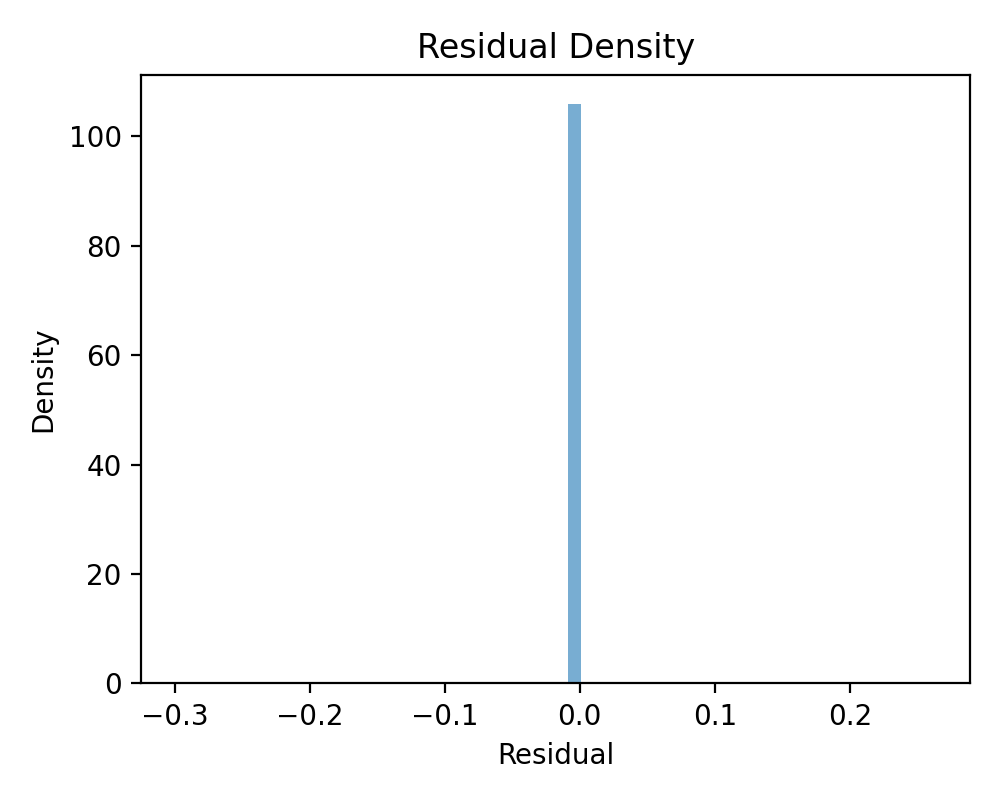}{Figure S30: residual density (\label{fig:supp-30})}\
\suppimage{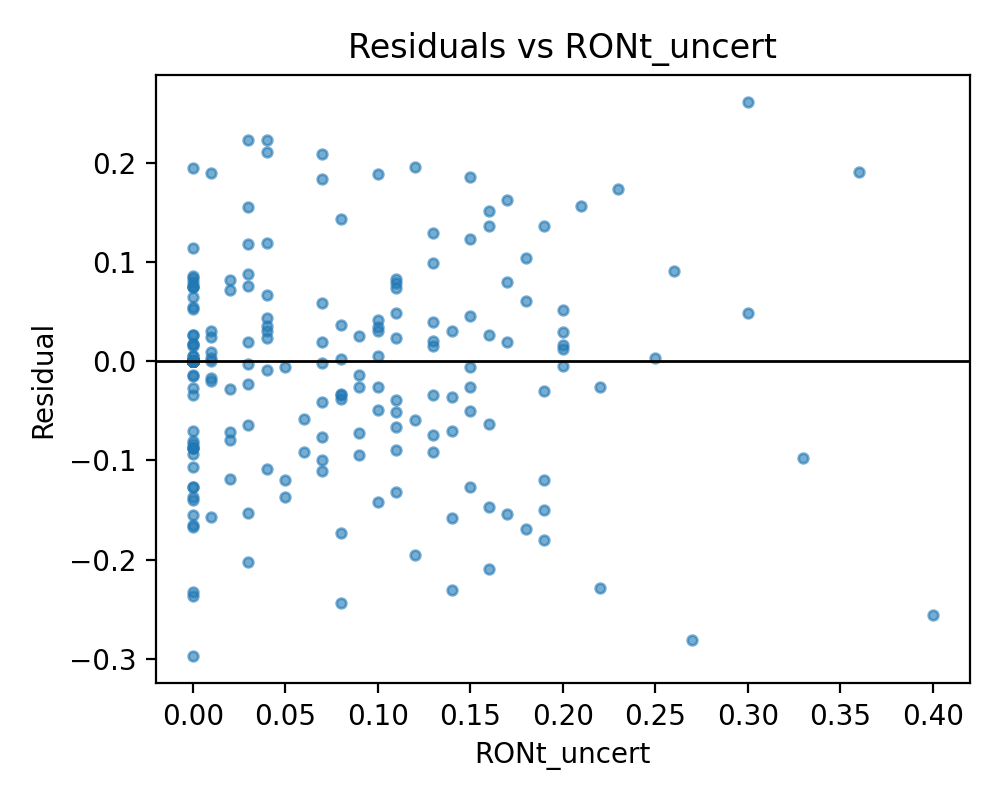}{Figure S31: residuals vs top feature RONt uncert (\label{fig:supp-31})}\hfill
\suppimage{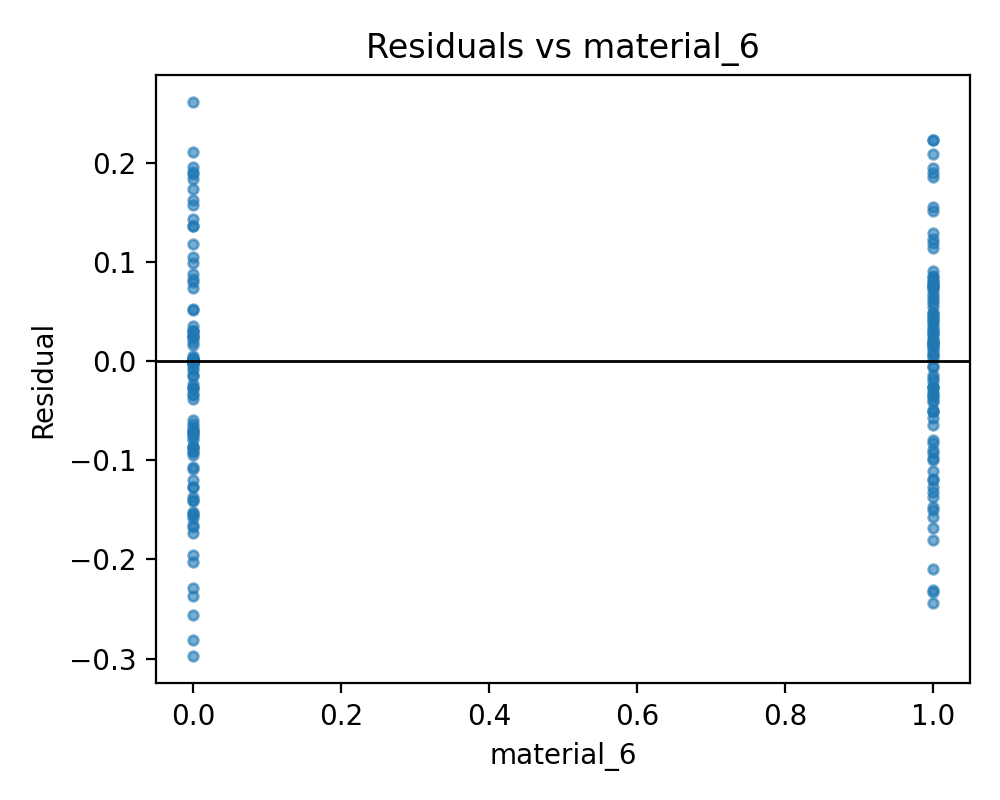}{Figure S32: residuals vs top feature material 6 (\label{fig:supp-32})}\
\suppimage{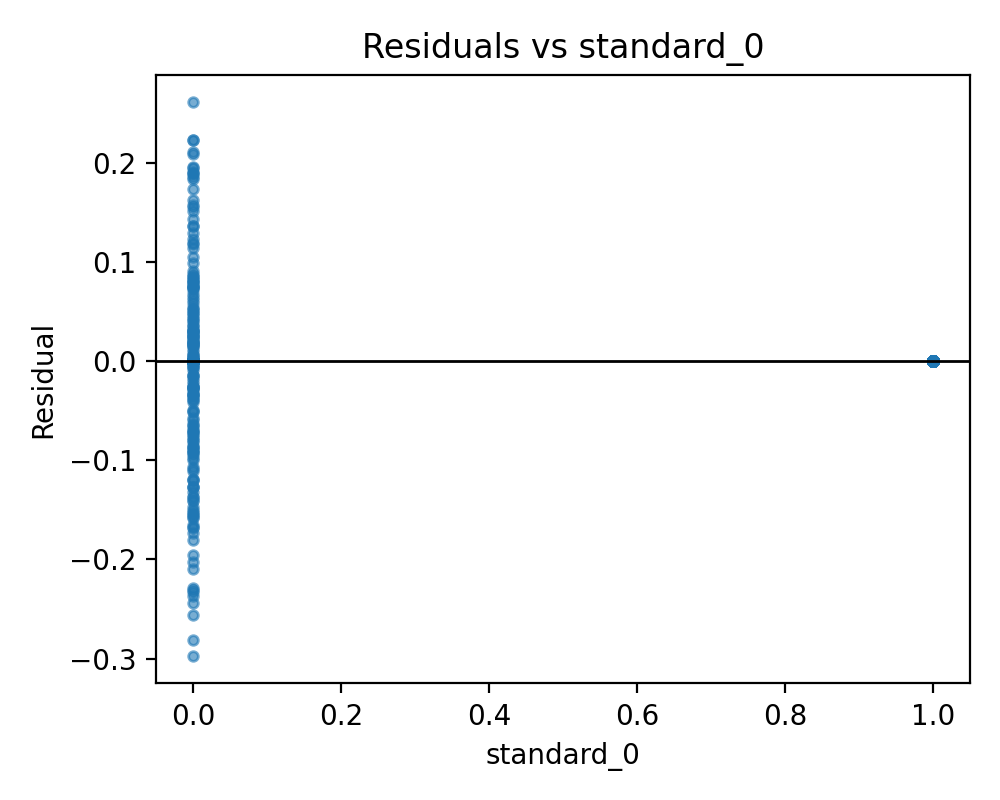}{Figure S33: residuals vs top feature standard 0 (\label{fig:supp-33})}\hfill
\suppimage{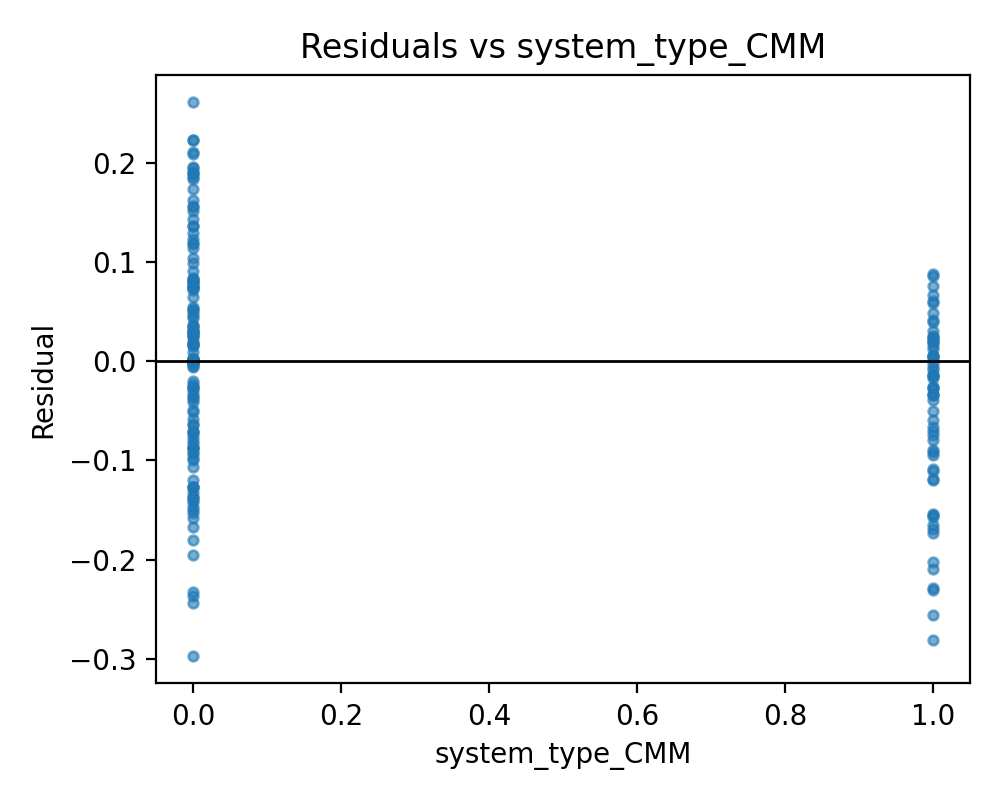}{Figure S34: residuals vs top feature system type CMM (\label{fig:supp-34})}\
\suppimage{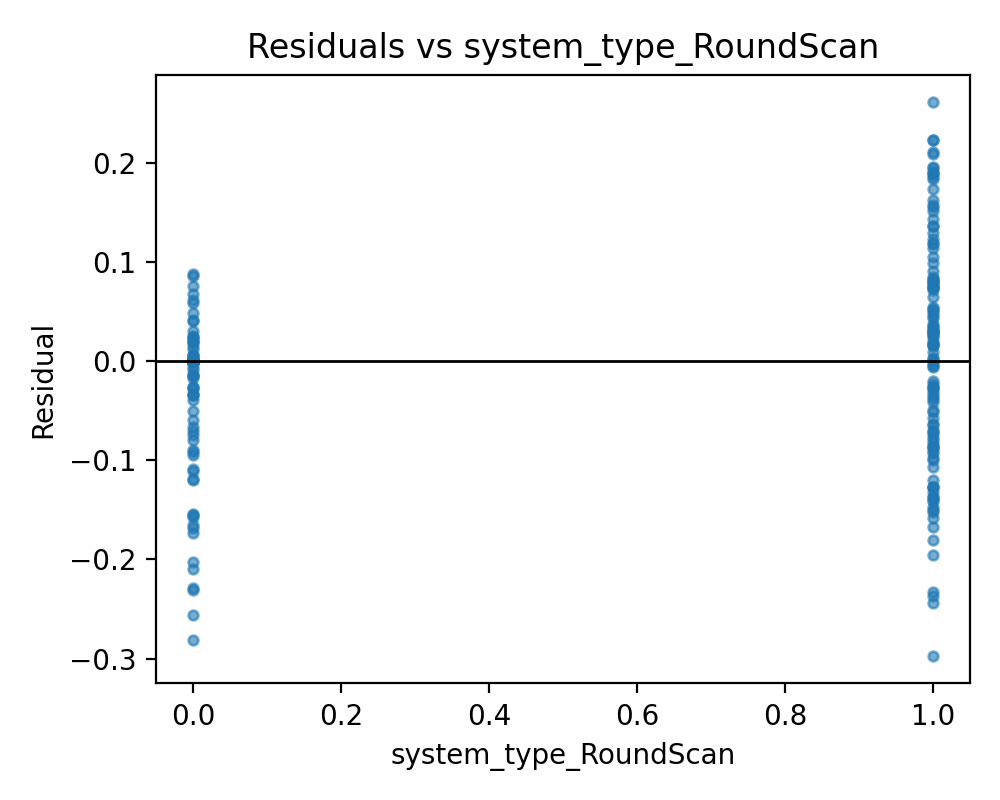}{Figure S35: residuals vs top feature system type RoundScan (\label{fig:supp-35})}\hfill
\suppimage{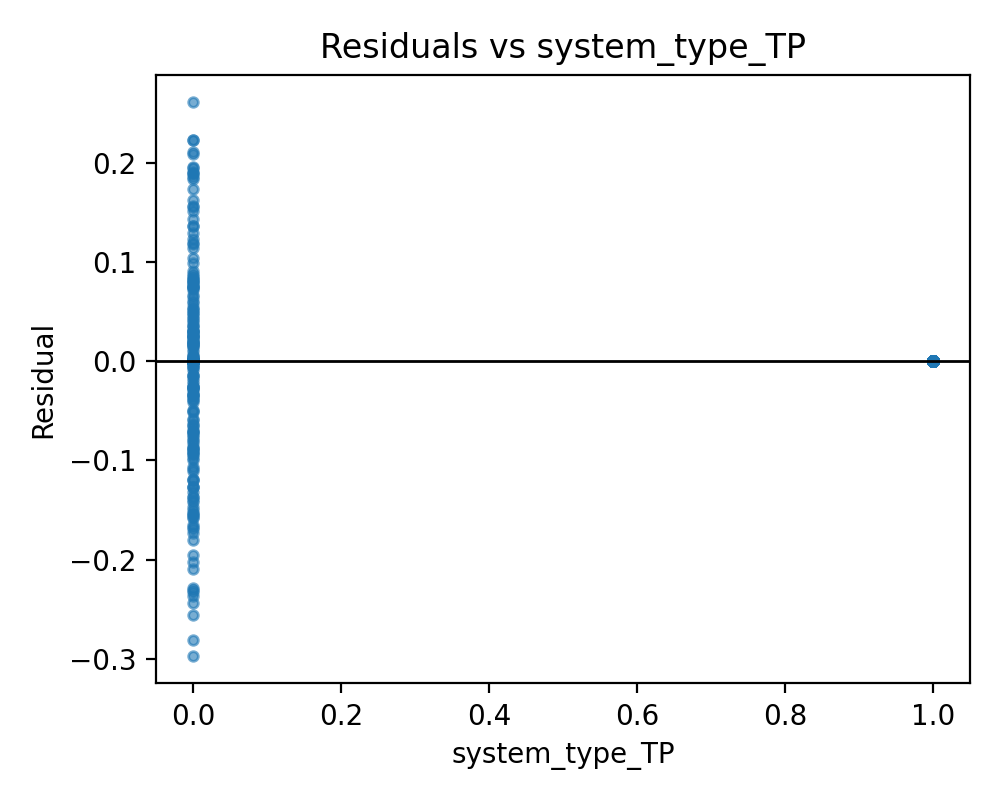}{Figure S36: residuals vs top feature system type TP (\label{fig:supp-36})}\
\suppimage{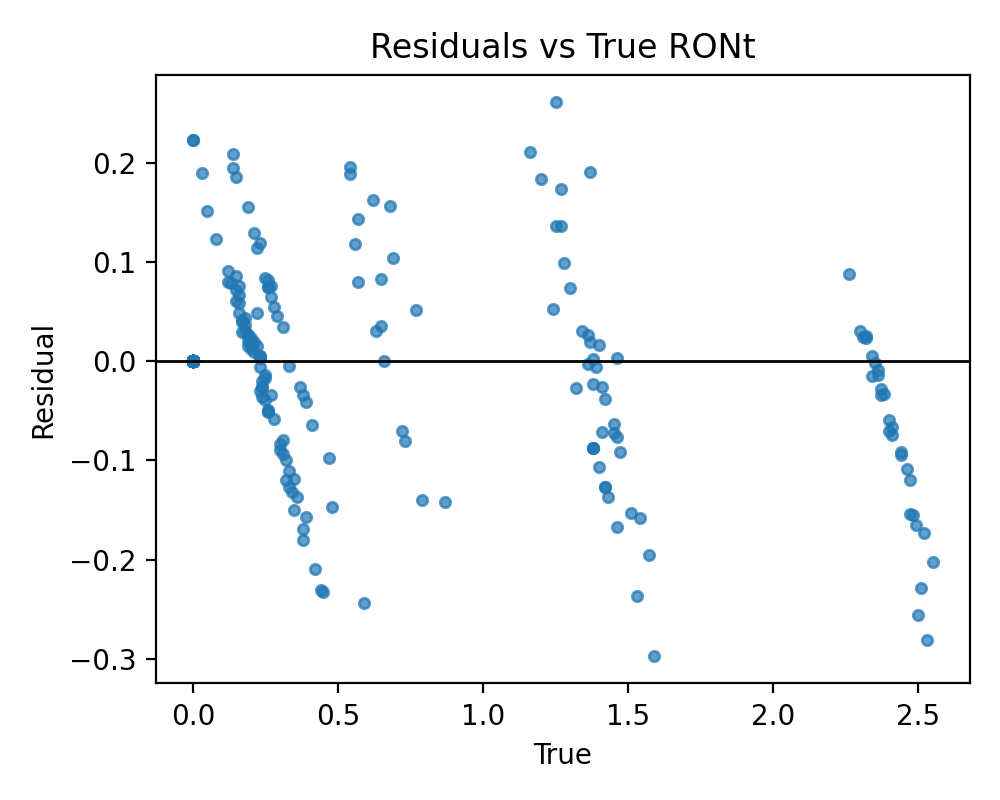}{Figure S37: residuals vs true (\label{fig:supp-37})}\hfill
\suppimage{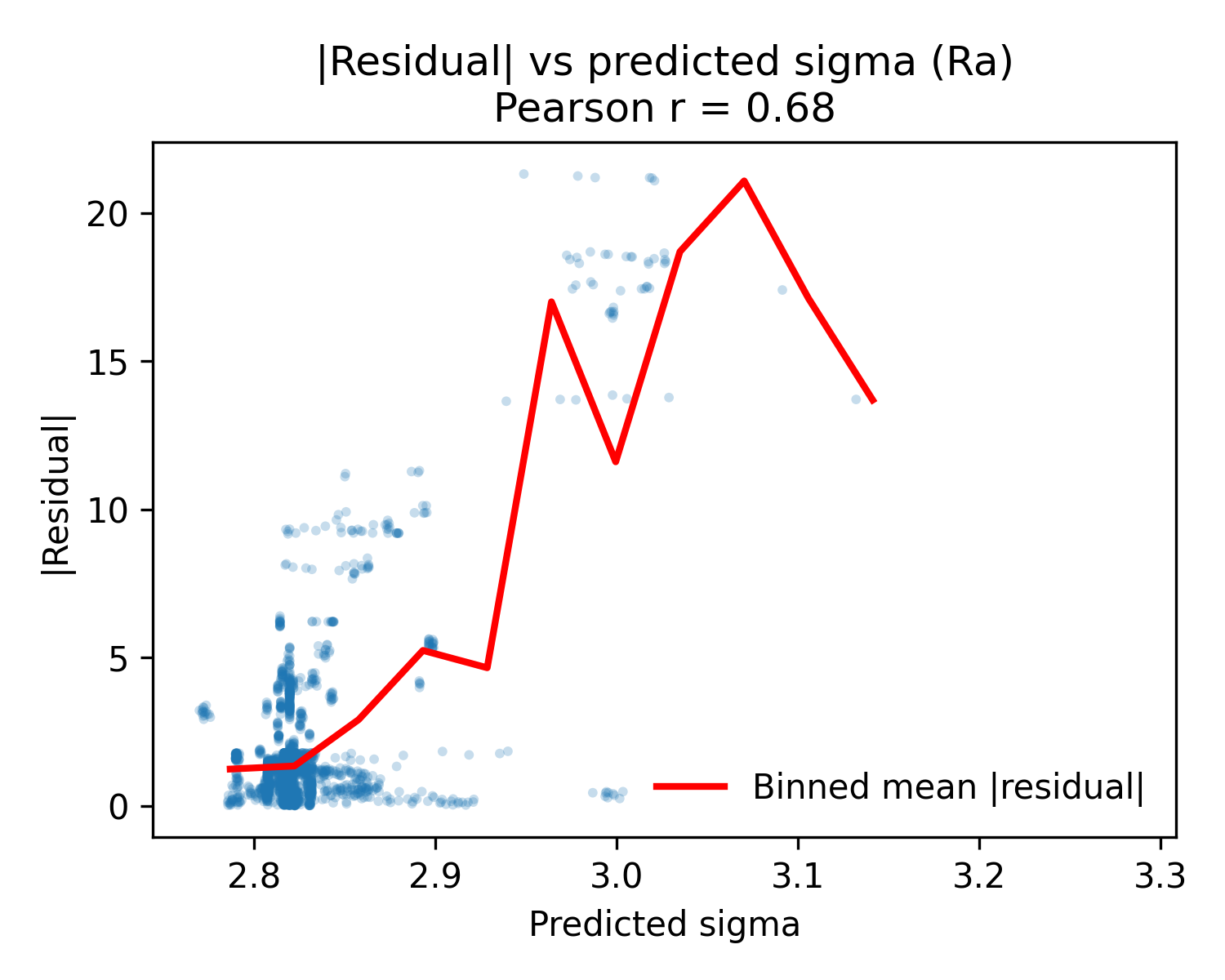}{Figure S38: abs residual vs sigma Ra (\label{fig:supp-38})}\
\suppimage{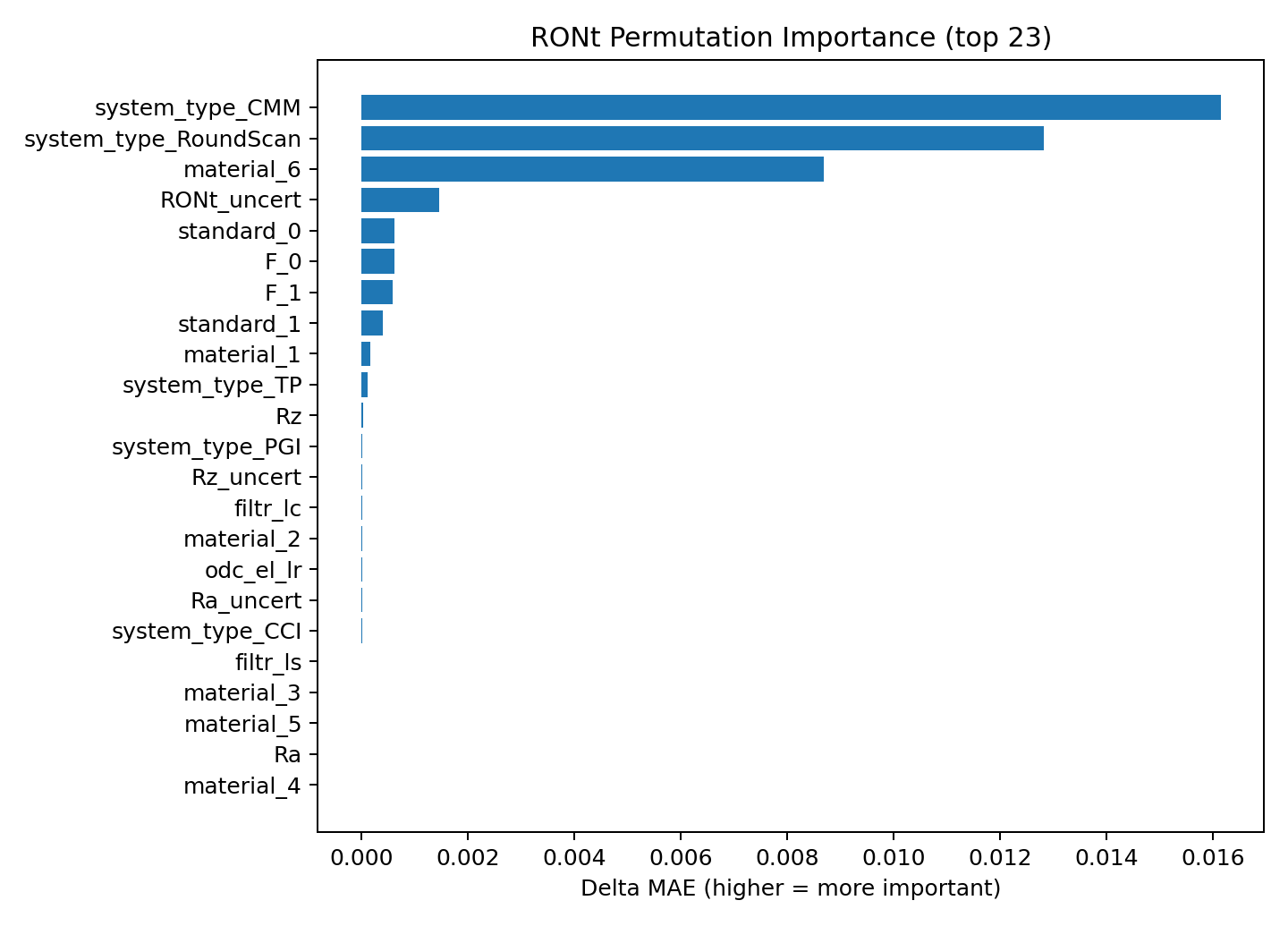}{Figure S39: permutation importance (\label{fig:supp-39})}\hfill
\suppimage{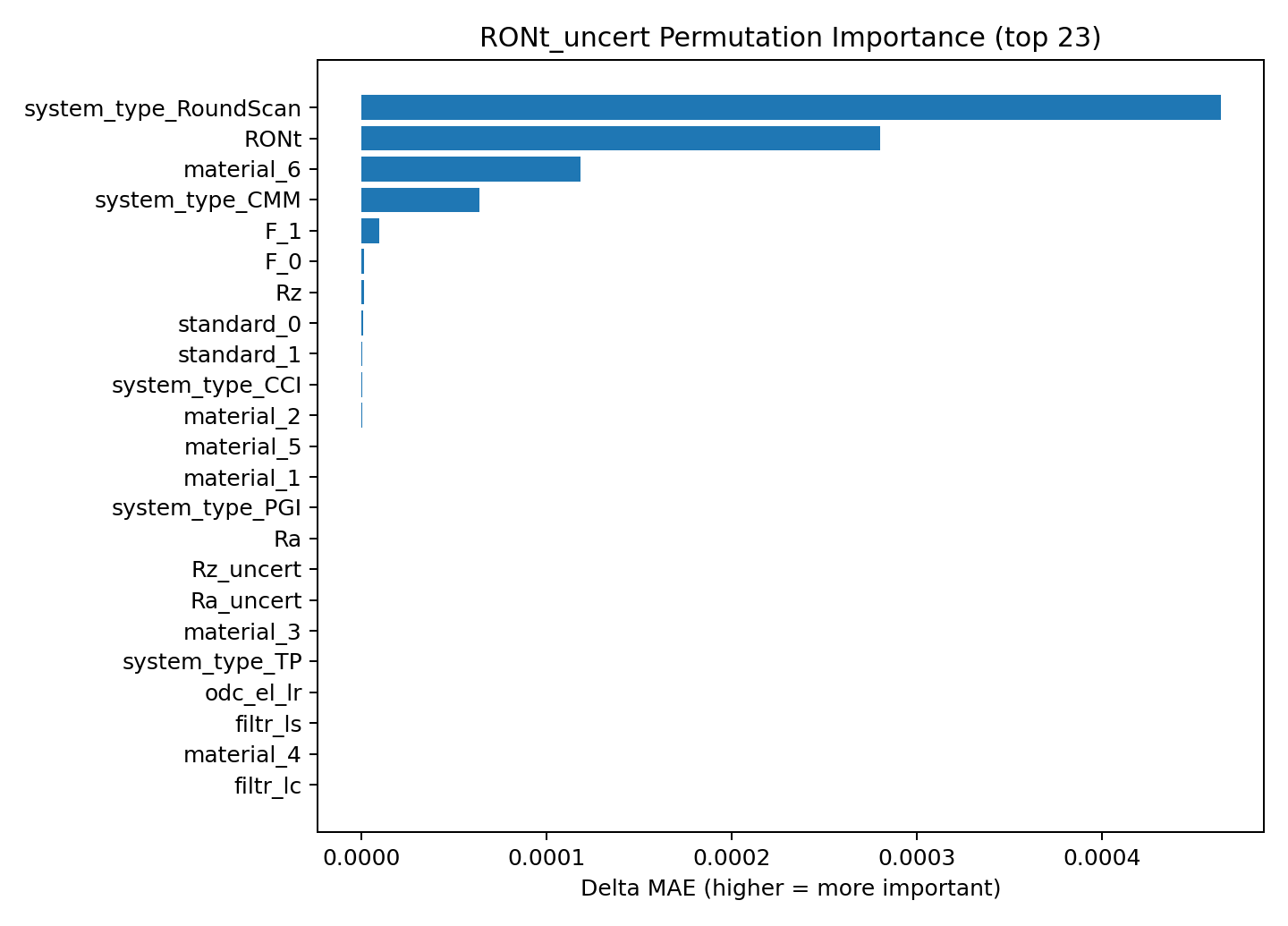}{Figure S40: permutation importance (\label{fig:supp-40})}\
\suppimage{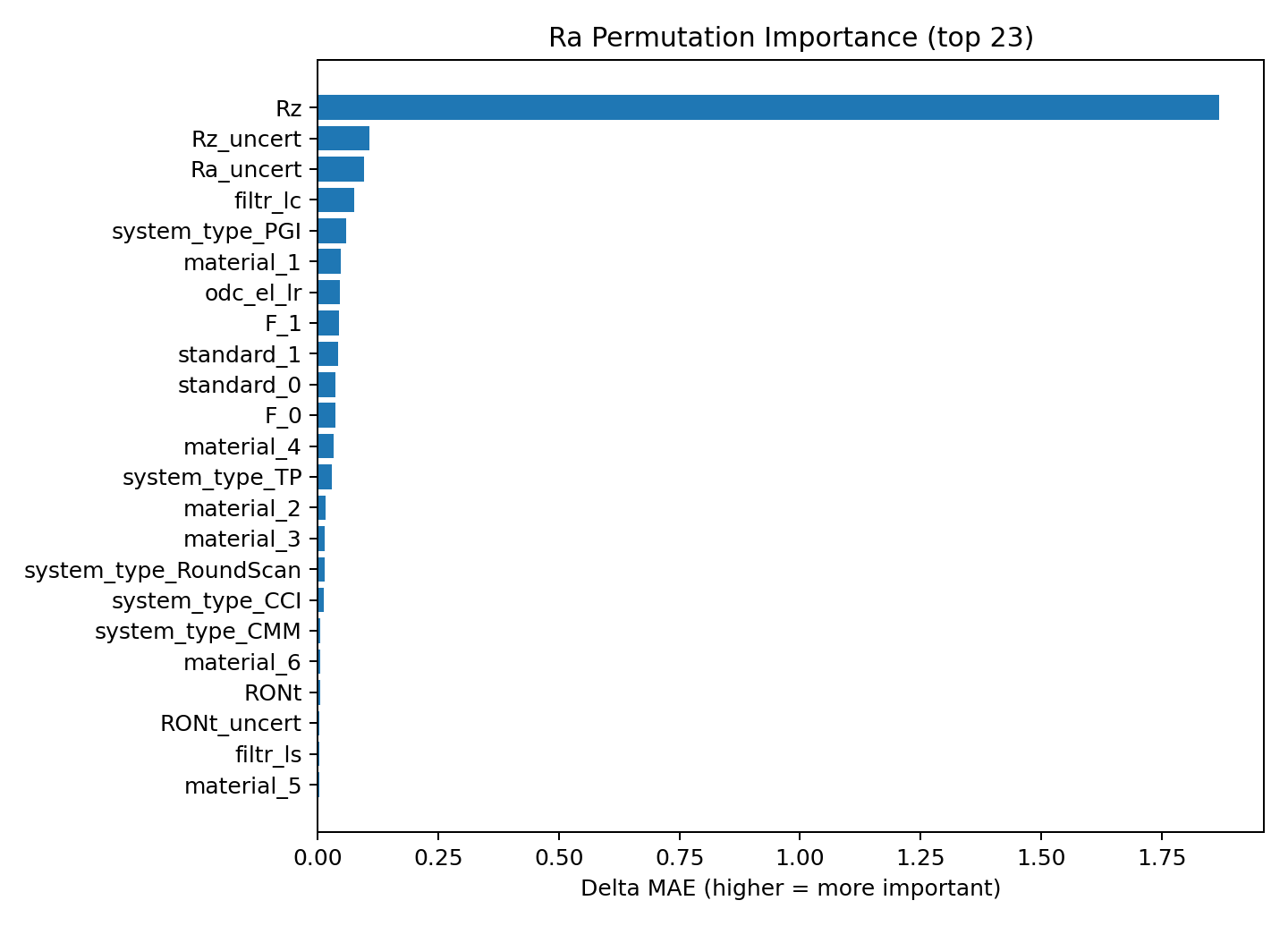}{Figure S41: permutation importance (\label{fig:supp-41})}\hfill
\suppimage{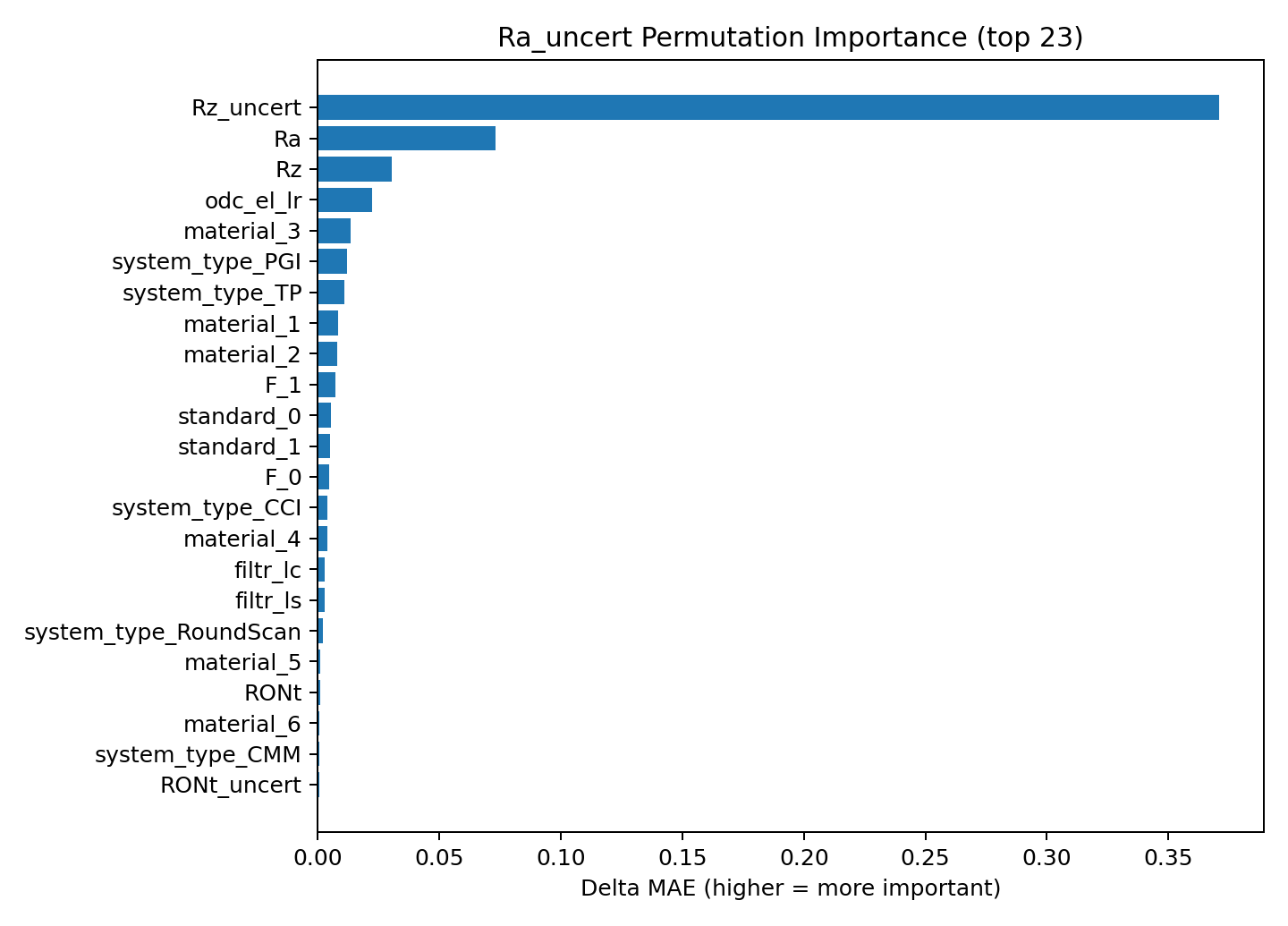}{Figure S42: permutation importance (\label{fig:supp-42})}\
\suppimage{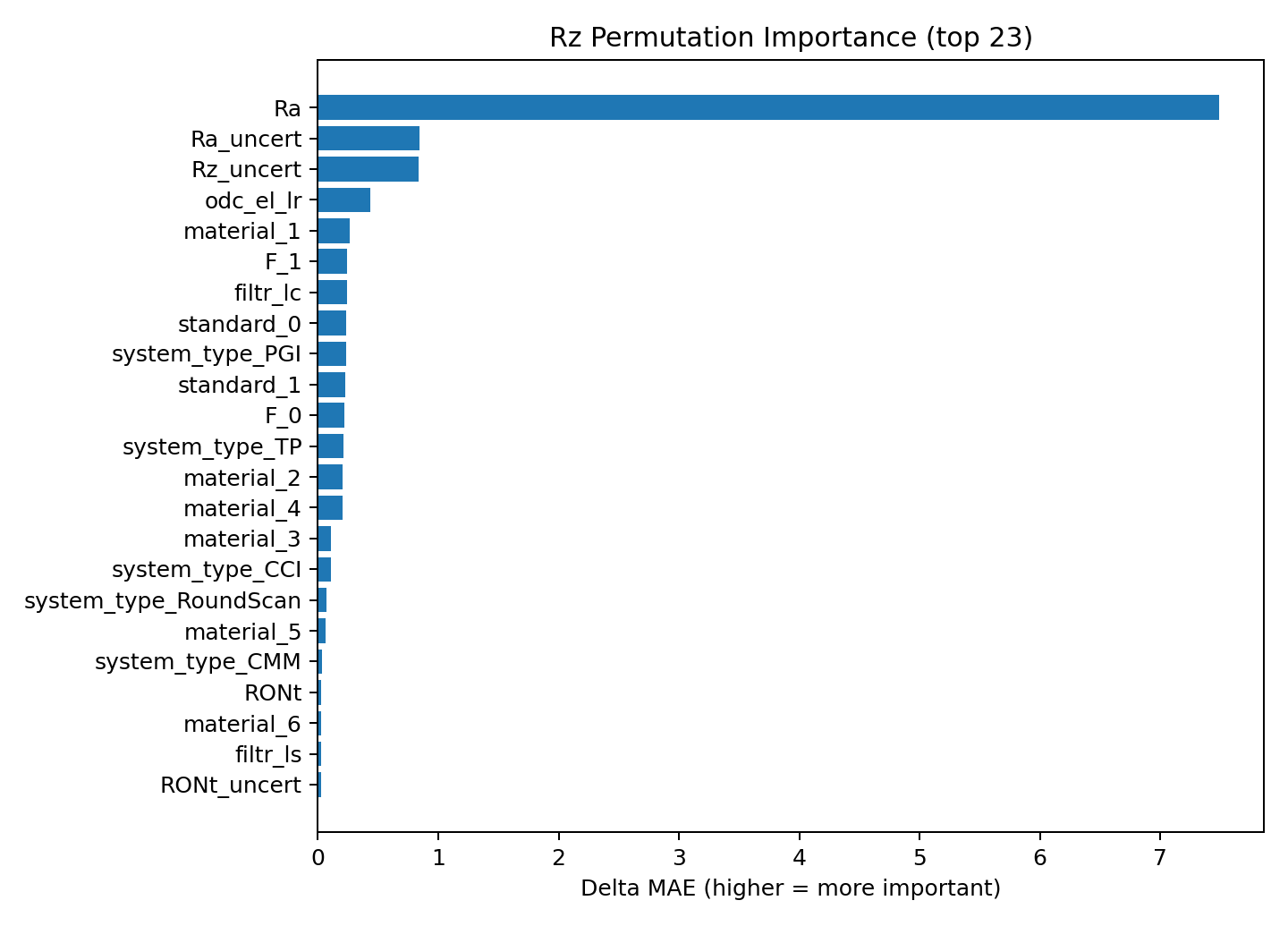}{Figure S43: permutation importance (\label{fig:supp-43})}\hfill
\suppimage{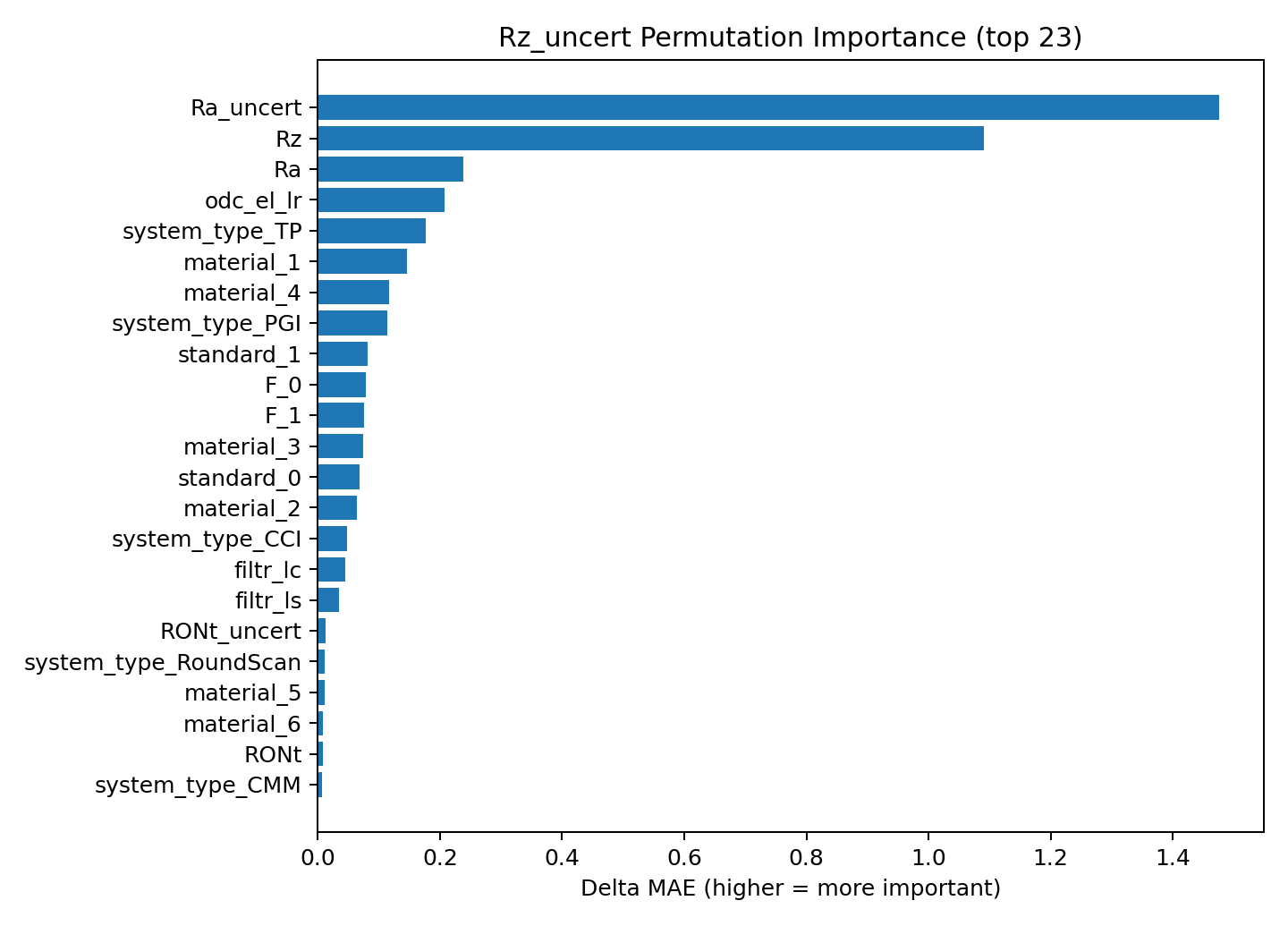}{Figure S44: permutation importance (\label{fig:supp-44})}\
\suppimage{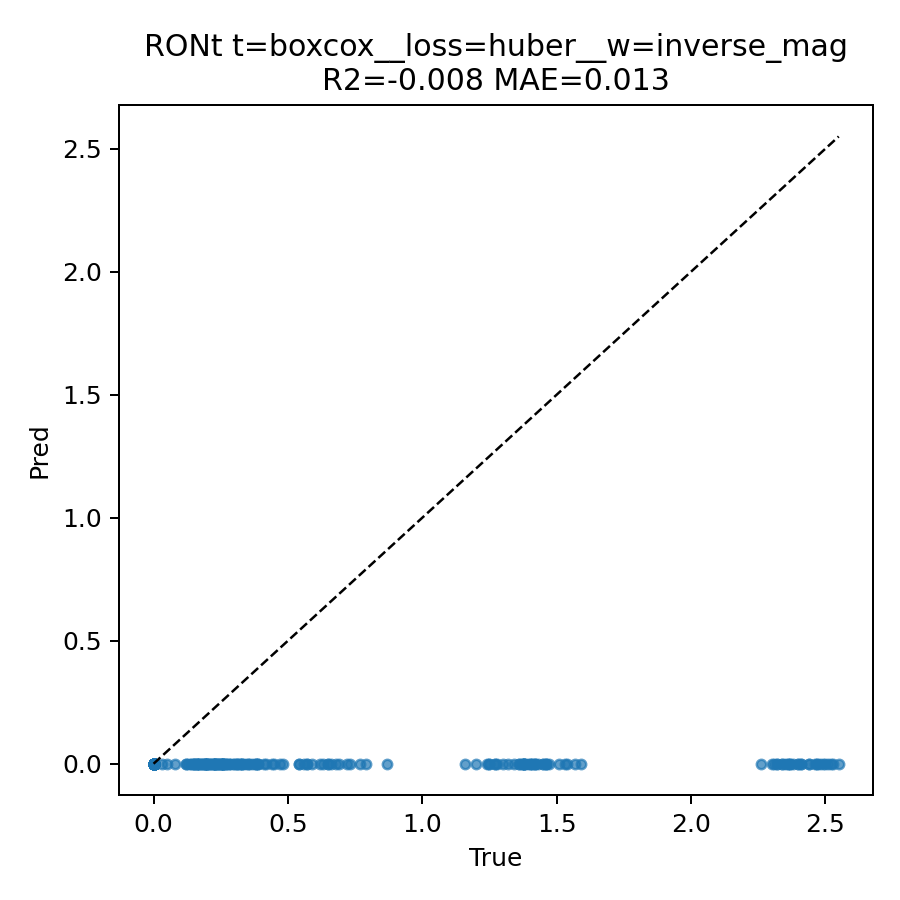}{Figure S45: pred vs true (\label{fig:supp-45})}\hfill
\suppimage{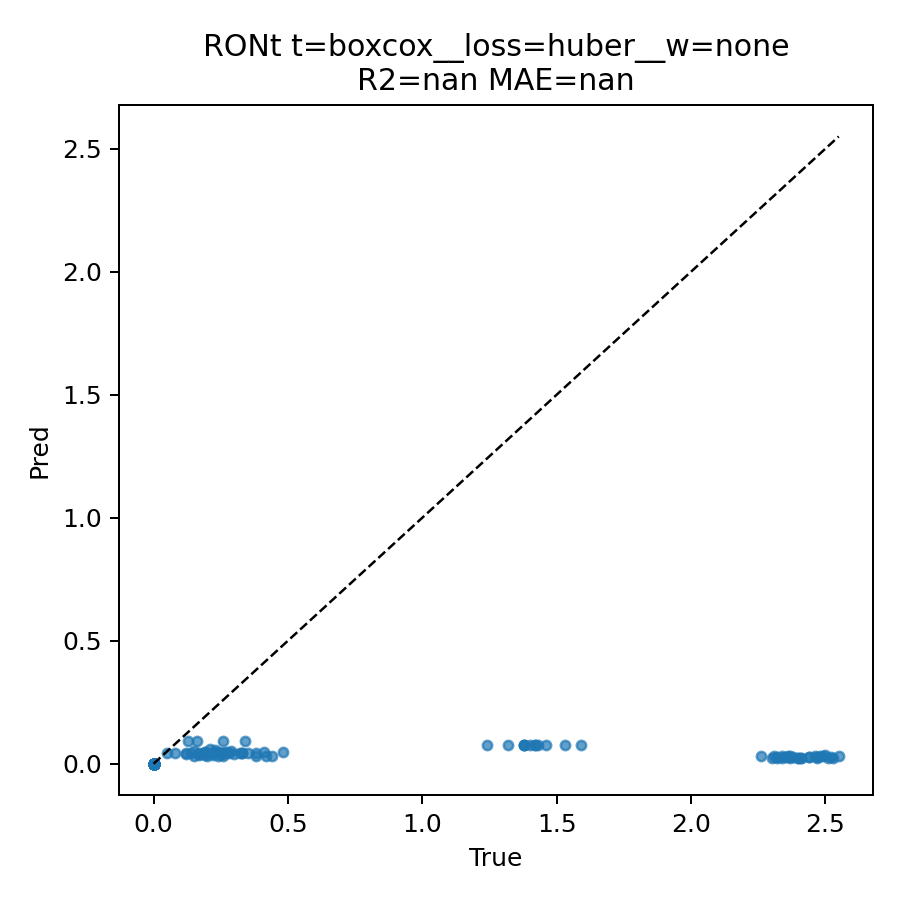}{Figure S46: pred vs true (\label{fig:supp-46})}\
\suppimage{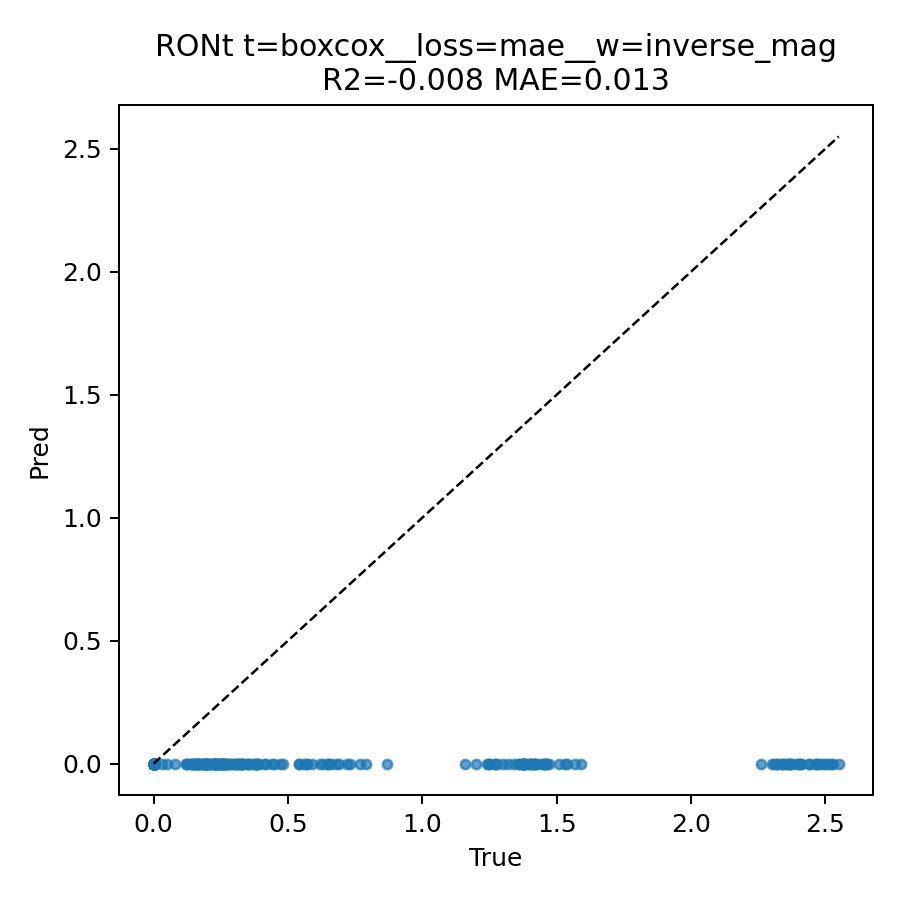}{Figure S47: pred vs true (\label{fig:supp-47})}\hfill
\suppimage{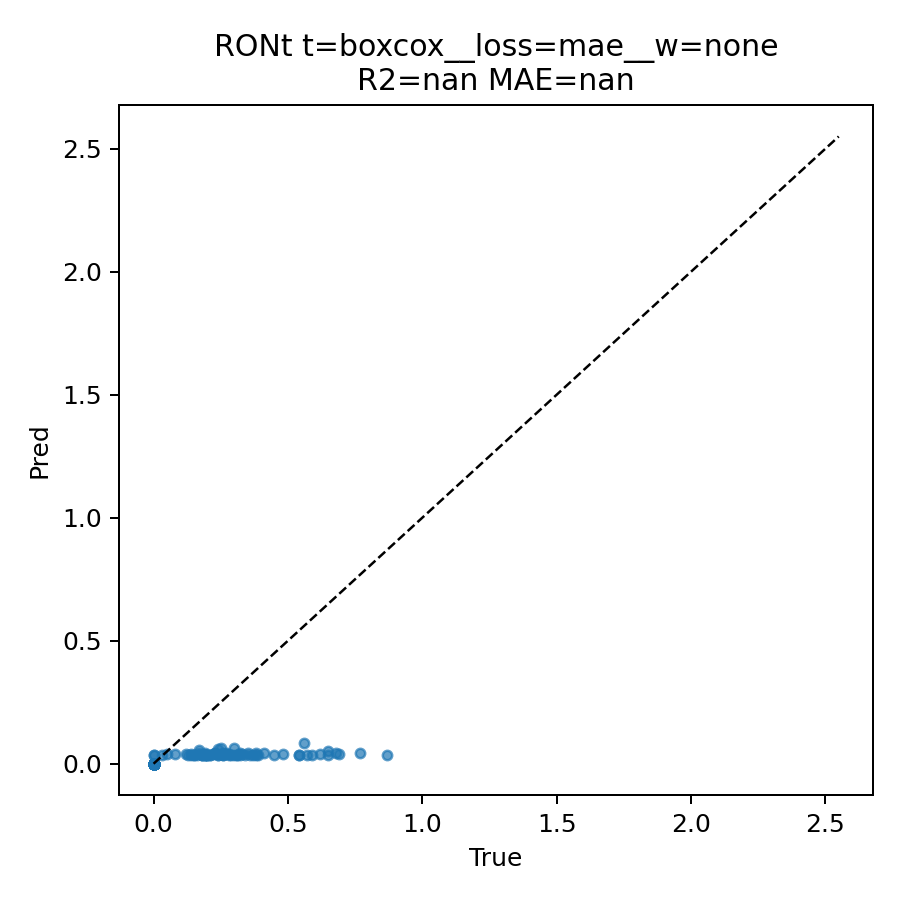}{Figure S48: pred vs true (\label{fig:supp-48})}\
\suppimage{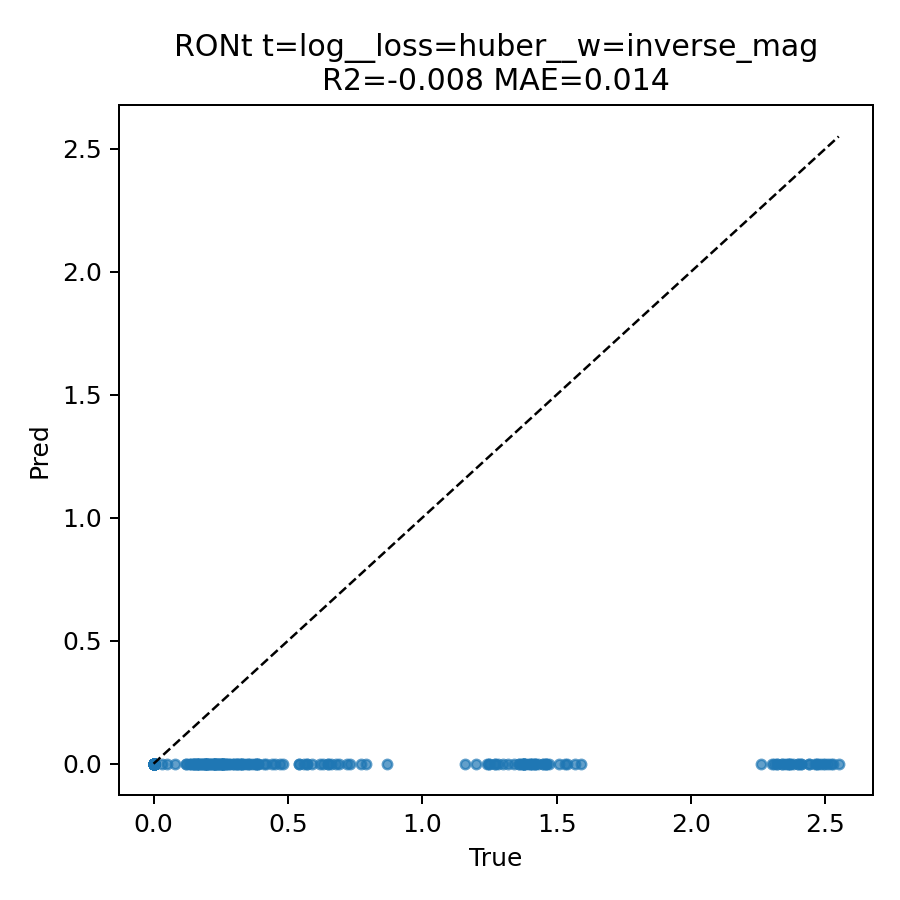}{Figure S49: pred vs true (\label{fig:supp-49})}\hfill
\suppimage{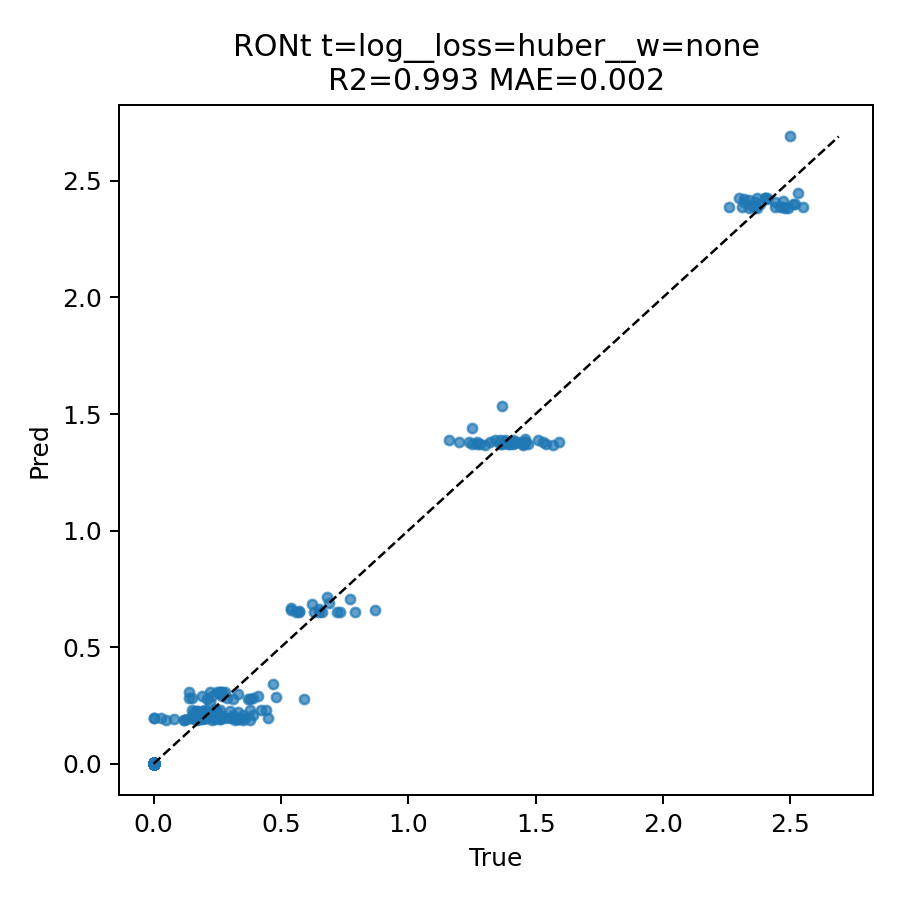}{Figure S50: pred vs true (\label{fig:supp-50})}\
\suppimage{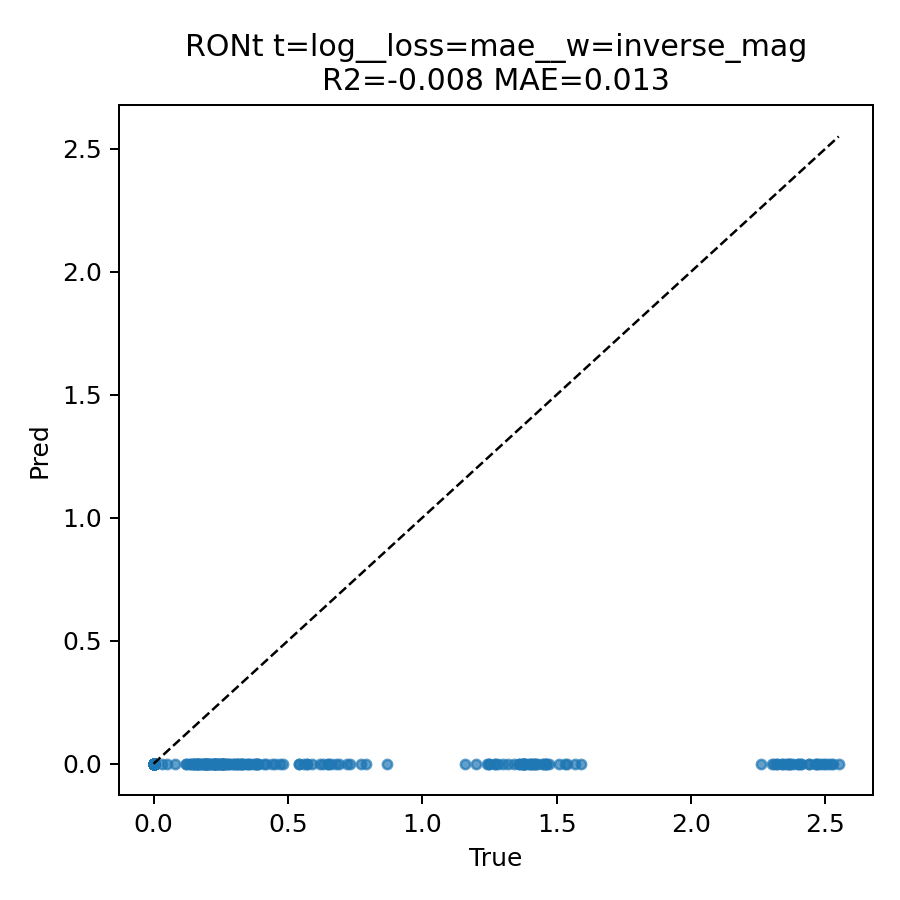}{Figure S51: pred vs true (\label{fig:supp-51})}\hfill
\suppimage{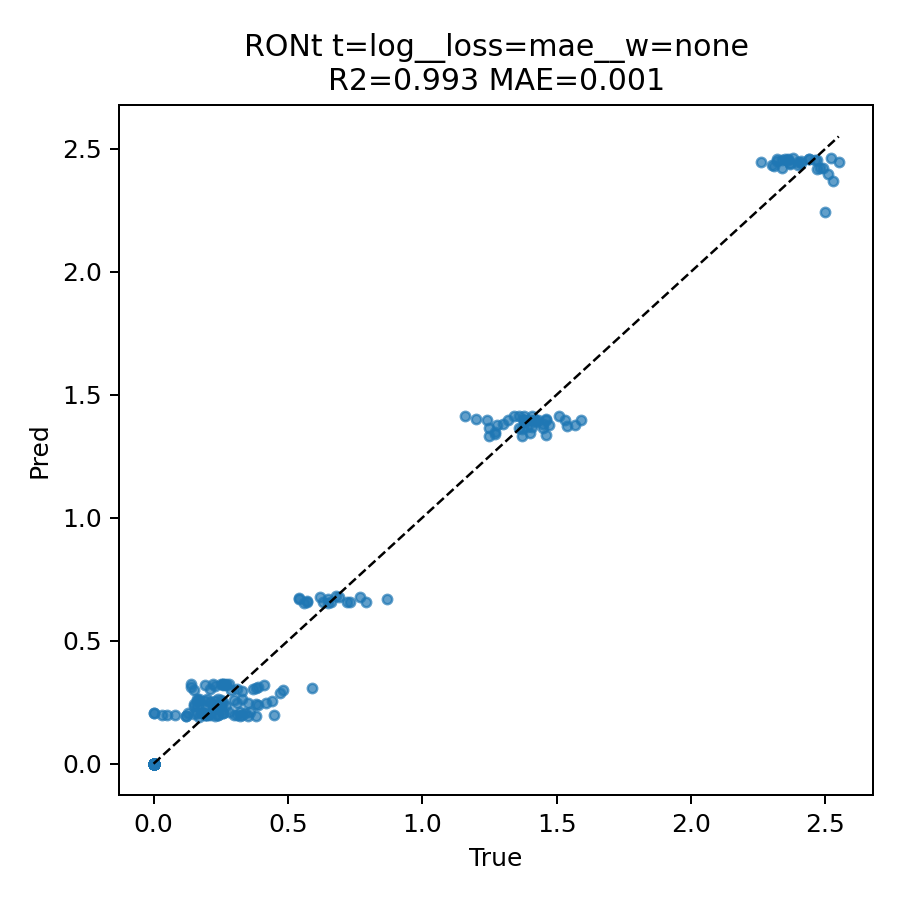}{Figure S52: pred vs true (\label{fig:supp-52})}\
\suppimage{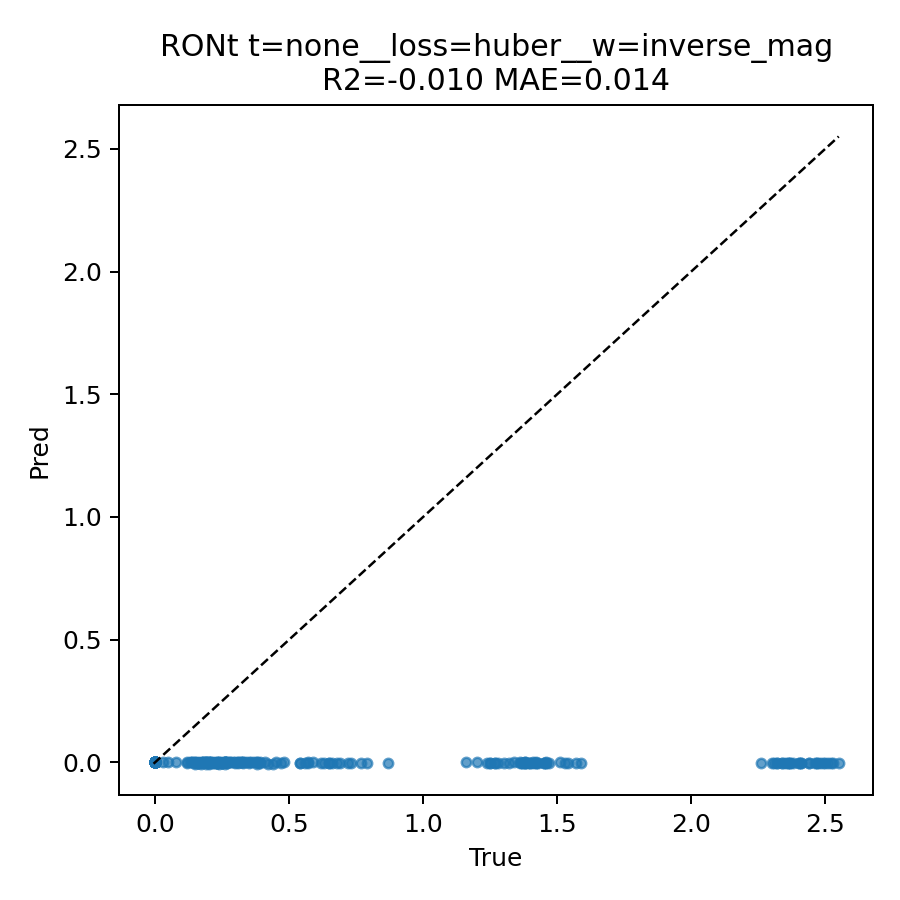}{Figure S53: pred vs true (\label{fig:supp-53})}\hfill
\suppimage{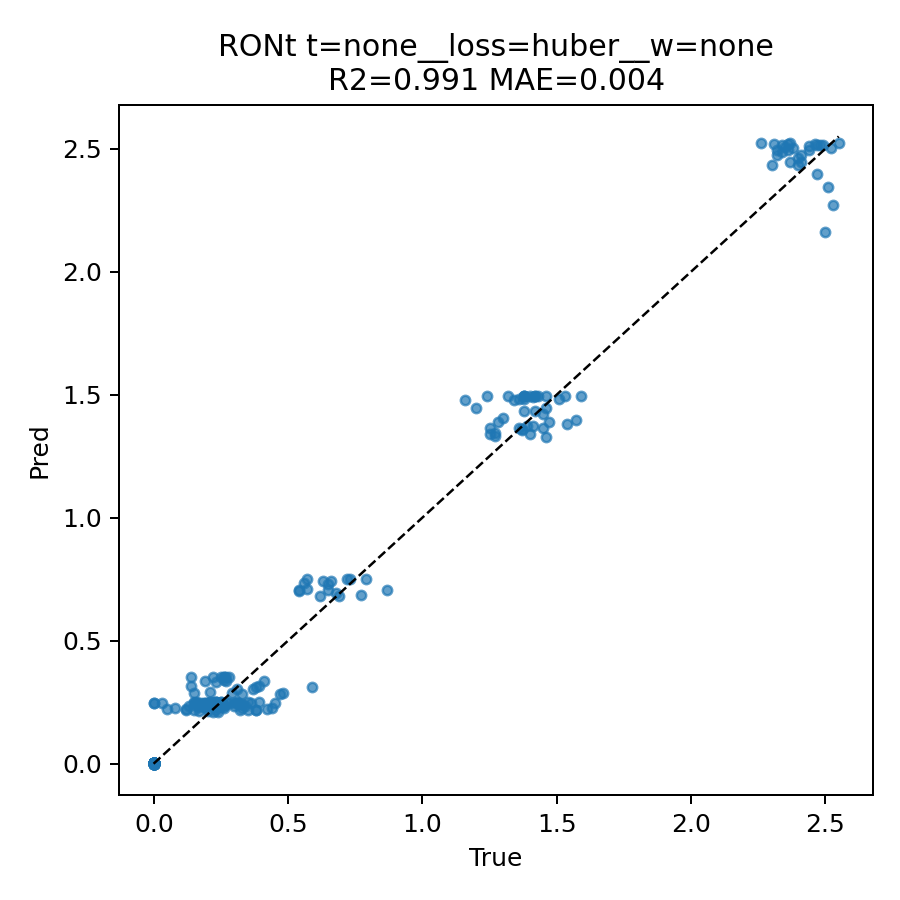}{Figure S54: pred vs true (\label{fig:supp-54})}\
\suppimage{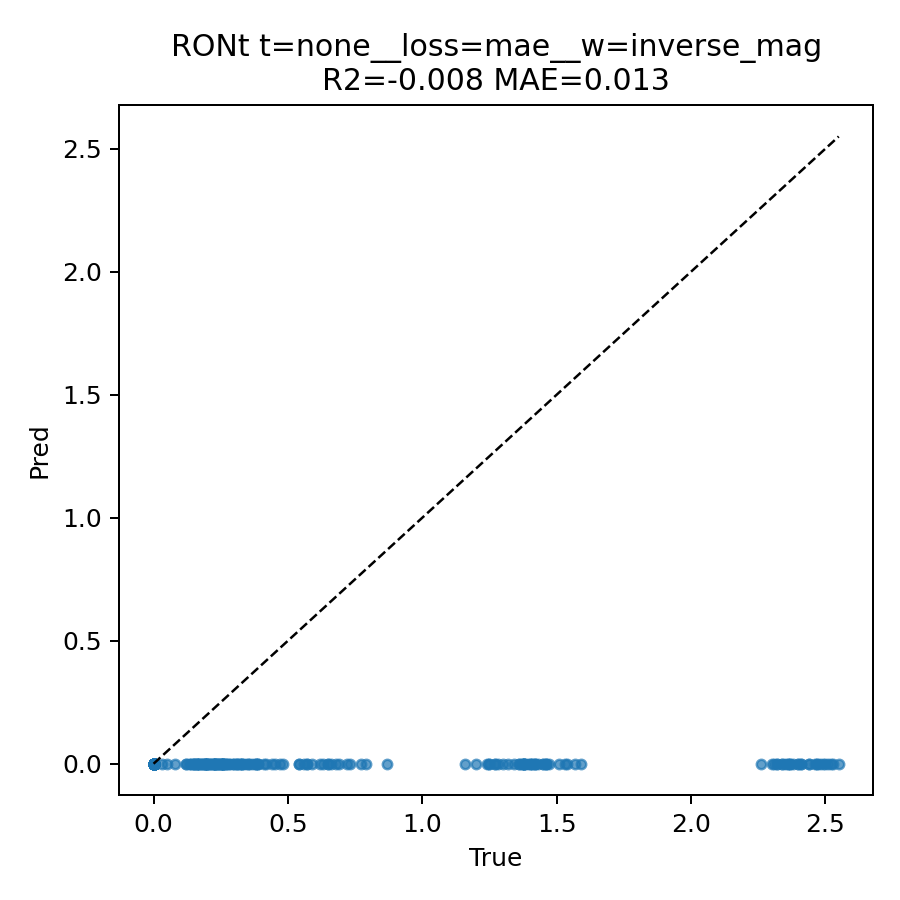}{Figure S55: pred vs true (\label{fig:supp-55})}\hfill
\suppimage{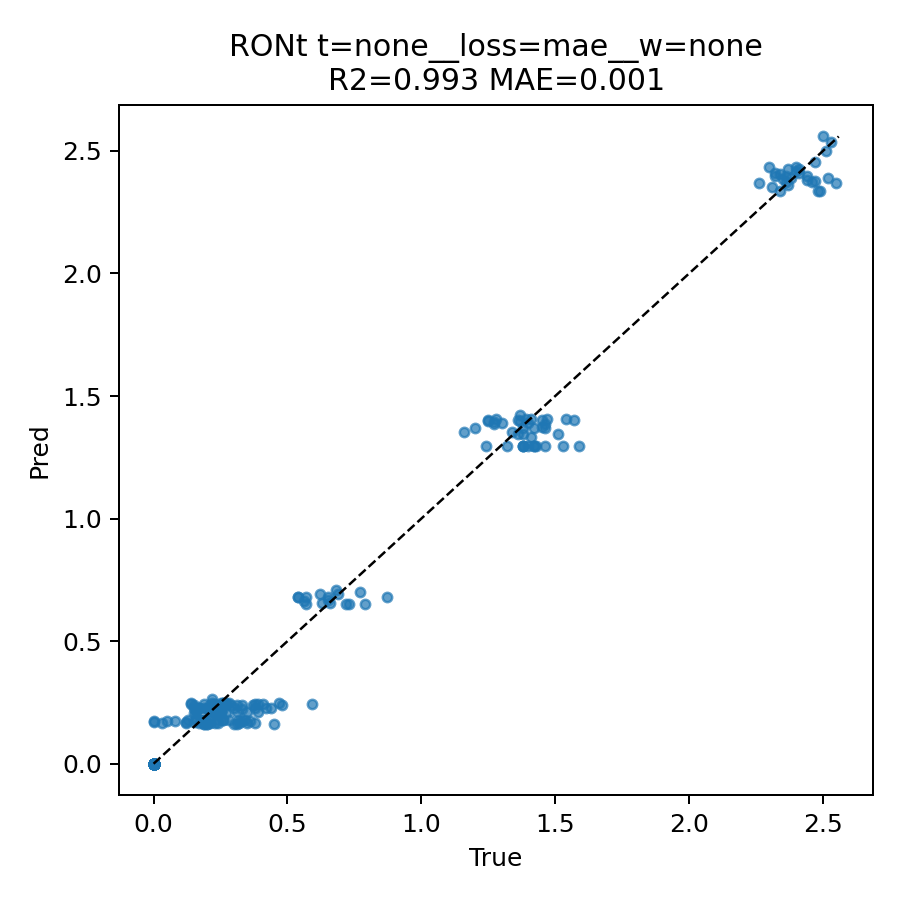}{Figure S56: pred vs true (\label{fig:supp-56})}\
\suppimage{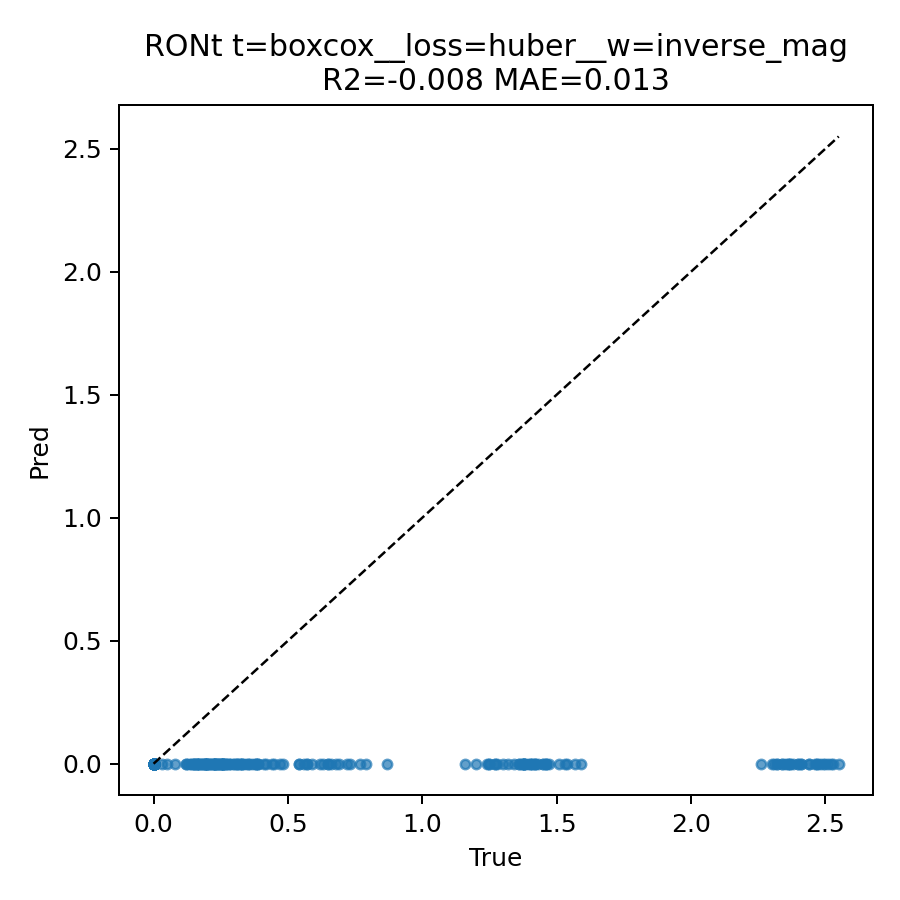}{Figure S57: pred vs true (\label{fig:supp-57})}\hfill
\suppimage{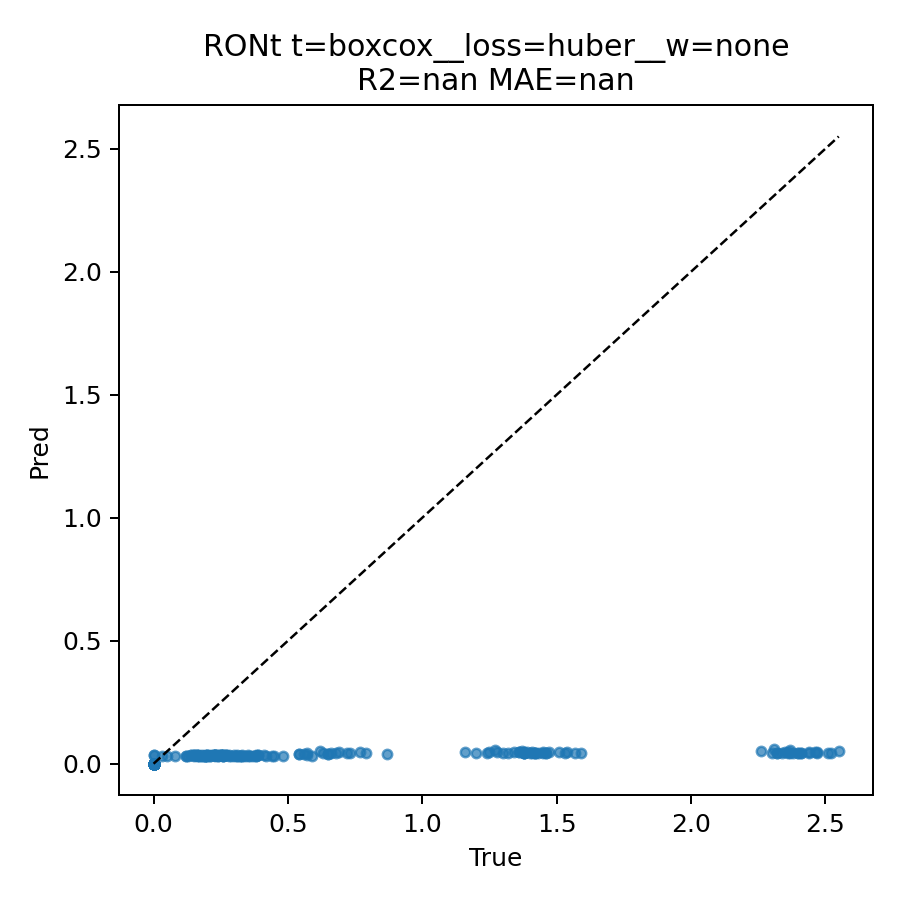}{Figure S58: pred vs true (\label{fig:supp-58})}\
\suppimage{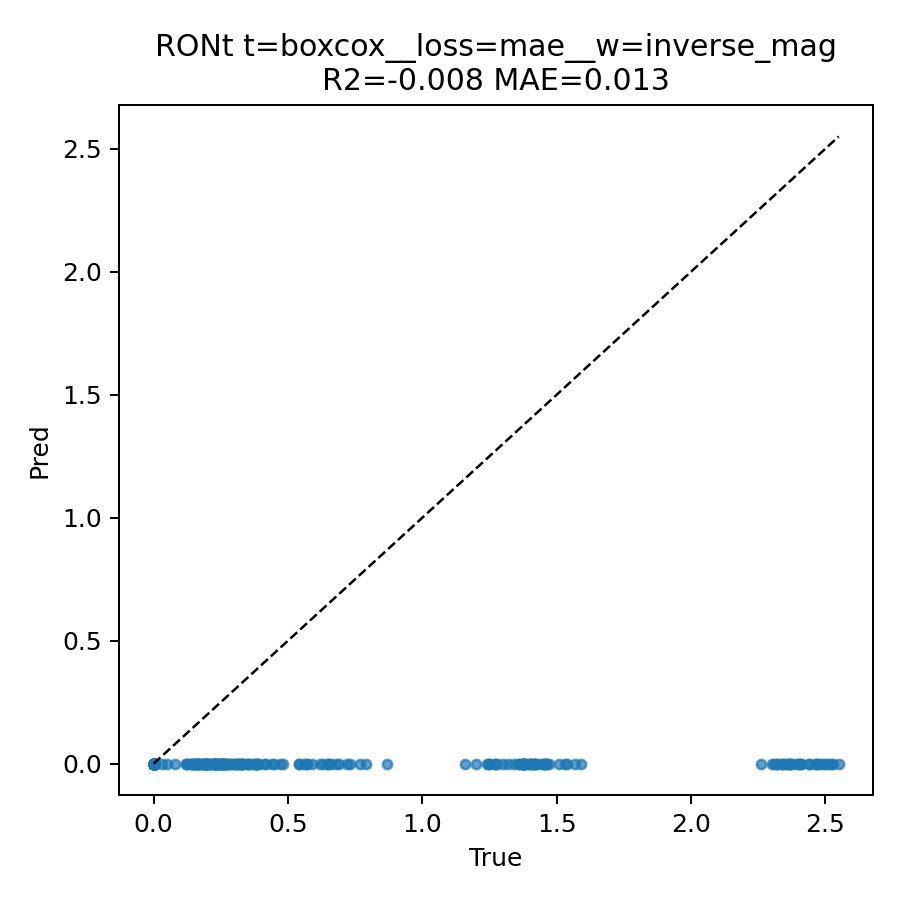}{Figure S59: pred vs true (\label{fig:supp-59})}\hfill
\suppimage{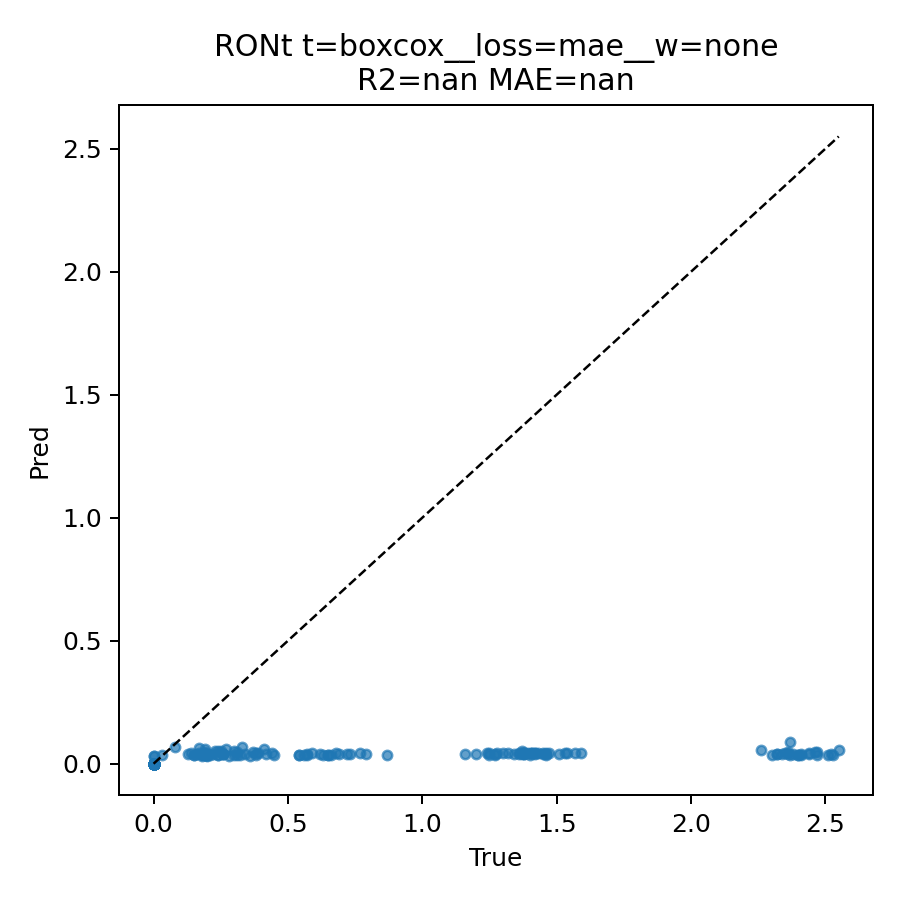}{Figure S60: pred vs true (\label{fig:supp-60})}\
\suppimage{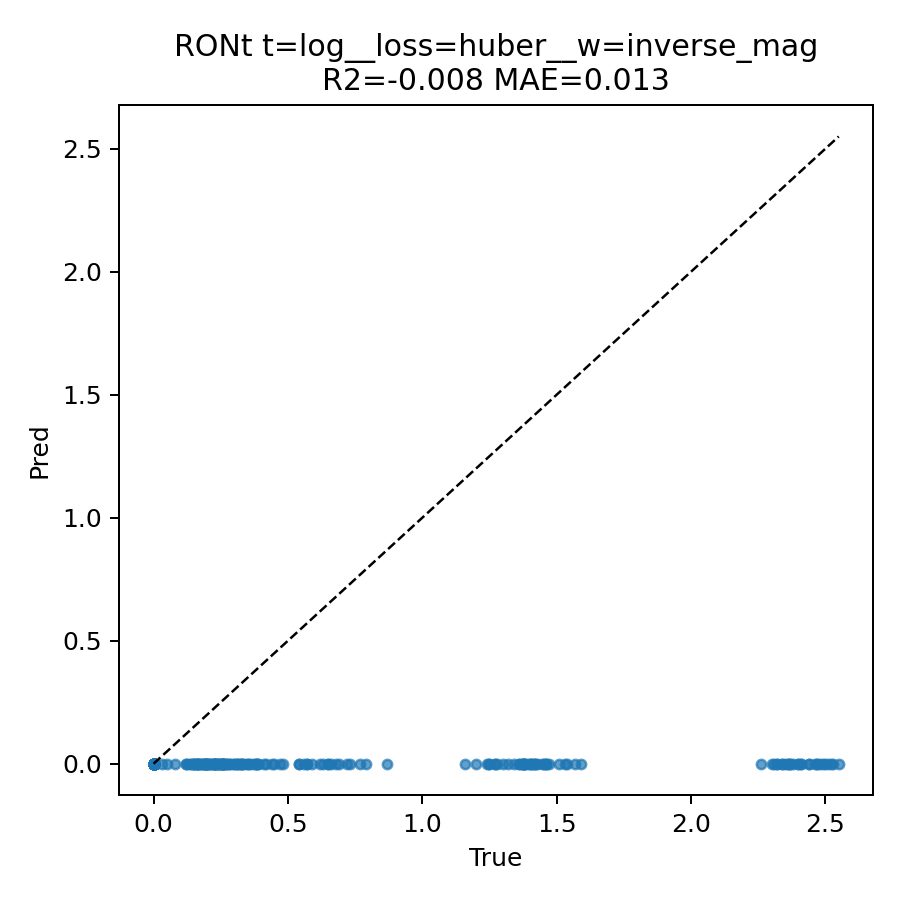}{Figure S61: pred vs true (\label{fig:supp-61})}\hfill
\suppimage{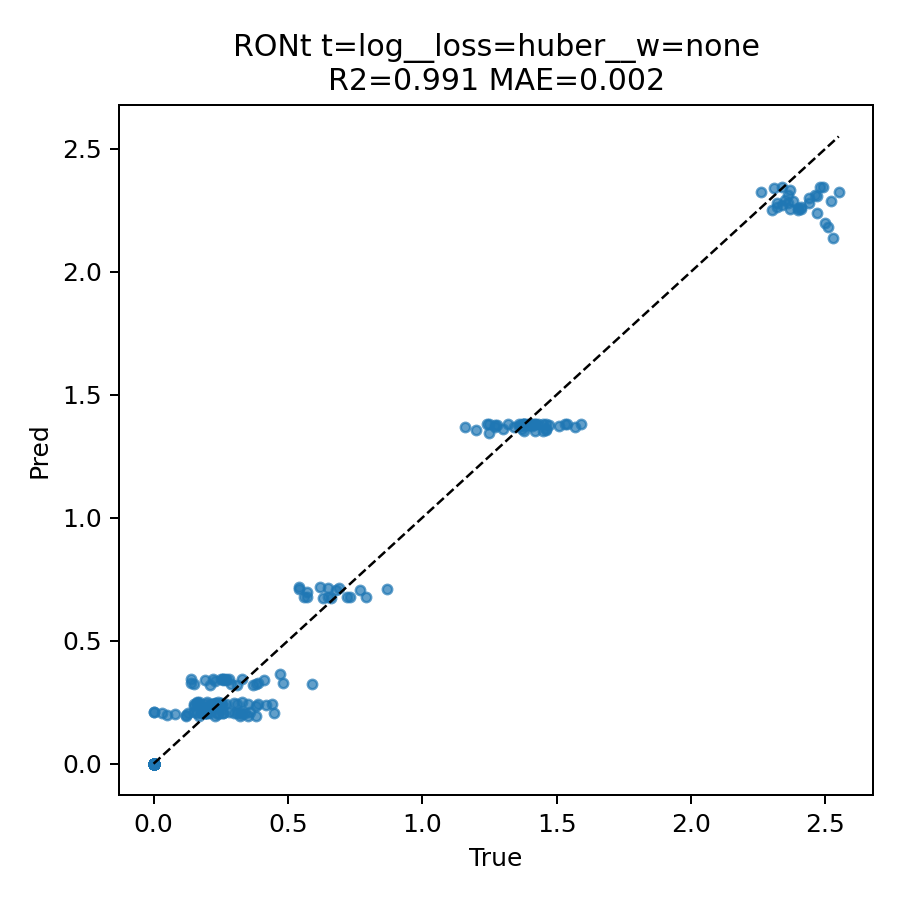}{Figure S62: pred vs true (\label{fig:supp-62})}\
\suppimage{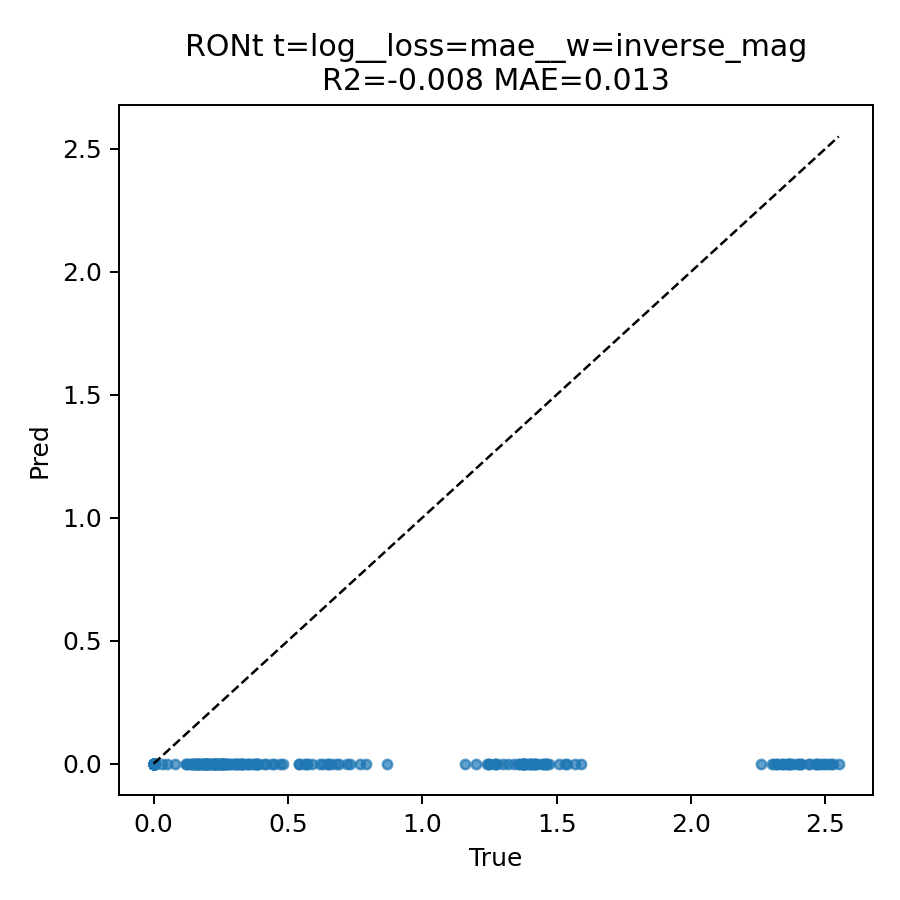}{Figure S63: pred vs true (\label{fig:supp-63})}\hfill
\suppimage{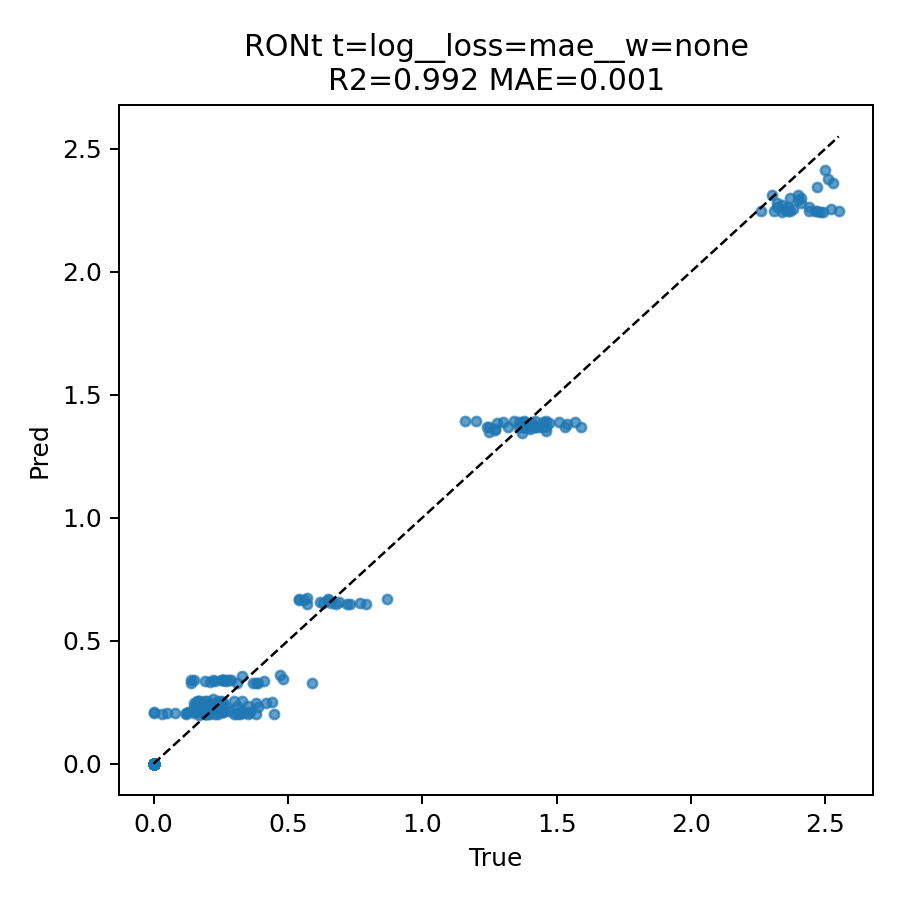}{Figure S64: pred vs true (\label{fig:supp-64})}\
\suppimage{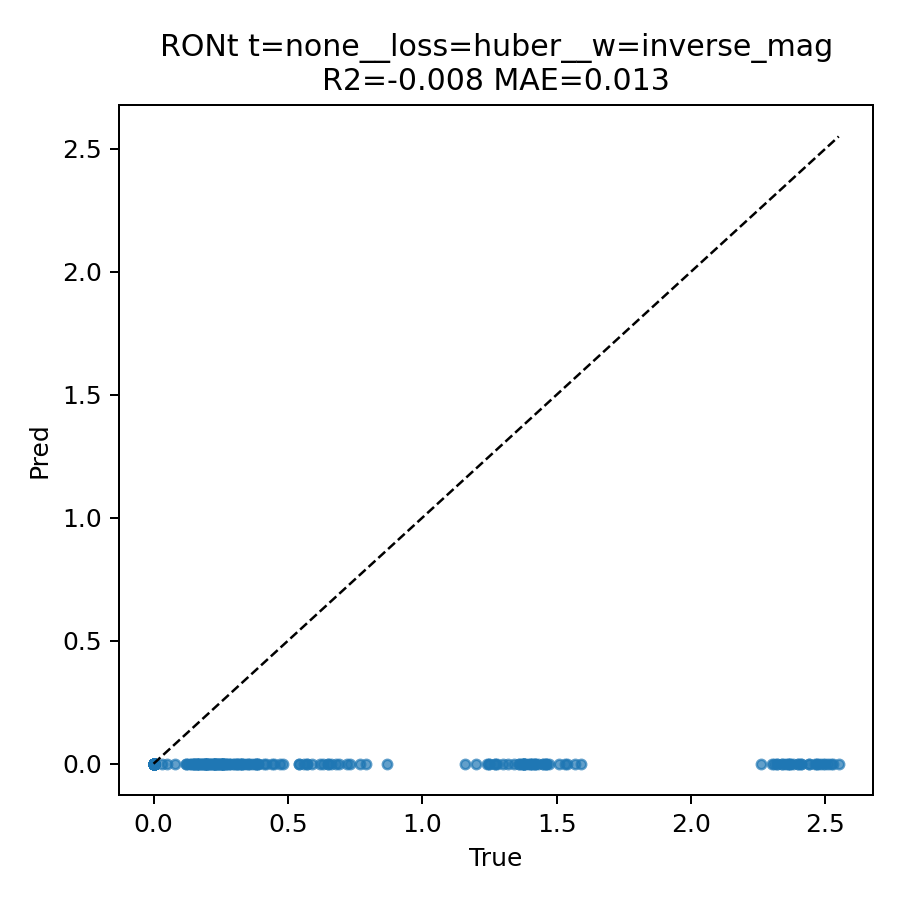}{Figure S65: pred vs true (\label{fig:supp-65})}\hfill
\suppimage{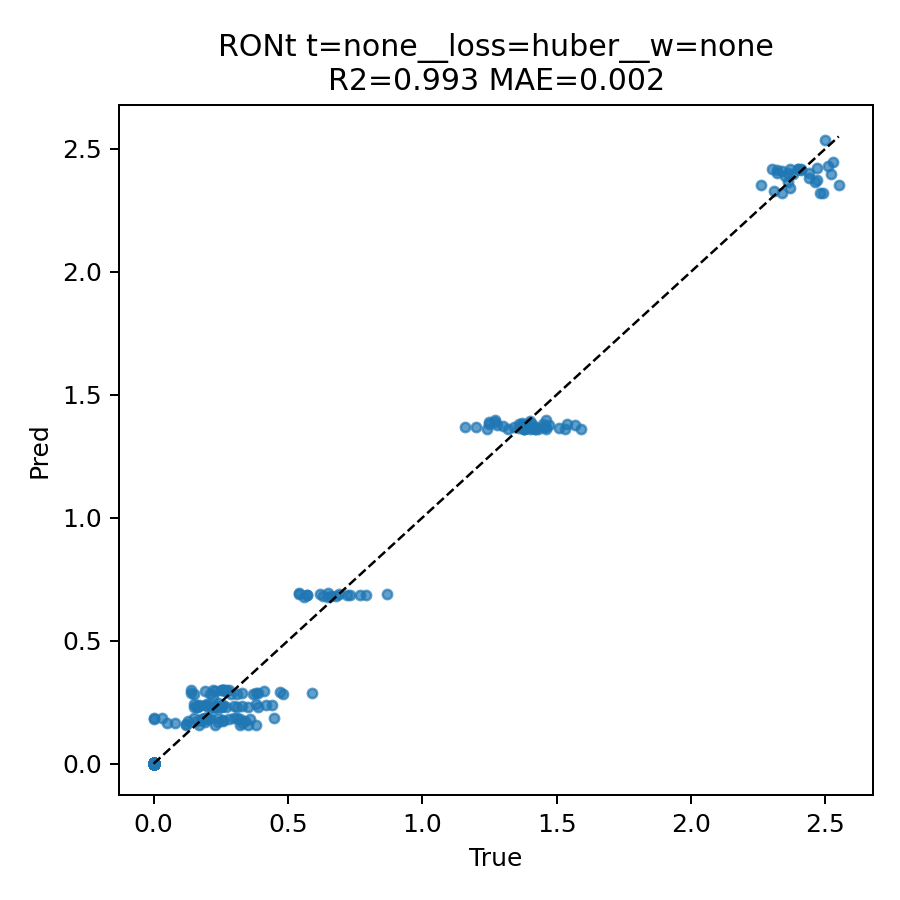}{Figure S66: pred vs true (\label{fig:supp-66})}\
\suppimage{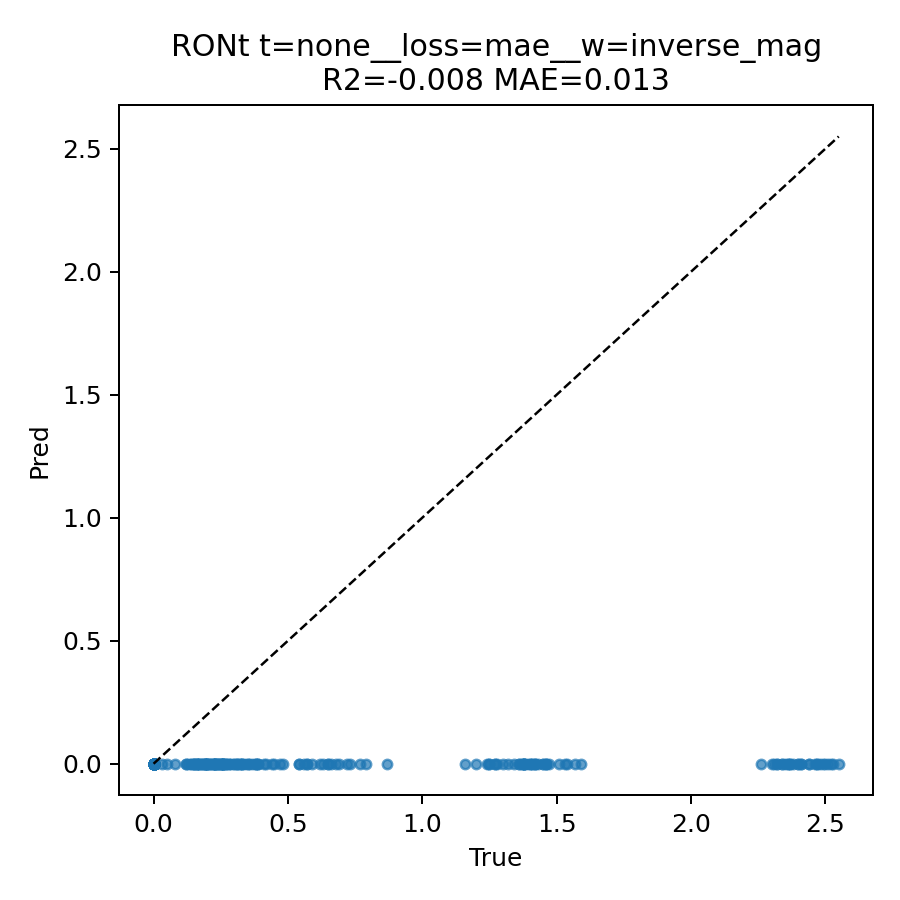}{Figure S67: pred vs true (\label{fig:supp-67})}\hfill
\suppimage{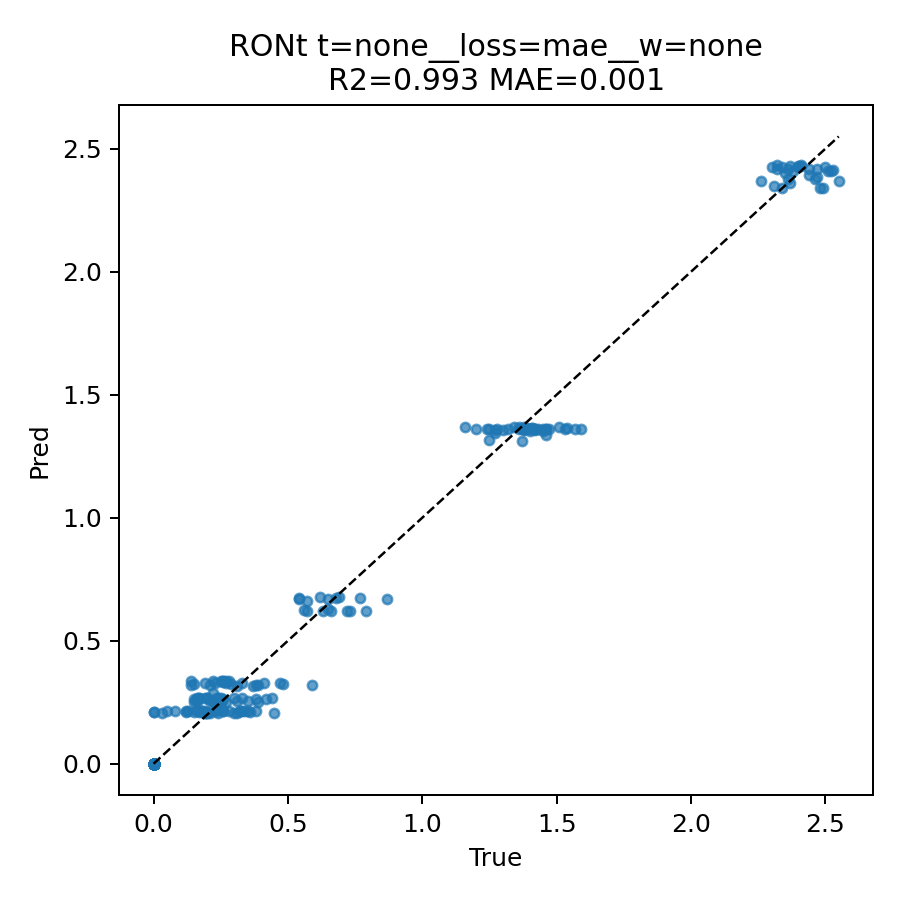}{Figure S68: pred vs true (\label{fig:supp-68})}\
\suppimage{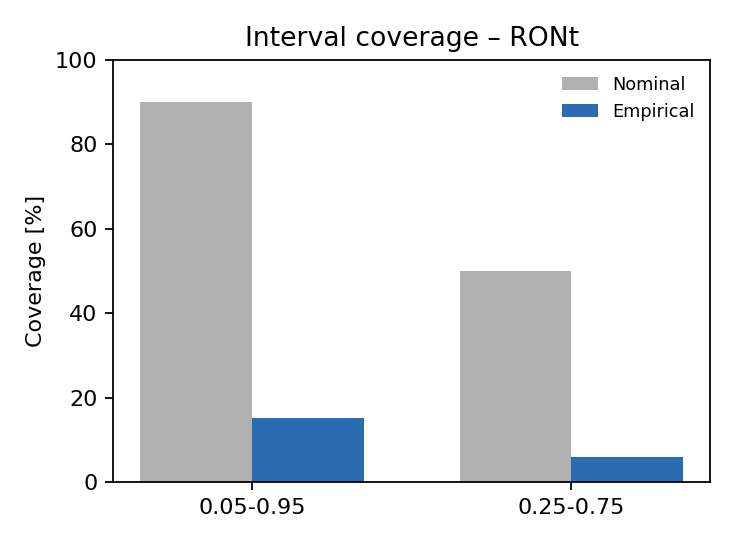}{Figure S69: coverage RONt (\label{fig:supp-69})}\hfill
\suppimage{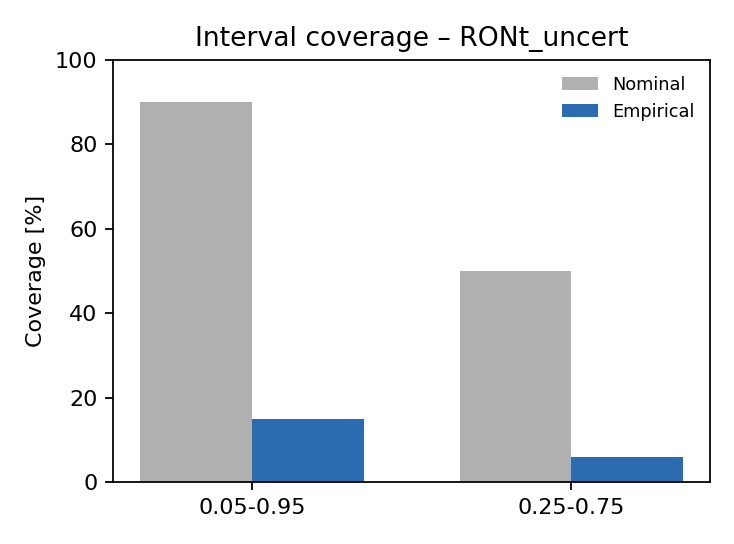}{Figure S70: coverage RONt uncert (\label{fig:supp-70})}\
\suppimage{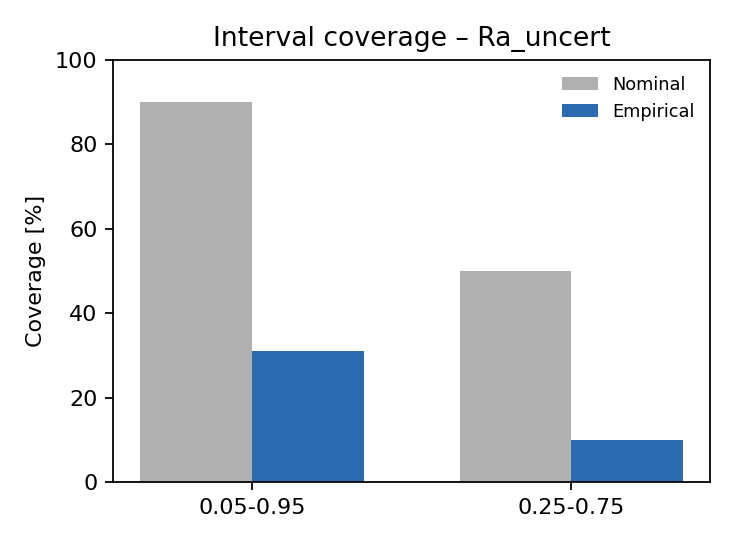}{Figure S71: coverage Ra uncert (\label{fig:supp-71})}\hfill
\suppimage{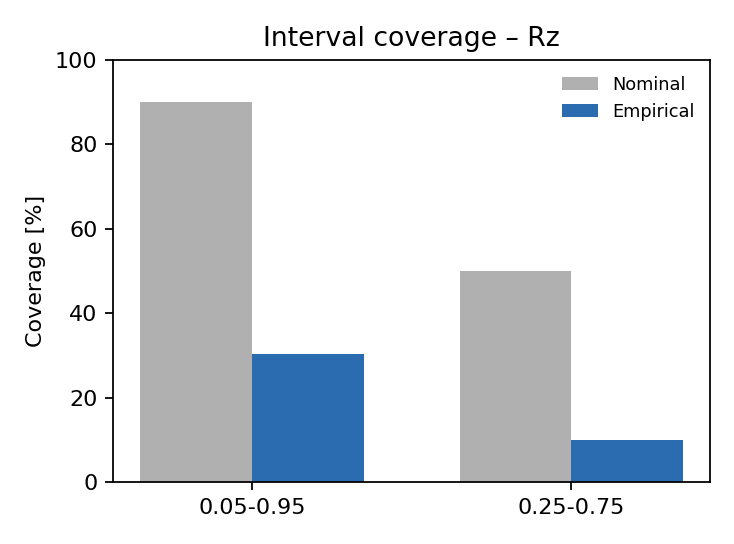}{Figure S72: coverage Rz (\label{fig:supp-72})}\
\suppimage{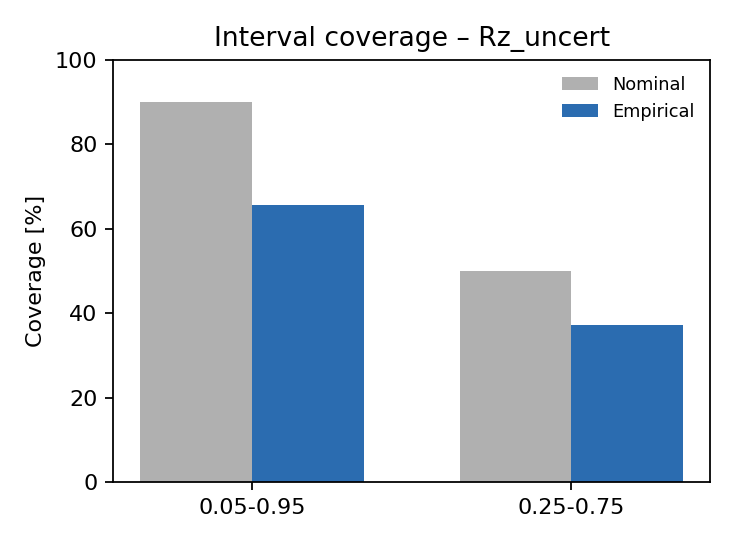}{Figure S73: coverage Rz uncert (\label{fig:supp-73})}\hfill
\suppimage{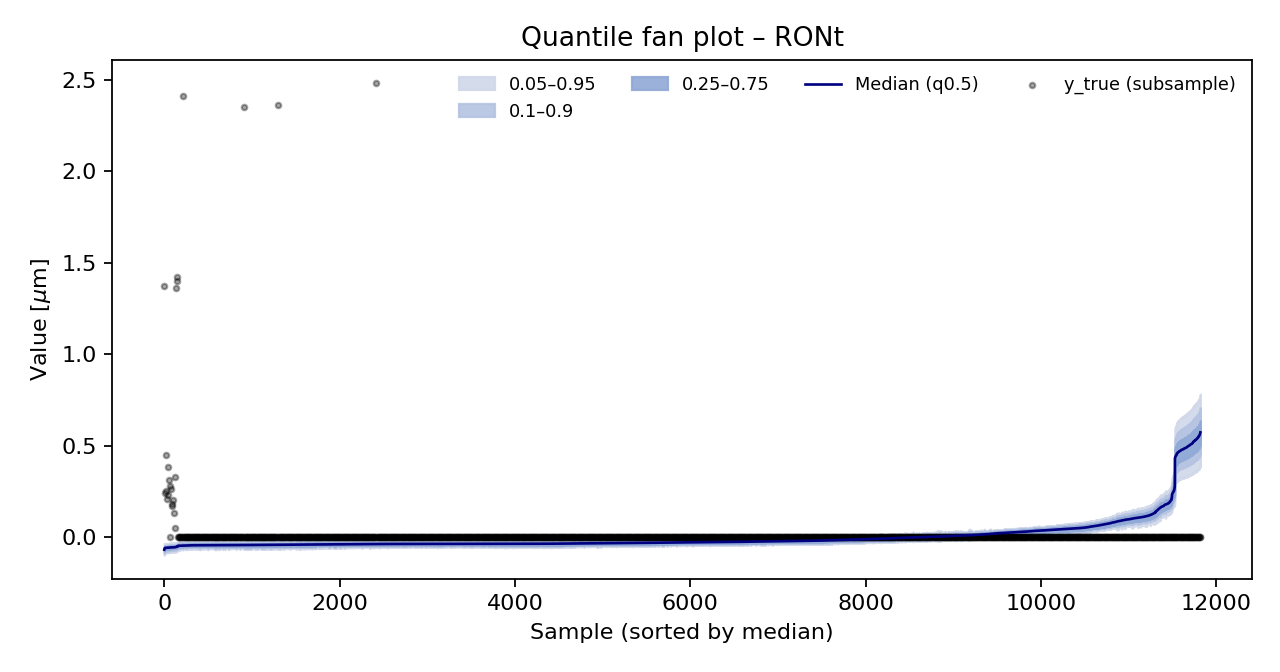}{Figure S74: fan RONt (\label{fig:supp-74})}\
\suppimage{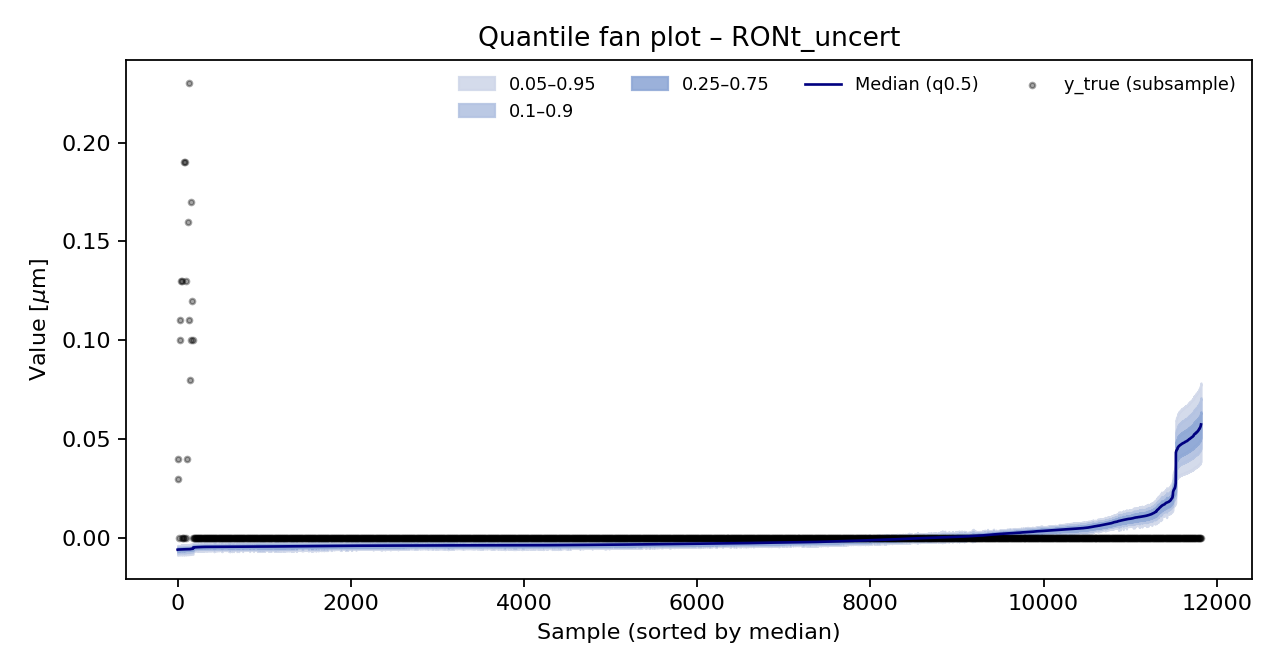}{Figure S75: fan RONt uncert (\label{fig:supp-75})}\hfill
\suppimage{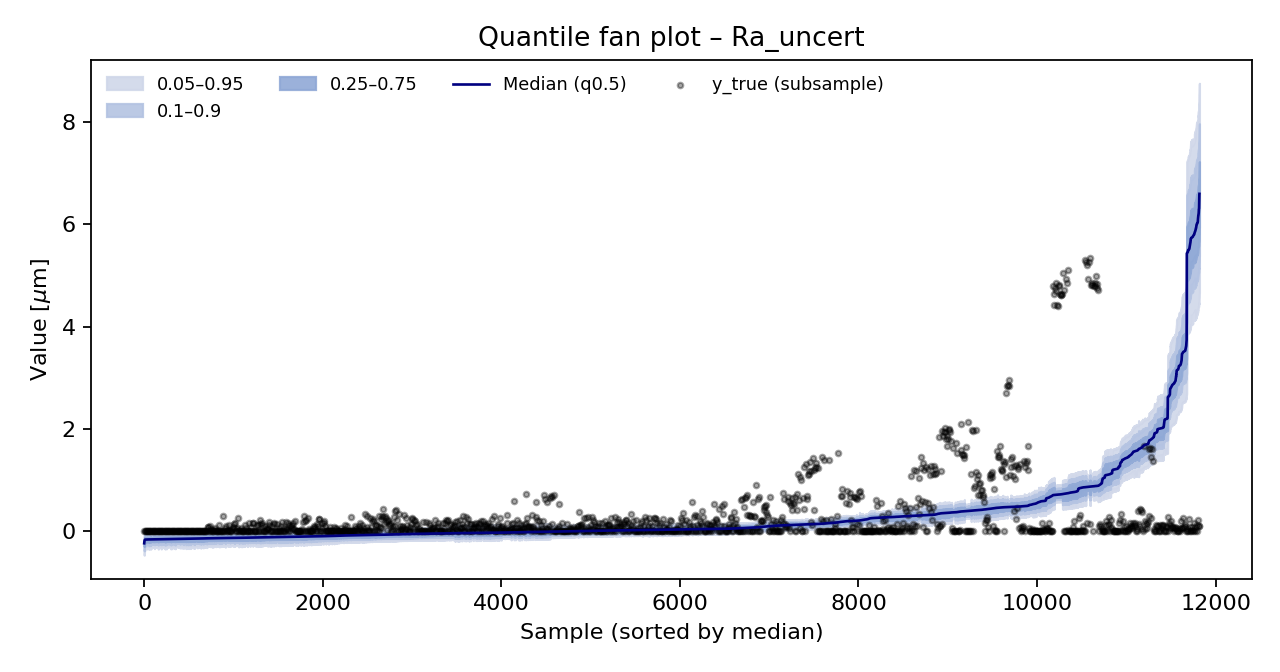}{Figure S76: fan Ra uncert (\label{fig:supp-76})}\
\suppimage{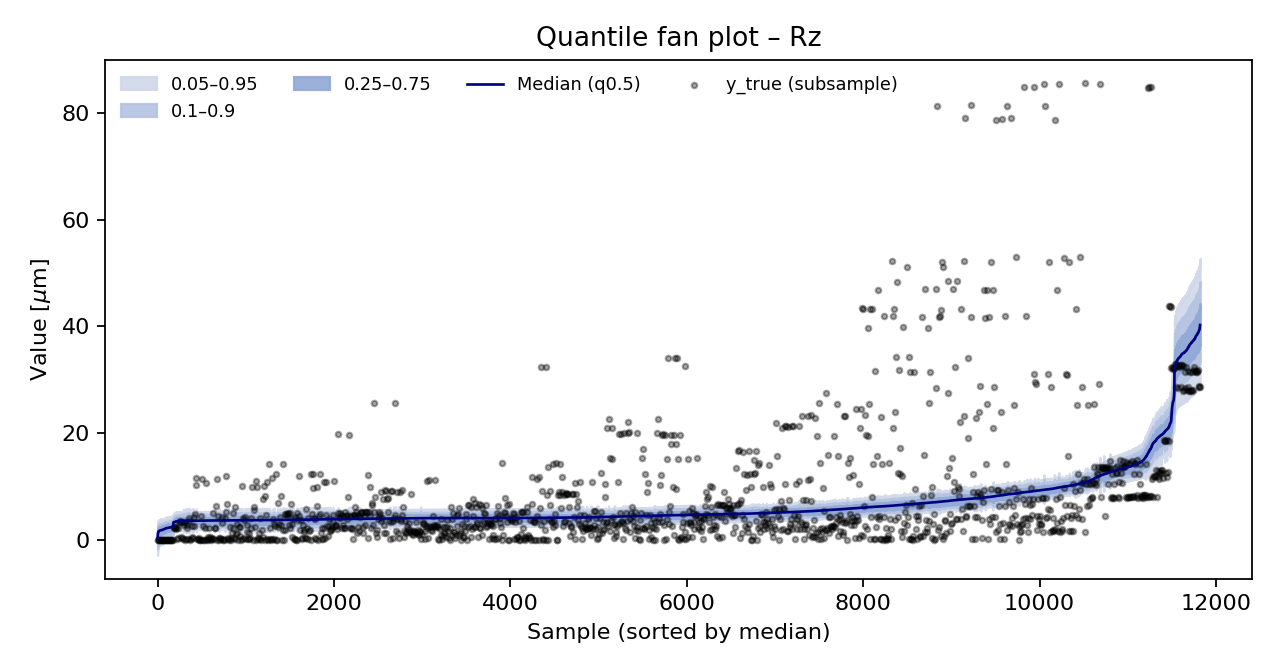}{Figure S77: fan Rz (\label{fig:supp-77})}\hfill
\suppimage{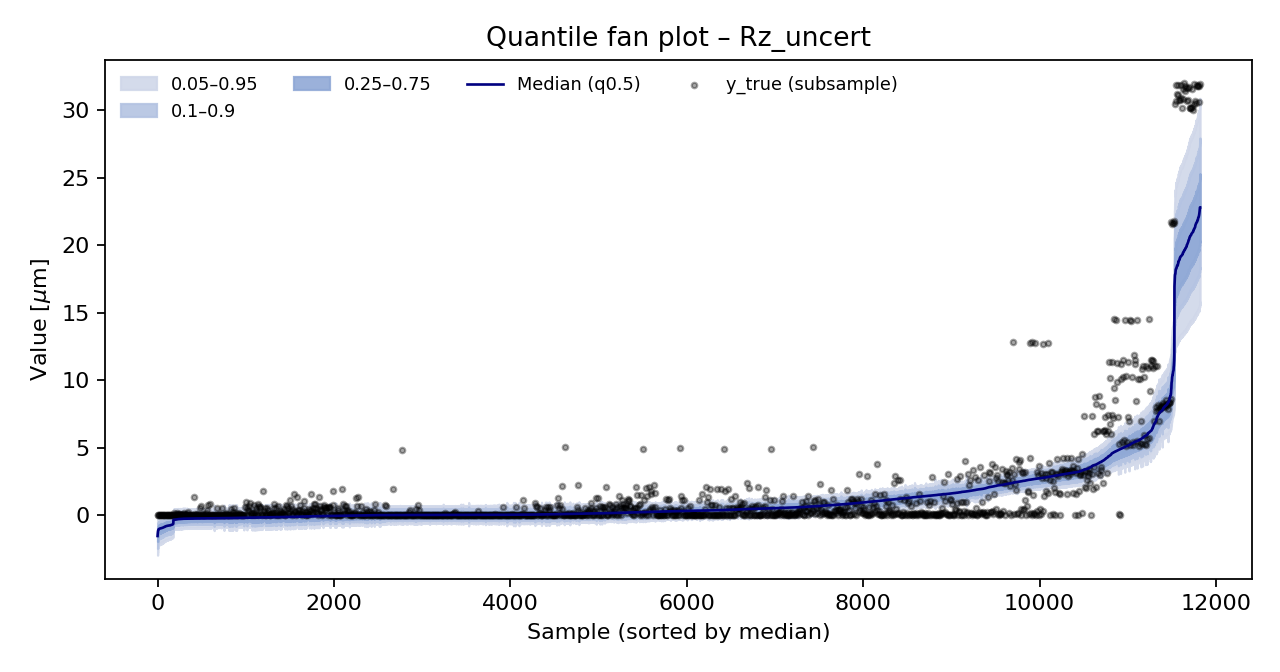}{Figure S78: fan Rz uncert (\label{fig:supp-78})}\
\suppimage{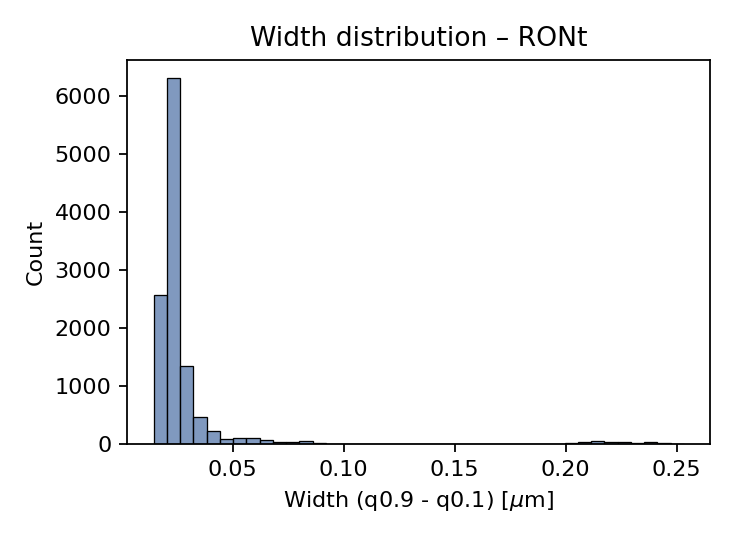}{Figure S79: width RONt (\label{fig:supp-79})}\hfill
\suppimage{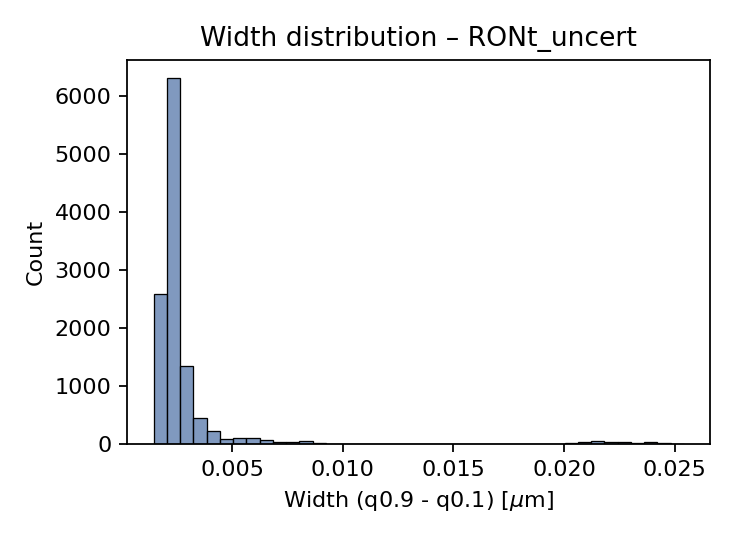}{Figure S80: width RONt uncert (\label{fig:supp-80})}\
\suppimage{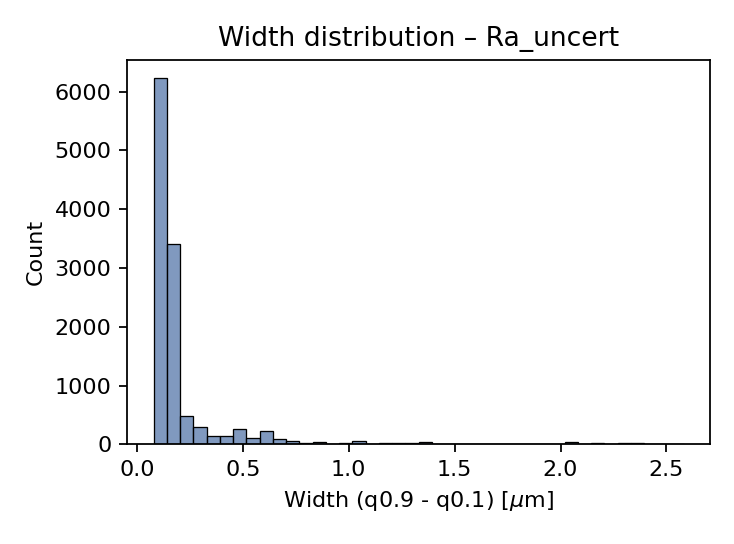}{Figure S81: width Ra uncert (\label{fig:supp-81})}\hfill
\suppimage{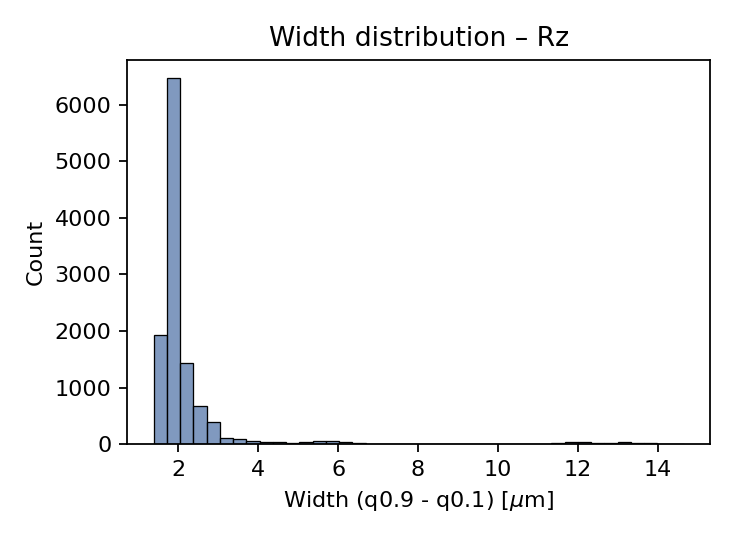}{Figure S82: width Rz (\label{fig:supp-82})}\
\suppimage{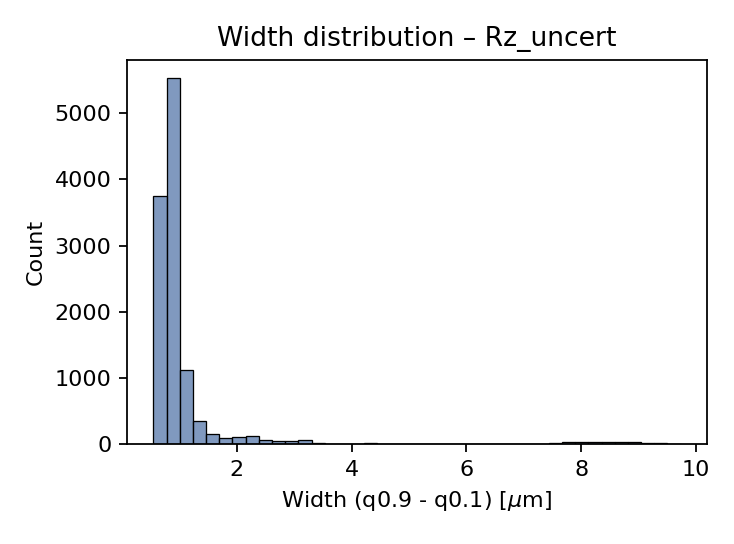}{Figure S83: width Rz uncert (\label{fig:supp-83})}\hfill
\suppimage{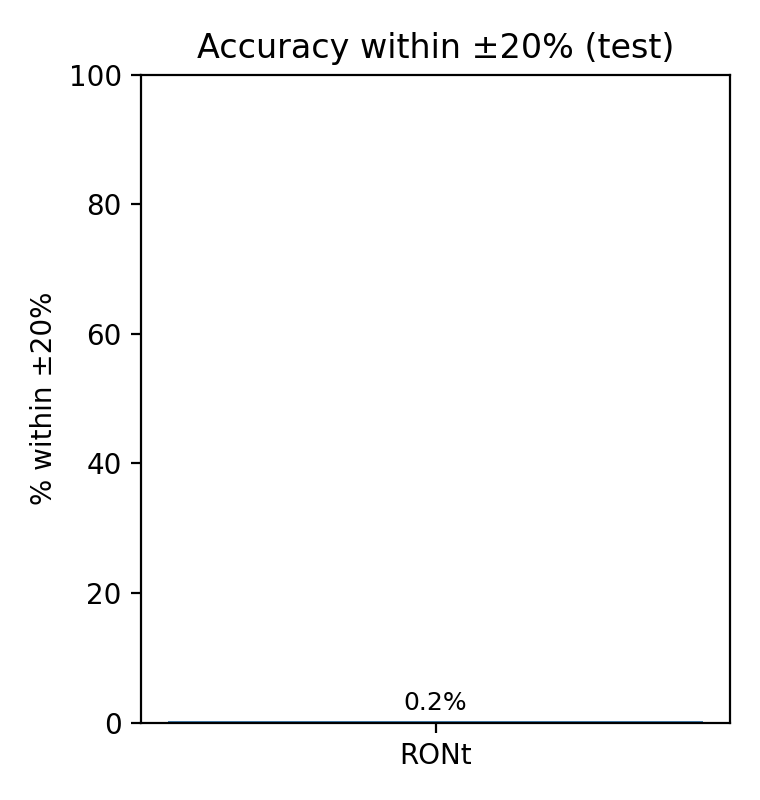}{Figure S84: accuracy within tol 20percent (\label{fig:supp-84})}\
\suppimage{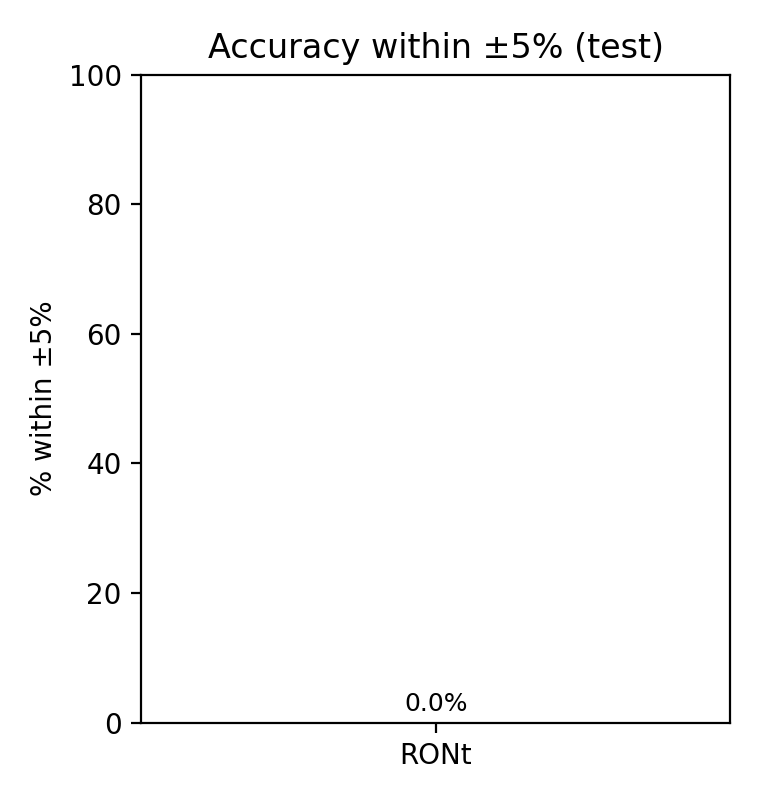}{Figure S85: accuracy within tol 5percent (\label{fig:supp-85})}\hfill
\suppimage{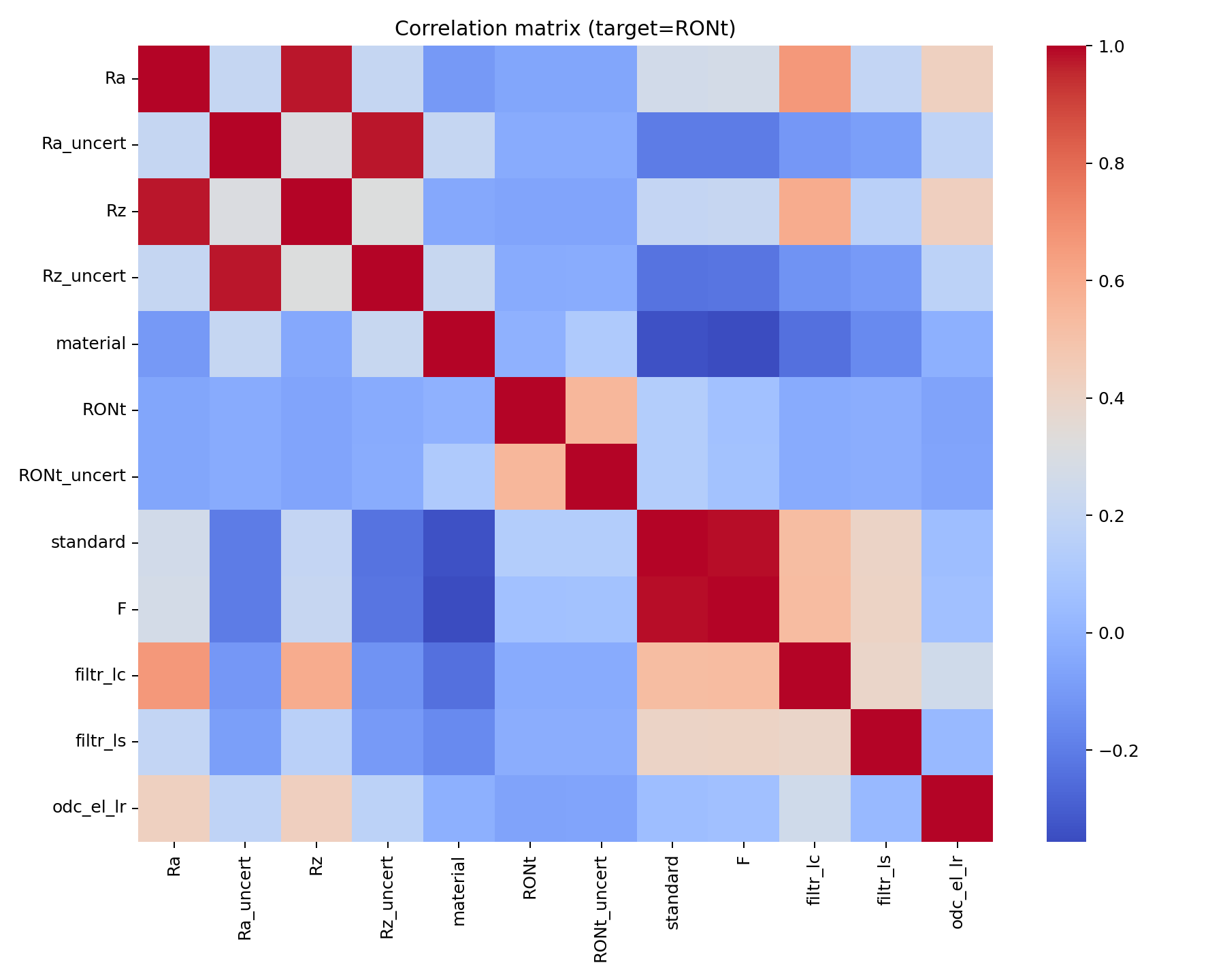}{Figure S86: correlation heatmap RONt (\label{fig:supp-86})}\
\suppimage{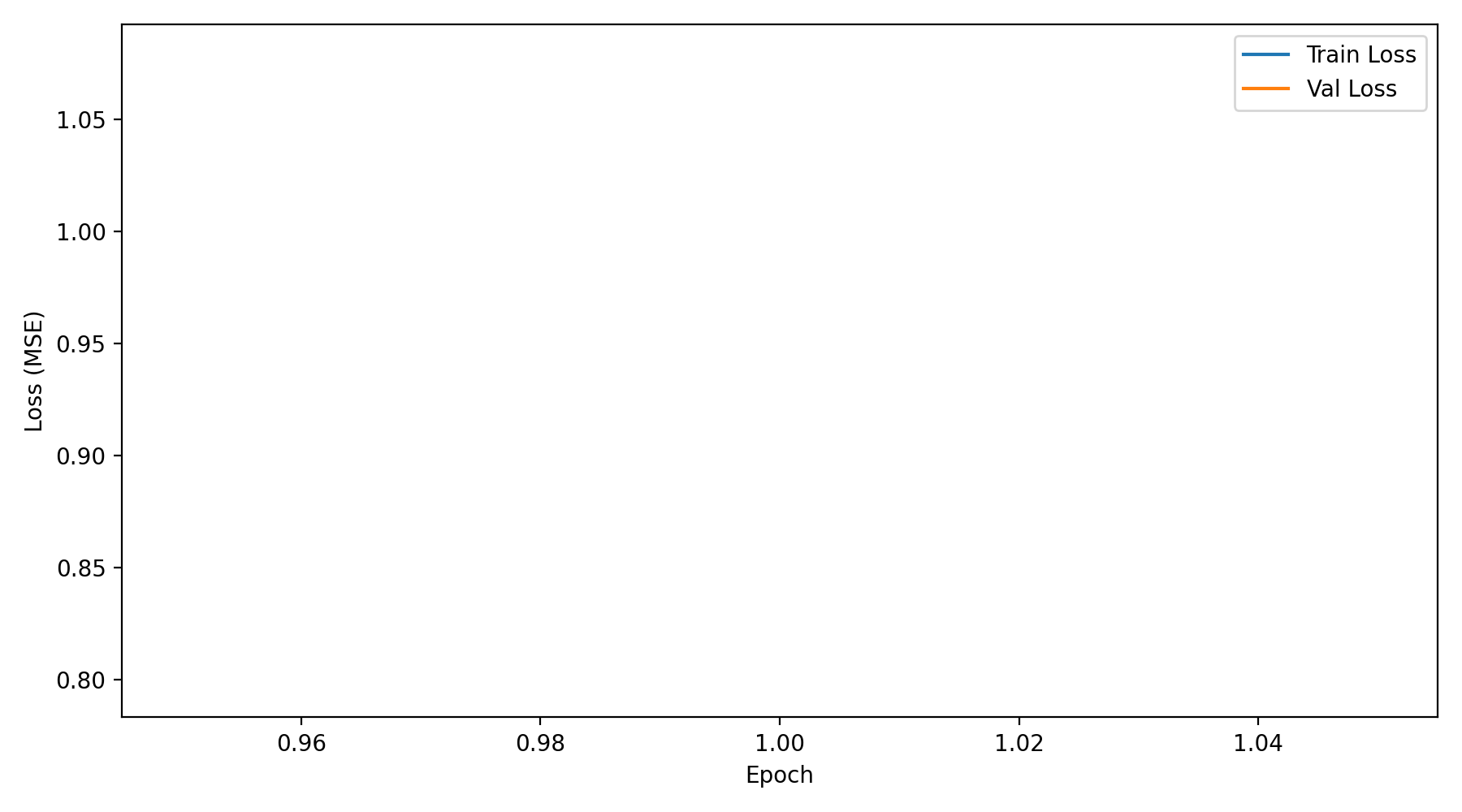}{Figure S87: loss curves (\label{fig:supp-87})}\hfill
\suppimage{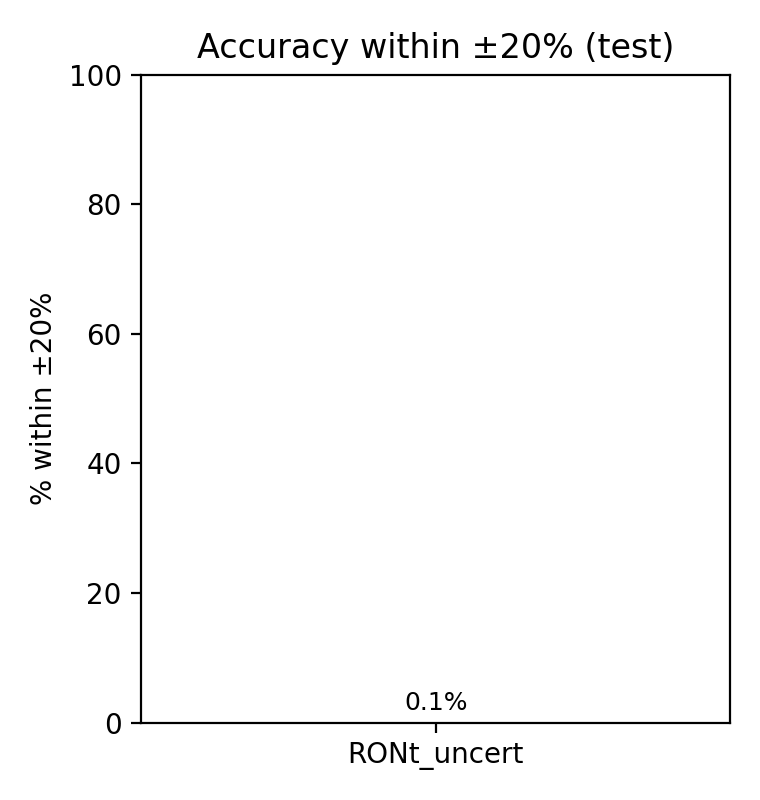}{Figure S88: accuracy within tol 20percent (\label{fig:supp-88})}\
\suppimage{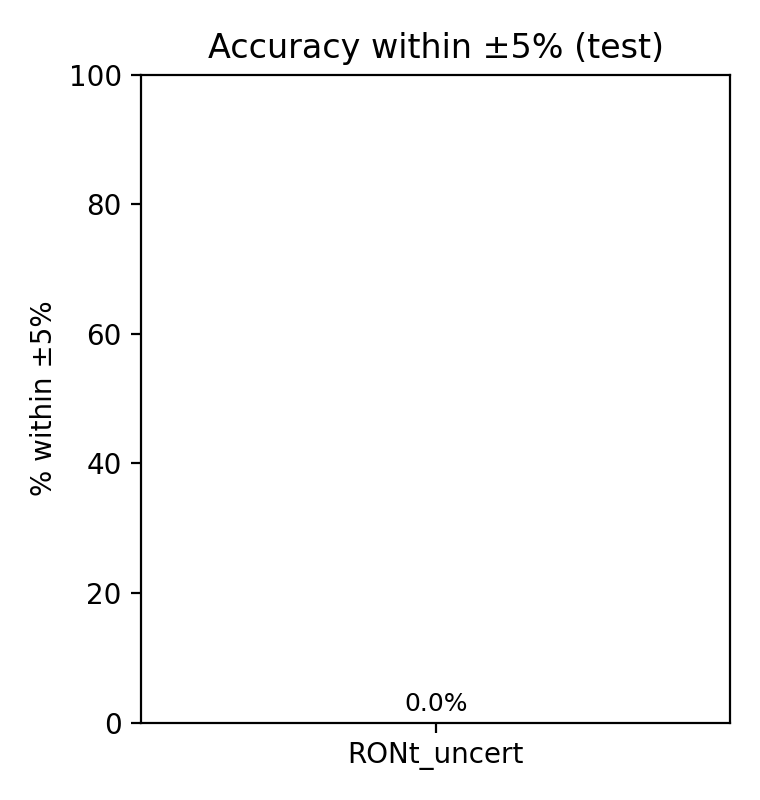}{Figure S89: accuracy within tol 5percent (\label{fig:supp-89})}\hfill
\suppimage{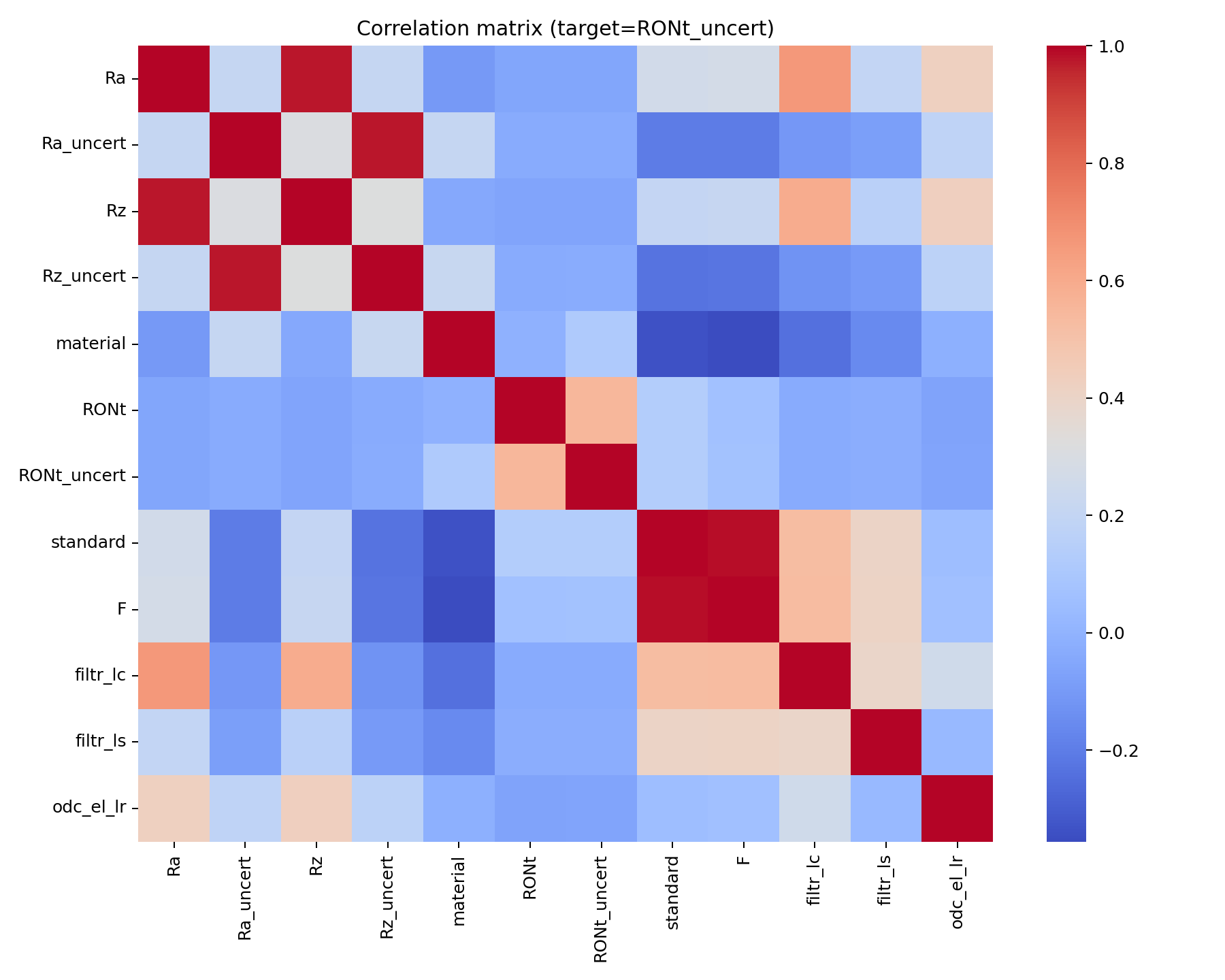}{Figure S90: correlation heatmap RONt uncert (\label{fig:supp-90})}\
\suppimage{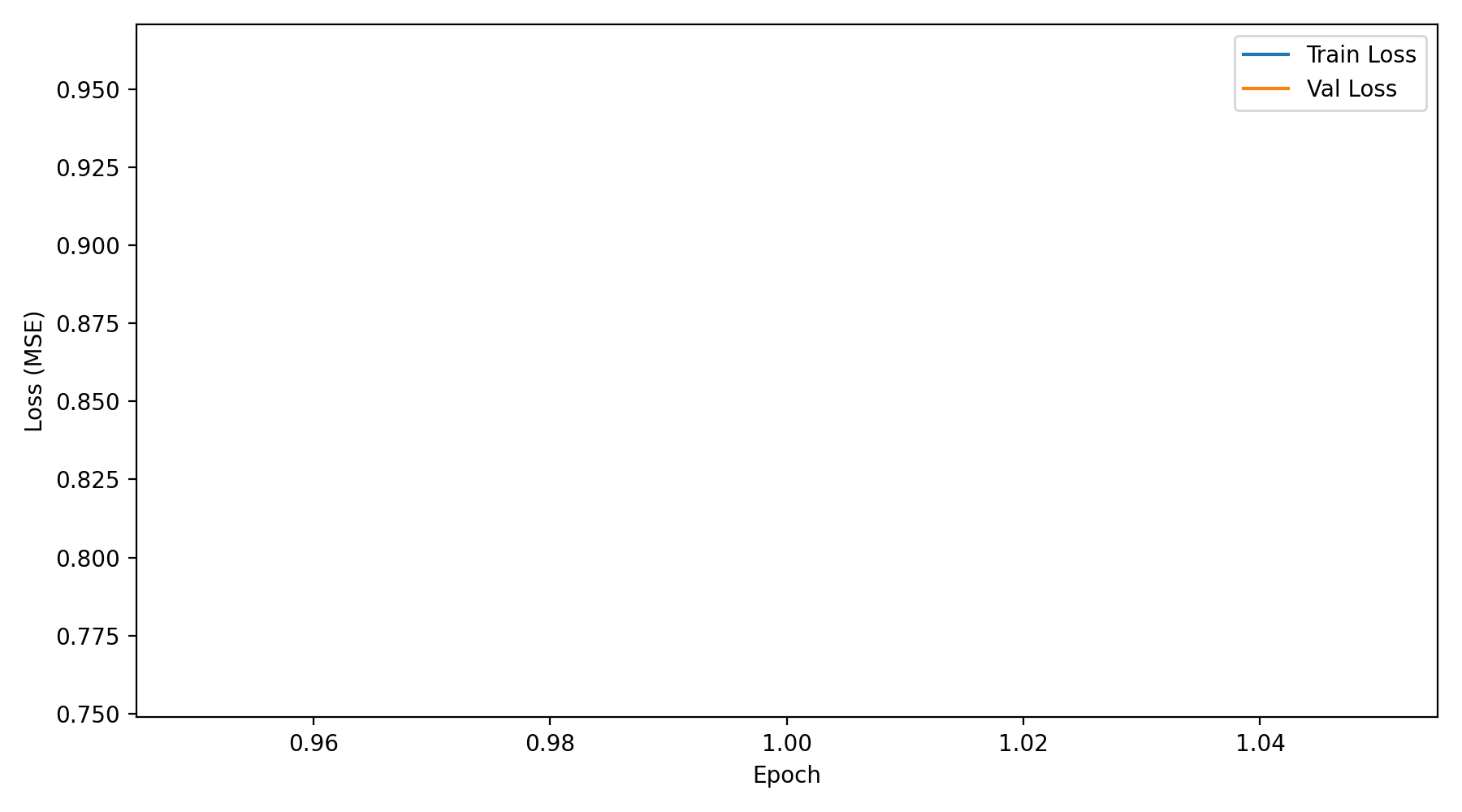}{Figure S91: loss curves (\label{fig:supp-91})}\hfill
\suppimage{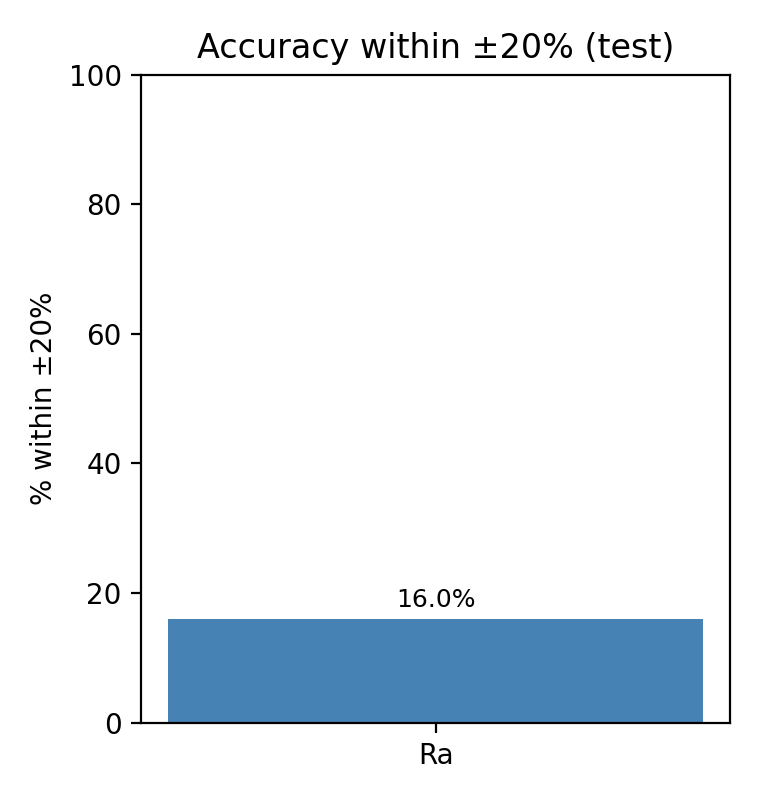}{Figure S92: accuracy within tol 20percent (\label{fig:supp-92})}\
\suppimage{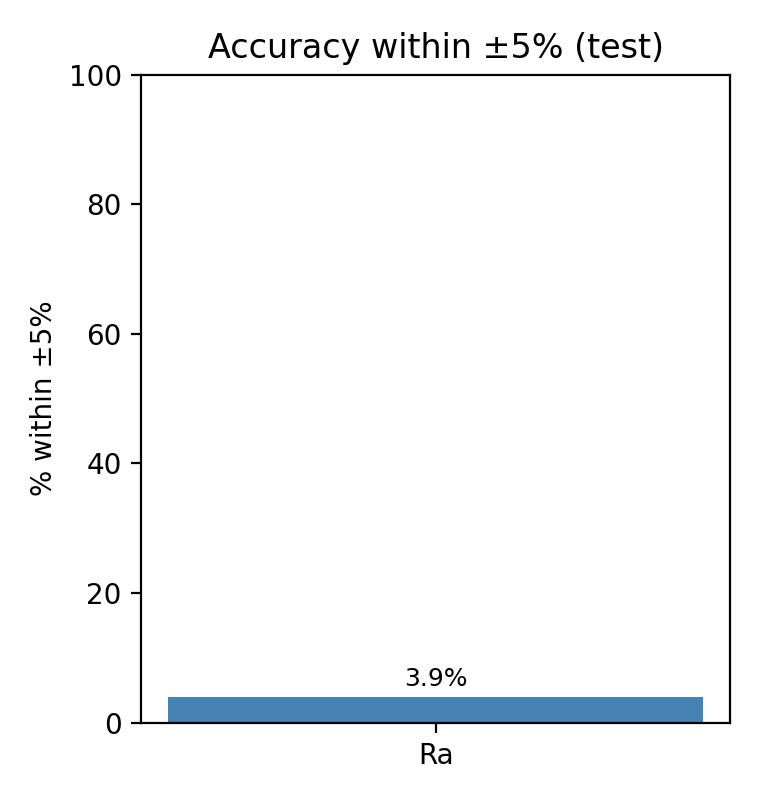}{Figure S93: accuracy within tol 5percent (\label{fig:supp-93})}\hfill
\suppimage{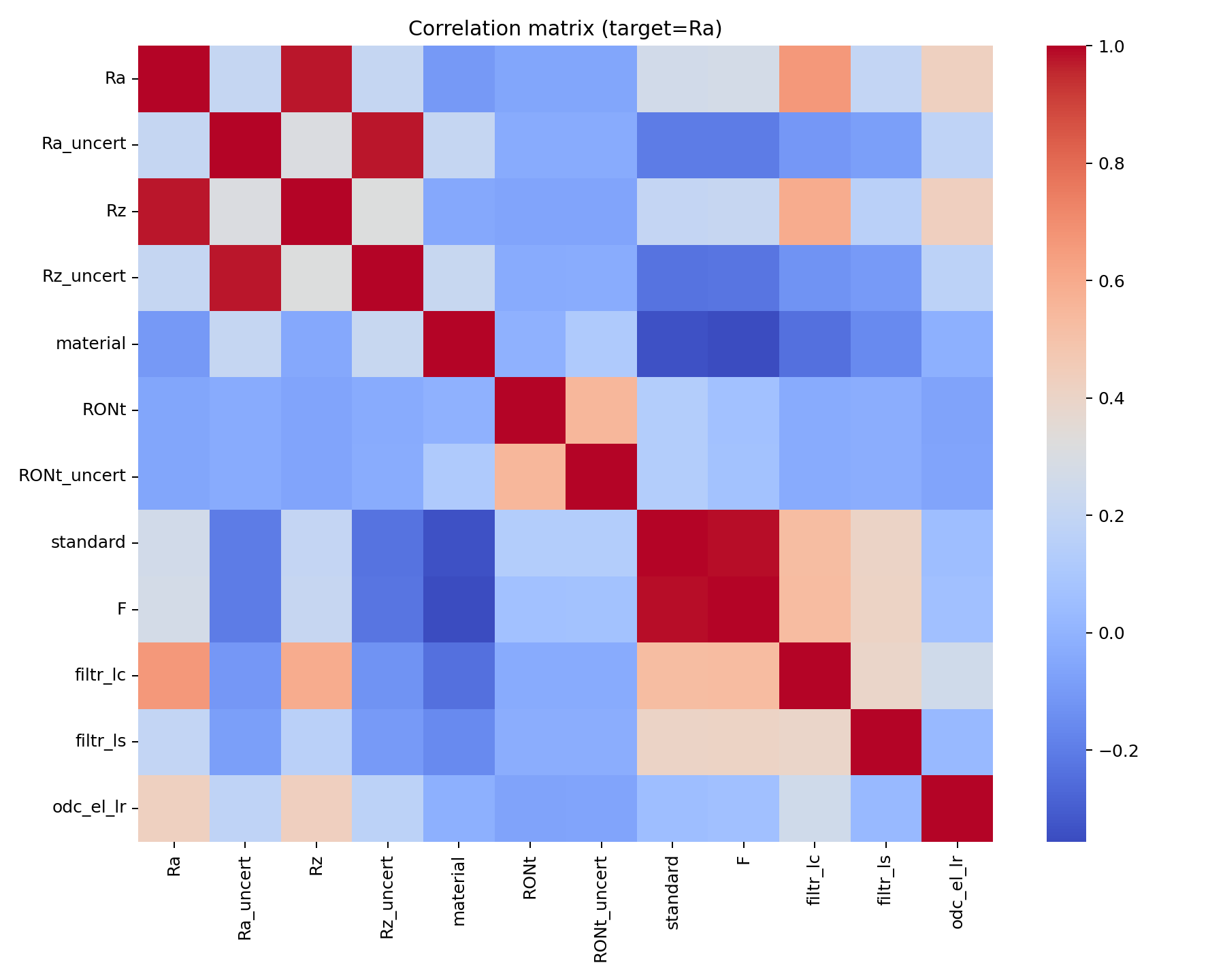}{Figure S94: correlation heatmap Ra (\label{fig:supp-94})}\
\suppimage{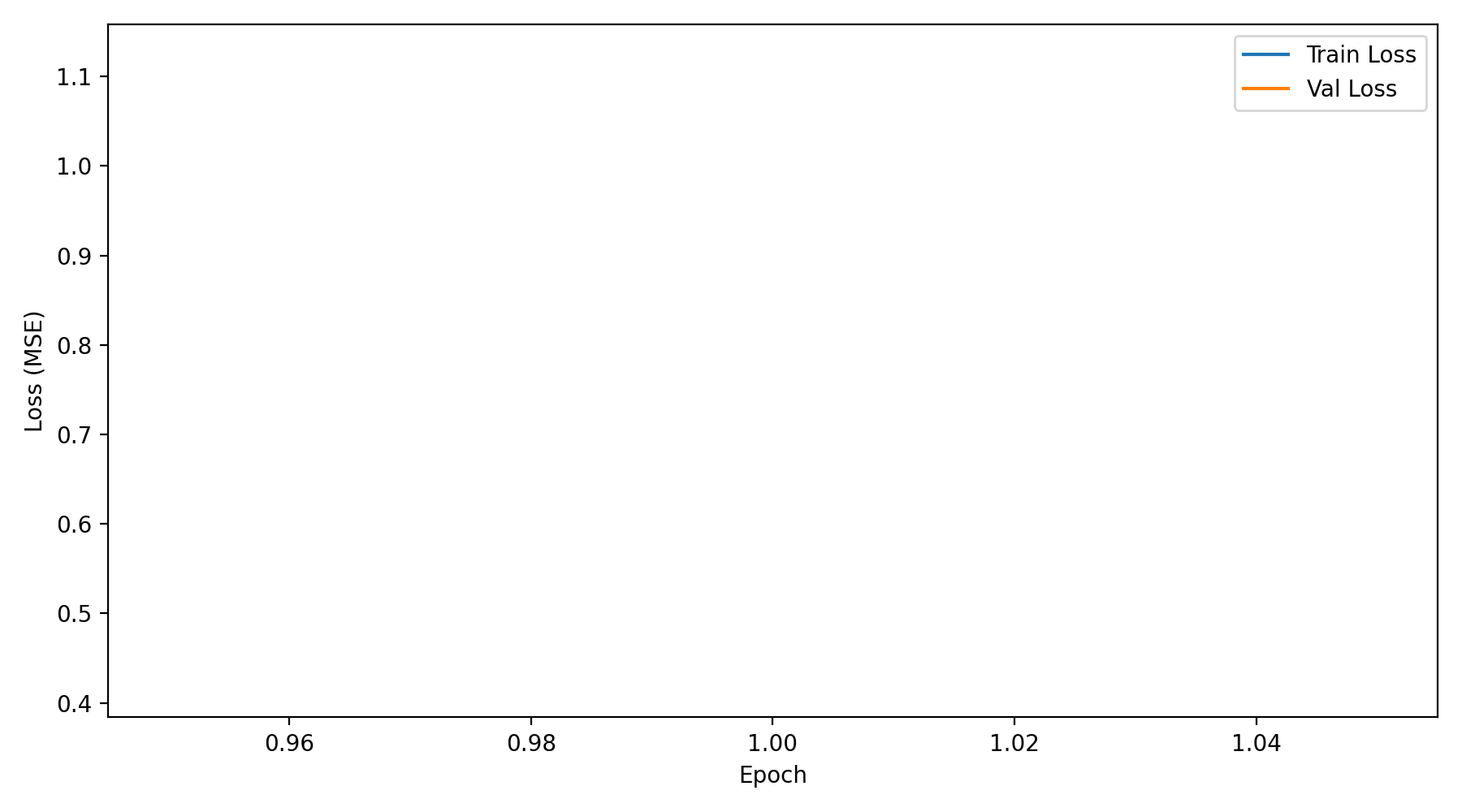}{Figure S95: loss curves (\label{fig:supp-95})}\hfill
\suppimage{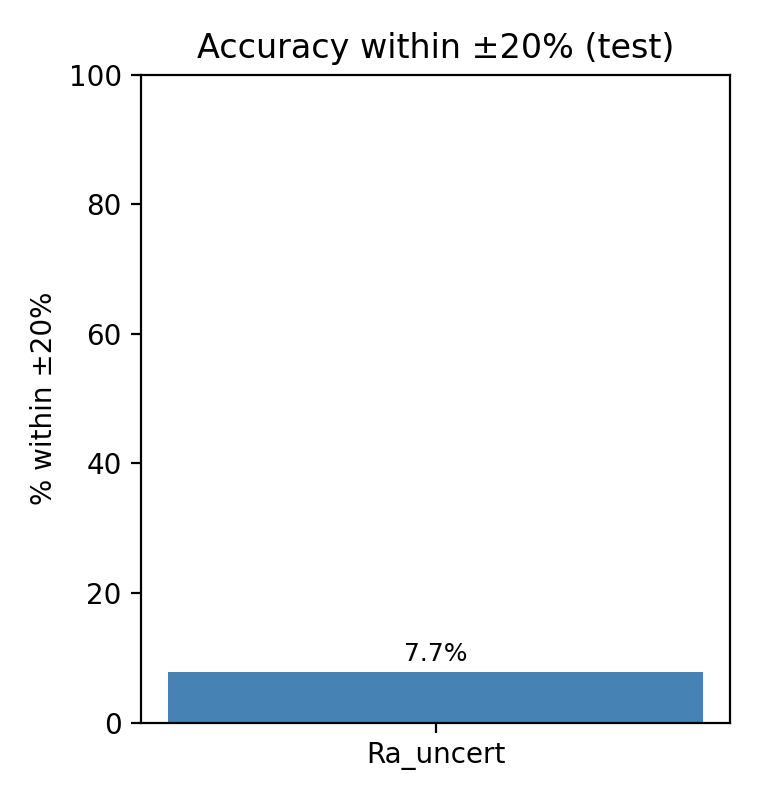}{Figure S96: accuracy within tol 20percent (\label{fig:supp-96})}\
\suppimage{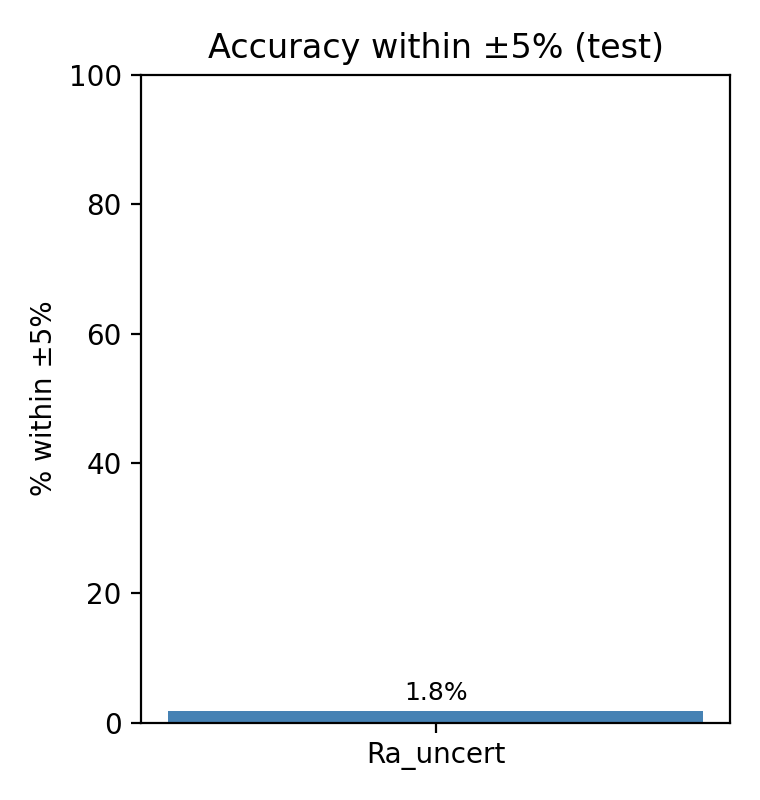}{Figure S97: accuracy within tol 5percent (\label{fig:supp-97})}\hfill
\suppimage{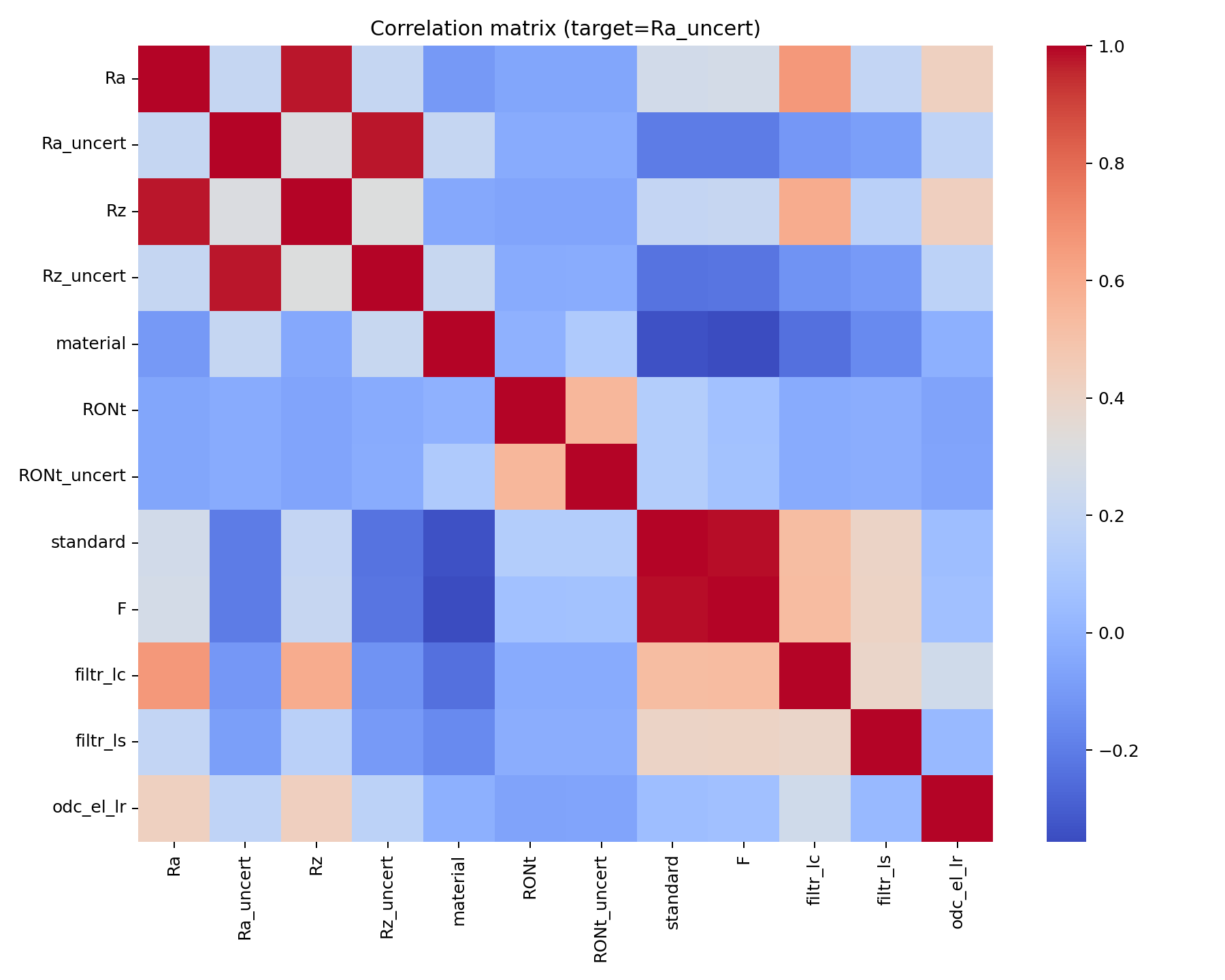}{Figure S98: correlation heatmap Ra uncert (\label{fig:supp-98})}\
\suppimage{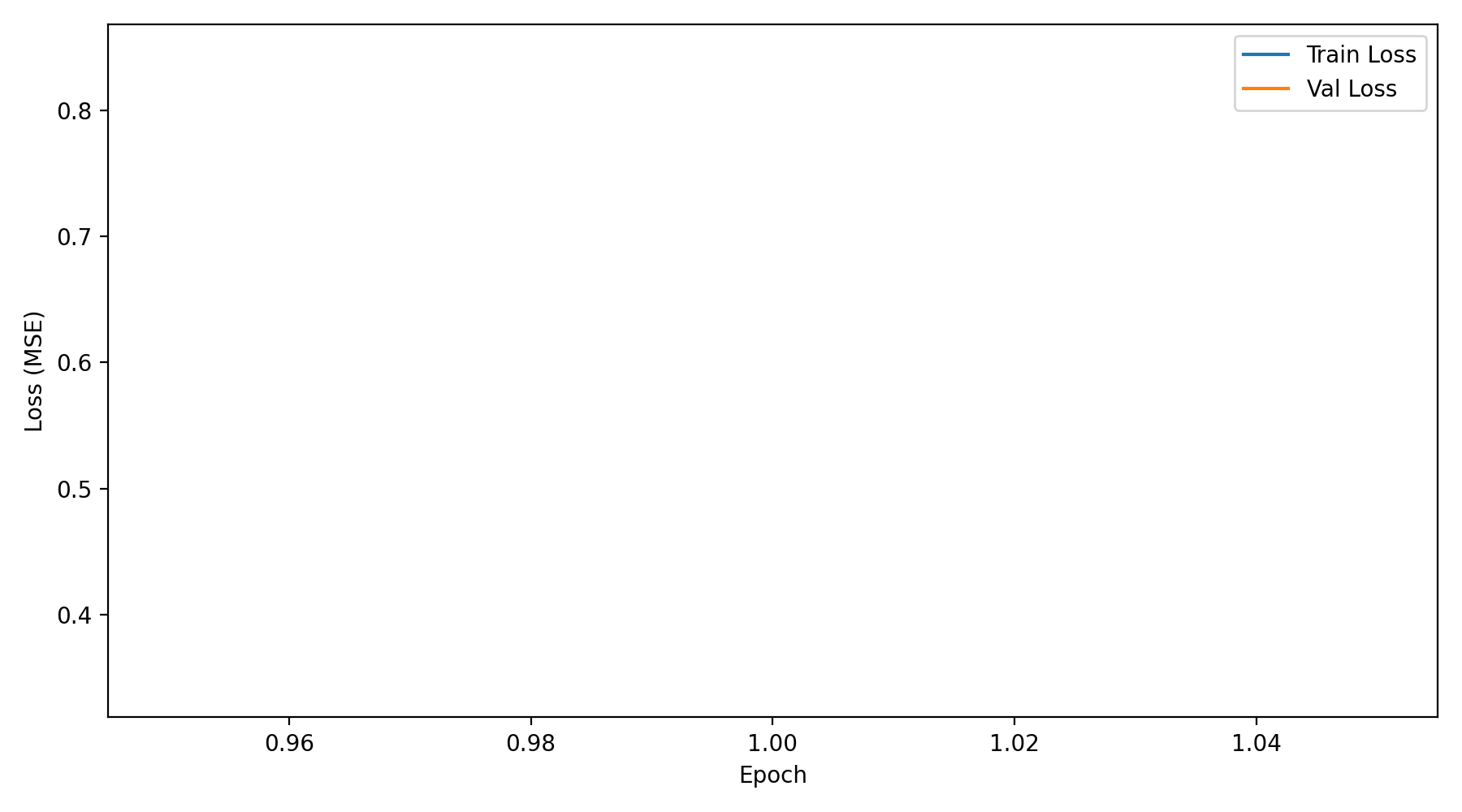}{Figure S99: loss curves (\label{fig:supp-99})}\hfill
\suppimage{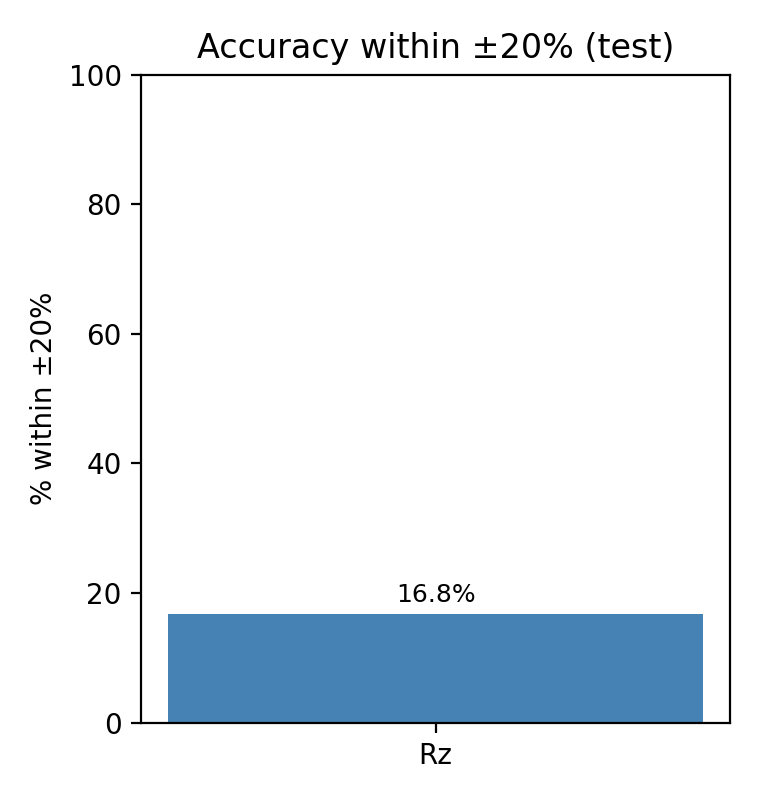}{Figure S100: accuracy within tol 20percent (\label{fig:supp-100})}\
\suppimage{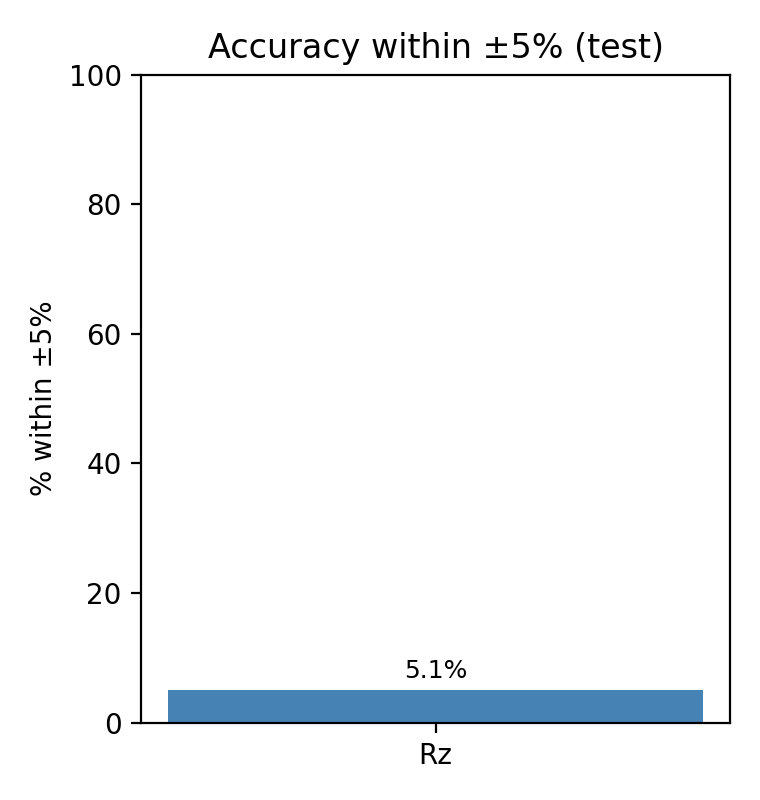}{Figure S101: accuracy within tol 5percent (\label{fig:supp-101})}\hfill
\suppimage{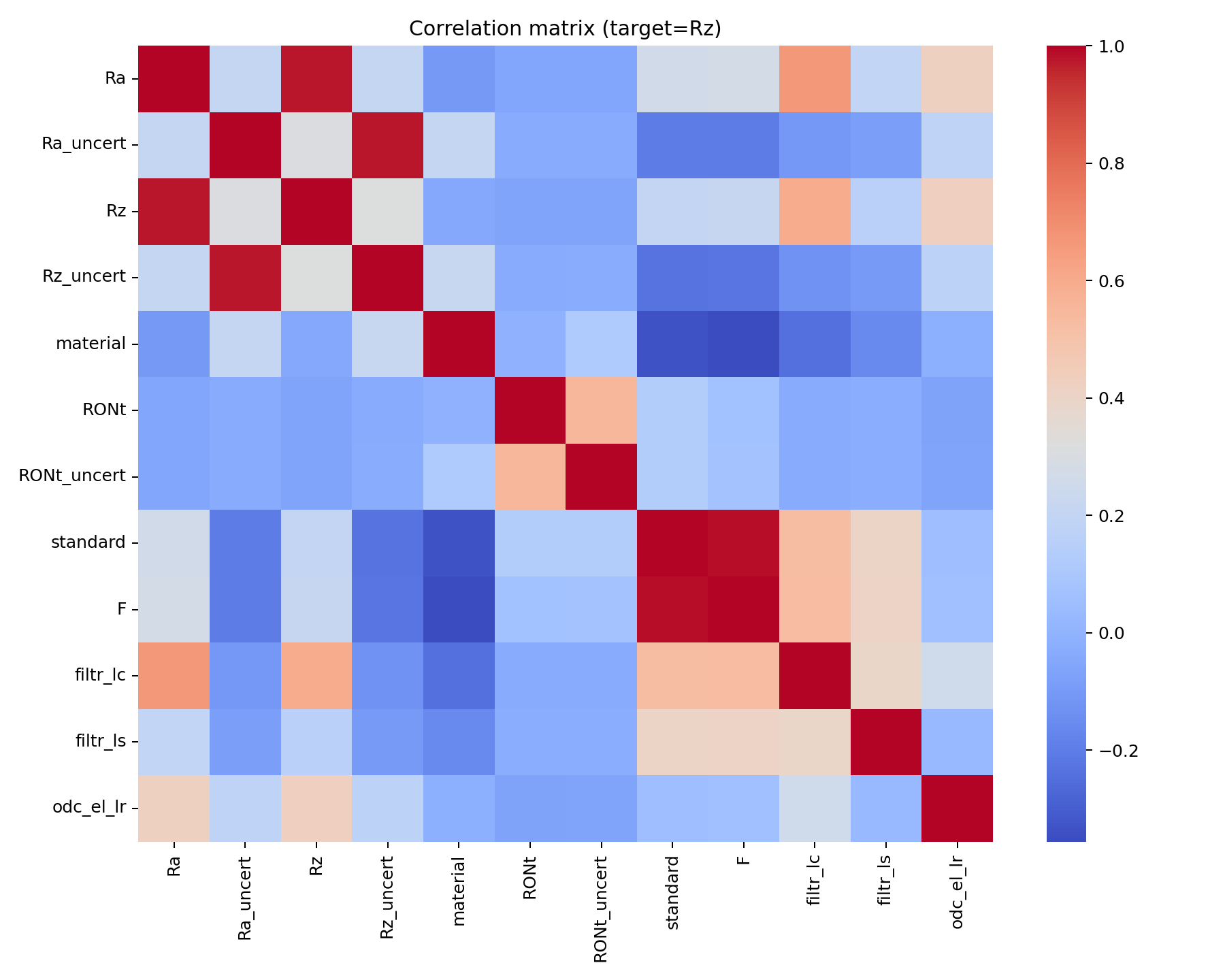}{Figure S102: correlation heatmap Rz (\label{fig:supp-102})}\
\suppimage{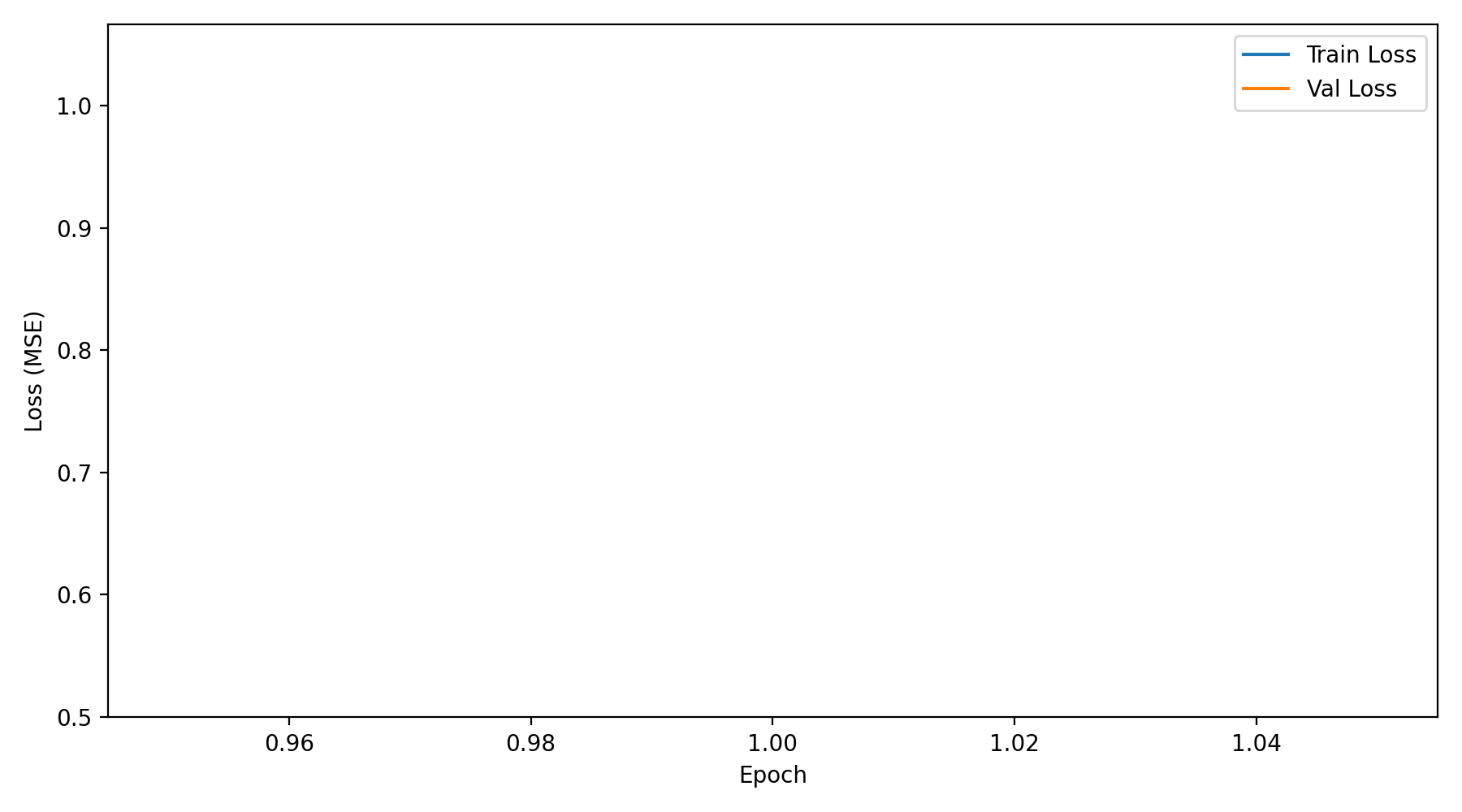}{Figure S103: loss curves (\label{fig:supp-103})}\hfill
\suppimage{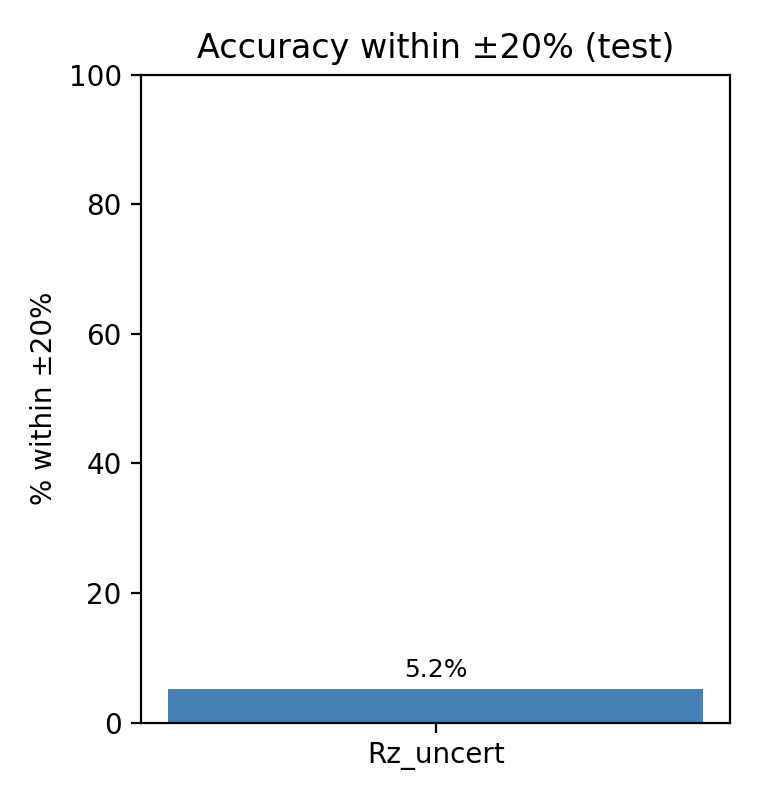}{Figure S104: accuracy within tol 20percent (\label{fig:supp-104})}\
\suppimage{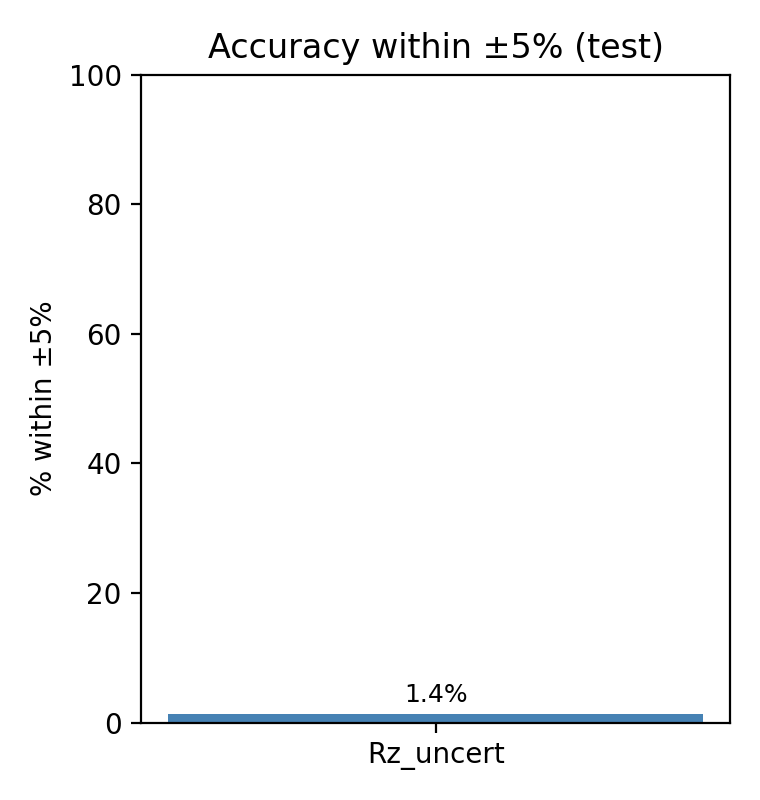}{Figure S105: accuracy within tol 5percent (\label{fig:supp-105})}\hfill
\suppimage{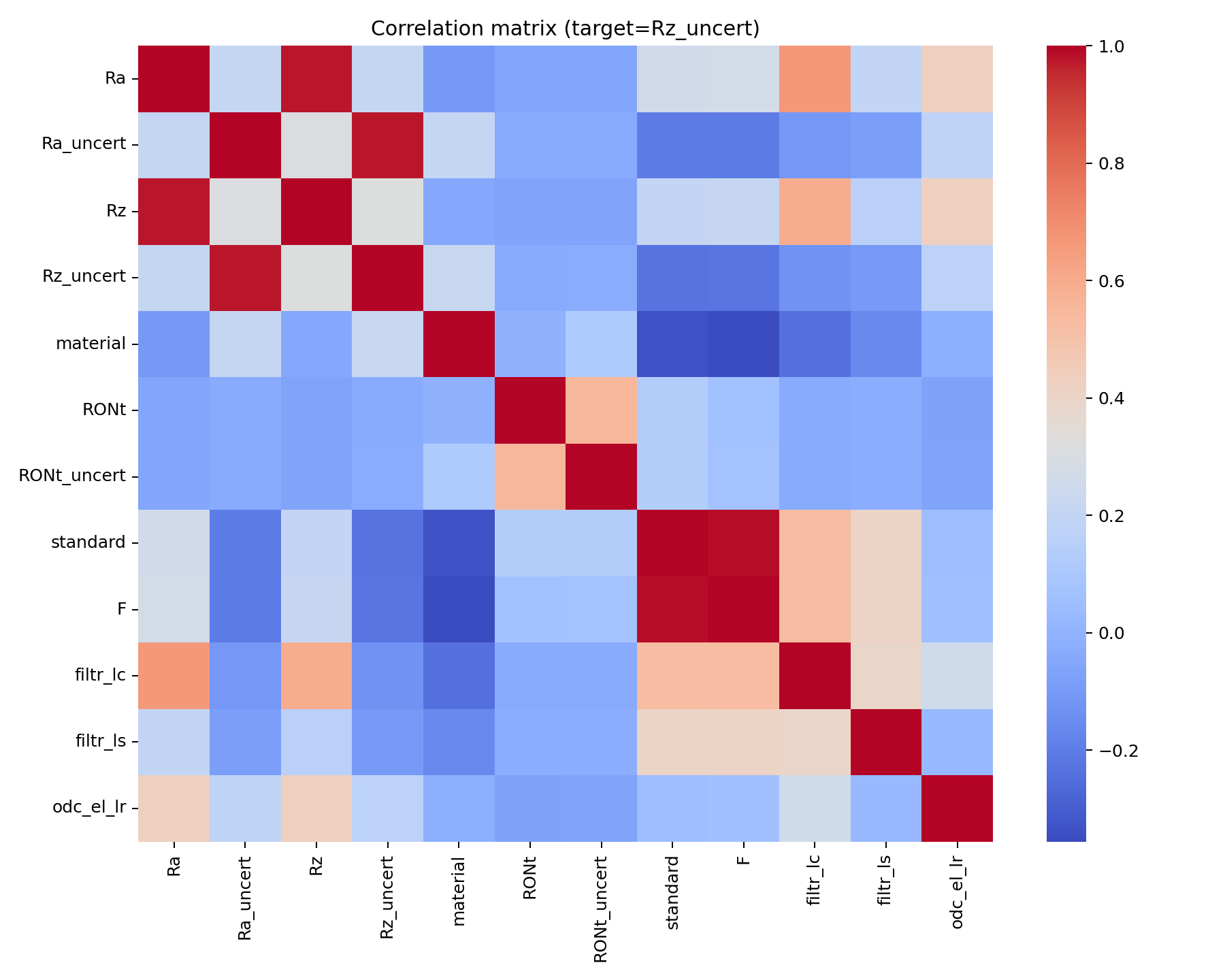}{Figure S106: correlation heatmap Rz uncert (\label{fig:supp-106})}\
\suppimage{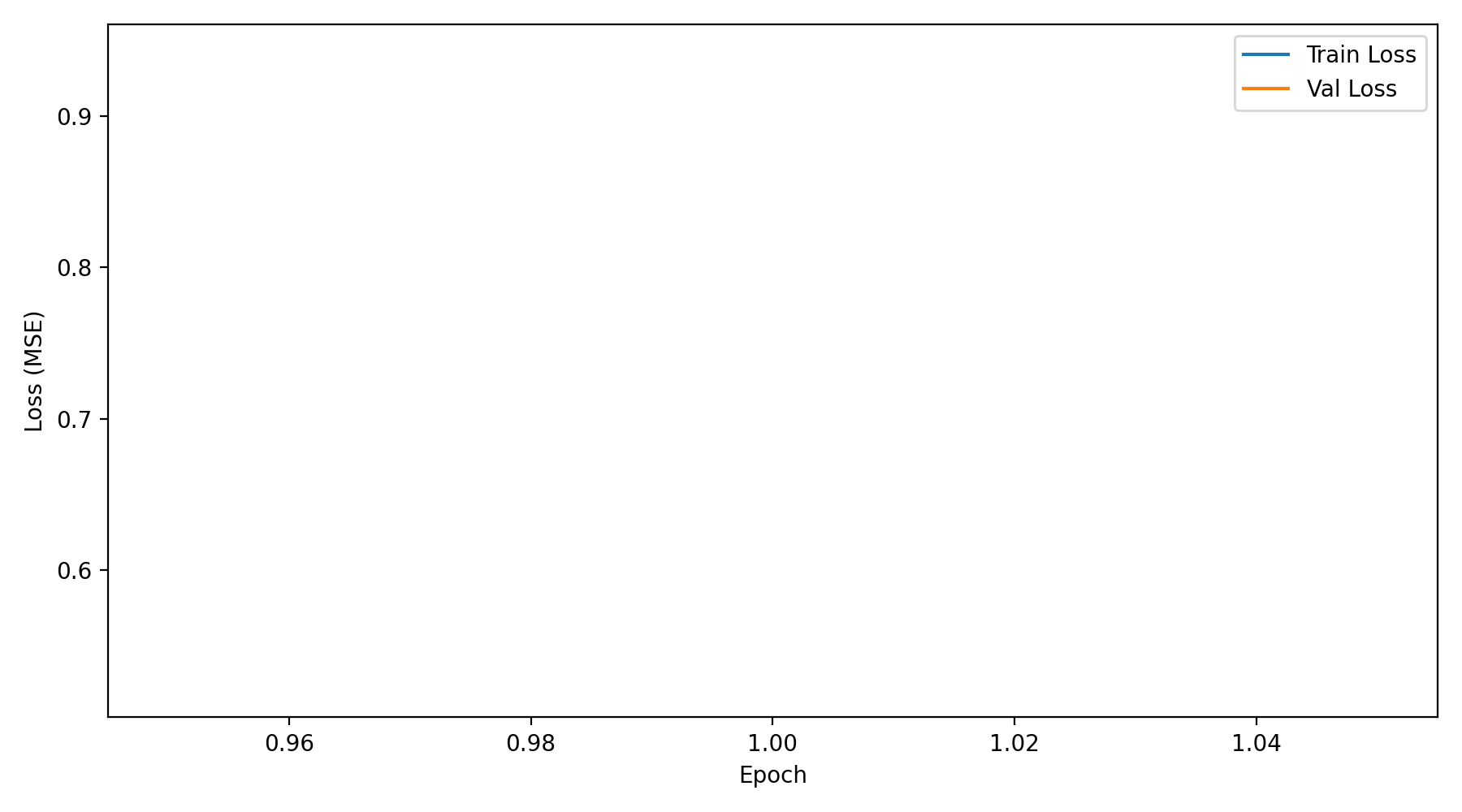}{Figure S107: loss curves (\label{fig:supp-107})}\hfill
\suppimage{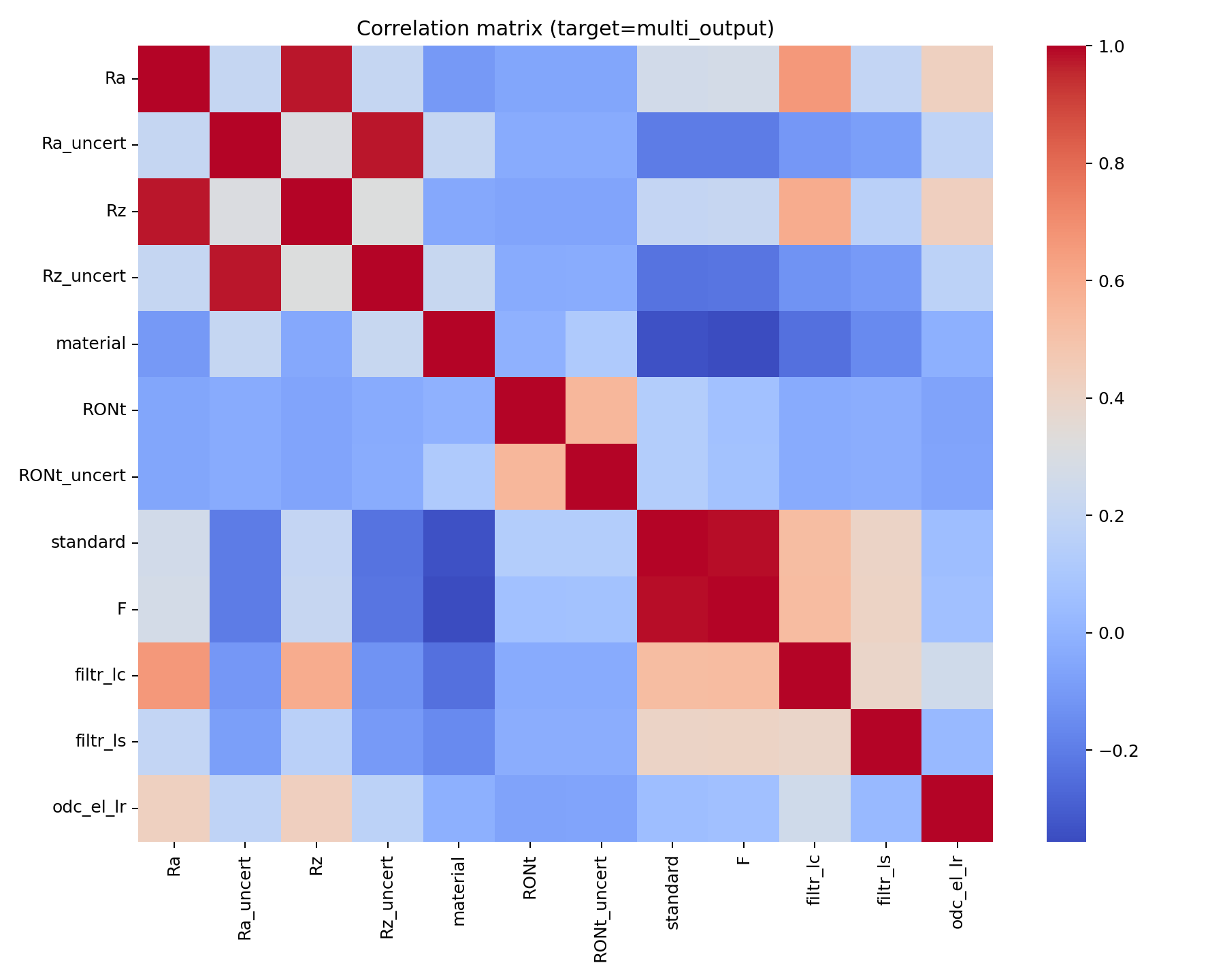}{Figure S108: correlation heatmap multi output (\label{fig:supp-108})}\
\suppimage{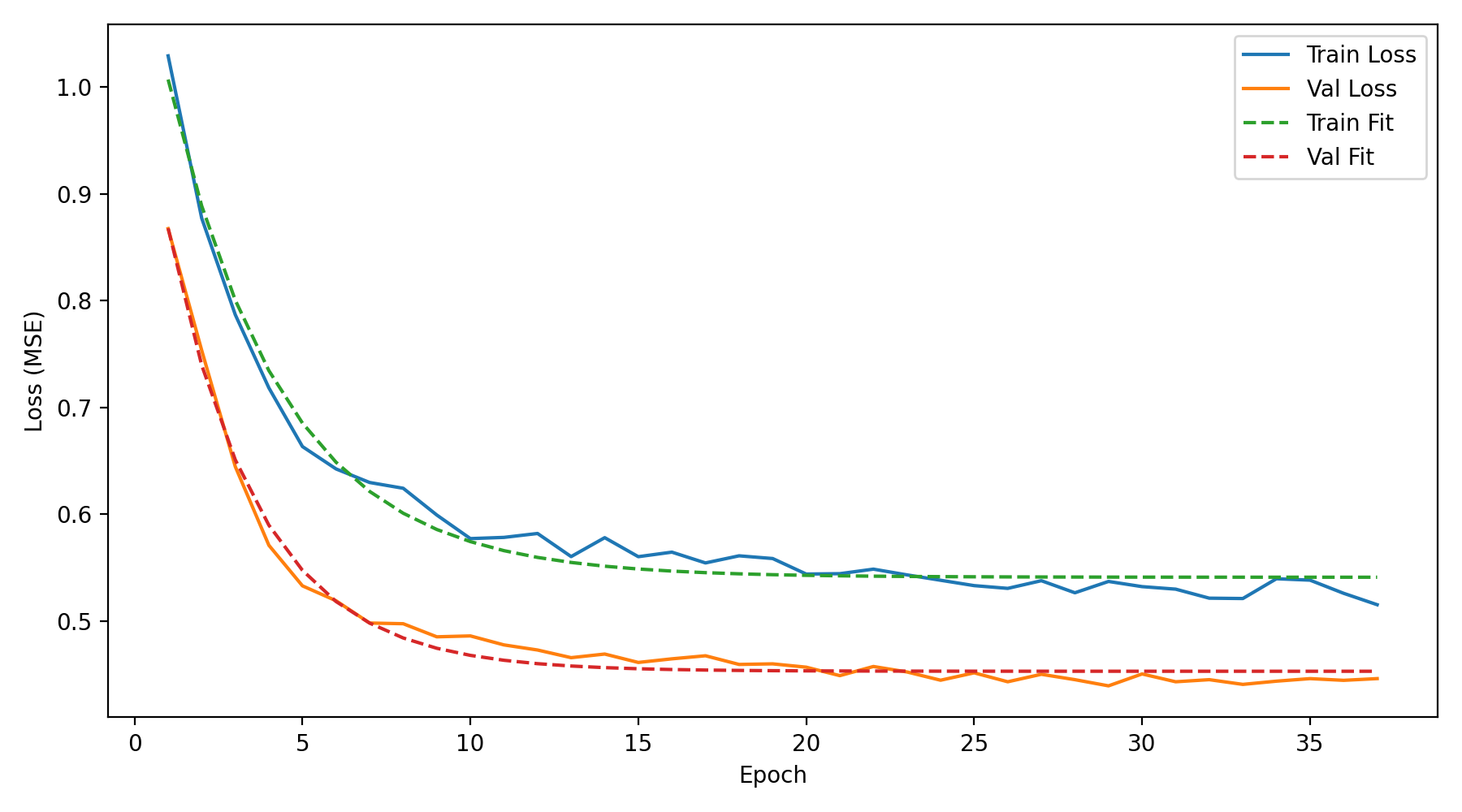}{Figure S109: loss curves (\label{fig:supp-109})}\hfill
\suppimage{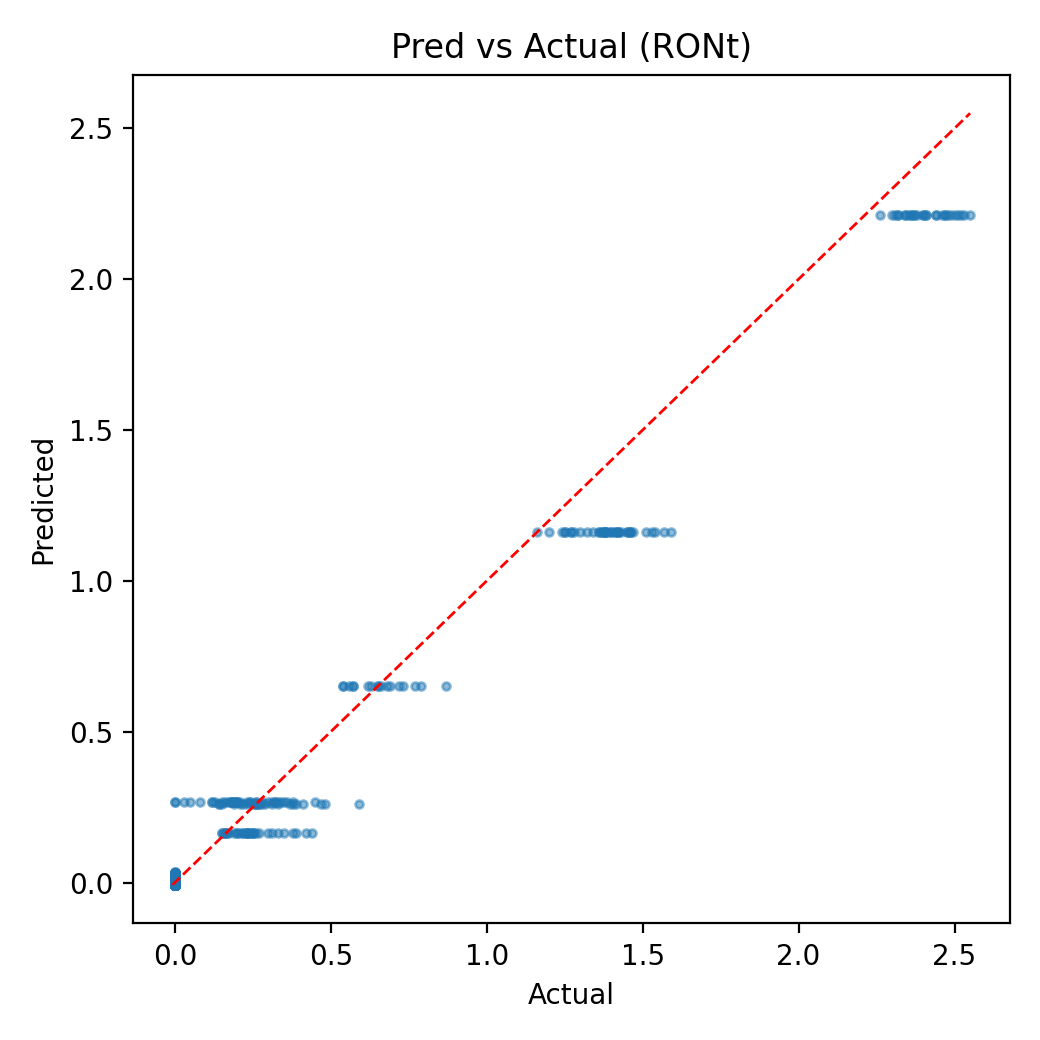}{Figure S110: pred vs actual RONt (\label{fig:supp-110})}\
\suppimage{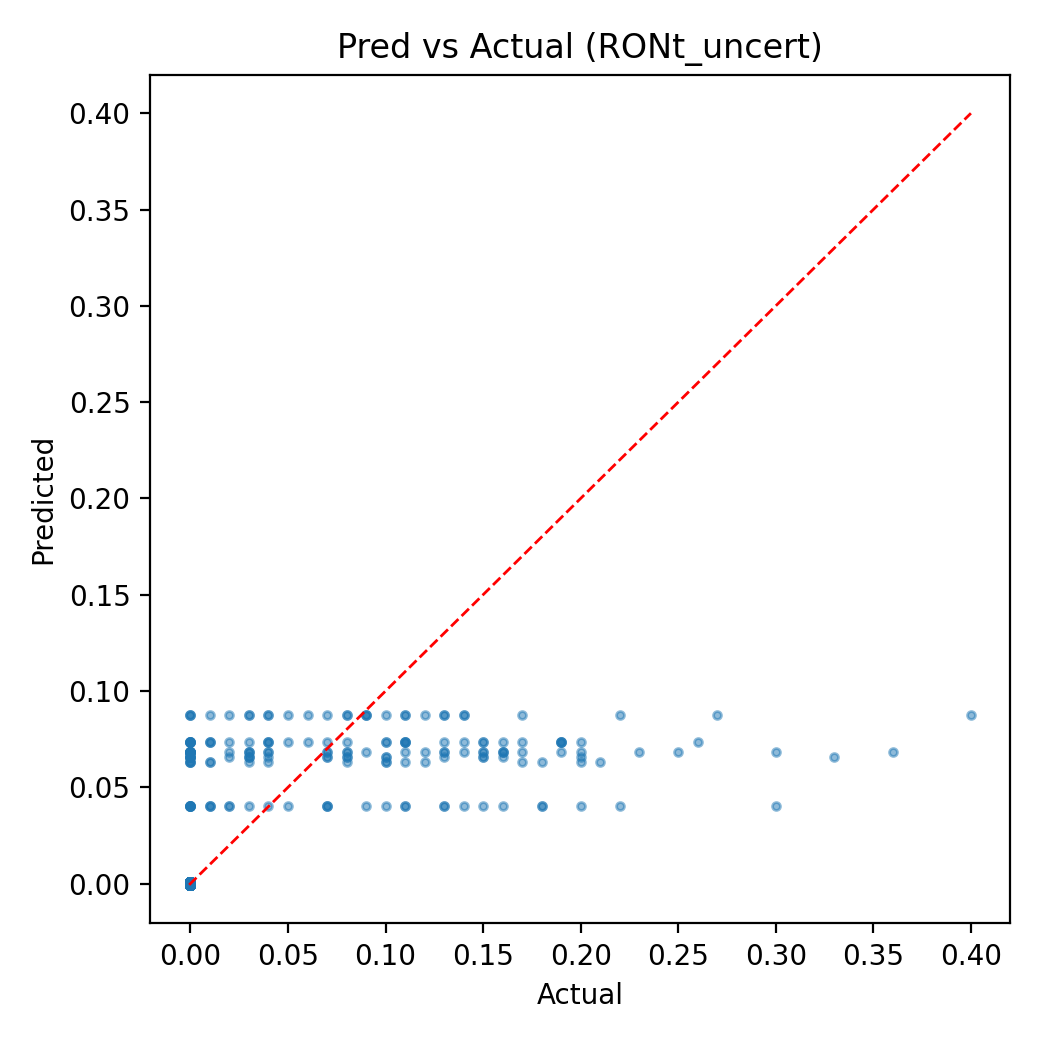}{Figure S111: pred vs actual RONt uncert (\label{fig:supp-111})}\hfill
\suppimage{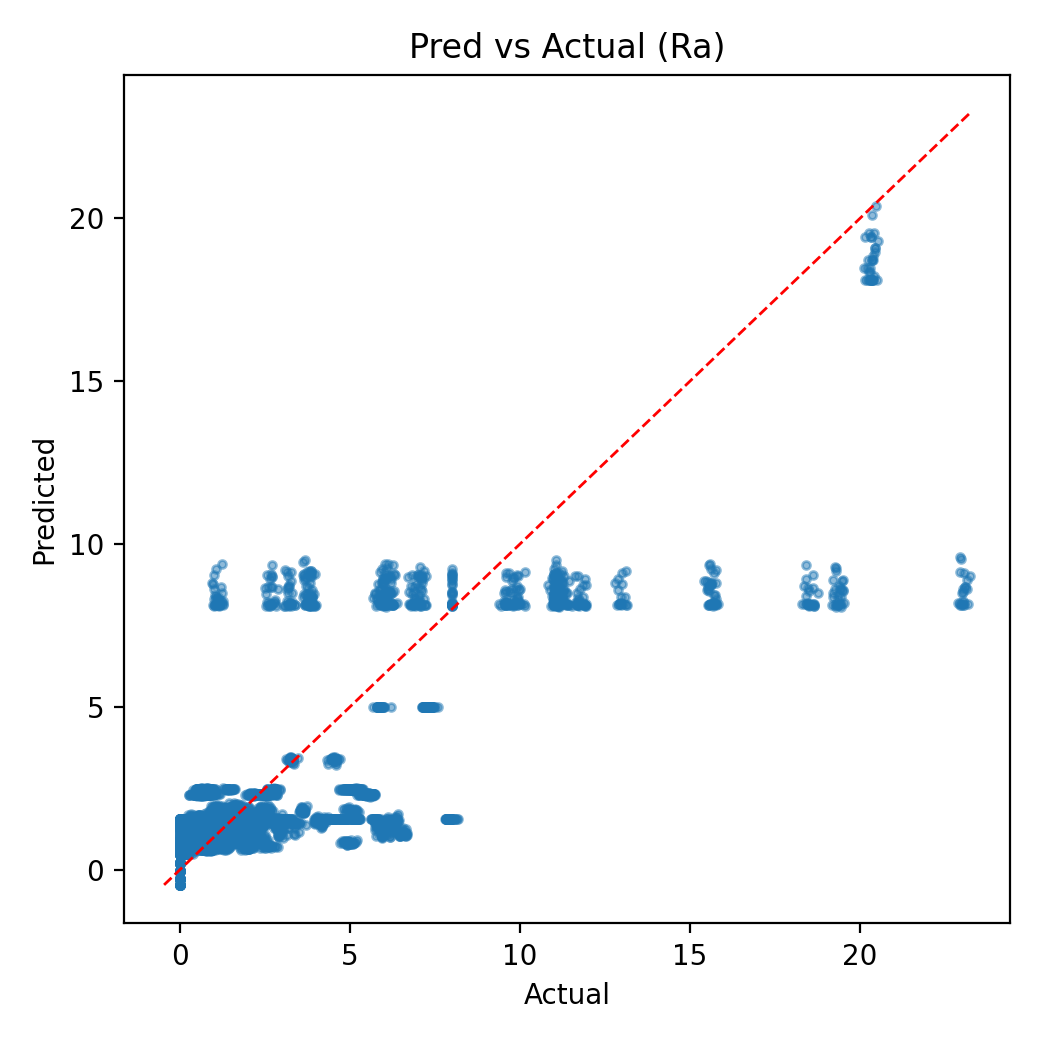}{Figure S112: pred vs actual Ra (\label{fig:supp-112})}\
\suppimage{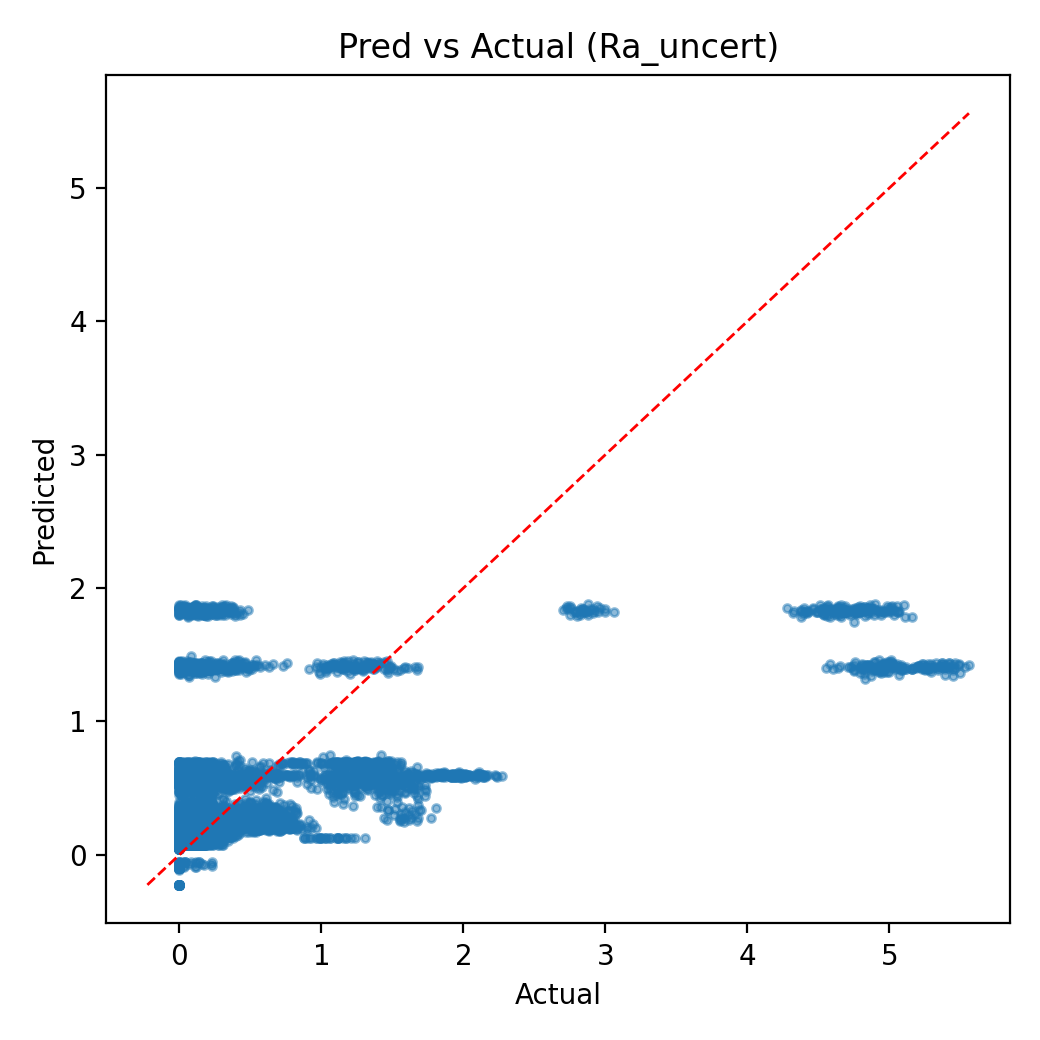}{Figure S113: pred vs actual Ra uncert (\label{fig:supp-113})}\hfill
\suppimage{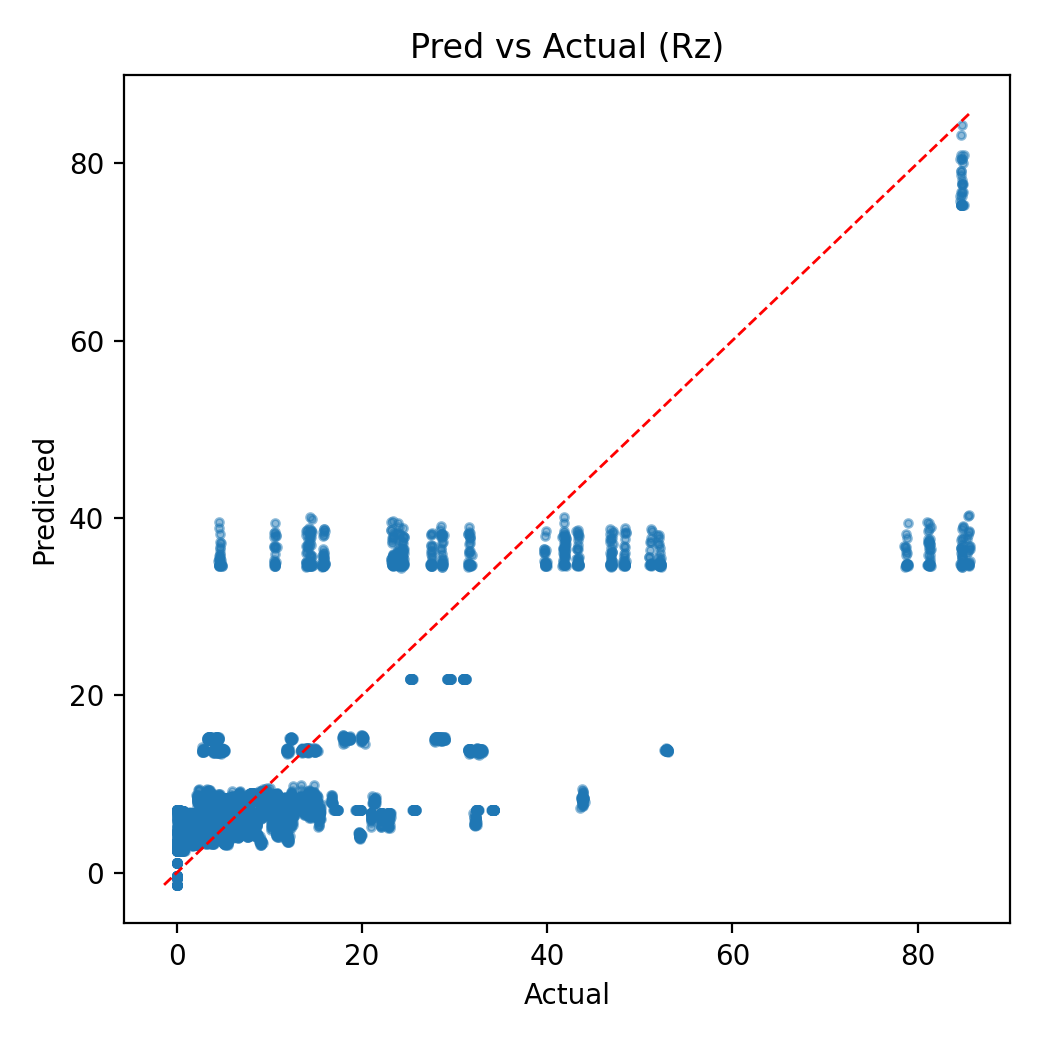}{Figure S114: pred vs actual Rz (\label{fig:supp-114})}\
\suppimage{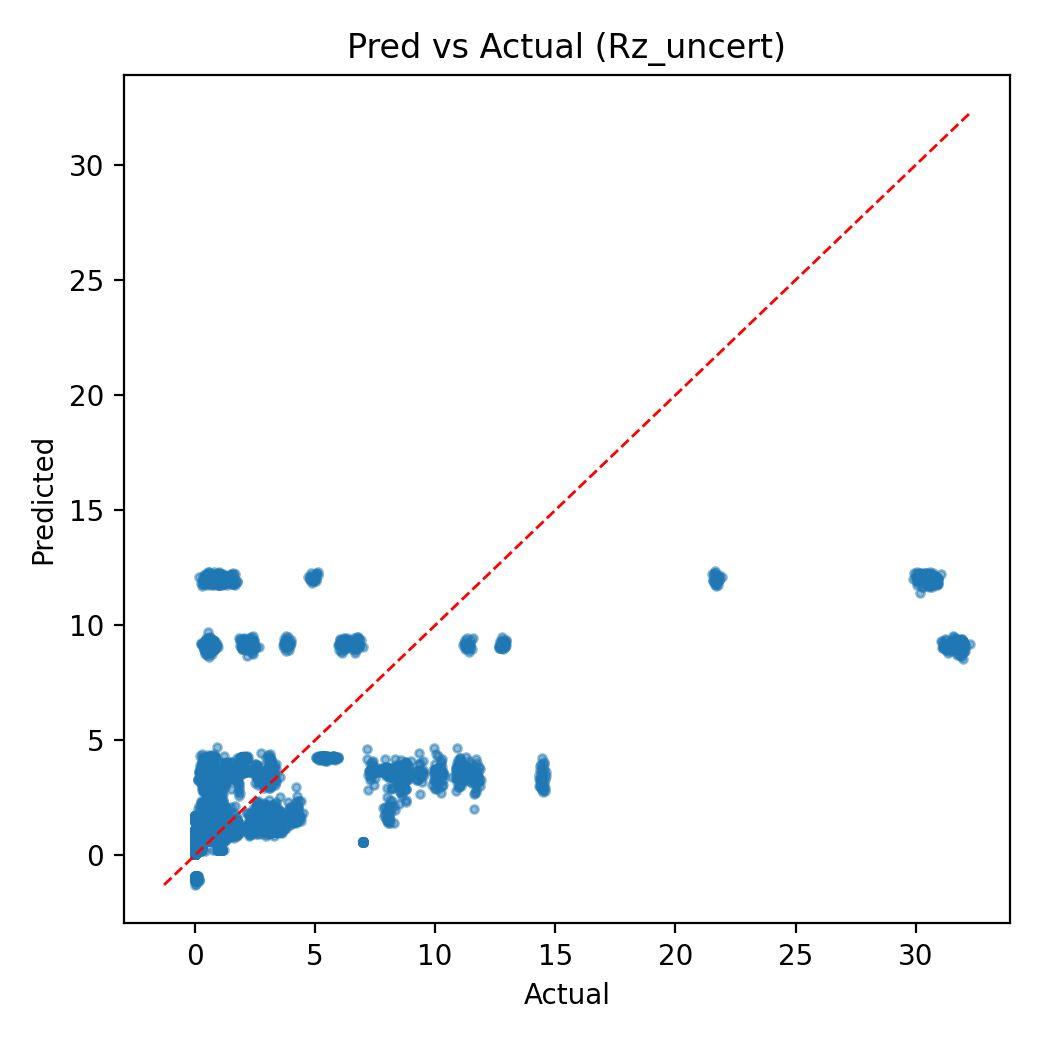}{Figure S115: pred vs actual Rz uncert (\label{fig:supp-115})}\hfill
\suppimage{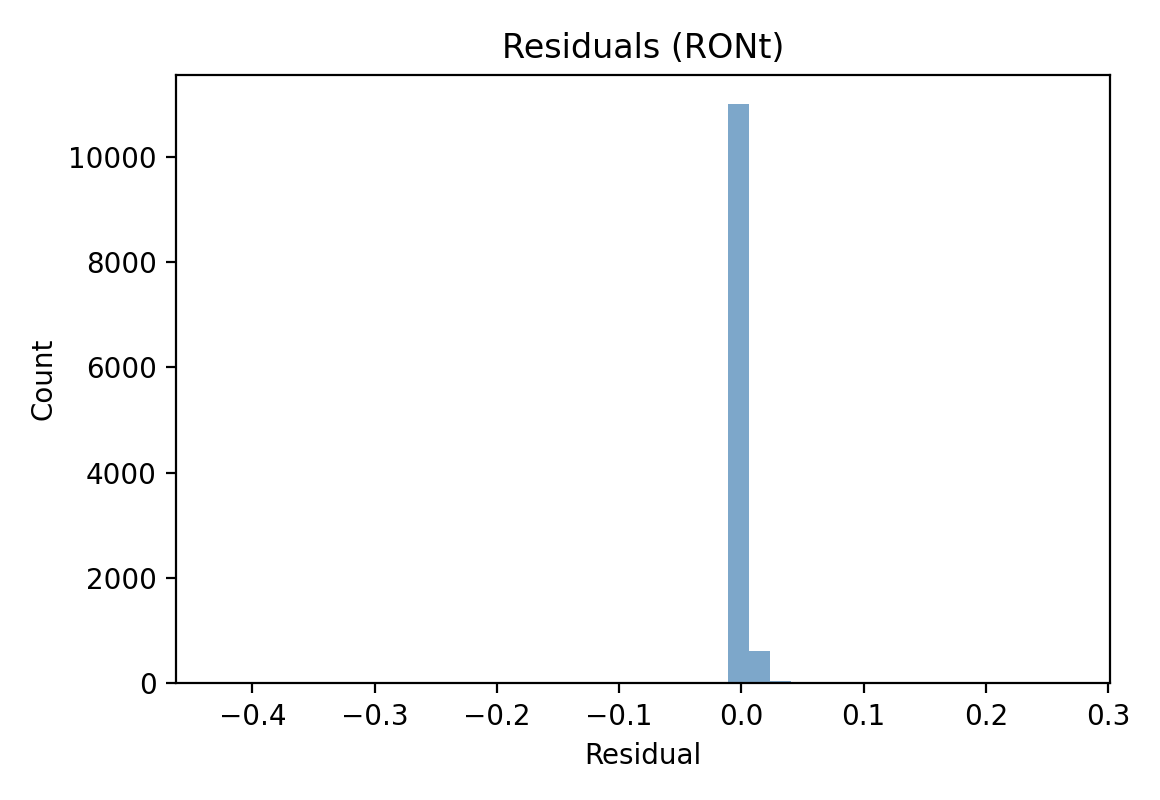}{Figure S116: residuals hist RONt (\label{fig:supp-116})}\
\suppimage{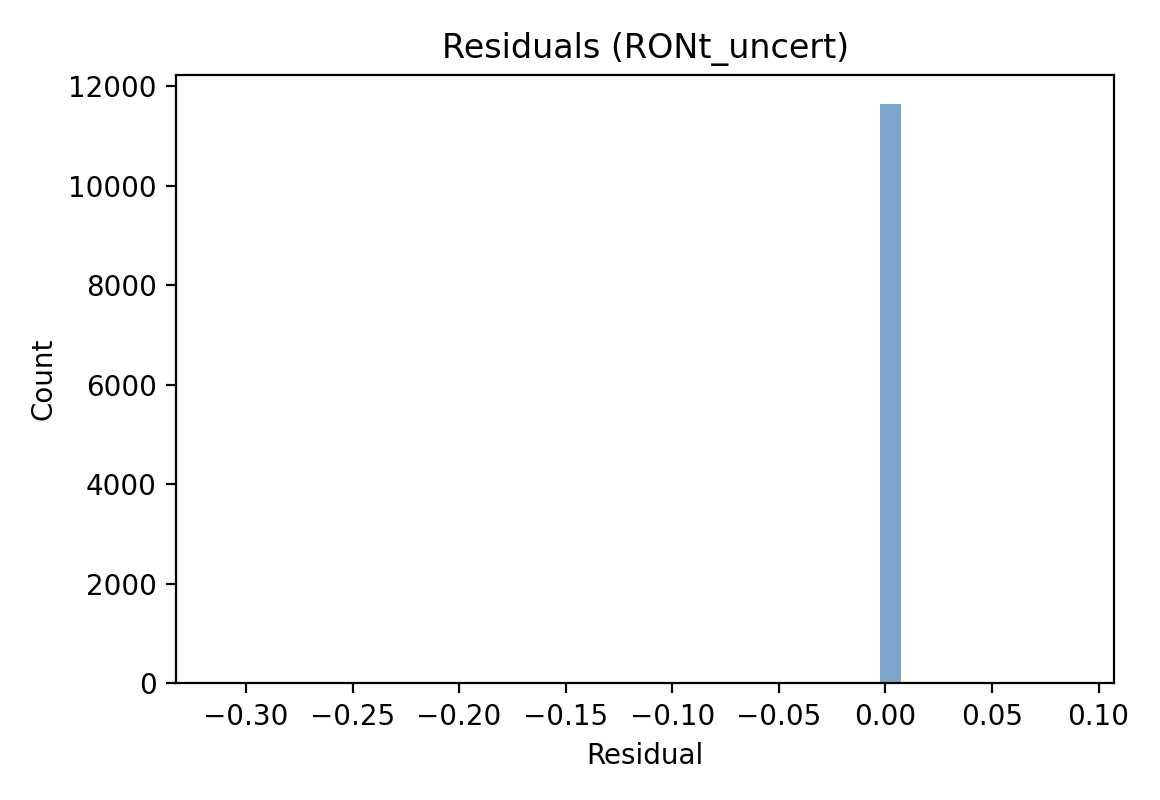}{Figure S117: residuals hist RONt uncert (\label{fig:supp-117})}\hfill
\suppimage{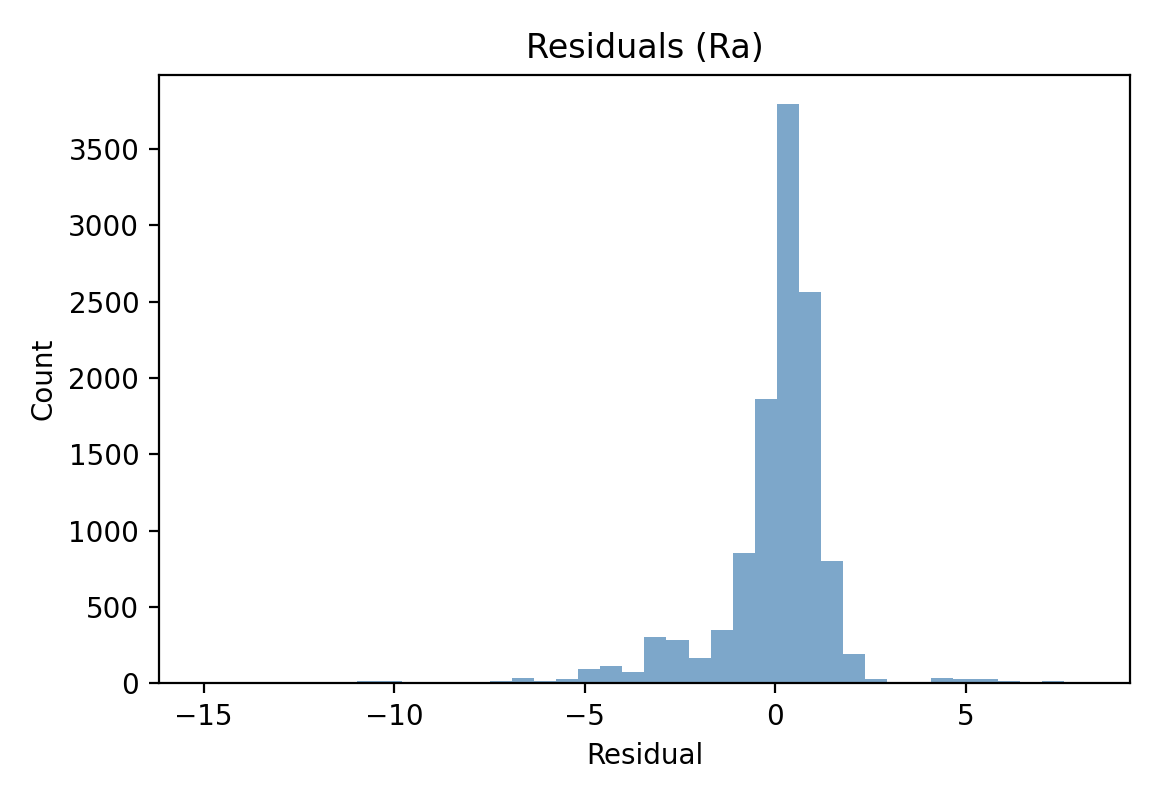}{Figure S118: residuals hist Ra (\label{fig:supp-118})}\
\suppimage{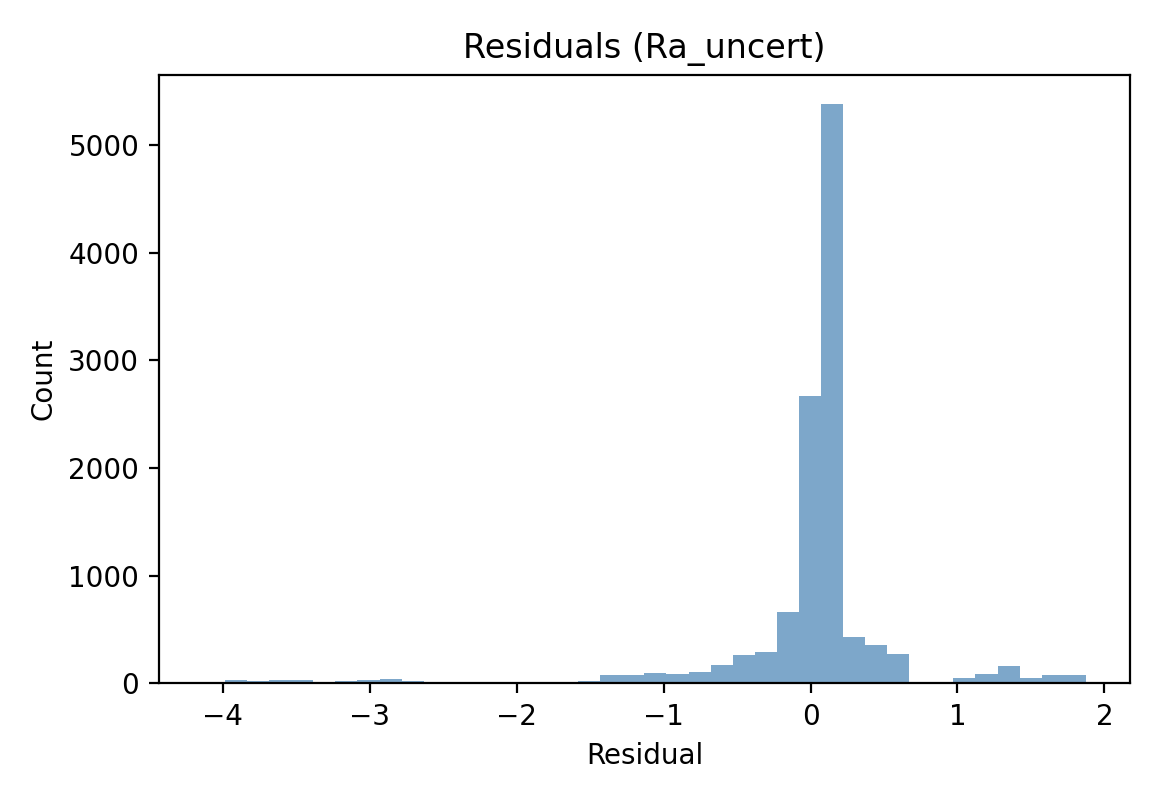}{Figure S119: residuals hist Ra uncert (\label{fig:supp-119})}\hfill
\suppimage{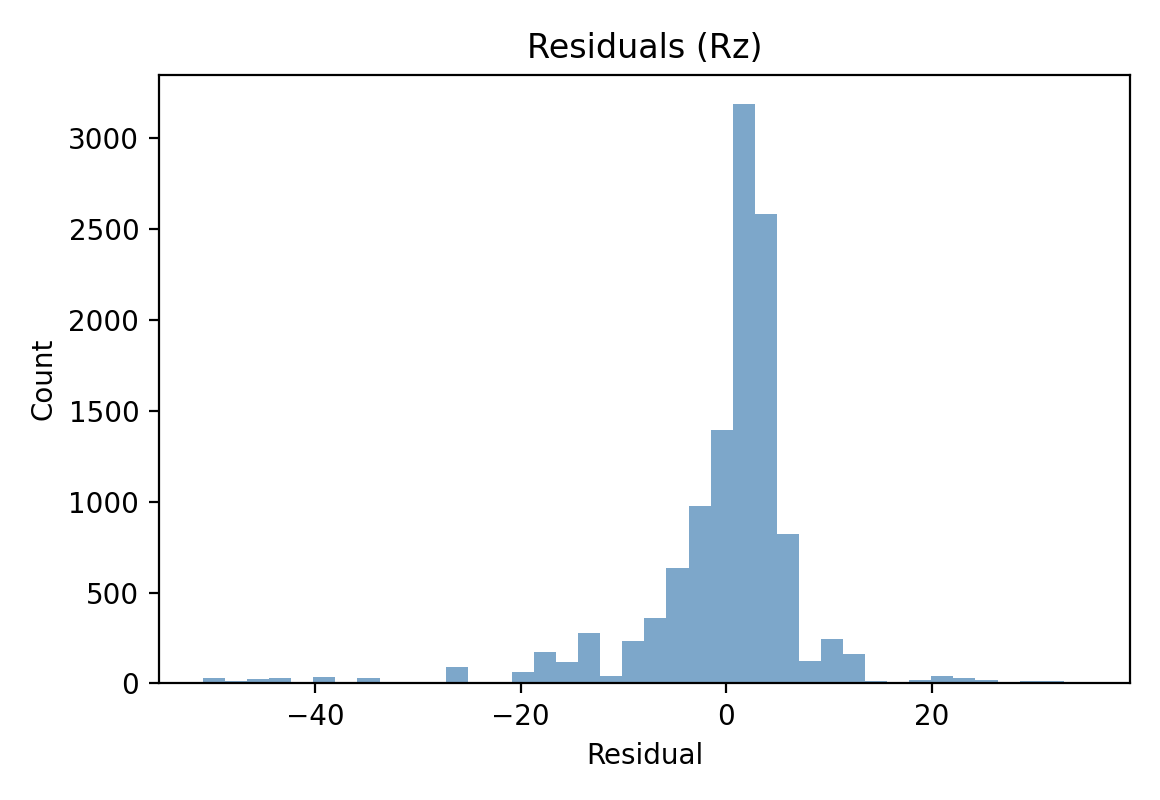}{Figure S120: residuals hist Rz (\label{fig:supp-120})}\
\suppimage{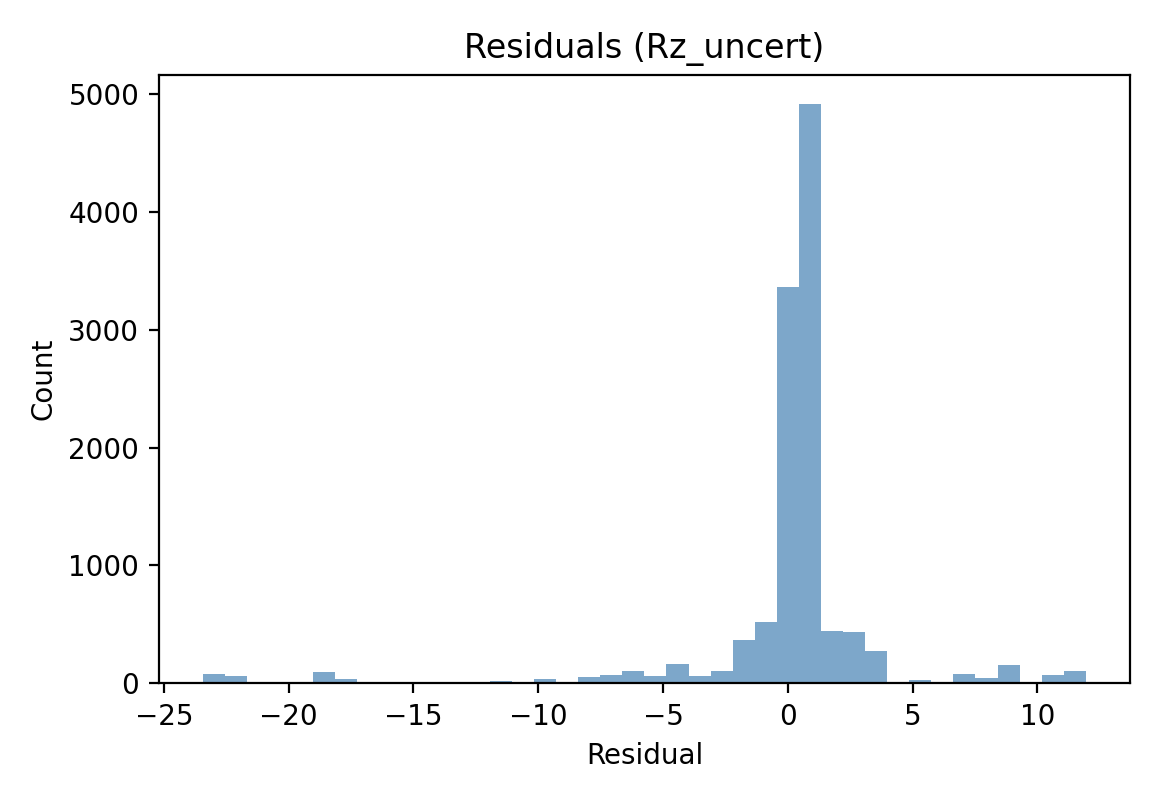}{Figure S121: residuals hist Rz uncert (\label{fig:supp-121})}\hfill
\suppimage{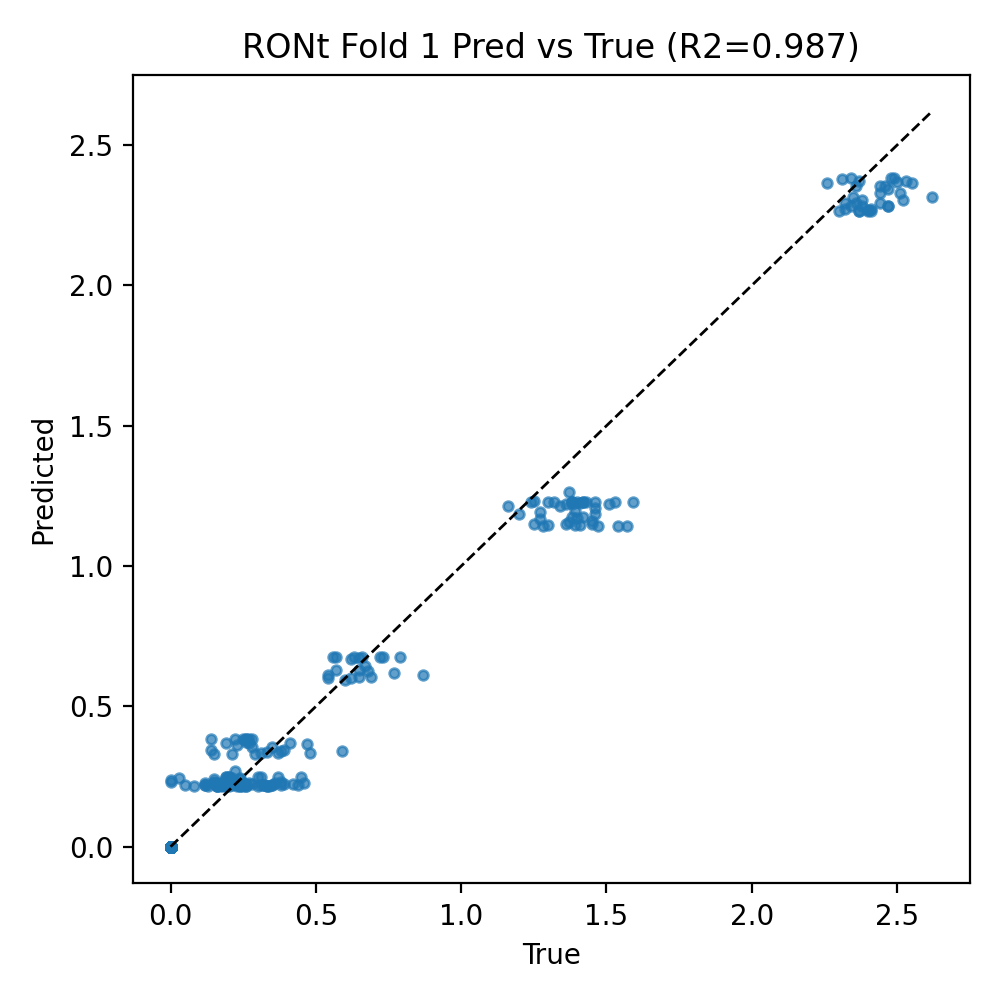}{Figure S122: pred vs true (\label{fig:supp-122})}\
\suppimage{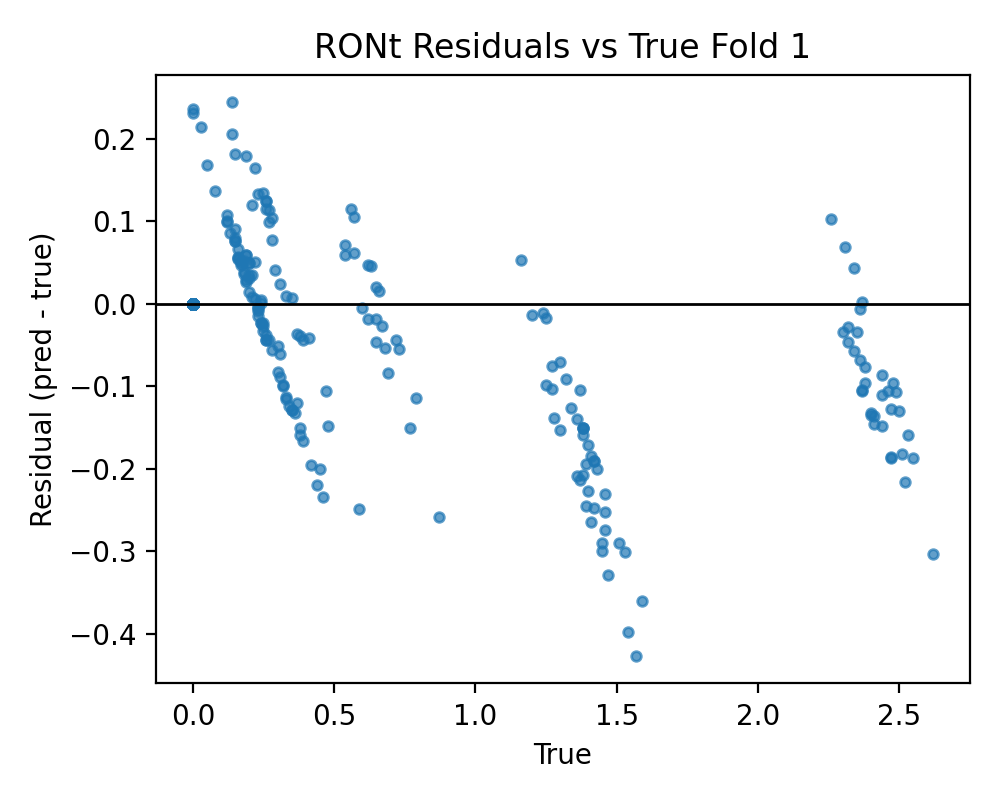}{Figure S123: residuals vs true (\label{fig:supp-123})}\hfill
\suppimage{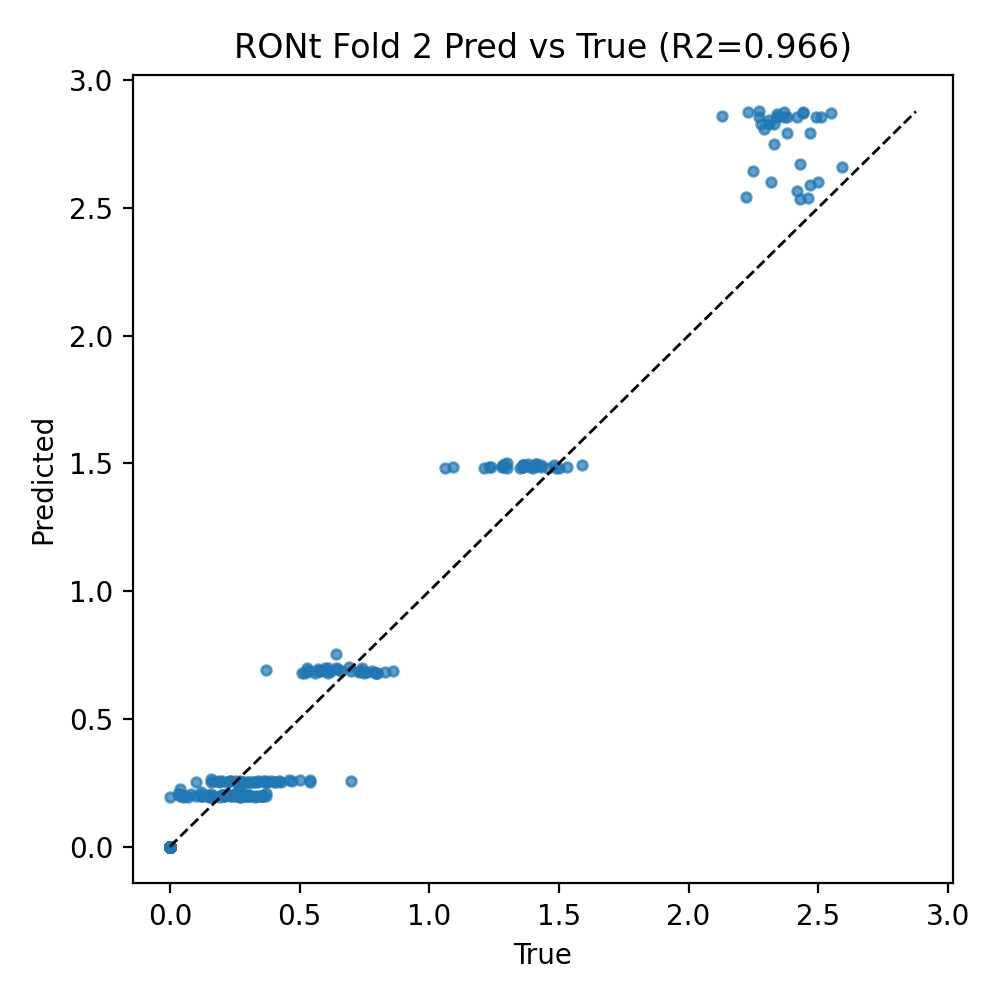}{Figure S124: pred vs true (\label{fig:supp-124})}\
\suppimage{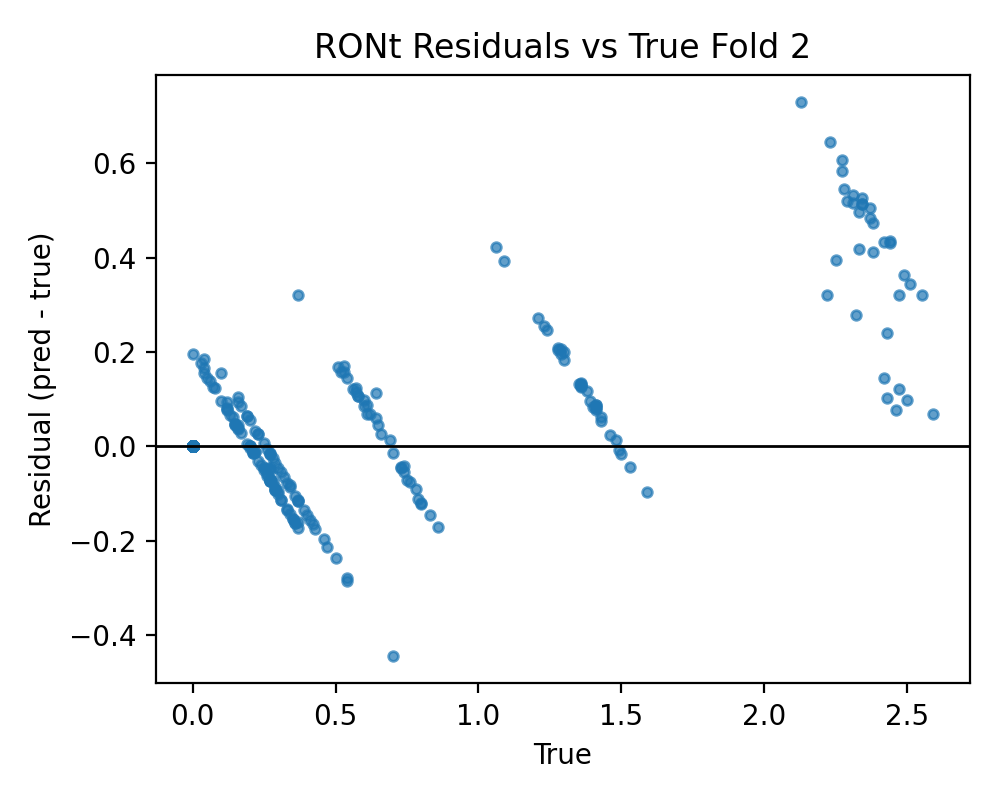}{Figure S125: residuals vs true (\label{fig:supp-125})}\hfill
\suppimage{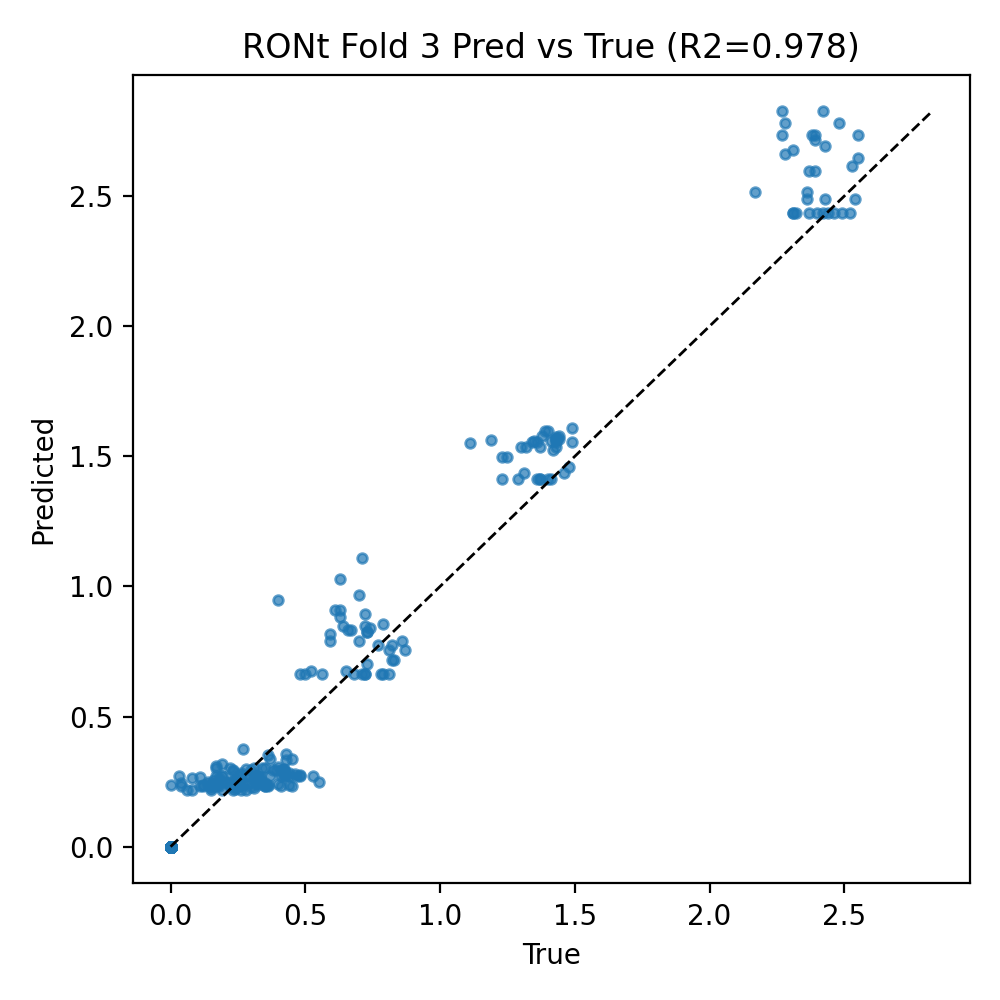}{Figure S126: pred vs true (\label{fig:supp-126})}\
\suppimage{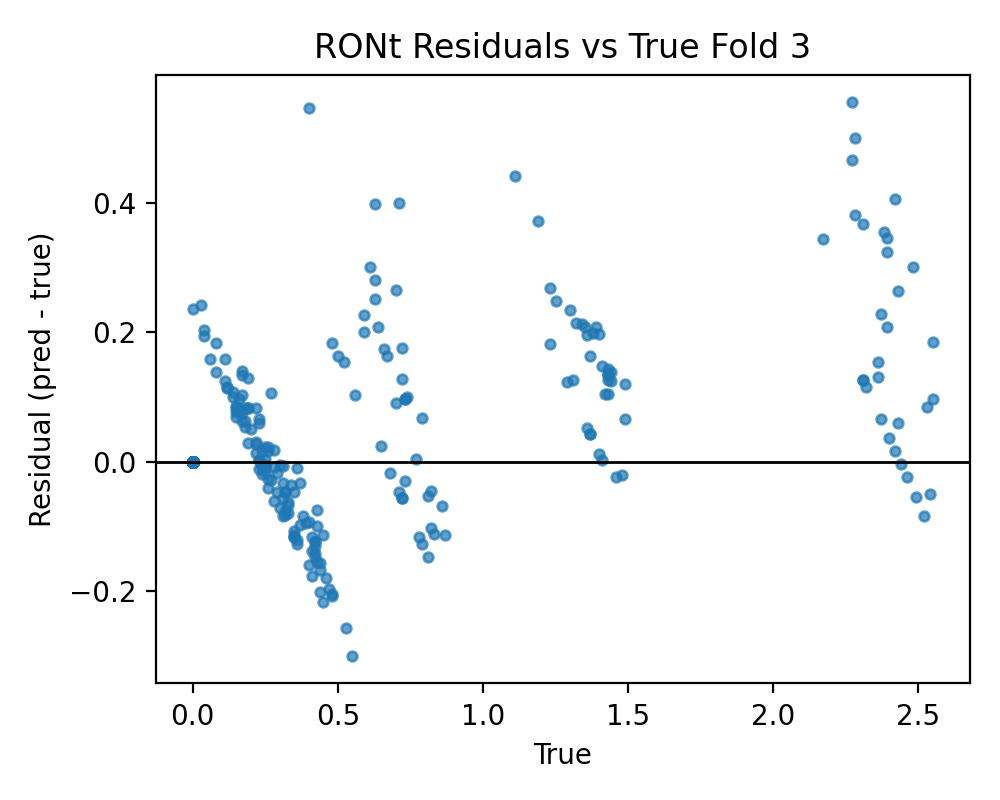}{Figure S127: residuals vs true (\label{fig:supp-127})}\hfill
\suppimage{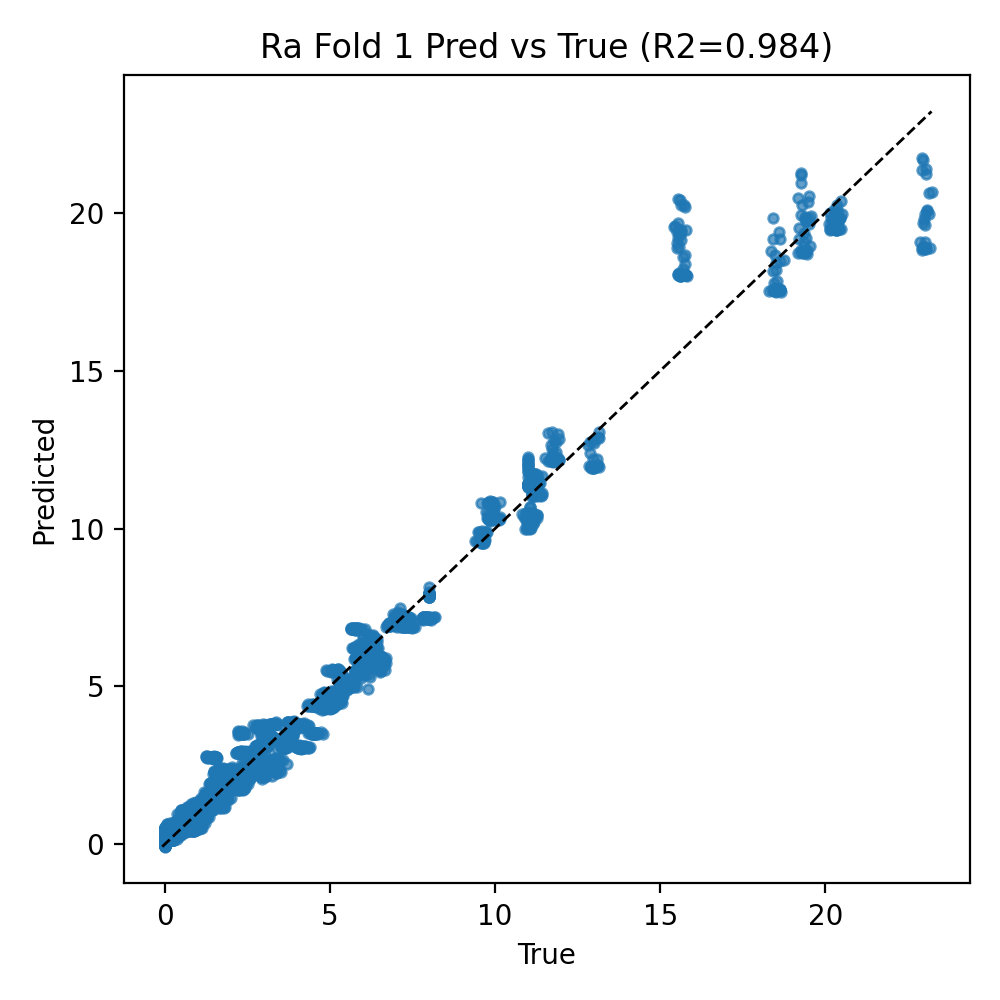}{Figure S128: pred vs true (\label{fig:supp-128})}\
\suppimage{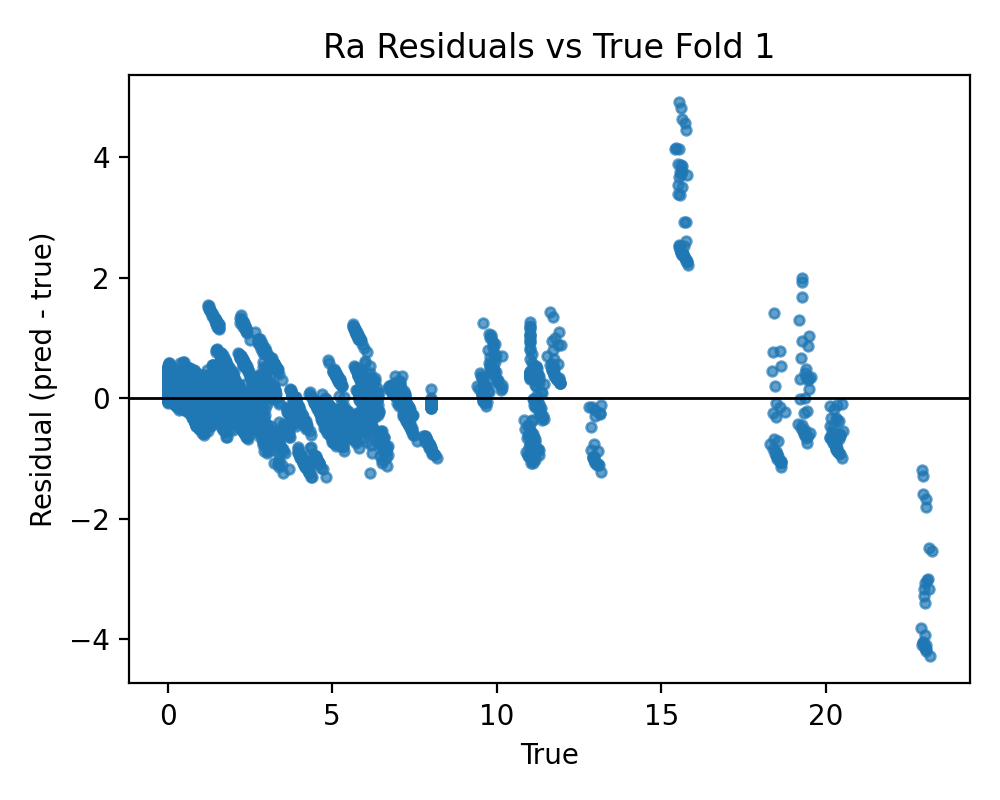}{Figure S129: residuals vs true (\label{fig:supp-129})}\hfill
\suppimage{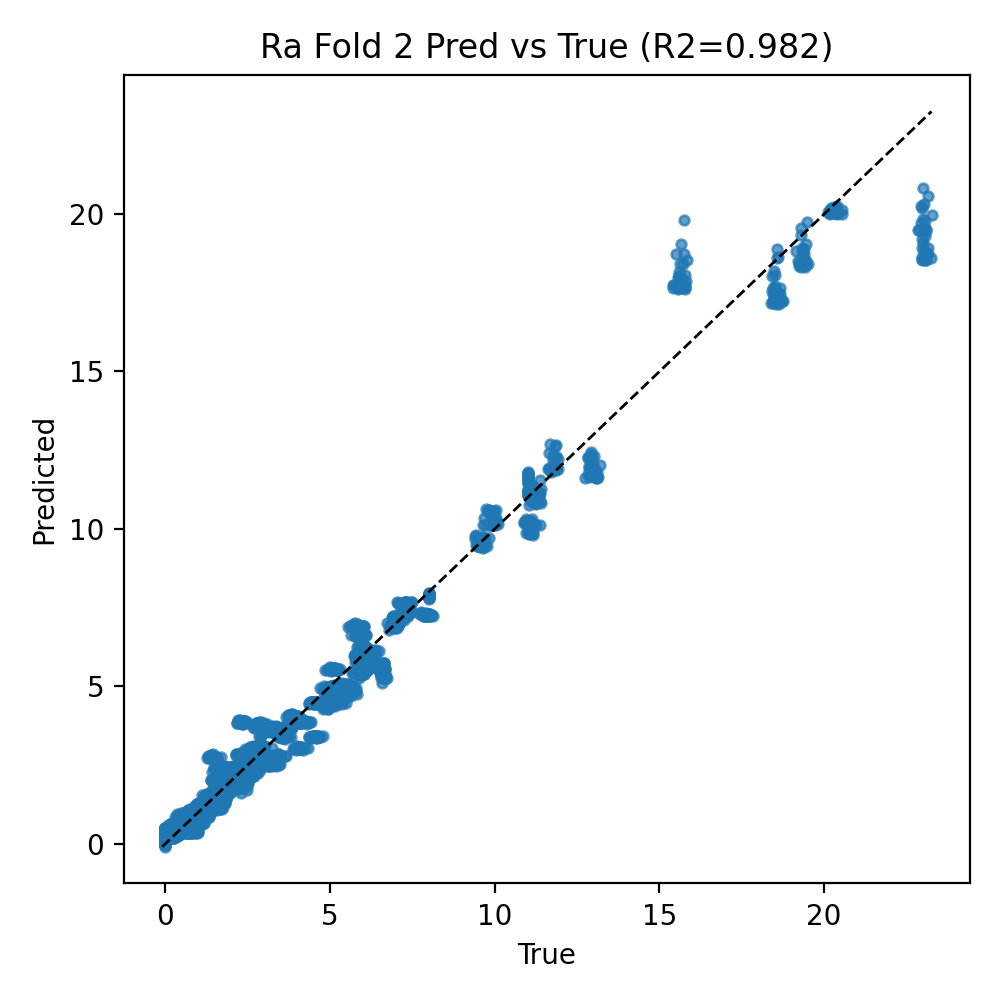}{Figure S130: pred vs true (\label{fig:supp-130})}\
\suppimage{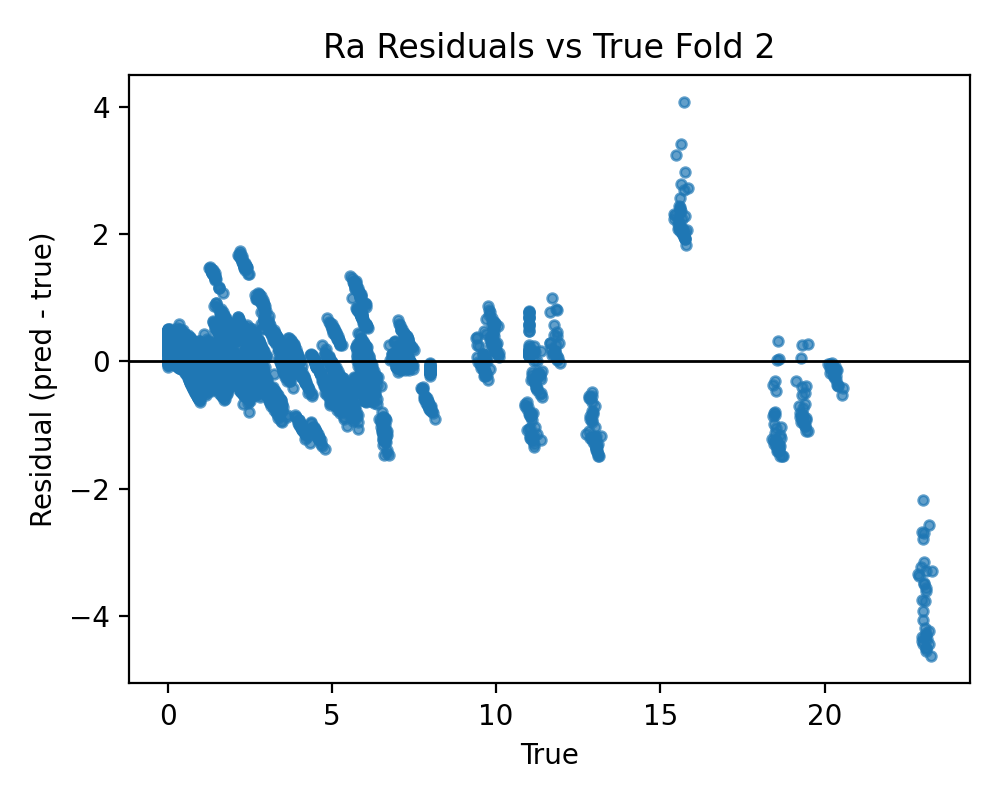}{Figure S131: residuals vs true (\label{fig:supp-131})}\hfill
\suppimage{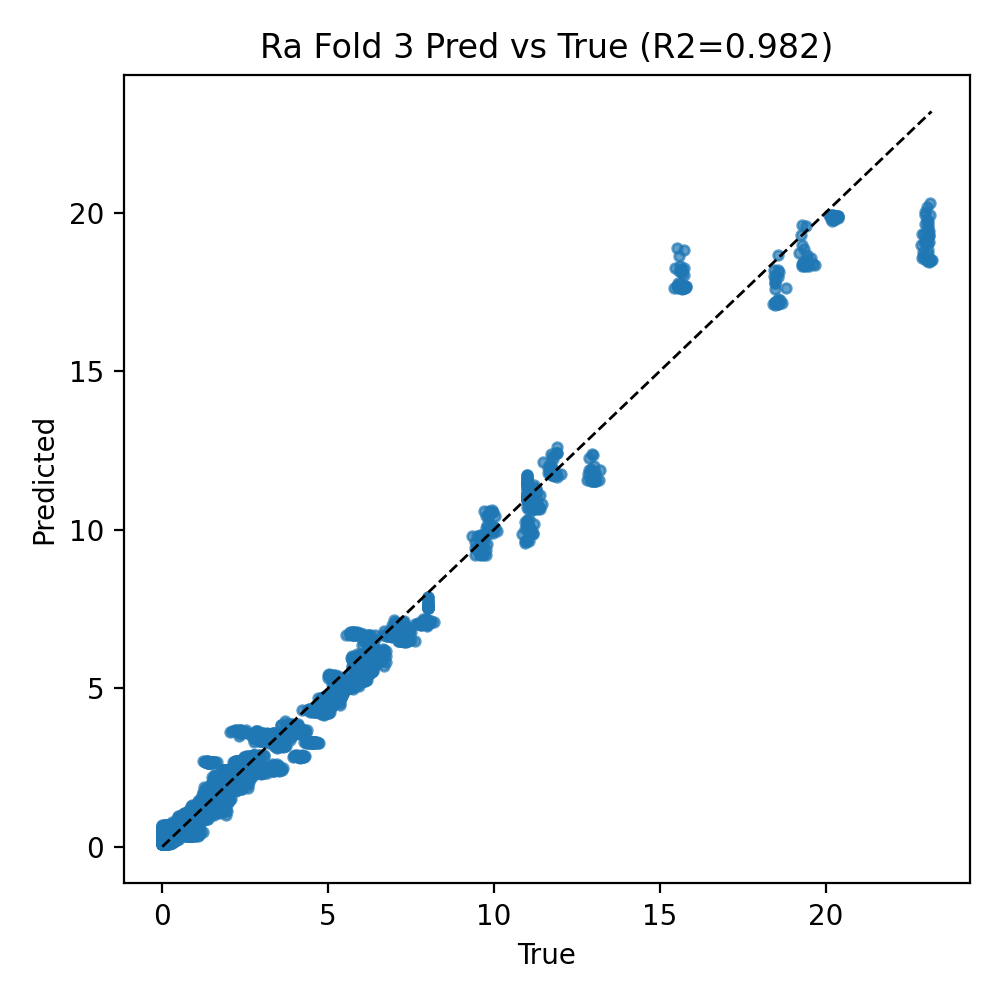}{Figure S132: pred vs true (\label{fig:supp-132})}\
\suppimage{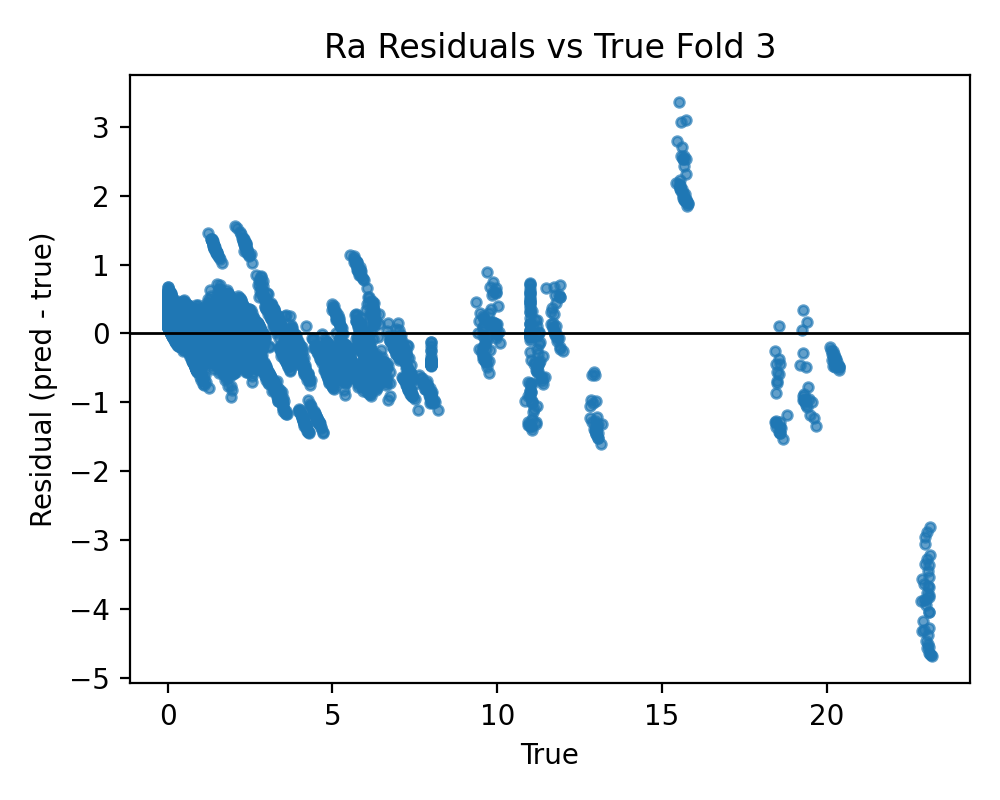}{Figure S133: residuals vs true (\label{fig:supp-133})}\hfill
\suppimage{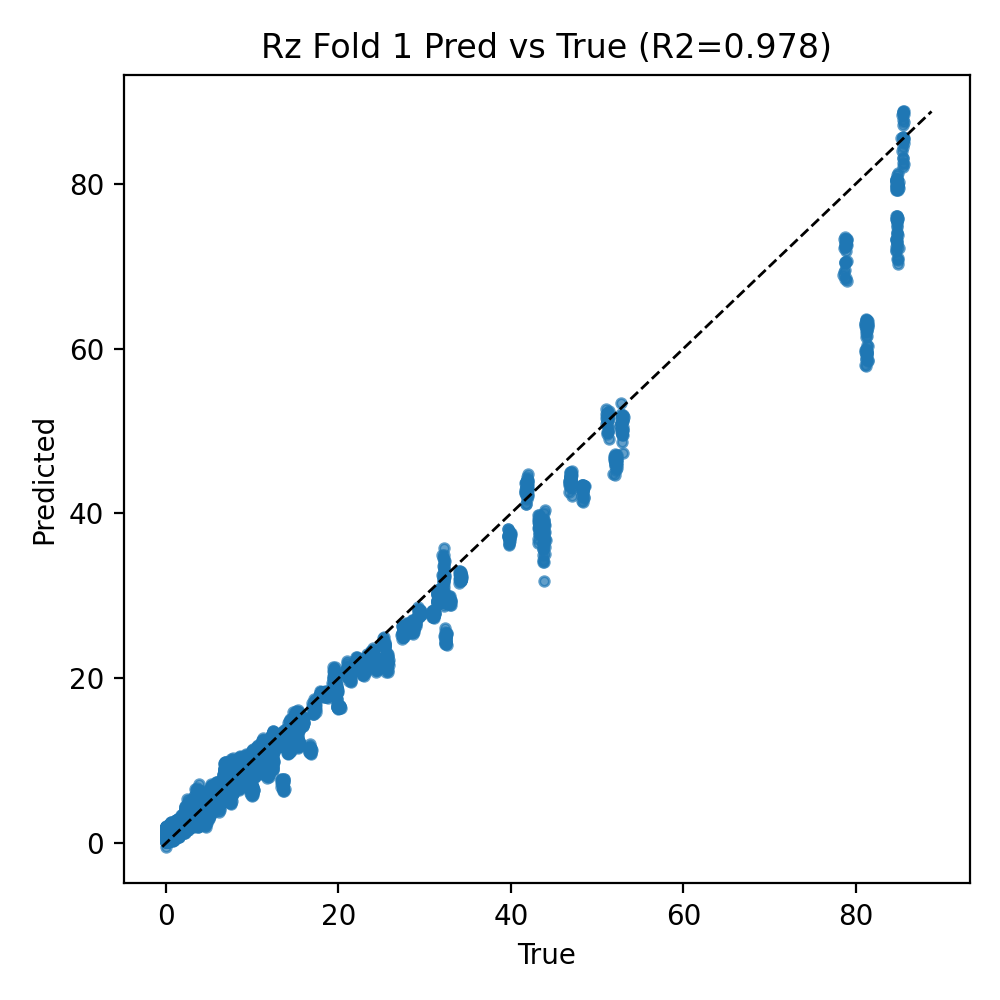}{Figure S134: pred vs true (\label{fig:supp-134})}\
\suppimage{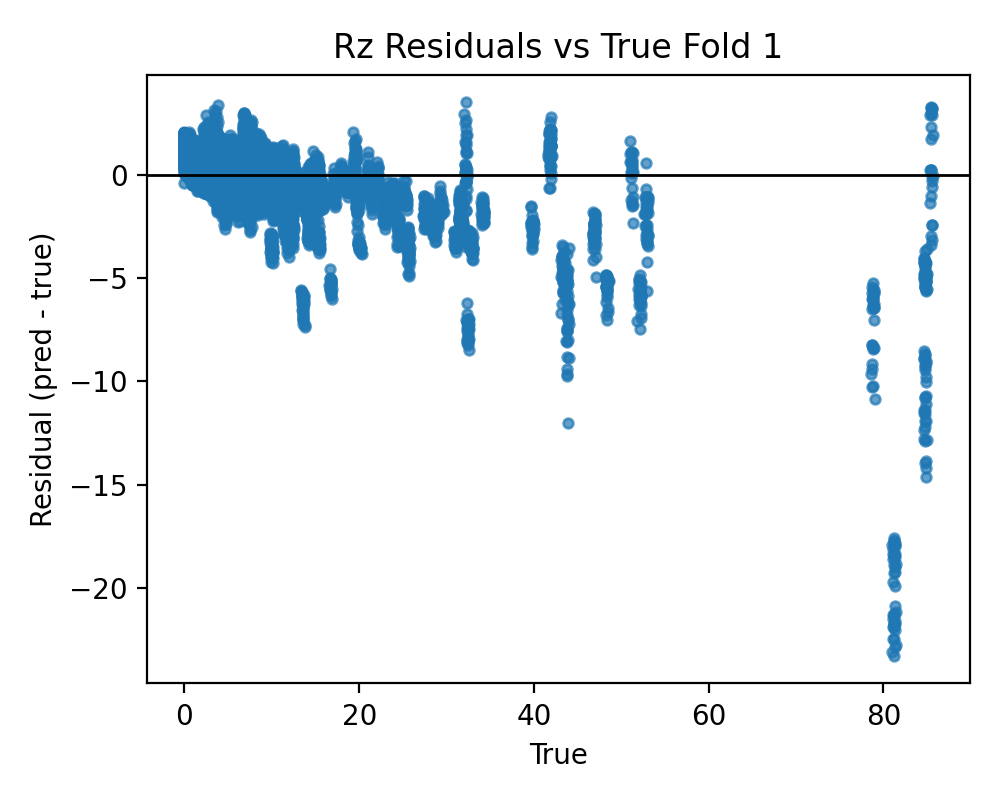}{Figure S135: residuals vs true (\label{fig:supp-135})}\hfill
\suppimage{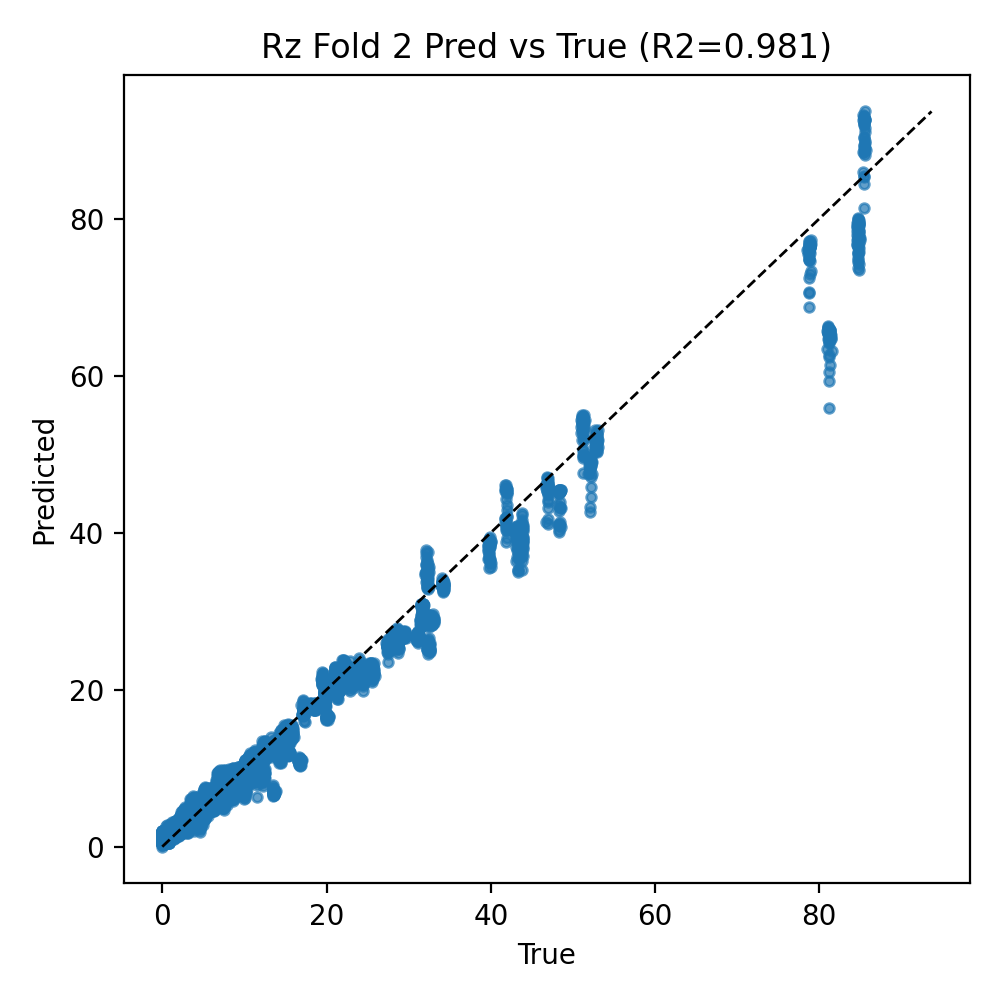}{Figure S136: pred vs true (\label{fig:supp-136})}\
\suppimage{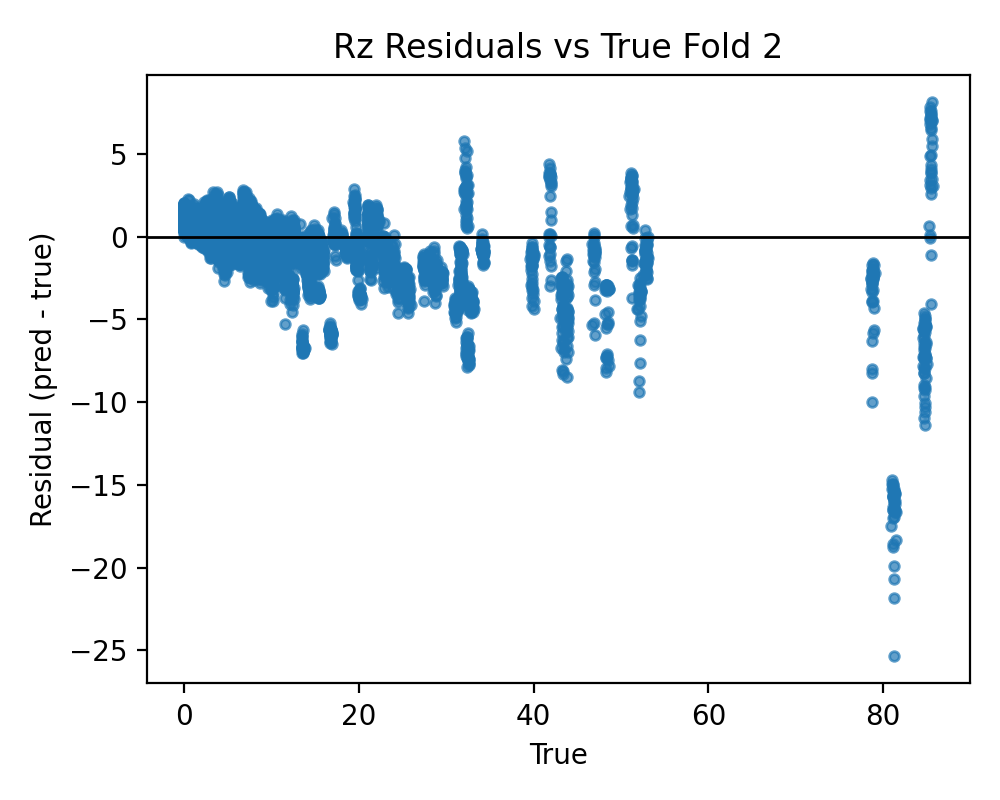}{Figure S137: residuals vs true (\label{fig:supp-137})}\hfill
\suppimage{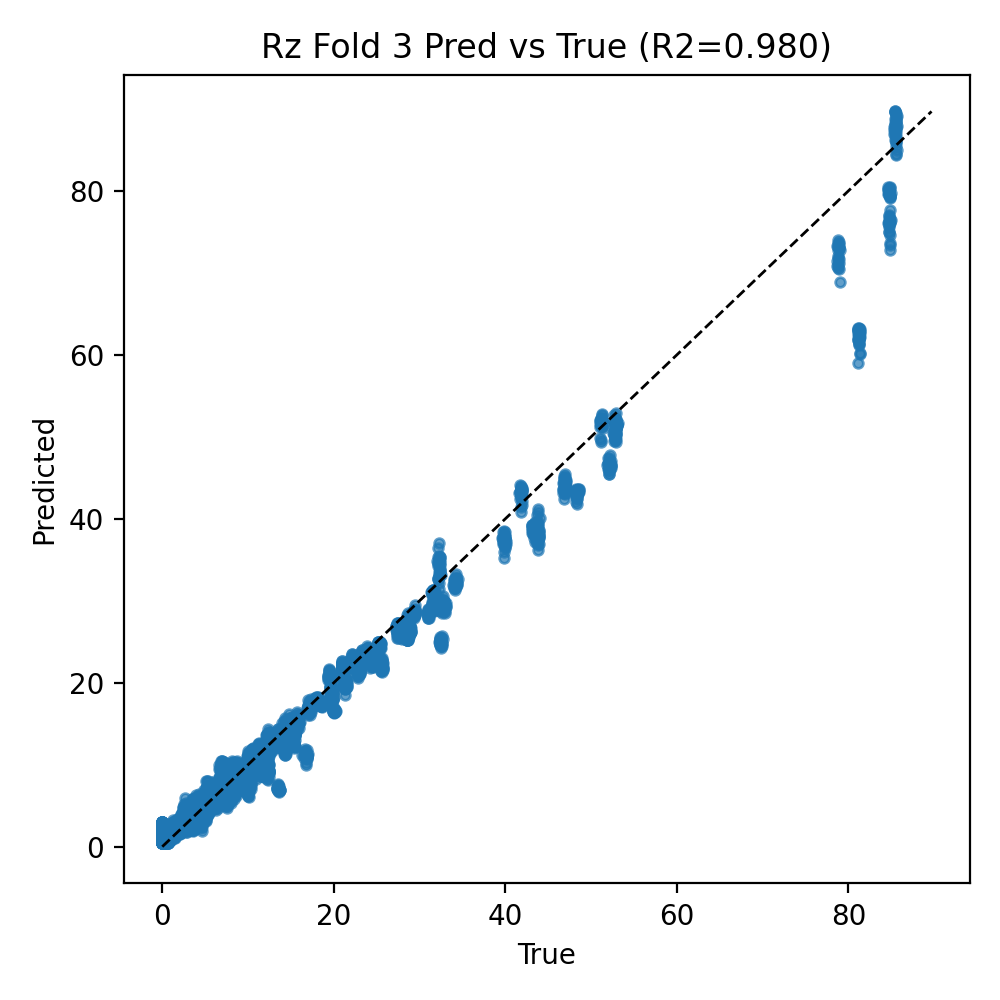}{Figure S138: pred vs true (\label{fig:supp-138})}\
\suppimage{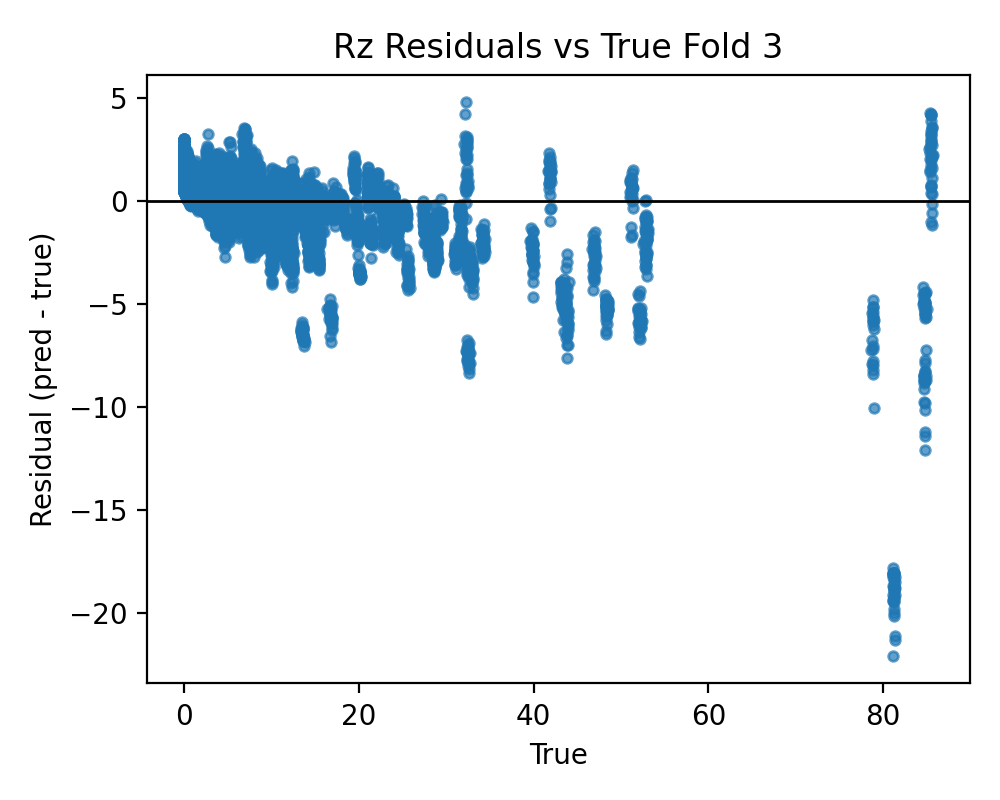}{Figure S139: residuals vs true (\label{fig:supp-139})}\hfill
\suppimage{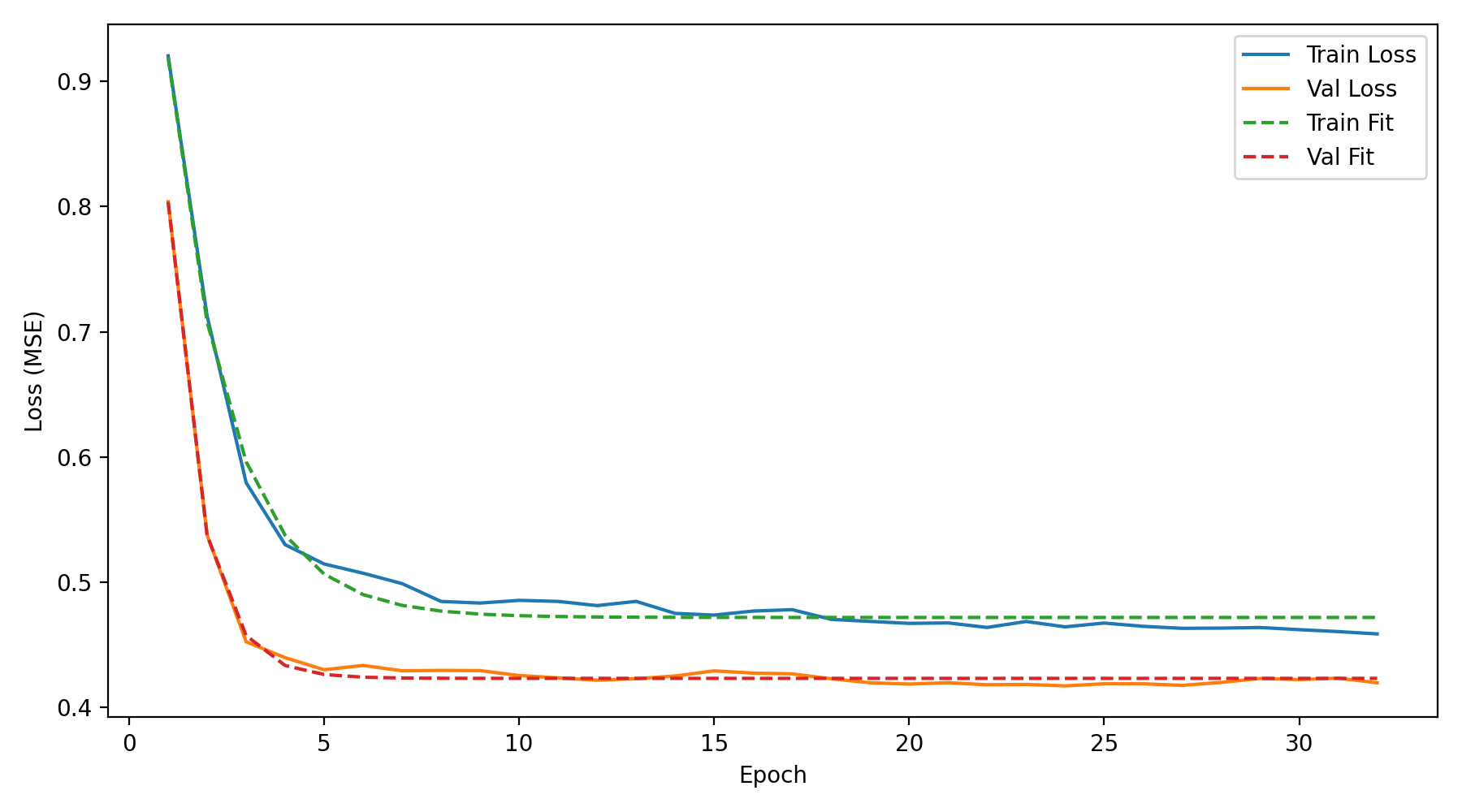}{Figure S140: loss curves (\label{fig:supp-140})}\
\suppimage{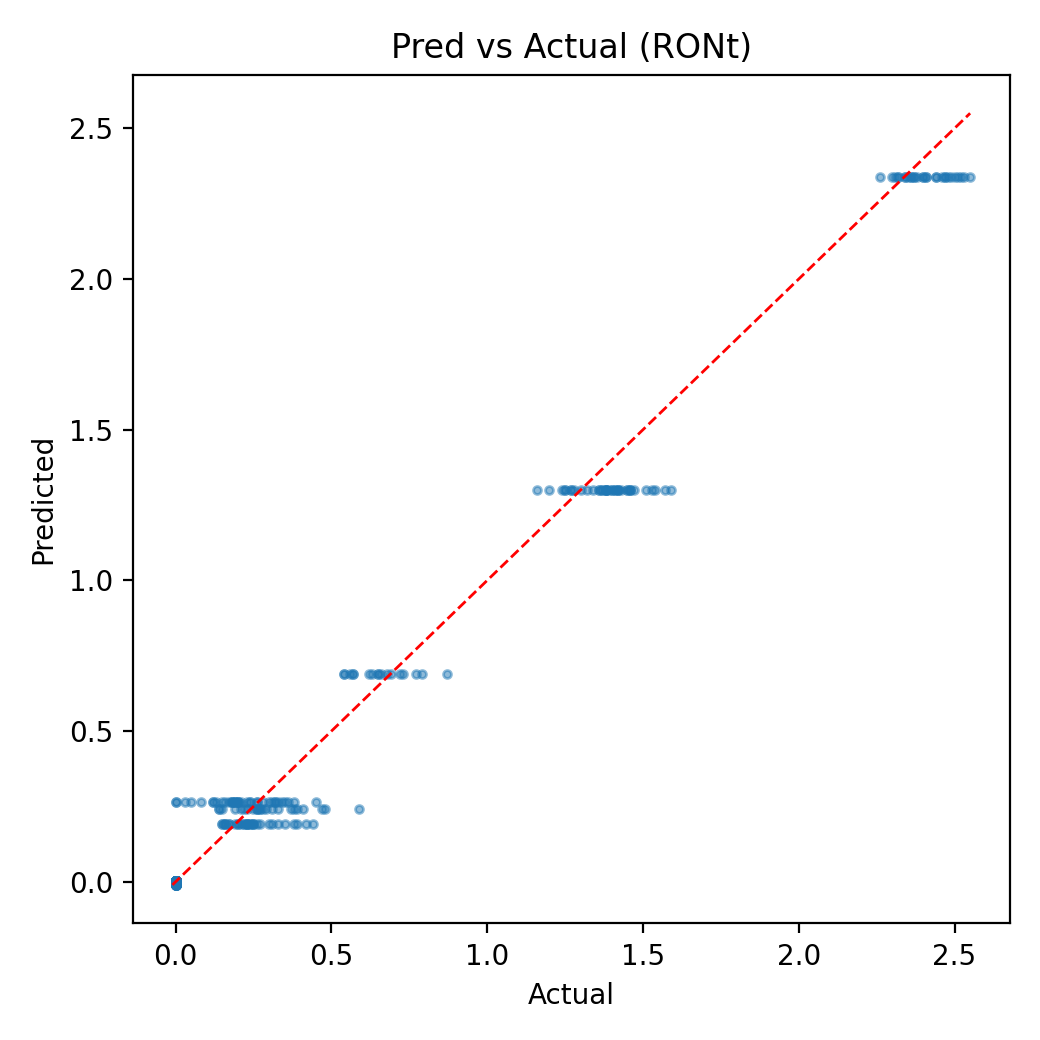}{Figure S141: pred vs actual RONt (\label{fig:supp-141})}\hfill
\suppimage{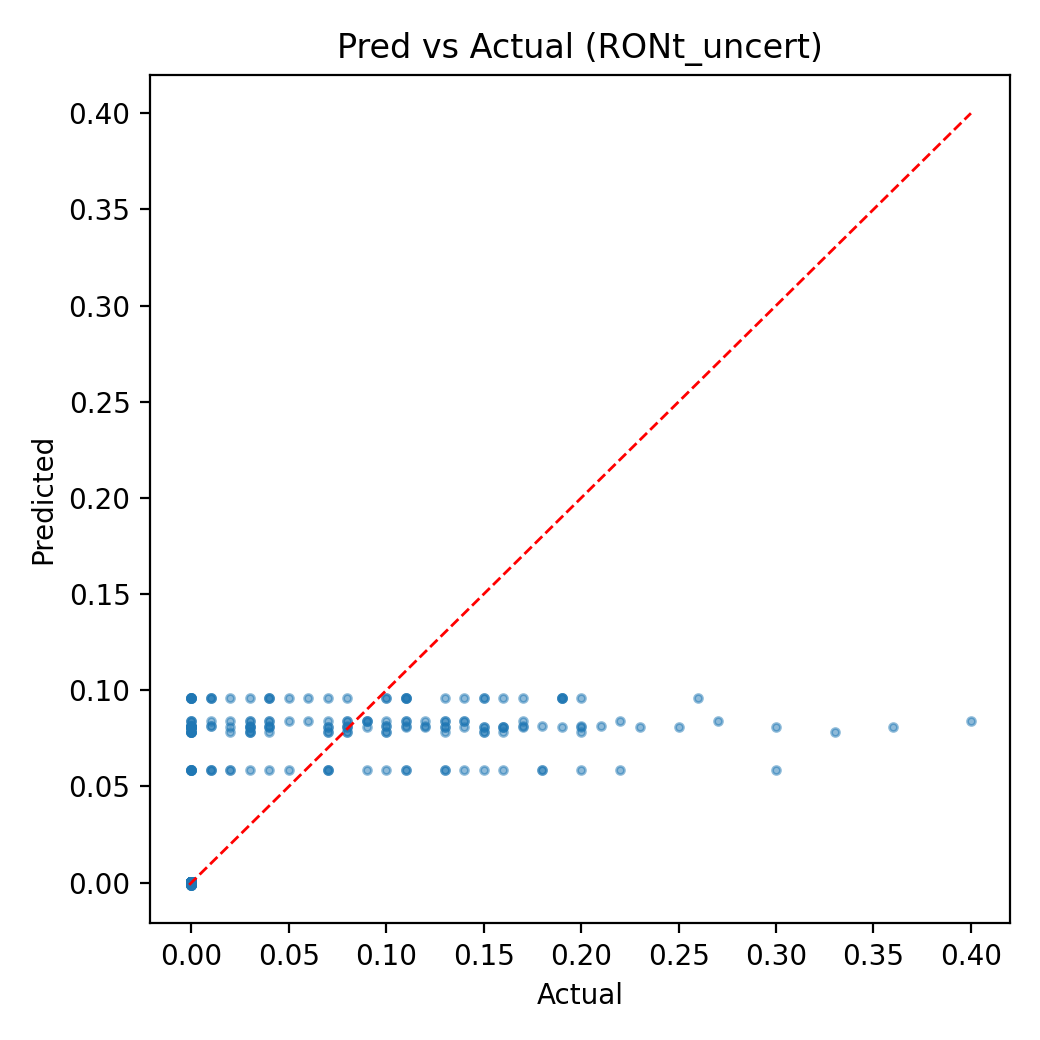}{Figure S142: pred vs actual RONt uncert (\label{fig:supp-142})}\
\suppimage{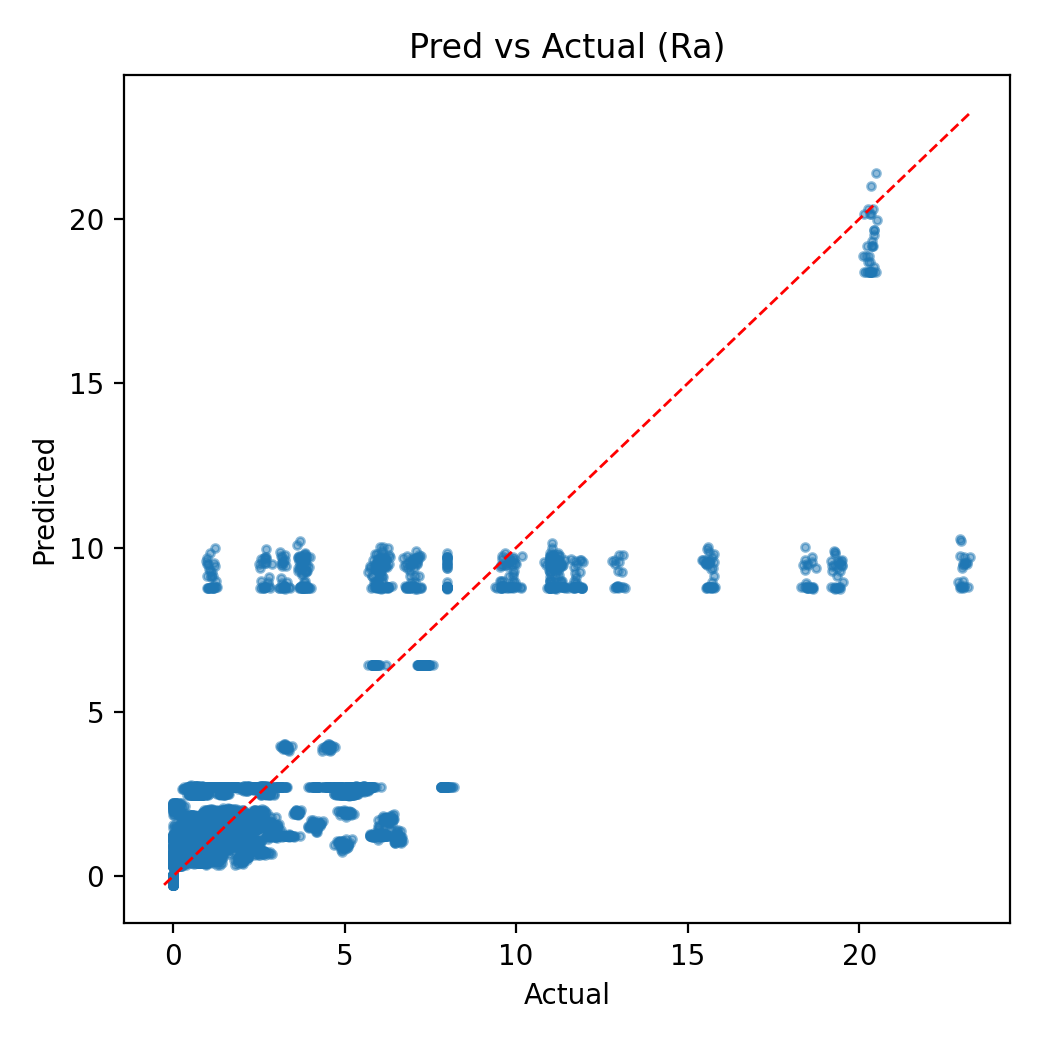}{Figure S143: pred vs actual Ra (\label{fig:supp-143})}\hfill
\suppimage{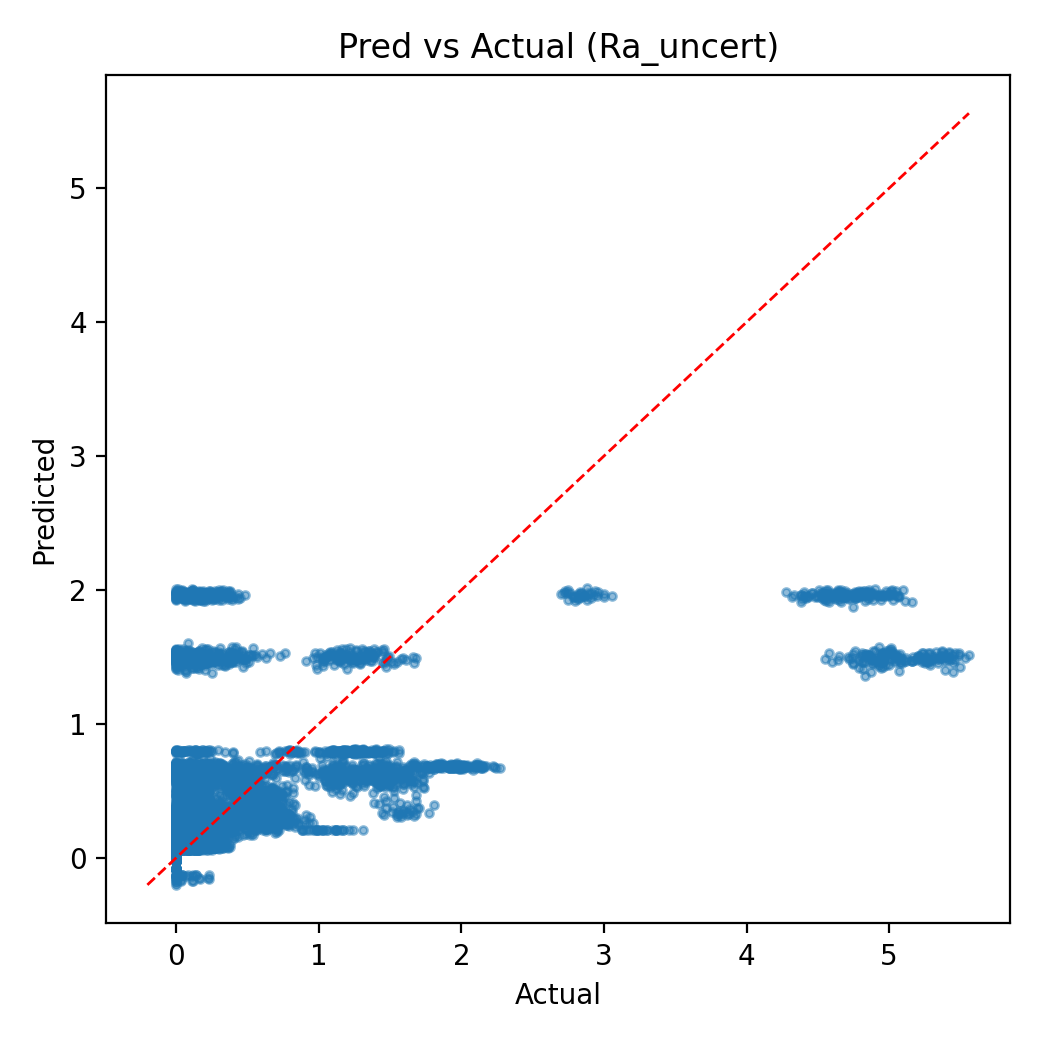}{Figure S144: pred vs actual Ra uncert (\label{fig:supp-144})}\
\suppimage{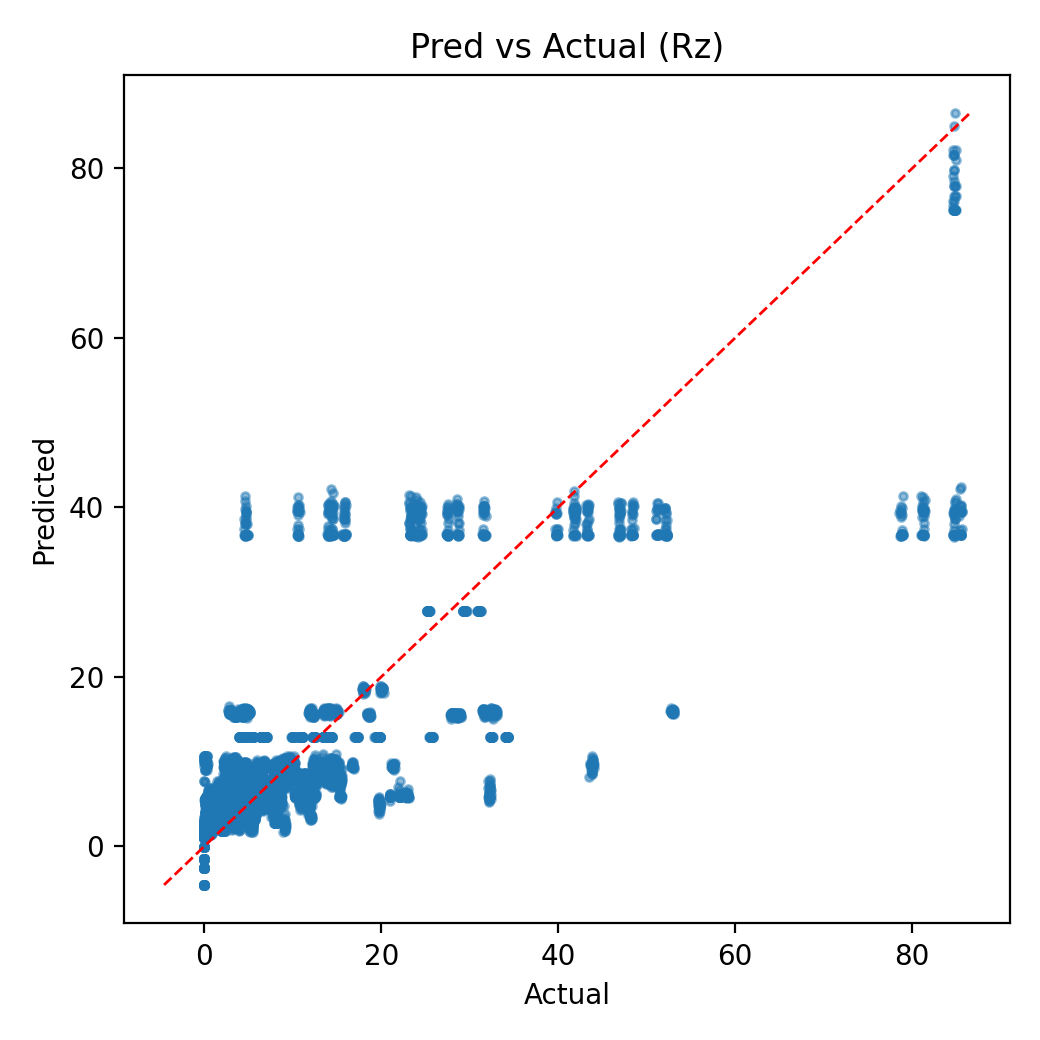}{Figure S145: pred vs actual Rz (\label{fig:supp-145})}\hfill
\suppimage{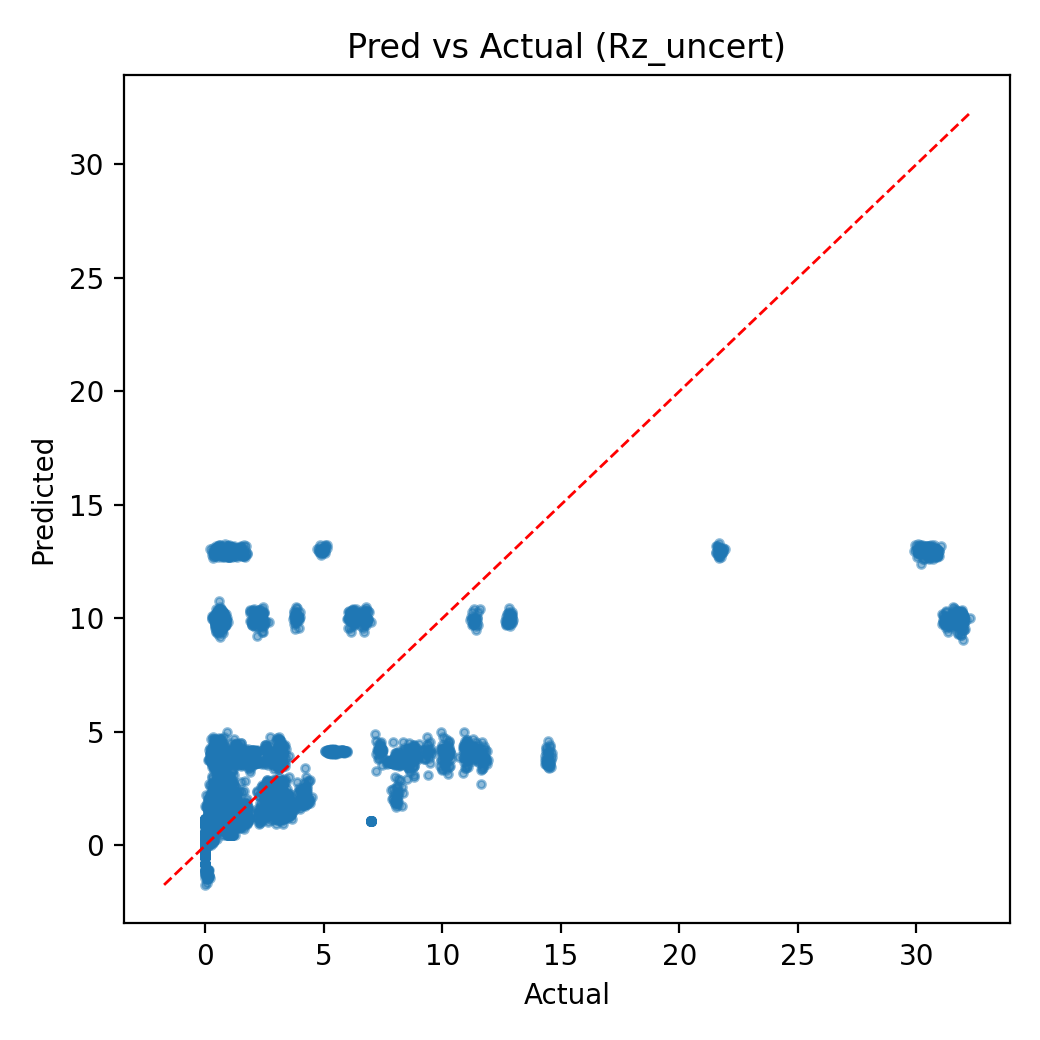}{Figure S146: pred vs actual Rz uncert (\label{fig:supp-146})}\
\suppimage{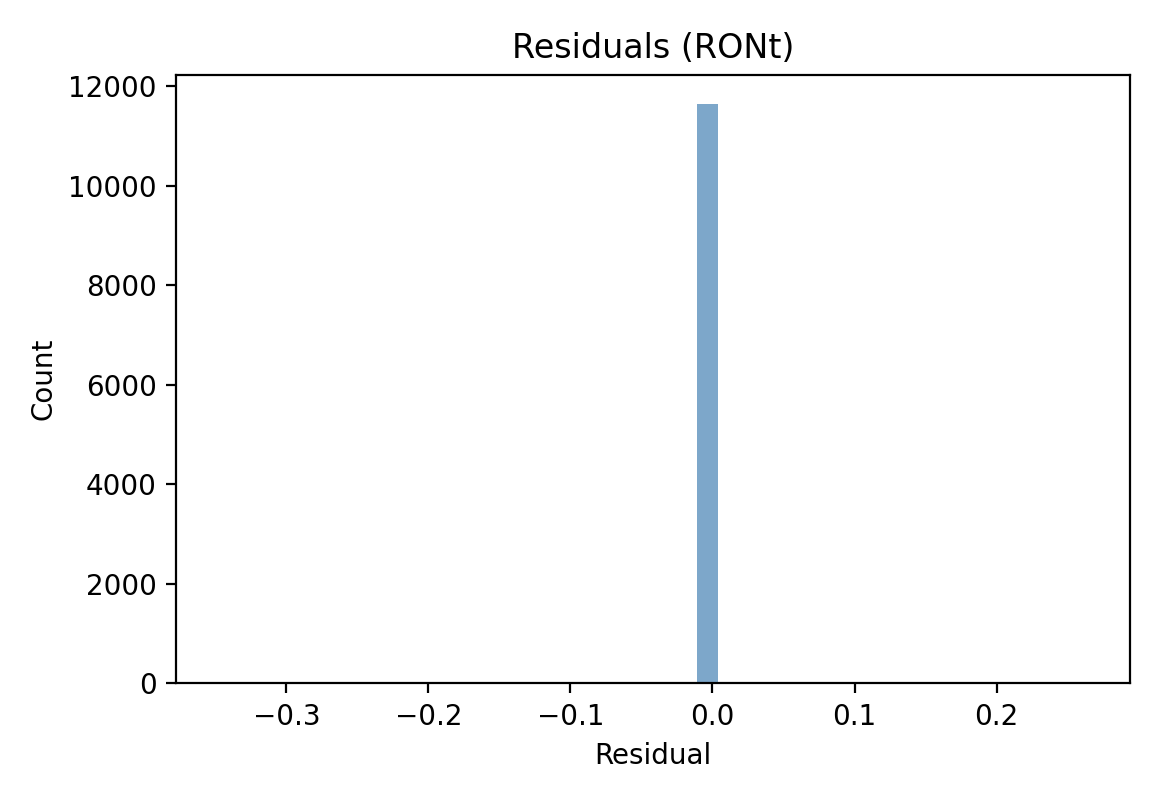}{Figure S147: residuals hist RONt (\label{fig:supp-147})}\hfill
\suppimage{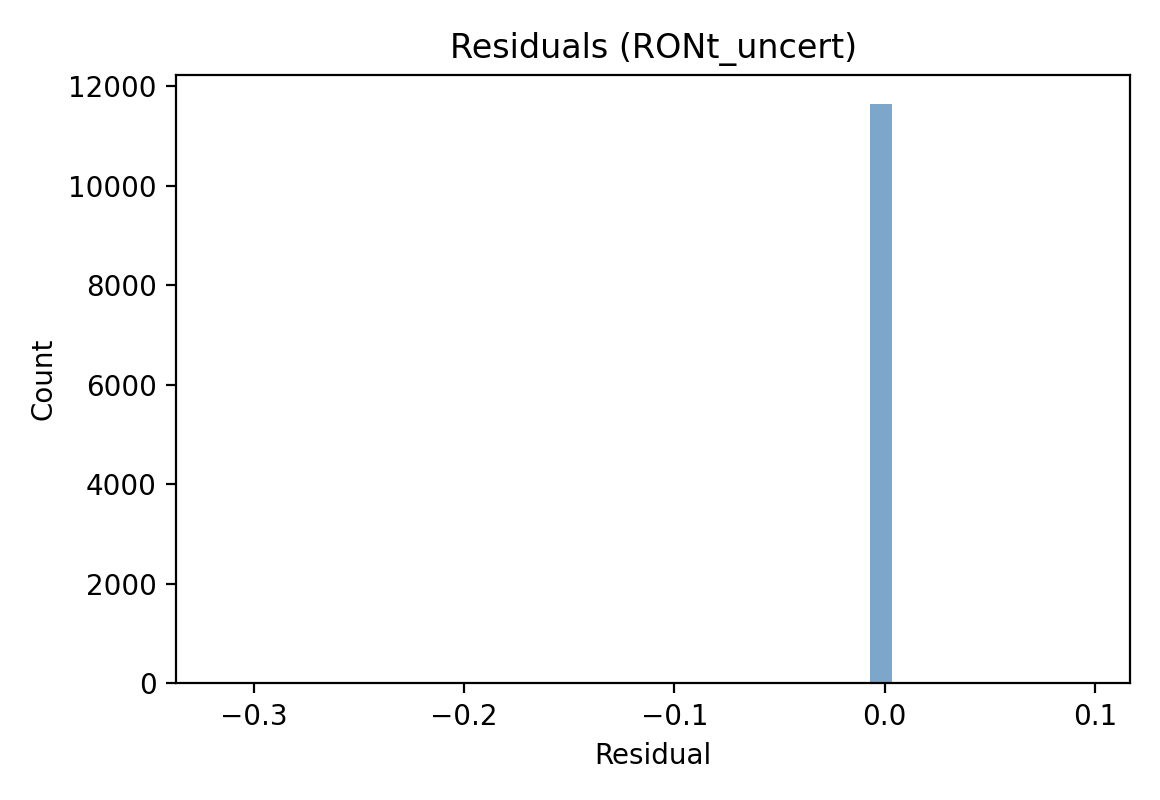}{Figure S148: residuals hist RONt uncert (\label{fig:supp-148})}\
\suppimage{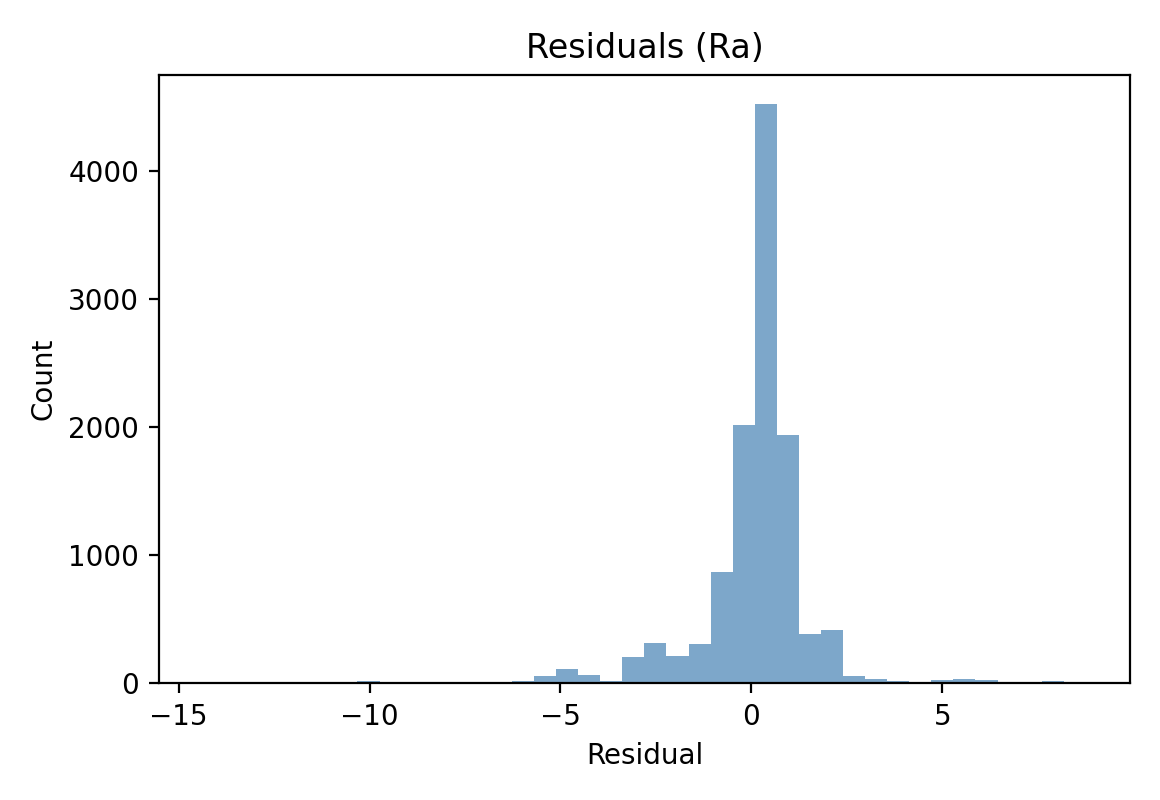}{Figure S149: residuals hist Ra (\label{fig:supp-149})}\hfill
\suppimage{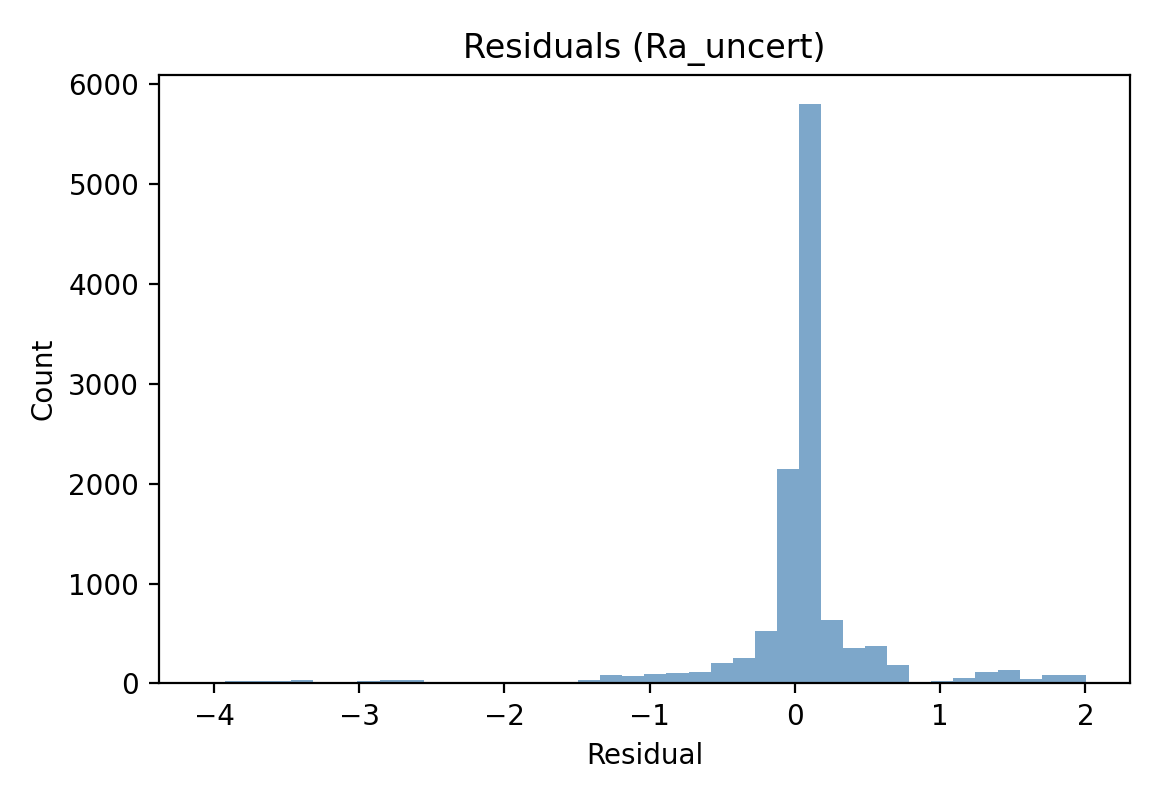}{Figure S150: residuals hist Ra uncert (\label{fig:supp-150})}\
\suppimage{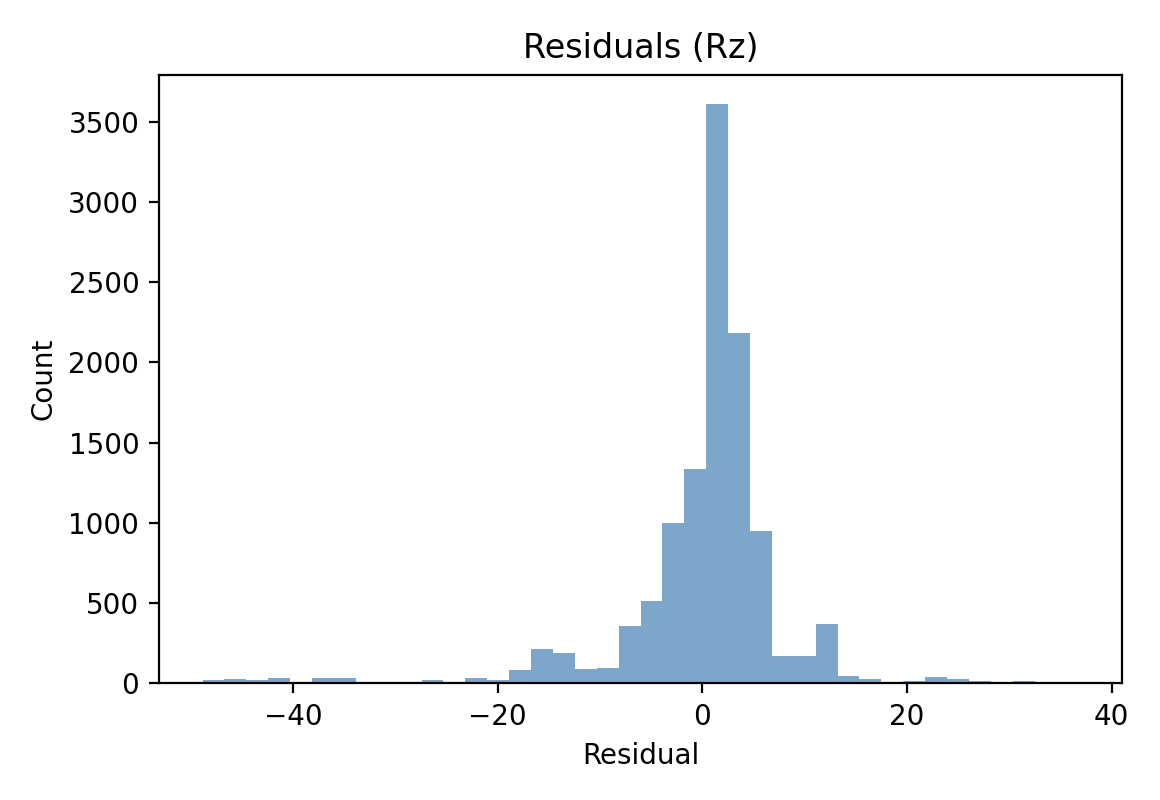}{Figure S151: residuals hist Rz (\label{fig:supp-151})}\hfill
\suppimage{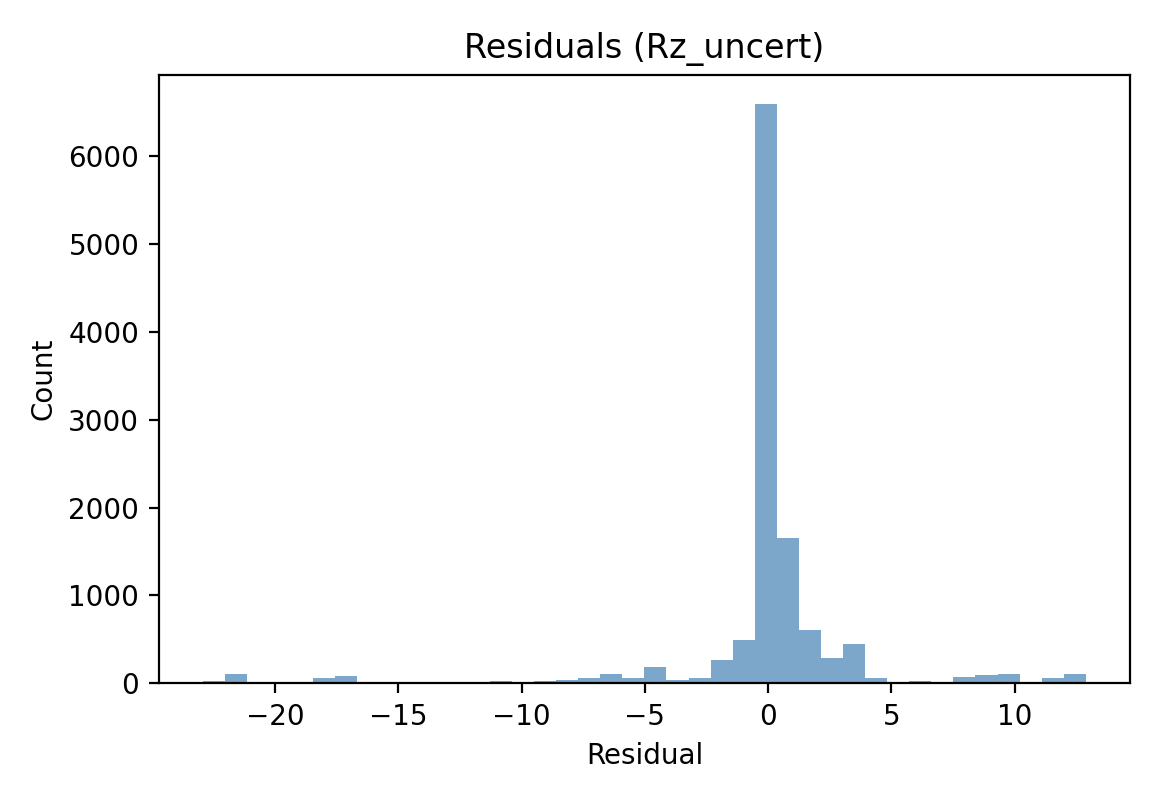}{Figure S152: residuals hist Rz uncert (\label{fig:supp-152})}\
\suppimage{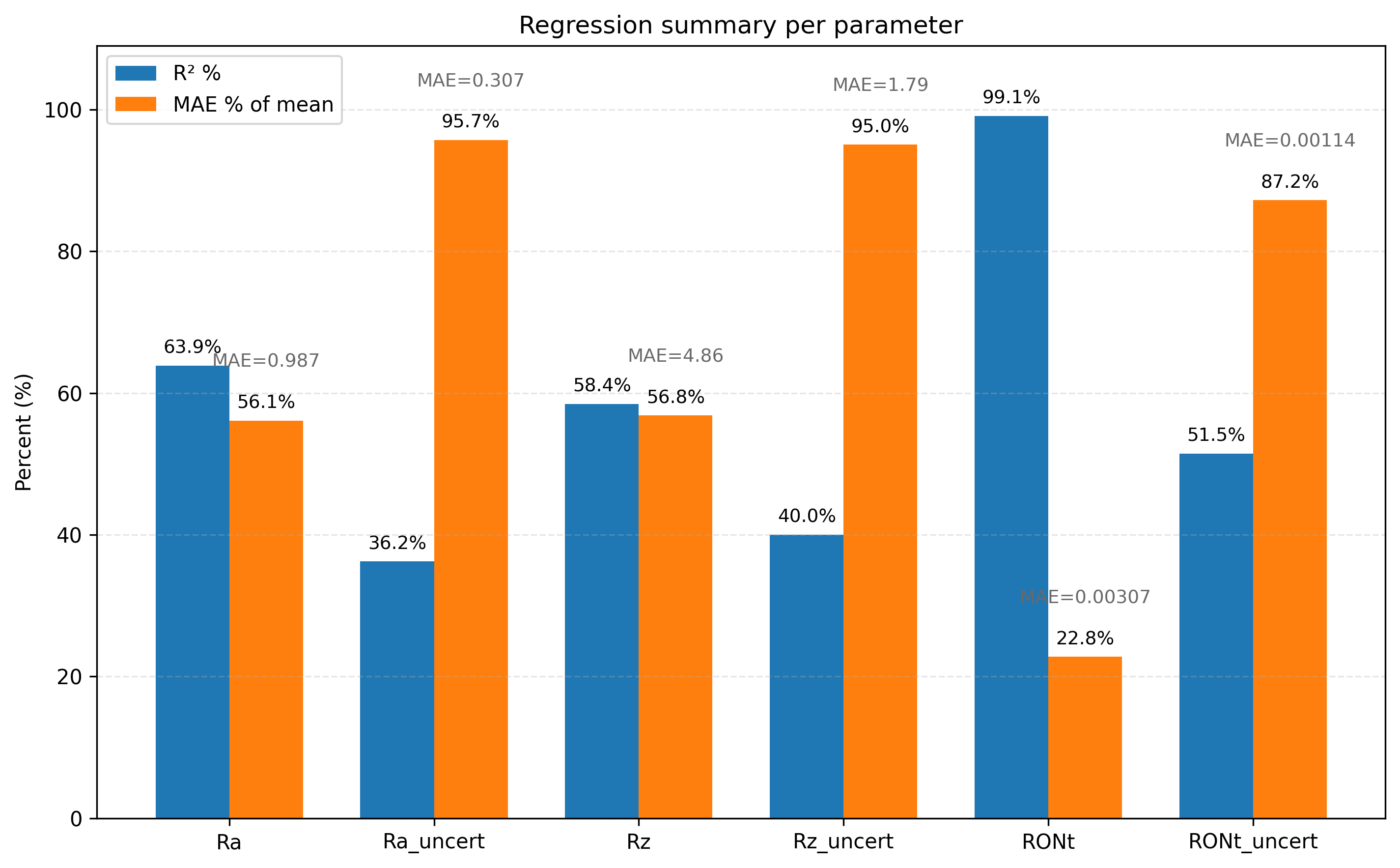}{Figure S153: regression summary bars (\label{fig:supp-153})}\hfill
\suppimage{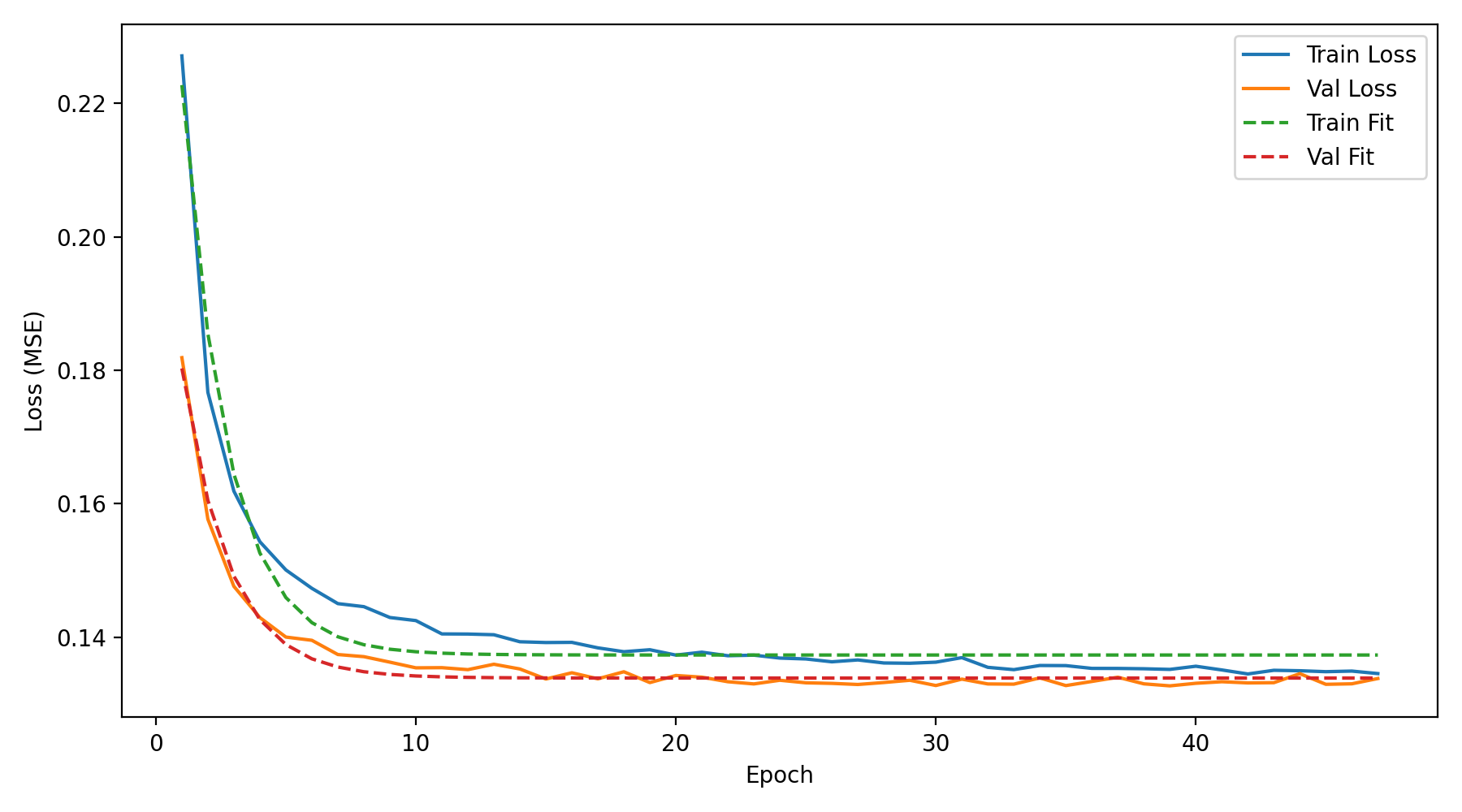}{Figure S154: loss curves (\label{fig:supp-154})}\
\suppimage{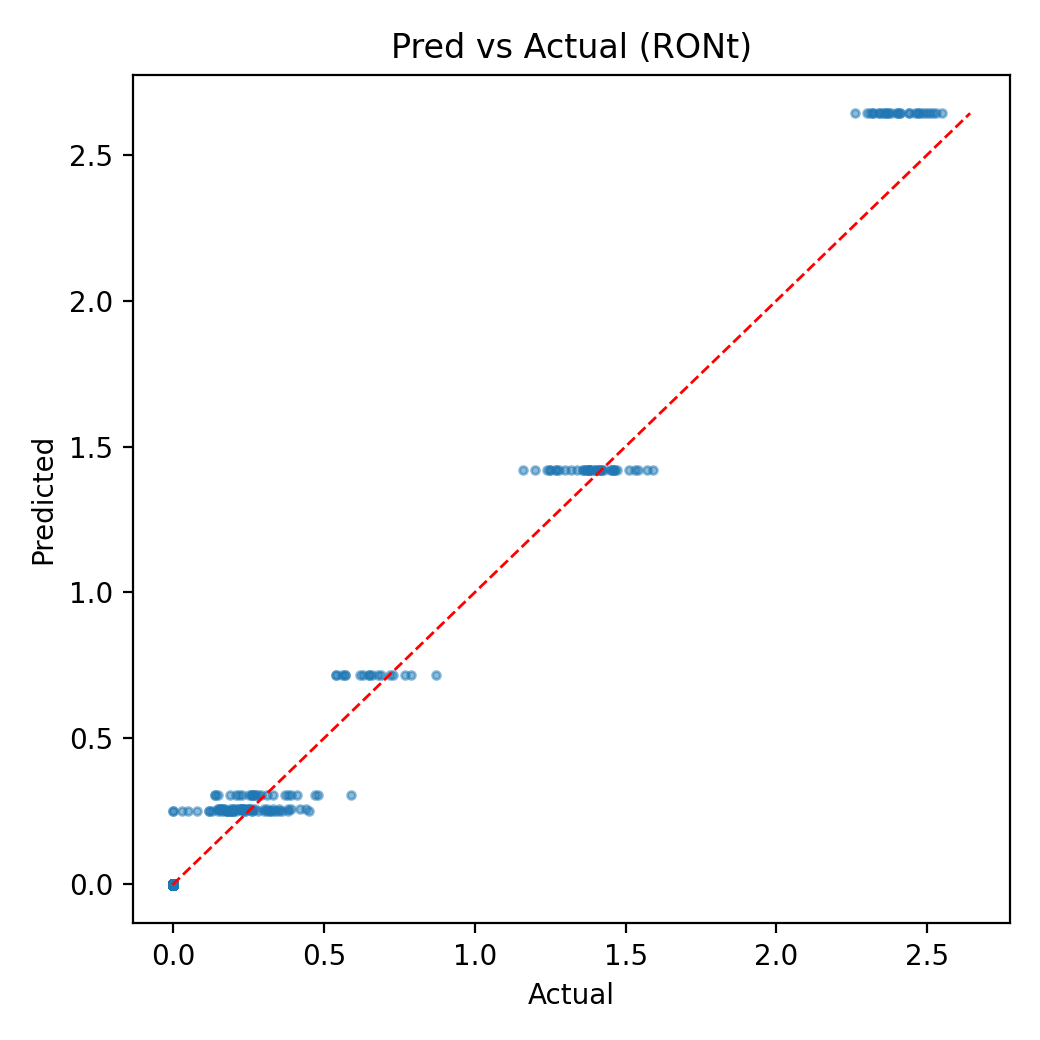}{Figure S155: pred vs actual RONt (\label{fig:supp-155})}\hfill
\suppimage{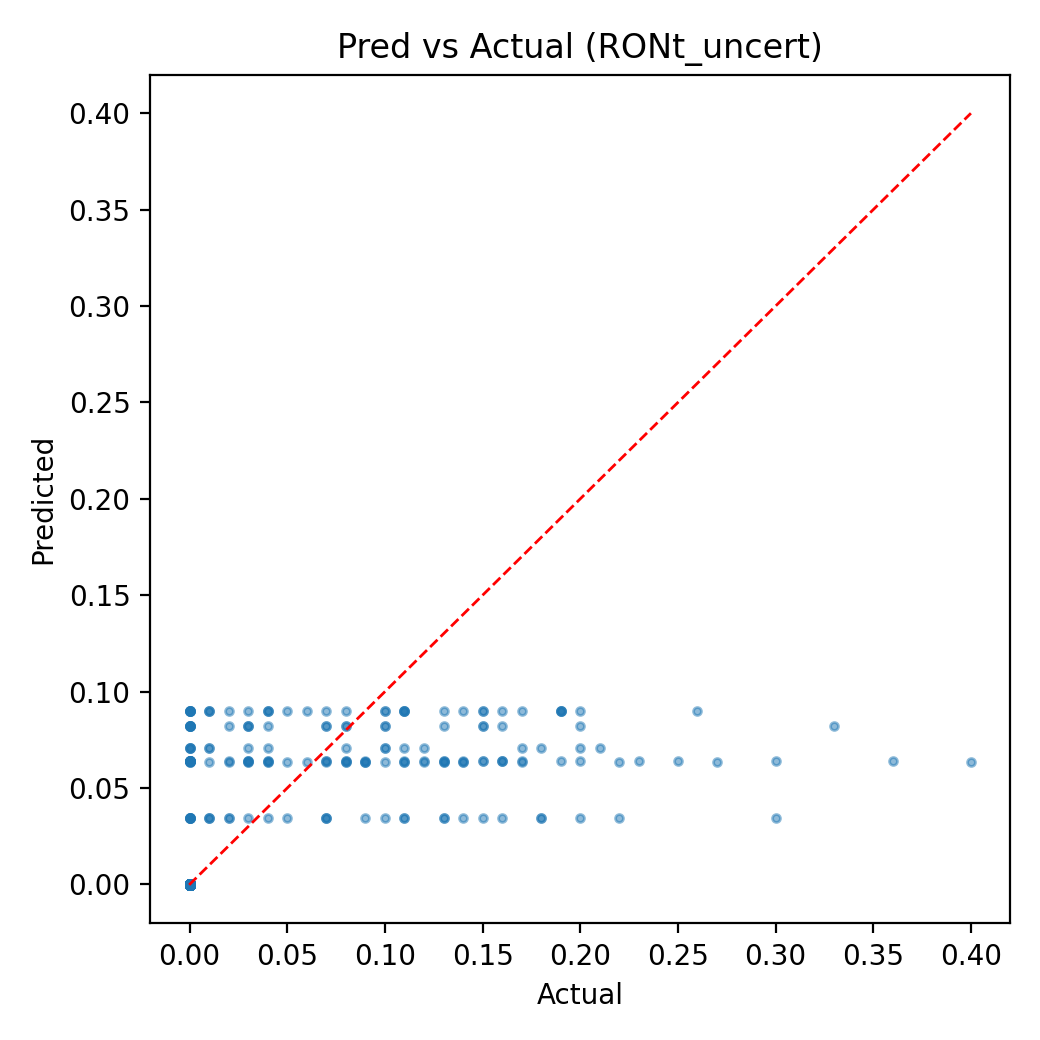}{Figure S156: pred vs actual RONt uncert (\label{fig:supp-156})}\
\suppimage{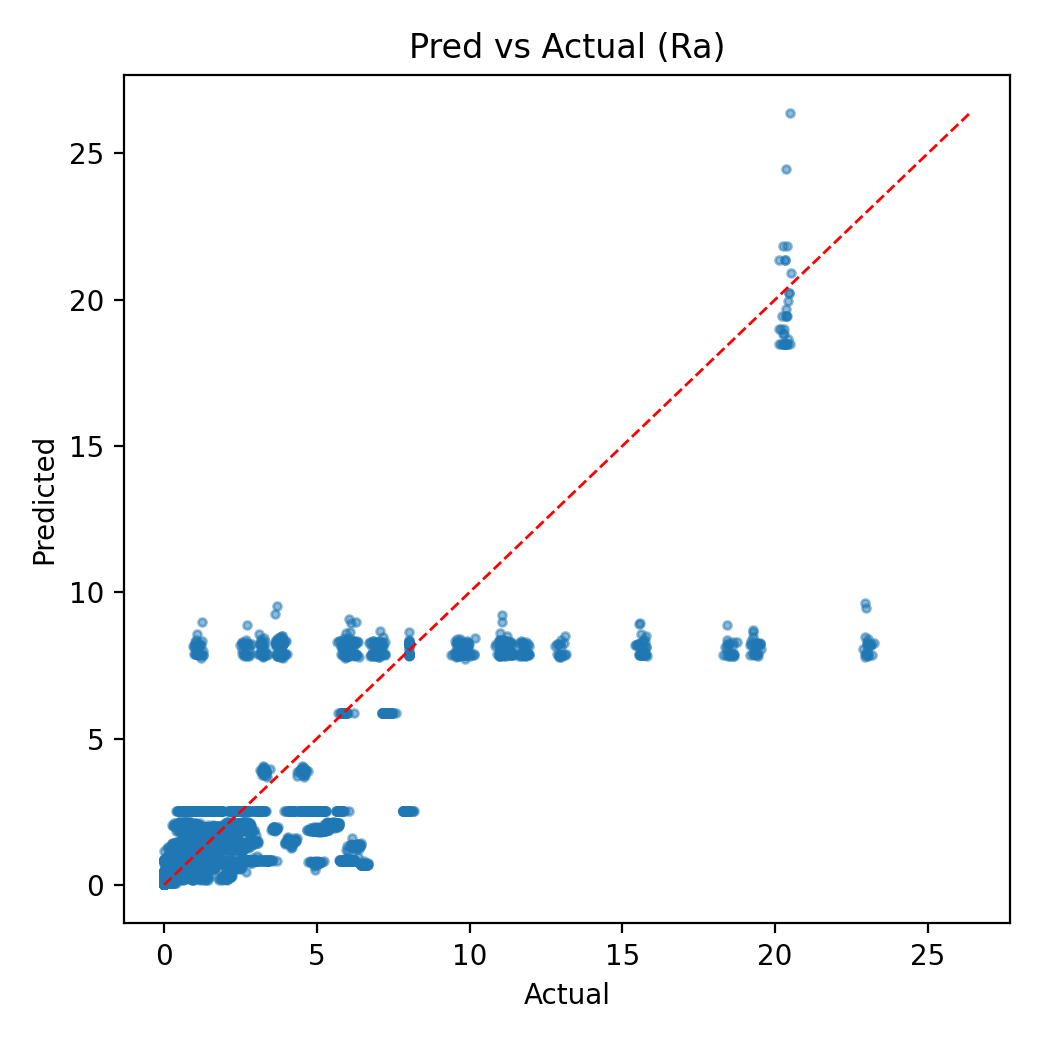}{Figure S157: pred vs actual Ra (\label{fig:supp-157})}\hfill
\suppimage{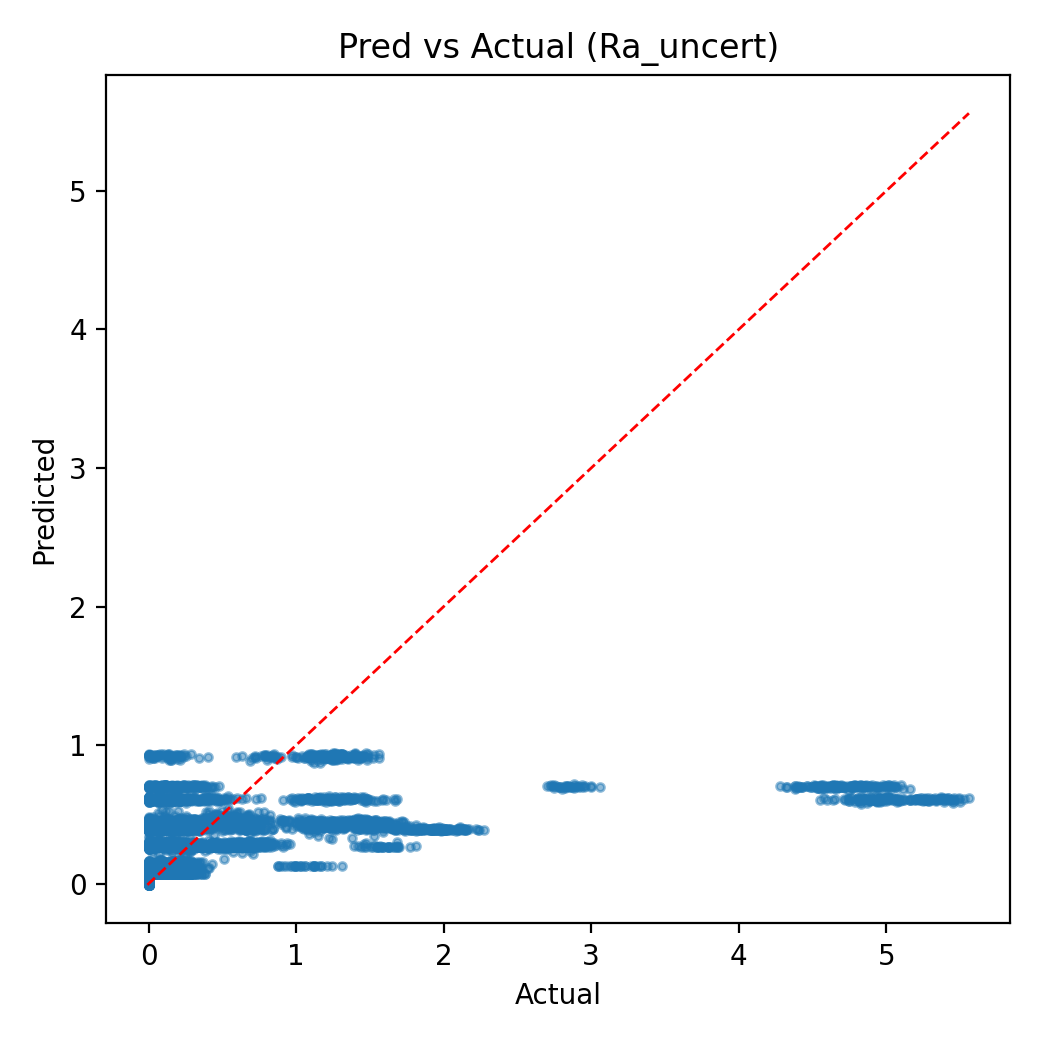}{Figure S158: pred vs actual Ra uncert (\label{fig:supp-158})}\
\suppimage{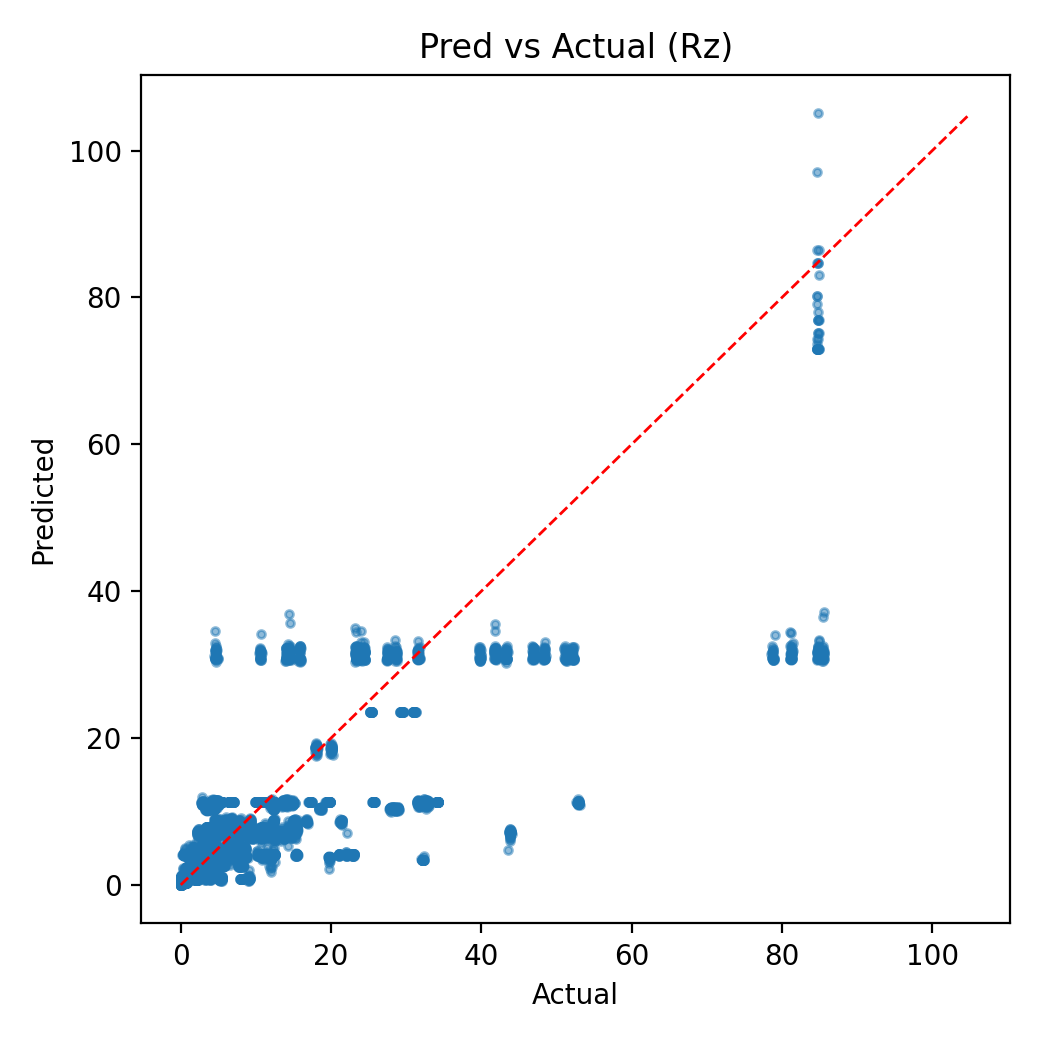}{Figure S159: pred vs actual Rz (\label{fig:supp-159})}\hfill
\suppimage{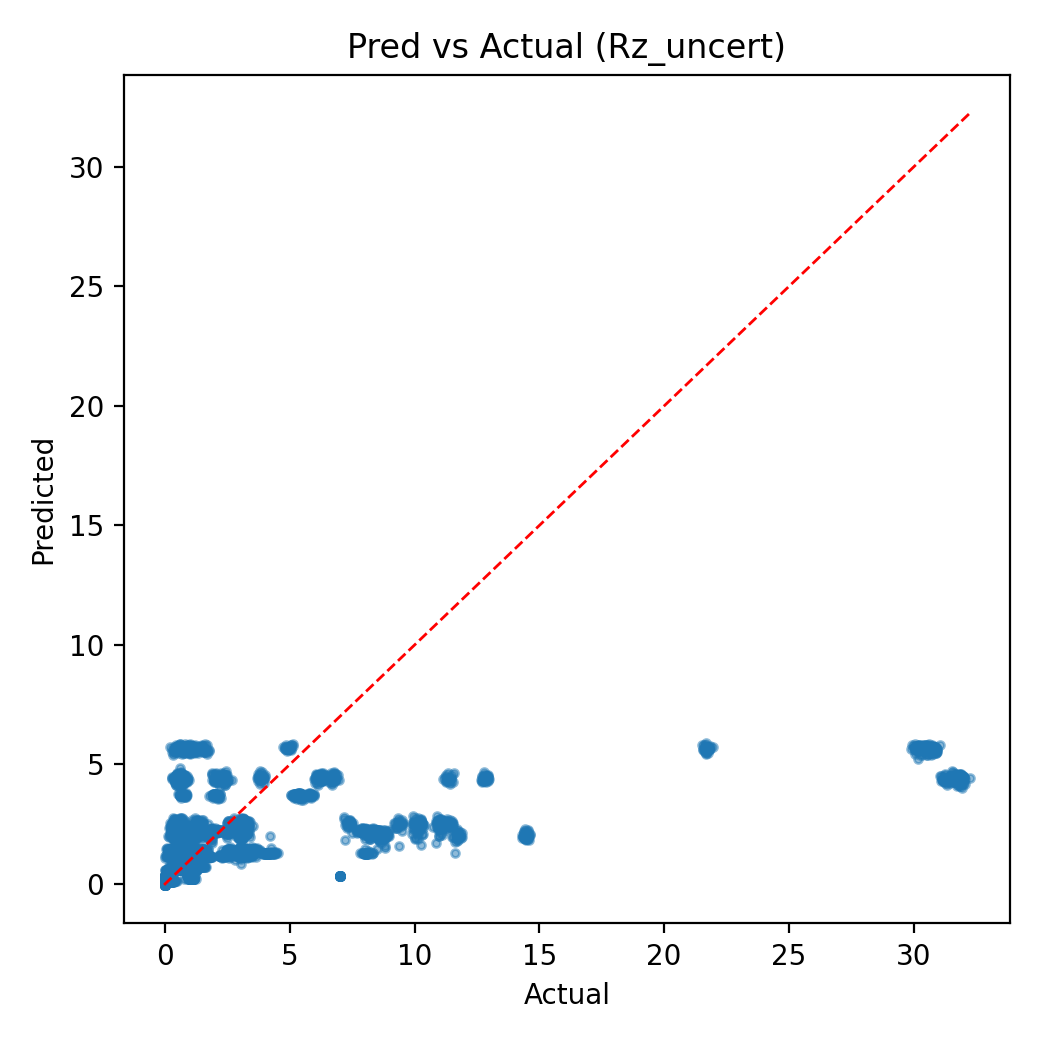}{Figure S160: pred vs actual Rz uncert (\label{fig:supp-160})}\
\suppimage{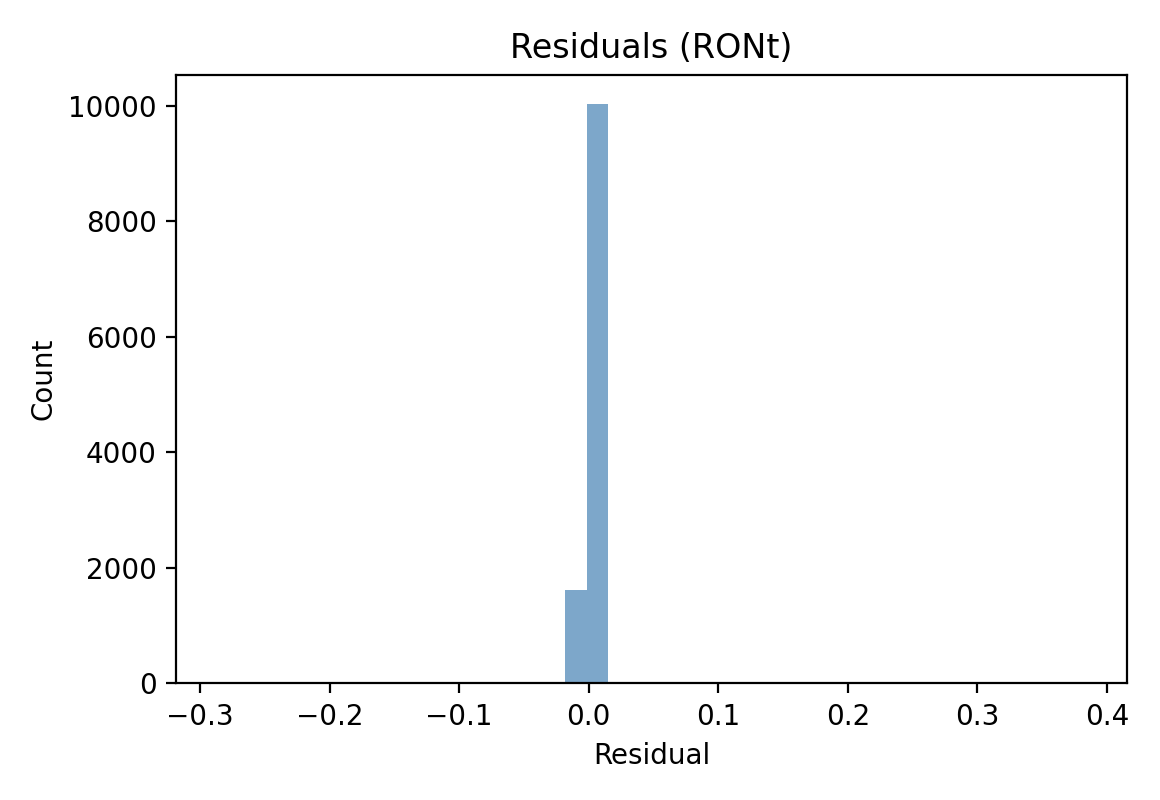}{Figure S161: residuals hist RONt (\label{fig:supp-161})}\hfill
\suppimage{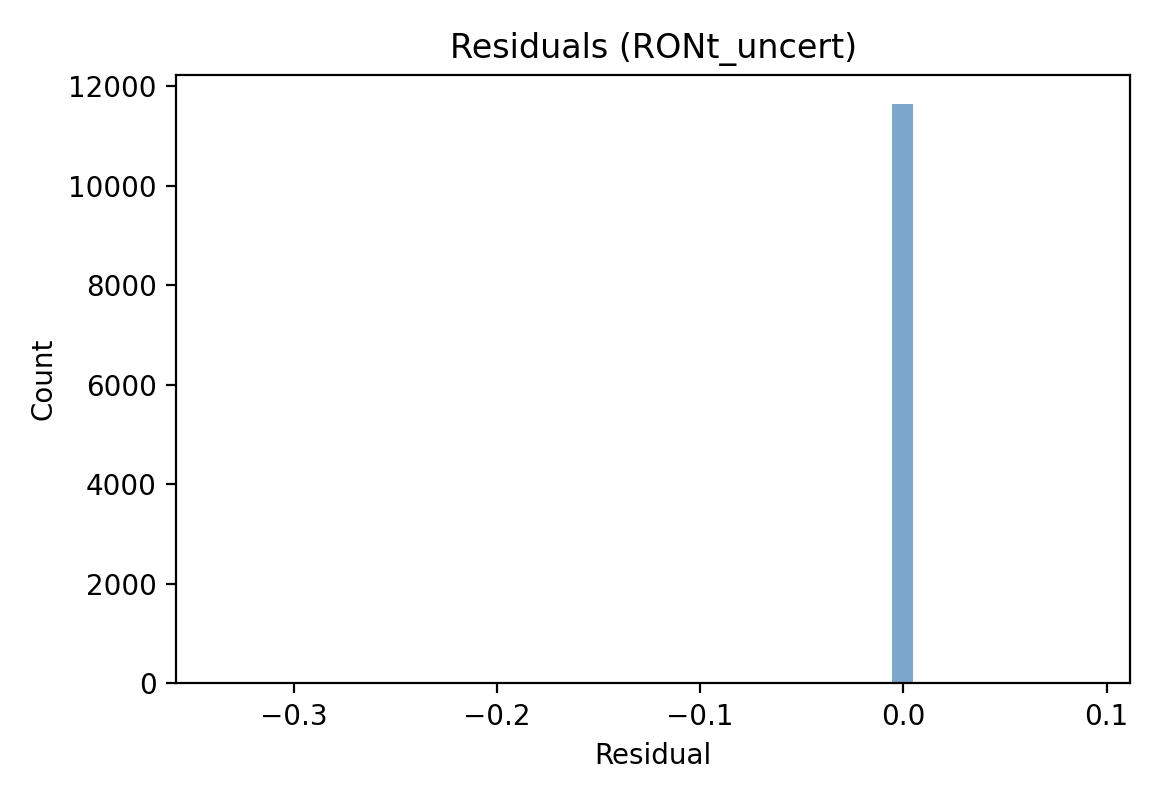}{Figure S162: residuals hist RONt uncert (\label{fig:supp-162})}\
\suppimage{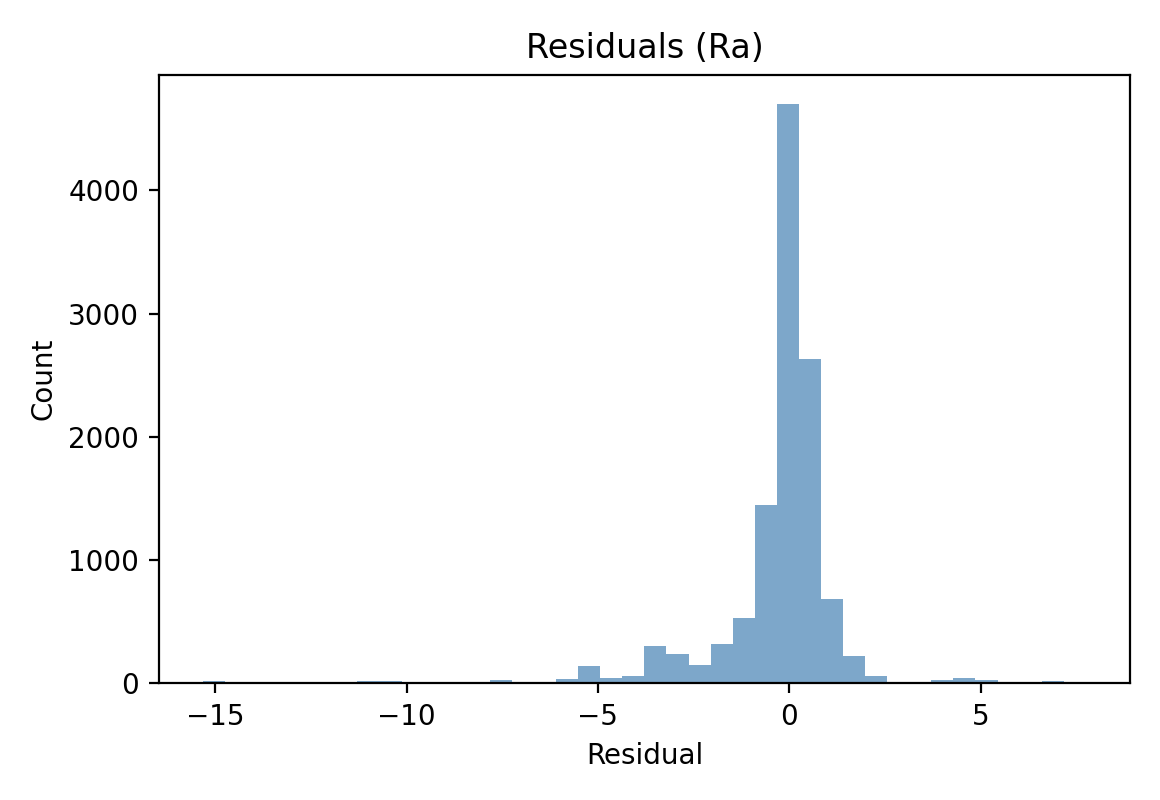}{Figure S163: residuals hist Ra (\label{fig:supp-163})}\hfill
\suppimage{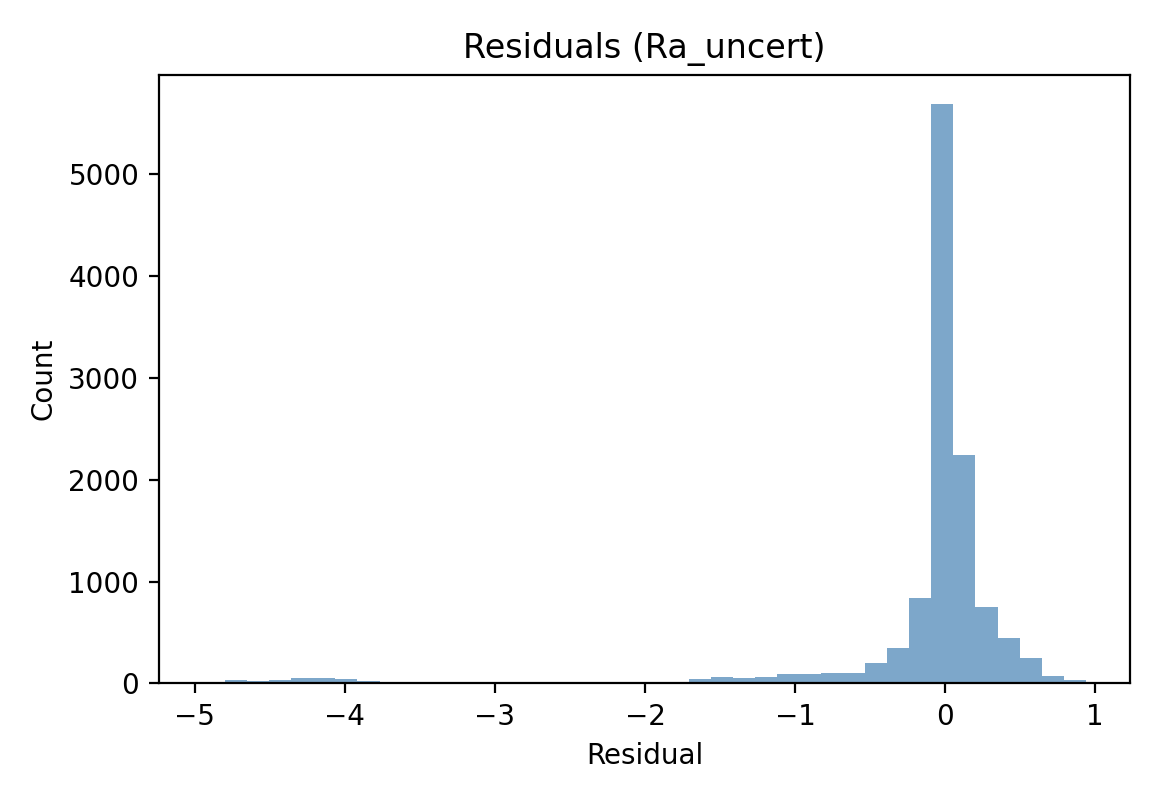}{Figure S164: residuals hist Ra uncert (\label{fig:supp-164})}\
\suppimage{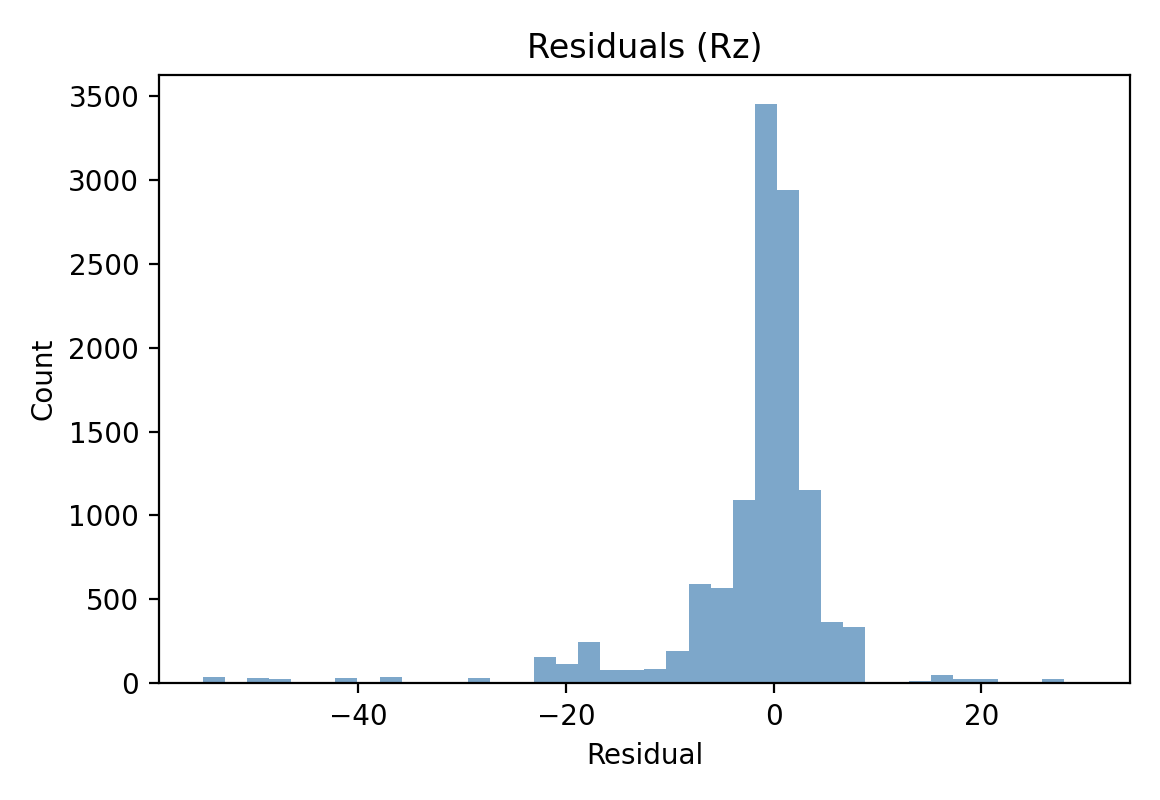}{Figure S165: residuals hist Rz (\label{fig:supp-165})}\hfill
\suppimage{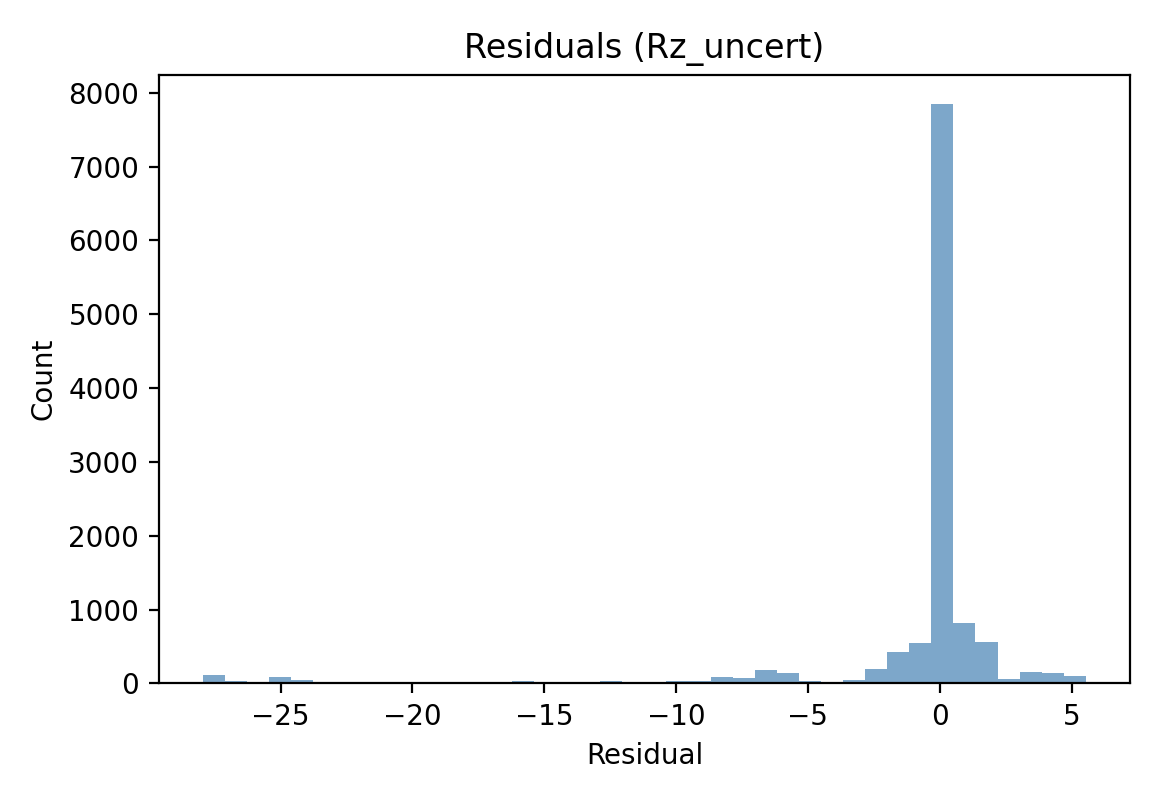}{Figure S166: residuals hist Rz uncert (\label{fig:supp-166})}\
\suppimage{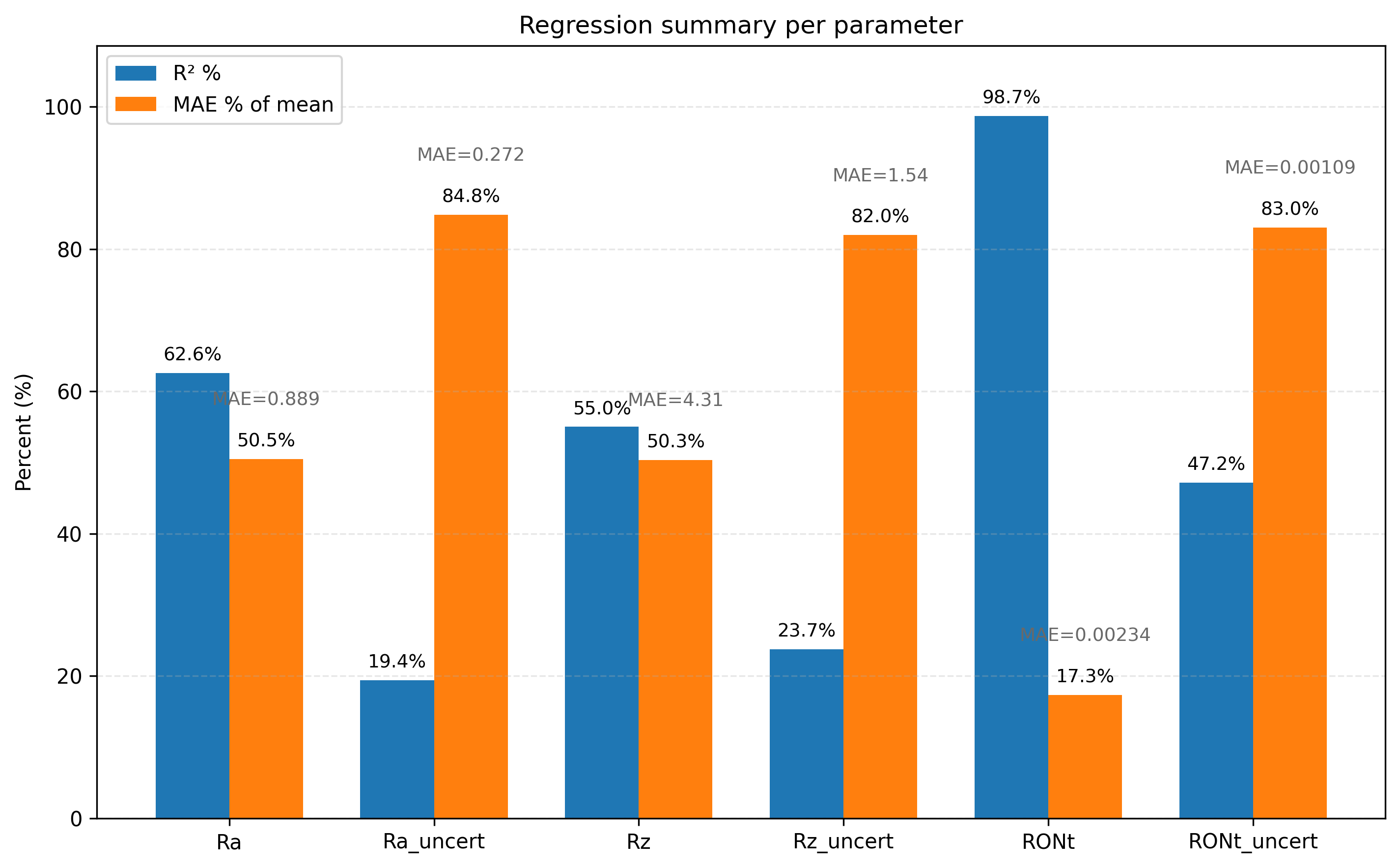}{Figure S167: regression summary bars (\label{fig:supp-167})}\hfill
\suppimage{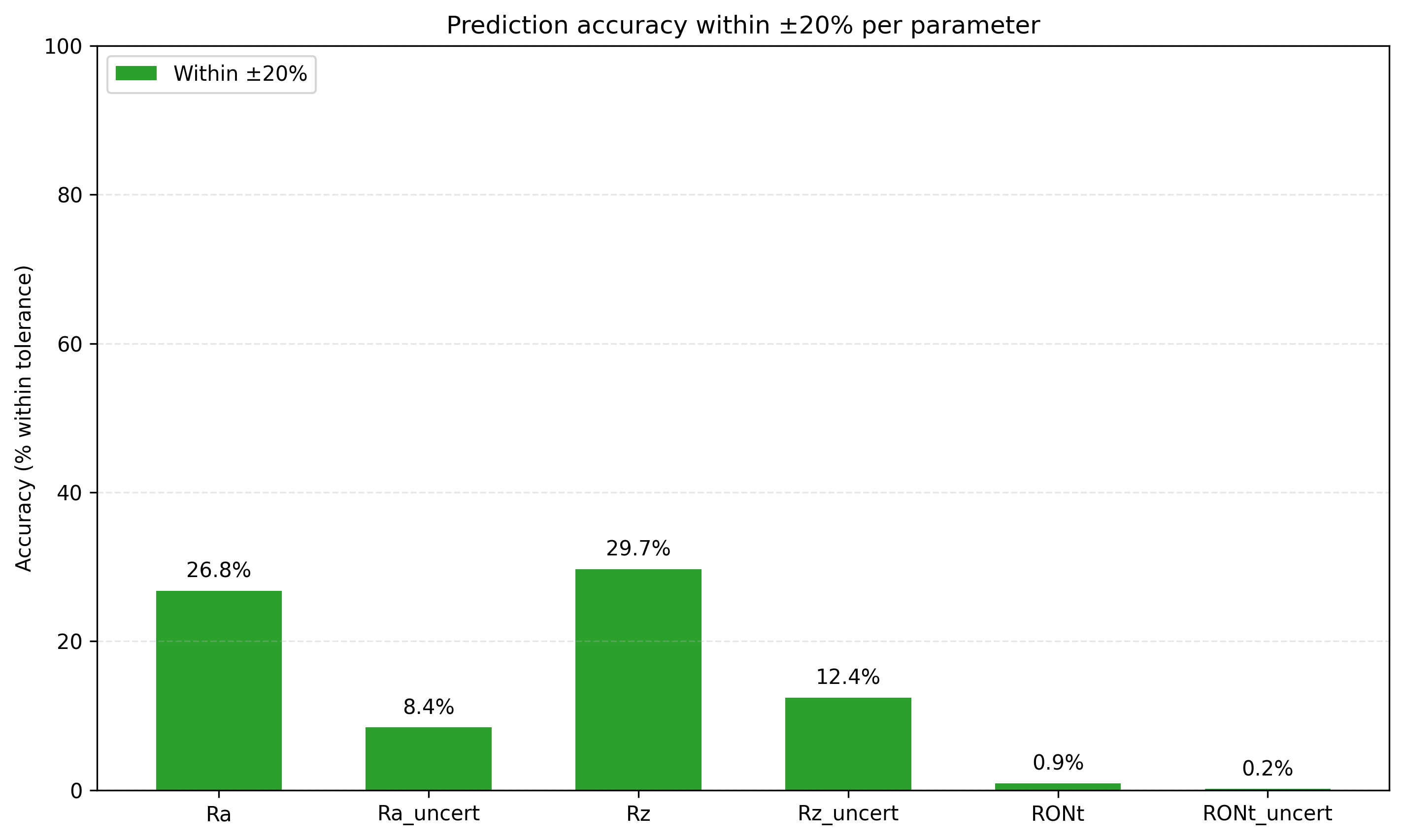}{Figure S168: accuracy within tol 20percent (\label{fig:supp-168})}\
\suppimage{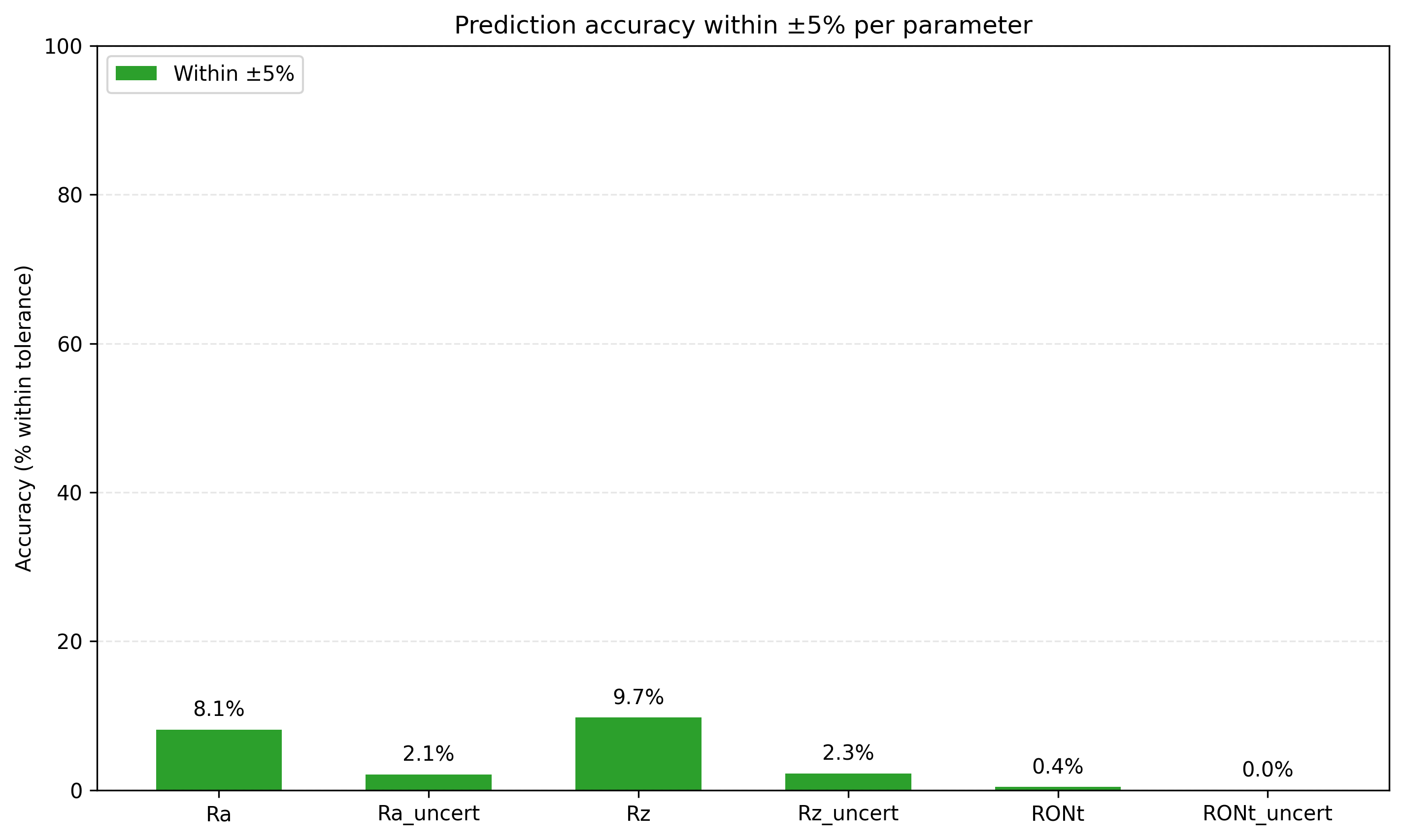}{Figure S169: accuracy within tol 5percent (\label{fig:supp-169})}\hfill
\suppimage{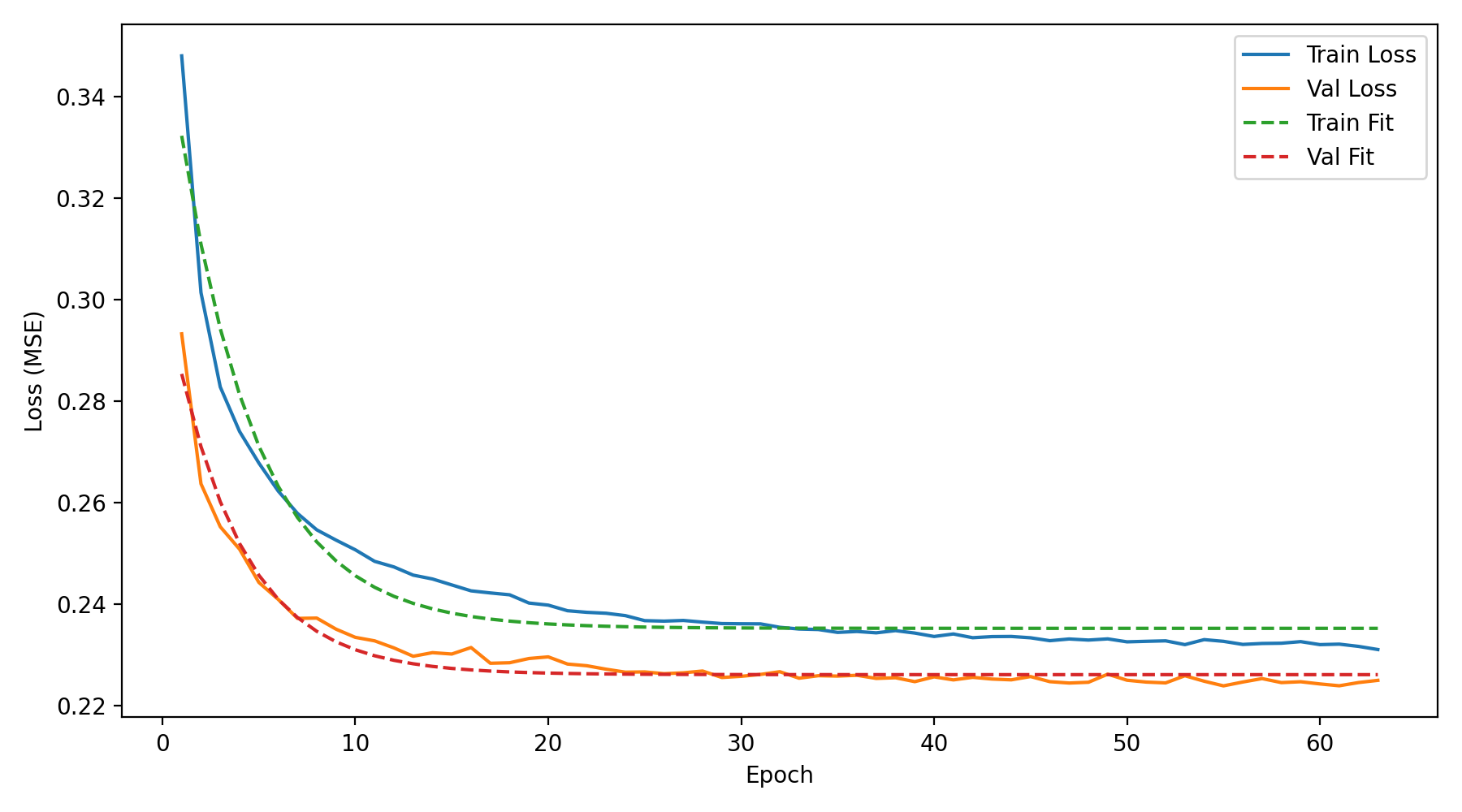}{Figure S170: loss curves (\label{fig:supp-170})}\
\suppimage{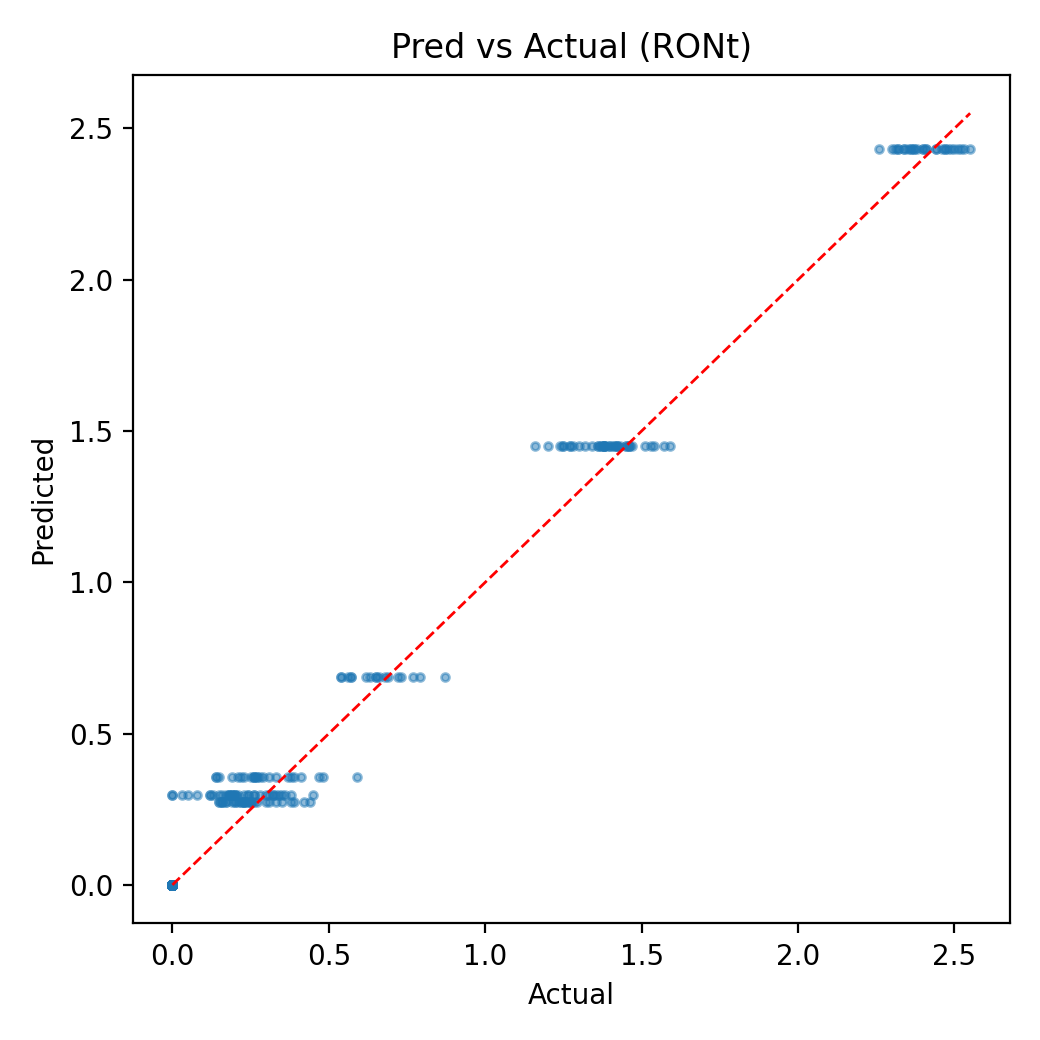}{Figure S171: pred vs actual RONt (\label{fig:supp-171})}\hfill
\suppimage{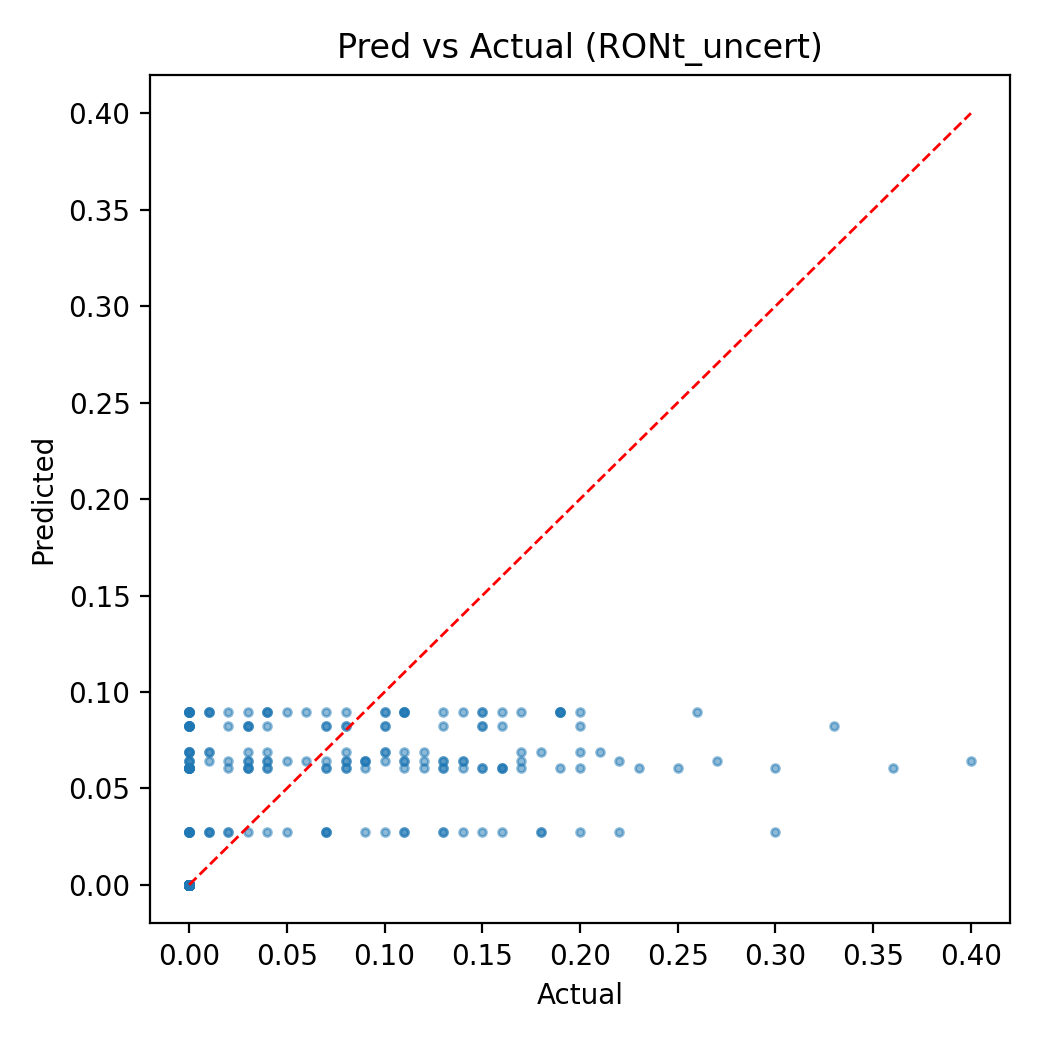}{Figure S172: pred vs actual RONt uncert (\label{fig:supp-172})}\
\suppimage{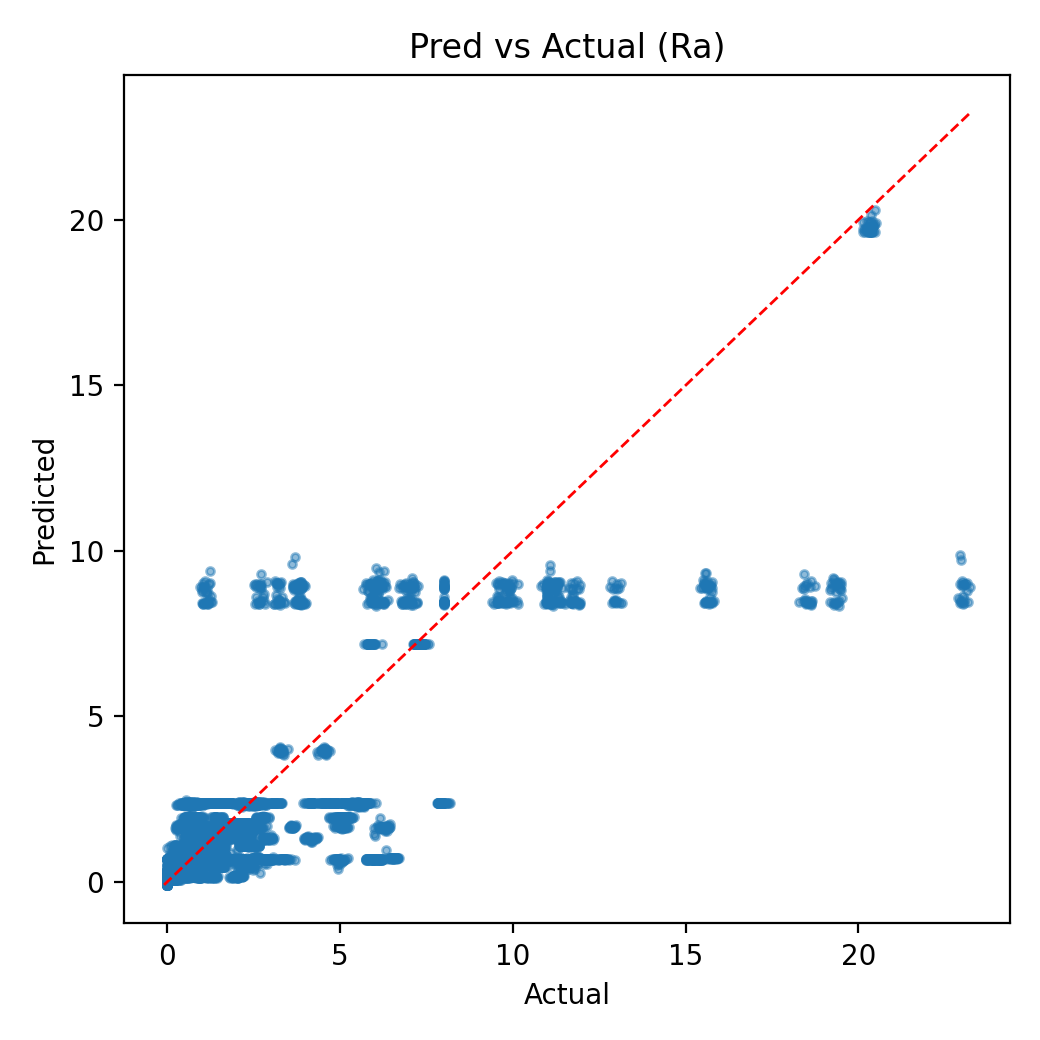}{Figure S173: pred vs actual Ra (\label{fig:supp-173})}\hfill
\suppimage{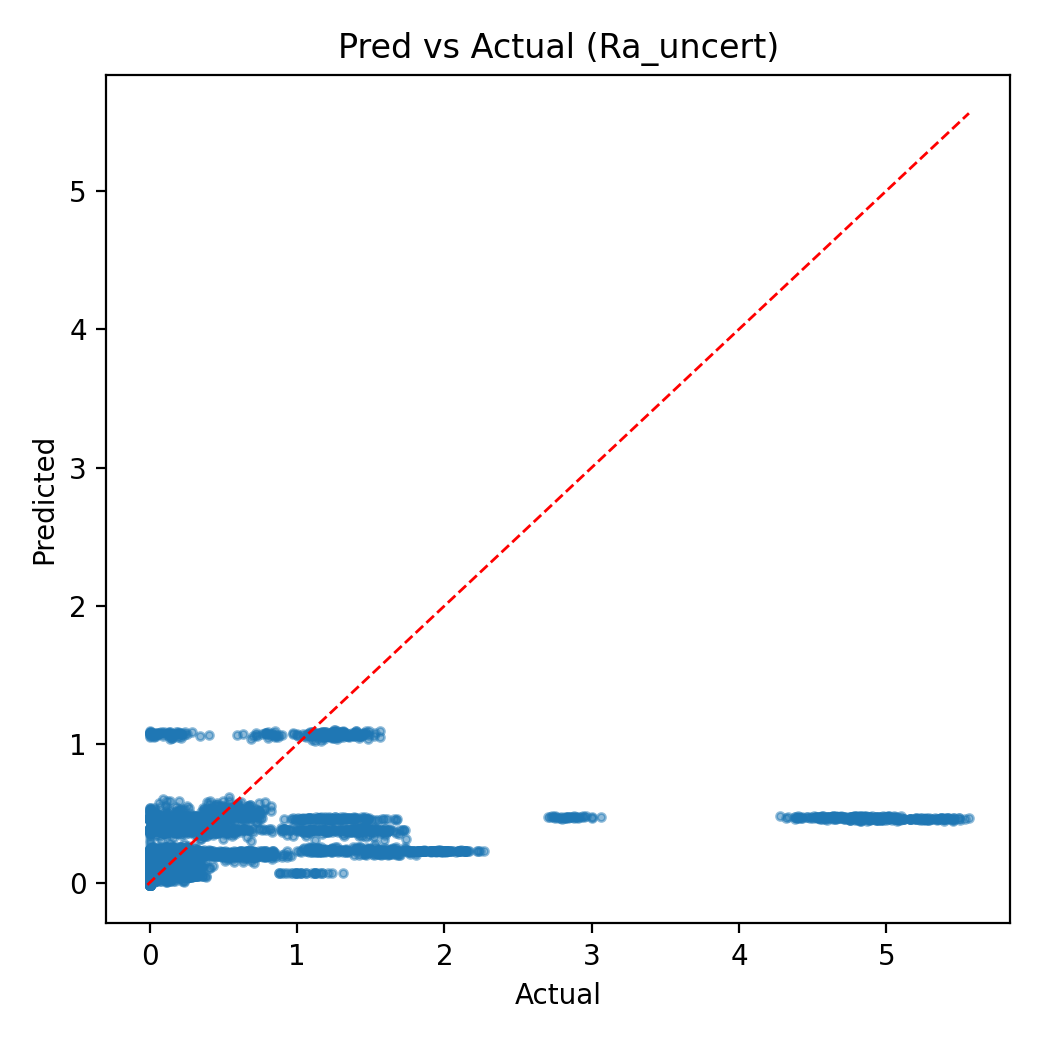}{Figure S174: pred vs actual Ra uncert (\label{fig:supp-174})}\
\suppimage{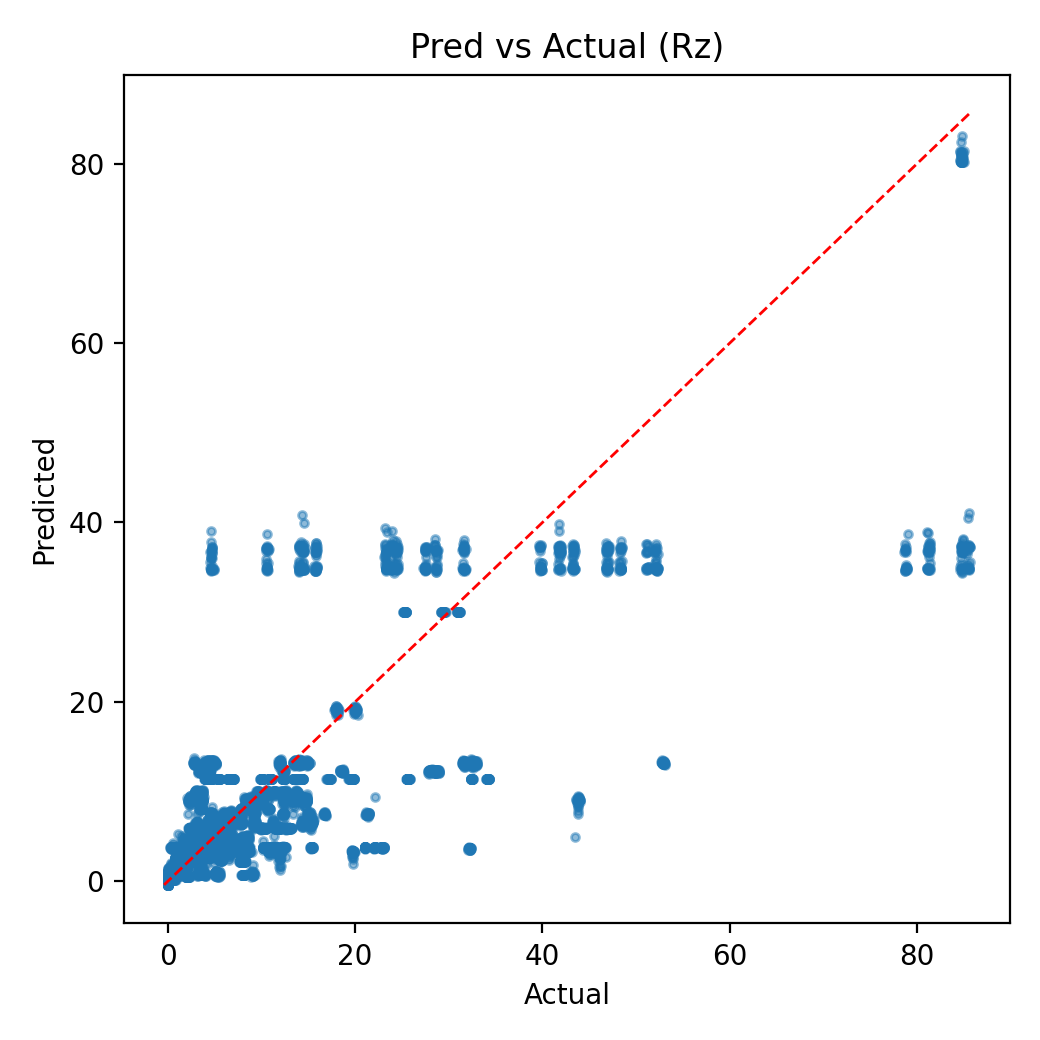}{Figure S175: pred vs actual Rz (\label{fig:supp-175})}\hfill
\suppimage{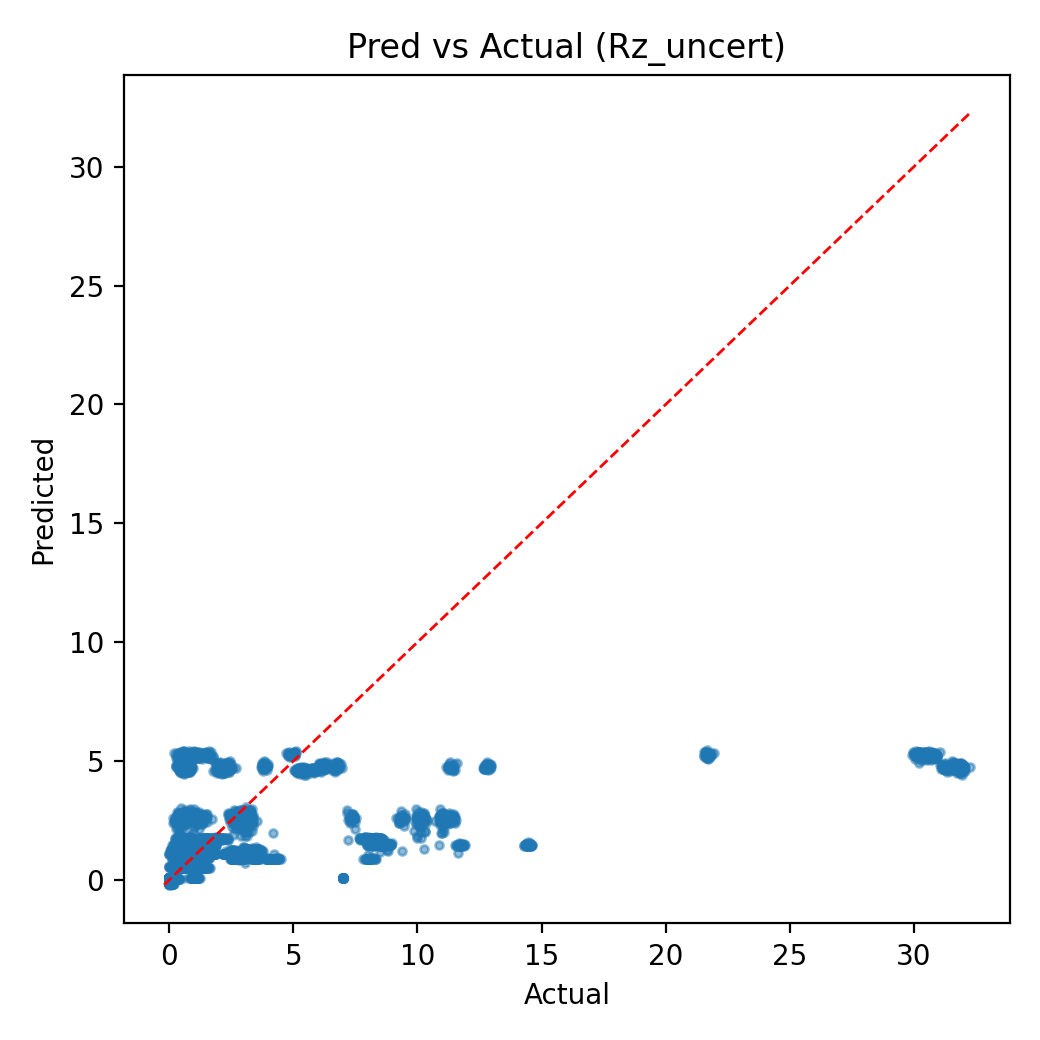}{Figure S176: pred vs actual Rz uncert (\label{fig:supp-176})}\
\suppimage{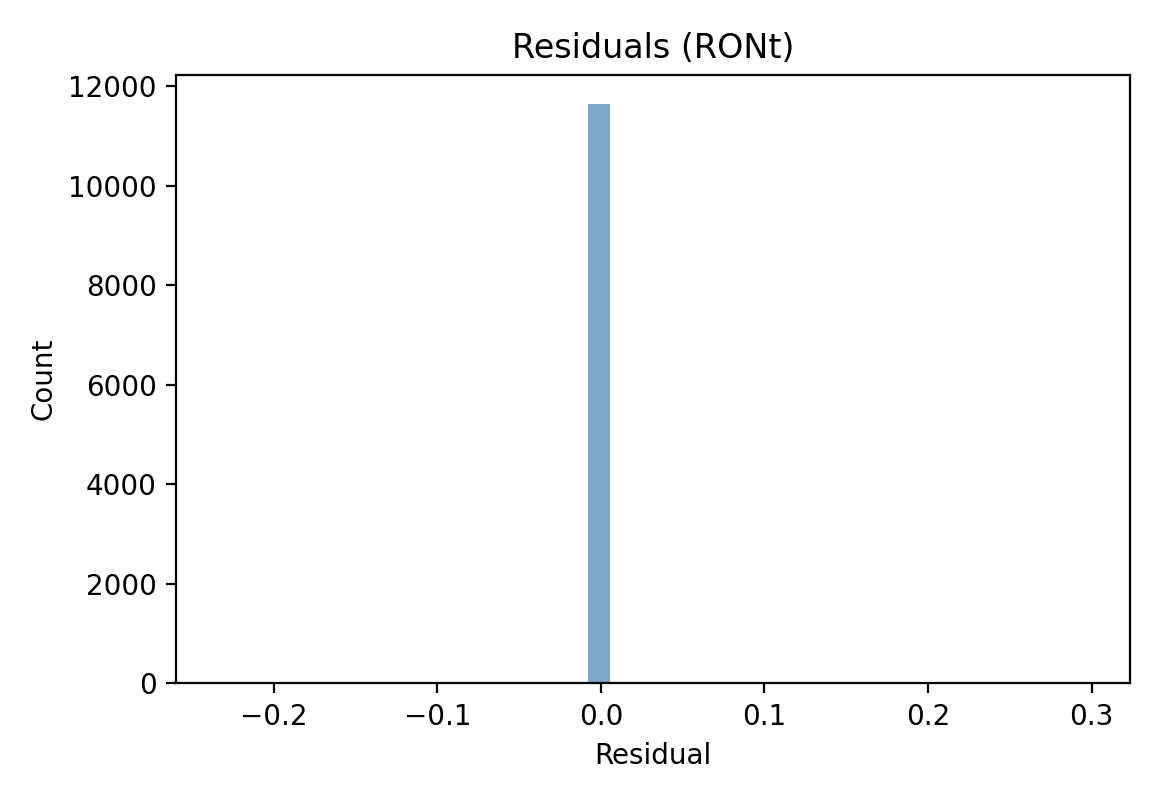}{Figure S177: residuals hist RONt (\label{fig:supp-177})}\hfill
\suppimage{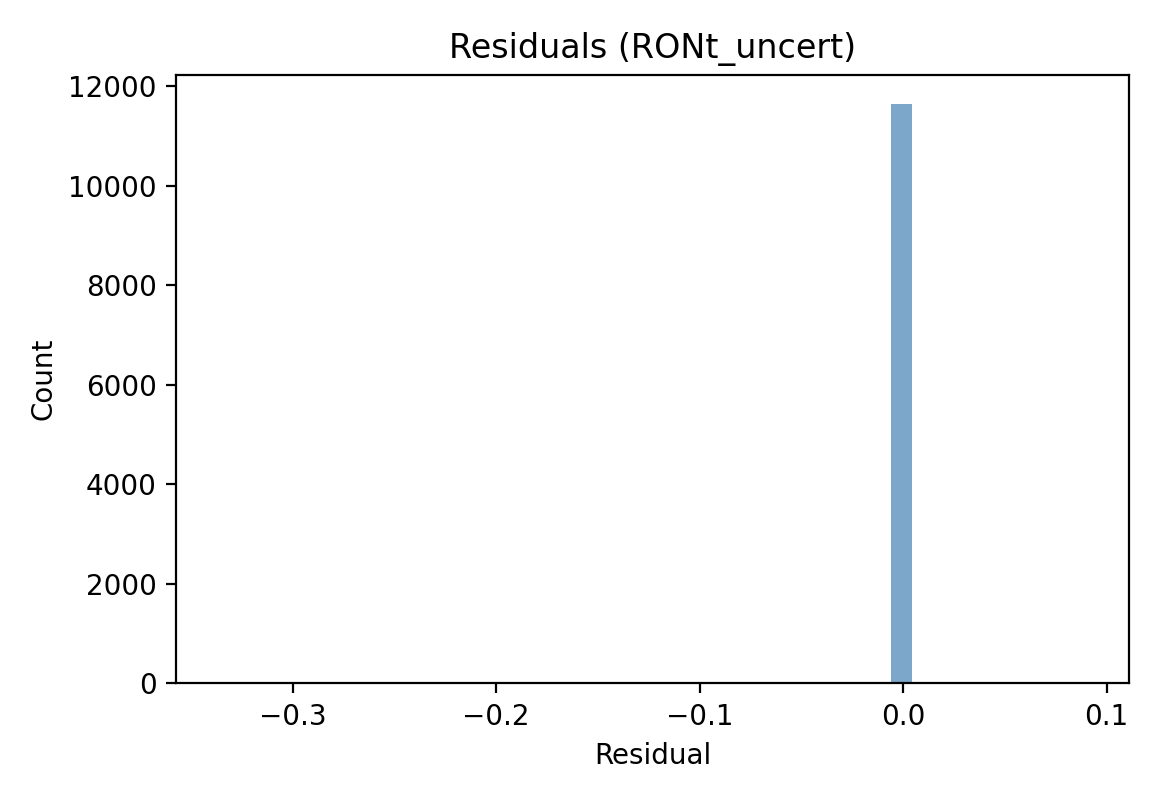}{Figure S178: residuals hist RONt uncert (\label{fig:supp-178})}\
\suppimage{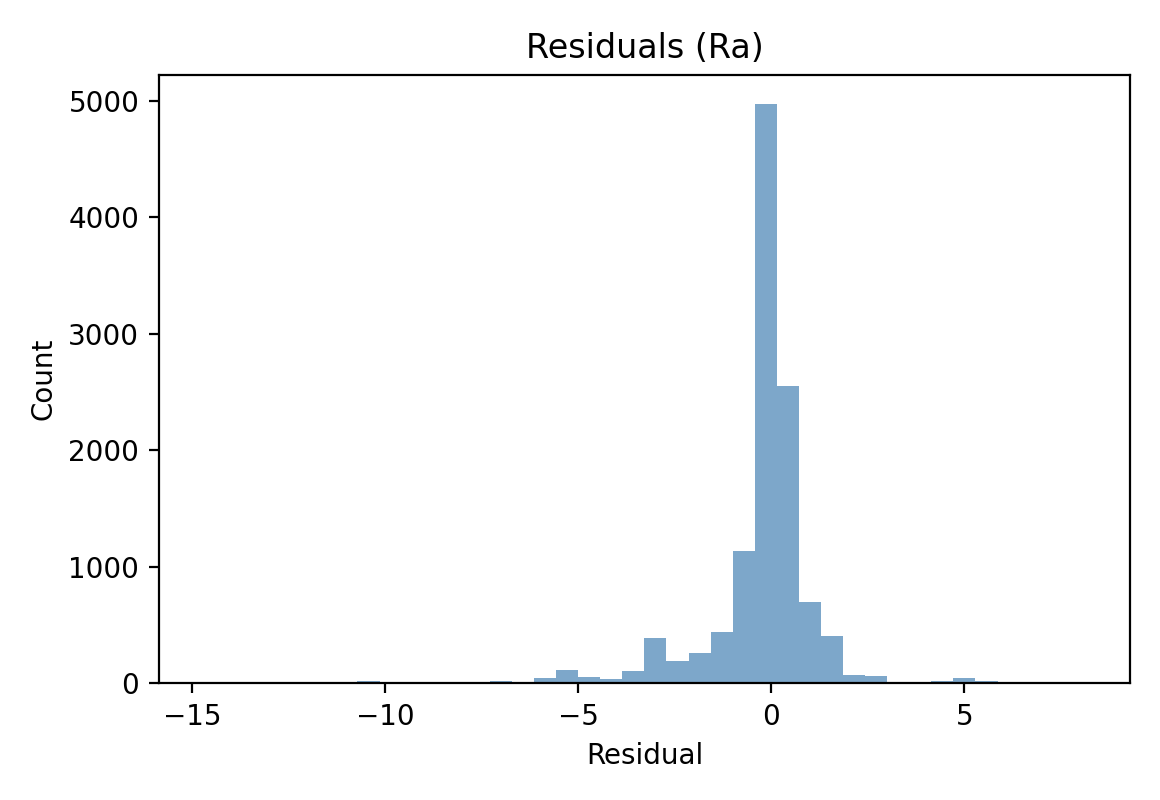}{Figure S179: residuals hist Ra (\label{fig:supp-179})}\hfill
\suppimage{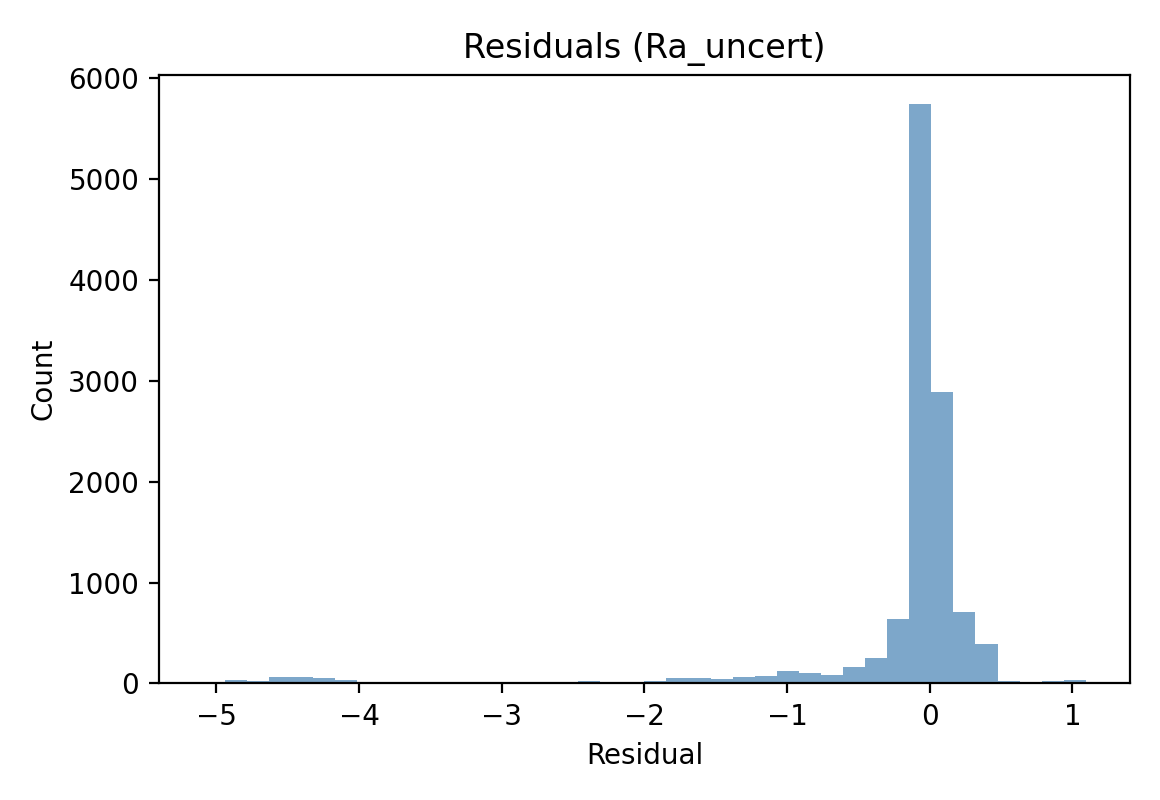}{Figure S180: residuals hist Ra uncert (\label{fig:supp-180})}\
\suppimage{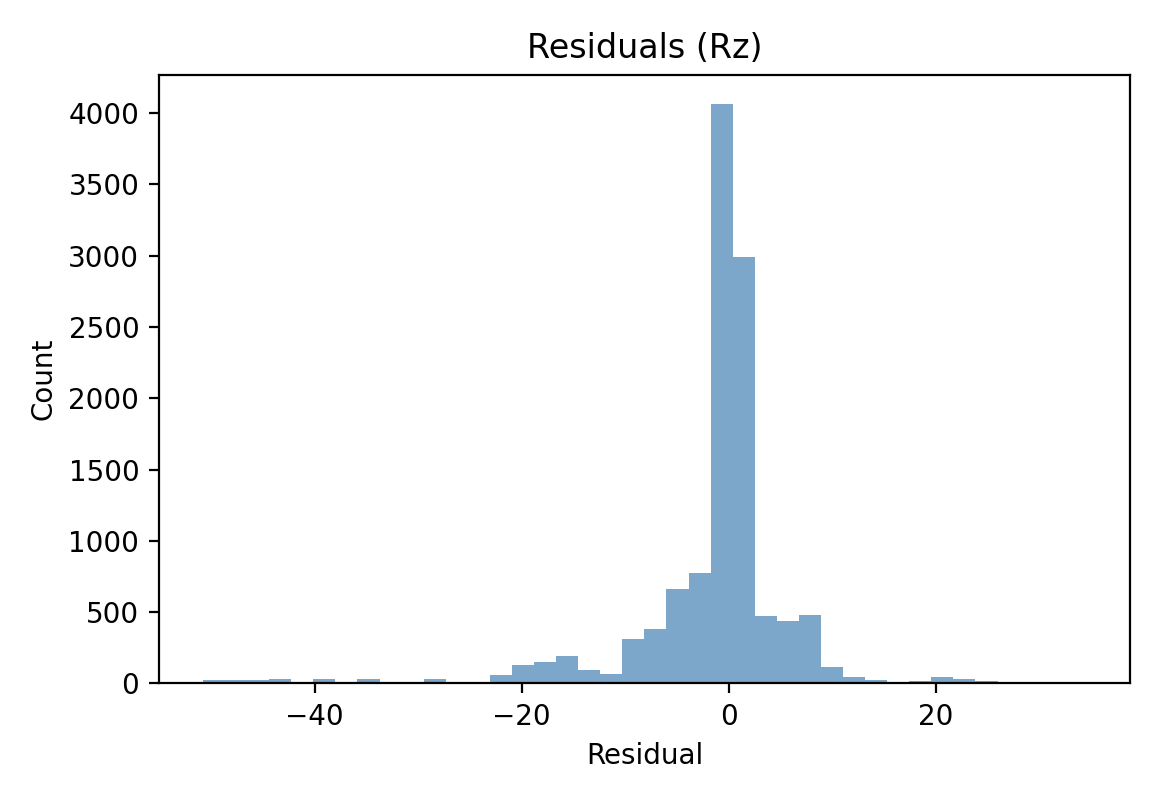}{Figure S181: residuals hist Rz (\label{fig:supp-181})}\hfill
\suppimage{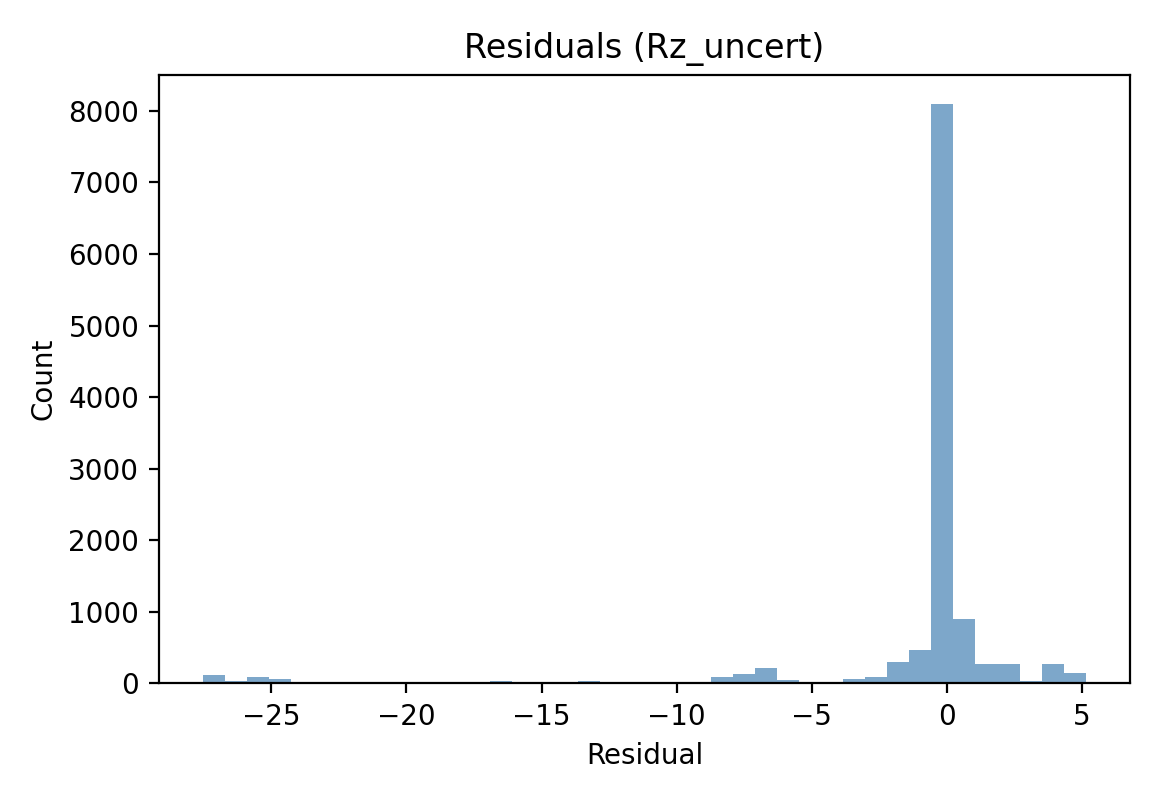}{Figure S182: residuals hist Rz uncert (\label{fig:supp-182})}\
\suppimage{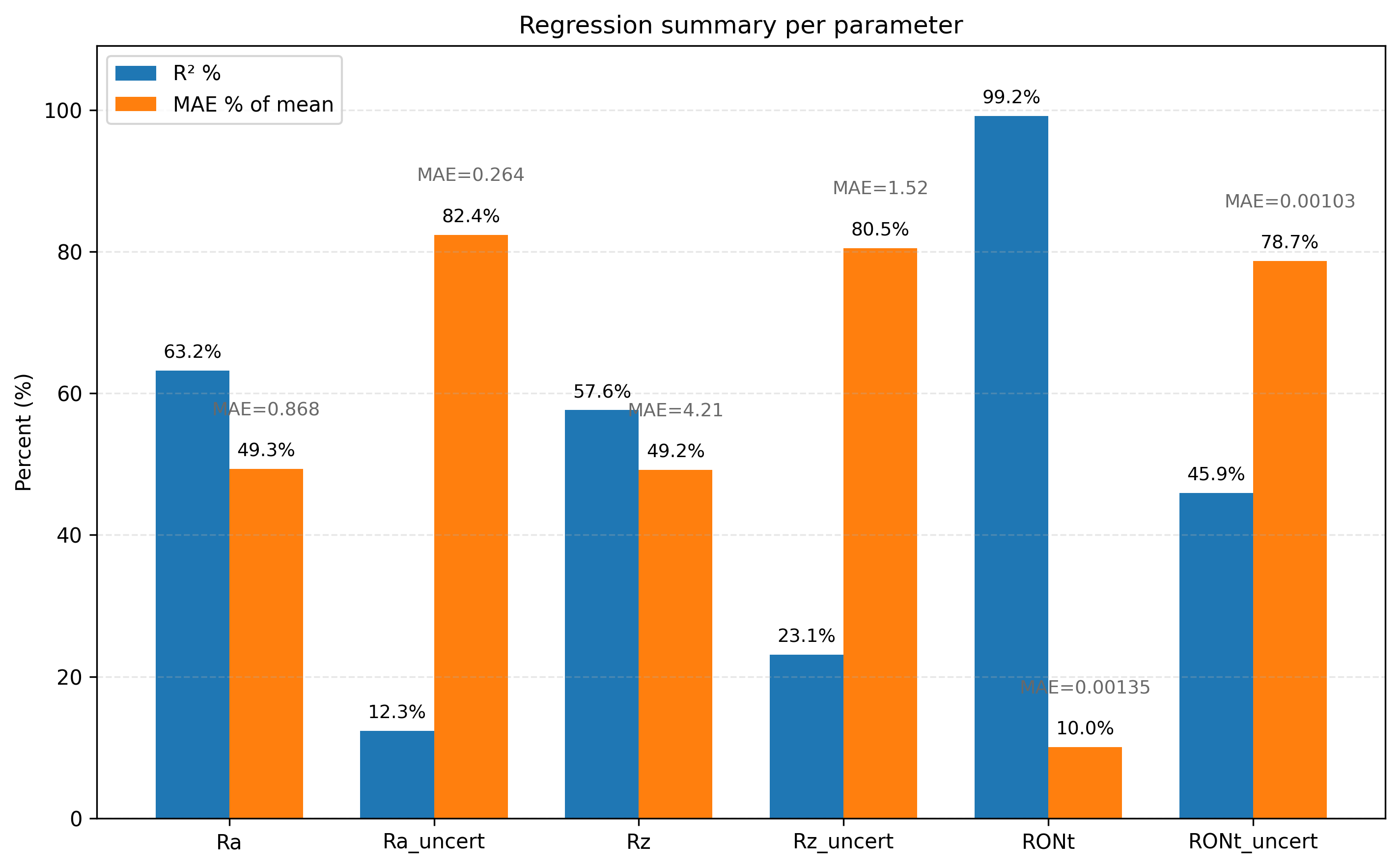}{Figure S183: regression summary bars (\label{fig:supp-183})}\hfill
\suppimage{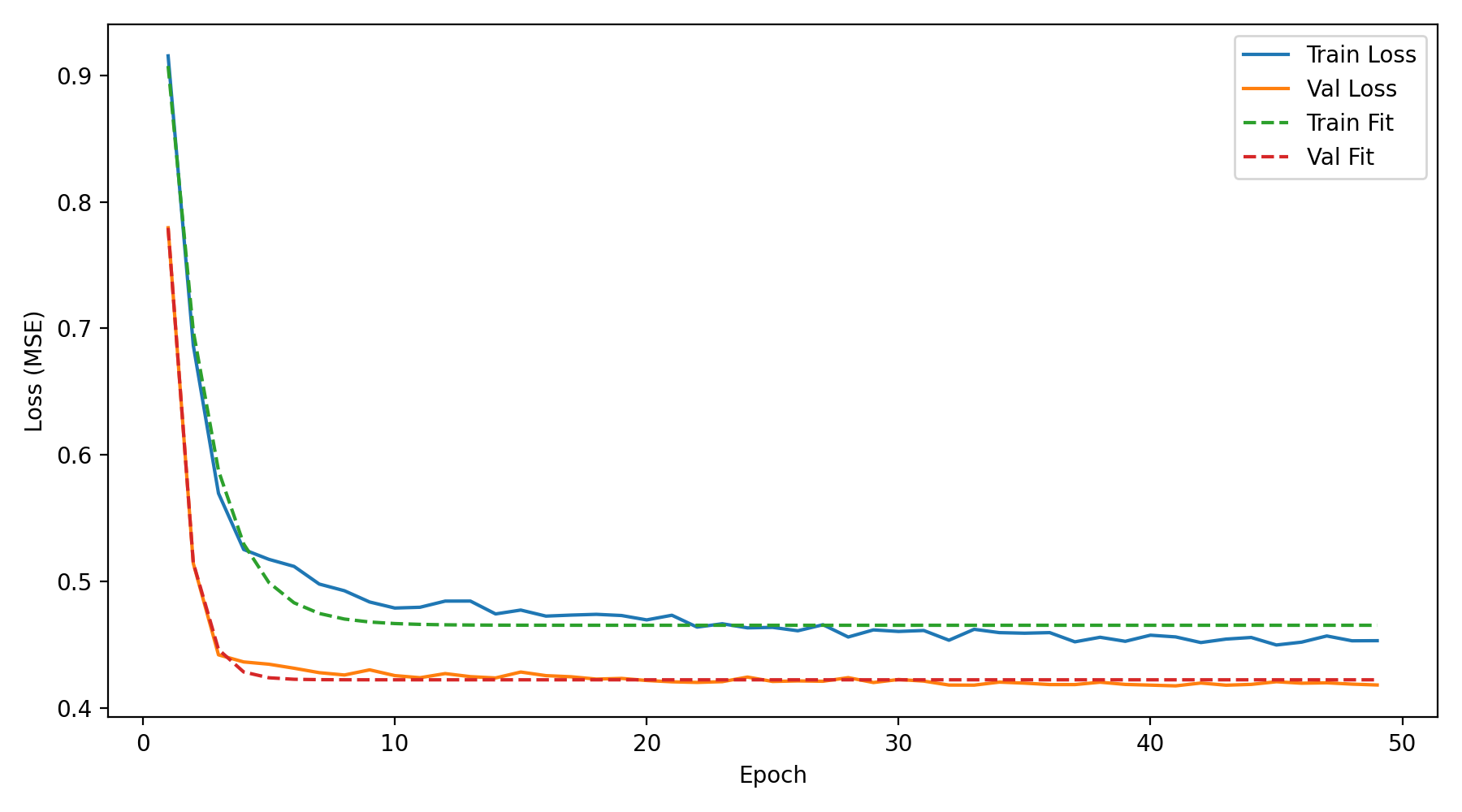}{Figure S184: loss curves (\label{fig:supp-184})}\
\suppimage{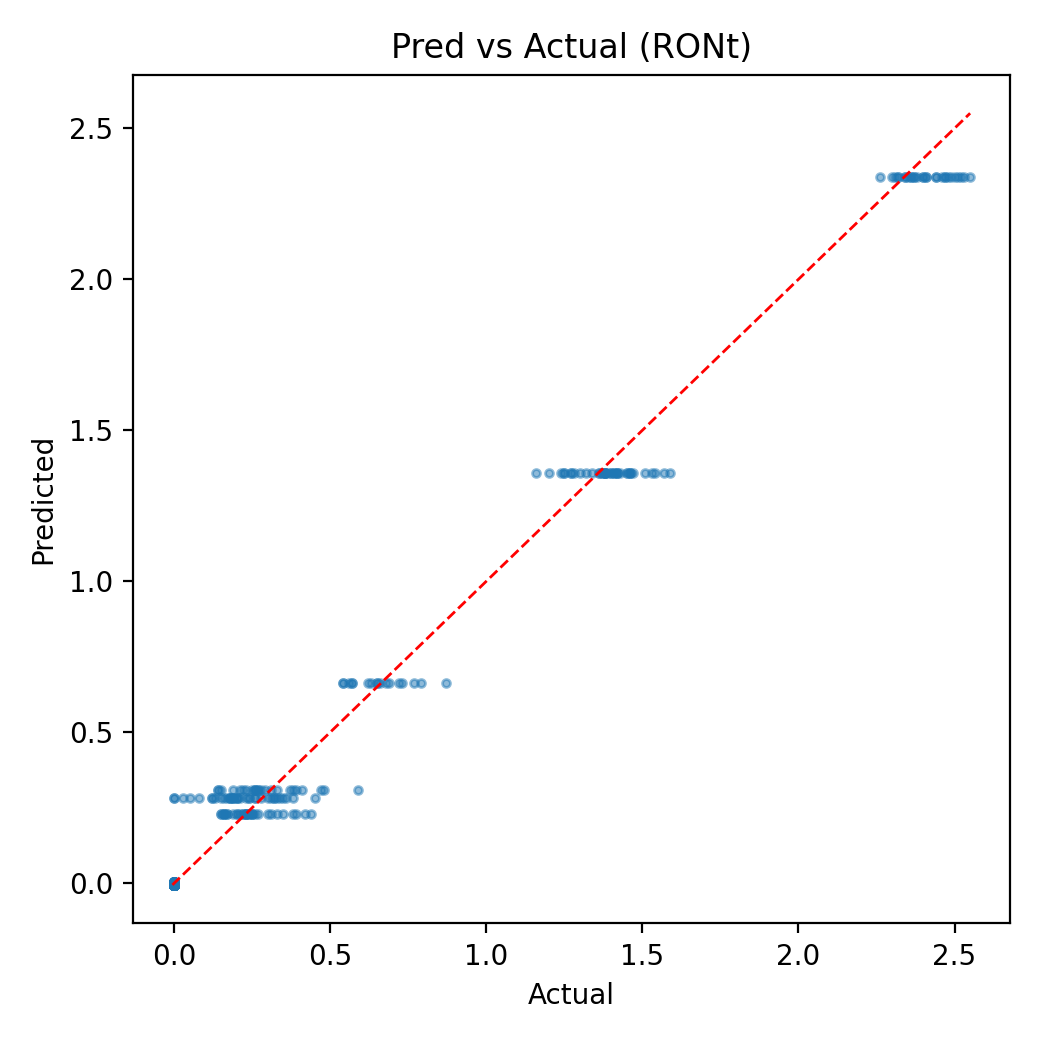}{Figure S185: pred vs actual RONt (\label{fig:supp-185})}\hfill
\suppimage{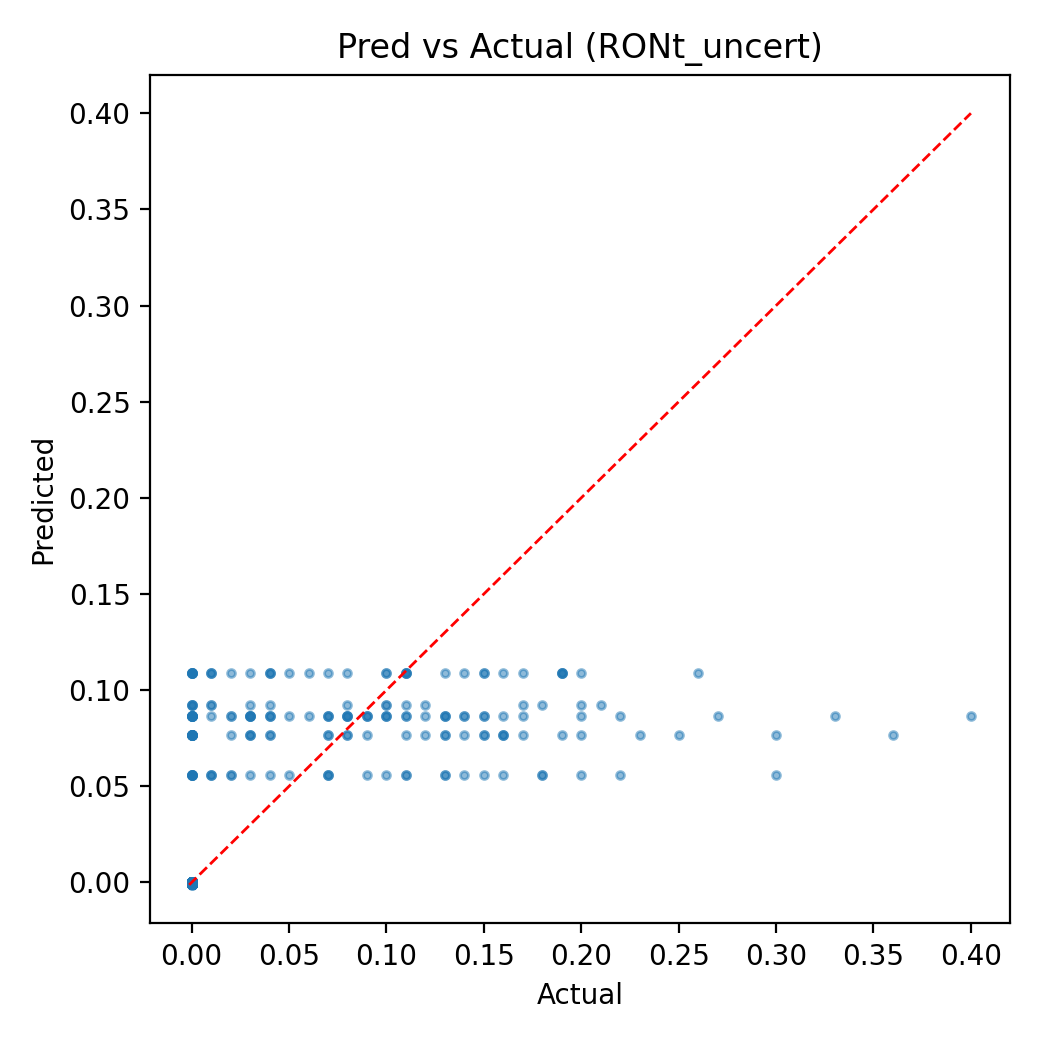}{Figure S186: pred vs actual RONt uncert (\label{fig:supp-186})}\
\suppimage{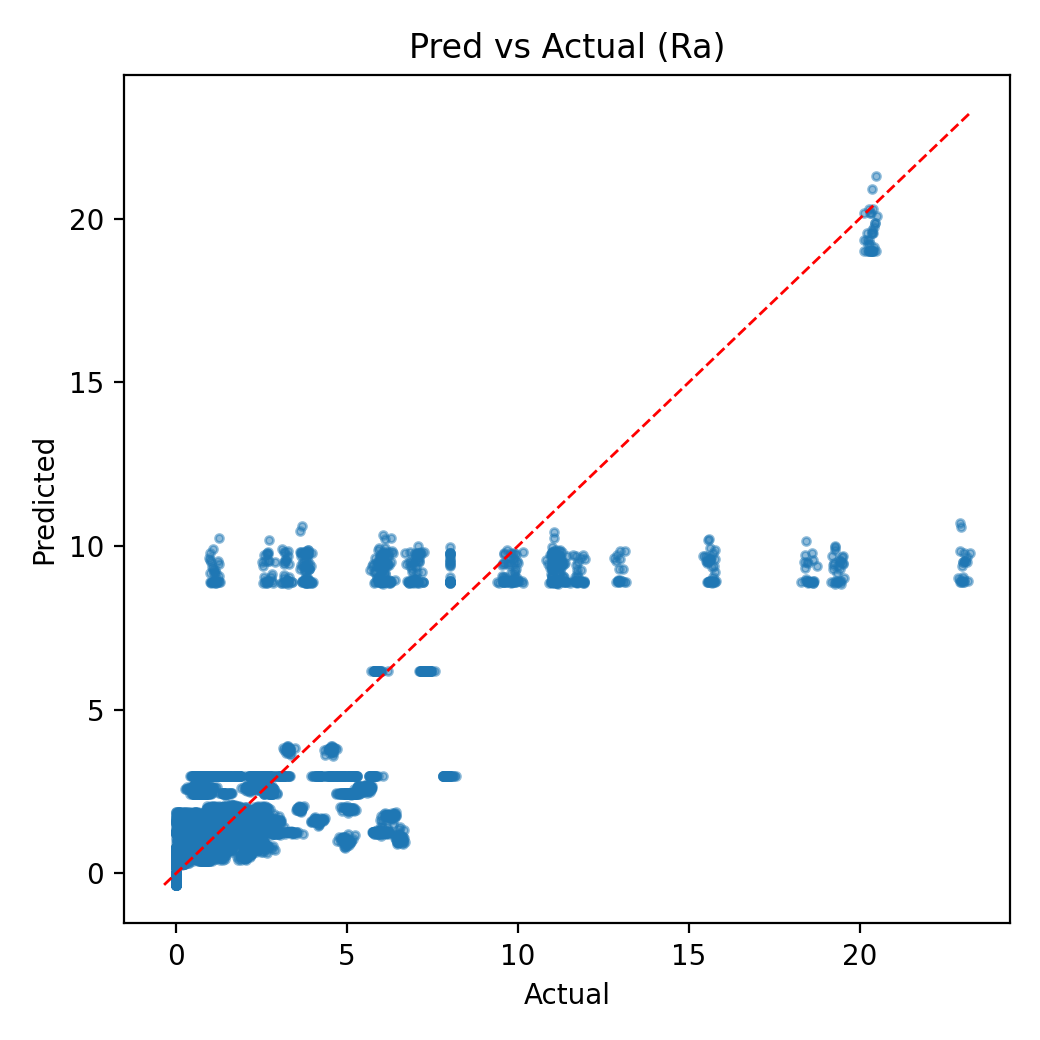}{Figure S187: pred vs actual Ra (\label{fig:supp-187})}\hfill
\suppimage{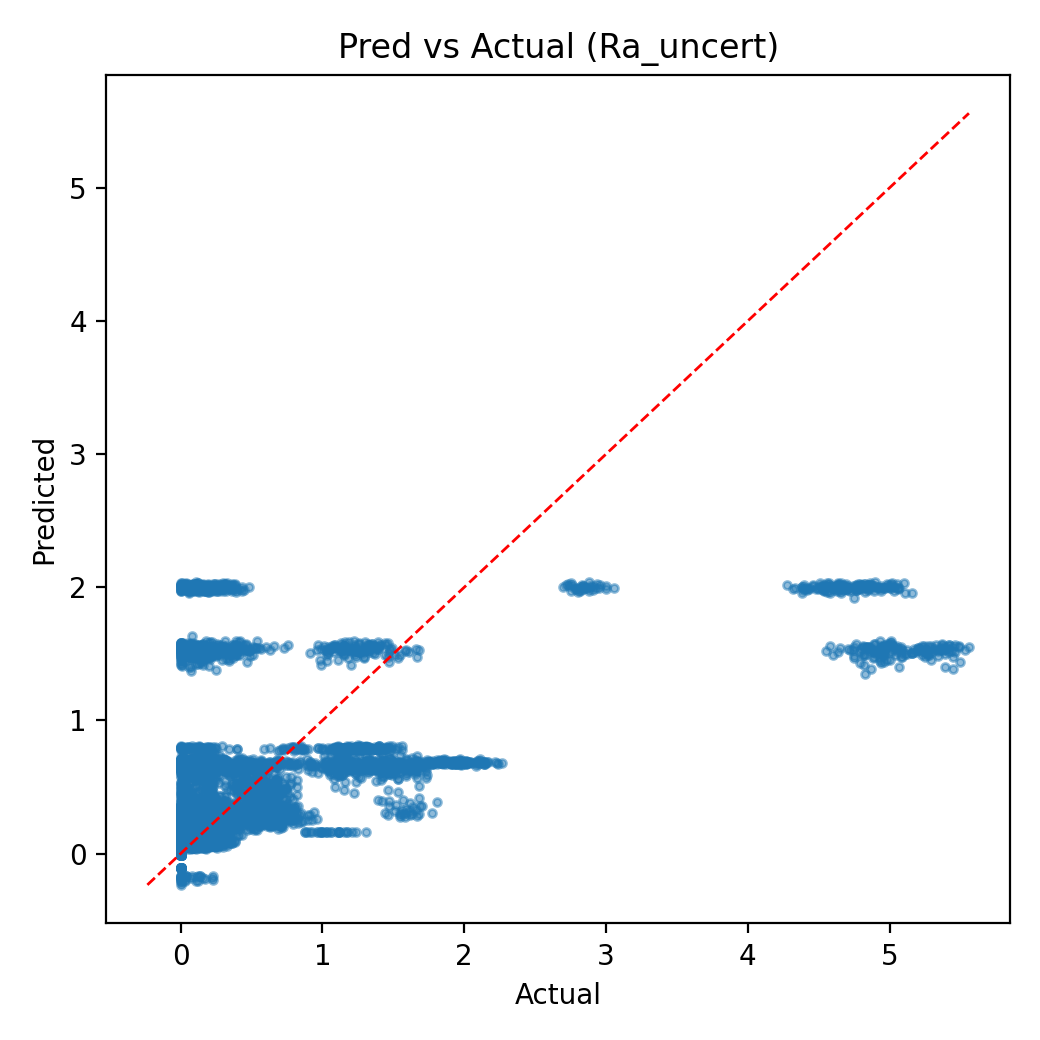}{Figure S188: pred vs actual Ra uncert (\label{fig:supp-188})}\
\suppimage{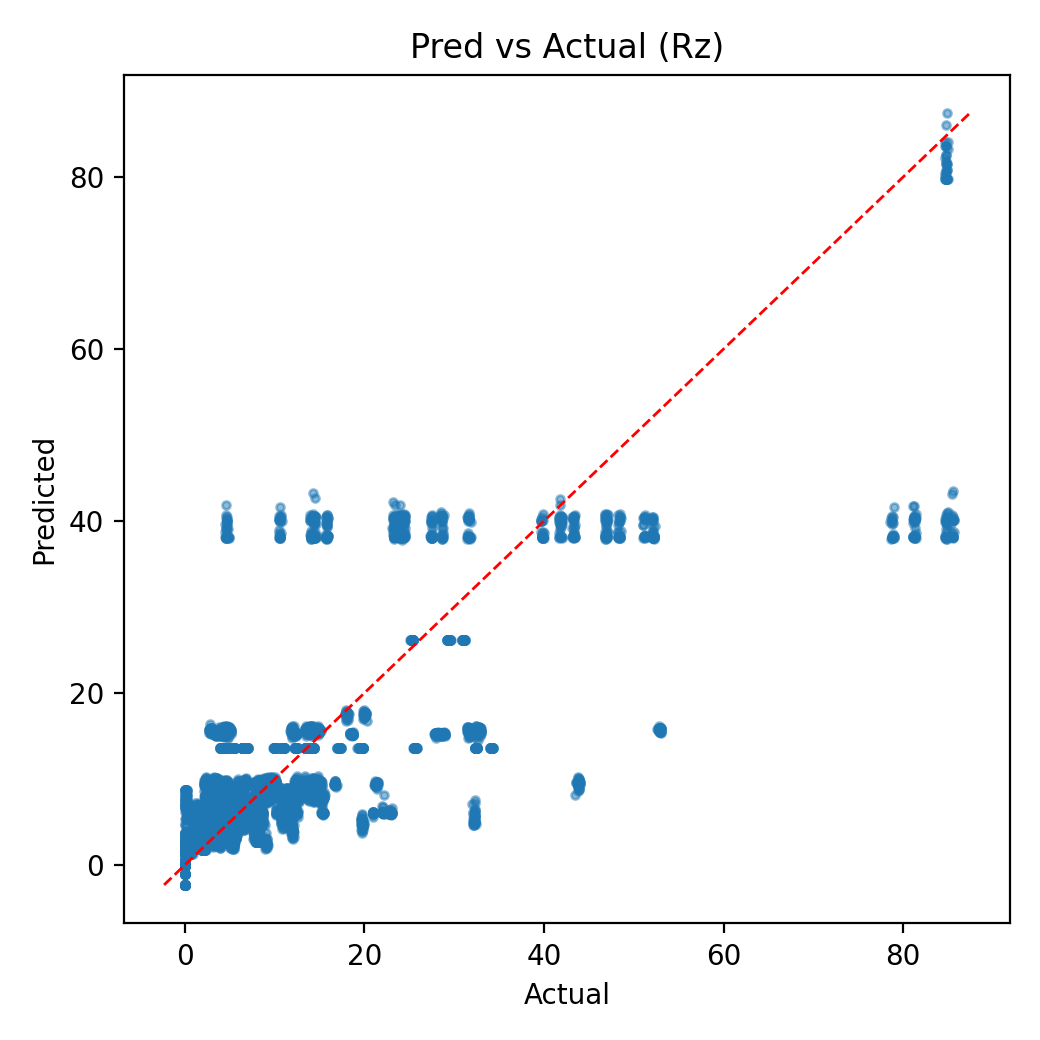}{Figure S189: pred vs actual Rz (\label{fig:supp-189})}\hfill
\suppimage{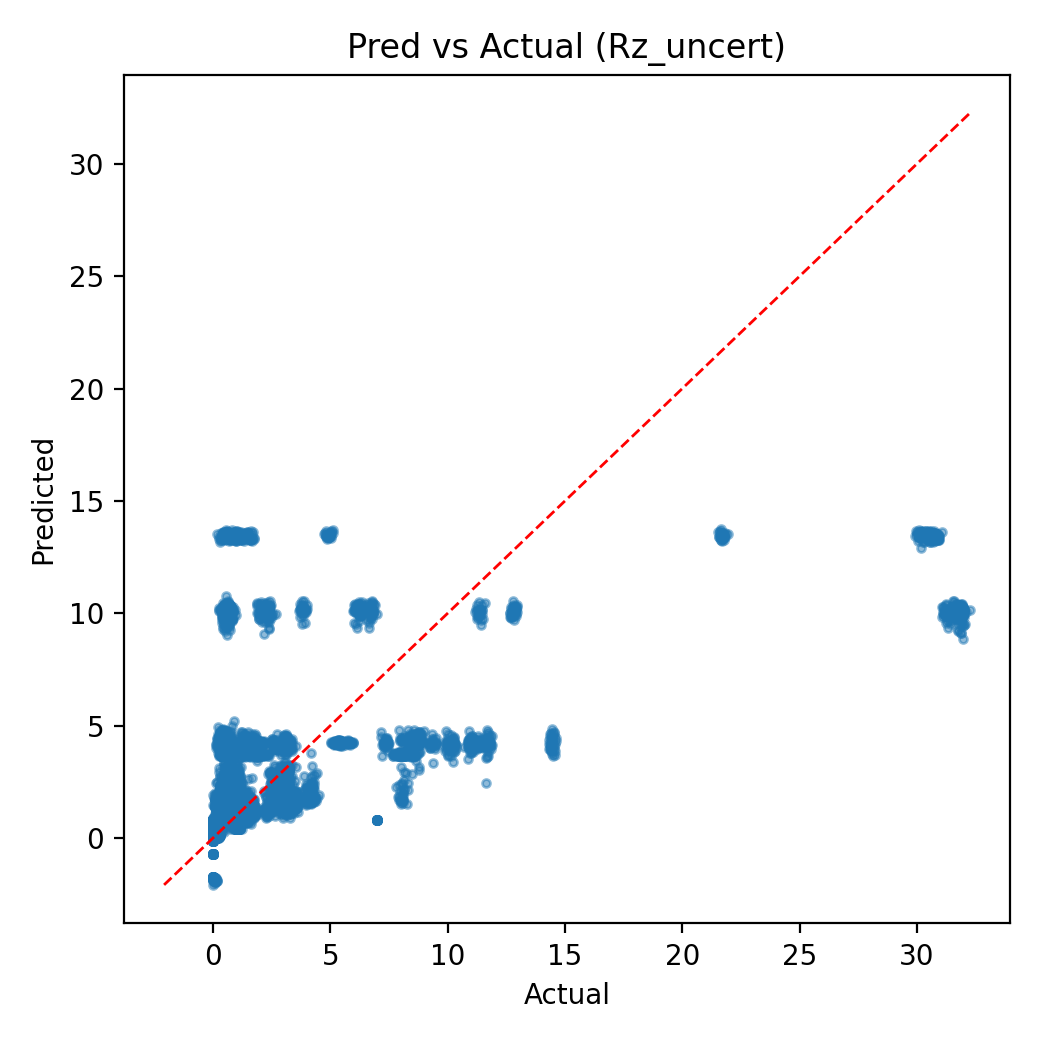}{Figure S190: pred vs actual Rz uncert (\label{fig:supp-190})}\
\suppimage{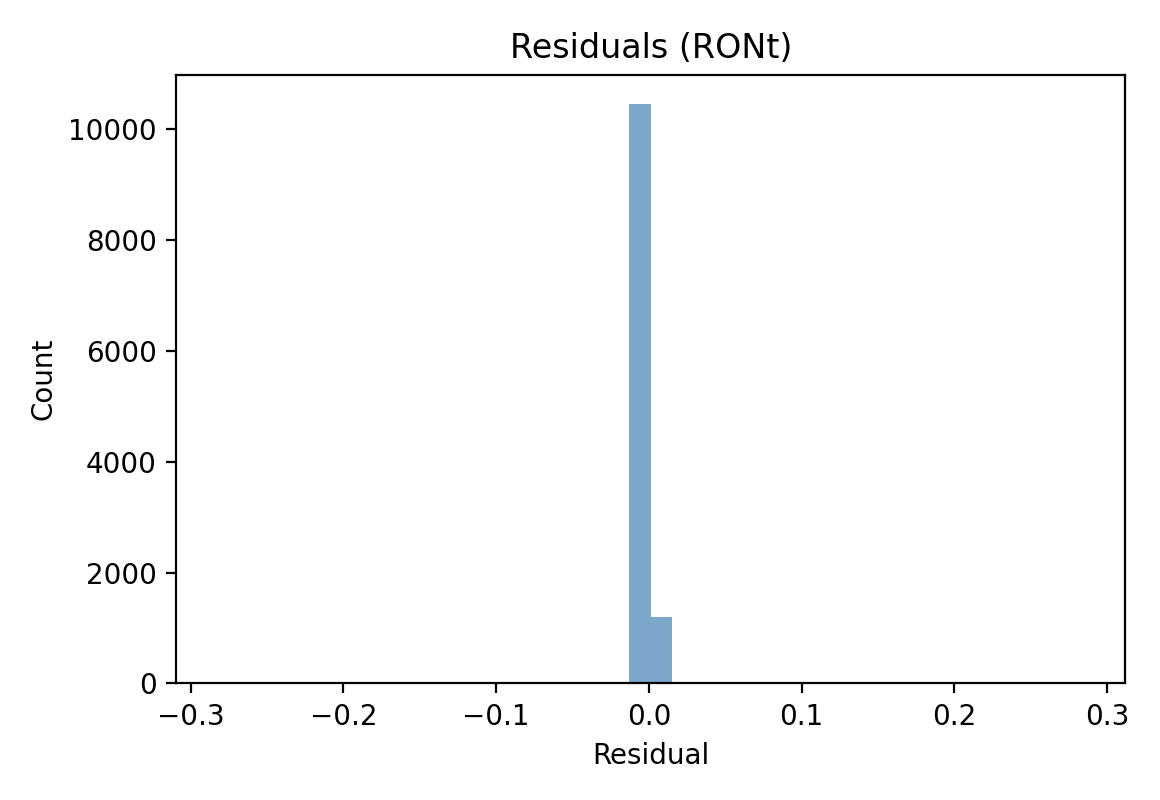}{Figure S191: residuals hist RONt (\label{fig:supp-191})}\hfill
\suppimage{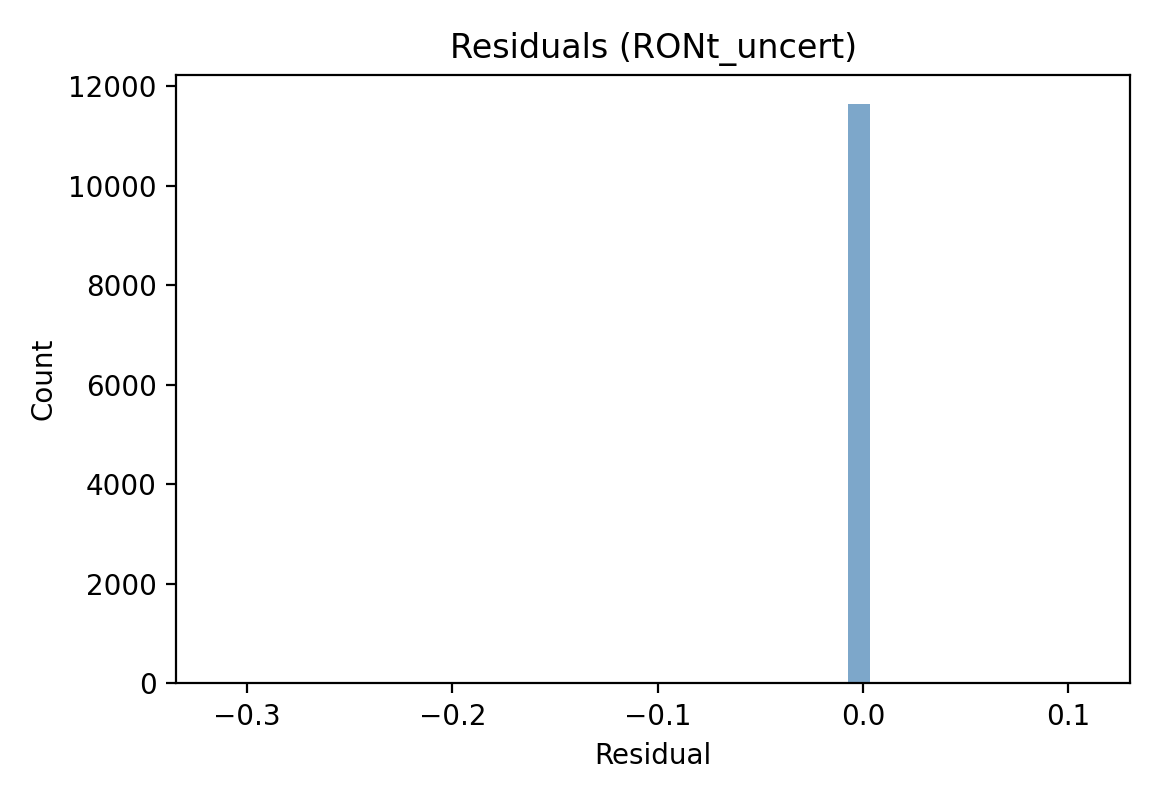}{Figure S192: residuals hist RONt uncert (\label{fig:supp-192})}\
\suppimage{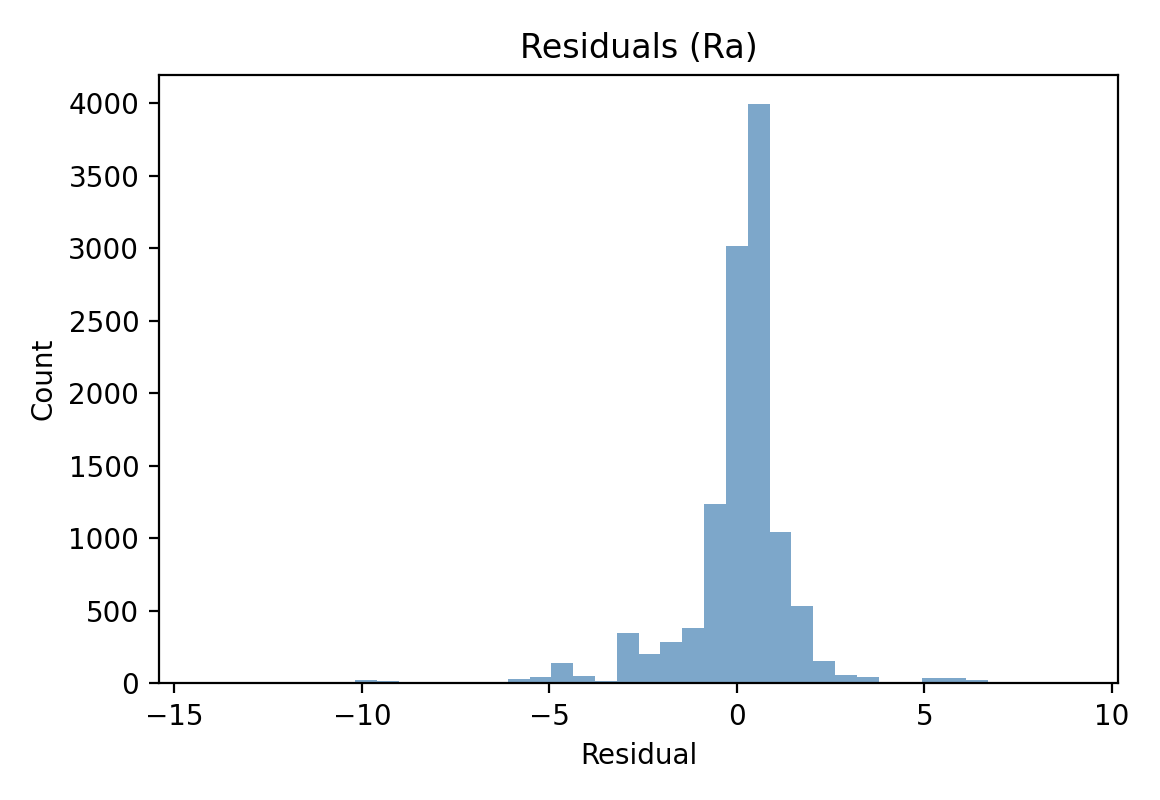}{Figure S193: residuals hist Ra (\label{fig:supp-193})}\hfill
\suppimage{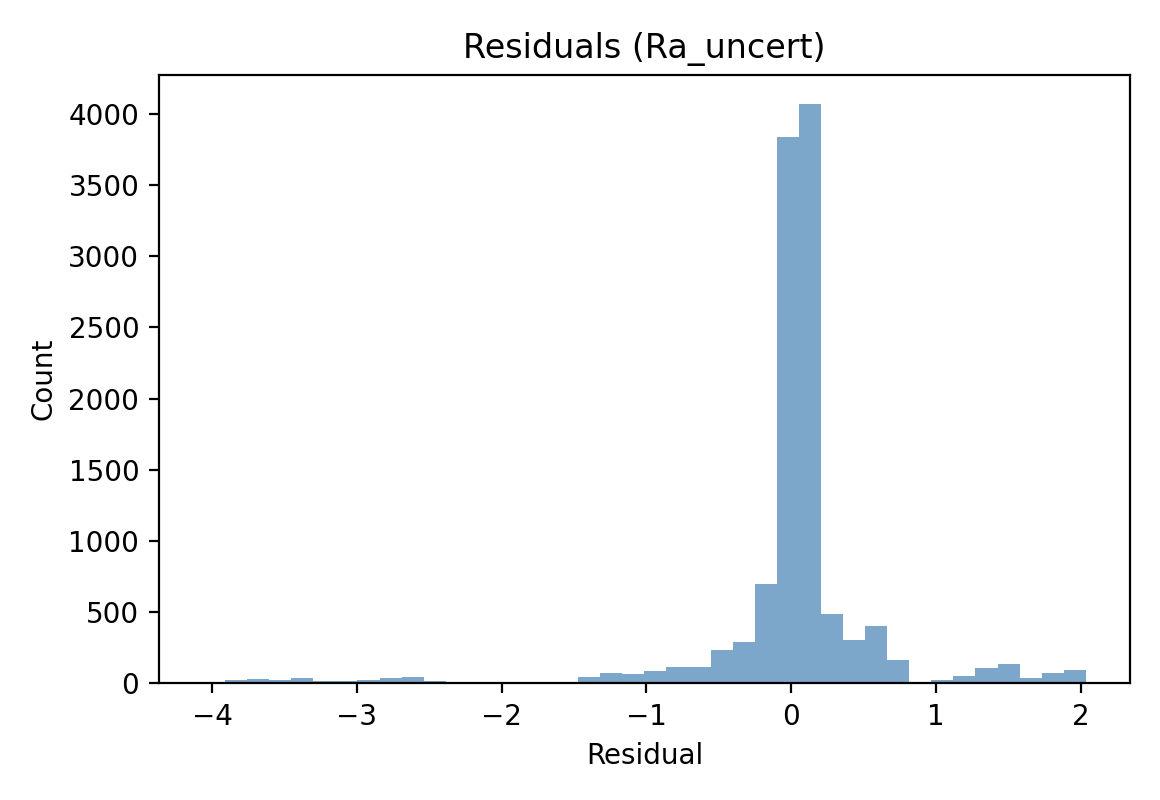}{Figure S194: residuals hist Ra uncert (\label{fig:supp-194})}\
\suppimage{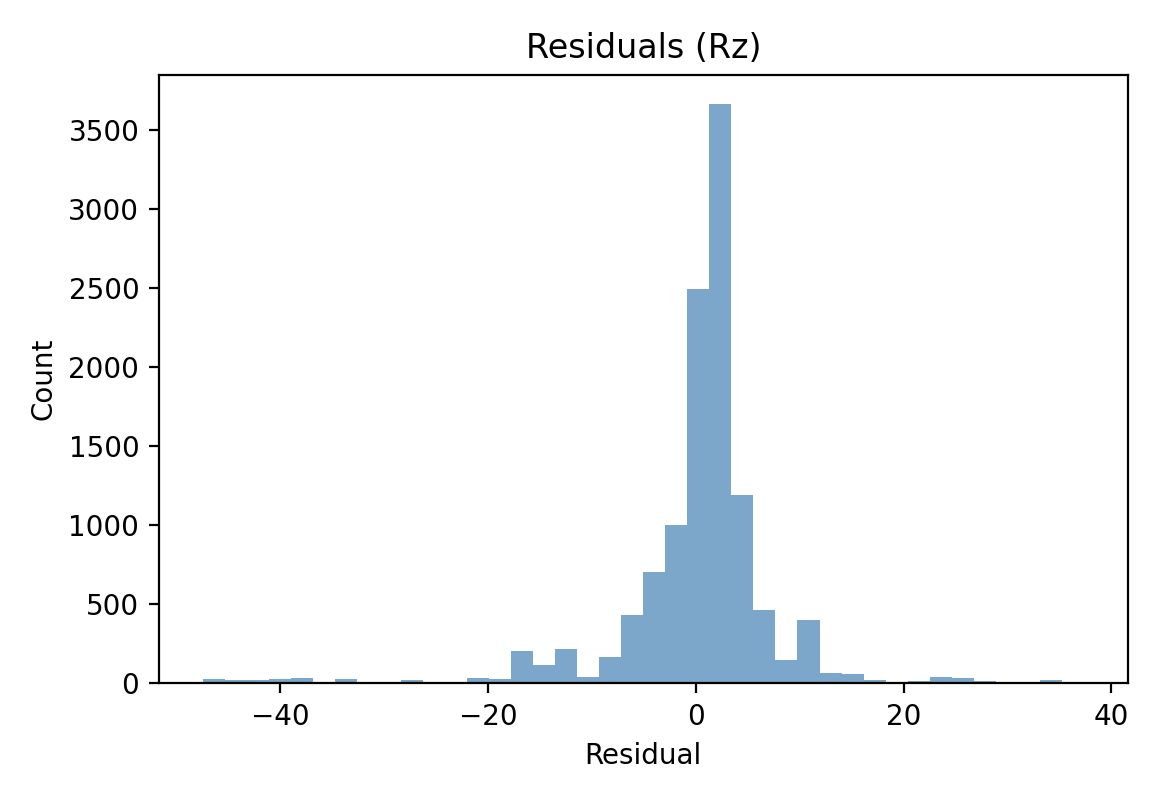}{Figure S195: residuals hist Rz (\label{fig:supp-195})}\hfill
\suppimage{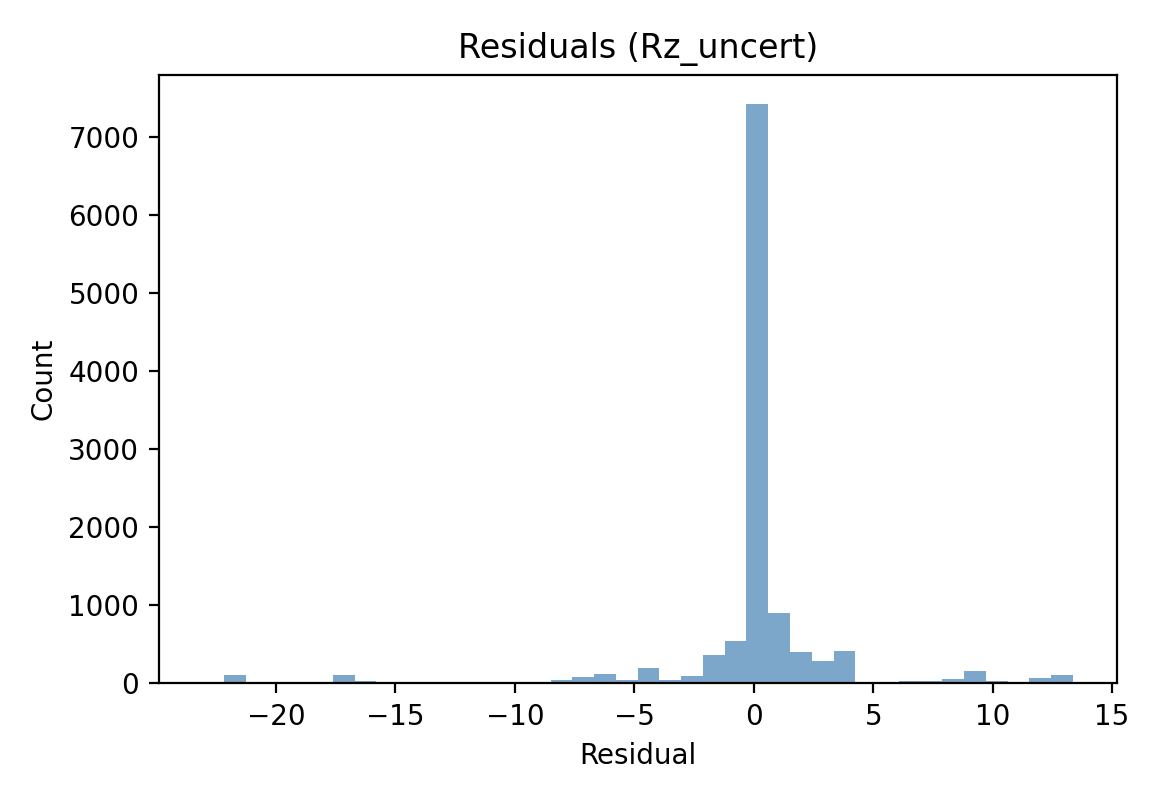}{Figure S196: residuals hist Rz uncert (\label{fig:supp-196})}\
\suppimage{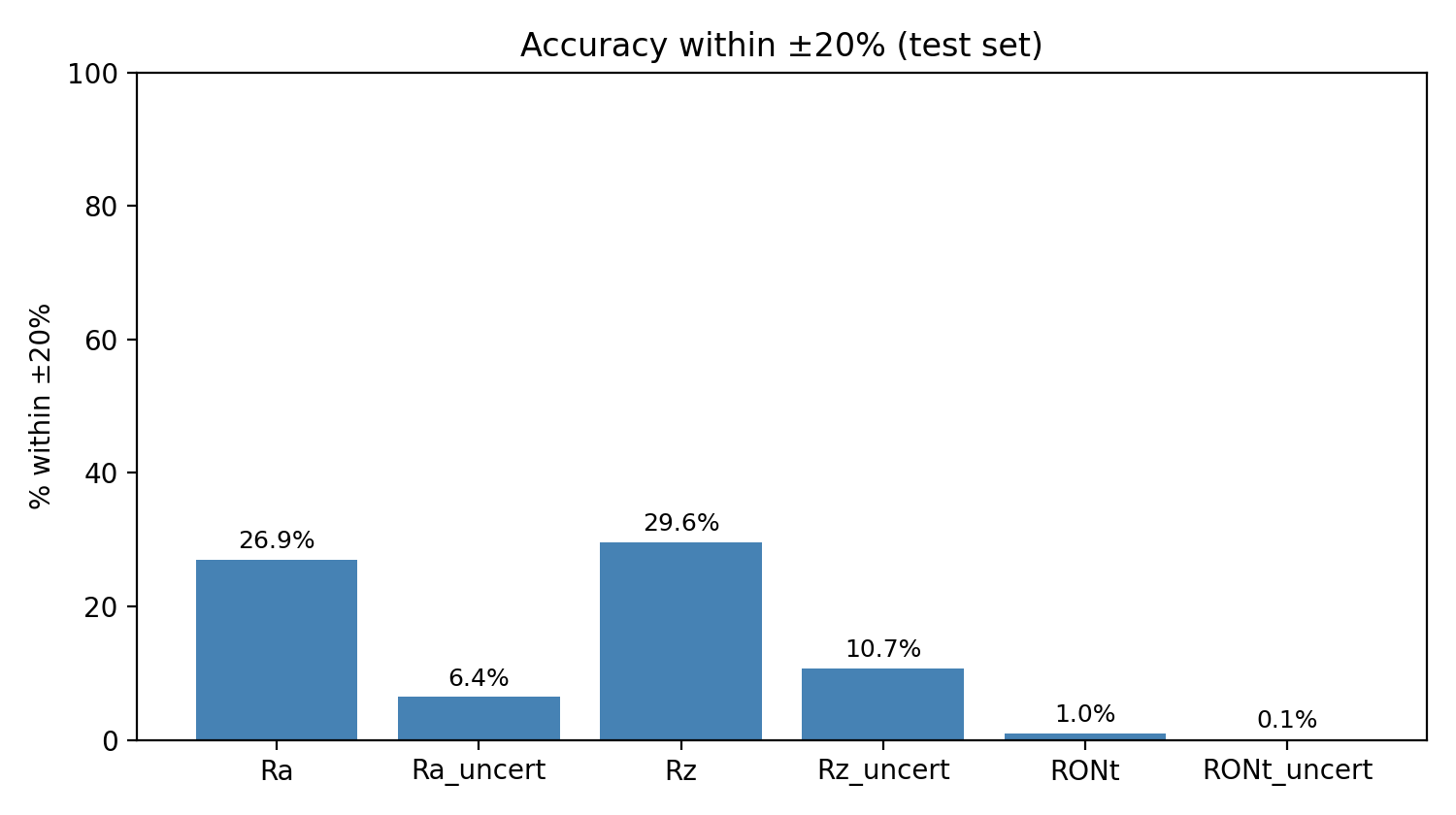}{Figure S197: accuracy within tol 20percent (\label{fig:supp-197})}\hfill
\suppimage{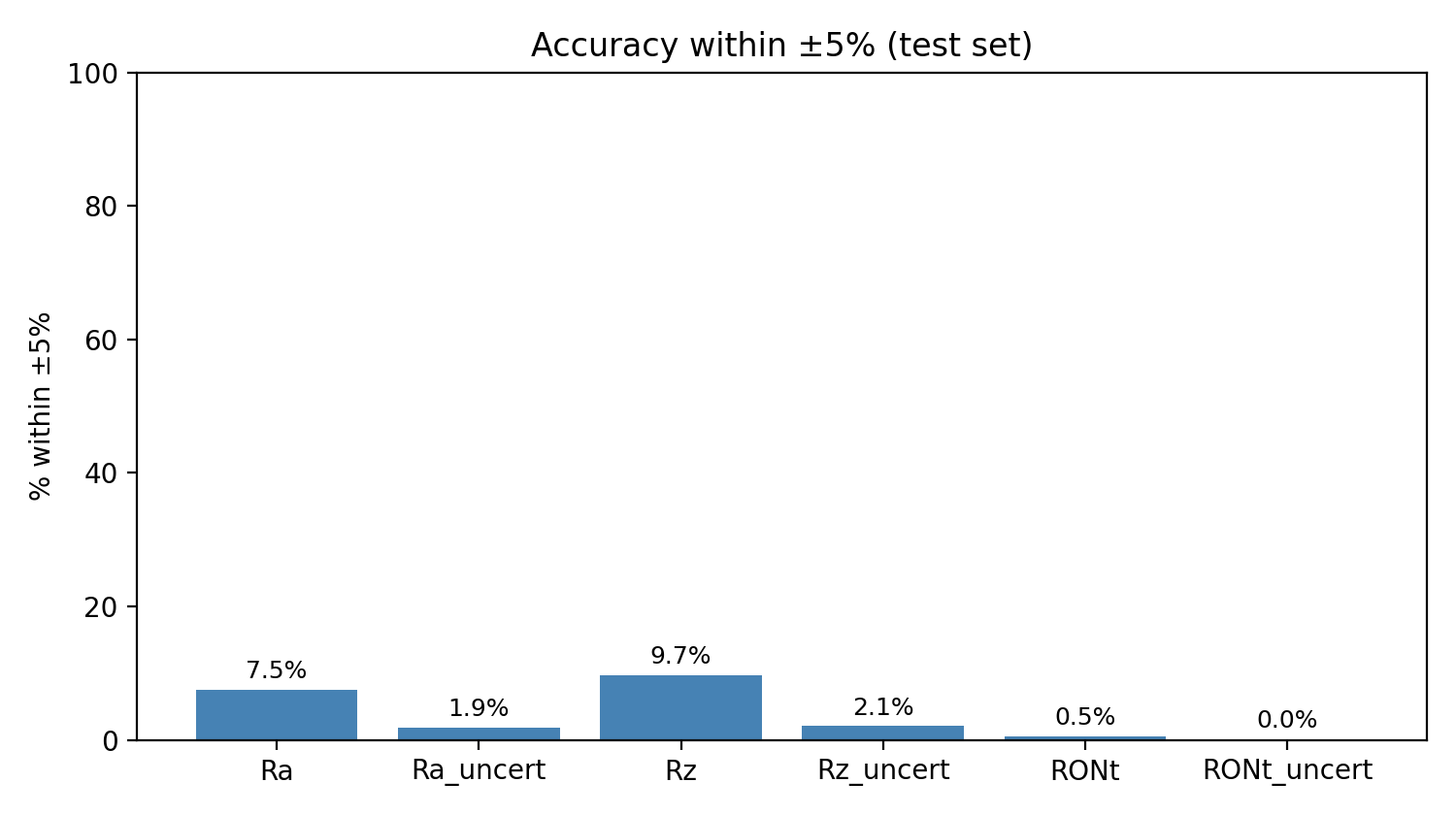}{Figure S198: accuracy within tol 5percent (\label{fig:supp-198})}\
\suppimage{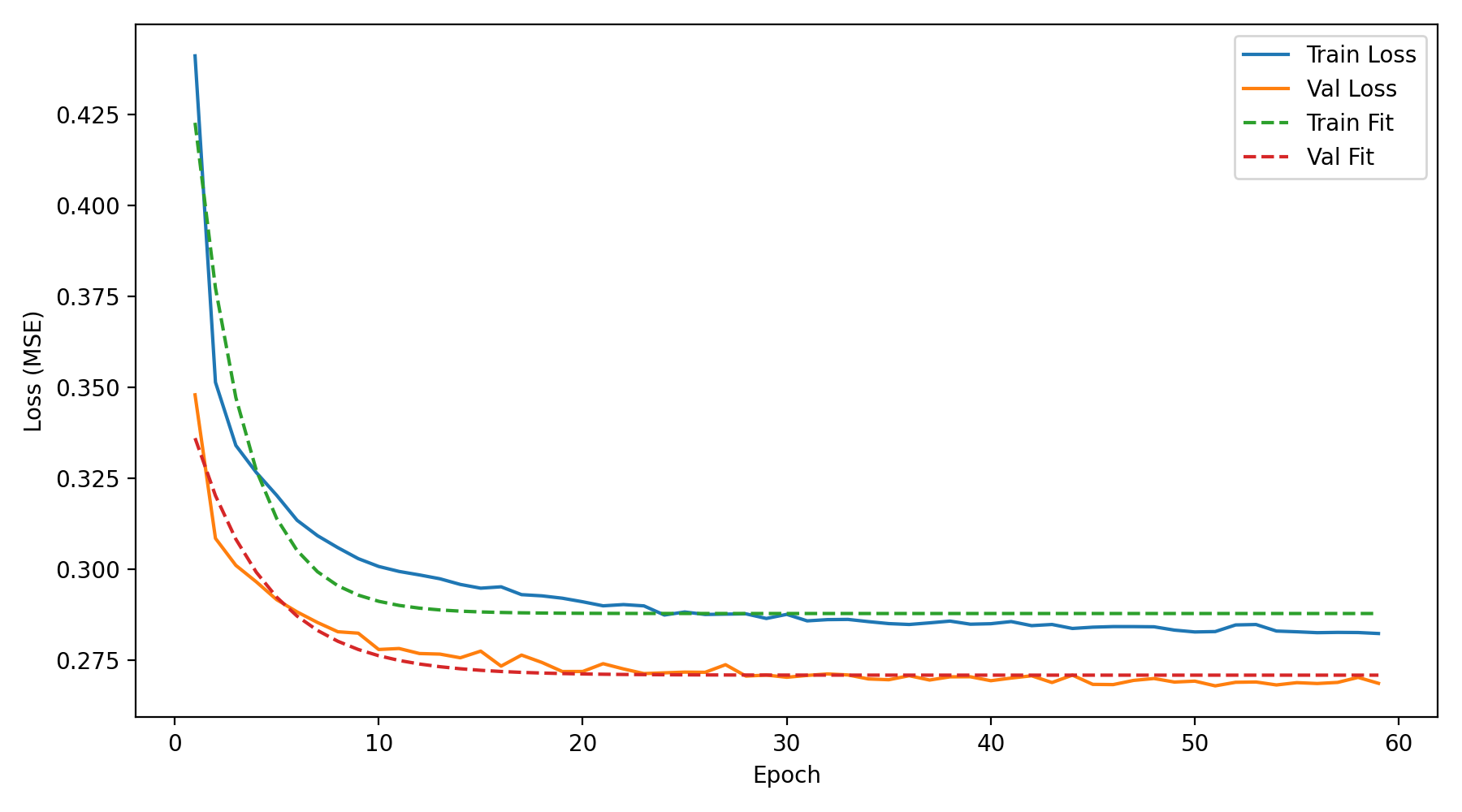}{Figure S199: loss curves (\label{fig:supp-199})}\hfill
\suppimage{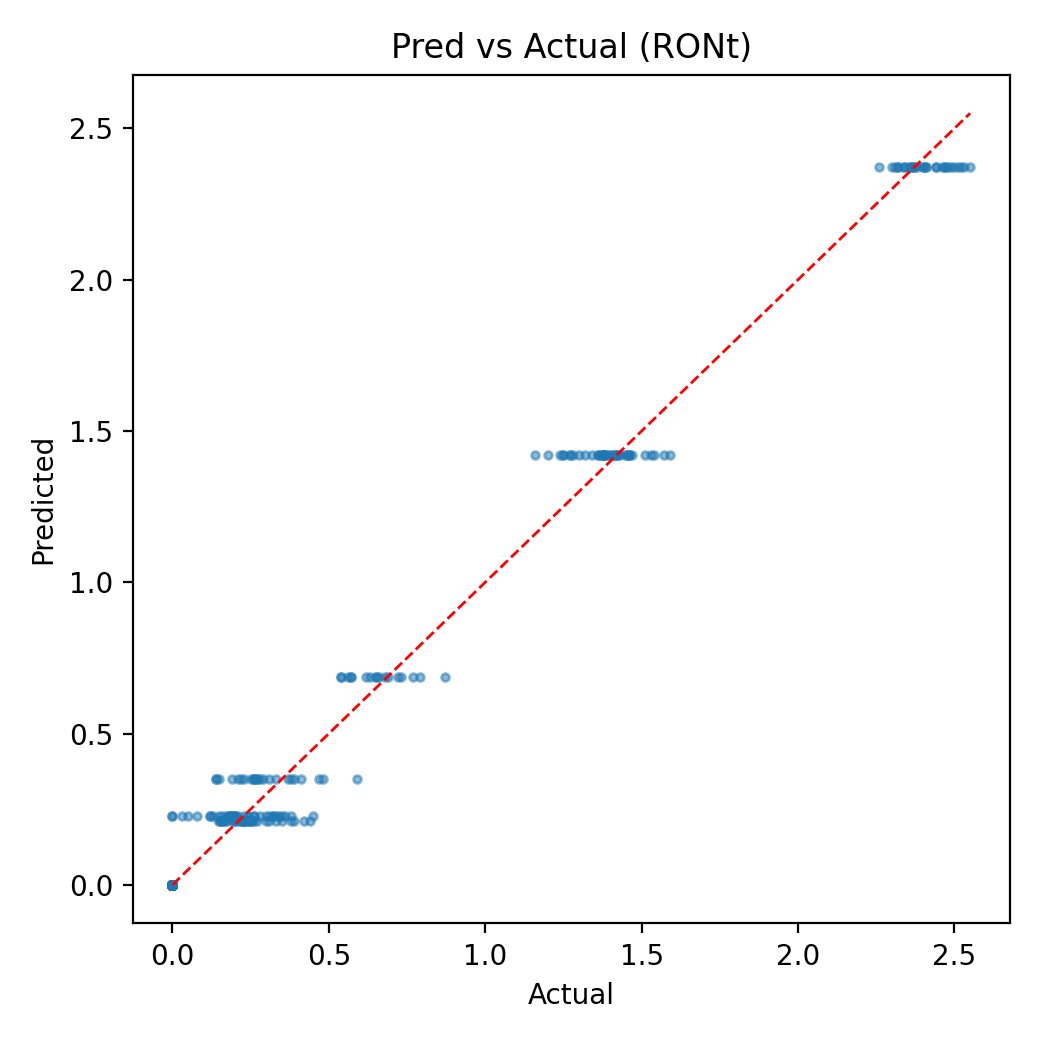}{Figure S200: pred vs actual RONt (\label{fig:supp-200})}\
\suppimage{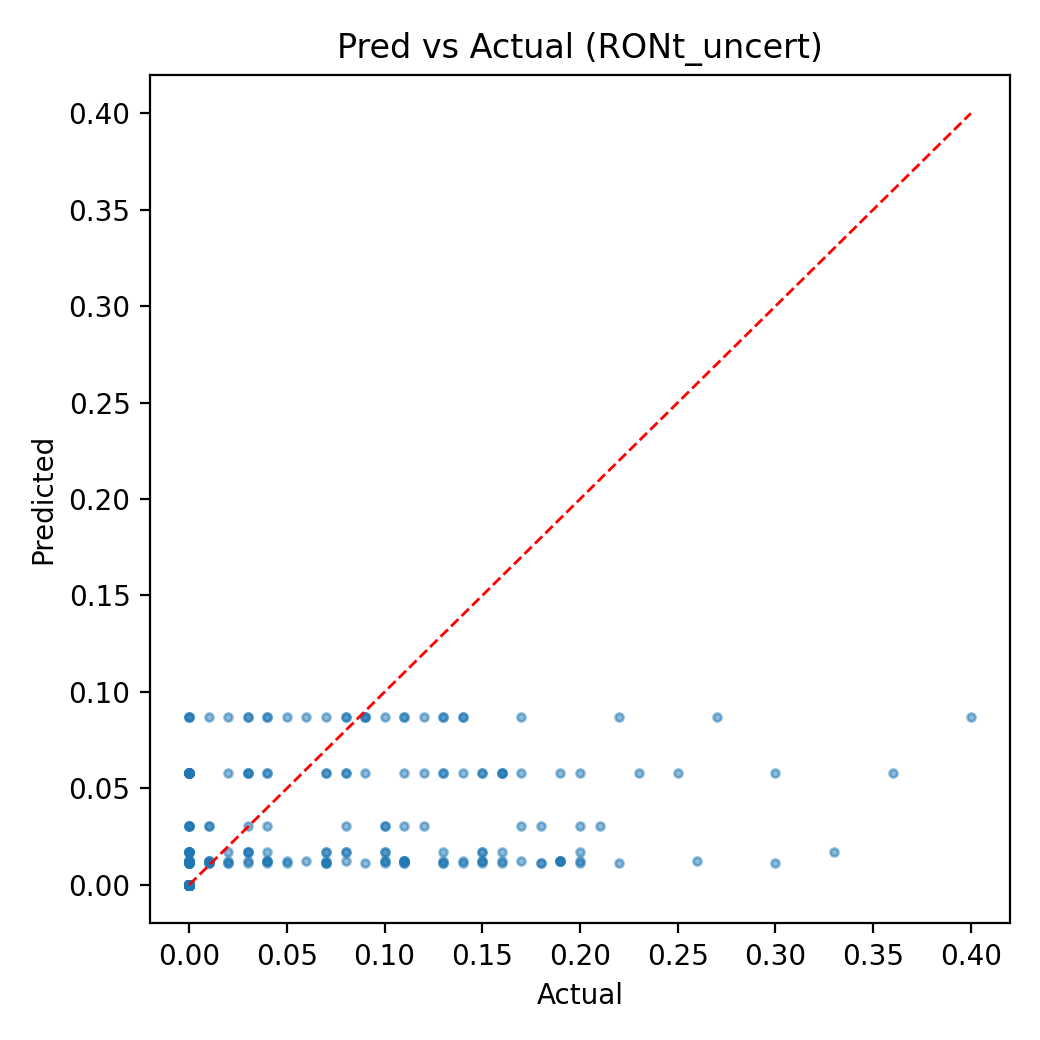}{Figure S201: pred vs actual RONt uncert (\label{fig:supp-201})}\hfill
\suppimage{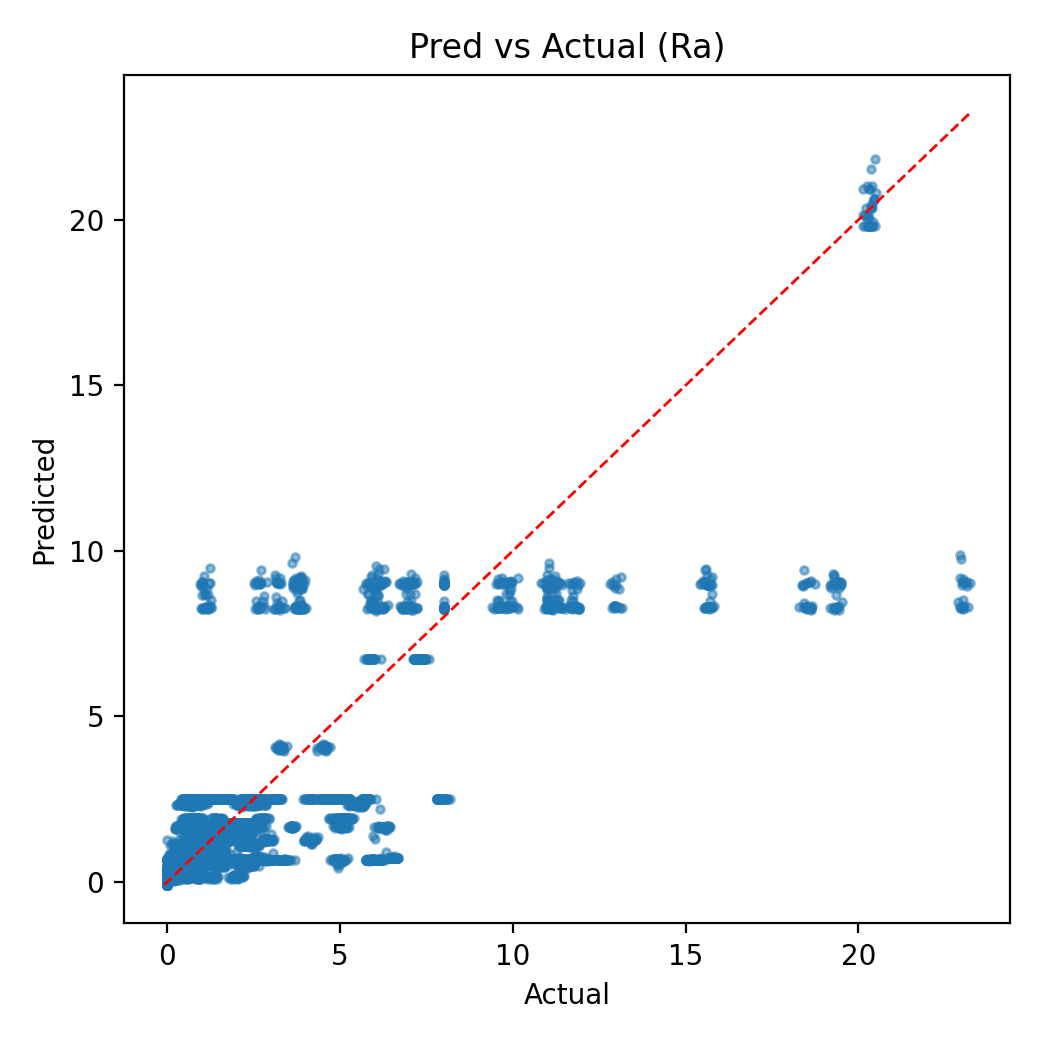}{Figure S202: pred vs actual Ra (\label{fig:supp-202})}\
\suppimage{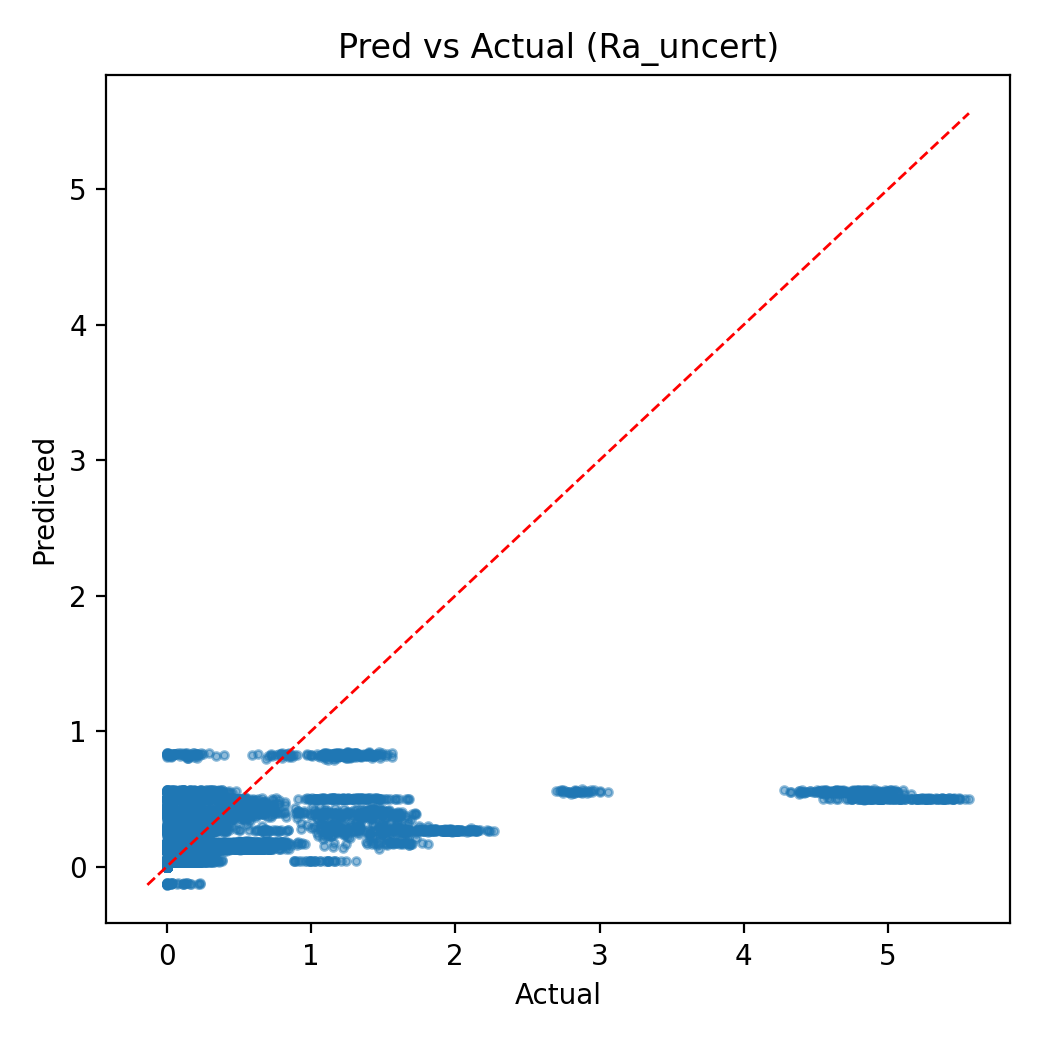}{Figure S203: pred vs actual Ra uncert (\label{fig:supp-203})}\hfill
\suppimage{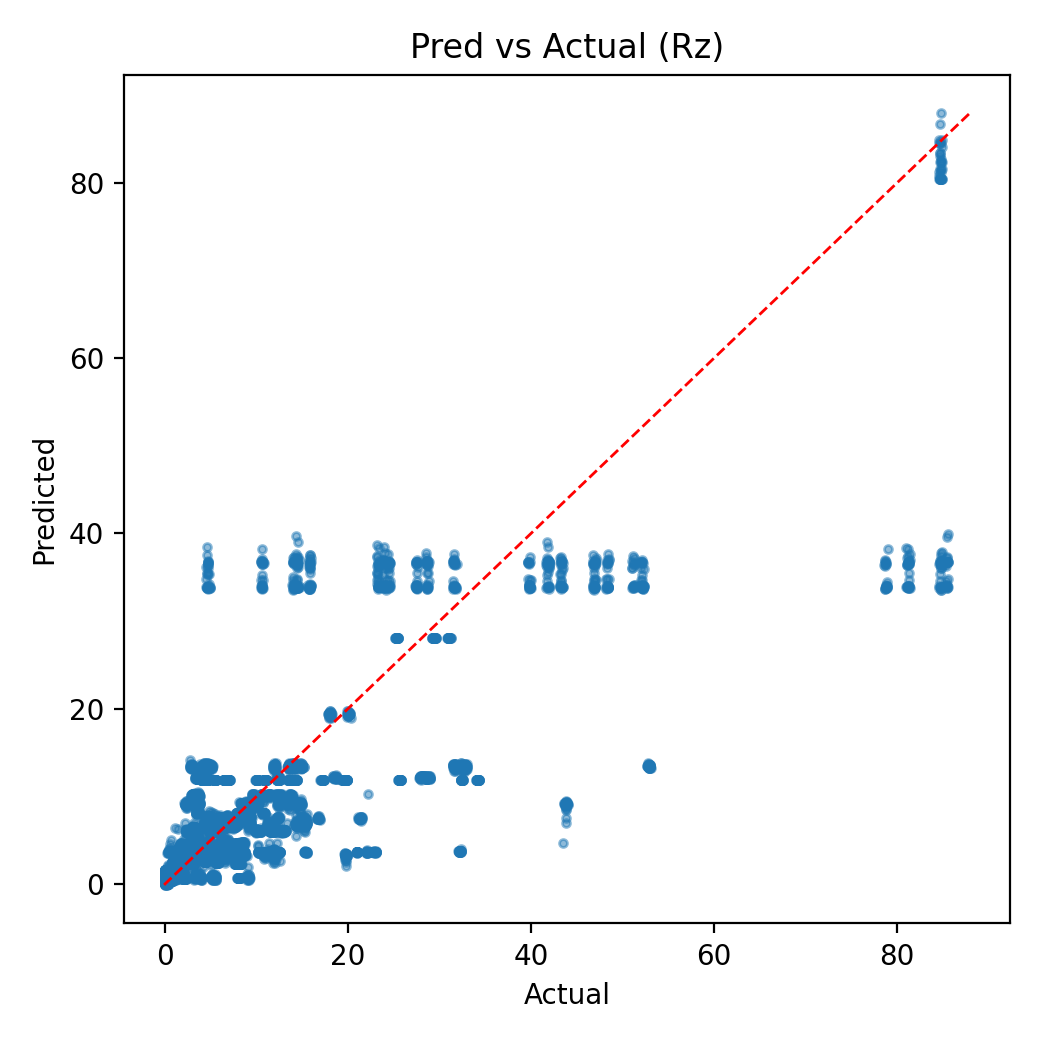}{Figure S204: pred vs actual Rz (\label{fig:supp-204})}\
\suppimage{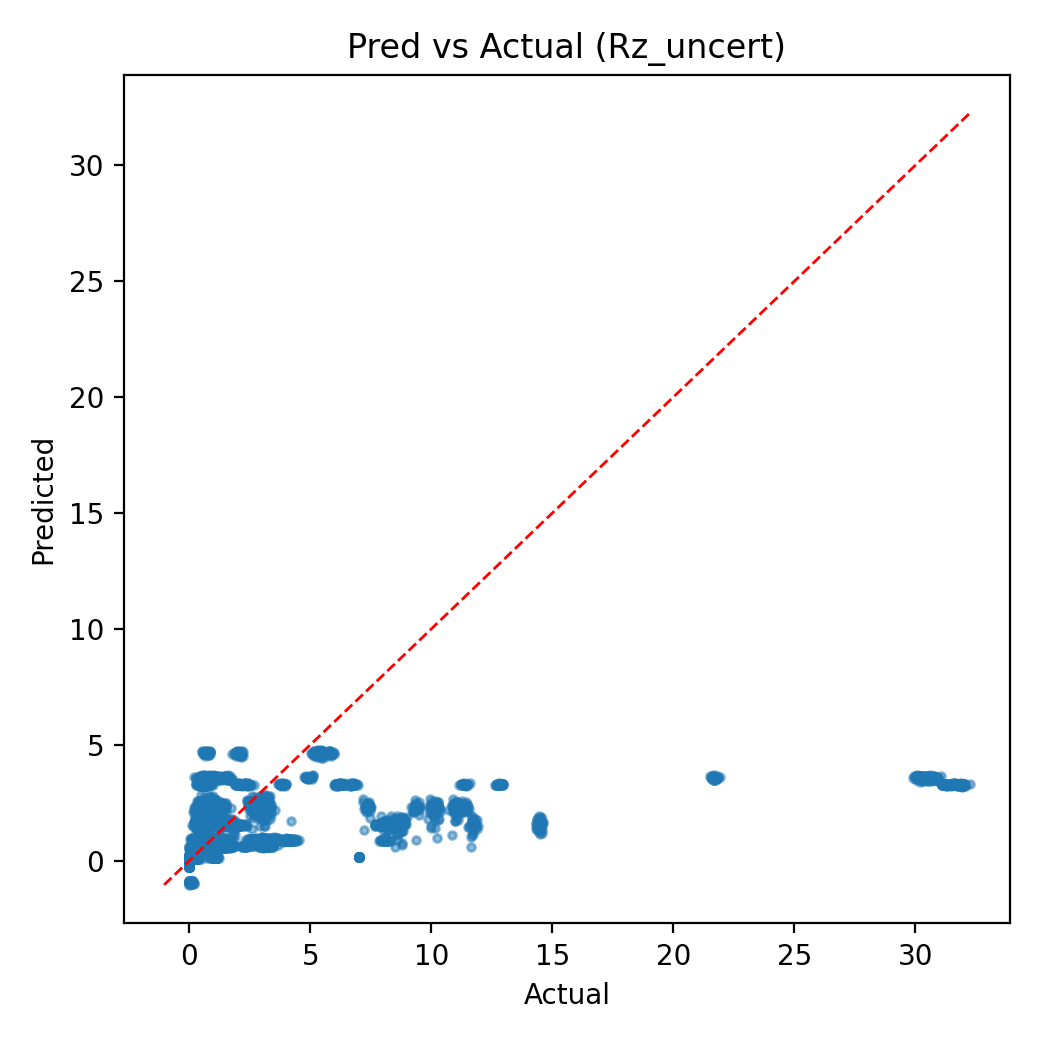}{Figure S205: pred vs actual Rz uncert (\label{fig:supp-205})}\hfill
\suppimage{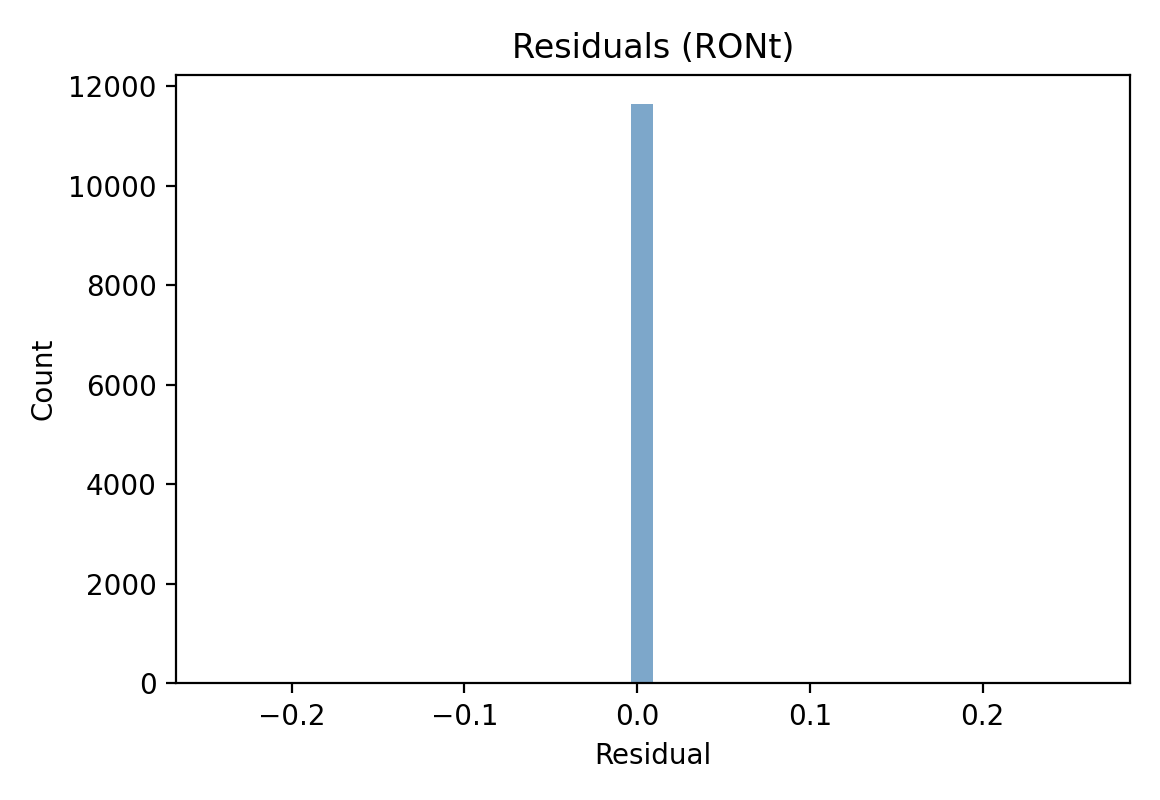}{Figure S206: residuals hist RONt (\label{fig:supp-206})}\
\suppimage{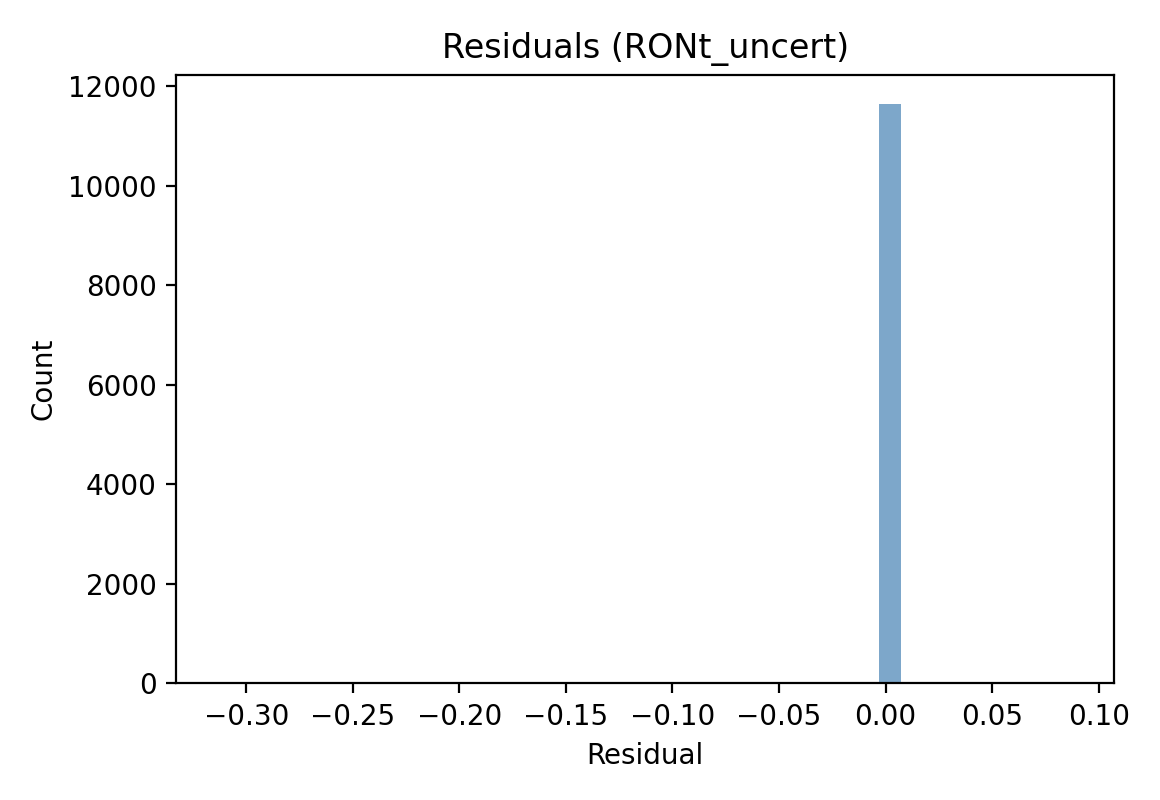}{Figure S207: residuals hist RONt uncert (\label{fig:supp-207})}\hfill
\suppimage{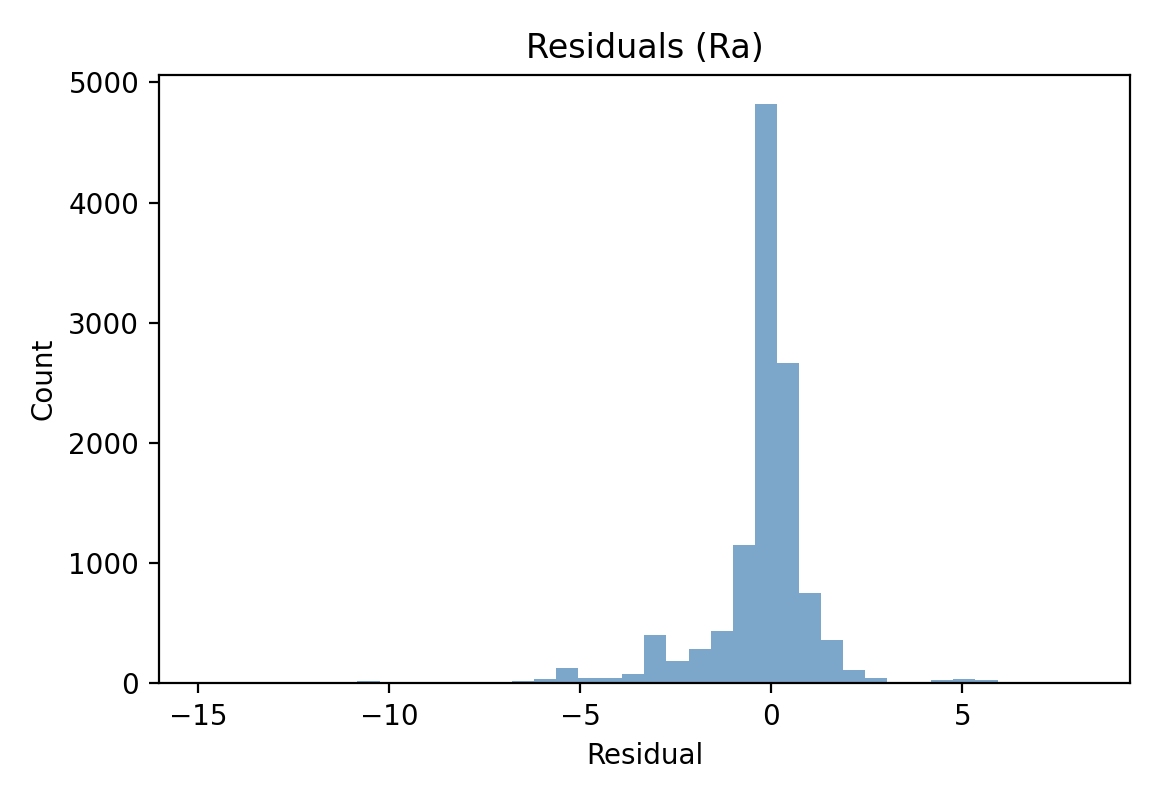}{Figure S208: residuals hist Ra (\label{fig:supp-208})}\
\suppimage{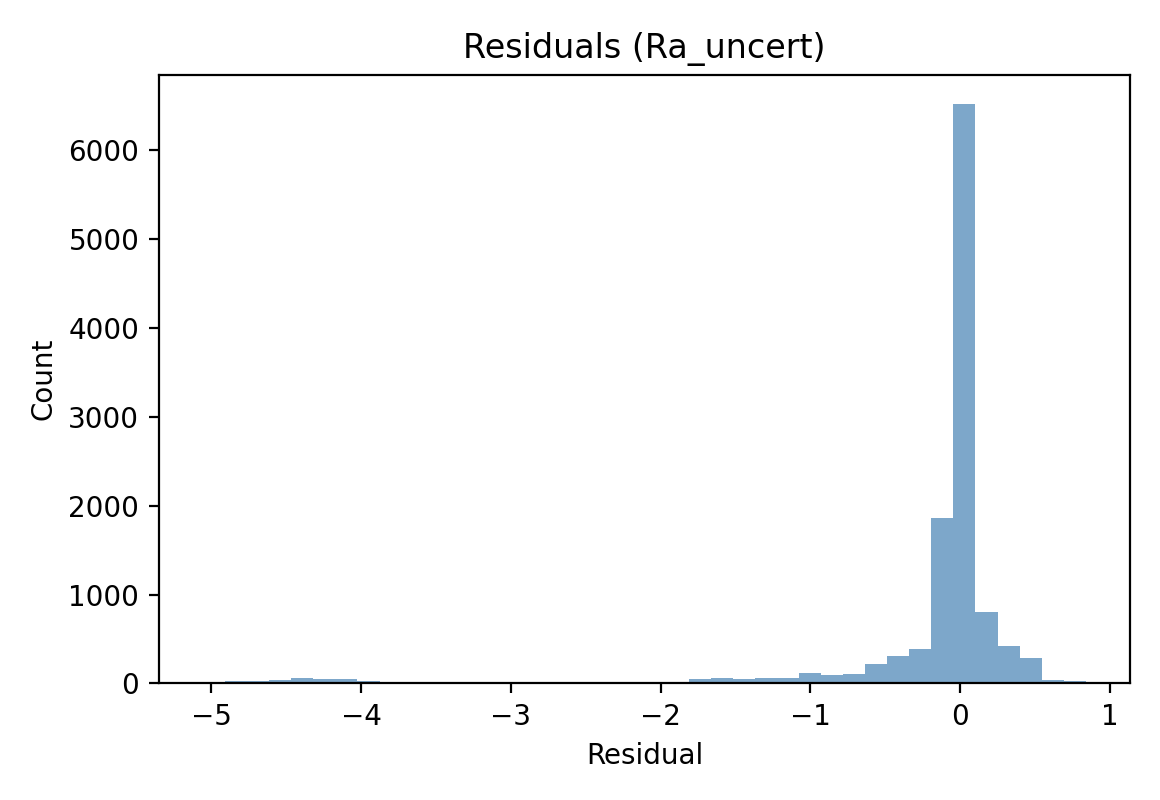}{Figure S209: residuals hist Ra uncert (\label{fig:supp-209})}\hfill
\suppimage{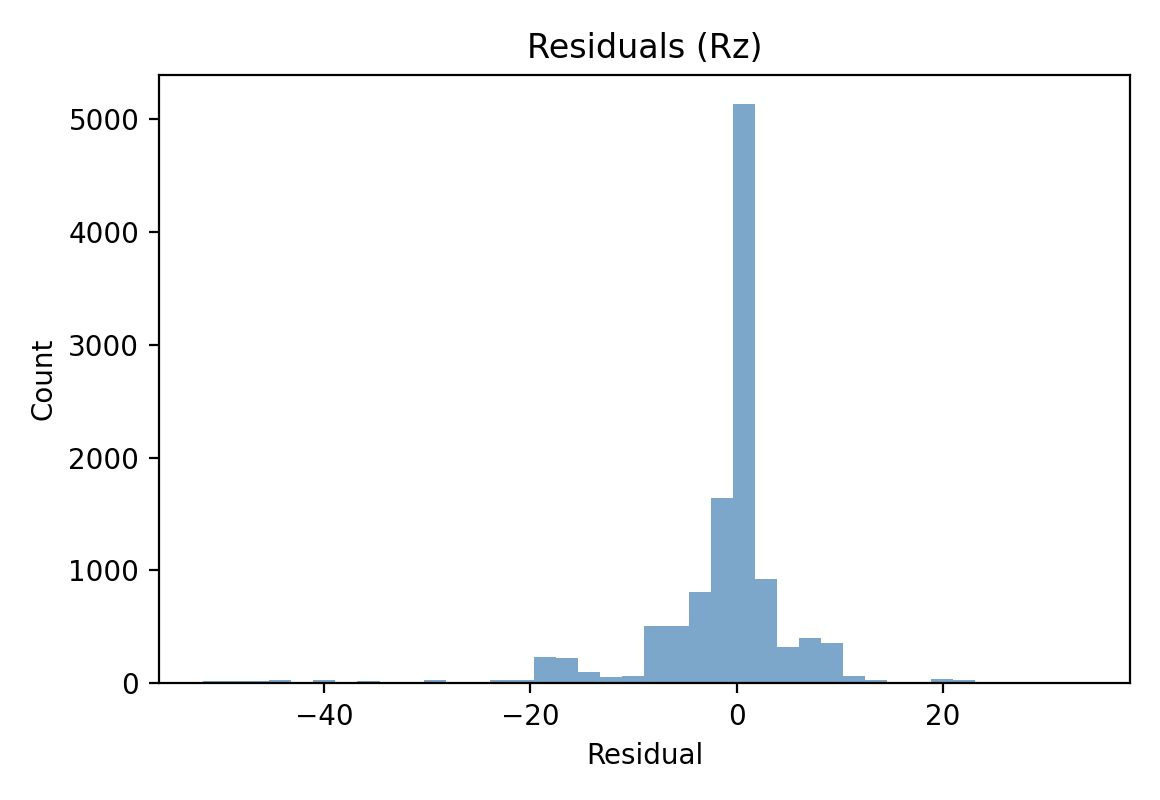}{Figure S210: residuals hist Rz (\label{fig:supp-210})}\
\suppimage{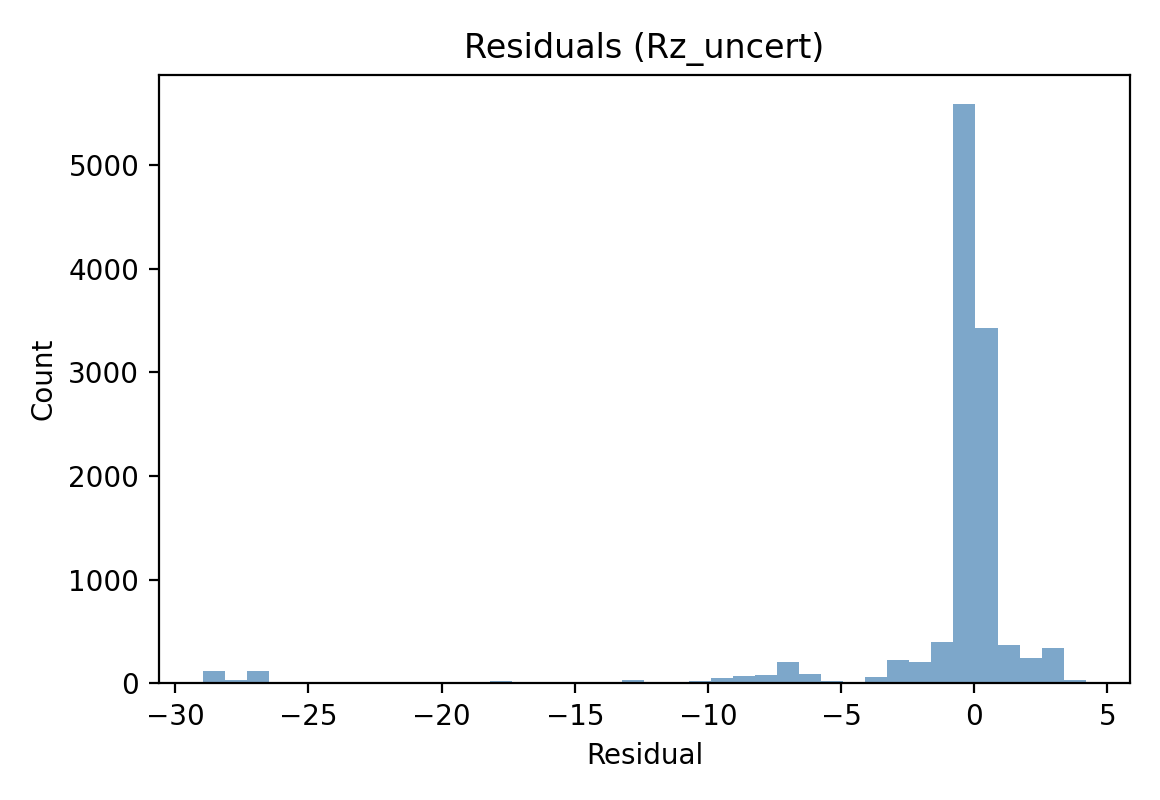}{Figure S211: residuals hist Rz uncert (\label{fig:supp-211})}\hfill
\suppimage{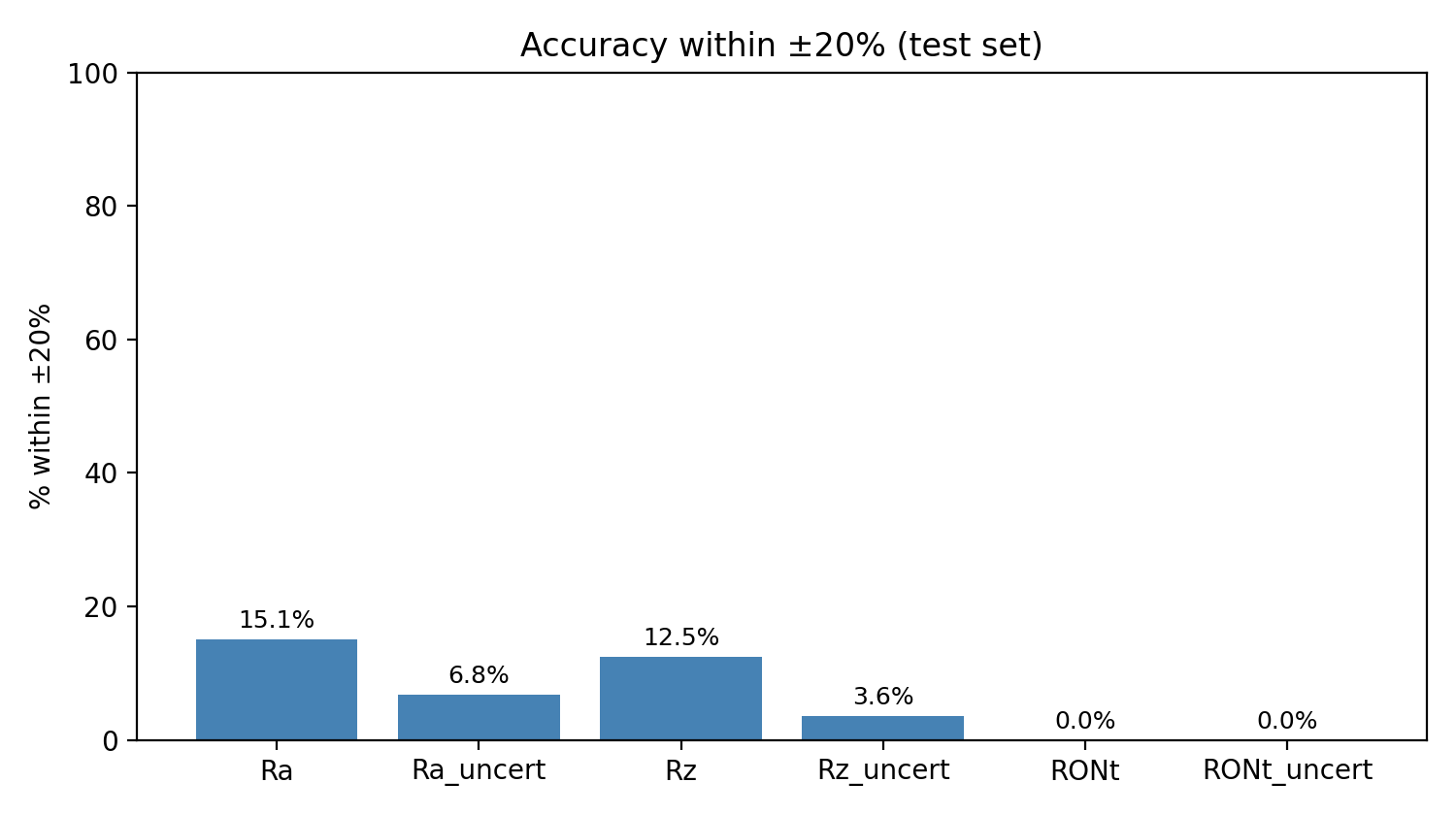}{Figure S212: accuracy within tol 20percent (\label{fig:supp-212})}\
\suppimage{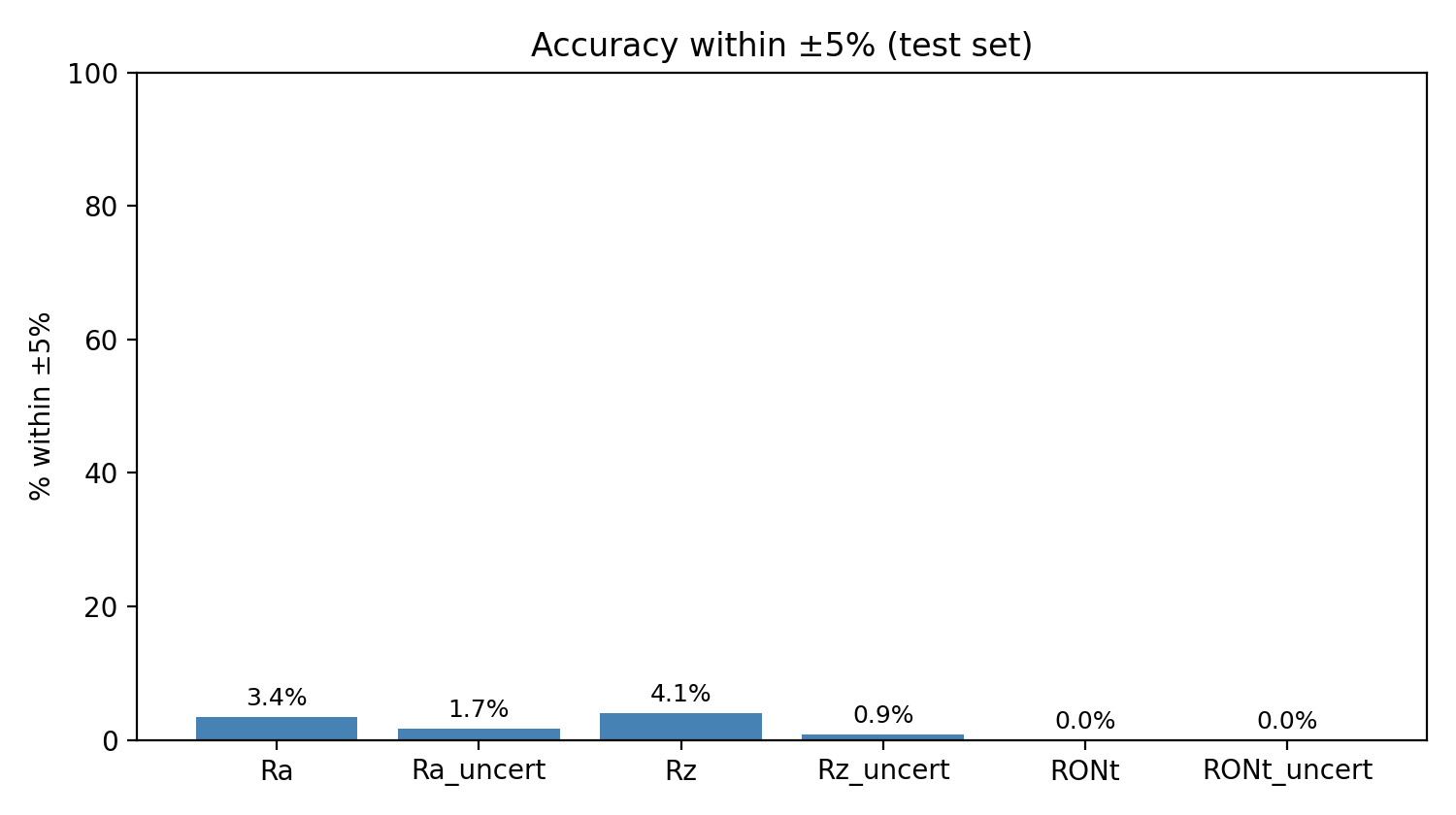}{Figure S213: accuracy within tol 5percent (\label{fig:supp-213})}\hfill
\suppimage{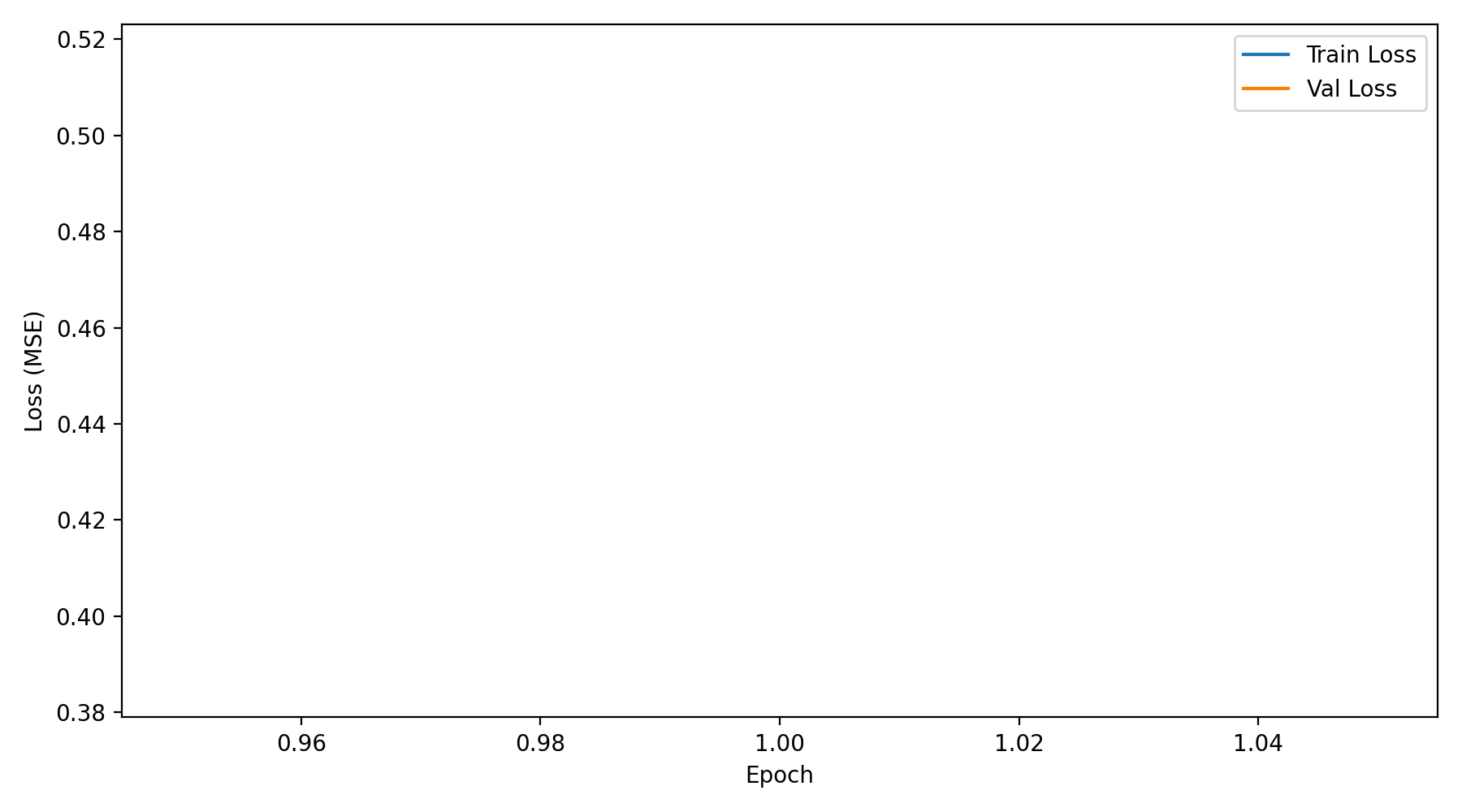}{Figure S214: loss curves (\label{fig:supp-214})}\
\suppimage{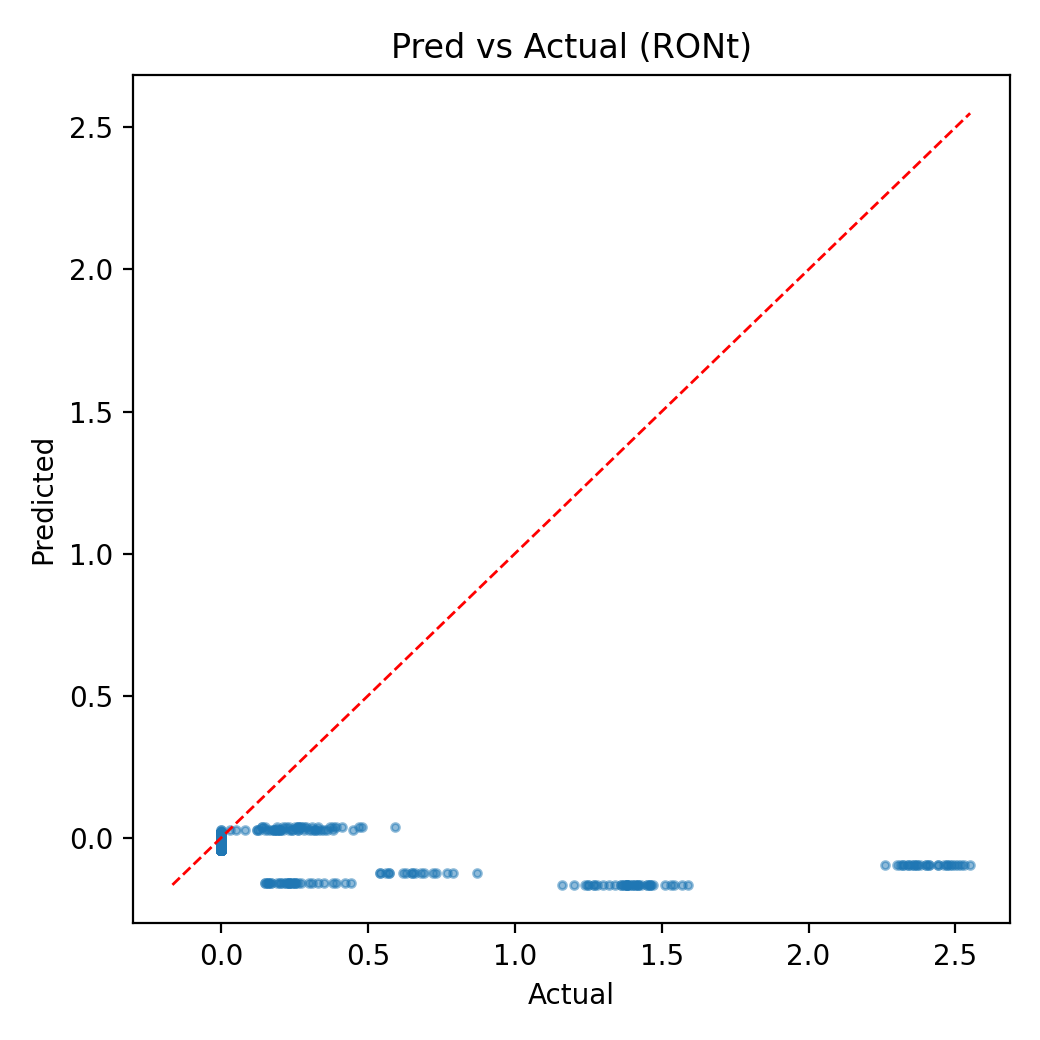}{Figure S215: pred vs actual RONt (\label{fig:supp-215})}\hfill
\suppimage{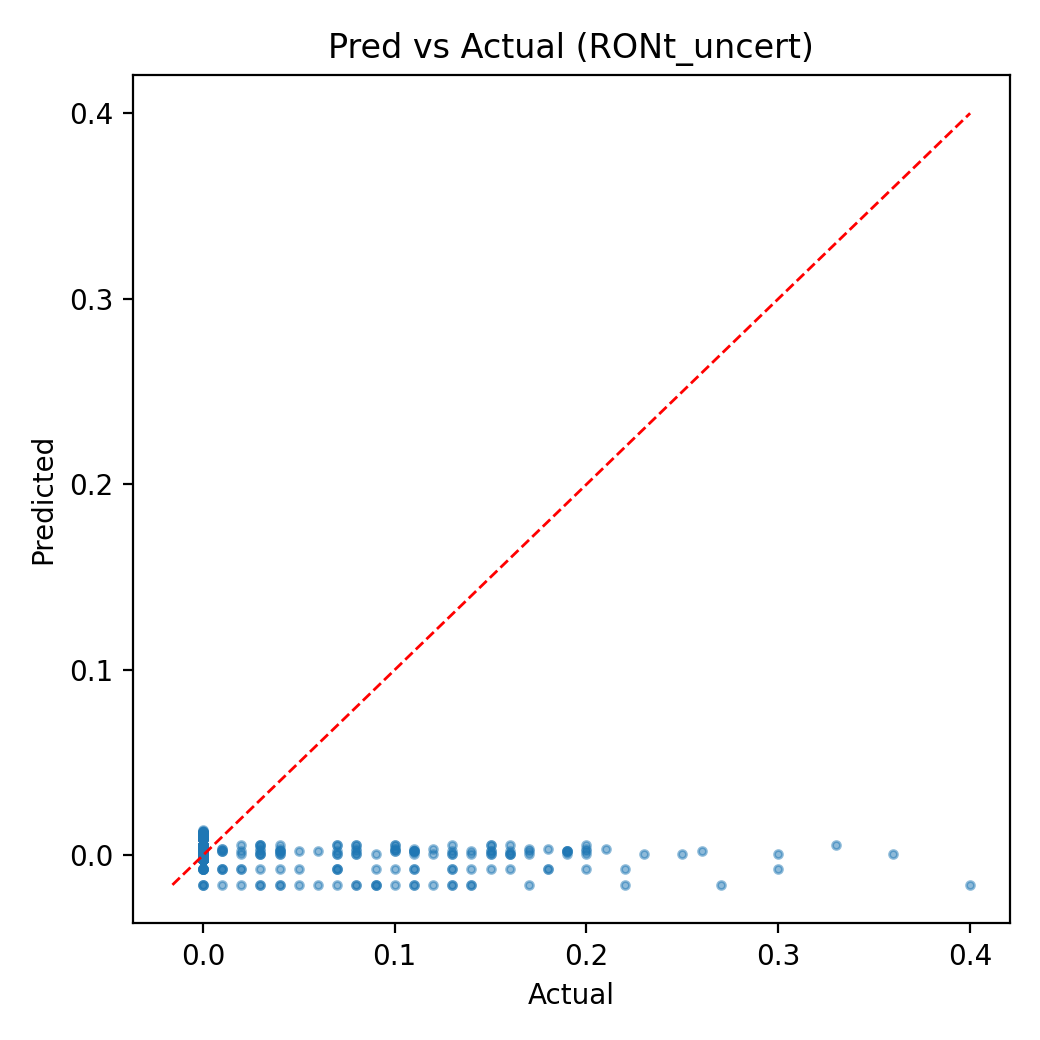}{Figure S216: pred vs actual RONt uncert (\label{fig:supp-216})}\
\suppimage{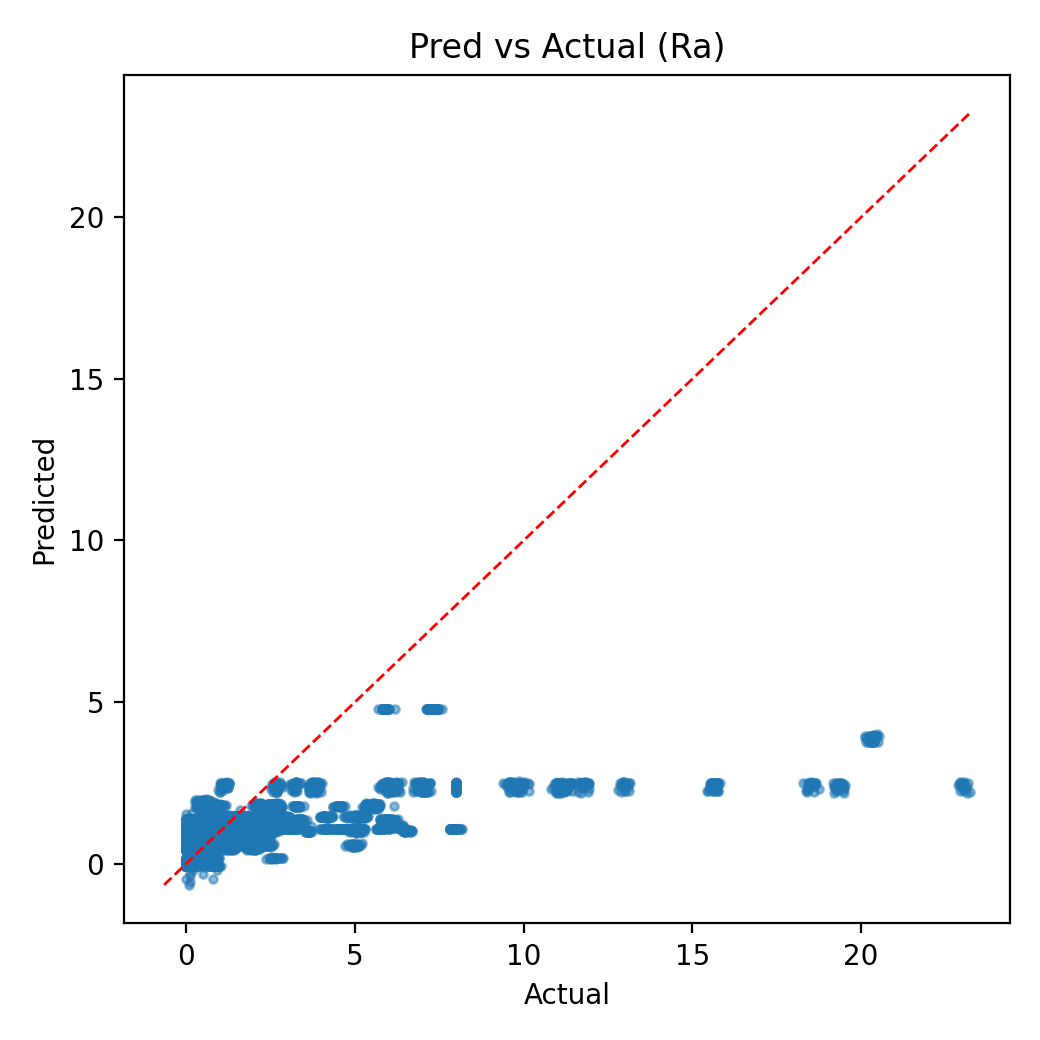}{Figure S217: pred vs actual Ra (\label{fig:supp-217})}\hfill
\suppimage{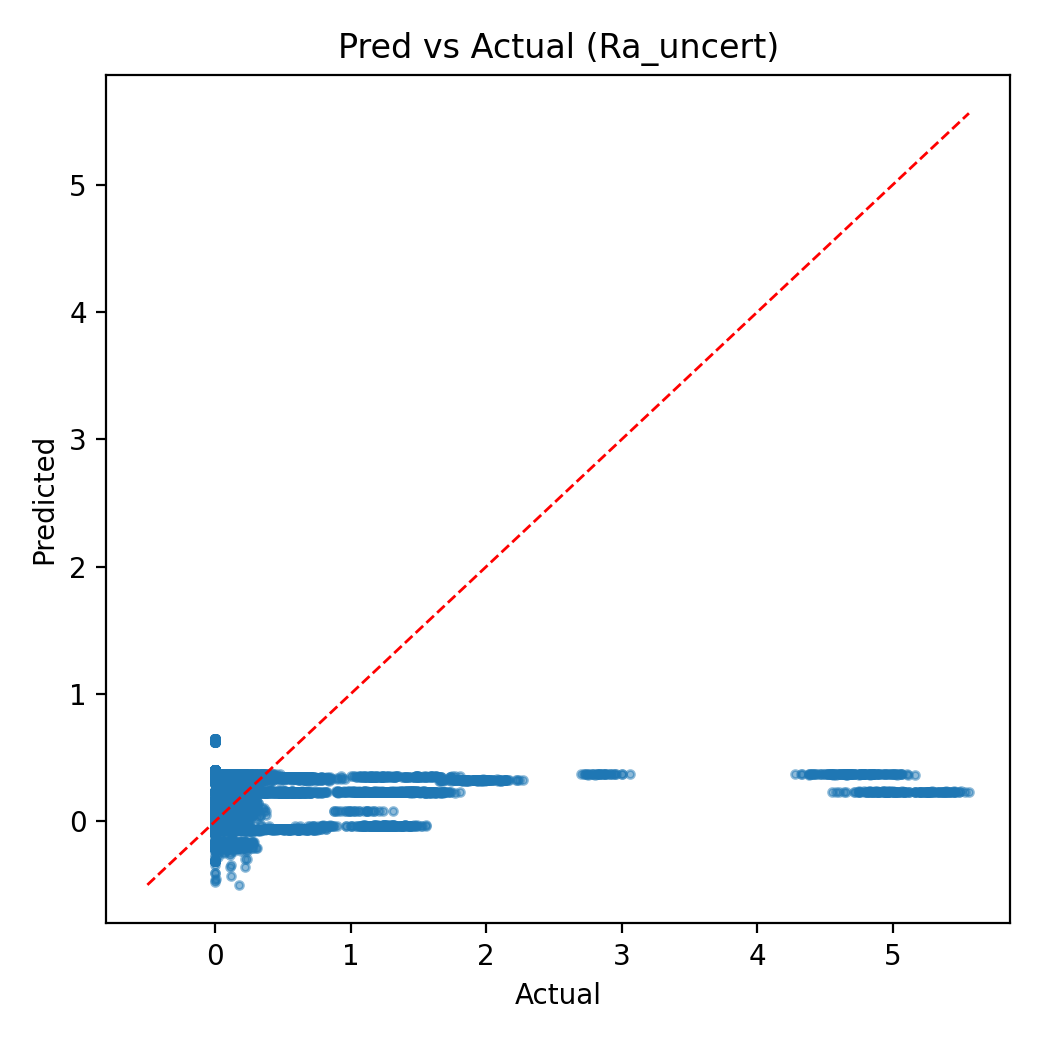}{Figure S218: pred vs actual Ra uncert (\label{fig:supp-218})}\
\suppimage{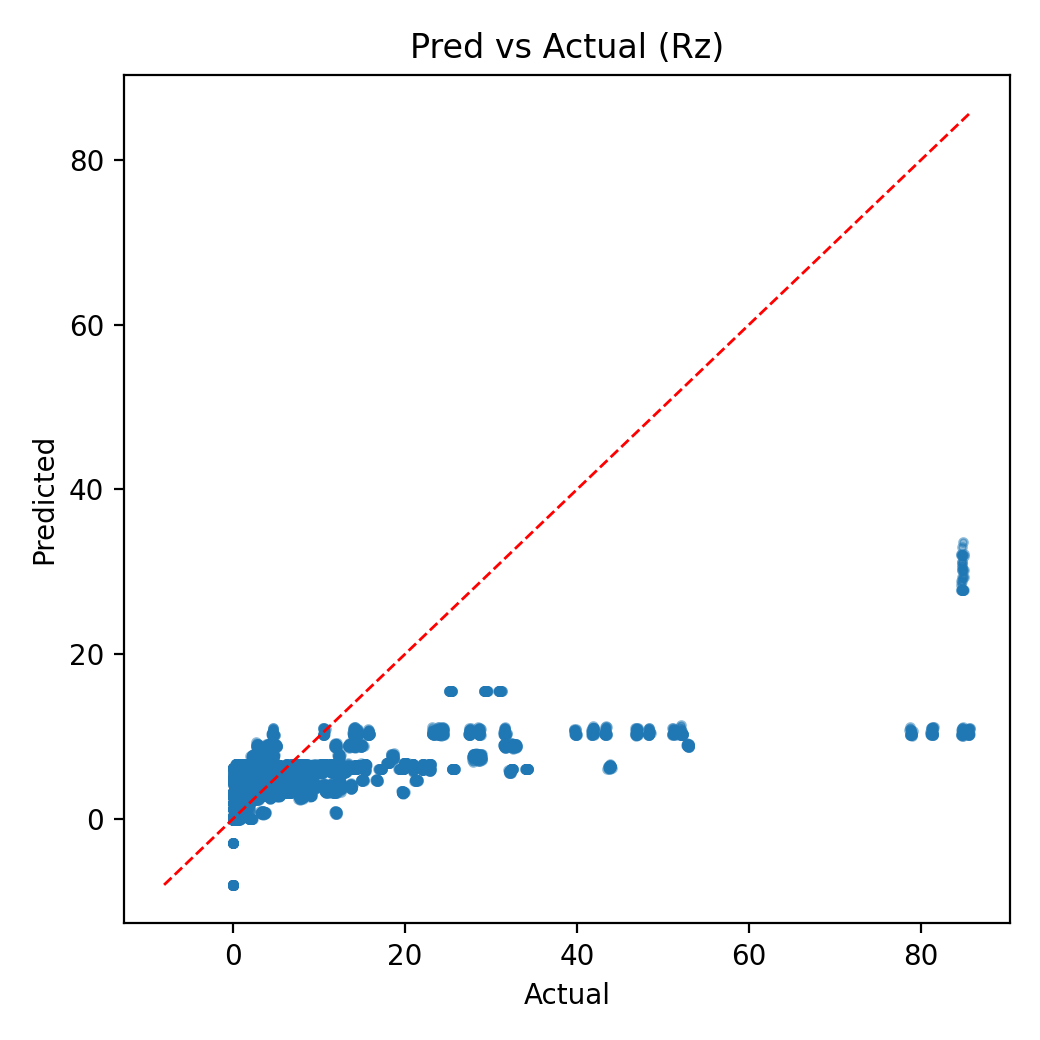}{Figure S219: pred vs actual Rz (\label{fig:supp-219})}\hfill
\suppimage{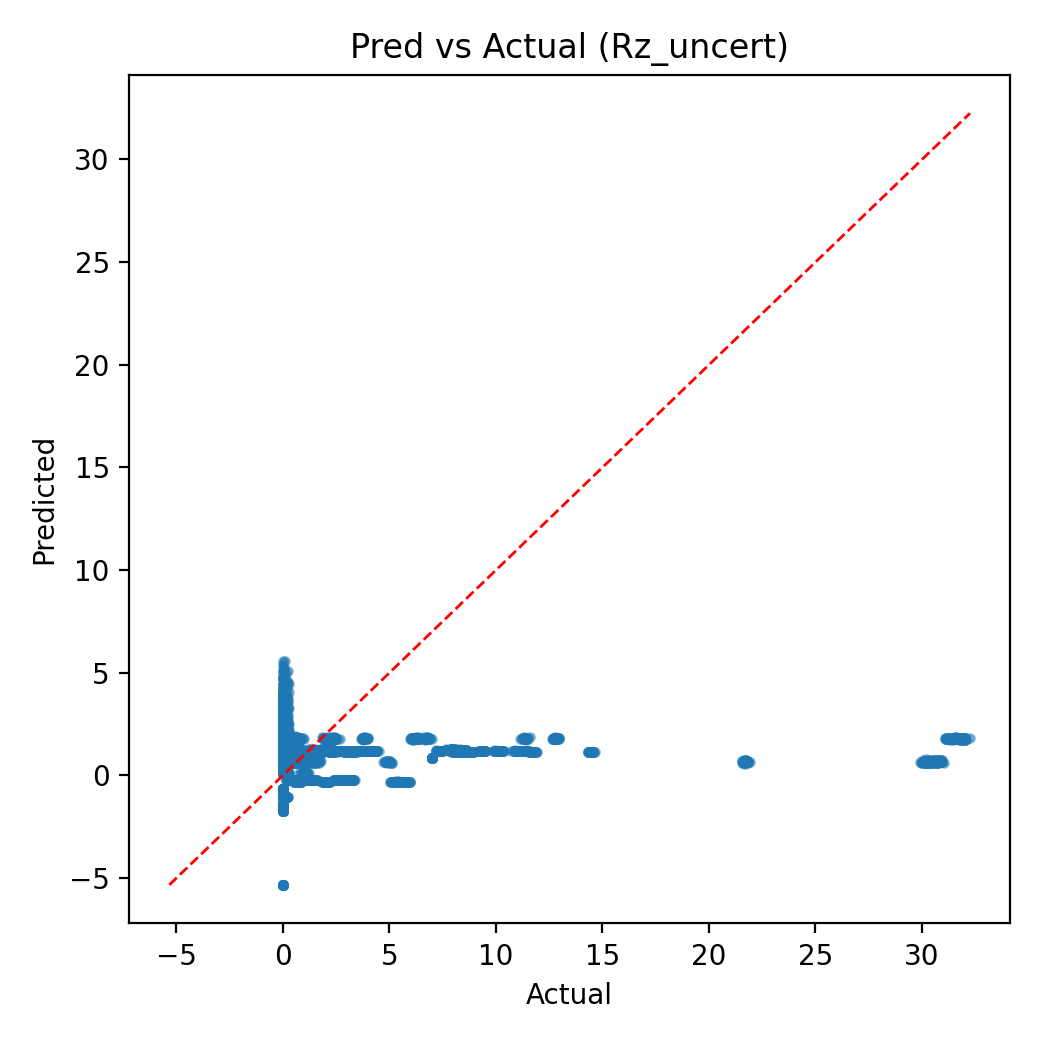}{Figure S220: pred vs actual Rz uncert (\label{fig:supp-220})}\
\suppimage{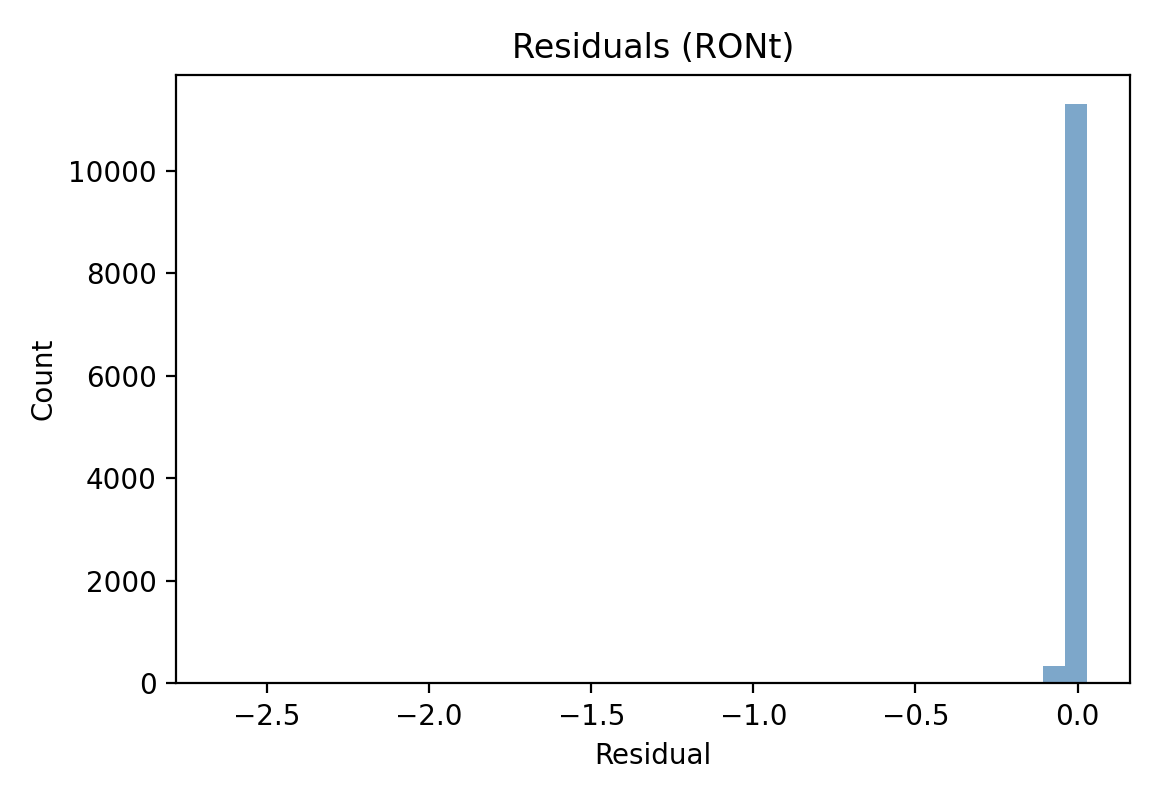}{Figure S221: residuals hist RONt (\label{fig:supp-221})}\hfill
\suppimage{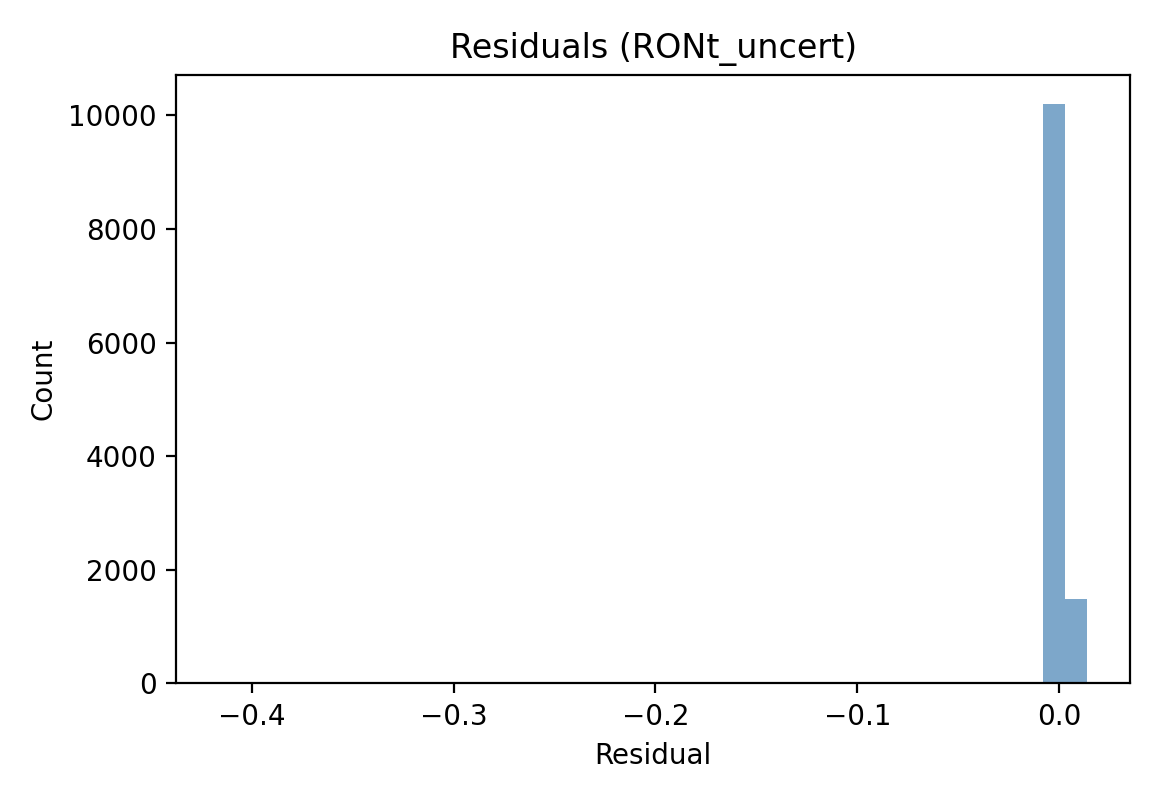}{Figure S222: residuals hist RONt uncert (\label{fig:supp-222})}\
\suppimage{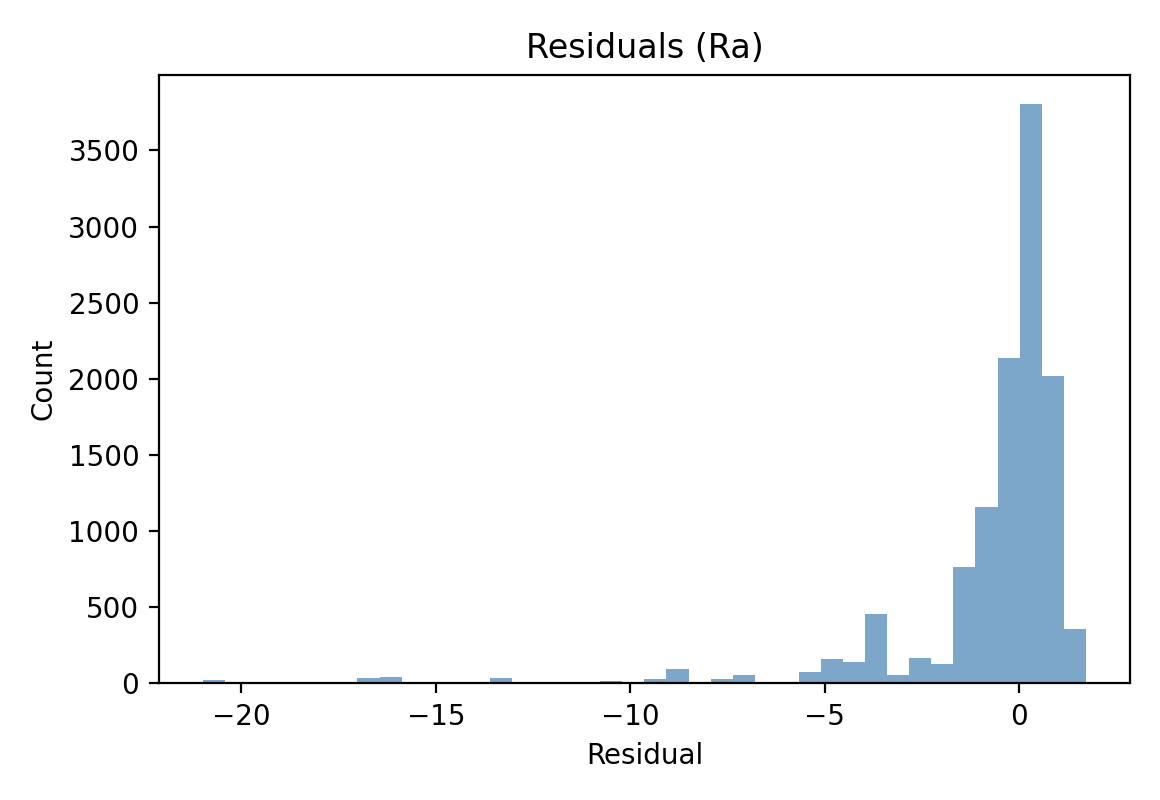}{Figure S223: residuals hist Ra (\label{fig:supp-223})}\hfill
\suppimage{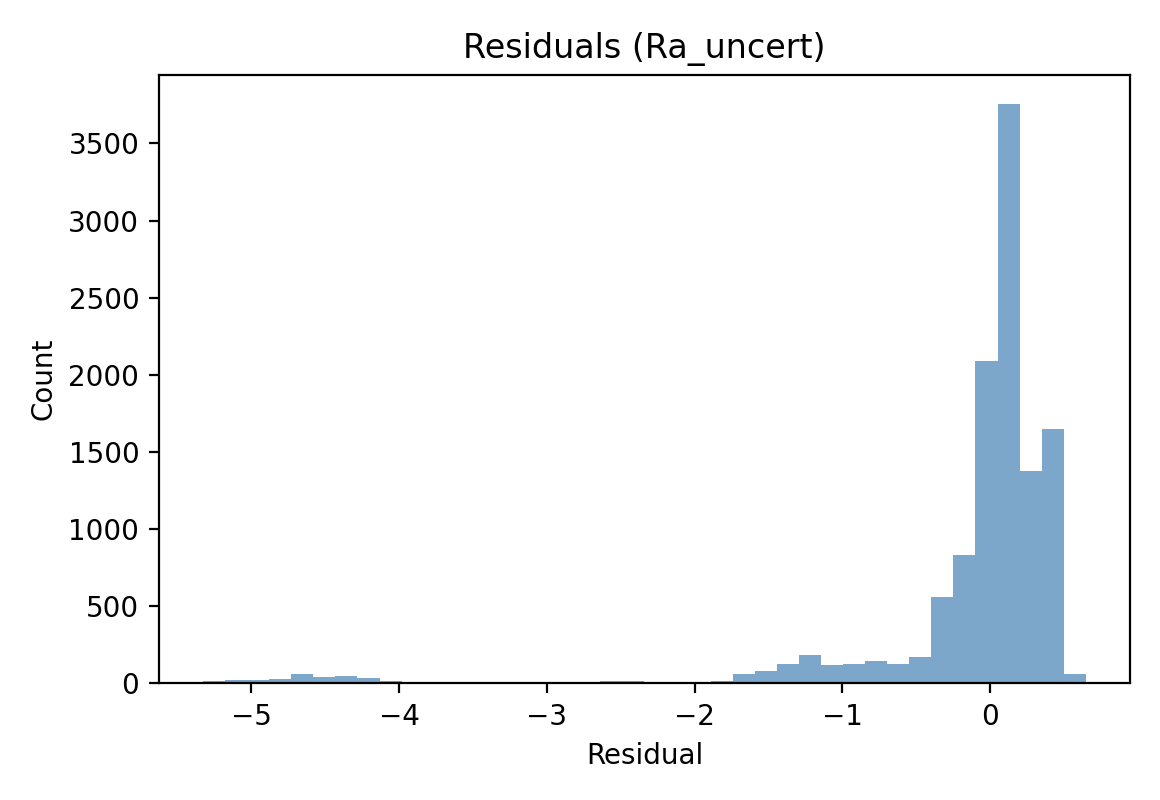}{Figure S224: residuals hist Ra uncert (\label{fig:supp-224})}\
\suppimage{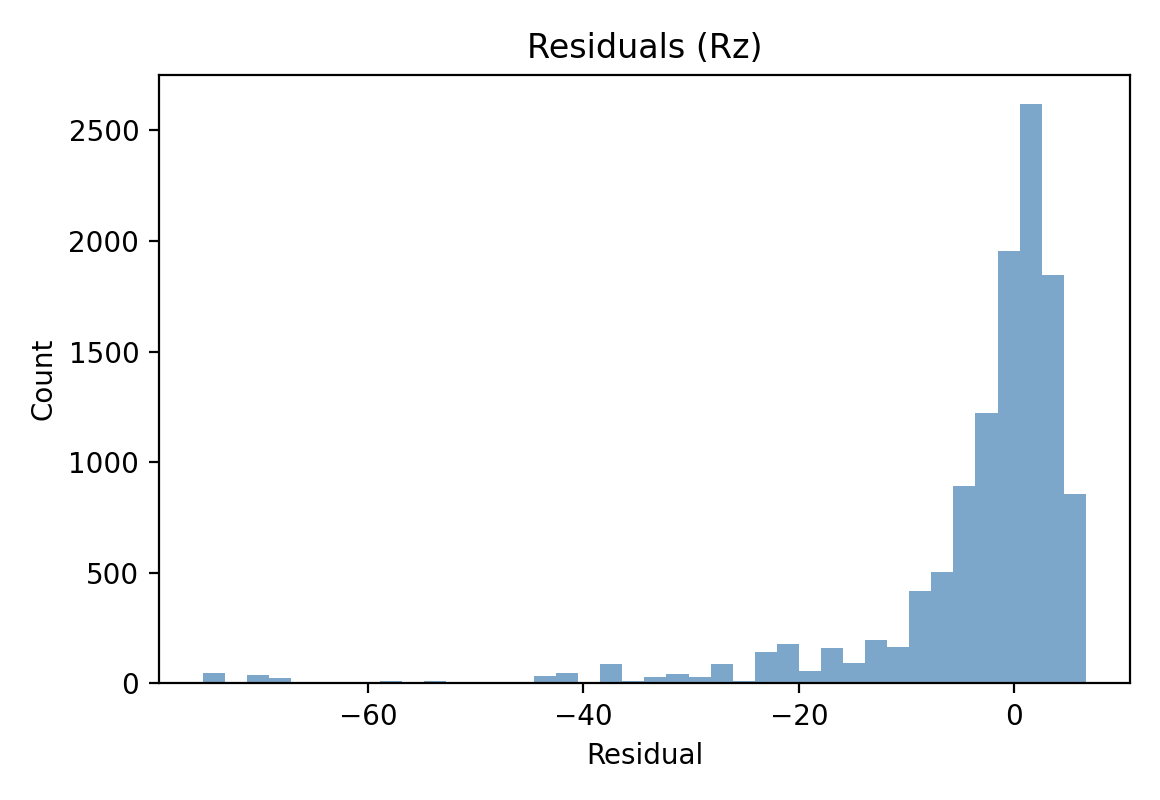}{Figure S225: residuals hist Rz (\label{fig:supp-225})}\hfill
\suppimage{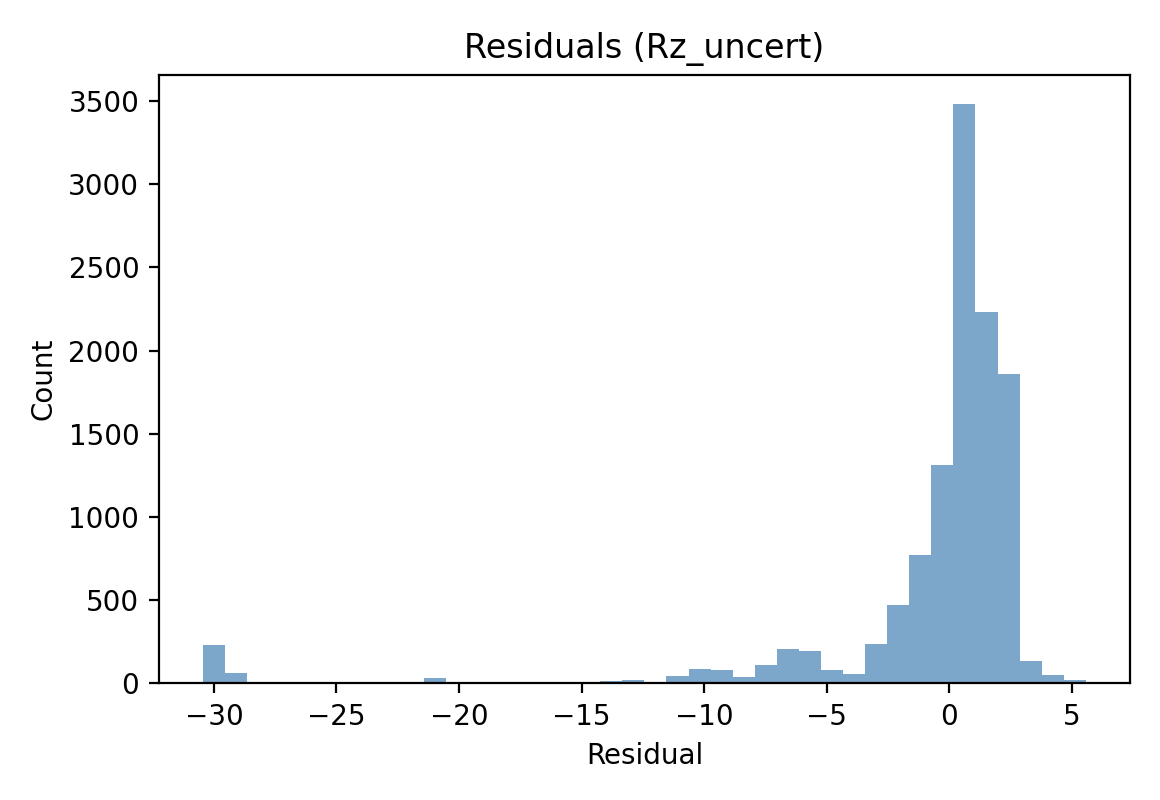}{Figure S226: residuals hist Rz uncert (\label{fig:supp-226})}\
\suppimage{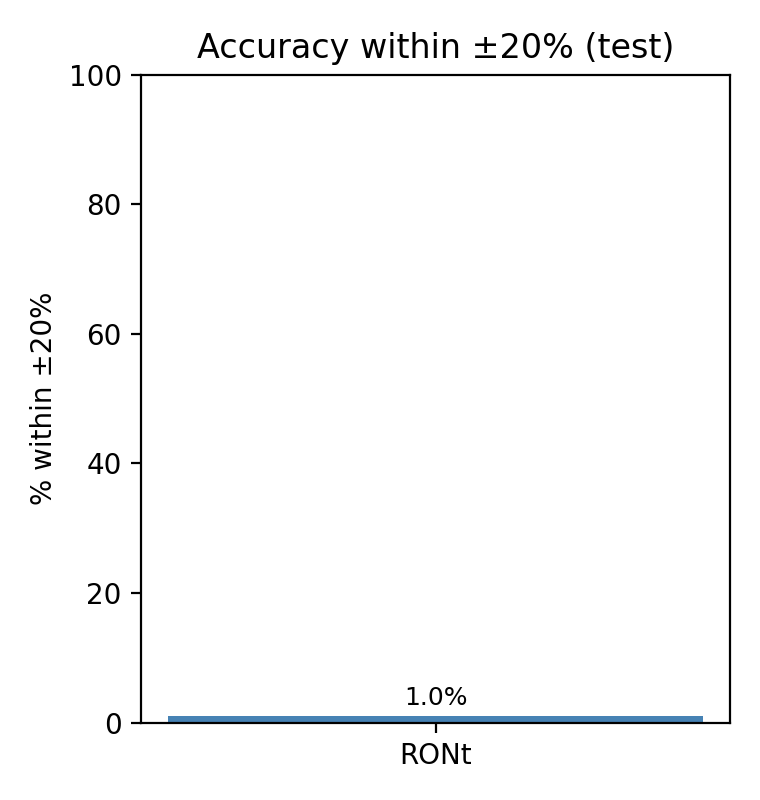}{Figure S227: accuracy within tol 20percent (\label{fig:supp-227})}\hfill
\suppimage{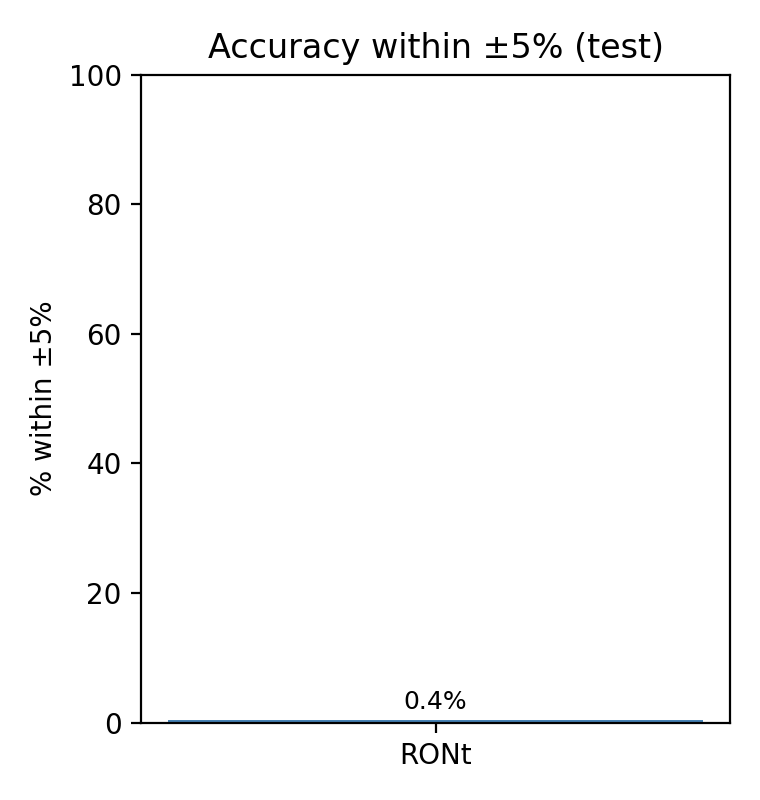}{Figure S228: accuracy within tol 5percent (\label{fig:supp-228})}\
\suppimage{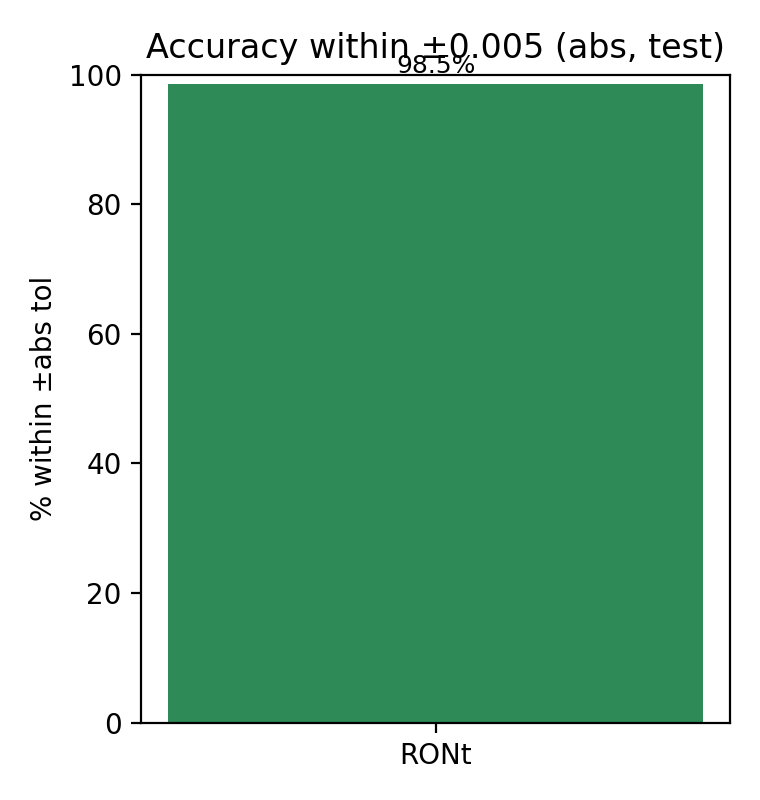}{Figure S229: accuracy within tol abs 0p005 (\label{fig:supp-229})}\hfill
\suppimage{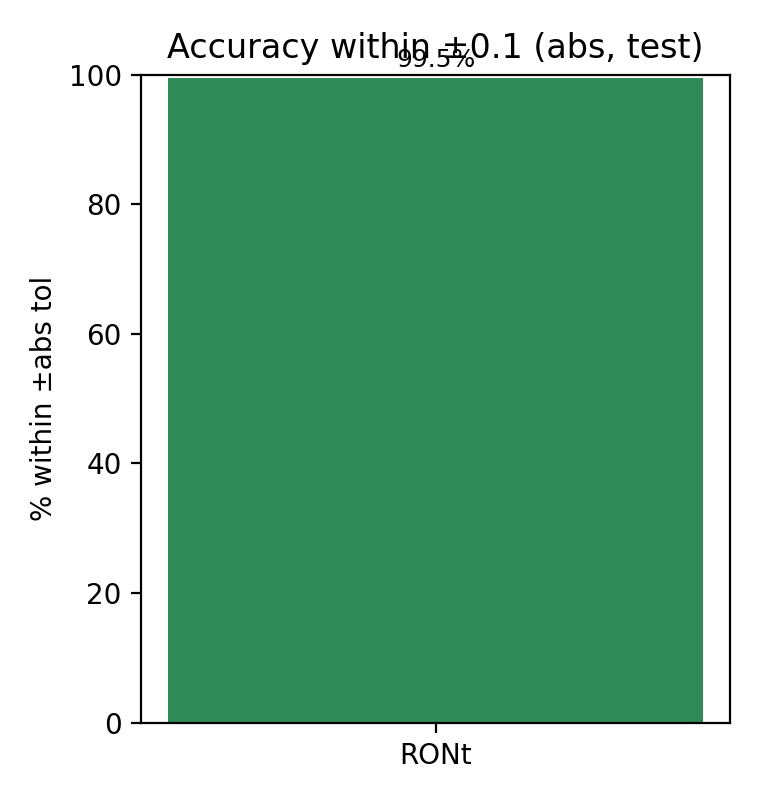}{Figure S230: accuracy within tol abs 0p1 (\label{fig:supp-230})}\
\suppimage{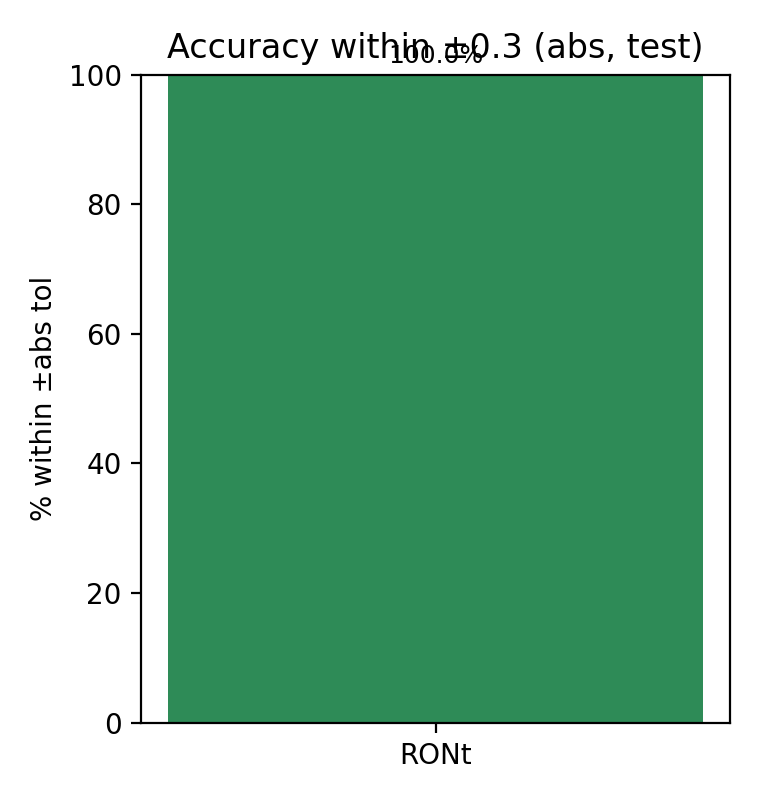}{Figure S231: accuracy within tol abs 0p3 (\label{fig:supp-231})}\hfill
\suppimage{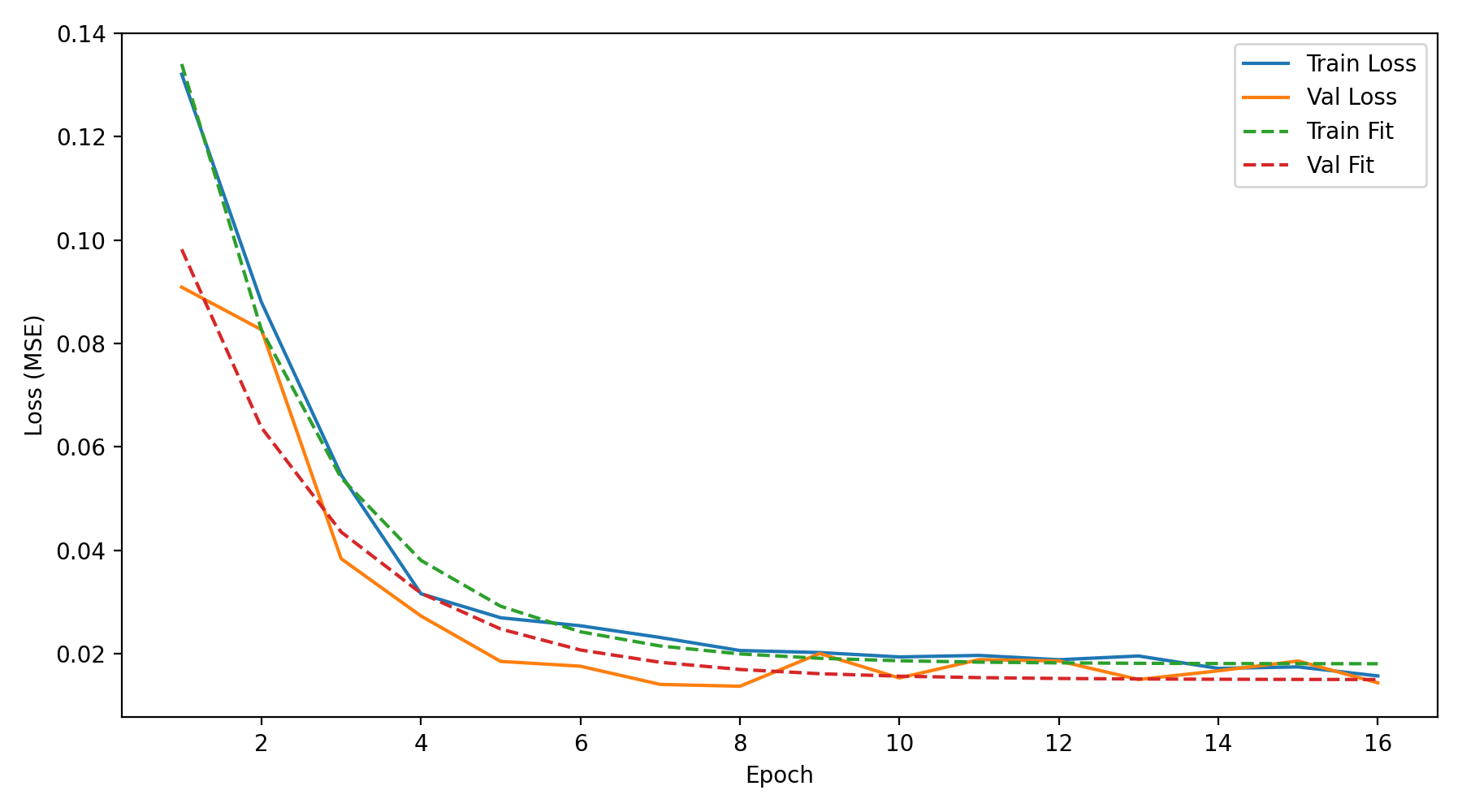}{Figure S232: loss curves (\label{fig:supp-232})}\
\suppimage{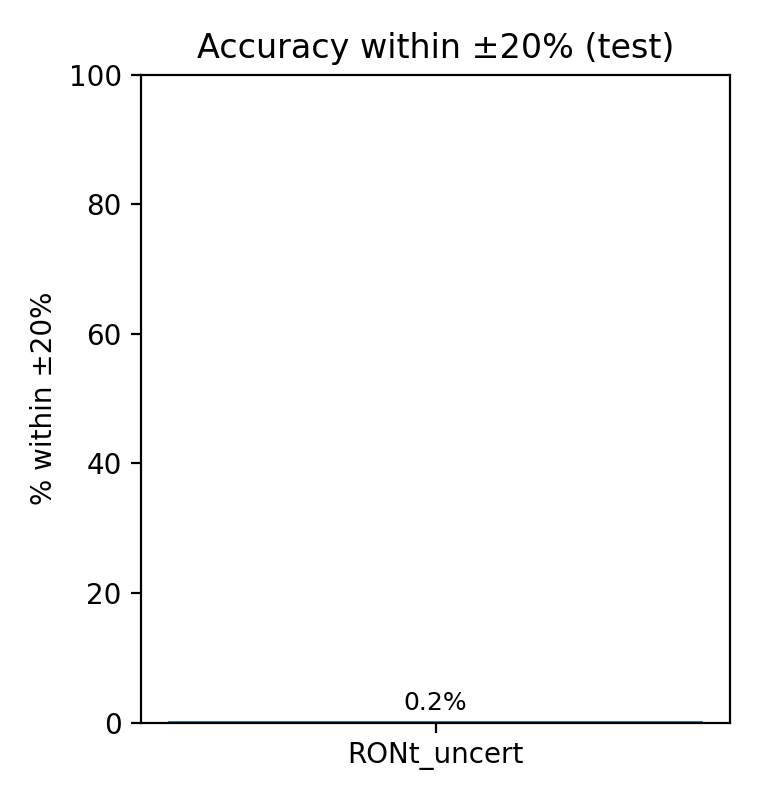}{Figure S233: accuracy within tol 20percent (\label{fig:supp-233})}\hfill
\suppimage{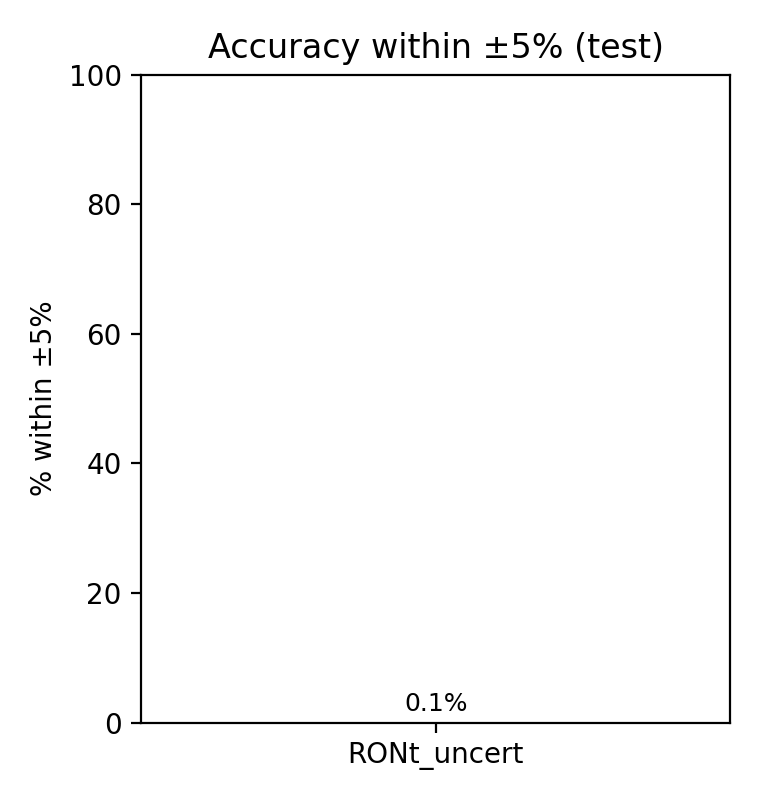}{Figure S234: accuracy within tol 5percent (\label{fig:supp-234})}\
\suppimage{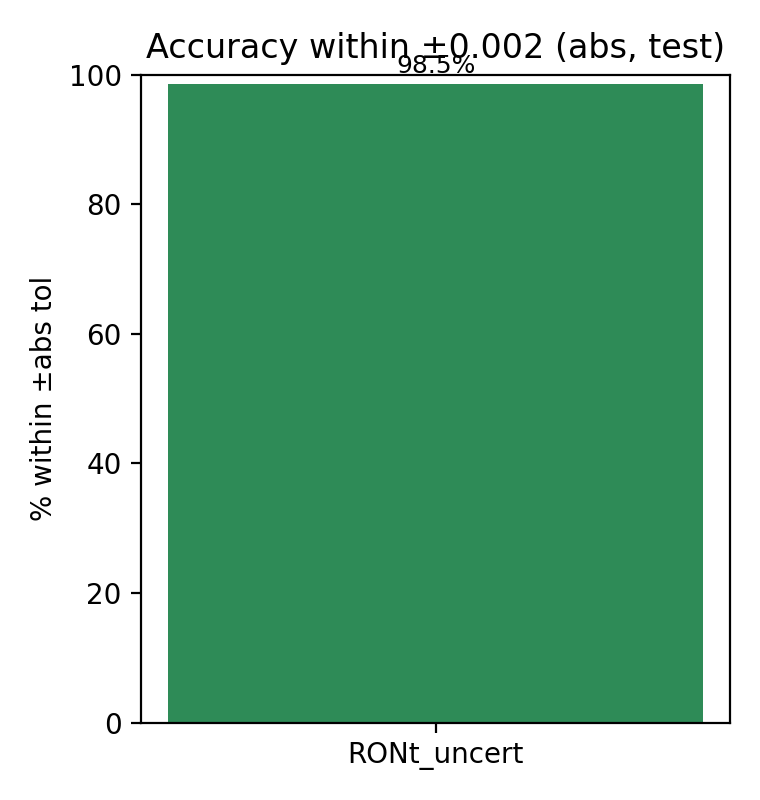}{Figure S235: accuracy within tol abs 0p002 (\label{fig:supp-235})}\hfill
\suppimage{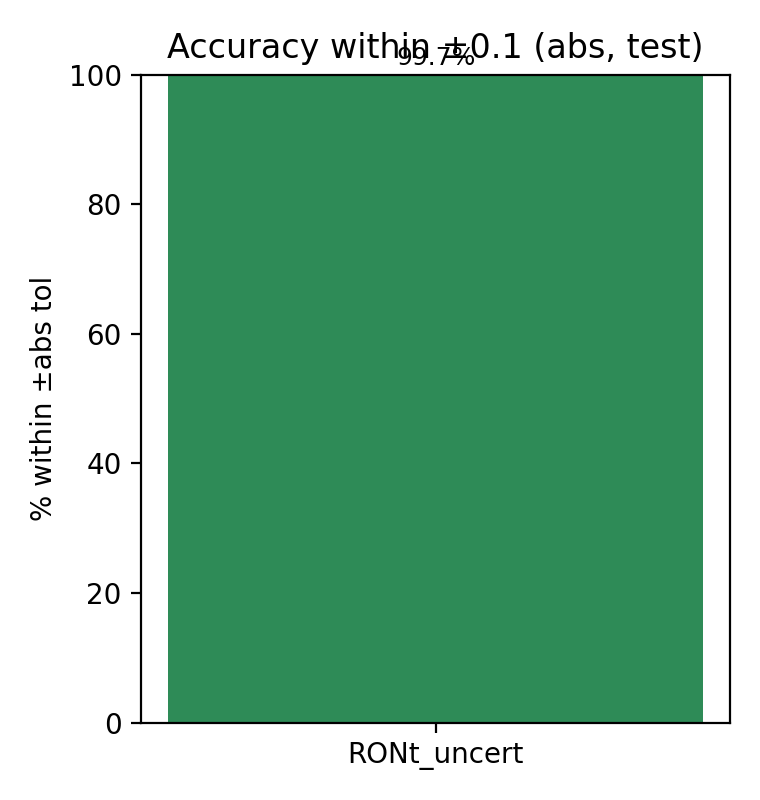}{Figure S236: accuracy within tol abs 0p1 (\label{fig:supp-236})}\
\suppimage{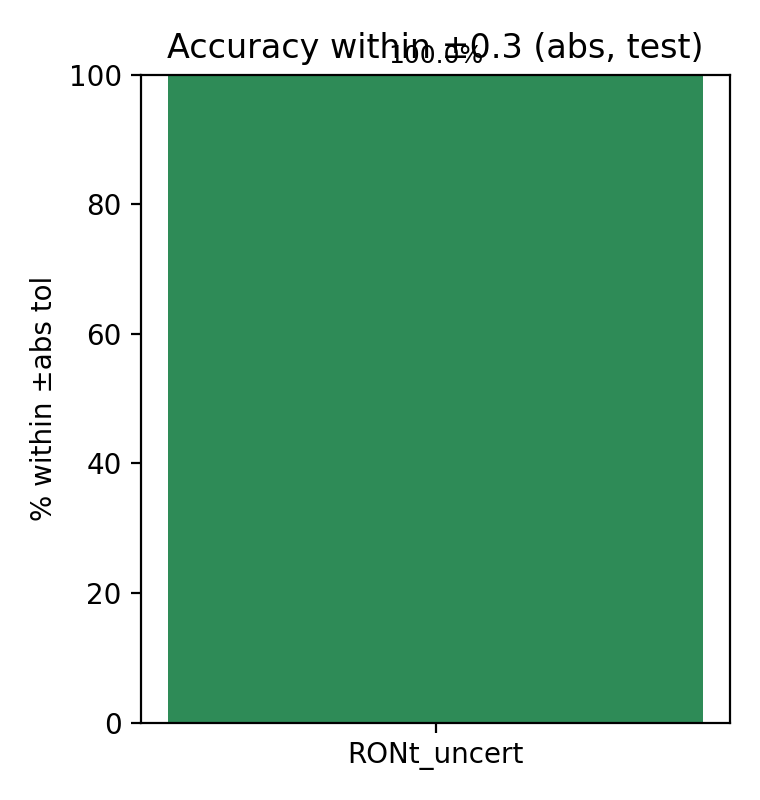}{Figure S237: accuracy within tol abs 0p3 (\label{fig:supp-237})}\hfill
\suppimage{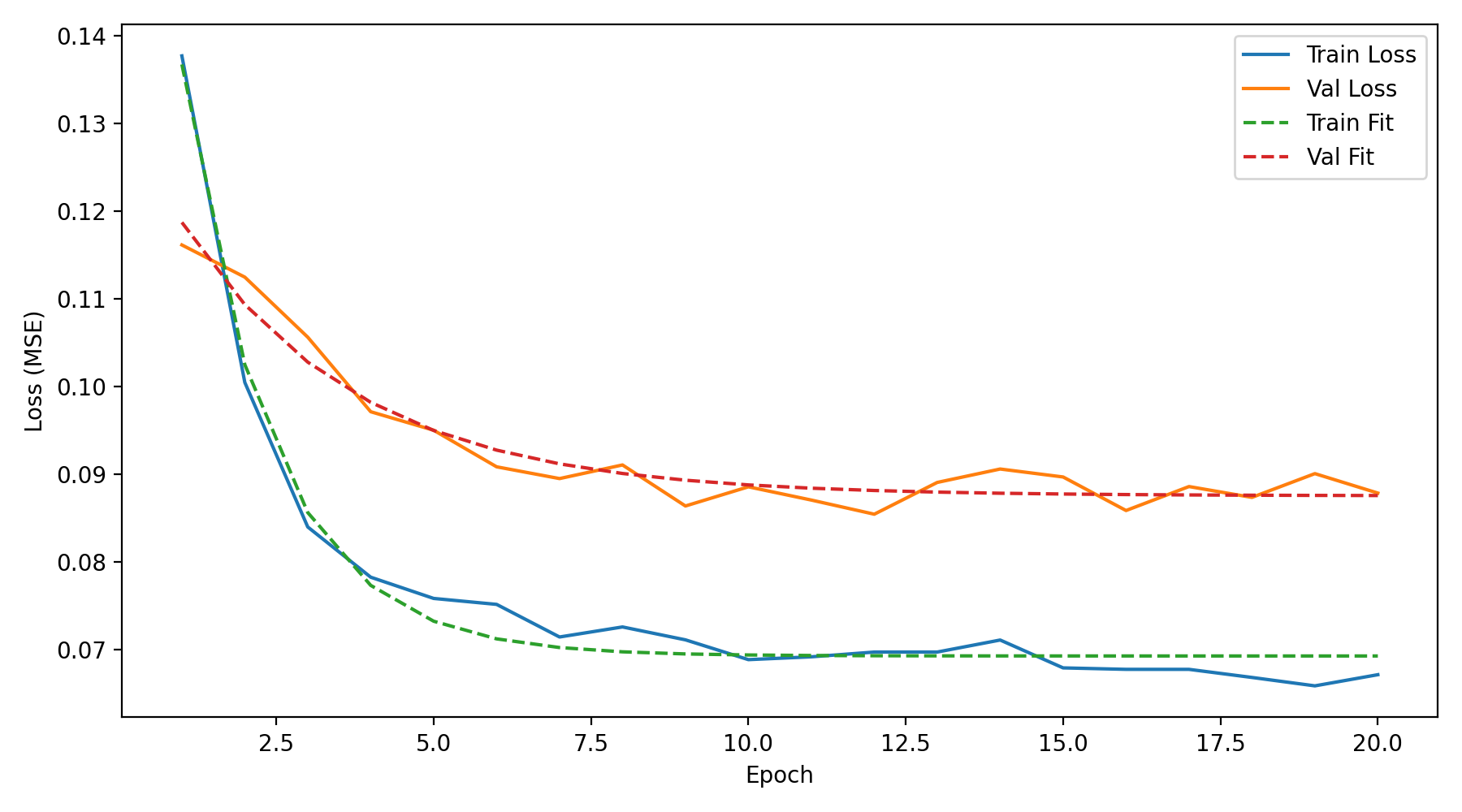}{Figure S238: loss curves (\label{fig:supp-238})}\
\suppimage{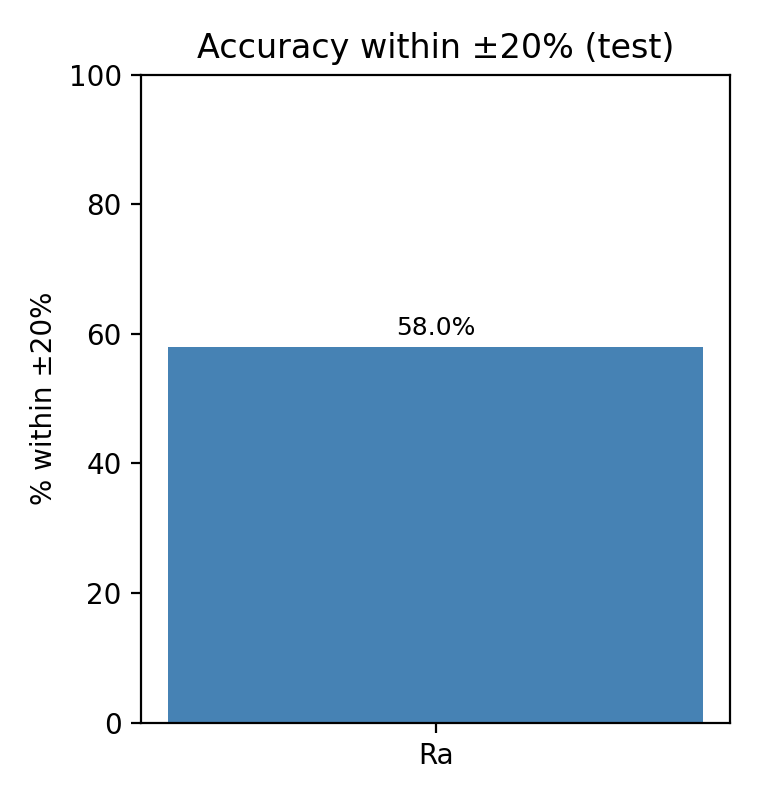}{Figure S239: accuracy within tol 20percent (\label{fig:supp-239})}\hfill
\suppimage{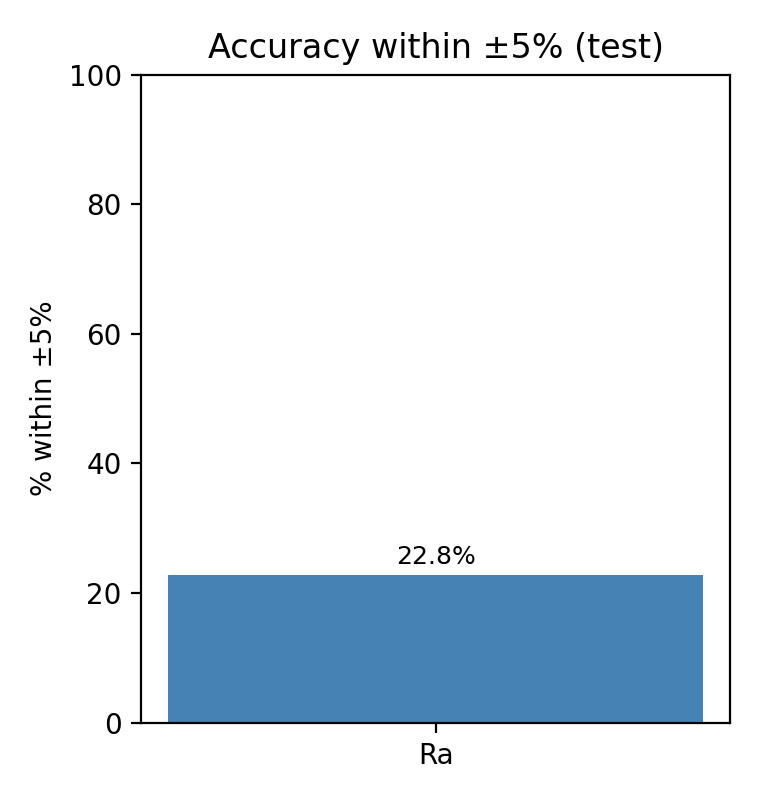}{Figure S240: accuracy within tol 5percent (\label{fig:supp-240})}\
\suppimage{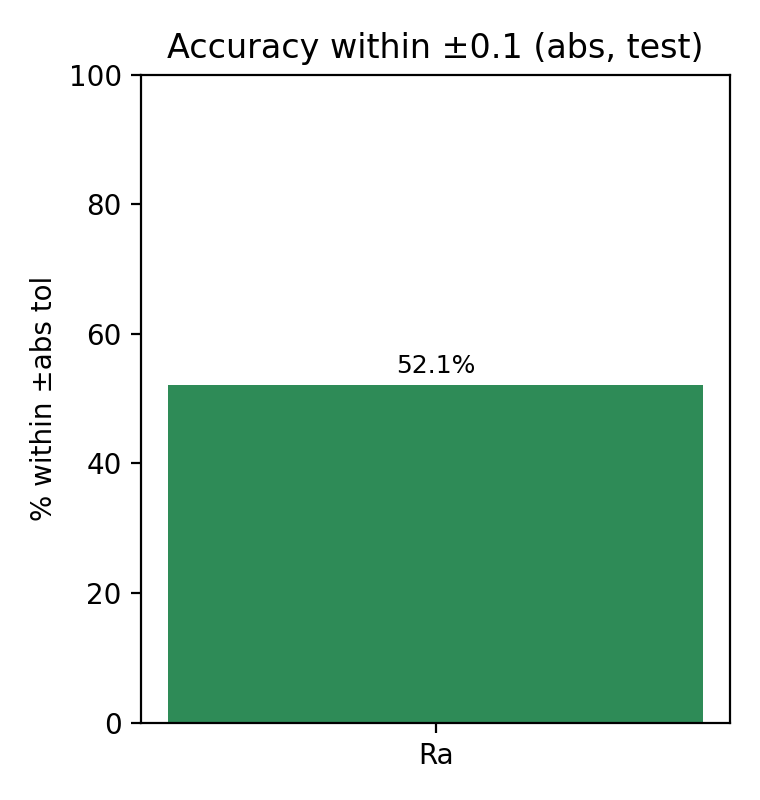}{Figure S241: accuracy within tol abs 0p1 (\label{fig:supp-241})}\hfill
\suppimage{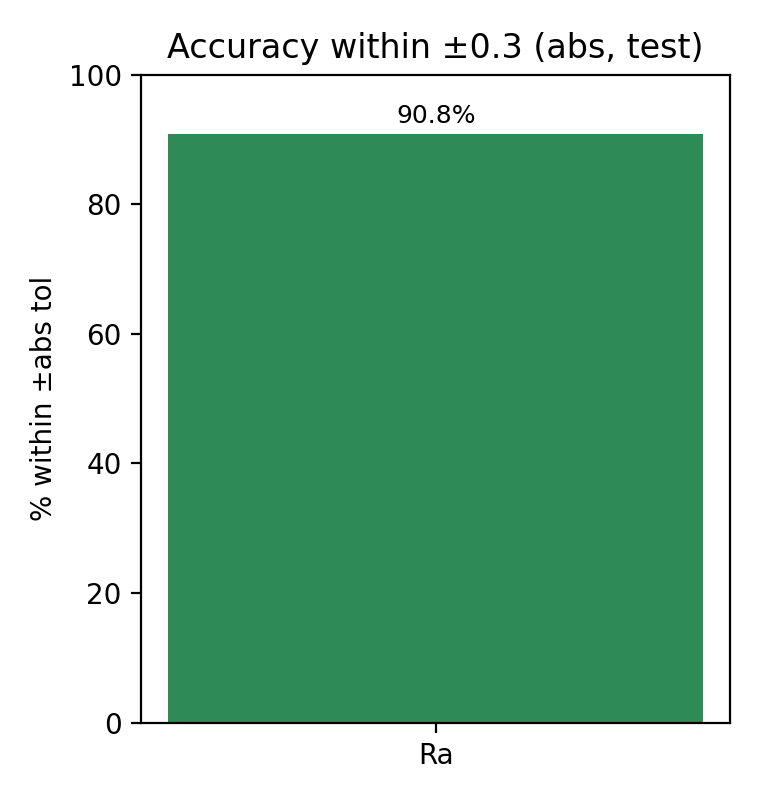}{Figure S242: accuracy within tol abs 0p3 (\label{fig:supp-242})}\
\suppimage{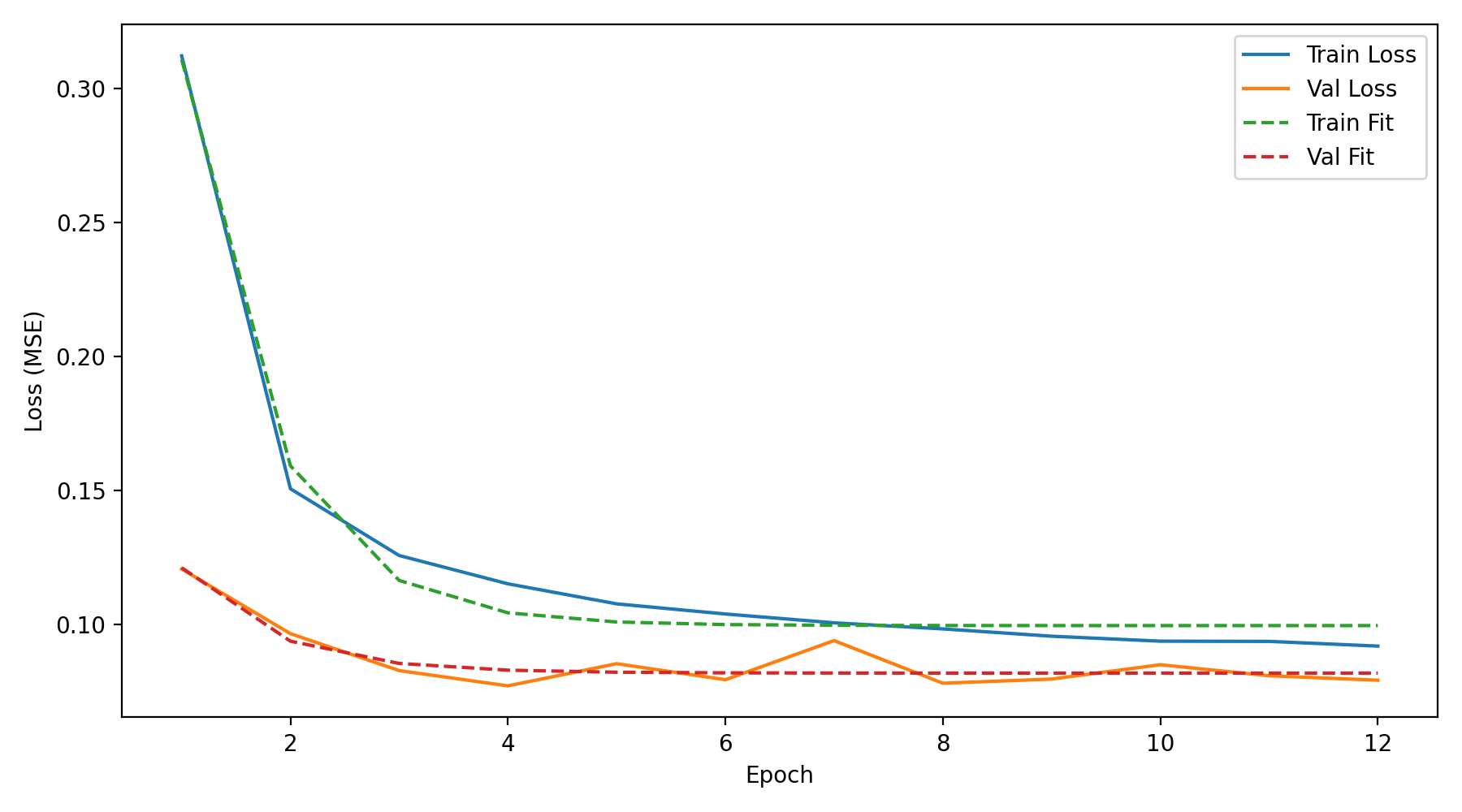}{Figure S243: loss curves (\label{fig:supp-243})}\hfill
\suppimage{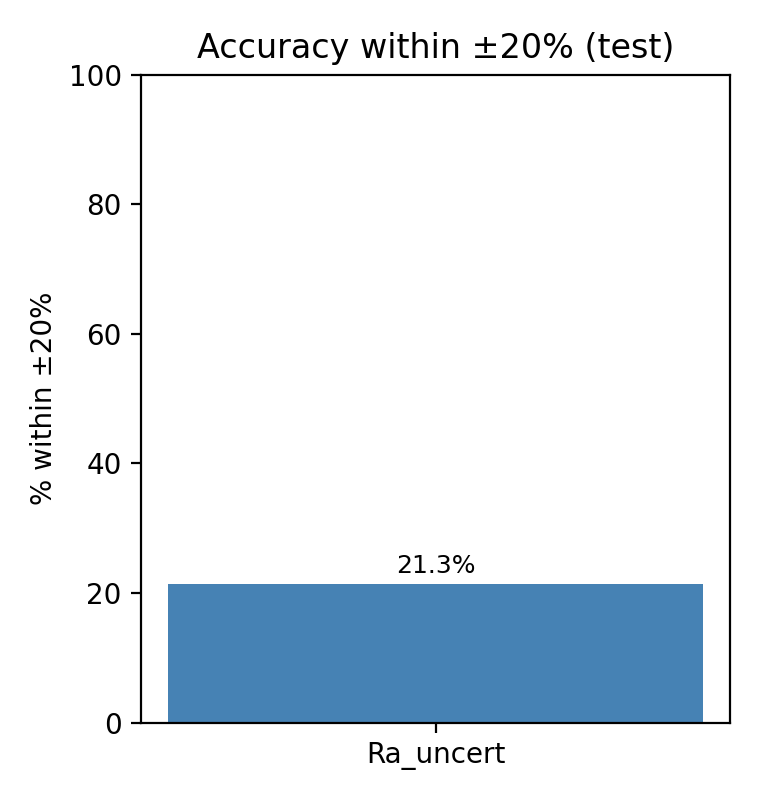}{Figure S244: accuracy within tol 20percent (\label{fig:supp-244})}\
\suppimage{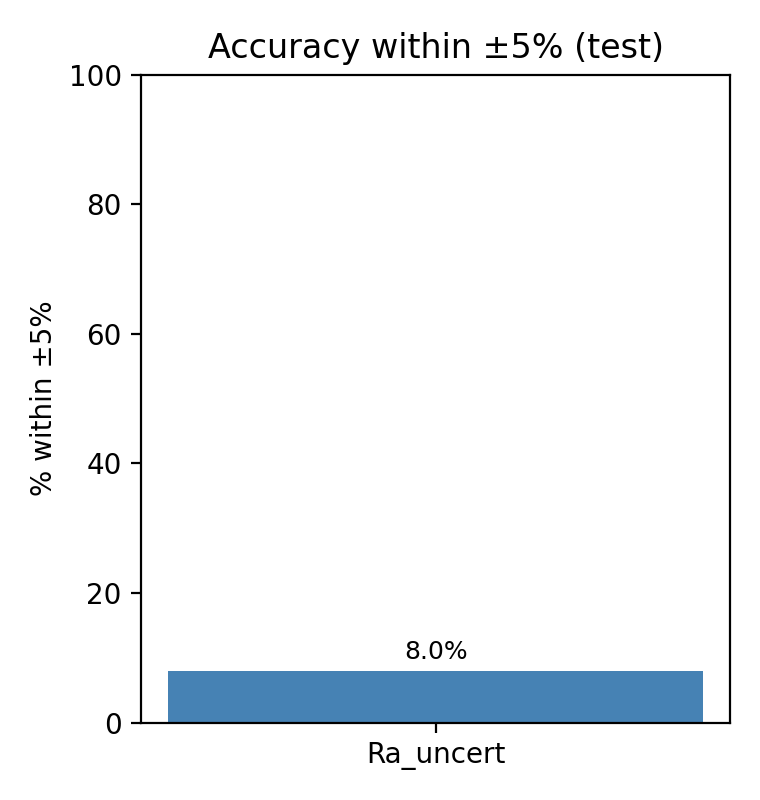}{Figure S245: accuracy within tol 5percent (\label{fig:supp-245})}\hfill
\suppimage{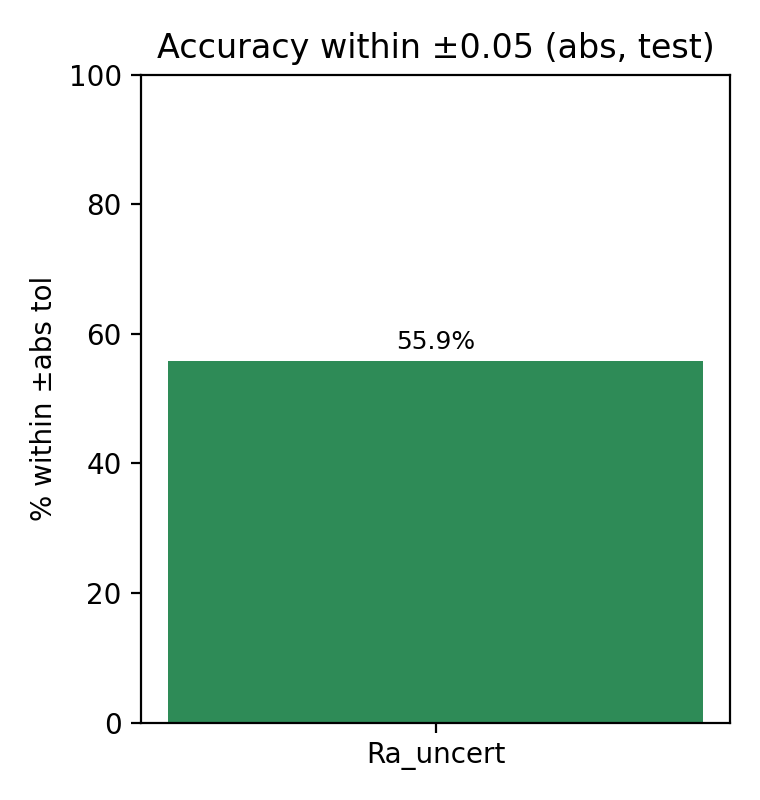}{Figure S246: accuracy within tol abs 0p05 (\label{fig:supp-246})}\
\suppimage{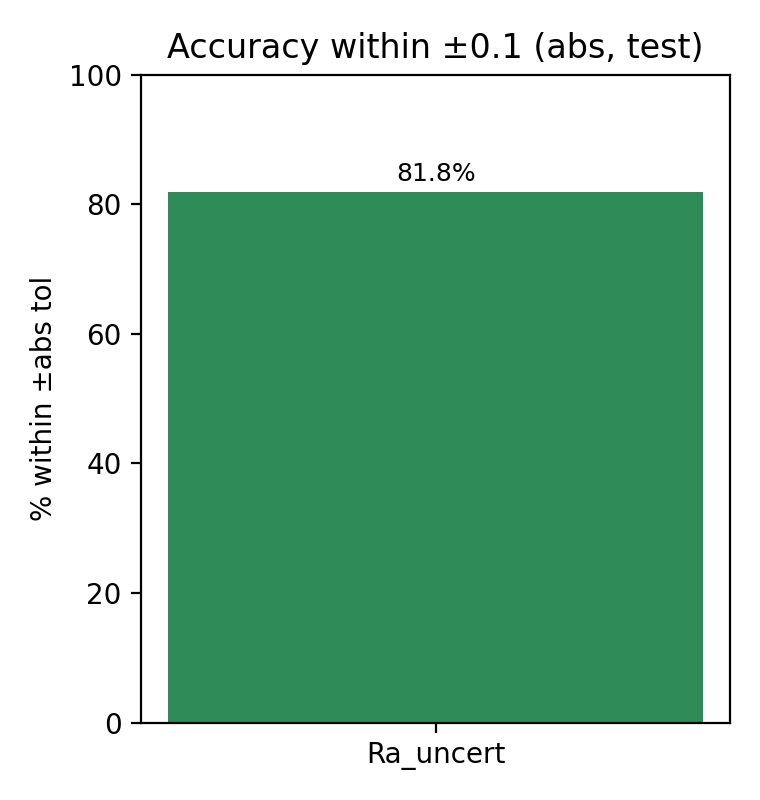}{Figure S247: accuracy within tol abs 0p1 (\label{fig:supp-247})}\hfill
\suppimage{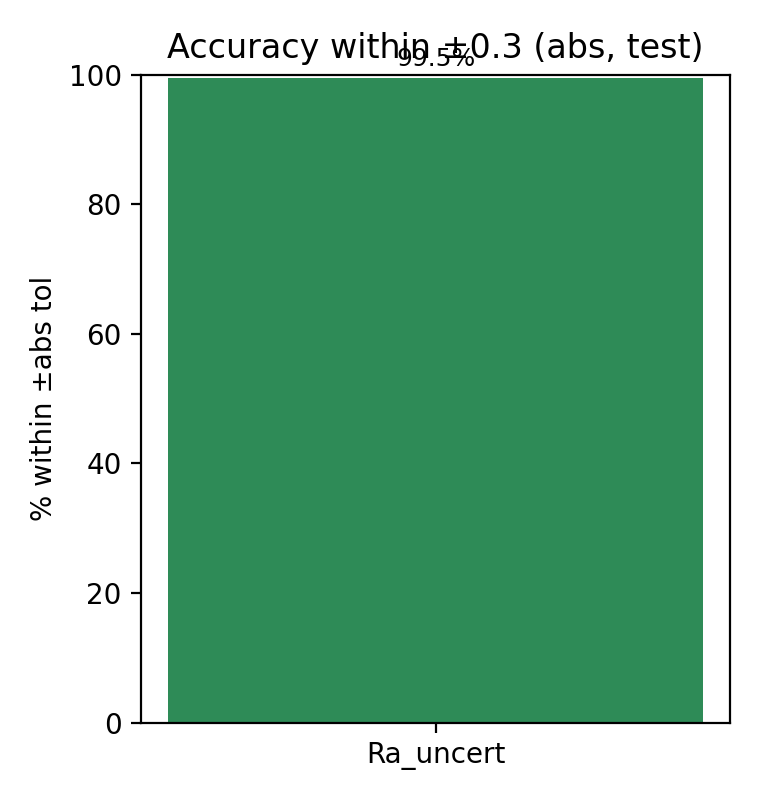}{Figure S248: accuracy within tol abs 0p3 (\label{fig:supp-248})}\
\suppimage{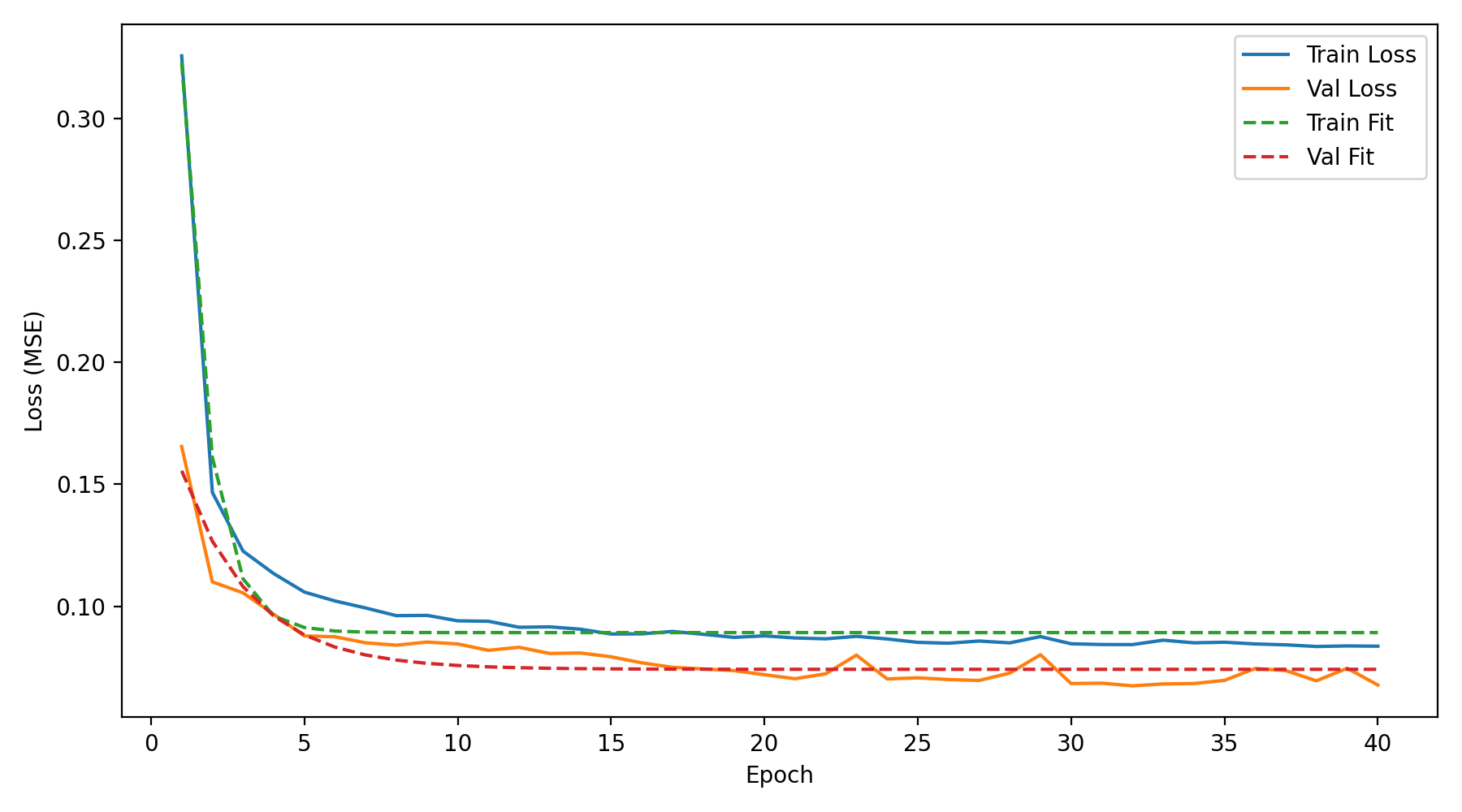}{Figure S249: loss curves (\label{fig:supp-249})}\hfill
\suppimage{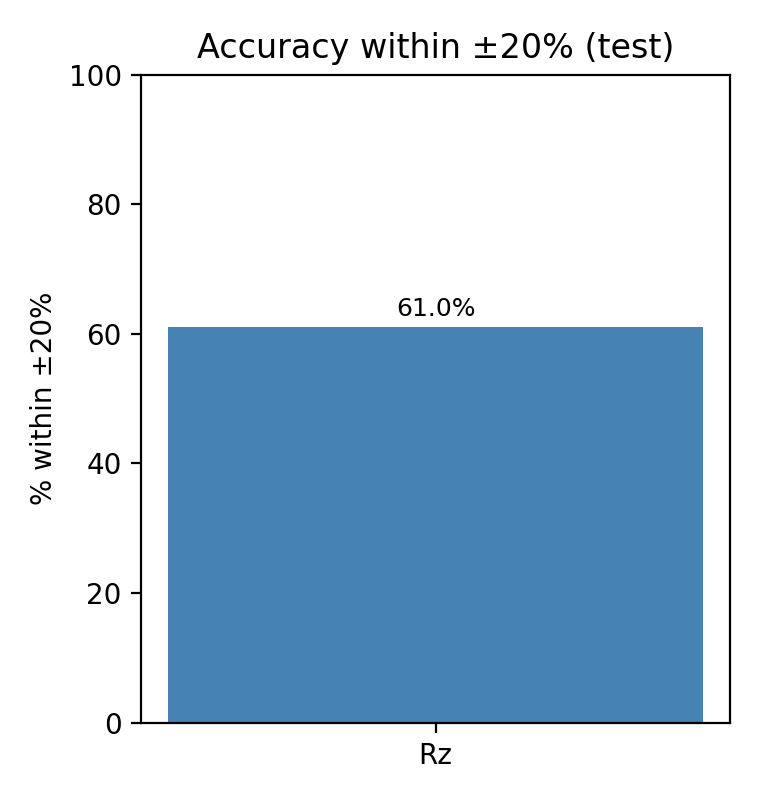}{Figure S250: accuracy within tol 20percent (\label{fig:supp-250})}\
\suppimage{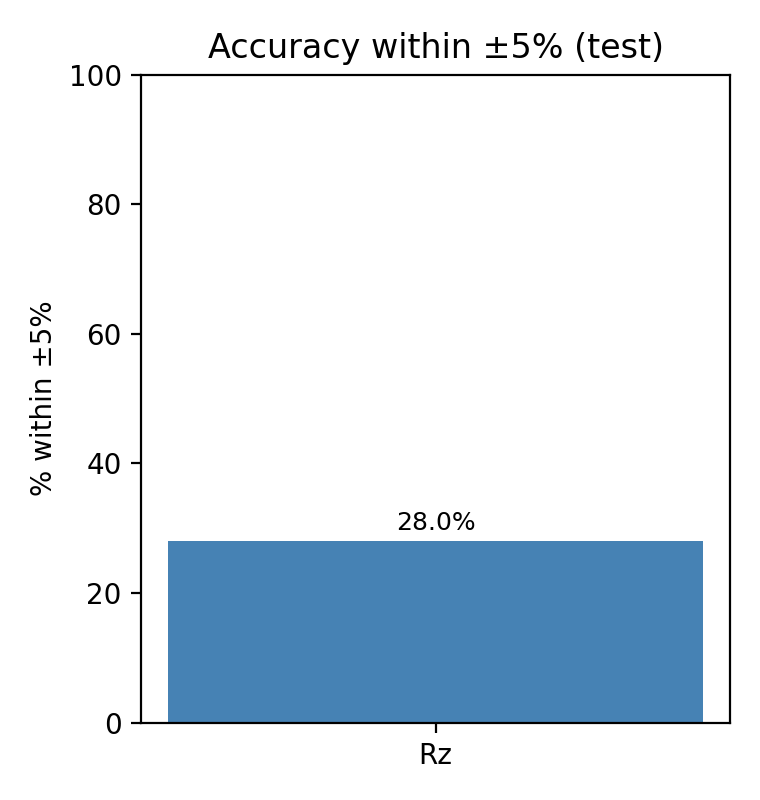}{Figure S251: accuracy within tol 5percent (\label{fig:supp-251})}\hfill
\suppimage{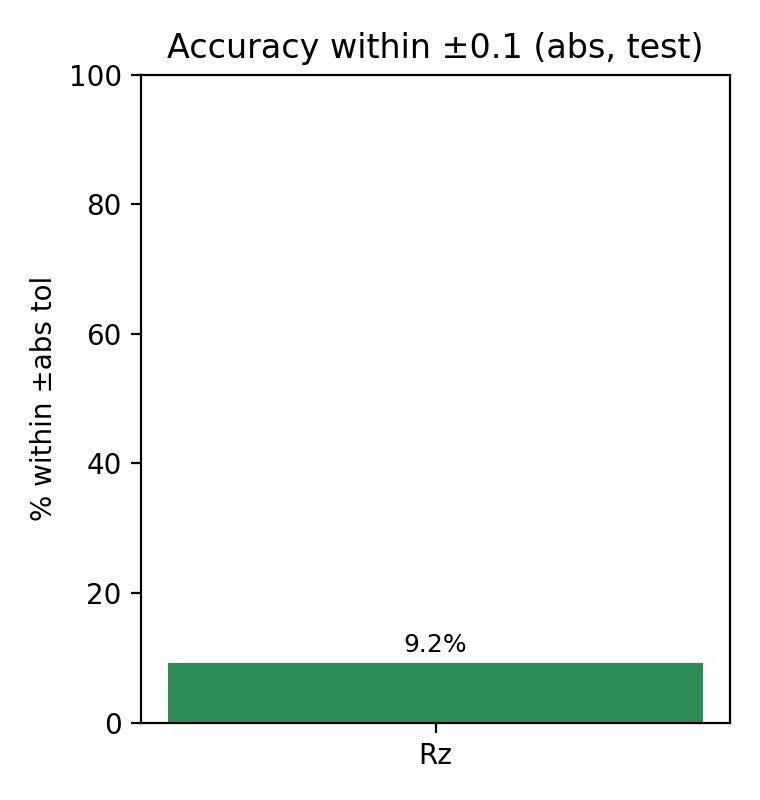}{Figure S252: accuracy within tol abs 0p1 (\label{fig:supp-252})}\
\suppimage{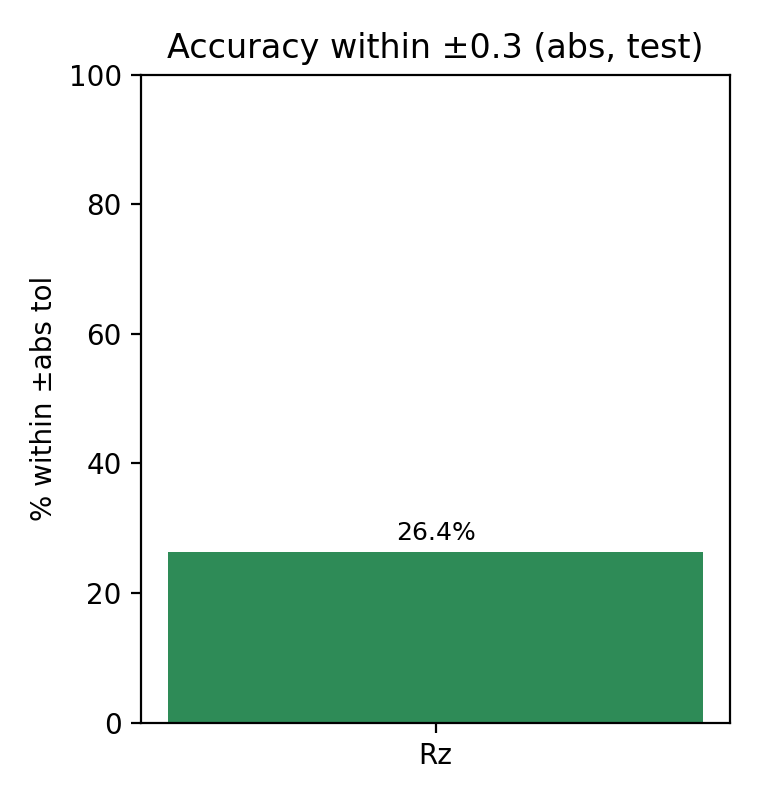}{Figure S253: accuracy within tol abs 0p3 (\label{fig:supp-253})}\hfill
\suppimage{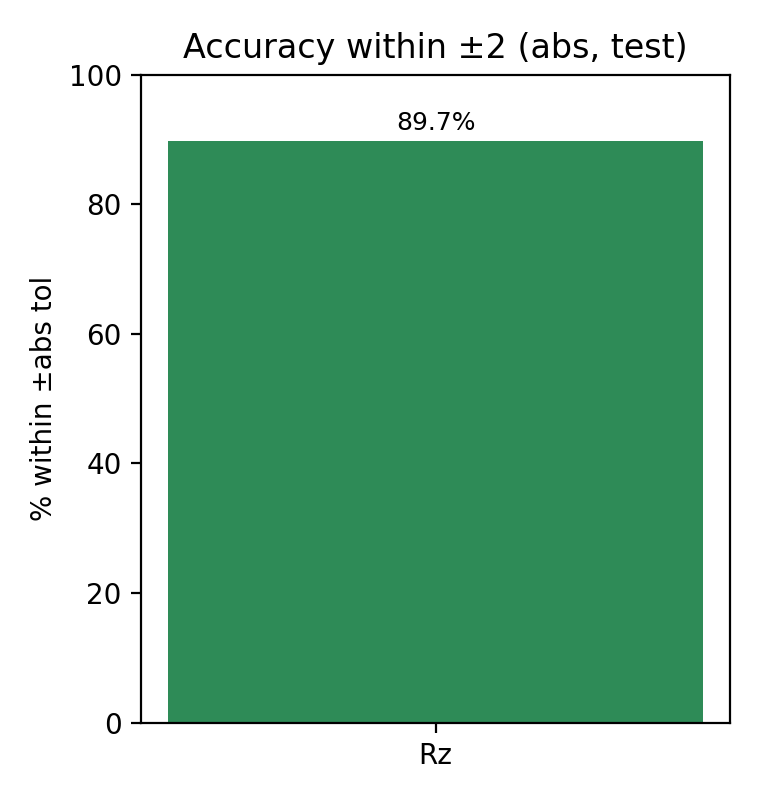}{Figure S254: accuracy within tol abs 2 (\label{fig:supp-254})}\
\suppimage{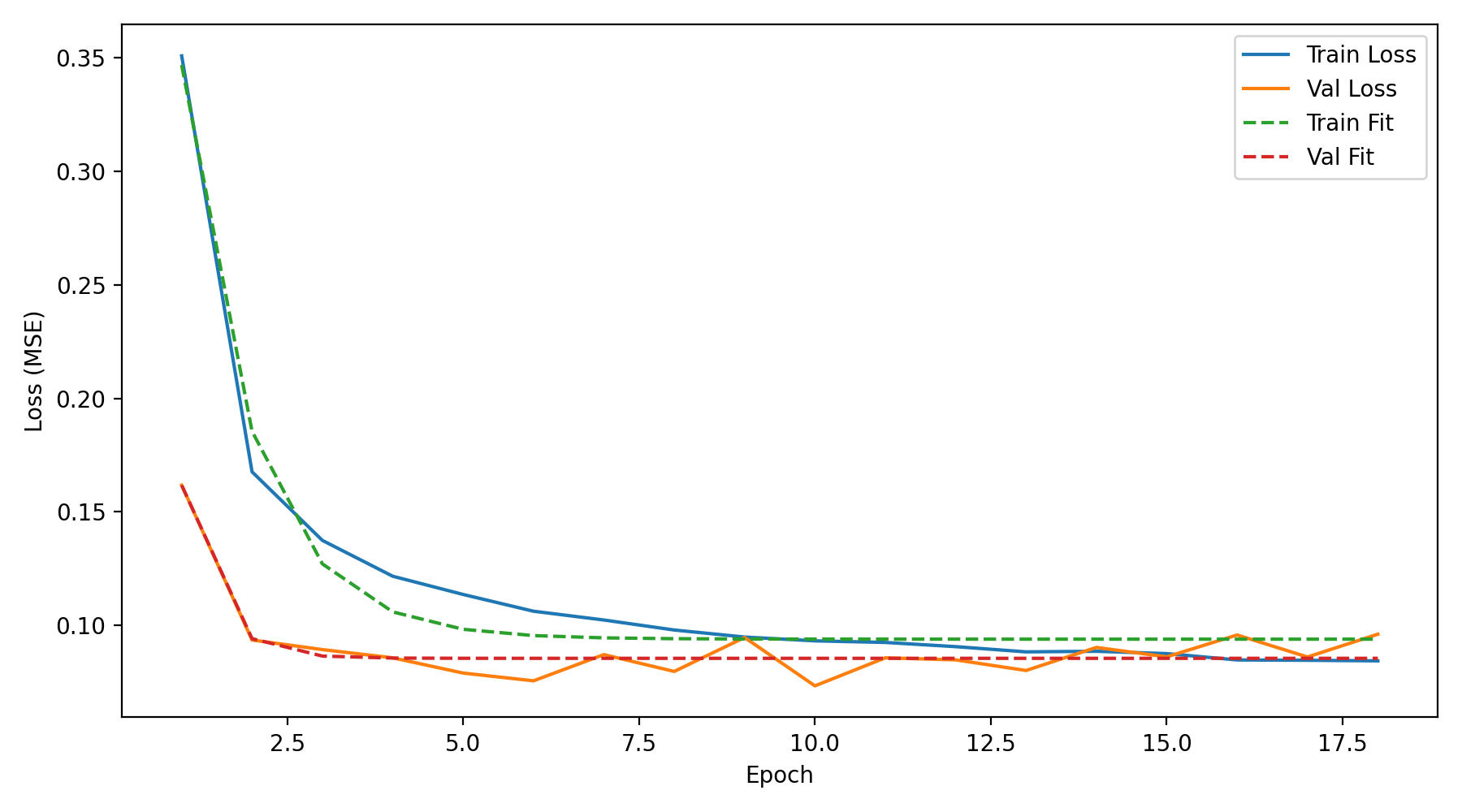}{Figure S255: loss curves (\label{fig:supp-255})}\hfill
\suppimage{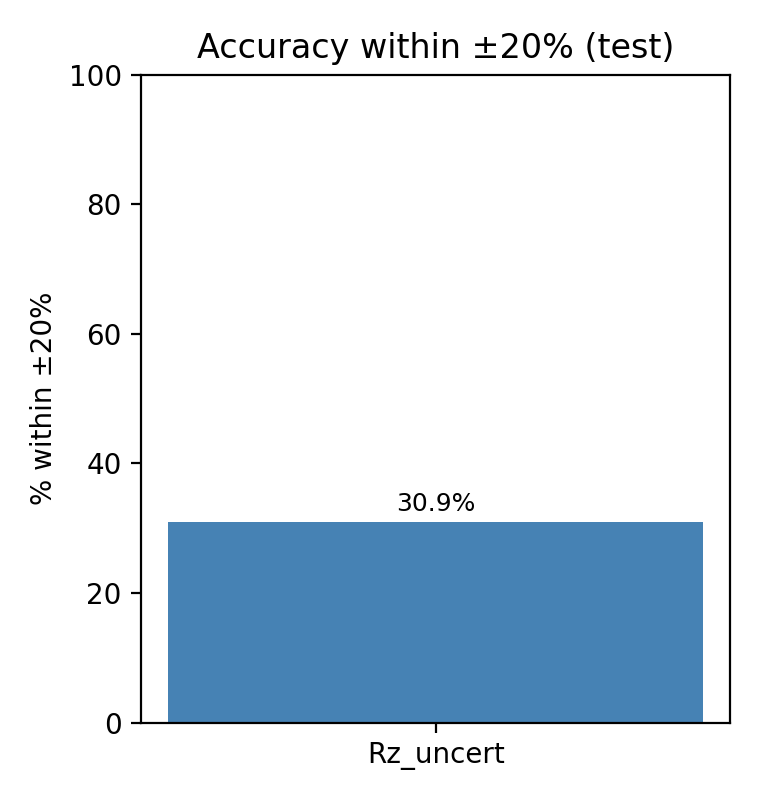}{Figure S256: accuracy within tol 20percent (\label{fig:supp-256})}\
\suppimage{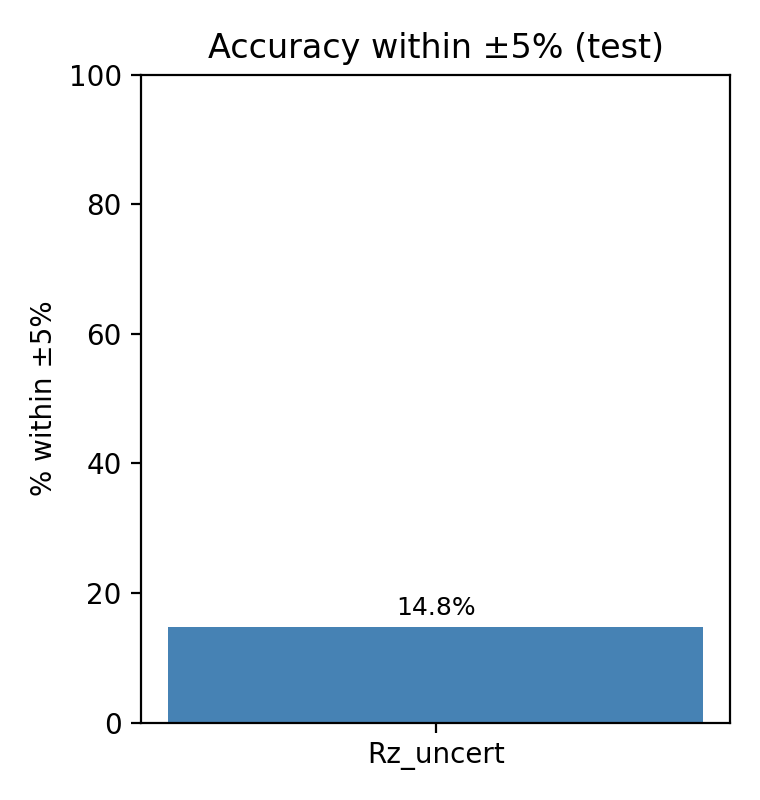}{Figure S257: accuracy within tol 5percent (\label{fig:supp-257})}\hfill
\suppimage{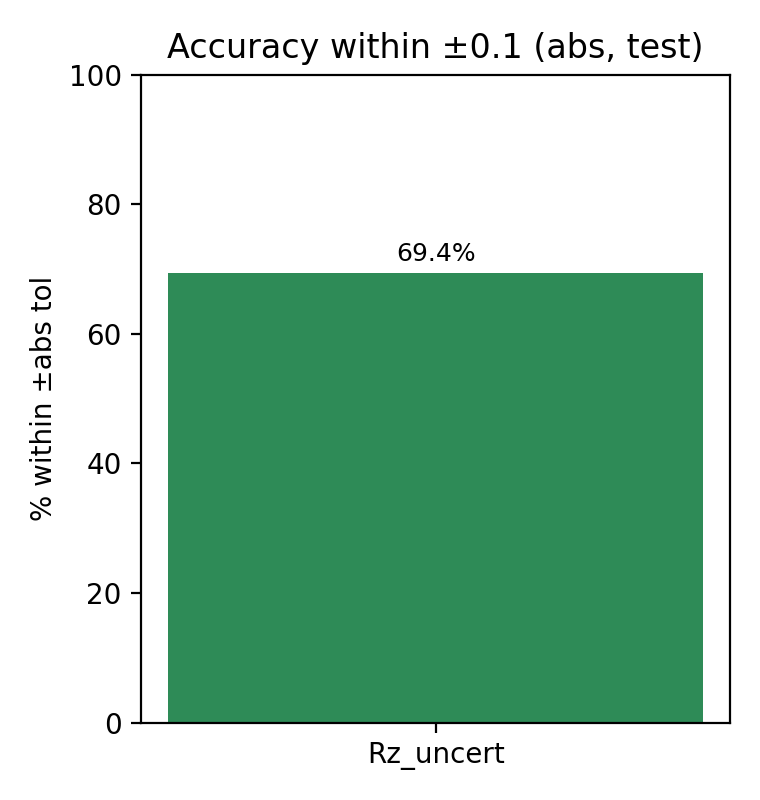}{Figure S258: accuracy within tol abs 0p1 (\label{fig:supp-258})}\
\suppimage{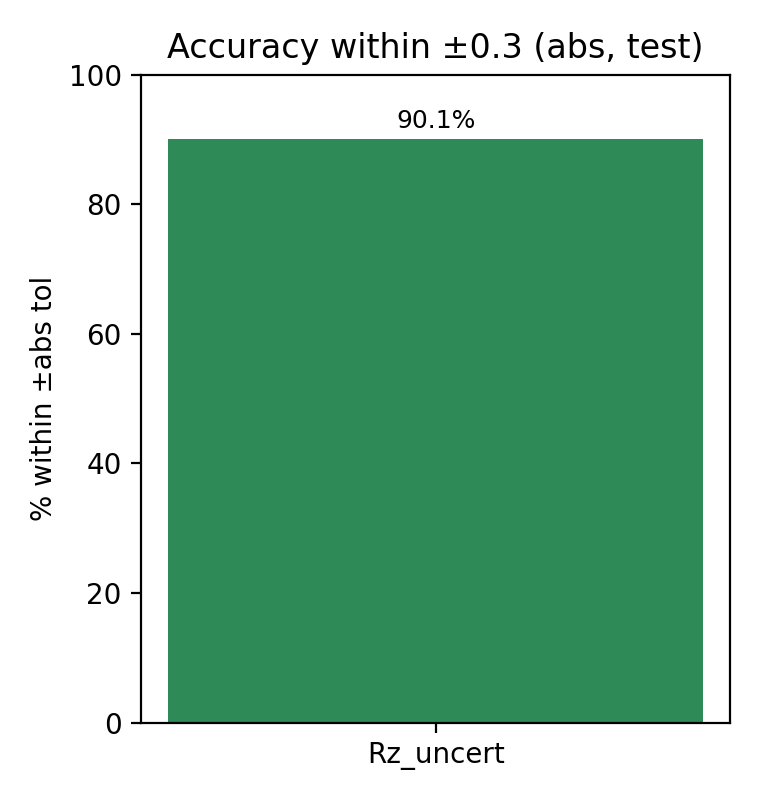}{Figure S259: accuracy within tol abs 0p3 (\label{fig:supp-259})}\hfill
\suppimage{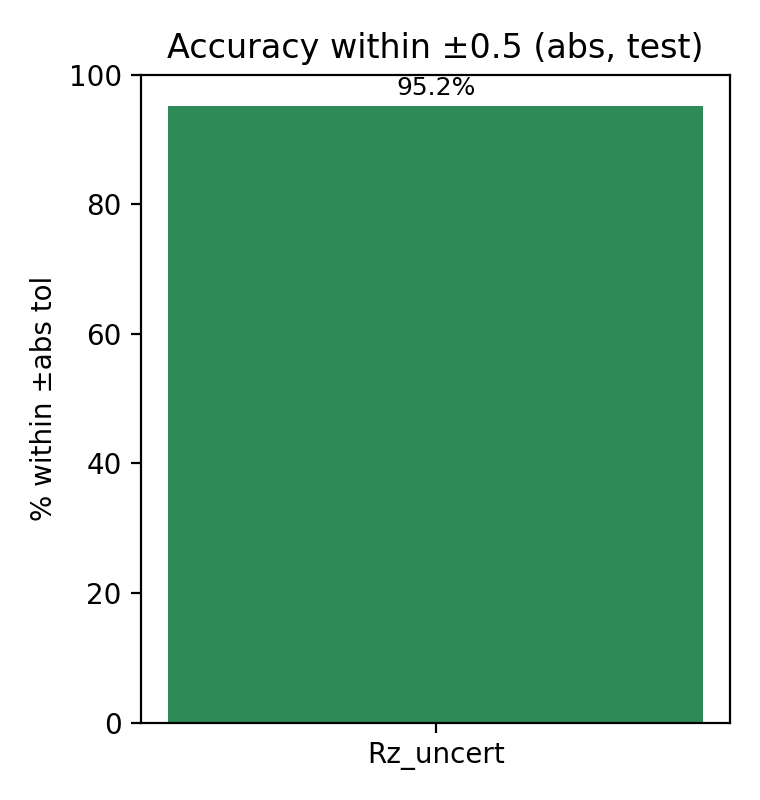}{Figure S260: accuracy within tol abs 0p5 (\label{fig:supp-260})}\
\suppimage{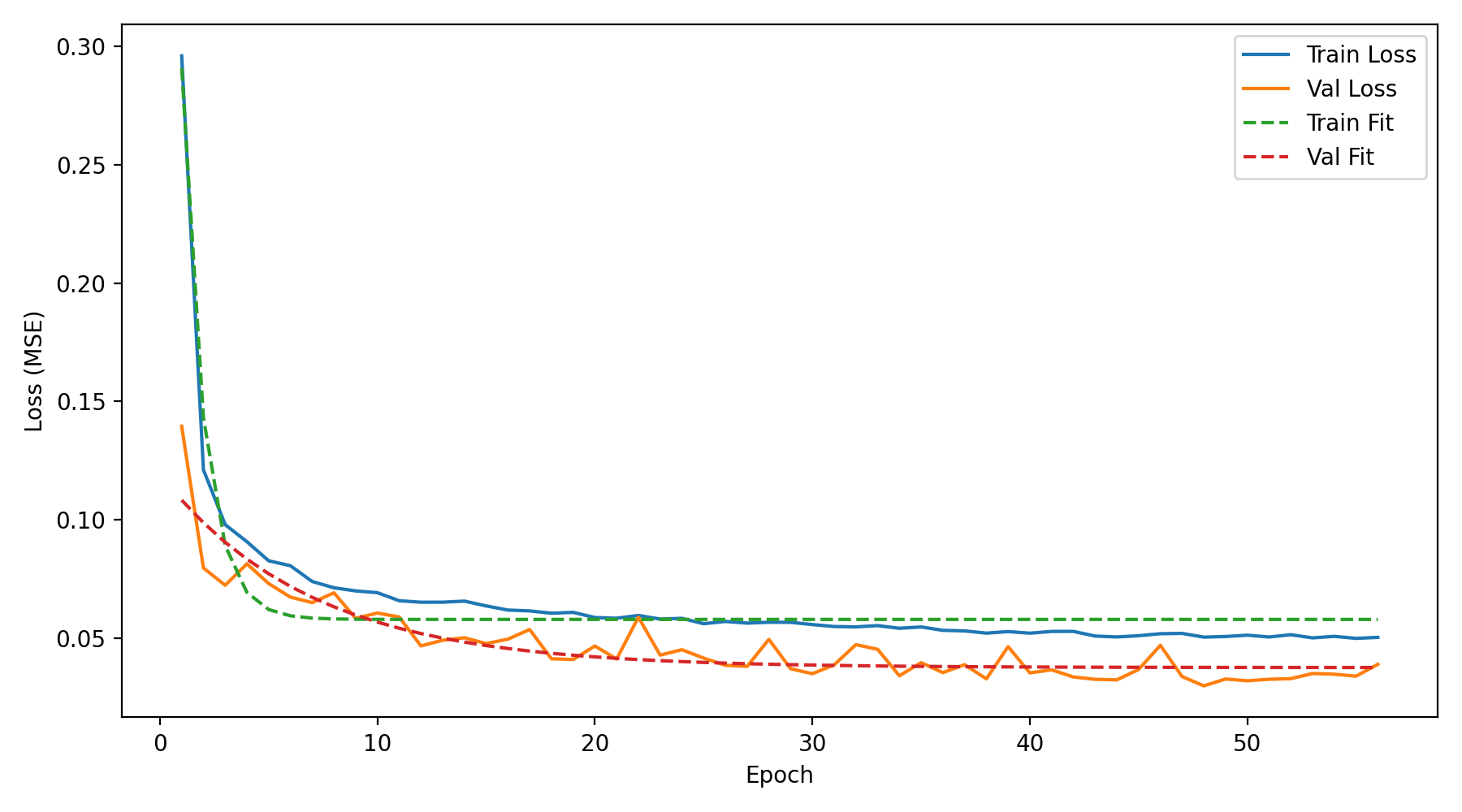}{Figure S261: loss curves (\label{fig:supp-261})}\hfill
\suppimage{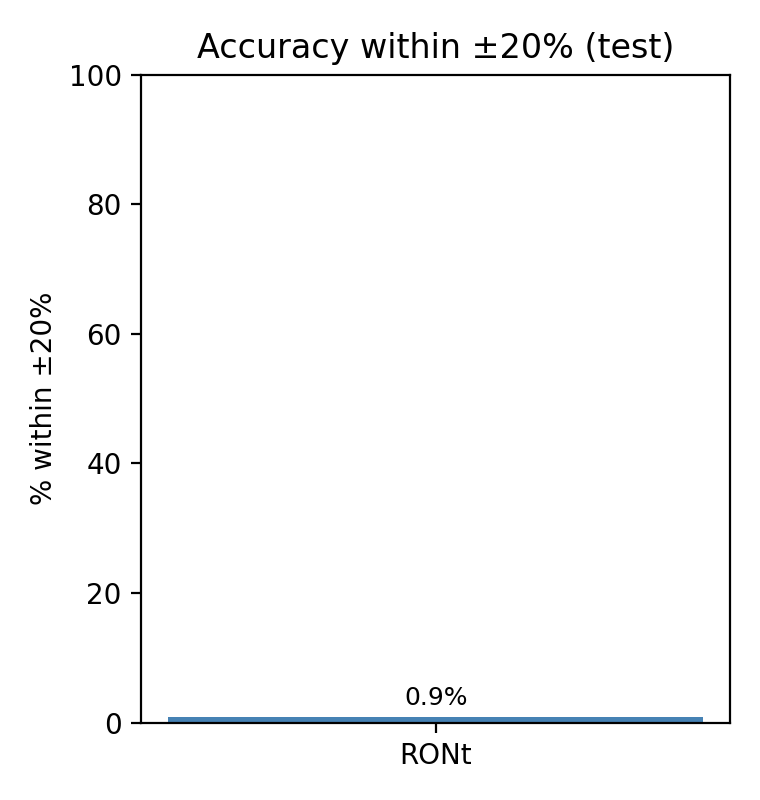}{Figure S262: accuracy within tol 20percent (\label{fig:supp-262})}\
\suppimage{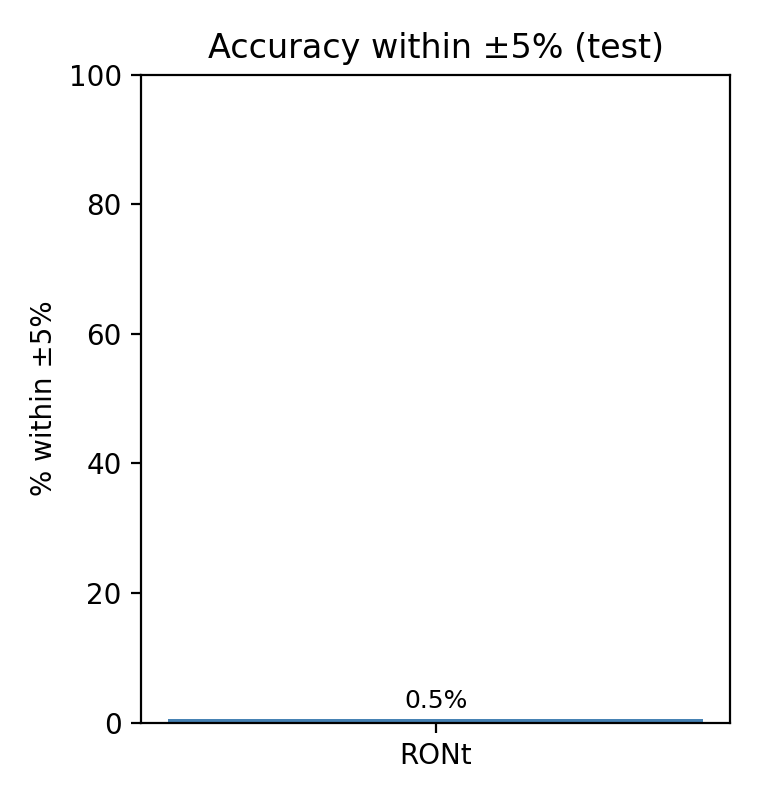}{Figure S263: accuracy within tol 5percent (\label{fig:supp-263})}\hfill
\suppimage{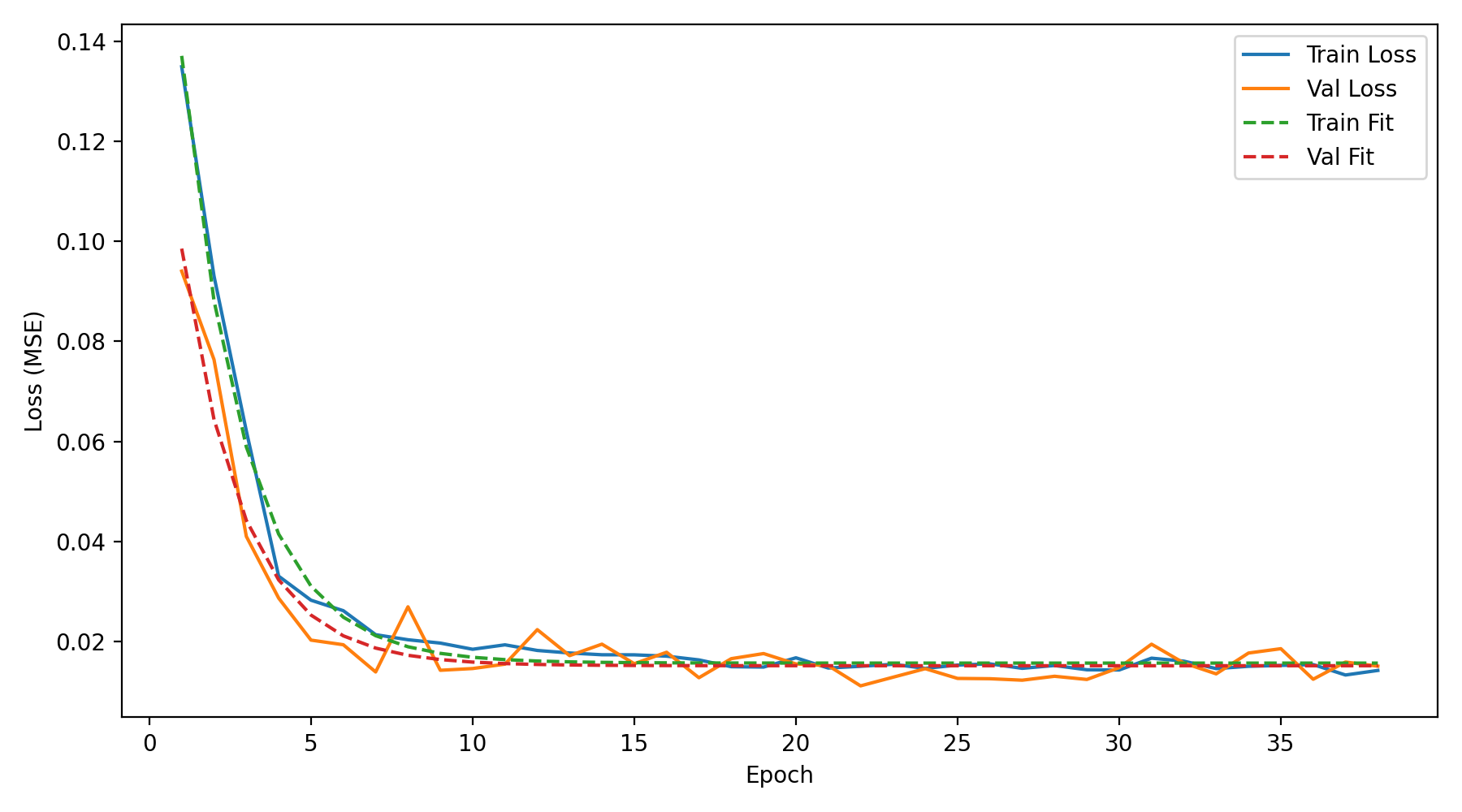}{Figure S264: loss curves (\label{fig:supp-264})}\
\suppimage{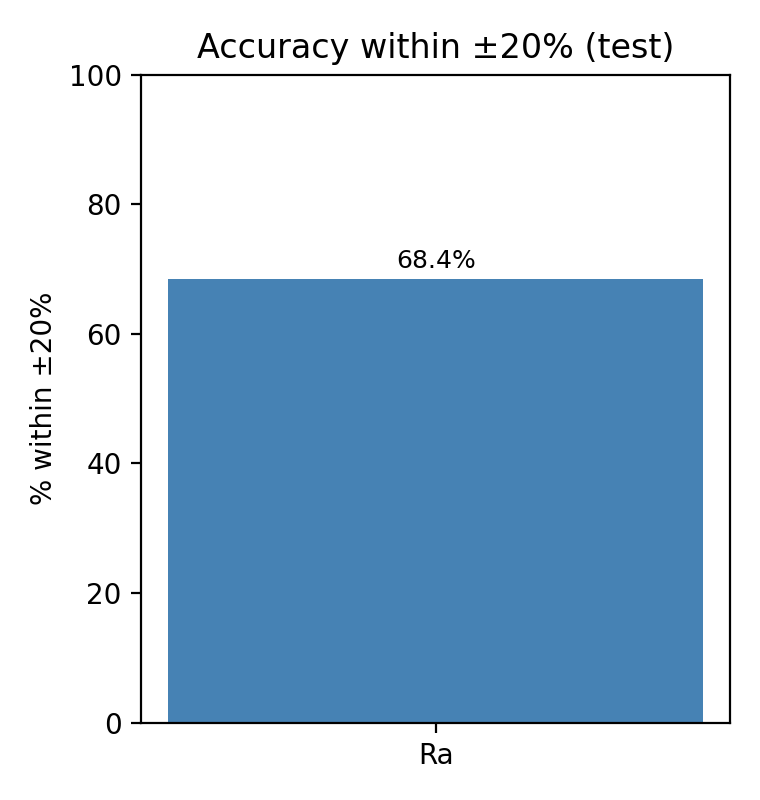}{Figure S265: accuracy within tol 20percent (\label{fig:supp-265})}\hfill
\suppimage{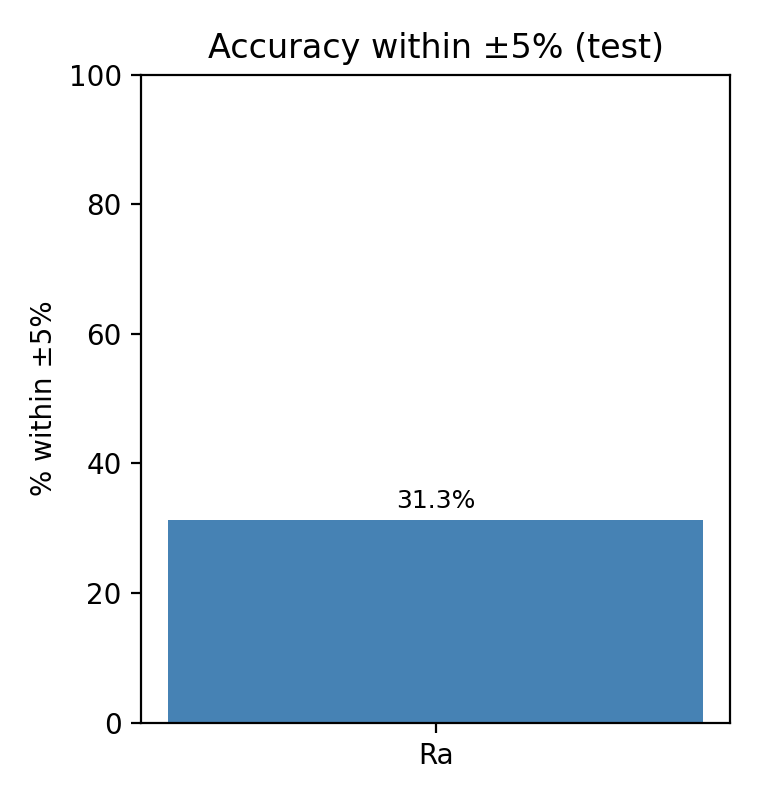}{Figure S266: accuracy within tol 5percent (\label{fig:supp-266})}\
\suppimage{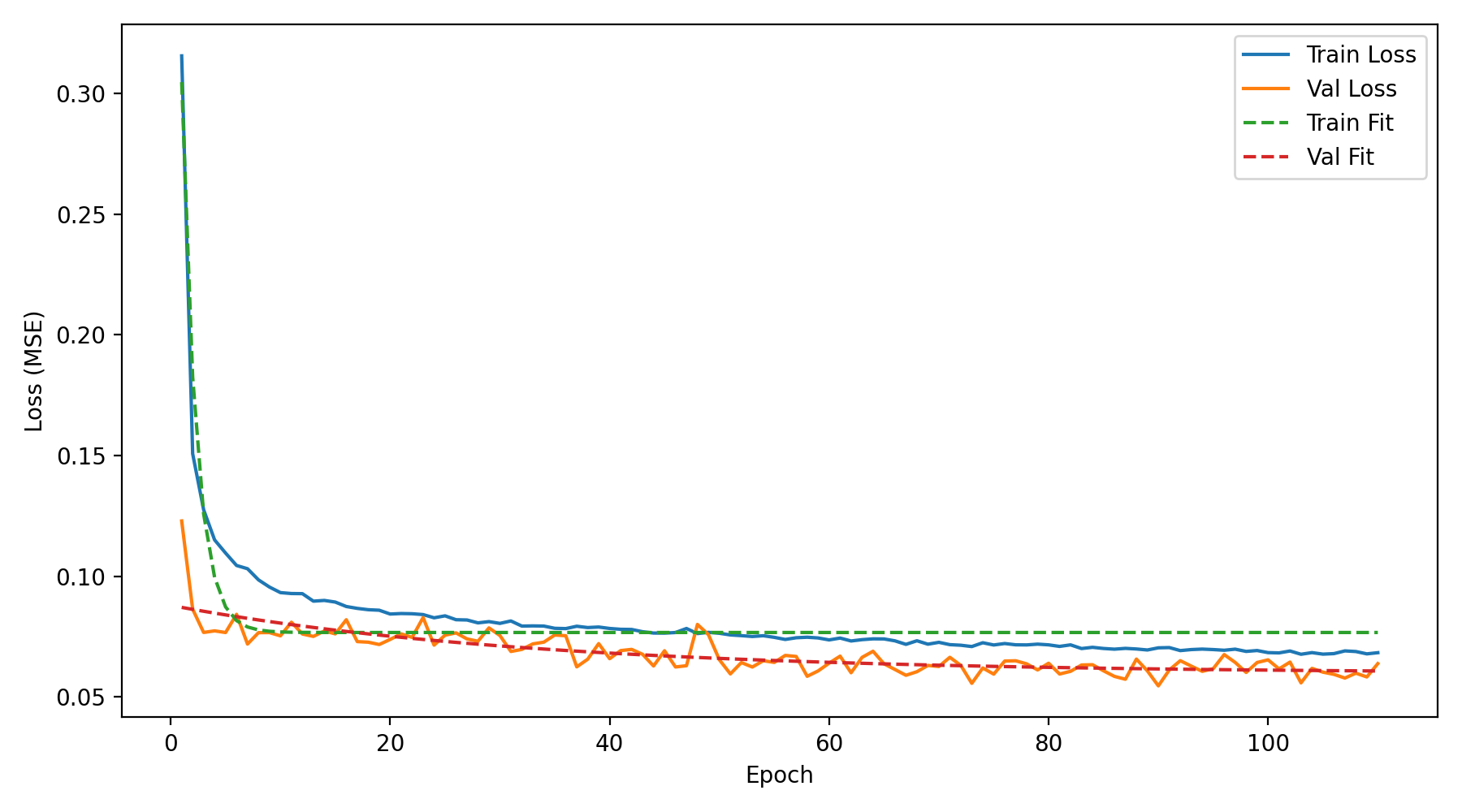}{Figure S267: loss curves (\label{fig:supp-267})}\hfill
\suppimage{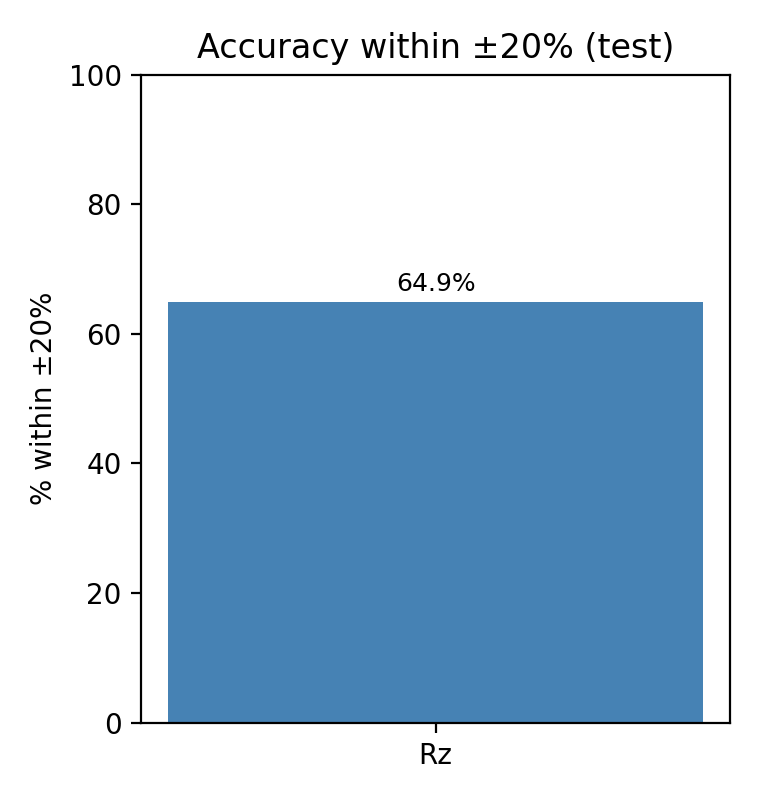}{Figure S268: accuracy within tol 20percent (\label{fig:supp-268})}\
\suppimage{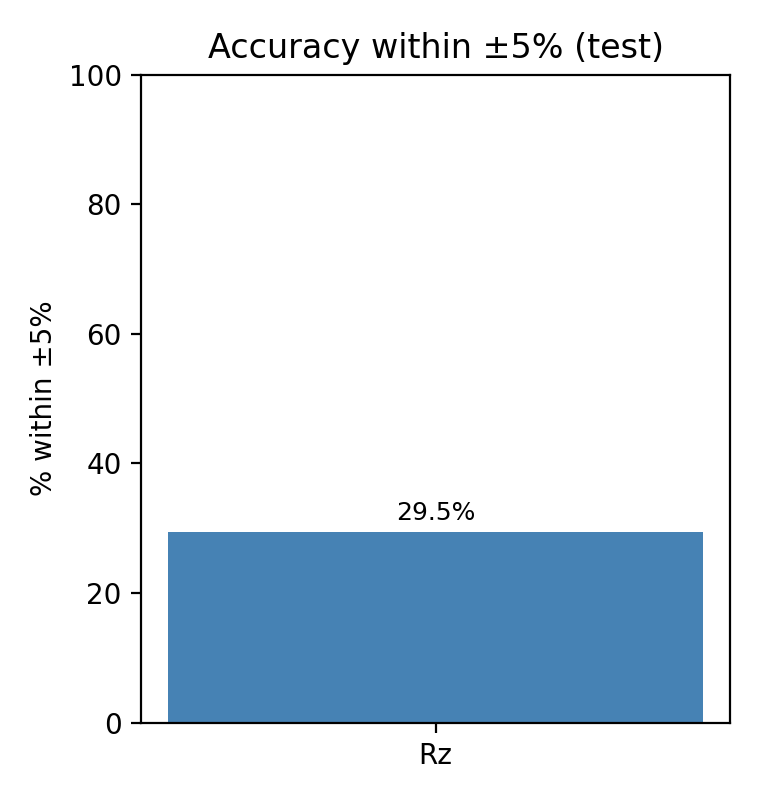}{Figure S269: accuracy within tol 5percent (\label{fig:supp-269})}\hfill
\suppimage{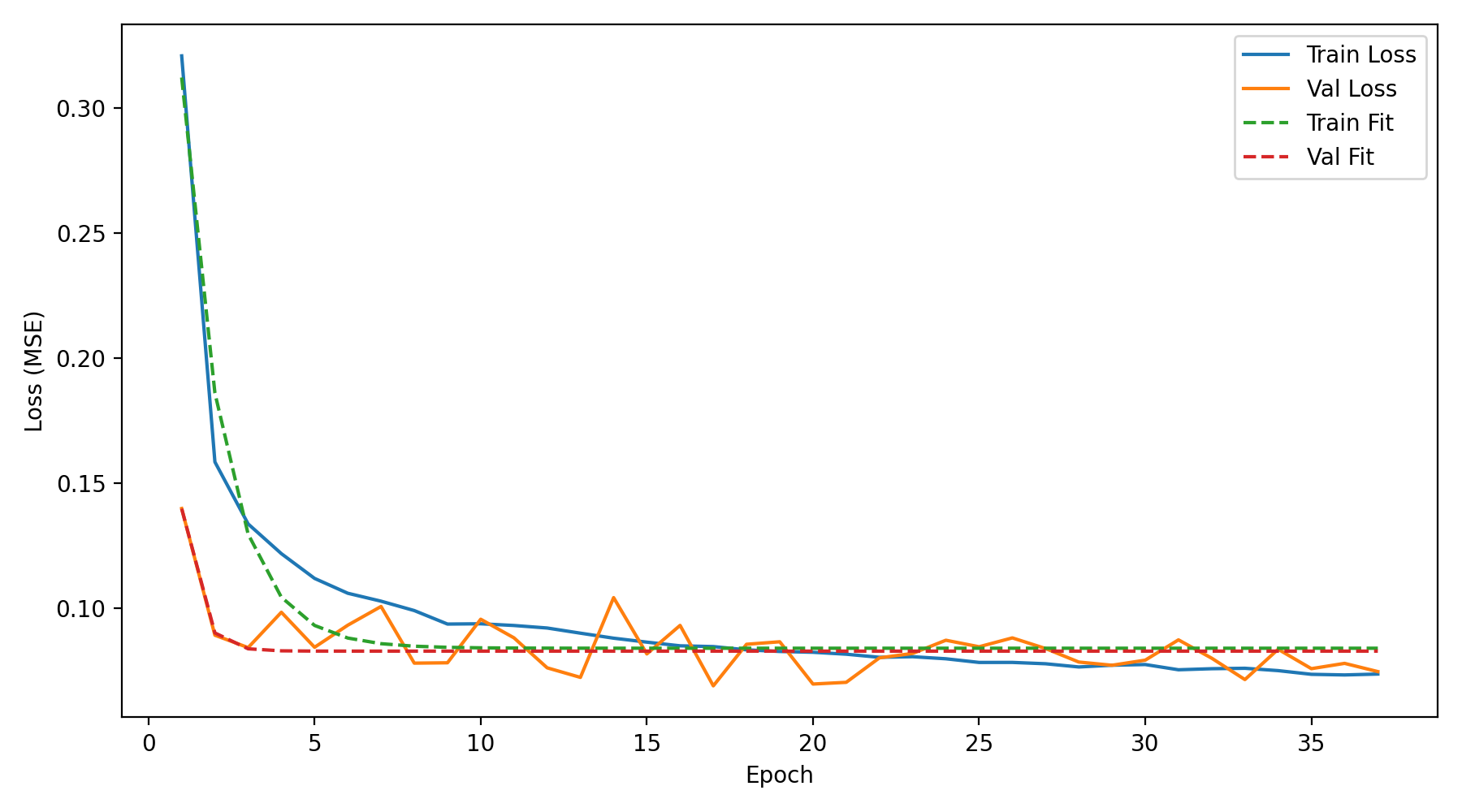}{Figure S270: loss curves (\label{fig:supp-270})}\
\suppimage{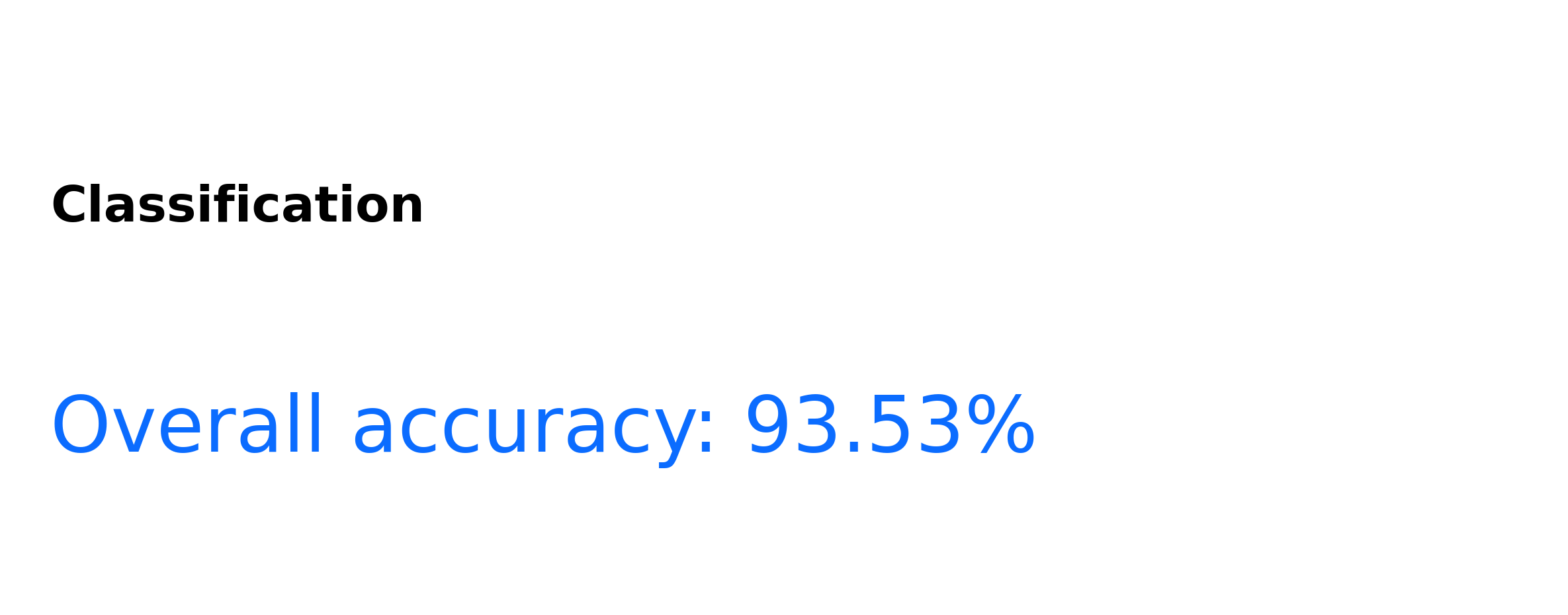}{Figure S271: headline classification (\label{fig:supp-271})}\hfill
\suppimage{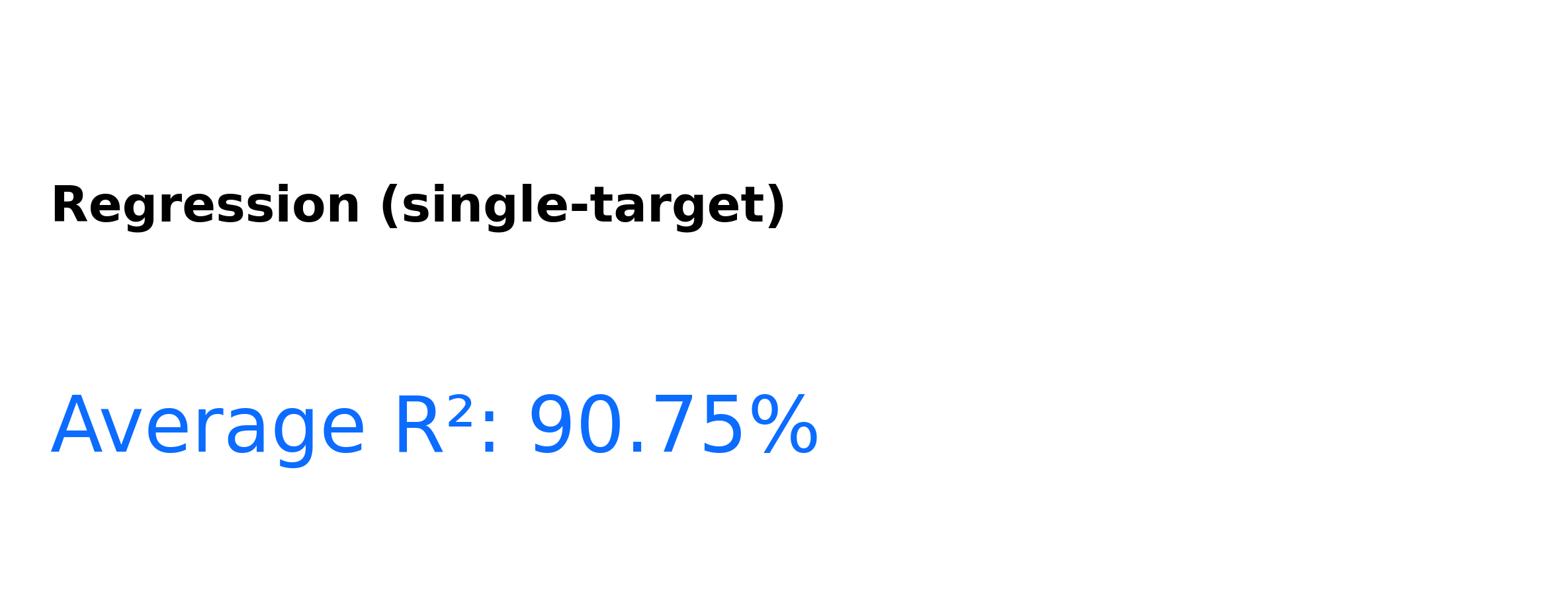}{Figure S272: headline regression (\label{fig:supp-272})}\\
\par
\endgroup

\section*{Supplementary Uncertainty Analysis}
This section expands the manuscript's uncertainty discussion with full numeric artefacts derived from the tabular outputs.

\subsection*{A. Interval Coverage and Widths}
Quantile empirical coverages and mean widths for central intervals (0.50, 0.80, 0.90 nominal) are summarised in Table~\ref{tab:supp_quantile_widths}. Conformal recalibration (Table~\ref{tab:supp_conformal_widths}) adjusts widths while restoring nominal coverage.

\begin{sidewaystable}[p]
  \centering
  \caption{Quantile-derived empirical coverage (EC) and mean width (MW) by target (pre-calibration).}
  \label{tab:supp_quantile_widths}
  \scriptsize
  \setlength{\tabcolsep}{5pt}
  \begin{tabular}{lccccc}
    	oprule
    Target & Interval & Nominal (NC) & Empirical (EC) & Mean Width [$\mu$m] & Notes \\
    \midrule
    Ra & 0.05--0.95 & 0.90 & 0.9832 & 1.2120 & Over-coverage \\
    Ra & 0.10--0.90 & 0.80 & 0.8849 & 0.7700 & Mild over-coverage \\
    Ra & 0.25--0.75 & 0.50 & 0.5832 & 0.4018 & Over-coverage \\
    Rz & 0.05--0.95 & 0.90 & 0.3022 & 3.6751 & Severe under-coverage \\
    Rz & 0.10--0.90 & 0.80 & 0.1890 & 2.3628 & Under-coverage \\
    Rz & 0.25--0.75 & 0.50 & 0.1003 & 1.2728 & Under-coverage \\
    RONt & 0.05--0.95 & 0.90 & 0.1507 & 0.04713 & Severe under-coverage \\
    RONt & 0.10--0.90 & 0.80 & 0.1034 & 0.03048 & Under-coverage \\
    RONt & 0.25--0.75 & 0.50 & 0.0602 & 0.01643 & Under-coverage \\
    Ra\_uncert & 0.05--0.95 & 0.90 & 0.3092 & 0.3535 & Under-coverage \\
    Ra\_uncert & 0.10--0.90 & 0.80 & 0.1825 & 0.2237 & Under-coverage \\
    Ra\_uncert & 0.25--0.75 & 0.50 & 0.0993 & 0.1166 & Under-coverage \\
    Rz\_uncert & 0.05--0.95 & 0.90 & 0.6554 & 1.7578 & Under-coverage \\
    Rz\_uncert & 0.10--0.90 & 0.80 & 0.5362 & 1.1324 & Under-coverage \\
    Rz\_uncert & 0.25--0.75 & 0.50 & 0.3725 & 0.6087 & Under-coverage \\
    RONt\_uncert & 0.05--0.95 & 0.90 & 0.1496 & 0.00473 & Severe under-coverage \\
    RONt\_uncert & 0.10--0.90 & 0.80 & 0.1029 & 0.00306 & Under-coverage \\
    RONt\_uncert & 0.25--0.75 & 0.50 & 0.0602 & 0.00165 & Under-coverage \\
    \bottomrule
  \end{tabular}
  \vspace{1mm}
  \begin{minipage}{0.9\linewidth}
    \footnotesize EC and NC are shown as fractions (0--1). Widths are reported in [$\mu$m]. Empirical coverage exceeding nominal indicates over-coverage; below nominal indicates under-coverage.
  \end{minipage}
\end{sidewaystable}

\begin{table}[H]
  \centering
  \caption{Conformal 90\% central interval coverage (COV) and mean width (MW) vs baseline 90\% width; width change $\Delta$ expressed as percent.}
  \label{tab:supp_conformal_widths}
  \begin{tabular}{lcccc}
    \hline
    Target & Quantile Width [$\mu$m] & Conformal Width [$\mu$m] & Width $\Delta$ [\%] & Conformal Coverage \\
    \hline
    Ra & 1.2120 & 0.6701 & -44.7 & 0.9055 \\
    Rz & 3.6751 & 3.0517 & -17.0 & 0.9010 \\
    RONt & 0.04713 & 2.7e-06 & -99.99 & 0.8987 \\
    \hline
  \end{tabular}
  \vspace{1mm}
  \begin{minipage}{0.9\linewidth}
    \footnotesize Conformal coverage is shown as a fraction (0--1). Negative width $\Delta$ denotes interval narrowing post conformal recalibration while maintaining nominal coverage.
  \end{minipage}
\end{table}

\subsection*{B. Interval Scoring Metrics}
Pinball, CRPS approximation and Winkler scores for the quantile model are detailed in Table~\ref{tab:supp_interval_scores}. Lower is better across metrics.

\begin{sidewaystable}[p]
  \centering
  \caption{Expanded interval scoring metrics (quantile model).}
  \label{tab:supp_interval_scores}
  \begin{tabular}{lrrrrr}
    \hline
    Target & Mean Pinball [$\mu$m] & CRPS (approx) [$\mu$m] & Winkler 80\% [$\mu$m] & Winkler 90\% [$\mu$m] & Note \\
    \hline
    Ra & 0.0561 & 0.1273 & 0.8701 & 1.2404 & High coverage 90\% \\
    Rz & 2.4961 & 4.7712 & 48.7100 & 87.2951 & Large scale \\
    RONt & 0.0263 & 0.0514 & 0.5111 & 0.8584 & Narrow scale \\
    Ra\_uncert & 0.2361 & 0.4536 & 4.6224 & 8.1124 & Learned uncertainty \\
    Rz\_uncert & 0.3978 & 0.8188 & 7.4755 & 11.2897 & Wider dispersion \\
    RONt\_uncert & 0.00262 & 0.00513 & 0.05094 & 0.08543 & Very small variance \\
    \hline
  \end{tabular}
  \vspace{1mm}
  \begin{minipage}{0.9\linewidth}
    \footnotesize All scores are in units of the response variable [$\mu$m]. Lower values indicate better probabilistic calibration and sharpness; Winkler penalises mis-coverage and width jointly.
  \end{minipage}
\end{sidewaystable}

\subsection*{C. Residual Tail Heaviness}
Excess kurtosis (Table~\ref{tab:supp_kurtosis}) highlights heavy-tailed error structure, especially for $RONt$ and $RONt\_uncert$. These motivate future adoption of robust likelihoods or quantile-local conformal adjustments.

\begin{table}[H]
  \centering
  \caption{Excess kurtosis of residuals (test set) for primary targets and direct uncertainty targets.}
  \label{tab:supp_kurtosis}
  \begin{tabular}{lrr}
    \hline
    Target & Excess Kurtosis & Comment \\
    \hline
    Ra & 33.89 & Heavy tail vs Gaussian (0) \\
    Rz & 34.98 & Heavy tail \\
    RONt & 176.14 & Extreme tail weight \\
    Ra\_uncert & 2.58 & Mild tail elevation \\
    Rz\_uncert & 39.83 & Heavy tail \\
    RONt\_uncert & 274.71 & Extreme tail / degeneracy \\
    \hline
  \end{tabular}
  \vspace{1mm}
  \begin{minipage}{0.9\linewidth}
    \footnotesize Gaussian reference excess kurtosis is 0; large positive values indicate heavy-tailed error distributions.
  \end{minipage}
\end{table}

\subsection*{D. Residual–Uncertainty Correlations}
Correlation between absolute residuals and predicted uncertainty targets ($|e|$ vs corresponding predicted uncertainty variable, e.g. \texttt{Ra\_uncert}) for aligned target pairs is shown in Table~\ref{tab:supp_correlations}. A stronger positive value indicates better heteroscedastic signal capture.

\begin{table}[H]
  \centering
  \caption{Absolute residual vs predicted uncertainty correlation coefficients.}
  \label{tab:supp_correlations}
  \begin{tabular}{lr}
    \hline
    Pair & r \\
    \hline
    |e(Ra)| vs Ra\_uncert & -0.054 \\
    |e(Rz)| vs Rz\_uncert & 0.031 \\
    |e(RONt)| vs RONt\_uncert & 0.789 \\
    \hline
  \end{tabular}
  \vspace{1mm}
  \begin{minipage}{0.9\linewidth}
    \footnotesize Positive correlation suggests predicted uncertainty scales with realised absolute errors (heteroscedastic signal capture).
  \end{minipage}
\end{table}


\section*{Acknowledgements}
A grant supported this work: project entitled: "Application of artificial intelligence in surface irregularities measurements", financed by the Ministry of Education and Science of the programme: Polish Metrology II PM-II/SP/0104/2024/02 of 01.02.2024\\
Projekt pt. „Zastosowanie sztucznej inteligencji w pomiarach nierówności powierzchni” finansowany przez Ministerstwo Nauki i~Szkolnictwa Wyższego w ramach programu Polska Metrologia 2 Nr PM-II/SP/0104/2024/02 z dnia 01.02.2024.

\vspace{6mm}
\begin{center}
  \IfFileExists{logo2.pdf}{\includegraphics[width=5cm]{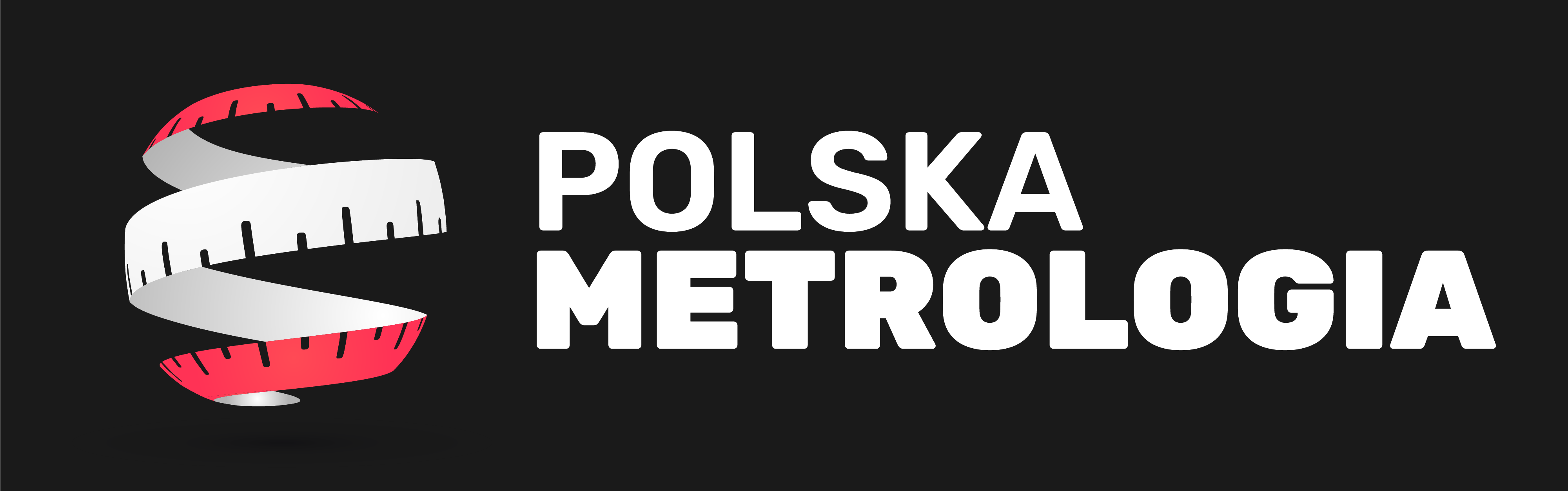}}{\fbox{\parbox{5cm}{\centering logo2.pdf missing}}}
\end{center}

\end{document}